\documentclass[twocolumn,nofootinbib,aps,amsmath,amssymb,10pt,floatfix,superscriptaddress,letterpaper,fleqn]{revtex4-1}

\usepackage{amssymb}
\usepackage{amsfonts}
\usepackage{amsmath}
\usepackage{makeidx}
\usepackage[bookmarksnumbered,colorlinks,bookmarks,breaklinks=true,linkbordercolor={1 1 1},linktocpage,citecolor=blue]{hyperref}
\usepackage{graphicx}
\usepackage{subfigure}
\usepackage{dcolumn}
\usepackage{bm}
\usepackage{fancyhdr}
\usepackage{floatflt}
\usepackage{color}
\usepackage[activate={true,nocompatibility},final,tracking=true,kerning=true,spacing=true,factor=1100,stretch=10,shrink=10]{microtype}

\usepackage{tabularx}
\usepackage[table]{xcolor}
\usepackage{longtable}
\usepackage{multirow}
\usepackage{afterpage}
\usepackage{booktabs}

\newcommand\T{\rule{0pt}{2.6ex}}       % Top strut
\newcommand\B{\rule[-1.2ex]{0pt}{0pt}} % Bottom strut
\def\etal{\textit{et al.~}}
\def\etals{\textit{et al.}}
\def\ql{``}
\def\qr{''\hspace{0.5mm}}
\def\qrs{''}

\def\a{\color{blue}}
\def\r{\color{red}}
\def\b{\color{black}}
\def\TH{$^{232}$Th}
\def\UT{$^{235}$U}

\def\CF{$^{252}$Cf}

\definecolor{lred}{RGB}{255,0,0}

\newcommand{\eg}{\textit{e.g.}}
\newcommand{\ie}{\textit{i.e.}}

\DeclareGraphicsExtensions{.pdf,.png,.jpg}

\begin{document}
\newcolumntype{L}[1]{>{\raggedright\arraybackslash}p{#1}}
\newcolumntype{C}[1]{>{\centering\arraybackslash}p{#1}}
\newcolumntype{R}[1]{>{\raggedleft\arraybackslash}p{#1}}

\title{IRDFF-II: A New Neutron Metrology Library}

\author{A.~Trkov}
\affiliation{NAPC--Nuclear Data Section, International Atomic Energy Agency, Vienna, Austria}

\author{P.J.~Griffin}
\affiliation{Sandia National Laboratories (SNL), Radiation Effects Science Center, Albuquerque, NM, USA}

\author{S.P.~Simakov}
\affiliation{Karlsruhe Institute for Technology (KIT), Eggenstein-Leopoldshafen, Germany}

\author{L.R.~Greenwood}
\affiliation{Pacific Northwest National Laboratory (PNNL), Richland, WA, USA}

\author{K.I.~Zolotarev}
\affiliation{Institute of Physics and Power Engineering (IPPE), Obninsk, Russia}

\author{R.~Capote}\email[Corresponding author: ]{r.capotenoy@iaea.org}
\affiliation{NAPC--Nuclear Data Section, International Atomic Energy Agency, Vienna, Austria}

\author{D.L.~Aldama}
\affiliation{Agencia de Energ\'ia Nuclear y Tecnolog\'ias de Avanzada, La Habana, Cuba}

\author{V.~Chechev\dag\thanks{deceased}}
\affiliation{V.G.~Khlopin Radium Institute, Saint Petersburg, Russia}

\author{C.~Destouches}
\affiliation{DER/CEA -- Centre d'$\acute{e}$tudes nucl$\acute{e}$aires de Cadarache, St. Paul Lez Durance, France}

\author{A.C.~Kahler}
\affiliation{Kahler Nuclear Data Services LLC, Jupiter, FL, USA}

\author{C.~Konno}
\affiliation{Japan Atomic Energy Agency, Tokai-mura, Naka-gun, Ibaraki-ken, Japan}

\author{M.~Ko\v{s}t\'al}
\affiliation{Research Centre \v{R}e\v{z} Ltd, Husinec-\v{R}e\v{z}, Czech Republic}

\author{M.~Majerle}
\affiliation{Nuclear Physics Institute, Czech Academy of Sciences, \v{R}e\v{z}, Czech Republic}

\author{E.~Malambu}
%\affiliation{SCK$\bullet$CEN, Boeretang 200, 2400 Mol, Belgium}
\affiliation{SCK$\bullet$CEN, Mol, Belgium}

\author{M.~Ohta}
\affiliation{National Institute for Quantum Radiological Science and Technology, Rokkasho-mura, Kamikita-gun, Aomori-ken, Japan}

\author{V.G.~Pronyaev}
\affiliation{Contractor, NAPC--Nuclear Data Section, International Atomic Energy Agency, Vienna, Austria}

\author{V.~Radulovi\'{c}}
\affiliation{Jo\v{z}ef Stefan Institute, Ljubljana, Slovenia}

\author{S.~Sato}
\affiliation{National Institute for Quantum Radiological Science and Technology, Rokkasho-mura, Kamikita-gun, Aomori-ken, Japan}

\author{M.~Schulc}
\affiliation{Research Centre \v{R}e\v{z} Ltd, Husinec-\v{R}e\v{z}, Czech Republic}

\author{E. \v{S}ime\v{c}kov\'{a}}
\affiliation{Nuclear Physics Institute, Academy of Sciences of the Czech Republic, \v{R}e\v{z}, Czech Republic}

\author{I.~Vavtar}
\affiliation{Jo\v{z}ef Stefan Institute, Ljubljana, Slovenia}

\author{J.~Wagemans}
\affiliation{SCK$\bullet$CEN, Mol, Belgium}

\author{M.~White}
\affiliation{Los Alamos National Laboratory (LANL), Los Alamos, NM, USA}

\author{H.~Yashima}
\affiliation{Institute for Integrated Radiation and Nuclear Science, Kyoto University, Kyoto, Japan}

\received{30 August 2019}
%\date{\today }

\begin{abstract}
\vspace*{2mm}
High quality nuclear data is the most fundamental underpinning for all neutron metrology applications. This paper describes the release of version II of the International Reactor Dosimetry and Fusion File (\mbox{IRDFF-II}) that contains a consistent set of nuclear data for fission and fusion neutron metrology applications up to 60~MeV neutron energy. The library is intended to support: a) applications in research reactors; b) safety and regulatory applications in the nuclear power generation in commercial fission reactors; and c) material damage studies in support of the research and development of advanced fusion concepts. The paper describes the contents of the library, documents the thorough verification process used in its preparation, and provides an extensive set of validation data gathered from a wide range of neutron benchmark fields.
The new \mbox{IRDFF-II} library includes 119 metrology reactions, four cover material reactions to support self-shielding corrections, five metrology metrics used by the dosimetry community, and cumulative fission products yields for seven fission products in three different neutron energy regions. In support of characterizing the measurement of the residual nuclei from the dosimetry reactions and the fission product decay modes, the present document lists the recommended decay data, particle emission energies and probabilities for 68 activation products.
It also includes neutron spectral characterization data for 29 neutron benchmark fields for the validation of the library contents. Additional six reference fields were assessed (four from plutonium critical assemblies, two measured fields for thermal-neutron induced fission on $^{233}$U and $^{239}$Pu targets) but not used for validation due to systematic discrepancies in C/E reaction rate values or lack of reaction-rate experimental data. Another ten analytical functions are included that can be useful for calculating average cross sections, average energy, thermal spectrum average cross sections and resonance integrals.
The \mbox{IRDFF-II} library and comprehensive documentation is available online at \href{https://www-nds.iaea.org/IRDFF/}{www-nds.iaea.org/IRDFF/}. Evaluated cross sections can be compared with experimental data and other evaluations at \href{https://www-nds.iaea.org/exfor/endf.htm}{www-nds.iaea.org/exfor/endf.htm}. The new library is expected to become the international reference in neutron metrology for multiple applications.
\end{abstract}

\vspace*{43mm}
\maketitle

%\addtocontents{toc}{\protect\clearpage}
\tableofcontents

\vspace*{12mm} \lhead{IRDFF-II: A New Neutron Dosimetry\ldots}
\chead{NUCLEAR DATA SHEETS} \rhead{A.~Trkov \etal} \lfoot{} \rfoot{} %

\vspace*{4mm}

\section{INTRODUCTION} \label{Sec_I_A}
High quality nuclear data are the most fundamental quantities underpinning all neutron metrology applications. The International Atomic Energy Agency (IAEA) Nuclear Data Section (NDS), through several research projects, sponsored an international collaboration to provide a consensus for the recommended evaluated nuclear data for such purposes. This new library, called version~II of the International Reactor Dosimetry and Fusion File (\mbox{IRDFF-II}), has undergone an exhaustive verification and validation process and is now being released to the metrology community. It supersedes the previous versions \mbox{IRDF-2002} \cite{IRDF2002} and \mbox{IRDFF-v1.05} \cite{IRDFF105}. This paper describes the content of the library and documents the thorough verification process used in its preparation. It also provides an extensive set of validation data for the library using data gathered from a wide range of neutron benchmark fields.

Well-validated nuclear data evaluation techniques are needed to produce a modern nuclear data evaluated library like IRDFF-II. An overview of nuclear data evaluation methodology can be found in Ref.~\cite{Smith:1991}; recent progress has been reviewed in Refs.~\cite{Capote:2010,Smith-Otuka:2012,WPEC-SG24}. Nowadays the generalized least-square method is the most common evaluation method, \eg, see~\cite{GMA,Smith:1993,GLUCS,KALMAN,GANDR}. A similar least-squares approach~\cite{Vinogradov:1987,Badikov:1992,Hermanne:2018} that is based on Pad\'e approximants~\cite{Pade:1892,Graves-Morris:1973,Baker:1975} has also been extensively employed by K.~Zolotarev to produce many of the new dosimetry evaluations for the \mbox{IRDFF-II} library.

Stochastic (Monte Carlo) evaluation methods have been also investigated by several researchers in the nuclear data field, \eg, see BFMC~\cite{BFMC}, UMC-B~\cite{UMCB}, UMC-G~\cite{UMCG,UMCG1}, and BMC~\cite{BMC} methods. These Bayesian approaches enable simultaneous consideration of both experimental data and model-calculated values as well as corresponding experimental and modeling correlations, and have been applied to large amounts of data (\eg, see Refs.~\cite{HFB-mass,TENDL-2017}).

The development of the \mbox{IRDFF-II} dosimetry library has greatly benefited from the availability of the comprehensive TENDL libraries~\cite{TENDL-2015,TENDL-2017}\footnote{The TENDL-2015.s60 file of the TENDL 2015 library contains exclusive representations of reactions and associated covariances up to 60~MeV. This file was extensively used for the extension of \mbox{IRDFF-II} evaluations up to 60 MeV.}, especially to extend the energy range of the existing evaluations for neutron-induced reactions on medium- and heavy-mass targets.

A general purpose neutron evaluated nuclear data file (ENDF) uses the evaluation techniques briefly discussed above to combine model calculations with experimental data to address the assembly of a complete and consistent set of neutron-induced reactions for a given target isotope. In this process an evaluator may need to make some compromises in the recommended nuclear data in order to satisfy consistency constraints on the nuclear data between different reaction channels, \eg, the total cross section must be exactly equal to the sum of the various reaction channel cross sections, while matching as much as possible the available experimental data.

In contrast to a general nuclear data evaluation, an entry in a metrology library is reaction-specific. The evaluation can start from a nuclear model calculation, but relies more heavily on experimental data to obtain recommended energy-dependent cross sections, including full covariance information for that particular reaction channel so that an accurate assessment of the uncertainty in reaction rates can be made for a given neutron spectrum field. Whereas information on the consistency of the evaluated data with integral measurements gathered in various reference neutron fields is not directly taken into account in the actual nuclear data evaluation process, it does influence the decisions to declare a reaction as \ql recommended\qr for use in metrology applications.

This paper presents a consistent set of nuclear data intended for neutron metrology that support: a) applications in research reactors; b) safety and regulatory applications in the nuclear power generation in commercial fission reactors; and c) material damage studies in research and development of advanced fusion concepts.

\section{OVERVIEW OF THE \mbox{IRDFF-II} LIBRARY} \label{Sec_II}
The new \mbox{IRDFF-II} library addresses incident neutron energies from 0 to 60~MeV. The library entries, enumerated in Table~\ref{Table_Long_I}, include 119 metrology reactions, four cover cross sections used to support shielding corrections, five metrology metrics used by the dosimetry community, seven cumulative fission products yields and corresponding decay data. In support of characterizing the measurement of the residual nuclei from the dosimetry reactions and the fission product decay modes, the library includes decay data, gamma and beta particle emission energies and probabilities for 68 residual nuclei. The library also includes neutron spectral characterization data for 45 neutron benchmark fields that were used to support the validation of the library contents. The following Secs. discuss these entries of the dosimetry library in more detail.

\subsection{Neutron Metrology Reactions}
There are several requirements for a given reaction to be accepted for use in metrology applications and included in the \mbox{IRDFF-II} library. These requirements include:
\begin{enumerate}
\item The existence of a high-fidelity nuclear data evaluation for the reaction channel that is documented and stored in the library;
\item A residual particle from the reaction that is measurable with a technique in use at multiple laboratories, \eg, gamma activation analysis, beta spectroscopy, mass spectrometry;
\item The existence of uncertainty data for the cross section that must come directly from the cross section evaluation process, captured in the form of a covariance matrix,
and stored in the library;
\item A reaction that is of interest to the commercial or research reactor community or to the fusion community (including metrology at high energy accelerators used to support fusion material damage studies, \eg, \href{https://www.ifmif.org/}{IFMIF}~\cite{IFMIF});
\item Evidence that the nuclear data evaluation took into consideration the available body of experimental data.
\end{enumerate}

Table~\ref{Table_Long_I} summarizes the 119 metrology reactions that meet these requirements and are included in the \mbox{IRDFF-II} library. The list includes all of the reactions that were present in the earlier \mbox{IRDF-2002}~\cite{IRDF2002} and IRDFF~v1.05~\cite{IRDFF105} libraries. The differences between library versions are indicated: column~8 indicates whether \mbox{IRDFF-v1.05} was updated (new) or not (old) compared to \mbox{IRDF-2002}. Likewise, column~9 indicates whether \mbox{IRDFF-II} was updated compared to \mbox{IRDFF-v1.05}. As seen from the table, several of the dosimetry reactions in the \mbox{IRDFF-II} were updated and many reactions were added, such as $^{27}$Al(n,2n)$^{26g}$Al, %$^{27}$Al(n,2n)$^{26m}$Al,
$^{28}$Si(n,p)$^{28}$Al, $^{29}$Si(n,X)$^{28}$Al, $^{113}$In(n,$\gamma$)$^{114m}$In, $^{238}$U(n,2n)$^{237}$U, and $^{209}$Bi(n,xn)$^{(210-x)}$Bi. The extensions were introduced to support specialized dosimetry applications.

High resolution atomic mass spectroscopy is often used to measure alpha production in materials such as $^{10}$B and $^{6}$Li. Measured alpha particles can originate from reactions other than the (n,$\alpha$) channel, hence the updated \mbox{IRDFF-II} library also includes reactions for the total alpha production cross section, \ie, $^{10}$B(n,X)$^{4}$He and $^{6}$Li(n,X)$^{4}$He reactions. Since boron and lithium are often used with a variety of enrichments in dosimetry applications, nuclear data for the $^{11}$B(n,X)$^{4}$He and $^{7}$Li(n,X)$^{4}$He reactions, as well as the composite elemental-based $^{\mbox{nat}}$B(n,X)$^{4}$He and $^{\mbox{nat}}$Li(n,X)$^{4}$He reactions, are included in the library. The availability of the separate $^{11}$B and $^{7}$Li alpha-production cross sections enables users to apply the \mbox{IRDFF-II} cross sections to variable enrichment dosimeters that might be employed at various metrology laboratories.

Previous input from users~\cite{CRP17} of the IRDFF-1.05 library~\cite{IRDFF105}, which was extended to address the energy range up to 60~MeV, indicated that natural elemental dosimeters, rather than isotopic ones, were usually used in dosimetry applications. Since, at the higher incident neutron energies on natural elemental dosimeters, several contributing reaction channels can produce the same residual nucleus that is used as a dosimetry monitor, the \mbox{IRDFF-II} library has also been extended to explicitly address the composite reactions that produce commonly used residual nuclei.

\LTcapwidth=\textwidth
\newpage
\begin{longtable*}{r | l | l | c | c c c |c c}
\caption{IRDFF-II nuclear data contents and evaluation sources. (r) denotes renormalization. Major sources of data are K.I. Zolotarev evaluations and
EAF-2010~\cite{EAF2010}, JENDL-4~\cite{JENDL4}, TENDL~\cite{TENDL-2015,TENDL-2017}, ENDF/B-VIII.0~\cite{ENDF8}, and JEFF-3.3~\cite{JEFF33} libraries.}
\label{Table_Long_I} \\
\hline \hline
\T  &              &             &        &                &                &          & \multicolumn{2}{c}{Consistency} \\
    &              &             &        &                &                &          &  v2002       & v-1.05           \\
 ID & Reaction     &  Label      & MAT    &  Energy        &  Evaluation    & Eval.    &  to          & to               \\
 No.&              &             &        &  Interval      &   source       & Date     &  v-1.05$^1$  & IRDFF-II$^2$     \\ \hline
\endfirsthead
\caption{(continued). IRDFF-II nuclear data contents and evaluation sources. (r) denotes renormalization.} \\
\hline \hline
\T  &              &             &        &                &                &          & \multicolumn{2}{c}{Consistency} \\
    &              &             &        &                &                &          &  v2002       & v-1.05           \\
 ID & Reaction     &  Label      & MAT    &  Energy        &  Evaluation    & Eval.    &  to          & to               \\
 No.&              &             &        &  Interval      &   source       & Date     &  v-1.05$^1$  & IRDFF-II$^2$     \\ \hline
\endhead
%Whatever you put before this command and below endfirsthead will be displayed at the top of the table in every page except the first one.
\hline
\addlinespace[1mm]
\multicolumn{9}{l}{~\protect{$^{1}$}~Column (v2002 to v-1.05) marks any change in IRDFF-v1.05 compared to IRDF-2002 as a reference up to 20~MeV.} \\
\multicolumn{9}{l}{~\protect{$^{2}$}~Column (v-1.05 to IRDFF-II) marks any change in IRDFF-II compared to IRDFF-v1.05 as a reference.} \\
\multicolumn{9}{l}{~Consistency legend: new = new evaluation; old = unchanged; -- = not present in reference.}\\
\hline \hline
\endfoot
\T
   1 & $^{\mbox{nat}}$Li(n,X)$^{3}$H                          & LiH3       &   300  &~0--60 MeV      &  IRDFF-II Li6H3,Li7H3                               & 2019      & --        & --        \\[1.25mm]

   2 & $^{\mbox{nat}}$Li(n,X)$^{4}$He                         & LiHe4      &   300  &~0--60 MeV      &  IRDFF-II Li6He4,Li7He4                             & 2019      & --        & --        \\[1.25mm]

   3 & $^{6}$Li(n,t)$^{4}$He                                  & Li6t       &   325  &~0--1~MeV       &  IAEA STD 2017~\cite{Car18}                         & 2018      & new       & new       \\
     &                                                        &            &   325  &~1--2.75~MeV    &  ENDF/B-VIII.0                                      & 2018      & new       & new       \\
     &                                                        &            &   325  &~2.75--60~MeV   &  EAF-2010(r)                                        & 2010      & new       & new       \\[1.25mm]

   4 & $^{6}$Li(n,X)$^{3}$H                                   & Li6H3      &   325  &~0--60~MeV      &  IRDFF-II Li6t                                      & 2019      & --        & --        \\[1.25mm]

   5 & $^{6}$Li(n,X)$^{4}$He                                  & Li6He4     &   325  &~0--20~MeV      &  MT24--ENDF/B-VIII.0                                & 2018      & --        & --        \\
     &                                                        &            &   325  &~20--60~MeV     &  EAF-2010(r) (n,2n$\alpha$)                         & 2010      & --        & --        \\
\hline
\end{longtable*}

These composite reactions include:
$^{\mbox{nat}}$Mg(n,X)$^{27}$Na, $^{\mbox{nat}}$Si(n,X)$^{28}$Al, $^{\mbox{nat}}$S(n,X)$^{32}$P,   $^{\mbox{nat}}$Ti(n,X)$^{45}$Ti, $^{\mbox{nat}}$Ti(n,X)$^{46}$Sc, $^{\mbox{nat}}$Ti(n,X)$^{47}$Sc, $^{\mbox{nat}}$Ti(n,X)$^{48}$Sc, $^{\mbox{nat}}$Cr(n,X)$^{51}$Cr, $^{\mbox{nat}}$Fe(n,X)$^{53}$Fe, $^{\mbox{nat}}$Fe(n,X)$^{51}$Cr, $^{\mbox{nat}}$Fe(n,X)$^{54}$Mn, $^{\mbox{nat}}$Fe(n,X)$^{56}$Mn, $^{\mbox{nat}}$Ni(n,X)$^{57}$Ni, $^{\mbox{nat}}$Ni(n,X)$^{58}$Co, $^{\mbox{nat}}$Ni(n,X)$^{60}$Co, $^{\mbox{nat}}$Cu(n,X)$^{60}$Co, $^{\mbox{nat}}$Cu(n,X)$^{62}$Cu, $^{\mbox{nat}}$Cu(n,X)$^{64}$Cu, $^{\mbox{nat}}$Zn(n,X)$^{64}$Cu, $^{\mbox{nat}}$Zn(n,X)$^{67}$Cu, $^{\mbox{nat}}$Zr(n,X)$^{89}$Zr, $^{\mbox{nat}}$Mo(n,X)$^{92m}$Nb, and $^{\mbox{nat}}$In(n,X)$^{114m}$In.

In the new \mbox{IRDFF-II} library the energy range was extended to consistently address incident neutron energies up to 60~MeV, so most of the library cross sections are a composite of separate existing nuclear data evaluations. Columns~5, 6, and~7 of Table~\ref{Table_Long_I} provide information on the source of the recommended nuclear data evaluations in the applicable incident neutron energy ranges.

%Nuclear data evaluations are provided in the \mbox{ENDF-6} format~\cite{ENDF-6}. In this format, documentation of the techniques used in a nuclear data evaluation and references are provided in File 1 (MF=1 in \mbox{ENDF-6} terminology), including all eigenvalues of the existing covaraince matrices.
Since clear documentation of the consistency of nuclear data used in an evaluation is important in the use of these data in metrology applications, Column~6 of Table~\ref{Table_Long_I} points to supplemental documentation for many of the entries in the \mbox{IRDFF-II} library.

\subsection{Reactions on Cover Materials}

Activation dosimeters can be used to characterize the neutron spectra through spectrum unfolding~\cite{McE67} or the least-squares-based spectrum adjustment approaches~\cite{Sta85}. Frequently they are shielded with cover materials to change the effective threshold energy and to move a dosimeter’s sensitivity from the thermal into the epithermal neutron energy region. Some commonly used approaches to model the effective cover material flux attenuation is to apply exponential attenuation, based on the energy-dependence of the total cross section of the cover material and the thickness of cover used~\cite{McE67}. Other approaches~\cite{Gri90} to the cover corrections use a more rigorous treatment based on radiation transport models of the neutron scattering within the cover. In order to support this variety of approaches to the treatment of the cover material, the \mbox{IRDFF-II} library provides the total, elastic and absorption cross sections for commonly used cover materials. Total, elastic, capture (and fission) cross sections are also provided for all monitor materials that exhibit resonance structure so that self-shielding corrections can be made, if needed. Table~\ref{Table_Long_I} lists the four cover materials that are addressed within the library.

\subsection{Dosimetry Metrics}

Various radiation dosimetry metrics are employed in the interpretation of radiation test results that supports various application areas. A dosimetry metric is the result of folding a calculated dosimetry-related energy-dependent response function with the incident neutron energy-dependent fluence. These energy-dependent response functions are treated within the \mbox{IRDFF-II} library in a manner similar to reaction-dependent cross sections. The dosimetry metrics used by the radiation effects community have been demonstrated to provide a correlation between the calculated quantity and a specific damage/failure mode of interest to an application area, \ie, iron embrittlement, that can result in failure of a critical weld in a reactor pressure vessel or gain degradation in a silicon bipolar junction transistor (BJT).  Monitor dosimetry measurements used  during the radiation testing are frequently combined with knowledge of the neutron spectrum during the irradiation to directly report the radiation dosimetry metrics of interest in the application. Since the \mbox{IRDFF-II} library is intended to support dosimetry applications for the fission and fusion communities, we have elected to include, for reference purposes, the primary radiation dosimetry metrics of interest to this community -- but to restrict this inclusion to just those dosimetry metrics that have been endorsed by a national nuclear regulator and/or by an international standards organization. Table~\ref{Table_Long_I} lists the five dosimetry metrics, called Damage cross sections, that are included in the \mbox{IRDFF-II} library.

%A wide range of radiation dosimetry metrics are employed in the interpretation of radiation test results that support various application areas. Monitor dosimetry measurements %fielded with the radiation testing are frequently combined with knowledge of the neutron spectrum during the irradiation to directly report the radiation dosimetry metrics of %interest in the application. Since the \mbox{IRDFF-II} library is intended to support dosimetry applications for the fission and fusion communities, we have elected to include, %for reference purposes, the primary radiation metrics of interest to this community -– but to restrict this inclusion to just those dosimetry metrics that have been endorsed by %a national nuclear regulator and/or by an international standards organization. Table~\ref{Table_Long_I} lists the five dosimetry metrics that are included in the %\mbox{IRDFF-II} library.
The primary dosimetry metric of interest to the commercial light water reactor (LWR) community is the displacement-per-atom (dpa) in iron. This metric is used to assess the embrittlement of structural materials near the critical structural areas, such as the beltline weld for the pressure vessel wall, where any loss of mechanical strength could result in safety concerns. The neutron energy-dependence of the iron dpa metric is set by the appropriate regulatory body for the nuclear reactor, thus, there are different iron dpa dosimetry metrics used by the community. The \mbox{IRDFF-II} contains the iron dpa damage metric from ASTM~E693-17 Standard Practice for Characterizing Neutron Exposures in Iron and Low Alloy Steels in Terms of Displacement Per Atom (DPA)~\cite{A693}. This damage metric was calculated by looking at the primary Frenkel pair production from each of the four naturally occurring iron isotopes and applying the Norgett-Robinson-Torrens (NRT) damage partition function~\cite{Nor75}. The \mbox{IRDFF-II} library also contains the \mbox{JEFF-3.3} library~\cite{JEFF33} NRT-based iron damage metric and the legacy iron damage metric from the \mbox{IRDF-2002} library.

In addition to the iron dpa metric, the radiation damage to electronics community has silicon and GaAs displacement damage standards defined in ASTM E722-14 Standard Practice for Characterizing Neutron Fluence Spectra in Terms of an Equivalent Monoenergetic Neutron Fluence for Radiation-Hardness Testing of Electronics~\cite{A722}. The same E722 silicon response functions are also supported in ASTM E1855-15 Standard Test Method for use of 2N2222A Silicon Bipolar Transistors and Neutron Spectrum Sensors and Displacement Damage Monitors~\cite{A1855}.

\subsection{Nuclear Decay Data}

The recommended source of nuclear decay data for the dosimetry community is the series of monographs by the Bureau International des Poids et Mesures (BIPM). This eight volume BIPM-5 Table of Radionuclides represents an international collaboration that captures primary recommended nuclear data for most reactions of interest to the dosimetry community. Updates to the \mbox{BIPM-5} are coordinated though the Decay Data Evaluation Project (DDEP)~\cite{DDEP} working group led by the Laboratoire National de Metrologie et d'Essais – Laboratoire National Henri Becquerel (\mbox{LNE-LNHB}). The recommended nuclear data are periodically updated in separate volumes of the \mbox{BIPM-5} and the ongoing recommended values are kept up to date at the \href{http://www.nucleide.org/DDEP_WG/DDEPdata.htm}{DDEP website}. Not all nuclides are addressed within the DDEP library. For nuclides that are not present, the recommended values are taken from the ENSDF \ql ADOPTED\qr dataset~\cite{ENSDF}\footnote{ENSDF nuclear structure and decay data can be easily extracted, understood and studied in an attractive user-friendly manner by means of LiveChart of Nuclides \cite{Livechart} and NuDat \cite{NUDAT}.}. While the DDEP/BIPM-5 represents an international consensus on the current recommended decay data to be used for metrology applications, it is critical in dosimetry applications that all nuclear data used in a given application be consistent. This means that it is critical that the nuclear decay data be consistent with the associated cross section data.
%Rephrased paragraph
An evaluator of nuclear reaction cross section data would normally use values that correspond to the BIPM-5 data at the time of the evaluation – but these values may not necessarily correspond to the BIPM-5 recommended values seen by a reader at a later time when the cross sections are being used in a metrology application. There are also cases where the evaluator, after consulting with members of the DDEP community, elected to weigh underlying measurements for the DDEP evaluation in a  different way and ended up with slightly different decay data. Thus, in order to assist the users of the \mbox{IRDFF-II} cross sections, the corresponding nuclear decay data that are consistent with the recommended cross sections are also documented within the \mbox{IRDFF-II} library files. Details are given in Sec.~ \ref{Sec_IV_C}.
%For most nuclear data evaluations, the evaluator has used data that corresponds to the BIPM-5 data at the time of the primary nuclear data evaluation – but this value may not necessarily correspond to the BIPM-5 recommended values seen by a reader at a later time when the cross section is being used for a metrology application. There are also cases where the evaluator of a given nuclear data file has, after consulting with members of the DDEP community, elected to weight individual nuclear data measurements used in their specific cross section evaluation in a slightly different way than DDEP has done in their recommendations for general applications. Thus, in order to assist the users of the \mbox{IRDFF-II} cross sections, the corresponding nuclear decay data that should be used in conjunction with the recommended cross sections are also documented within the \mbox{IRDFF-II} library files. Details are given in Sec.~\ref{Sec_V_B}.

\subsection{Atomic Mass Data}

Metrology applications also require data on the elemental and isotopic mass used for the target materials. The most common metrological application for these data is the determination of the number of target atoms in an activation foil of a given mass.

\subsubsection{Elemental}

The recommended elemental atomic weights are taken from the International Union of Pure and Applied Chemistry (IUPAC) Commission on Isotopic Abundances and Atomic Weights (CIAAW)~\cite{CIAAW}. These data represents revisions made by the CIAAW in 2017 and published in the Pure and Applied Chemistry.

\subsubsection{Isotopic}

The recommended atomic weights for individual isotopes are taken from the recommendations of the Atomic Mass Data Center and are represented by the Atomic Mass evaluation 2016 (AME2016)~\cite{AME2016}. It is noted that geological materials are known for which the element has an isotopic composition that is outside the uncertainties cited here for normal materials. We should also note that many dosimetry applications may use enriched or depleted materials that are the result of isotopic separation techniques, in which case the user has to resort to the isotopic data that are provided.

\subsection{Natural Abundance Data}

The natural isotopic abundance data are also required in order to convert the densities of target materials into the number densities of the constituting isotopes. The recommended natural abundance data come from the CIAAW recommendations that refer to the 2016 IUPAC Technical Report~\cite{Abn16}. These values represent the best measurement of isotopic abundances from a single terrestrial source.

Some of the dosimetry reactions addressed in the \mbox{IRDFF-II} are for natural materials and include the production of a residual isotope from all contributing reactions. The abundance data used in the compilation of these multi-reaction combinations are given in Table~\ref{Table_Long_XV} of Sec.~\ref{Sec_IV_A}.
%details the isotopic abundance data used to describe the naturally occurring material.

\subsection{Fission Product Yields}

Since many important dosimetry metrics are for fission reactions, it is also important the fission yields used to support the dosimetry measurements be clearly documented. Thus, the recommended nuclear data contained in the \mbox{IRDFF-II} library include the cumulative fission yields for all the relevant fission products used for metrology applications which are listed in  Tables~\ref{Table_Long_XVIII}, \ref{Table_Long_XIX}, and \ref{Table_Long_XX} in the three generic incident energy regions.
\begin{table}[t]
\vspace{-4mm}
\caption{IRDFF-II thermal-neutron fission yields ($E=0.0253$~eV).}
\label{Table_Long_XVIII}
\begin{tabular}{r r c }
\hline \hline
\T
   Target     & Reaction     & Cumulative \\
              & Product      & Yield      \\ \hline
\\[1mm]
%  $^{232}$Th & $^{ 95}$Zr & 0.0000E+00~$\pm$~ 0.00~\% \\
%             & $^{ 99}$Mo & 0.0000E+00~$\pm$~ 0.00~\% \\
%             & $^{103}$Ru & 0.0000E+00~$\pm$~ 0.00~\% \\
%             & $^{106}$Ru & 0.0000E+00~$\pm$~ 0.00~\% \\
%             & $^{137}$Cs & 0.0000E+00~$\pm$~ 0.00~\% \\
%             & $^{137}$Ba & 0.0000E+00~$\pm$~ 0.00~\% \\
%             & $^{140}$Ba & 0.0000E+00~$\pm$~ 0.00~\% \\
%             & $^{140}$La & 0.0000E+00~$\pm$~ 0.00~\% \\
%             & $^{144}$Ce & 0.0000E+00~$\pm$~ 0.00~\% \\ \hline
   $^{235}$U  & $^{ 95}$Zr & 6.5042E-02~$\pm$~~1.00~\% \\
              & $^{ 99}$Mo & 6.1399E-02~$\pm$~~1.30~\% \\
              & $^{103}$Ru & 3.1118E-02~$\pm$~~2.10~\% \\
              & $^{106}$Ru & 4.0958E-03~$\pm$~~2.30~\% \\
              & $^{137}$Cs & 6.0897E-02~$\pm$~~1.04~\% \\
%             & $^{137}$Ba & 5.7492E-02~$\pm$~~1.38~\% \\
              & $^{140}$Ba & 6.3444E-02~$\pm$~~1.00~\% \\
%             & $^{140}$La & 6.3450E-02~$\pm$~~1.00~\% \\
              & $^{144}$Ce & 5.4781E-02~$\pm$~~0.90~\% \\[1mm] \hline
%  $^{238}$U  & $^{ 95}$Zr & 0.0000E+00~$\pm$~ 0.00~\% \\
%             & $^{ 99}$Mo & 0.0000E+00~$\pm$~ 0.00~\% \\
%             & $^{103}$Ru & 0.0000E+00~$\pm$~ 0.00~\% \\
%             & $^{106}$Ru & 0.0000E+00~$\pm$~ 0.00~\% \\
%             & $^{137}$Cs & 0.0000E+00~$\pm$~ 0.00~\% \\
%             & $^{137}$Ba & 0.0000E+00~$\pm$~ 0.00~\% \\
%             & $^{140}$Ba & 0.0000E+00~$\pm$~ 0.00~\% \\
%             & $^{140}$La & 0.0000E+00~$\pm$~ 0.00~\% \\
%             & $^{144}$Ce & 0.0000E+00~$\pm$~ 0.00~\% \\ \hline
\T
   $^{237}$Np & $^{ 95}$Zr & 5.9364E-02~$\pm$~11.80~\% \\
              & $^{ 99}$Mo & 6.6813E-02~$\pm$~~6.80~\% \\
              & $^{103}$Ru & 5.8640E-02~$\pm$~~7.30~\% \\
              & $^{106}$Ru & 1.5421E-02~$\pm$~19.40~\% \\
              & $^{137}$Cs & 7.0334E-02~$\pm$~11.71~\% \\
%             & $^{137}$Ba & 6.6401E-02~$\pm$~11.76~\% \\
              & $^{140}$Ba & 5.8591E-02~$\pm$~~9.90~\% \\
%             & $^{140}$La & 5.8661E-02~$\pm$~~9.90~\% \\
              & $^{144}$Ce & 4.2716E-02~$\pm$~12.76~\% \\[1mm] \hline
\T
   $^{239}$Pu & $^{ 95}$Zr & 4.8805E-02~$\pm$~~1.10~\% \\
              & $^{ 99}$Mo & 6.1410E-02~$\pm$~~0.70~\% \\
              & $^{103}$Ru & 6.8709E-02~$\pm$~~1.20~\% \\
              & $^{106}$Ru & 4.3404E-02~$\pm$~~2.30~\% \\
              & $^{137}$Cs & 6.5796E-02~$\pm$~~1.22~\% \\
%             & $^{137}$Ba & 6.2163E-02~$\pm$~~1.46~\% \\
              & $^{140}$Ba & 5.2880E-02~$\pm$~~1.10~\% \\
%             & $^{140}$La & 5.2989E-02~$\pm$~~1.10~\% \\
              & $^{144}$Ce & 3.7623E-02~$\pm$~~0.90~\% \\[1mm] \hline
\T
   $^{241}$Am & $^{ 95}$Zr & 3.9643E-02~$\pm$~~3.00~\% \\
              & $^{ 99}$Mo & 6.5724E-02~$\pm$~~3.20~\% \\
              & $^{103}$Ru & 7.1423E-02~$\pm$~~6.10~\% \\
              & $^{106}$Ru & 5.0867E-02~$\pm$~16.60~\% \\
              & $^{137}$Cs & 7.2069E-02~$\pm$~~6.60~\% \\
%             & $^{137}$Ba & 6.8107E-02~$\pm$~~6.66~\% \\
              & $^{140}$Ba & 5.8045E-02~$\pm$~~2.50~\% \\
%             & $^{140}$La & 5.8550E-02~$\pm$~~2.50~\% \\
              & $^{144}$Ce & 3.3905E-02~$\pm$~~3.70~\% \\[1mm] \hline\hline
\end{tabular}
%\vspace{+5mm}
\vspace{-3mm}
\end{table}

\begin{table}[hbtp]
\vspace{5mm}
\caption{IRDFF-II fast-neutron fission yields ($E\approx 400-500$~keV).}
\label{Table_Long_XIX}
\begin{tabular}{r r c }
\hline \hline
\T
 Target     & Reaction & Cumulative \\
            & Product  & Yield      \\ \hline
\\[1mm]
 $^{232}$Th & $^{ 95}$Zr &    5.4494E-02~$\pm$~~2.90~\% \\
            & $^{ 99}$Mo &    2.8740E-02~$\pm$~~2.80~\% \\
            & $^{103}$Ru &    1.5179E-03~$\pm$~~6.30~\% \\
            & $^{106}$Ru &    5.3236E-04~$\pm$~~5.70~\% \\
            & $^{137}$Cs &    6.1790E-02~$\pm$~~5.12~\% \\
%           & $^{137}$Ba &    5.8329E-02~$\pm$~~5.22~\% \\
            & $^{140}$Ba &    7.6222E-02~$\pm$~~3.19~\% \\
%           & $^{140}$La &    7.6222E-02~$\pm$~~3.19~\% \\
            & $^{144}$Ce &    7.6334E-02~$\pm$~~6.12~\% \\[1mm] \hline
\T
 $^{235}$U  & $^{ 95}$Zr &    6.4589E-02~$\pm$~~1.30~\% \\
            & $^{ 99}$Mo &    5.8957E-02~$\pm$~~1.90~\% \\
            & $^{103}$Ru &    3.2809E-02~$\pm$~~1.40~\% \\
            & $^{106}$Ru &    4.6597E-03~$\pm$~~7.60~\% \\
            & $^{137}$Cs &    5.8572E-02~$\pm$~~1.92~\% \\
%           & $^{137}$Ba &    5.5295E-02~$\pm$~~2.13~\% \\
            & $^{140}$Ba &    6.0586E-02~$\pm$~~1.10~\% \\
%           & $^{140}$La &    6.0594E-02~$\pm$~~1.10~\% \\
            & $^{144}$Ce &    5.1578E-02~$\pm$~~1.80~\% \\[1mm] \hline
\T
 $^{238}$U  & $^{ 95}$Zr &    5.2506E-02~$\pm$~~1.60~\% \\
            & $^{ 99}$Mo &    6.2147E-02~$\pm$~~1.60~\% \\
            & $^{103}$Ru &    6.0331E-02~$\pm$~~1.80~\% \\
            & $^{106}$Ru &    2.5063E-02~$\pm$~~5.30~\% \\
            & $^{137}$Cs &    6.0045E-02~$\pm$~~2.33~\% \\
%           & $^{137}$Ba &    5.6682E-02~$\pm$~~2.53~\% \\
            & $^{140}$Ba &    6.0457E-02~$\pm$~~1.29~\% \\
%           & $^{140}$La &    6.0457E-02~$\pm$~~1.29~\% \\
            & $^{144}$Ce &    4.6916E-02~$\pm$~~2.25~\% \\[1mm] \hline
\T
 $^{237}$Np & $^{ 95}$Zr &    5.6715E-02~$\pm$~~2.70~\% \\
            & $^{ 99}$Mo &    7.7238E-02~$\pm$~16.90~\% \\
            & $^{103}$Ru &    5.3778E-02~$\pm$~11.20~\% \\
            & $^{106}$Ru &    2.2333E-02~$\pm$~10.80~\% \\
            & $^{137}$Cs &    6.2129E-02~$\pm$~~3.62~\% \\
%           & $^{137}$Ba &    5.8651E-02~$\pm$~~3.74~\% \\
            & $^{140}$Ba &    5.7593E-02~$\pm$~~2.00~\% \\
%           & $^{140}$La &    5.7674E-02~$\pm$~~2.00~\% \\
            & $^{144}$Ce &    4.1743E-02~$\pm$~~4.79~\% \\[1mm] \hline
\T
 $^{239}$Pu & $^{ 95}$Zr &    4.6909E-02~$\pm$~~2.50~\% \\
            & $^{ 99}$Mo &    5.8366E-02~$\pm$~~2.30~\% \\
            & $^{103}$Ru &    6.5709E-02~$\pm$~~2.50~\% \\
            & $^{106}$Ru &    4.0375E-02~$\pm$~~7.80~\% \\
            & $^{137}$Cs &    6.3098E-02~$\pm$~~2.21~\% \\
%           & $^{137}$Ba &    5.9624E-02~$\pm$~~2.35~\% \\
            & $^{140}$Ba &    5.2916E-02~$\pm$~~1.50~\% \\
%           & $^{140}$La &    5.3067E-02~$\pm$~~1.50~\% \\
            & $^{144}$Ce &    3.4994E-02~$\pm$~~1.70~\% \\[1mm] \hline
\T
 $^{241}$Am & $^{ 95}$Zr &    4.1632E-02~$\pm$~~8.20~\% \\
            & $^{ 99}$Mo &    5.3596E-02~$\pm$~~5.50~\% \\
            & $^{103}$Ru &    6.5234E-02~$\pm$~~6.70~\% \\
            & $^{106}$Ru &    4.8254E-02~$\pm$~~6.90~\% \\
            & $^{137}$Cs &    6.2281E-02~$\pm$~18.80~\% \\
%           & $^{137}$Ba &    5.8866E-02~$\pm$~18.84~\% \\
            & $^{140}$Ba &    4.8853E-02~$\pm$~~4.50~\% \\
%           & $^{140}$La &    4.9310E-02~$\pm$~~4.50~\% \\
            & $^{144}$Ce &    3.4353E-02~$\pm$~~6.80~\% \\[1mm] \hline\hline
\end{tabular}
\end{table}

For the measurements of time-integrated fluence rates with fission products up to periods of about six weeks, the most common measured fission product used for fission research reactor dosimetry applications is $^{140}$Ba (half-life of 12.7527~days). In order to avoid isotopic separation techniques, the $^{140}$Ba activity is often determined by directly counting its daughter product, $^{140}$La (half-life of 1.6781~days). The independent fission yields are required to treat the transient equilibrium between the $^{140}$La and $^{140}$Ba.

Other fission products are commonly used to support reactor irradiations of different periods~\cite{A704}. The isotope $^{95}$Zr (half-life 64.032~days) is commonly used for shorter irradiation durations, up to 6 months while $^{103}$Ru (half-life of 39.247~days) is used for irradiations up to 4 months. The isotope $^{144}$Ce (half-life 284.91~days) is also a commonly measured fission product in reactor applications and can be used for irradiations periods of 2 to 3 years. The isotope $^{137}$Cs (half-life of 30.08~years) is used to support reactor irradiations of 30 to 40 years. The $^{137}$Cs activity is often measured by counting the activity in its $^{137m}$Ba decay product~\cite{A1005} (half-life of~2.552 minutes). The $^{106}$Ru fission product (half-life 371.8~days) is also often counted~\cite{A1005}. The transient equilibrium of its $^{106}$Rh decay product (half-life of 30.07~seconds) with the $^{106}$Rh independent yield can be ignored for most applications.

In summary, the relevant fission products currently addressed in dosimetry-related standards applications include the nine isotopes: $^{95}$Zr, $^{99}$Mo, $^{103}$Ru, $^{106}$Ru, $^{137}$Cs, $^{137m}$Ba, $^{140}$Ba, $^{140}$La, and $^{144}$Ce, but cumulative fission yields are needed only for seven of them.
\begin{table}[hbtp]
\vspace{5mm}
\caption{IRDFF-II fusion-peak fission yields ($E\approx 14$~MeV).}
\label{Table_Long_XX}
\begin{tabular}{r r c }
\hline \hline
\T
 Reaction   & Reaction & Cumulative \\
 Label      & Product  & Yield      \\ \hline
\\[1mm]
 $^{232}$Th & $^{ 95}$Zr & 5.0470E-02~$\pm$~13.60~\% \\
            & $^{ 99}$Mo & 1.9747E-02~$\pm$~~5.30~\% \\
            & $^{103}$Ru & 8.9252E-03~$\pm$~~7.20~\% \\
            & $^{106}$Ru & 1.0873E-02~$\pm$~10.40~\% \\
            & $^{137}$Cs & 6.0482E-02~$\pm$~~8.81~\% \\
%           & $^{137}$Ba & 5.7138E-02~$\pm$~~8.86~\% \\
            & $^{140}$Ba & 5.7087E-02~$\pm$~~4.00~\% \\
%           & $^{140}$La & 5.7143E-02~$\pm$~~4.00~\% \\
            & $^{144}$Ce & 4.2355E-02~$\pm$~~8.66~\% \\[1mm] \hline
\T
 $^{235}$U  & $^{ 95}$Zr & 5.0620E-02~$\pm$~~4.30~\% \\
            & $^{ 99}$Mo & 5.0301E-02~$\pm$~~2.80~\% \\
            & $^{103}$Ru & 3.1210E-02~$\pm$~~4.40~\% \\
            & $^{106}$Ru & 1.8000E-02~$\pm$~16.80~\% \\
            & $^{137}$Cs & 5.7013E-02~$\pm$~24.00~\% \\
%           & $^{137}$Ba & 5.4064E-02~$\pm$~24.01~\% \\
            & $^{140}$Ba & 4.4929E-02~$\pm$~~1.90~\% \\
%           & $^{140}$La & 4.5283E-02~$\pm$~~1.90~\% \\
            & $^{144}$Ce & 3.1575E-02~$\pm$~~1.30~\% \\[1mm] \hline
\T
 $^{238}$U  & $^{ 95}$Zr & 4.6389E-02~$\pm$~~1.22~\% \\
            & $^{ 99}$Mo & 5.7713E-02~$\pm$~~0.80~\% \\
            & $^{103}$Ru & 4.4881E-02~$\pm$~~2.30~\% \\
            & $^{106}$Ru & 2.4945E-02~$\pm$~~5.30~\% \\
            & $^{137}$Cs & 5.4359E-02~$\pm$~~8.21~\% \\
%           & $^{137}$Ba & 5.1345E-02~$\pm$~~8.26~\% \\
            & $^{140}$Ba & 4.6437E-02~$\pm$~~0.70~\% \\
%           & $^{140}$La & 4.6481E-02~$\pm$~~0.70~\% \\
            & $^{144}$Ce & 3.6147E-02~$\pm$~~3.57~\% \\[1mm] \hline
\T
 $^{237}$Np & $^{ 95}$Zr & 5.5453E-02~$\pm$~16.00~\% \\
            & $^{ 99}$Mo & 4.8632E-02~$\pm$~11.00~\% \\
            & $^{103}$Ru & 4.3312E-02~$\pm$~16.00~\% \\
            & $^{106}$Ru & 2.9975E-02~$\pm$~23.00~\% \\
            & $^{137}$Cs & 5.1076E-02~$\pm$~23.00~\% \\
%           & $^{137}$Ba & 4.8375E-02~$\pm$~23.00~\% \\
            & $^{140}$Ba & 4.8131E-02~$\pm$~11.00~\% \\
%           & $^{140}$La & 4.8281E-02~$\pm$~11.00~\% \\
            & $^{144}$Ce & 2.9997E-02~$\pm$~16.00~\% \\[1mm] \hline
\T
 $^{239}$Pu & $^{ 95}$Zr & 4.3837E-02~$\pm$~~6.00~\% \\
            & $^{ 99}$Mo & 5.6556E-02~$\pm$~~5.00~\% \\
            & $^{103}$Ru & 5.8641E-02~$\pm$~~6.00~\% \\
            & $^{106}$Ru & 3.5345E-02~$\pm$~~8.00~\% \\
            & $^{137}$Cs & 5.0107E-02~$\pm$~~8.00~\% \\
%           & $^{137}$Ba & 4.8059E-02~$\pm$~~8.00~\% \\
            & $^{140}$Ba & 4.0556E-02~$\pm$~~5.00~\% \\
%           & $^{140}$La & 4.1866E-02~$\pm$~~6.00~\% \\
            & $^{144}$Ce & 2.9598E-02~$\pm$~~5.00~\% \\[1mm] \hline
\T
 $^{241}$Am & $^{ 95}$Zr & 3.1194E-02~$\pm$~~6.00~\% \\
            & $^{ 99}$Mo & 4.3358E-02~$\pm$~~6.00~\% \\
            & $^{103}$Ru & 5.0404E-02~$\pm$~~6.00~\% \\
            & $^{106}$Ru & 4.0805E-02~$\pm$~~6.00~\% \\
            & $^{137}$Cs & 4.2719E-02~$\pm$~~6.00~\% \\
%           & $^{137}$Ba & 4.1374E-02~$\pm$~~6.00~\% \\
            & $^{140}$Ba & 3.3620E-02~$\pm$~~6.00~\% \\
%           & $^{140}$La & 3.5299E-02~$\pm$~~8.00~\% \\
            & $^{144}$Ce & 2.4968E-02~$\pm$~~6.00~\% \\[1mm] \hline\hline
\end{tabular}
\end{table}

\clearpage
The cumulative yields are used because they are generally known with a higher precision than the independent yields. However, care must be taken to ensure that the cumulative yields reach equilibrium. The typical cooling times that are necessary are listed below:
\noindent
\begin{tabular}{ p{0.25\columnwidth} p{0.72\columnwidth} }
%\\[-2mm]
\T
$^{95}$Zr               & The longest lived member of the chain is $^{95}$Y at 10.3 minutes.  All the others are less than 1 minute.  If we wait for 6 half-lives that would be 1 hour. \\[1mm]
$^{99}$Mo               & Longest half-life has $^{99m}$Nb at 2.6 m, requiring cooling time of about 15 minutes. \\[1mm]
$^{103}$Ru - $^{103}$Tc & Half-life 54 s, $^{103}$Mo half life 1.13 m, requiring cooling time of about 13 minutes. \\[1mm]
$^{106}$Ru - $^{106}$Tc & Half-life 36 s, requiring cooling time of about 3.6 minutes. \\[1mm]
$^{137}$Cs - $^{137}$Xe & Half-life 3.82 m, requiring cooling time of about 23 minutes. \\[1mm]
$^{140}$Ba - $^{140}$Cs & Half-life 1.06 m, requiring cooling time of about 6 minutes. \\[1mm]
$^{144}$Ce - $^{144}$La & Half-life 40.7 s, requiring cooling time of about 4.1 minutes. \\[3mm]
\end{tabular}
It should also be noted that several radionuclides of these chains have Kr or  Xe precursors.  This means that part or all of the chain yield can be lost as Kr or Xe gas unless the fissionable material is irradiated in a gas tight container.  This is a very well-known effect and actinides are commonly sealed in quartz to avoid losing any Kr or Xe, which would lead to a low reading of the cumulative fission product.

Some activities, such as nonproliferation monitoring~\cite{NASn,HDBK}, have a wider range of application-specific fission yield products that are monitored. This application area has independently standardized the underlying fission products yields to be used, hence this application area is not addressed by these \mbox{IRDFF-II} fission-yield recommended values.

Work continues in refining the recommended yields and the treatment of their dependence on the energy of the incident neutron. There is an ongoing international collaboration, sponsored through an IAEA coordinated research project (CRP) (project number F42007 approved 19 December 2018), to provide recommended fission yields and to address recommended approaches for handling the energy-dependence of the fission yields. The results of this CRP will not be ready in time for inclusion in the present release of IRDFF-II.

For present dosimetry applications, pending more investigation and validation of the energy-dependent modeling, the recommendation is to map the application need into one of the three generic energy regions supported by the commonly available nuclear data libraries: thermal, fast, and 14~MeV. The thermal energy generally equates to evaluated data for 0.0253~eV incident neutrons. The fast energy equates to evaluated data gathered using a range of broad-spectrum fast fission spectra. Within the commonly available tabulated nuclear data files, the fast region is characterized with a reference energy of 500~keV in the \mbox{ENDF/B-VIII.0} library and with an energy of 400~keV in the \mbox{JEFF-3.3} library. The 14~MeV energy equates to evaluated data validated using data gathered from DT fusion neutron sources.

Pending results from ongoing international collaboration on updating the fission product yields, the \mbox{IRDFF-II} library has adopted a set of recommended fission yields. These recommended fission yields are taken from the \mbox{JEFF-3.3} library, if available. Alternatively, \eg, in the case of the 14~MeV fission yields for $^{237}$Np, $^{239}$Pu and $^{241}$Am, the recommended yields are taken from the \mbox{ENDF/B-VIII.0} library.

\subsection{Neutron Benchmark Fields}

As part of the library evaluation and acceptance process for reactions accepted for inclusion within the \mbox{IRDFF-II} library, data validation was addressed using experimental measurements in integral neutron benchmark fields. Such measurements are independent of the differential data (measured as a function of incident energy), used by the evaluators in conjunction with nuclear data models to generate the recommended  nuclear reaction cross section data files. Since these neutron benchmark fields are tied to the validation evidence for the \mbox{IRDFF-II} library, the decision was made to include the neutron spectral characterizations for these integral benchmark fields as part of the \mbox{IRDFF-II} package. Table~\ref{Table_Long_XIII} summarizes the 45 neutron fields that were considered in the \mbox{IRDFF-II} validation and provides brief descriptions of the selected 29 neutron benchmark fields. These  benchmark fields are addressed in further detail in Secs.~\ref{Sec_VII} and~\ref{Sec_VIII} on sensitivity, data validation and data consistency in this paper. The \mbox{IRDFF-II} library provides recommended spectra for these neutron fields as well as for several other spectra not selected as benchmarks, but useful in making data comparisons, \eg, a constant spectrum, a linear spectrum, and a 1/$E$ spectrum in the region 0.5~eV~$<$~$E$~$<$~20~MeV. Maxwellian spectra corresponding to different temperatures are also included that cover a range of temperatures (in the laboratory coordinate system) from 25~keV to 60~keV, which are of interest for astrophysics. The spectra are part of the \mbox{IRDFF-II} distribution.
%, stored in pseudo-ENDF format as MF~3 MT~261 ENDF entries. The covariance information is in the corresponding ENDF files MF~33.
Note that the spectra calculated with detailed computational models of the experiments by the Monte Carlo technique (\eg, reactor spectra) are used as \textit{prior} spectra in the least-squares fit. The uncertainty in the \textit{prior} spectrum, in addition to being a subjective evaluation at this time, includes large uncertainty contributors which make direct comparison with measured data very difficult due to the large associated uncertainties. The total uncertainty of the calculated \textit{prior} obviously includes the statistical uncertainty of the calculations, which usually is less than 10~\%, except in the tails, where the spectrum approaches zero. The full spectrum covariance matrices from computational models being subjective and not available for all cases, the covariances in \mbox{IRDFF-II} only include the statistical component for cross-checking the uncertainties in the calculated reaction rates of monitor reactions with median energies close to the tails of the spectrum.

\begin{table*}[thbp]
\vspace{-5mm}
\caption{List of IRDFF-II benchmark neutron fields} \label{Table_Long_XIII}
\begin{tabular}{r | l | l | c | p{11.2cm} } \hline \hline
\T
No. & Name & MAT & $E_{\mbox{aver}}[MeV]$ & ~~~~~~~~~~~~~Description \B \\
\hline
\addlinespace[1mm]
\multicolumn{5}{c}{Measured by Time-of-Flight neutron fields including the \CF(sf) standard}\\
\addlinespace[1mm]
\hline
\T
 1 & $^{252}$Cf(sf)        & 9861 & 2.121 & Spontaneous fission neutron spectrum from $^{252}$Cf \\
 2 & $^{235}$U PFNS        & 9228 & 2.000 & Thermal-neutron induced prompt fission spectrum from $^{235}$U \\
 3 & $^{9}$Be(d,n)~16MeV   & 9408 & 5.608 & Spectrum of neutrons from 16~MeV deuterons incident on a beryllium target \\
 4 & $^{9}$Be(d,n)~40MeV   & 9409 & 15.58 & Spectrum of neutrons from 40~MeV deuterons incident on a beryllium target \\
\hline
\addlinespace[1mm]
\multicolumn{5}{c}{Measured by Time-of-Flight neutron fields not accepted as benchmark fields}\\
\addlinespace[1mm]
\hline
\T
 1 & $^{233}$U PFNS        & 9222 & 2.030 & Thermal-neutron induced prompt fission spectrum from $^{233}$U \\
 2 & $^{239}$Pu PFNS       & 9437 & 2.073 & Thermal-neutron induced prompt fission spectrum from $^{239}$Pu \\
\hline
\addlinespace[1mm]
\multicolumn{5}{c}{Neutron benchmark fields from detailed computational models}\\
\addlinespace[1mm]
\hline
\T
 1 & ACRR-FF-32            & 9010 & 0.575 & ACRR-FF-32 Reactor Extended Cavity Spectrum 640-group \\
 2 & ACRR-CdPoly           & 9011 & 0.657 & ACRR-CdPoly Reactor Bucket Spectrum 640-group \\
 3 & ACRR-PLG              & 9012 & 0.439 & ACRR-PLG    Reactor Bucket Spectrum 640-group \\
 4 & ACRR-LB44             & 9013 & 0.715 & ACRR-LB44   Reactor Bucket Spectrum 640-group \\
 5 & FREC-II               & 9015 & 0.545 & FREC-II Spectrum (external cavity attached to ACRR) 640-group \\
 6 & SPR-III               & 9014 & 1.251 & SPR-III Reactor Central Cavity Spectrum 640-group \\
 7 & Mol BR1 Mark-III      & 9020 & 1.864 & Mol BR1 Mark-III, \UT~converter in Cd and Graphite cavity, 640-groups \\
 8 & LR0-Rez               & 9032 & 0.646 & Rez-LR0 Reactor spectrum, 640-group \\
 9 & TRIGA-JSI             & 9041 & 0.389 & TRIGA Mark-II Pneumatic tube (bare), 640-group \\
10 & TRIGA-JSI/BN          & 9042 & 0.848 & TRIGA Mark-II boron nitride cover, 640-group \\
11 & TRIGA-JSI/B4C         & 9043 & 0.923 & TRIGA Mark-II boron carbide cover, 640-group \\
12 & TRIGA-JSI/10B4C       & 9044 & 1.090 & TRIGA Mark-II enriched boron carbide cover, 640-group \\
13 & ISNF                  & 9004 & 1.058 & ISNF Reactor Spectrum 725-group \\
14 & CFRMF                 & 9005 & 0.741 & CFRMF Reactor Spectrum from IRDF-2002 \\
15 & Sigma-Sigma           & 9007 & 0.763 & Sigma-Sigma facility in $^{nat}$U and BC spheres inside Graphite column, 725g \\
\T
16 & HMF001                & 9101 & 1.433 & Godiva, central region, 725-group  \\
17 & HMF028                & 9102 & 1.343 & Flattop-25, central region, 725-group \\
18 & IMF007                & 9103 & 0.570 & Big-Ten 725-group  \\
19 & FMR001                & 9110 & 1.483 & IPPE-BR1, central region, 725-group \\
\T
20 & FNS-Grph-096mm        & 9201 & 5.267 & FNS-Graphite block with a D-T source and monitors at  96 mm, 725 group \\
21 & FNS-Grph-293mm        & 9202 & 1.957 & FNS-Graphite block with a D-T source and monitors at 293 mm, 725 group \B \\
\hline
\addlinespace[1mm]
\multicolumn{5}{c}{ICSBEP spectra not accepted as benchmark fields}\\
\addlinespace[1mm]
\hline
\T
1  & PMF001                & 9104 & 1.797 & Jezebel, central region, 725-group  \\
2  & PMF002                & 9105 & 1.747 & Jezebel-240, central region, 725-group \\
3  & PMF006                & 9106 & 1.589 & Flattop-Pu, central region, 725-group  \\
4  & PMF008                & 9107 & 1.681 & Thor, central region, 725-group  \\
\hline
\addlinespace[1mm]
\multicolumn{5}{c}{Analytical spectrum functions accepted as benchmark fields}\\
\addlinespace[1mm]
\hline
\T
 1 & Thermal Maxw.         & 9901 &       & Thermal Maxwellian at 293.6 K  \\
 2 & 1/E [0.55eV--2MeV]    & 9902 &       & Pure 1/E between Ecd and E2 ($0.55$~eV$<E<2$~MeV) \\
 3 & Maxw.(25~keV)         & 9925 &       & Maxwellian at 25 keV\\
 4 & Maxw.(30~keV)         & 9930 &       & Maxwellian at 30 keV \\
\hline
\addlinespace[1mm]
\multicolumn{5}{c}{Analytical spectrum functions not used as benchmark fields}\\
\addlinespace[1mm]
\hline
\T
 1 & Const.                & 9900 &       & Constant spectrum Phi=1 \\
 2 & 1/E [0.5eV--20MeV]     & 9904 &       & Pure 1/E between Ecd and E2 (0.5 eV $<$ E $<$ 20 MeV) \\
 3 & Maxw.Fiss.            & 9905 &       & Pure Maxwellian fission spectrum at temperature 2.03 MeV \\
 4 & Linear                & 9910 &       & Linear spectrum Phi=E (1.E-5 eV $<$ E $<$ 20 MeV) \\
 5 & Maxw.(32~keV)         & 9932 &       & Maxwellian at 32 keV \\
 6 & Maxw.(35~keV)         & 9935 &       & Maxwellian at 35 keV \\
 7 & Maxw.(40~keV)         & 9940 &       & Maxwellian at 40 keV \\
 8 & Maxw.(45~keV)         & 9945 &       & Maxwellian at 45 keV \\
 9 & Maxw.(50~keV)         & 9950 &       & Maxwellian at 50 keV \\
10 & Maxw.(60~keV)         & 9960 &       & Maxwellian at 60 keV \\
\hline\hline
\end{tabular}
\vspace{-4mm}
\end{table*}

\section{IRDFF-II FORMAT AND DATA VERIFICATION} \label{Sec_III}

In order to support the user community, the cross sections in the \mbox{IRDFF-II} library are distributed in several different formats, described in the sections below.

The baseline library is in the \mbox{ENDF-6} format, where the full library data file contains several logical files for different nuclides, identified by the so-called MAT numbers. Each nuclide file is further subdivided into subfiles identified by the MF numbers that designate the type of data and the MT numbers that designate the reaction type. A brief extract is given below for convenience. Full details are given in the ENDF-6 manual~\cite{ENDF-6}.

\begin{itemize}
  \item[MAT] Material/nuclide identifier is a four-digit number that uniquely identifies the target nuclide. \\
  \item[MF] Data type identifier:
  \item[1] General information. The first section in this subfile is identified by MT=451 and contains evaluation description in plain text, including eigenvalues of the  covarince matrices.
  \item[3] Cross sections and other parameters as a function of incident particle energy.
  \item[10] Cross sections for the excitation of discrete states, or radionuclide production cross sections.
  \item[33] Covariance data of the quantities in MF=3.
  \item[40] Covariance data for the quantities in MF=10.
  \item[MT] Reaction type number (\eg, 16$\equiv$(n,2n), 102$\equiv$(n,$\gamma$), 103$\equiv$(n,p), \textit{etc}.). For full details see Appendix~B of the \mbox{ENDF-6} Manual~\cite{ENDF-6}.
\end{itemize}

The derived libraries in other formats are generated by processing baseline libraries in point-wise cross-section representation with codes such as NJOY-2016~\cite{NJOY2016} or PREPRO~\cite{PREPRO}. At the request of the users, the recommended cross sections are also available in three column format (energy, cross section and the total uncertainty) to be used as reference activation cross sections for relative activation measurements\footnote{Note that a full uncertainty propagation of these cross sections for a broad spectrum requires the use of the complete covariance matrix, which is only given in ENDF-6 format.}. If a user requires the cross sections at a different temperature, the threshold reactions would be practically unaffected. Reaction cross sections with resonance structure  might need additional processing with codes like NJOY or PREPRO.

Note that residual-production covariance matrices in ENDF subfile MF=40 that contain more than one product cannot be processed with the current version of NJOY~\cite{NJOY2016}. This is applicable to many of the natural element evaluations in IRDFF-II. A way around this problem is to split the evaluated file in such a way that only one product is present at a time, and process the partial files separately. Hopefully, an update to the processing code will be available soon that will support legitimate ENDF-6 format covariance representation.

\subsection{ENDF-6 Format Point Cross Sections} \label{Sec_III_A}
The baseline library contains point cross sections at 293.6~K in \mbox{ENDF-6} format representation that is well-documented and has the flexibility to capture the highest fidelity characterization of neutron cross sections, including the resonance region, as well as the associated covariance files. This format can also be used to capture the nuclear decay data and the neutron spectral characterization data. The \mbox{ENDF-6} file structure places reaction cross sections in MF=3 and production cross sections in MF=10. Both of these file types are used to capture the cross sections within the \mbox{IRDFF-II} library. The covariance data that correspond to data found in these sections are stored in MF=33 and MF=40, respectively. Radioactive decay data are given in MF=8. The file structure permits the inclusion of in-line documentation using MF=1, MT=451 entries.
%Whereas the \mbox{ENDF-6} file structure permits the energy distribution of secondary particles to be presented in MF=5 and MF=6, this field does not fit the \mbox{IRDFF-II} characterization of the benchmark neutron spectra, so, these data appear in a special \mbox{IRDFF-II} file in the MF=3/33 with MT=261, an unassigned reaction type number within the formal \mbox{ENDF-6} file structure.

\subsection{ENDF-6 Format Neutron Spectra} \label{Sec_III_B}
Whereas the \mbox{ENDF-6} file structure permits the energy distributions of secondary particles to be presented in MF=5 and MF=6, this field does not fit the \mbox{IRDFF-II} characterization for the benchmark neutron spectra. These data appear in a separate \mbox{IRDFF-II} library file in pseudo-ENDF format as MF=3, MT=261 sub-files. The spectra are differentiated by the MAT number. The covariance information is stored in the corresponding ENDF subfiles MF=33, MT=261. This is a non-standard use of the ENDF format, which uses MT=261 that is formally undefined in the ENDF-6 manual~\cite{}. Storing the spectra in this format is convenient because for many operations on the data and for generating the comparison plots the same software can be used as for the cross sections.

\subsection{ENDF-6 Format 640 and 725 Group Representation} \label{Sec_III_C}

Whereas a point cross section format is best for capturing the physics-based structure behind the cross section representation, this format does not readily support post-processing of the cross sections and the extraction of spectrum-averaged cross sections, reaction rates from integral neutron fields, or many other relevant material dosimetry metrics. For these applications, a multigroup representation is generally used. An important consideration with multigroup representations is the energy group structure. Whereas few-group representation may be sufficient for some applications, other applications require a fine-group representation.

A common fine-group structure that was found to be sufficient for capturing the energy-dependent shape of most responses relevant to the fission reactor community is the extended SAND-II group structure~\cite{McE67}. The extension here represented an increase from the original 18~MeV \mbox{SAND-II} maximum energy up to 20 MeV. Thus, this 640-group energy group structure starts at 10$^{-4}$~eV and goes up to 20~MeV. Since the \mbox{IRDFF-II} library goes up to 60~MeV, in this representation the \mbox{IRDFF-II} data are essentially truncated at 20~MeV and collapsed into the 640-group structure.

The \mbox{IRDFF-II} application addresses the needs of the fusion and low temperature material science communities as well as the fission reactor community, hence a second 725-group representation that extends up to 60~MeV is also used to characterize the data. The energy group structure for this 725-group representation is formed by:
\begin{itemize}
\item starting at 10$^{-5}$~eV and adding 45 groups to cover the energy range from 10$^{-5}$~eV up to the 10$^{-4}$~eV, where the 640-group SAND-II structure starts;
\item using the same 640-groups between 10$^{-4}$~eV and 20~MeV;
\item adding 20 groups of 0.5~MeV width in energy to extend the range up to 30~MeV;
\item adding 10 groups of 1~MeV width in energy to further extend the range up to 40~MeV;
\item adding 10 groups of 2~MeV width in energy to further extend the range up to 60-MeV.
\end{itemize}

In this collapsing of the point cross sections to the 640 or 725 group structure, an energy-dependent weighting function can be used. The two most common weighting functions are:
\begin{enumerate}
\item A thermal Maxwellian shape in the thermal region, a 1/$E$ shape in the reactor neutron slowing down region, a fission spectrum shape in the fast fission reactor region, and a fusion peak at high energies of the fusion blanket region. The GROUPR module of the NJOY-2012 code with the IWT=8 weighting function using the default weighting function shape parameters (thermal breakpoint of 0.1 eV, thermal temperature of 0.025 eV, fission breakpoint of 0.1 MeV, fission temperature of 1.4 MeV) can be used to collapse the data that appears in the file. The disadvantage of this weighting function is that it might not be representative of reactor spectra at epithermal energies and could even have a wrong gradient.
\item A flat weighting across the energy bin. Given the fine group nature of the two energy bin structures, this flat weighting is more general at high energies and mostly sufficient at lower energies as well.
\end{enumerate}
In the preparation of the \mbox{IRDFF-II} multigroup data the second option, flat weighting, was used. This collapsing was performed with the GROUPIE module of the PREPRO-2018 data processing package~\cite{PREPRO}. The same procedures are applicable also to the \mbox{IRDFF-II} file of neutron spectra in \mbox{ENDF-6} format.

\subsection{MCNP ACE Format}\label{Sec_III_D}

The ACE (A Compact ENDF) format is a common data format that is used in continuous-energy neutron-photon Monte Carlo codes like MCNP~\cite{MCNP} and SERPENT~\cite{SERPENT}. The virtue of this format is that it can capture the continuous energy nature of the underlying cross section without resorting to a multi-group representation. Since many fission and fusion community applications use the MCNP code to model the neutron spectrum at locations of interest for dosimetry measurements, the inclusion of this format increases the fidelity supported by the application and makes it easy for the user to interface with the recommended \mbox{IRDFF-II} library.

\subsection{Data Verification} \label{Sec_III_E}

A critical step in the verification of the formal correctness of the \mbox{IRDFF-II} library files was to ensure that the \mbox{ENDF-6} format rules were obeyed. For this purpose, the ENDF Utility Codes were employed~\cite{ENDF_utility}, namely:

\begin{itemize}
\setlength{\itemindent}{1.0cm}
\item[CHECKR] checks the syntax;
\item[FIZCON~] checks the internal consistency of the data;
\item[PSYCHE~] does an in-depth checking, particularly regarding the covariance data.
\end{itemize}

In addition, the COVEIG code~\cite{COVEIG} was used to check for symmetry, large correlation coefficients and negative eigenvalues in the covariance matrices.

The ability to process the files in \mbox{ENDF-6} format was checked by running the PREPRO codes~\cite{PREPRO}. The following modules are applicable:
\begin{itemize}
\setlength{\itemindent}{1.2cm}
\item[LINEAR~~~] makes all cross sections linearly interpolable;
\item[GROUPIE~] generates group-averaged cross sections;
\item[COMPLOT] makes graphical comparison with other evaluated data files.
\end{itemize}
%\end{description}

Finally, the NJOY code~\cite{NJOY2016} was employed to prepare plots of the covariance matrices and to generate the \mbox{IRDFF-II} library in ACE format. Quality assurance procedure~\href{https://www-nds.iaea.org/publications/indc/indc-sec-0107/}{INDC(SEC)-0107}~\cite{INDC(SEC)-0107} was employed to minimize the possibility of data processing errors when generating the files in ACE format.

\section{RECOMMENDED NUCLEAR DATA FOR NEUTRON METROLOGY} \label{Sec_IV}

This Sec. provides a more in-depth examination of the various nuclear data components within the \mbox{IRDFF-II} library and provides details on the process used to arrive at the recommendations.

\subsection{Atomic Mass and Isotopic Natural Abundance Data} \label{Sec_IV_A}

%\LTcapwidth=\textwidth
%\begin{longtable*}[t]{r | c | c | c | c}
\begin{table*}[htb]
\vspace{-2mm}
\caption{IRDFF-II isotopic abundances, nuclear and elemental mass data.
Numbers in parentheses indicate the absolute uncertainties, \textit{e.g.}, \protect{$92.411(24)\equiv 92.411\pm 0.024$}.}\label{Table_Long_XV}
\begin{tabular}{r | c | c | c | c}
\hline \hline
\T
Target     & Abundance & Isotopic Atomic   & Isotopic Atomic & Elemental weight value \\
nucleus    & [atom \%] & Mass Excess [keV] & Mass [$\mu$amu] &    or range [amu]      \\ \hline
\T $^{6}$Li                          &  7.589 (24)    & +14086.8789 (14)     & 6015122.8874 (15)      & [6.938, 6.997]      \\
 $^{7}$Li                            &  92.411 (24)   & +14907.105 (4)       & 7016003.437 (5)        & [6.938, 6.997]      \\
									
 $^{10}$B                            &  19.82 (2)     & +12050.609 (15)      & 10012936.862 (16)      & [10.806, 10.821]    \\
 $^{11}$B                            &  80.18 (2)     & +8667.707 (12)       & 11009305.167 (13)      & [10.806, 10.821]    \\
									
 $^{19}$F                            &  100.  (0)     & -1487.4442 (09)      & 18998403.1629 (9)      & 18.998403 163(6)    \\
									
 $^{23}$Na                           &  100.  (0)     & -9529.8525 (18)      & 22989869.2820 (19)     & 22.98976928 (2)     \\
									
 $^{24}$Mg                           &  78.951 (12)   & -13933.569 (13)      & 23985041.697 (14)      & [24.304, 24.307]    \\
									
 $^{27}$Al                           &  100. (0)      & -17196.86 (5)        & 26981538.41 (5)        & 26.9815385 (7)      \\
									
$^{28}$Si                            & 92.22968 (44)  & -21492.7943 (5)      & 27.9769265350 (5)      & [28.084, 28.086]    \\
$^{29}$Si                            & 4.68316 (32)   & -21895.0784 (6)      & 28.9764946653 (6)      & [28.084, 28.086]    \\
									
 $^{31}$P                            &  100. (0)      & -24440.5410 (7)      & 30973761.9986 (7)      & 30.973761 998 (5)   \\
									
 $^{32}$S                            &  95.04074 (88) & -26015.5336 (13)     & 31972071.1744 (14)     & [32.059, 32.076]    \\
									
 $^{45}$Sc                           &  100. (0)      & -41071.9 (7)         & 44955907.5 (7)         & 44.955908 (5)       \\
									
 $^{46}$Ti                           &  8.249 (21)    & -44127.80 (16)       & 45952626.86 (18)       & 47.867 (1)          \\
 $^{47}$Ti                           &  7.437 (14)    & -44937.36 (12)       & 46951757.75 (12)       & 47.867 (1)          \\
 $^{48}$Ti                           &  73.720 (22)   & -48492.71 (11)       & 47947940.93 (12)       & 47.867 (1)          \\
									
 $^{51}$V                            &  99.7503 (6)   & -52203.8 (4)         & 50 943956.9 (4)        & 50.9415 (1)         \\
									
 $^{50}$Cr                           &  0.2497 (6)    & -50262.1 (4)         & 49 946041.4 (5)        & 51.9961 (6)         \\
 $^{52}$Cr                           &  83.7895 (117) & -55419.2 (3)         & 51 940505.0 (4)        & 51.9961 (6)         \\
									
 $^{55}$Mn                           &  100. (0)      & -57712.4 (3)         & 54938043.2 (3)         & 54.938 044 (3)      \\
									
 $^{54}$Fe                           &  5.8450 (230)  & -56254.5 (4)         & 53939608.3 (4)         & 55.845 (2)          \\
 $^{54}$Fe                           &  5.8450 (230)  & -56254.5 (4)         & 53939608.3 (4)         & 55.845 (2)          \\
 $^{56}$Fe                           &  91.7540 (240) & -60607.1 (3)         & 55934935.6 (3)         & 55.845 (2)          \\
 $^{58}$Fe                           &  0.2819 (27)   & -62155.1 (3)         & 57933273.7 (4)         & 55.845 (2)          \\
									
 $^{59}$Co                           &  100. (0)      & -62229.7 (4)         & 58933193.7 (4)         & 58.933194 (4)       \\
									
 $^{58}$Ni                           &  68.0769 (59)  & -60228.7 (4)         & 57935341.8 (4)         & 58.6934 (4)         \\
									
 $^{63}$Cu                           &  69.174 (20)   & -65579.8 (4)         & 62929597.2 (5)         & 63.546 (3)          \\
 $^{65}$Cu                           &  30.826 (20)   & -67263.7 (6)         & 64927789.5 (7)         & 63.546 (3)          \\
									
 $^{64}$Zn                           &  49.1704 (83)  & -66004.0 (6)         & 63929141.8 (7)         & 65.38 (2)           \\
 $^{67}$Zn                           &  4.0401 (18)   & -67880.3 (8)         & 66 927127.5 (8)        & 65.38 (2)           \\
									
 $^{75}$As                           &  100. (0)      & -73034.2 (9)         & 74 921594.6 (9)        & 74.921595 (6)       \\
									
 $^{89}$Y                            &  100. (0)      & -87708.4 (16)        & 88 905841.2 (17)       & 88.90584 (2)        \\
									
 $^{90}$Zr                           &  51.452 (9)    & -88772.54 (12)       & 89904698.76 (13)       & 91.224 (2)          \\
									
 $^{93}$Nb                           &  100. (0)      & -87212.8 (15)        & 92906373.2 (16)        & 92.90637 (2)        \\
									
 $^{92}$Mo                           &  14.649 (17)   & -86808.58 (16)       & 91906807.16 (17)       & 95.95 (1)           \\
									
 $^{103}$Rh                          &  100. (0)      & -88031.7 (23)        & 102905494.1 (25)       & 102.90550 (2)       \\
									
 $^{109}$Ag                          &  48.1608 (51)  & -88719.4 (13)        & 108904755.8 (14)       & 107.8682 (2)        \\
									
 $^{113}$In                          &  4.281 (17)    & -89367.12 (19)       & 112904060.45 (20)      & 114.818(1)          \\
 $^{115}$In                          &  95.719 (17)   & -89536.346 (12)      & 114903878.774 (13)     & 114.818(1)          \\
									
 $^{127}$I                           &  100.          & -88984 (4)           & 126904472 (4)          & 126.90447 (3)       \\
									
 $^{139}$La                          &  99.91119 (24) & -87226.2 (20)        & 138906358.8 (22)       & 138.90547 (7)       \\
									
 $^{141}$Pr                          &  100.0 (0)     & -86015.6 (17)        & 140907658.4 (18)       & 140.90766 (2)       \\
									
 $^{169}$Tm                          &  100.0 (0)     & -61275.2 (08)        & 168934218.4 (09)       & 168.93422 (2)       \\
									
 $^{181}$Ta                          &  99.98799 (8)  & -48438.3 (14)        & 180947999.3 (15)       & 180.94788 (2)       \\
									
 $^{186}$W                           &  28.4259 (62)  & -42508.5 (12)        & 185954365.2 (13)       & 183.84 (1)          \\
									
 $^{197}$Au                          &  100.0 (0)     & -31139.7 (05)        & 196966570.1 (06)       & 196.966569 (5)      \\
									
 $^{199}$Hg                          &  16.938 (9)    & -29546.1 (05)        & 198968281.0 (06)       & 200.592 (3)         \\
									
 $^{204}$Pb                          &  1.4245 (12)   & -25109.9 (11)        & 203973043.4 (12)       & 207.2 (1)           \\
									
 $^{209}$Bi                          &  100.0 (0)     & -18258.7 (14)        & 208980398.5 (15)       & 208.98040 (1)       \\
									
 $^{232}$Th                          &  99.998862 (2) & +35446.8 (14)        & 232038053.7 (15)       & 232.0377 (4)        \\
									
 $^{235}$U                           &  0.72041 (36)  & +40918.8 (11)        & 235043928.2 (12)       & 238.02891 (3)       \\
									
 $^{238}$U                           &  99.27417 (36) & +47307.8 (15)        & 238050787.0 (16)       & 238.02891 (3)       \\

 $^{237}$Np                          &  ---           & +44871.7 (11)        & 237048171.7 (12)       & ---                 \\
									
 $^{239}$Pu                          &  ---           & +48588.3 (11)        & 239052161.7 (12)       & ---                 \\
									
 $^{241}$Am                          &  ---           & +52934.4 (11)        & 241056827.4 (12)       & ---                 \\[1mm]
\hline\hline
\end{tabular}
%\vspace{-2mm}
\end{table*}
%\end{longtable*}

The atomic weights for the target elements are given in Table~\ref{Table_Long_XV} in atomic mass units. The reported uncertainties correspond to those for isotopes in the normal terrestrial materials.
Column~2 of Table~\ref{Table_Long_XV} gives the recommended natural abundance value for all targets in the \mbox{IRDFF-II} library.
Column~3 of Table~\ref{Table_Long_XV} shows the isotopic weights – represented as the mass excess in energy units and Column~4 gives the corresponding values in micro-atomic mass units, as quoted in the original publication~\cite{AME2016}.

\subsection{Dosimetry Cross Sections} \label{Sec_IV_B}

The \mbox{IRDFF-II} recommended cross sections are chosen from available \mbox{ENDF-6}~\cite{ENDF-6} formatted point cross sections from different sources as discussed below. Whereas some applications, such as radiation transport calculations, require the use of a complete nuclear data evaluation that includes all potential reaction channels, the \mbox{IRDFF-II} neutron metrology library only requires the specification of a cross section for the channels used within the metrology application. However, unlike many applications, it requires a consistent uncertainty characterization for the cross section in the form of an energy-dependent covariance matrix. Because of the demanding requirements on neutron metrology cross sections and corresponding uncertainties, the recommended evaluations, rather than being taken from existing nuclear data evaluations, are often the result of a specifically funded metrology-related activity. Table~\ref{Table_Long_I} provides a list of the cross sections addressed within the \mbox{IRDFF-II} library and identifies the source evaluations. Columns~8 of Table~\ref{Table_Long_I} also shows how the \mbox{IRDFF-v1.05} library differs from the previous \mbox{IRDF-2002}. Note that \ql old\qr means no change in the data, symbol "new" means the data were newly evaluated. Otherwise, Column~9 indicates how the current \mbox{IRDFF-II} compares with \mbox{IRDFF-v1.05}. Sometimes the recommended evaluation represents a combination where different energy regions are derived from different evaluated data sources. Column~6 provides reference to auxiliary documentation that describes the derivation of some of the recommended evaluations.

\subsection{Nuclear Decay Data}  \label{Sec_IV_C}

This Sec. addresses the \mbox{IRDFF-II} recommended nuclear decay data for activation products and for fission products.
Tables~\ref{Table_Long_XIVa} (beta emitters) and \ref{Table_Long_XIV} (gamma emitters) give the half-lives and emission probabilities associated with the residual nuclei (activation product) for all of the radionuclides according to the detected decay radiation. It also shows the most recent recommended  gamma and beta decay data, namely the emission energies and emission probabilities for all of the residual nuclei associated with the reactions addressed within \mbox{IRDFF-II}.
\begin{table}[t]
\vspace{-4mm}
\caption{IRDFF-II recommended nuclear decay data from BIPM for pure beta emitters. Numbers in parentheses indicate the absolute uncertainties, \textit{e.g.}, \protect{$12.312(25)\equiv 12.312\pm 0.025$}.}
\label{Table_Long_XIVa}
\begin{tabular}{c | c c | c | c | c}
\hline \hline
\multicolumn{4}{c}{ }              & \multicolumn{2}{c}{Beta Data}      \\
Reaction & Half-life& Half-life    & Time & Aver.$E_{\beta}$ & Emiss.   \\
Product  & (recomm.)& (DD eval.)   & unit &   [keV]          & Prob. [\%] \\
\hline
\T
$^{3}$H    & 12.32(2)   & \a 12.312(25)\b & a & 5.68(1)  & 100. \\
$^{32}$P   & 14.263(36) & 14.284(36)      & d & 695.5(3) & 100. \\
$^{106}$Ru & 371.8(18)  & \a 371.5(21)\b  & d & 10.03(6) & 100. \\ \hline \hline
\end{tabular}
\vspace{-3mm}
\end{table}

\newpage
Note that recommended data are those used in the \mbox{IRDFF-II} cross-section evaluation. These data agree within uncertainties with the latest decay data evaluations (labeled \textbf{DD eval.} in Table headers). The latest decay data evaluations have been compiled into associated \mbox{IRDFF-II} decay data file that is released together with the \mbox{IRDFF-II} neutron metrology library.

%\newpage
\LTcapwidth=\textwidth
%\vspace*{6mm}
\begin{longtable*}[!t]{r | c c c | c | c | c | c }
\caption{IRDFF-II recommended nuclear decay data content for gamma emitters.
(DD eval.) refers to the latest Chechev compilation at \url{https://www-nds.iaea.org/IRDFFtest/RCM3/Chechev-RCM3.pdf}.
(*) indicates data taken from ENSDF. New DDEP evaluations undertaken within this project are highlighted in \a blue\b. 
Numbers in parentheses indicate the absolute uncertainties, \textit{e.g.}, \protect{$109.734(14)\equiv 109.734\pm 0.014$}.}
\label{Table_Long_XIV} \\ \hline \hline
\addlinespace[0.5mm]
\multicolumn{5}{c}{ }          & \multicolumn{2}{c}{Gamma/Xray}        &                 \\
Reaction & Half-life& Half-life & Time  &~Gamma/X-ray    & Emiss.Prob. [\%] & Emiss.Prob. [\%] &  Source \\
Product  & (recomm.)&(DD eval.) & unit  &~Energy[keV]    & (recomm.)        &(DD eval.)& Document        \\ \hline
\endfirsthead
\caption{(continued). IRDFF-II recommended nuclear decay data content for gamma emitters.
(DD eval.) refers to the latest Chechev compilation at \url{https://www-nds.iaea.org/IRDFFtest/RCM3/Chechev-RCM3.pdf}.
(*) indicates data taken from ENSDF. New DDEP evaluations undertaken within this project are highlighted in \a blue\b.} \\
\hline \hline
\addlinespace[0.5mm]
\multicolumn{5}{c}{ }          & \multicolumn{2}{c}{Gamma/Xray}        &                 \\
Reaction & Half-life& Half-life & Time  &~Gamma/X-ray    & Emiss.Prob. [\%] & Emiss.Prob. [\%] &  Source \\
Product  & (recomm.)&(DD eval.) & unit  &~Energy[keV]    & (recomm.)        &(DD eval.)& Document        \\ \hline
\hline
\endhead
%Whatever you put before this command and below endfirsthead will be displayed at the top of the table in every page except the first one.
\hline \hline
\endfoot
\hline \hline
\endlastfoot
\T $^{18}$F &   109.77(5)    &\a 109.734(14)\b & m & 511.                              &  193.72(38)   &                  & BIPM  \\
            &                &                 &   & 0.525 X$K_\alpha1$                &  0.013(4)     &                  &       \\[0.5mm]
            &                &                 &   & 0.525 X$K_\alpha2$                &  0.007(2)     &                  &       \\[0.5mm]

 $^{22}$Na  &   2.6018(22)   & \a 2.6020\b     & a & 1274.537(7)                       &  99.94(13)    &    99.940(14)    & BIPM  \\
            &                &                 &   &  511.                             & 180.7(2)      &    180.76(4)     &       \\[0.5mm]

 $^{24}$Na  &\r 14.997(12)\b & \a 14.958(2)\b  & h & 1368.630(5)                       &  99.9934(5)   &    99.9936(15)   & BIPM  \\
            &                &                 &   & 2754.049(13)                      &  99.862(3)    &    99.855(5)     &       \\[0.5mm]

 $^{27}$Mg  &  9.458(12)     &                 & m &  843.76(10)                       &  71.800(20)   &    71.8(4)       & ENSDF \\
\end{longtable*}

\clearpage
\subsection{Fission Yield Data} \label{Sec_IV_D}

\begin{figure*}[!bhtp]
\vspace{-3mm}
 \includegraphics[width=0.98\textwidth]{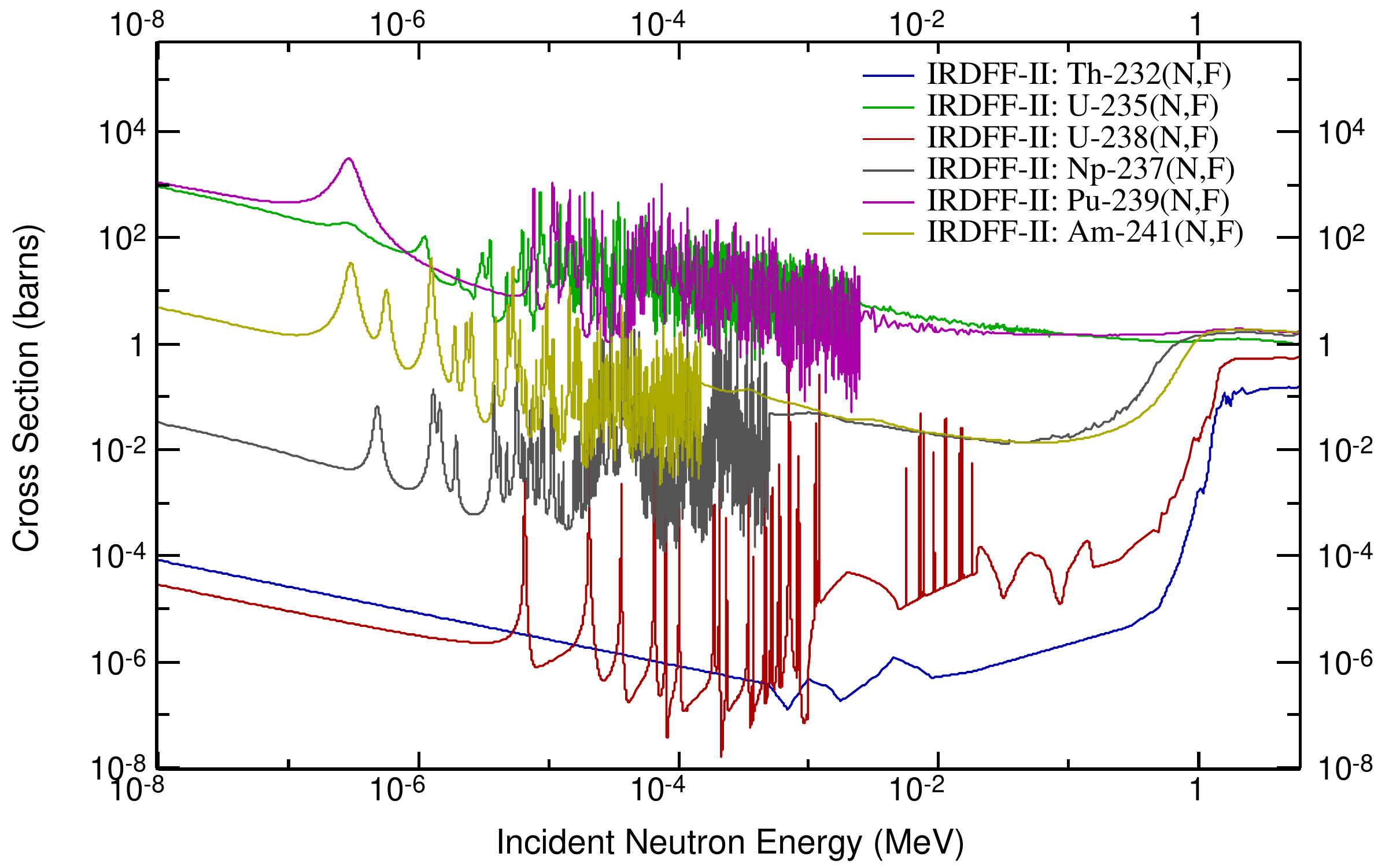}
\vspace{-2mm}
\caption{(Color online) Comparison of the neutron-induced fission cross-section data of the actinides in the \mbox{IRDFF-II} Library.}
\label{Fig:Actin_fis_xs}
\vspace{-2mm}
\end{figure*}

The fission cross sections of the actinides included in the \mbox{IRDFF-II} library differ significantly in the thermal and the epithermal neutron energy range, which determines the relative importance of the fission yield data. Comparison of the fission cross sections is given in Fig.~\ref{Fig:Actin_fis_xs} for $^{232}$Th, $^{235,238}$U, $^{237}$Np, $^{239}$Pu, and $^{241}$Am targets. %Thermal fission yields are listed for fissile isotopes as well as for $^{237}$Np and $^{241}$Am.

The baseline recommended fission yields for dosimetry applications are taken from the \mbox{JEFF-3.3} library. These data are modifications to the \mbox{JEFF-3.1.1} library and are also identified as \mbox{UKFY3.7}~\cite{Kel09}. When the fission yields do not appear in the \mbox{JEFF-3.3} library, \eg, the 14~MeV fission yields for $^{239}$Pu and $^{237}$Np, the recommended values are taken from the \mbox{ENDF/B-VIII.0} library.

While the dosimetry community looks forward to the future use of a more refined energy-dependence to the recommended fission yields, the current \mbox{IRDFF-II} library recommendation is based on a selection of the appropriate yield from one of the three energy-dependent groups: thermal, fast (average energy in the range 400~keV to 500~keV), and 14~MeV.  %

\subsection{Thermal Neutron Cross Sections and Resonance Integrals} \label{Sec_IV_E}
Neutron dosimetry is the inverse of neutron activation analysis (NAA). Independent data sets exist like the Kayzero library for NAA by the k$_0$ standardization method~\cite{Kayzero}, for example. There is a direct relation between the constants in the Kayzero data-base and the differential cross sections~\cite{Trkov:NAA}. Since the constants were measured independently of the differential cross sections, the k$_0$ and Q$_0$ constants contained in the Kayzero library can be used for the validation of the thermal capture cross sections and resonance integrals in the \mbox{IRDFF-II} library, in addition to the comparison with other compilations like the Atlas of Resonance Parameters~\cite{Mug16}. The comparison is discussed and presented in Sec.~\ref{Sec_VI_G}.

\section{CROSS-SECTION COMPARISON BETWEEN AVAILABLE EVALUATIONS FOR SELECTED REACTIONS} \label{Sec_V}

The following Secs. provide a commented review and details on some of the recommended cross sections that appear within the \mbox{IRDFF-II} library. The reactions addressed below focus on the new and updated content of the library. Some of the reaction-specific verification evidence is summarized below and comparisons are made with other evaluations.
Note that all reaction evaluations present in the \mbox{IRDFF-II} library feature both the evaluated mean values and the corresponding covariance matrices. The definition of the covariance matrix requires that it be positive semi-definite. The eigenvalues of all covariance matrices have been checked to be non-negative in the evaluator's energy grids; all calculated eigenvalues are listed in the \mbox{IRDFF-II} files in ENDF-6 format in the comment section (MT=451) for each dosimetry reaction. Users should be aware that numerical processing of covariance matrices may introduce very small negative eigenvalues due to round-off errors; processed covariance matrices require to be checked for negative eigenvalues, and regularized if needed by adding a very small number to the diagonal elements.

Whereas the ordering of the nuclear data reaction content in Table~\ref{Table_Long_I} is with increasing atomic number of the target nuclei, the ordering of Secs. here is different. This new library includes composite reactions and extends earlier evaluation up to 60~MeV. The ordering of Secs. below is designed to most easily illustrate the verification steps taken in the preparation of the library. It starts with a simple high energy reaction, the $^{58}$Ni(n,p)$^{58}$Co reaction, then proceeds to address the related composite reaction, $^{\mbox{nat}}$Ni(n,X)$^{58}$Co, that is now added to the library. The following Secs. then address more complicated high energy composite dosimetry reactions such as $^{\mbox{nat}}$Fe(n,X)$^{54}$Mn, resonance reactions and new reactions that have a reaction threshold above 20~MeV. The survey of reactions then ends with a discussion of the very complex composite reactions, such as gas-production reactions $^{\mbox{nat}}$Li(n,X)$^{4}$He and $^{\mbox{nat}}$B(n,X)$^{4}$He, where there can be multiple target isotopes and multiple contributing reaction channels.

\subsection{Reaction Q-values and Thresholds}
Reaction threshold is directly related to its Q-value. As a rule, exothermic reactions do not have a threshold, but in some cases there is an effective threshold due to the Coulomb barrier in the outgoing charged-particle emission, and the probability of the reaction is very small. However, \ql small\qr is actually case-dependent, since in \ql well-thermalised systems \qr the neutron flux at thermal energies can be many orders of magnitude higher than the fast flux, where the cross section is significant. The special cases in the \mbox{IRDFF-II} library are listed in Table~\ref{Q-values}, together with the Q-values and the adopted effective threshold energies.
\begin{table}[!hbtp]
\vspace{-2mm}
\caption{Reaction Q-values and thresholds} \label{Q-values}
\begin{tabular}{r | l l | c r r} \hline \hline
\T ID  & Reaction                       & Label   &   MAT & Q-Value & Eff. Thr.    \\
    &                                &         &       & [keV]   & [keV]        \\ \hline
\T 1& $^{10}$B(n,$\alpha$)$^{7}$Li   & B10a    & 525   & 2789.323&  0.0         \\
2   & $^{47}$Ti(n,p)$^{47}$Sc        & Ti47p   & 2228  & 181.580 &  800.0       \\
3   & $^{54}$Fe(n,p)$^{54}$Mn        & Fe54p   & 2625  & 85.2127 &  0.0         \\
4   & $^{54}$Fe(n,$\alpha$)$^{51}$Cr & Fe54a   & 2625  & 842.760 &  0.0         \\
5   & $^{58}$Ni(n,p)$^{58}$Co        & Ni58p   & 2825  & 400.800 &  400.0       \\
6   & $^{59}$Co(n,$\alpha$)$^{56}$Mn & Co59a   & 2725  & 328.400 &  334.016     \\
7   & $^{63}$Cu(n,$\alpha$)$^{60}$Co & Cu63a   & 2925  & 1714.400&  2250.0      \\
8   & $^{64}$Zn(n,p)$^{64}$Cu        & Zn64p   & 3025  & 203.500 &  500.0       \\
9   & $^{67}$Zn(n,p)$^{67}$Cu        & Zn67p   & 3034  & 220.690 &  0.0         \\
10  & $^{92}$Mo(n,p)$^{92m}$Nb       & MoNb92pm& 4225  & 425.680 &  1000.0      \\ \hline \hline
\end{tabular}
\vspace{-6mm}
\end{table}

\subsection{Description of Selected Cross-Section Evaluations}

\subsubsection{$^{58}\!$Ni(n,p)$^{58}\!$Co}

The zero to 20~MeV portion of the selected cross section comes from an IAEA-sponsored evaluation performed by K. Zolotarev in December 2002, documented as part of the \mbox{RRDF-2002} library, and released within the IAEA-generated \mbox{IRDF-2002} library. The IAEA sponsored the extension of this library from 20~MeV to 60~MeV, which was incorporated into the evaluation in February 2014. This extension used the model-based sub-library \mbox{TENDL-2013}(s60) of the \mbox{TENDL-2013} library with explicit representation of reaction cross sections up to 60~MeV, while renormalizing (labelled in this paper as (r)) the calculations in the energy range 20~MeV to 60~MeV to ensure continuity of the data at 20 MeV. The associated \mbox{IRDFF-II} covariance matrices correctly reflect the derivation of the evaluation. These development details can be seen in the entry in Table~\ref{Table_Long_I} for this reaction.

\begin{figure}[!thb]
\vspace{-2mm}
\subfigure[~Comparison to selected experimental data from EXFOR \cite{EXF08}.]
{\includegraphics[width=\columnwidth]{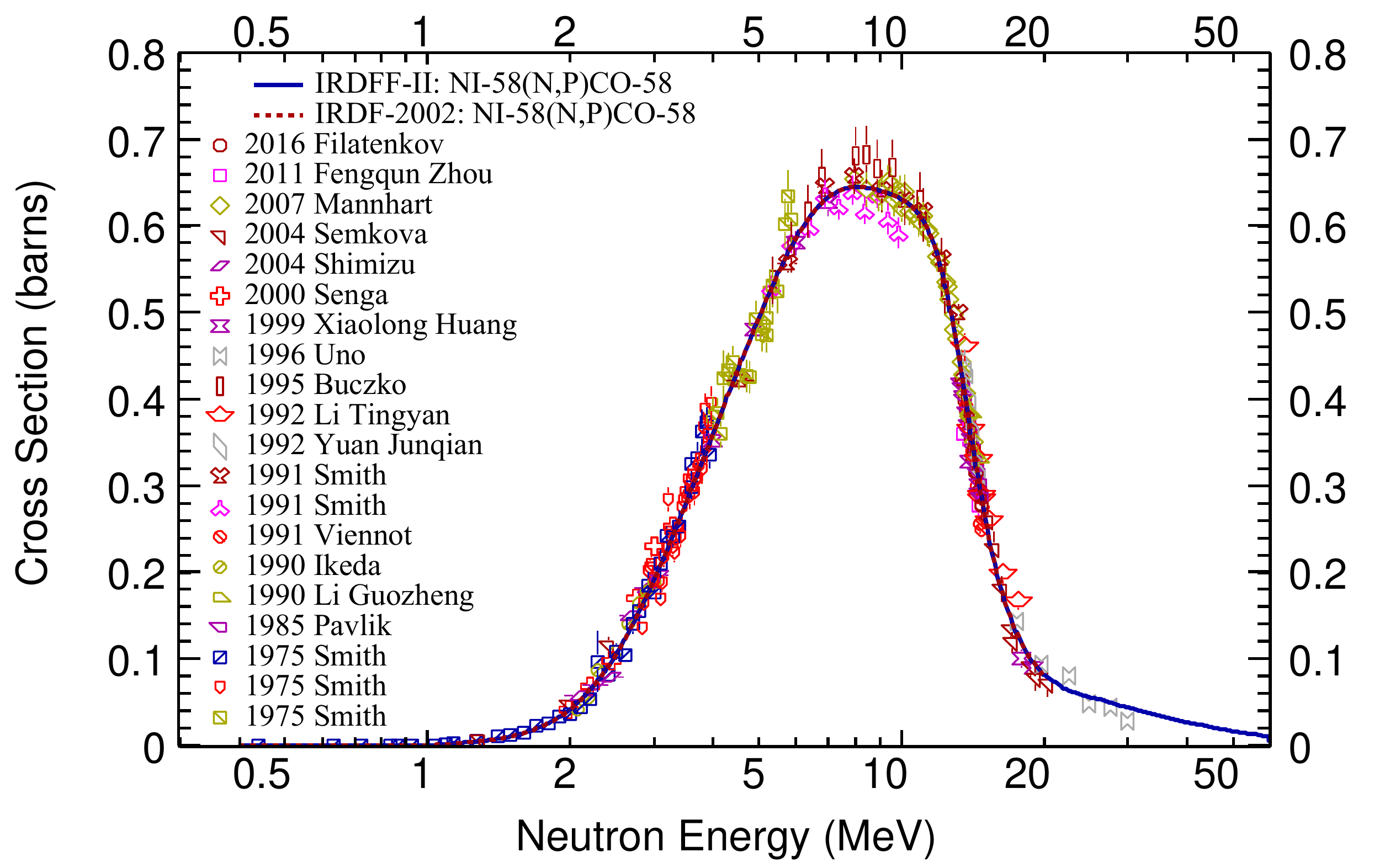}}
\subfigure[~Ratio of cross-section evaluations to \mbox{IRDFF-II} evaluation.]
{\includegraphics[width=\columnwidth]{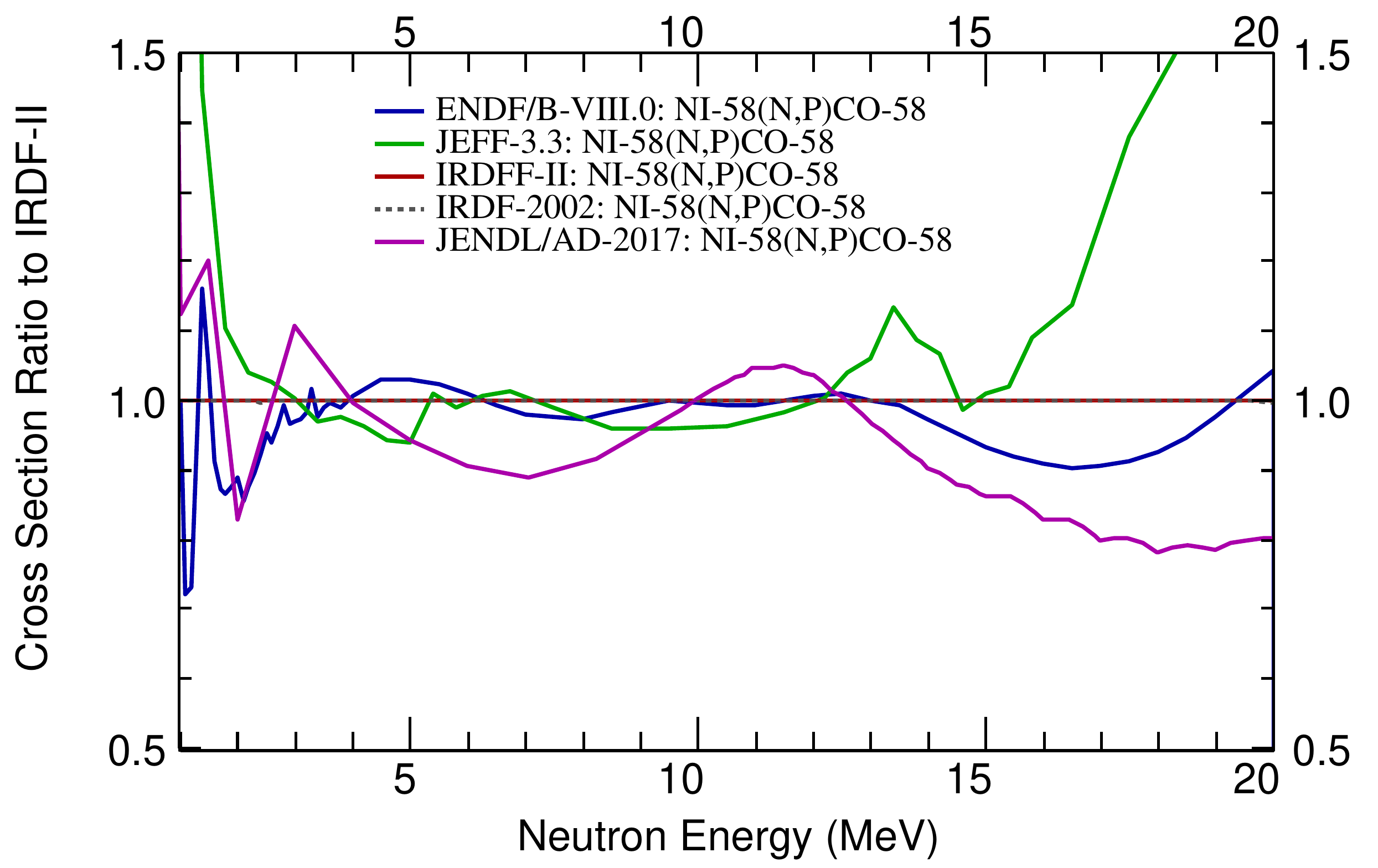}}
\vspace{-3mm}
\caption{(Color online) $^{58}$Ni(n,p)$^{58}$Co cross-section evaluation in \mbox{IRDFF-II} library relative to other data.}
\label{Fig_VI_A_12}
\vspace{-3mm}
\end{figure}
As noted in Sec.~\ref{Sec_I_A}, it is critical that recommended cross sections for dosimetry applications have evidence that the evaluation took into consideration the available body of experimental data. No experimental data exists for neutron energies greater than 30~MeV. Fig.~\ref{Fig_VI_A_12}(a) shows good agreement of the selected EXFOR data with the \mbox{IRDFF-II} cross sections. This comparison provides strong verification evidence for the consistency of the evaluation with measured differential data.

Fig.~\ref{Fig_VI_A_12}(a) also shows that the cross sections for this reaction are practically equal to those in the older \mbox{IRDF-2002}. For most fission reactor applications, the relevant 10~\%-90~\% sensitivity region for this reaction is between 2~MeV and 7~MeV. In order to also illustrate the variations between evaluations within this region, Fig.~\ref{Fig_VI_A_12}(b) shows the ratio of the various evaluations to the \mbox{IRDFF-II} recommended cross section, namely \mbox{ENDF/B-VIII.0}, \mbox{JEFF-3.3}, and \mbox{JENDL/AD-2017}. Variations on the order of 10~\% are observed. The curve for the \mbox{IRDFF-II} ratio is unity by definition in this figure. The \mbox{JEFF-3.3} evaluation shows significant differences near threshold and above 15~MeV.

\begin{figure}[!thb]
\vspace{-4mm}
\subfigure[~Cross-section correlation matrix.]
{\includegraphics[width=\columnwidth]{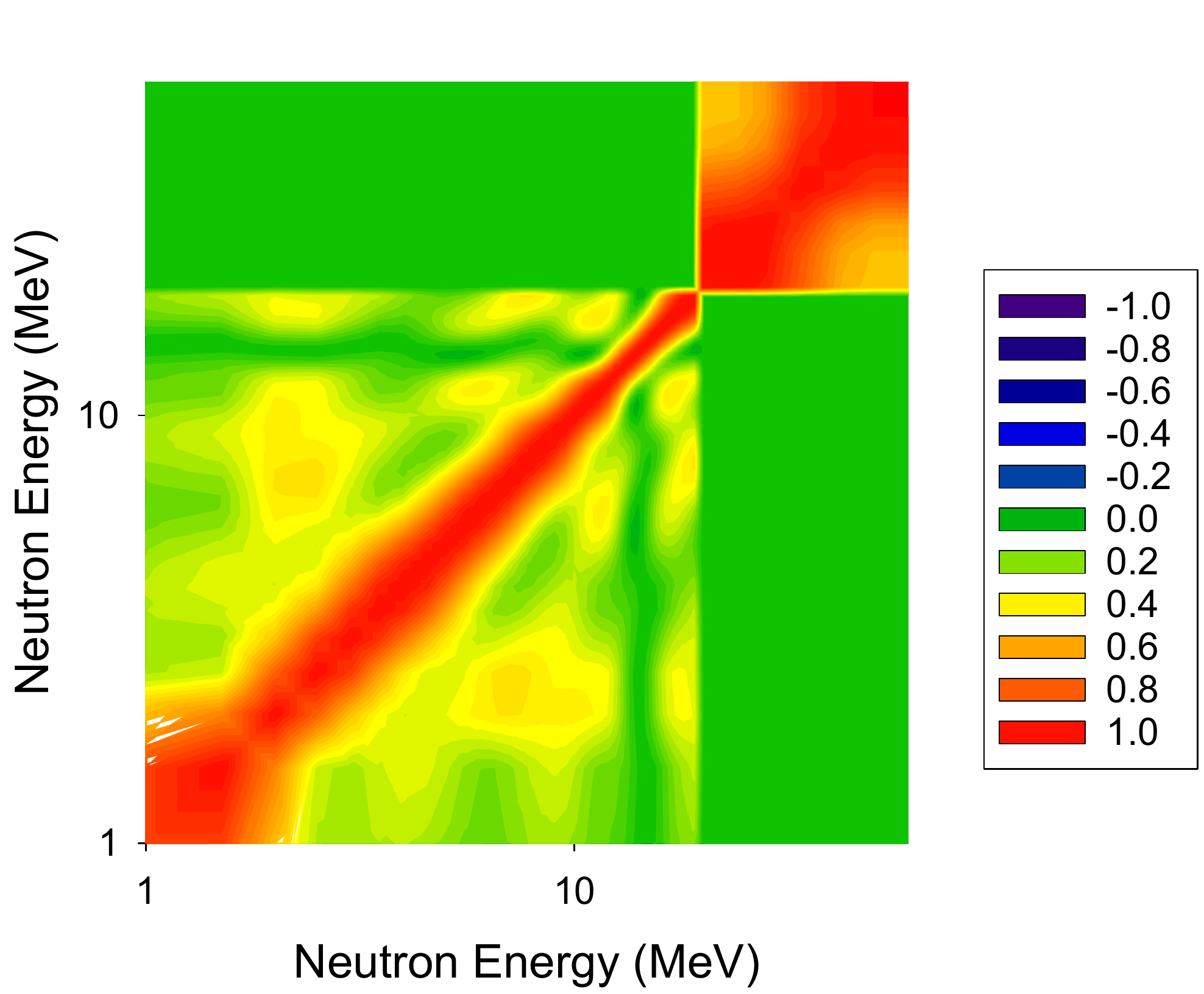}}
\subfigure[~One-sigma uncertainties in \% ($\equiv 100\times \frac{\sqrt{cov(i,i)}}{\mu_i}$), being $\mu_i$ the corresponding cross-section mean value.]
{\includegraphics[width=\columnwidth]{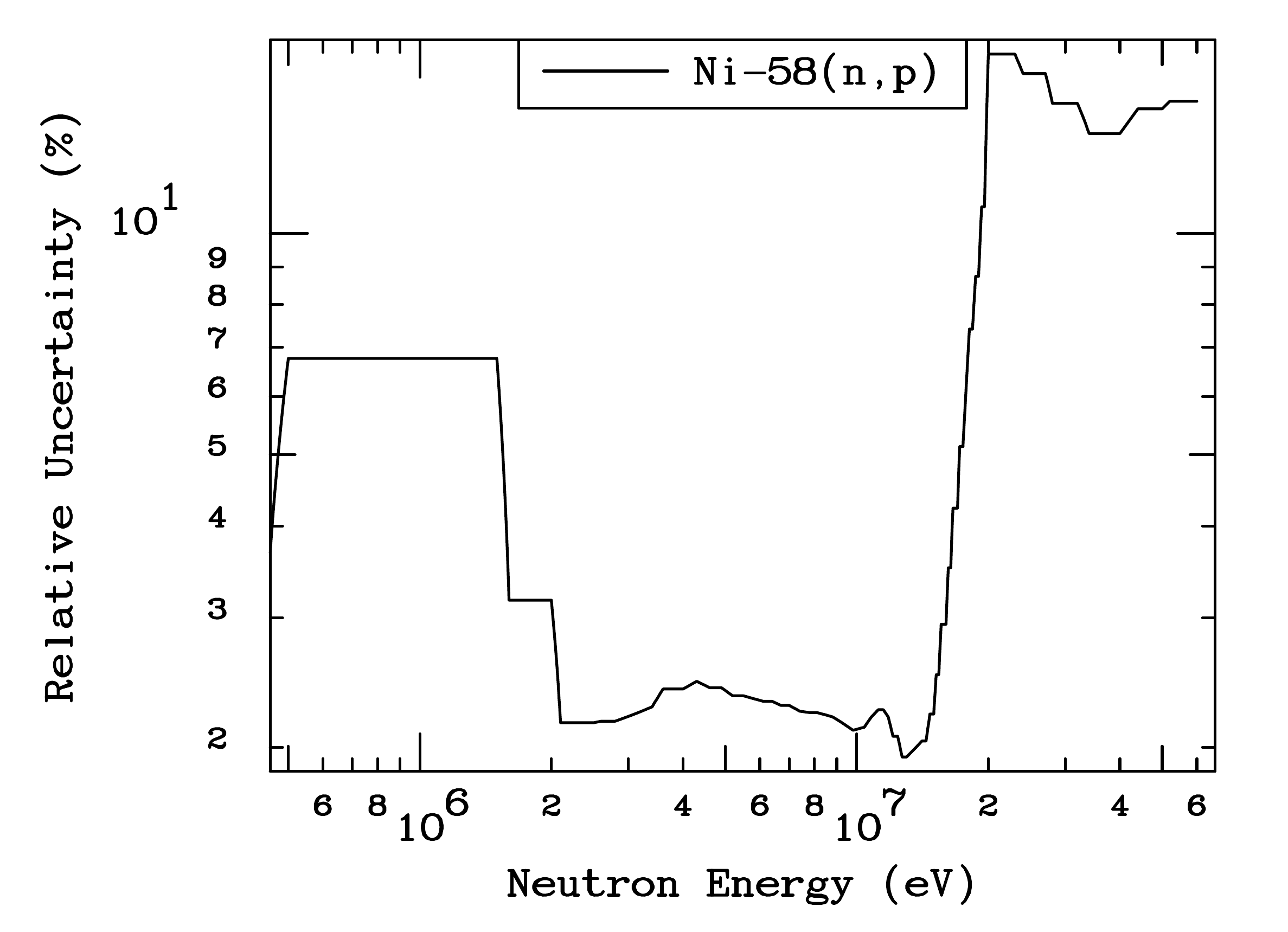}}
\vspace{-3mm}
\caption{(Color online) Uncertainties and correlations of the $^{58}$Ni(n,p)$^{58}$Co cross-section evaluation in \mbox{IRDFF-II} library.}
\label{Fig_VI_A_34}
\vspace{-3mm}
\end{figure}
The inclusion of uncertainty data, in the form of a covariance matrix, is a required feature of dosimetry cross sections. Fig.~\ref{Fig_VI_A_34}(b) shows the energy-dependent 1-sigma uncertainty in the cross section using the energy group structure provided by the evaluator. The energy structure used in the uncertainty characterization is seen to only have a single large-width bin in the 0.4~MeV to 1.5~MeV region, so, due to the correlation of the cross section within this region, the energy-grouped uncertainty in Fig.~\ref{Fig_VI_A_34}(b) does not exhibit the large uncertainty suggested by the variation between different evaluations in Fig.~\ref{Fig_VI_A_12}(b). Fig.~\ref{Fig_VI_A_34}(a) shows the correlation matrix for the evaluation. The covariance matrix elements in the $>$20~MeV region are seen to be uncorrelated with those at lower energies -– as expected based on the evaluation methodology, since the two regions originate from different evaluations.

As discussed above, the definition of the covariance matrix requires that it be positive semi-definite. In order to verify this mathematical property, the smallest and the largest eigenvalues of the covariance matrix are shown in Table~\ref{Table_VI_A_1} for this reaction; all eigenvalues are positive numbers.
\begin{table}
\vspace{-2mm}
\caption{List of lowest eigenvalues in ascending order of the $^{58}$Ni(n,p)$^{58}$Co covariance matrix.} \label{Table_VI_A_1}
\begin{tabular}{r r r}
\hline \hline
    No. & Row   & Eigenvalue  \\ \hline
 \T  1  &   1   & 0.00000E+00 \\
     2  &  11   & 2.98127E-08 \\
     3  &  17   & 3.02008E-08 \\
     4  &  24   & 3.03488E-08 \\
     5  &  20   & 3.17274E-08 \\
     6  &   8   & 3.23051E-08 \\
     7  &  21   & 3.27524E-08 \\
     8  &  22   & 3.37126E-08 \\
     9  &  12   & 3.53861E-08 \\
    10  &  14   & 3.71190E-08 \\
  ...   &       &             \\
    44  &  39   & 1.16138E-01 \\ \hline\hline
\end{tabular}
\vspace{-3mm}
\end{table}

Another verification step is the comparison of the calculated-to-experimental ratio of the spectrum-averaged cross section (SACS) in the $^{252}$Cf spontaneous fission and $^{235}$U(n$_{th}$,f) prompt fission neutron benchmark fields. We emphasize that this, within the context of the evaluation, is treated purely as a verification step since the nuclear data evaluator also commonly examines this metric as part of their verification process for the evaluation. The treatment of validation in these benchmark fields is addressed later in Sec.~\ref{Sec_VII}. Table~\ref{Table_VI_A_2} compares the spectrum-averaged cross sections produced with the \mbox{IRDFF-II} library to that reported by the nuclear data evaluator in the evaluation documentation.

\begin{table}
\caption{C/E of SACS for $^{58}$Ni(n,p)$^{58}$Co reaction and data used by the nuclear data evaluator.} \label{Table_VI_A_2}
\begin{tabular}{r | r c l | r c}
\hline \hline
Spectrum      & \multicolumn{5}{c}{Experiment}                     \mbox{IRDFF-II}        \\
              & \multicolumn{4}{c}{Cross Section}   Source       & Calculated x.s. \\
              & \multicolumn{4}{c}{(mb)}                         &  (mb)           \\ \hline
\T $^{235}$U(n$_{th}$,f) &  108.9 & $\pm$ & 4.78~\%   & \cite{Man85}  & 107.44          \\
                         &  106.0 & $\pm$ & 6.60~\%   & \cite{Hor89}  &                 \\
                         &  108.5 & $\pm$ & 1.29~\%   & \cite{Man99}  &                 \\
$^{252}$Cf(s.f.)         &  117.6 & $\pm$ & 1.28~\%   & \cite{Man87a} & 117.36          \\
                         &  117.5 & $\pm$ & 1.30~\%   & \cite{Man02}  &                 \\ \hline\hline
\end{tabular}
\vspace{-3mm}
\end{table}

The residual nucleus produced in this reaction is $^{58}$Co – which, in addition to its 70.86~day half-life for the ground state, also has a 9.04~hour half-life isomer. It is critical to note that, consistent with adopted notation, this evaluation captures the complete $^{58}$Ni(n,p)$^{58(m+g)}$Co cross section, that is, it represents the combined (n,p) reaction that populates the $^{58}$Co ground state and the meta-stable states. An important consideration in the use of this reaction for dosimetry in commercial fission reactors is the consideration of the thermal neutron burn-up correction that can be a factor in long irradiations in the presence of a thermal neutron environment. The effective burn-up cross section, via the $^{58}$Co(n,$\gamma$) reaction, is 1650~barns for $^{58g}$Co while that for the $^{58m}$Co residual nucleus is 1.4x10$^{5}$ barns. The caveats on the need to make thermal neutron burn-up corrections for applications in light water reactors are found in ASTM E264~\cite{A264}. The cross sections and the decay data of the metastable isotope $^{58m}$Co are not included in the present version of \mbox{IRDFF-II}.

\begin{figure}[!hbtp]
\vspace{-2mm}
\includegraphics[width=\columnwidth]{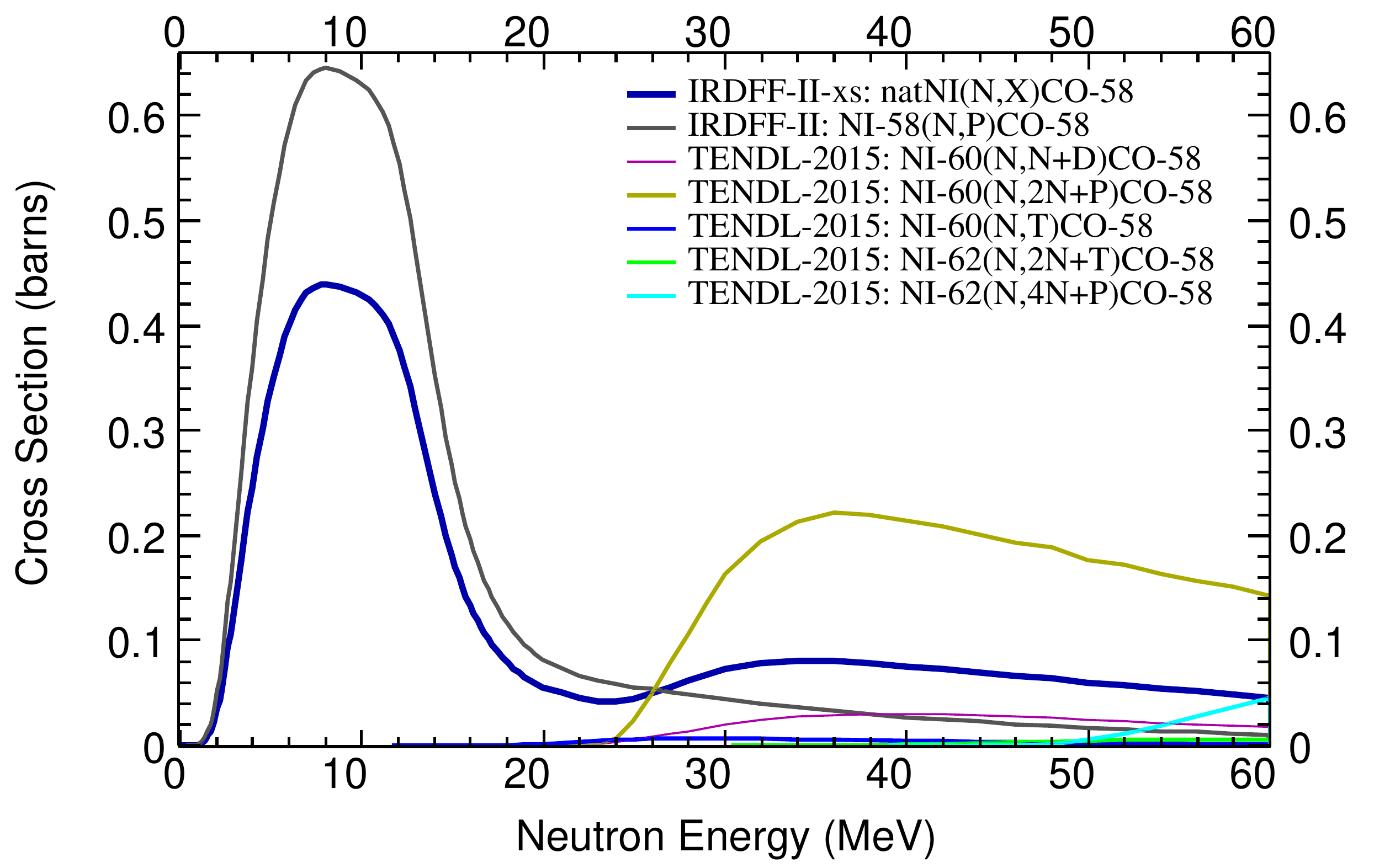}
\vspace{-2mm}
\caption{(Color online) Contributing reaction cross sections to the $^{58}$Co production from neutrons incident on a $^{\mbox{nat}}$Ni target. Note that the cross sections are not scaled by the isotopic abundance.}
\label{xNi-Co58-major-isot}
\vspace{-2mm}
\end{figure}

\subsubsection{$^{\mbox{nat}}$Ni(n,X)$^{58}\!$Co}
Nickel is an element that has several naturally occurring isotopes and many reactions on different isotopes may lead to the same residual. Monitor samples are usually made of natural elements, so it is advantageous to provide the data for the residual production from the element, since production paths may include a multitude of minor channels which a user would need to consider. The cross sections and the covariance data for the $^{\mbox{nat}}$Ni(n,X)$^{58}$Co reaction were assembled from the \mbox{IRDFF-II} evaluation for the $^{58}$Ni(n,p) reaction (described in the previous section), which is the dominant contribution at low energies. The contributing reactions at higher energies were taken from the \mbox{TENDL-2015} library, as shown in Fig.~\ref{xNi-Co58-major-isot}.

\begin{figure}[!hbtp]
\vspace{-2mm}
\includegraphics[width=\columnwidth]{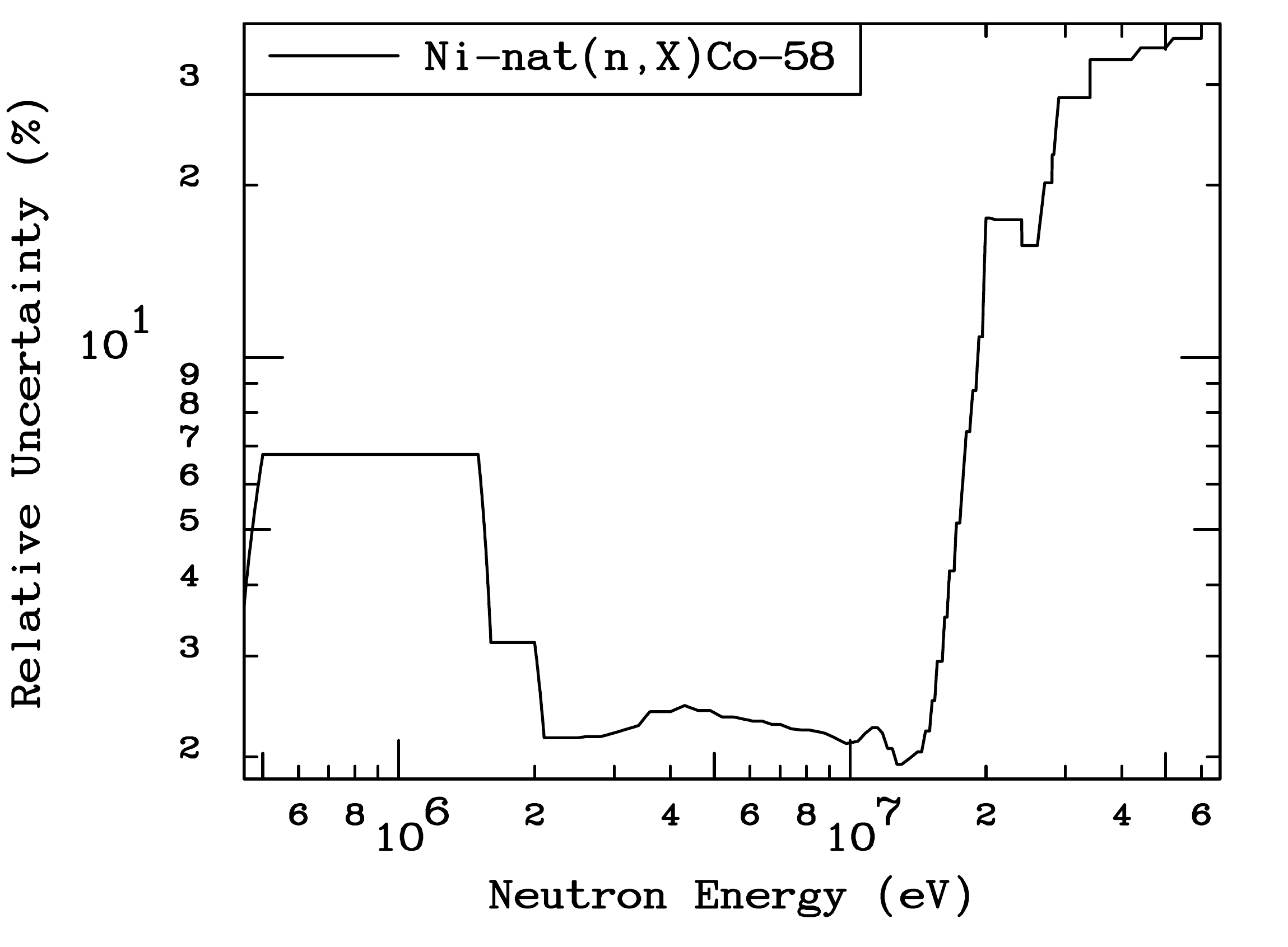}
\caption{(Color online) Uncertainties in the $^{\mbox{nat}}$Ni(n,X)$^{58}$Co reaction cross sections.}
\label{Ni-nat_IRDFF_Co58_stddev_format}
\vspace{-2mm}
\end{figure}
The covariance plot and the uncertainty are practically the same as for the $^{58}$Ni(n,p)$^{58}$Co reaction except for the high energy part.
The 1-sigma standard deviation is shown in Fig.~\ref{Ni-nat_IRDFF_Co58_stddev_format}.
%The plot of the correlation matrix for the $^{\mbox{nat}}$Ni(n,X)$^{58}$Co reaction is shown in Fig.~\ref{Fig:Ni(n,X)Co58_cov}. The plot of the 1-sigma uncertainty is shown in Fig.~\ref{Fig:Ni(n,X)Co58_unc}.
%
%\begin{figure}[hbtp]
%  \vspace{5cm} \r Covariance plot to be included \b
%% \includegraphics[width=\columnwidth]{Figures/Nip_IRDFF-II_cor_lsl_gen_log.pdf}
%\caption{Correlation matrix of the $^{\mbox{nat}}$Ni(n,X)$^{58}$Co reaction cross sections.}
%\label{Fig:Ni(n,X)Co58_cov}
%\end{figure}
%
\subsubsection{$^{60}\!$Ni(n,p)$^{60}\!$Co}
\begin{figure}[!thb]
\vspace{-2mm}
\subfigure[~Comparison to selected experimental data from EXFOR \cite{EXF08}.]
{\includegraphics[width=1.02\columnwidth]{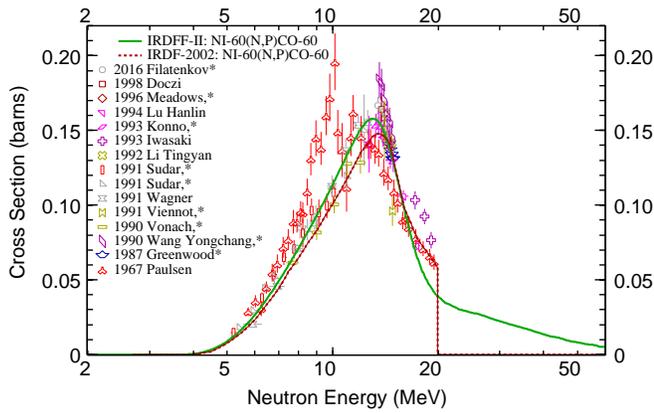}}
\subfigure[~Ratio of cross-section evaluations to \mbox{IRDFF-II} evaluation.]
{\includegraphics[width=\columnwidth]{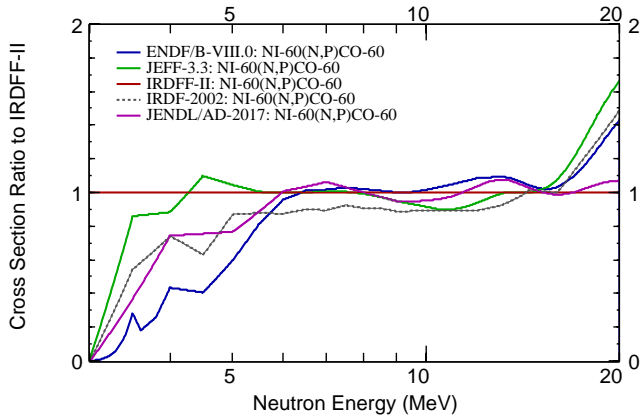}}
\vspace{-3mm}
\caption{(Color online) $^{60}$Ni(n,p)$^{60}$Co cross-section evaluation in \mbox{IRDFF-II} library relative to other data.}
\label{pNi60}
\vspace{-3mm}
\end{figure}

The cross sections and the covariance data for the $^{60}$Ni(n,p)$^{60}$Co reaction in the energy range from threshold to 21~MeV were evaluated by Zolotarev~\cite{Zol08} in 2008. Extension to 60~MeV was done at the IAEA based on TENDL-2011, renormalizing the cross sections at 21~MeV for continuity. Comparison of the cross sections with experimental data is shown in Fig.~\ref{pNi60}(a). The figure also shows the comparison between IRDF-2002 and IRDFF-II, which remained unchanged from version IRDFF-v1.05.

\begin{figure}[!thb]
\vspace{-4mm}
\subfigure[~Cross-section correlation matrix.]
{\includegraphics[width=\columnwidth]{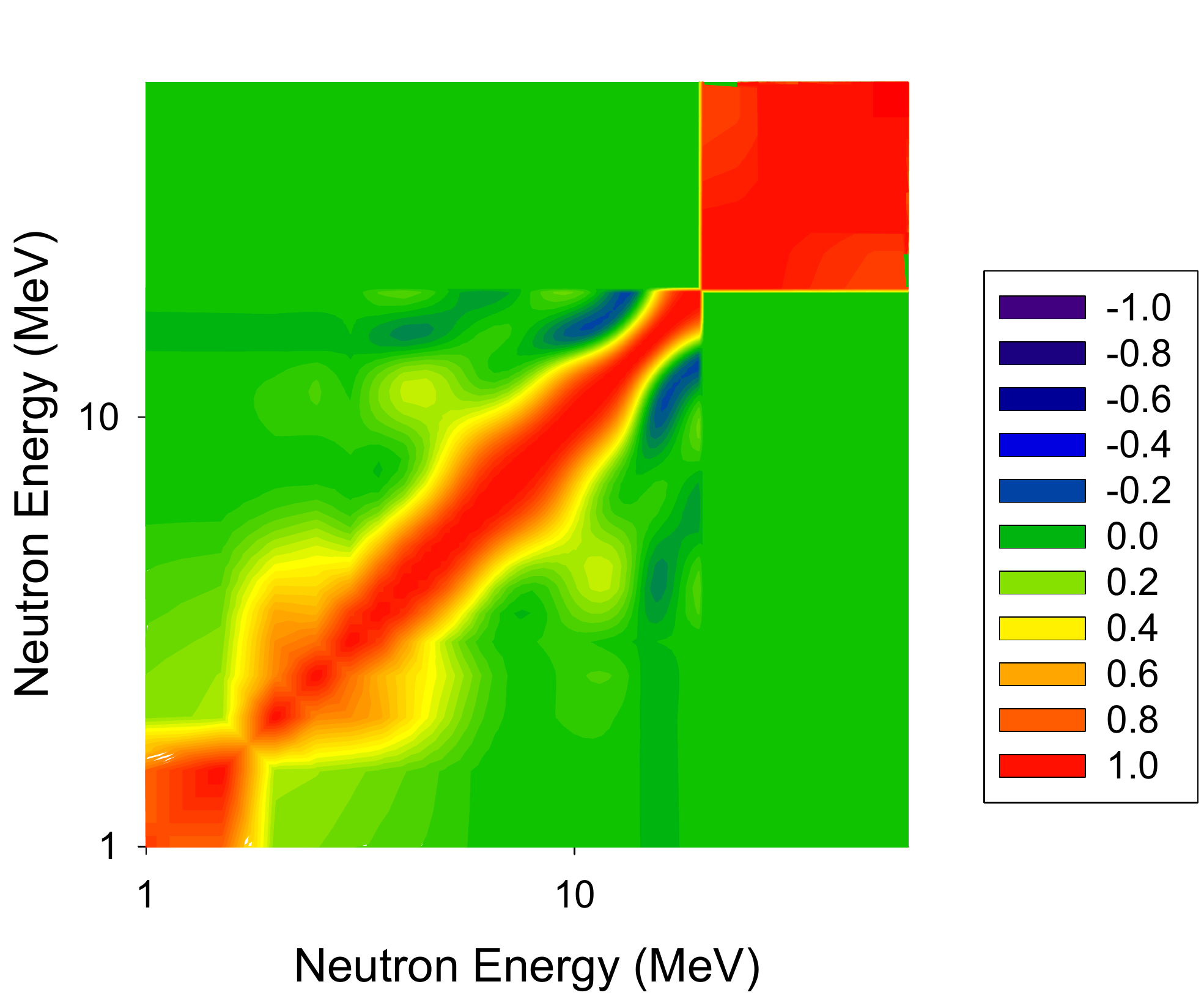}}
\subfigure[~One-sigma uncertainties ($\equiv \sqrt{cov(i,i)}$).]
{\includegraphics[width=\columnwidth]{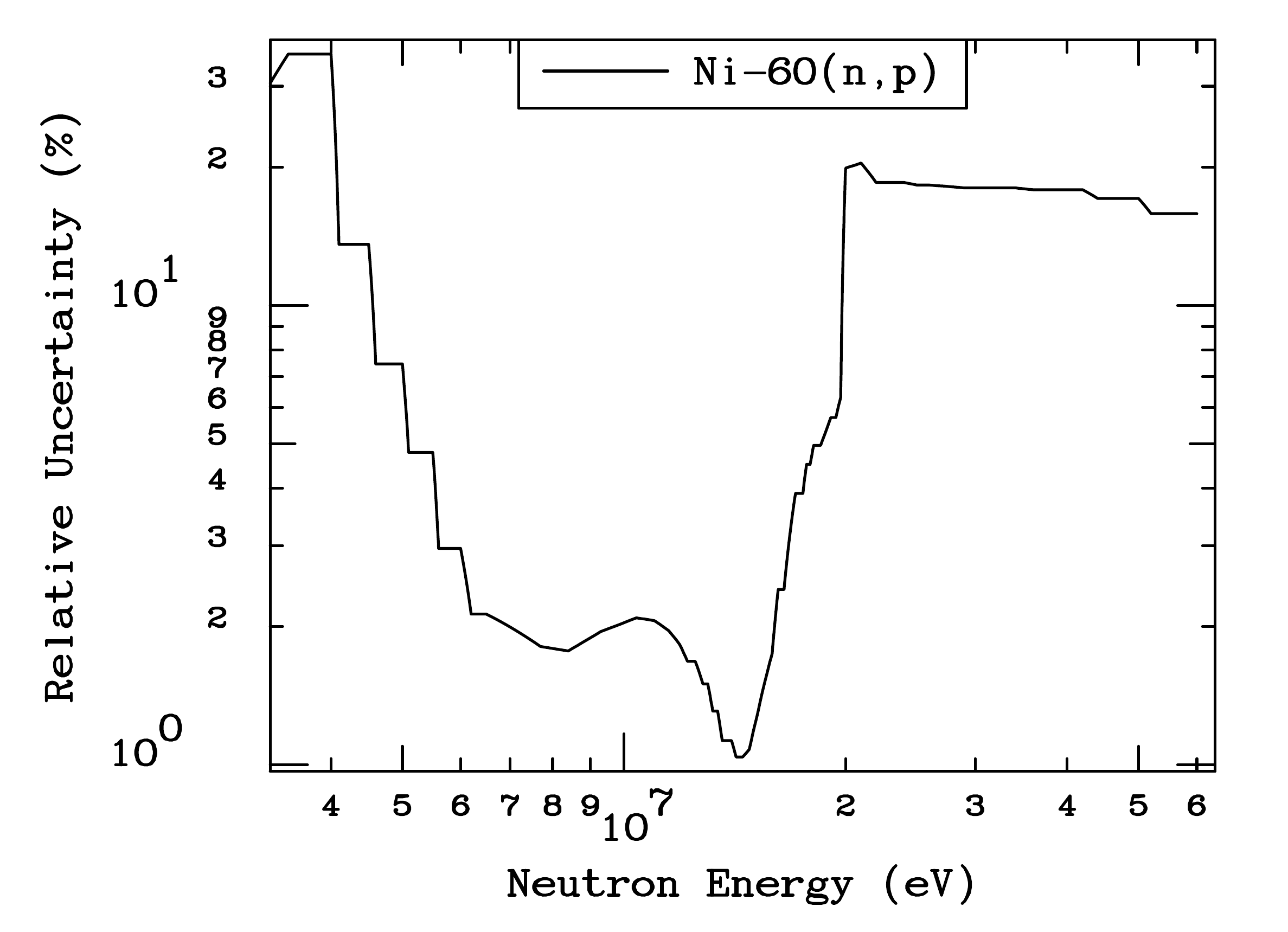}}
\vspace{-3mm}
\caption{(Color online) Uncertainties and correlations of the $^{60}$Ni(n,p)$^{60}$Co cross-section evaluation in \mbox{IRDFF-II} library.}
\label{pNi60-unc}
\vspace{-3mm}
\end{figure}

Cross sections in different evaluated libraries differ significantly near threshold as shown in Fig.~\ref{pNi60}(b), where the ratios of the various evaluations to the \mbox{IRDFF-II} recommended cross section are given. These differences may be important if this reaction monitor is used in reactor spectra. It is interesting that the cross section curves differ also at the plateau of the cross section near 13~MeV. The plot of the 1-sigma uncertainty is shown in Fig.~\ref{pNi60-unc}(b), and the corresponding correlation matrix for the $^{60}$Ni(n,p)$^{60}$Co reaction is shown in Fig.~\ref{pNi60-unc}(a).

\begin{figure}[!hbtp]
\vspace{-2mm}
 \includegraphics[width=0.99\columnwidth]{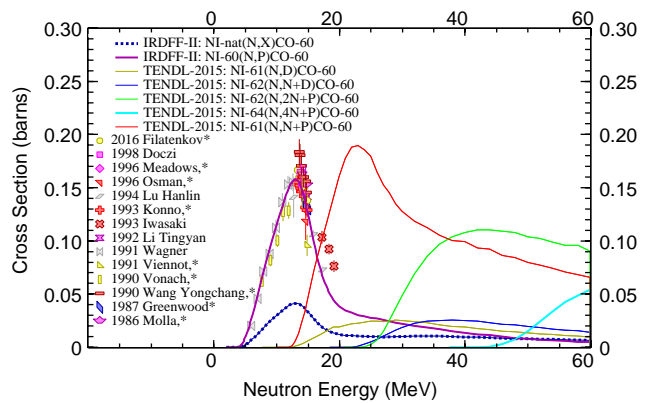}
\vspace{-4mm}
\caption{(Color online) Contributing reactions to the $^{60}$Co production from neutrons incident on a $^{\mbox{nat}}$Ni target. Note that the cross sections are not scaled by the isotopic abundance.}
% \r If included, this plot needs to be re-drawn \b}
\label{Fig:Ni(n,X)Co60}
\end{figure}

The $^{60}$Ni(n,p)$^{60}$Co cross sections are traditionally included in dosimetry libraries, but the users should note that several neutron induced reactions from the heavier nickel isotopes lead to the same product, as shown in Fig.~\ref{Fig:Ni(n,X)Co60}. Although the abundance of $^{61}$Ni is about twenty times smaller than the abundance of $^{60}$Ni, the latter may start contributing to the  $^{60}$Co production at energies above 12~MeV through the $^{61}$Ni(n,np)$^{60}$Co reaction. The user is advised to use the $^{\mbox{nat}}$Ni(n,X)$^{60}$Co cross sections that are newly introduced into the \mbox{IRDFF-II} library, noting that the cross sections are not validated for dosimetry purposes in neutron fields with a significant contribution of the flux above 12~MeV.

\clearpage
\subsubsection{$^{54}\!$Fe(n,p)$^{54}\!$Mn}
\begin{figure}[!thb]
\vspace{-2mm}
\subfigure[~Comparison to selected experimental data from EXFOR \cite{EXF08}.]
{\includegraphics[width=0.99\columnwidth]{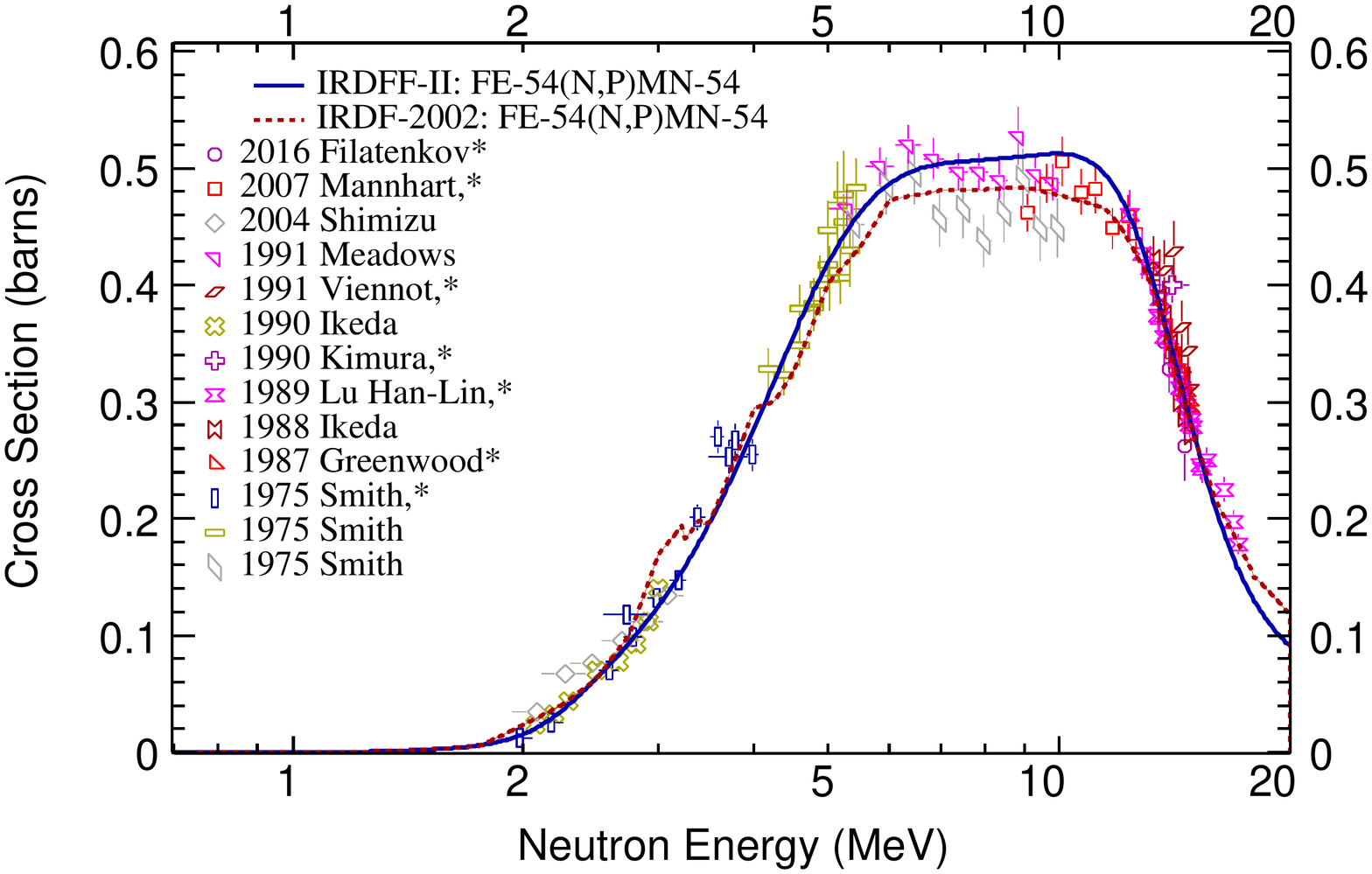}}
\subfigure[~Ratio of cross-section evaluations to \mbox{IRDFF-II} evaluation.]
{\includegraphics[width=0.99\columnwidth]{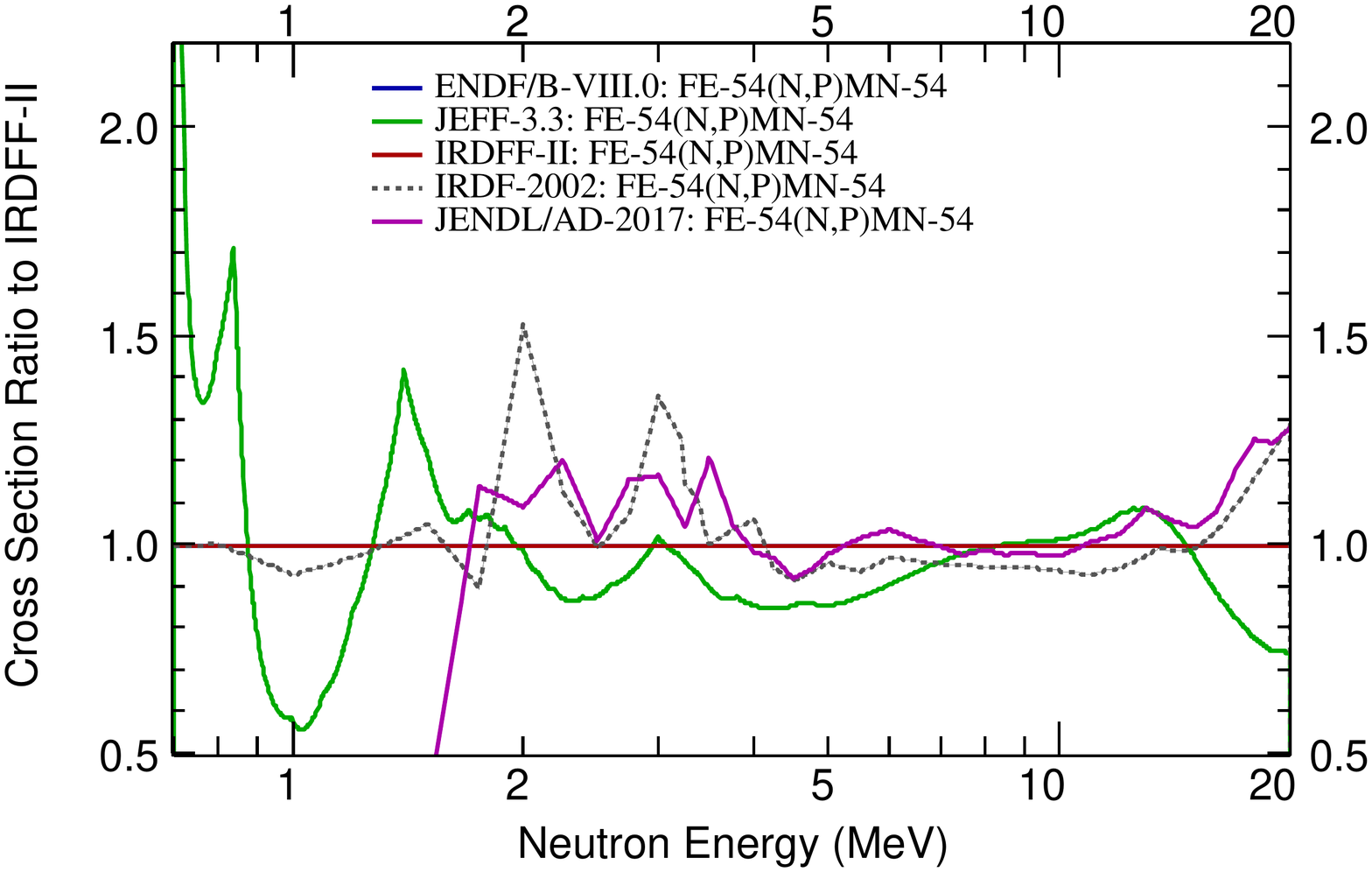}}
\vspace{-3mm}
\caption{(Color online) $^{54}$Fe(n,p)$^{54}$Mn cross-section evaluation in \mbox{IRDFF-II} library relative to other data.}
\label{pFe54}
\vspace{-3mm}
\end{figure}

The $^{54}$Fe(n,p)$^{54}$Mn reaction cross sections and covariances  below 20~MeV were evaluated by Zolotarev~\cite{Zol13} in 2013. Extension to 60~MeV was done at the IAEA based on TENDL-2013, renormalizing the cross sections at 20~MeV for continuity. Comparison of the cross sections with experimental data is shown in Fig.~\ref{pFe54}(a). The figure also shows the comparison between IRDF-2002 and IRDFF-II, which remained unchanged from version IRDFF-v1.05. Cross sections in different evaluated libraries differ significantly near threshold as shown in Fig.~\ref{pFe54}(b), where the ratios of the various evaluations to the \mbox{IRDFF-II} recommended cross section are given. These differences may be important if this reaction monitor is used in reactor spectra.

\begin{figure}[t]
\vspace{-4mm}
\subfigure[~Cross-section correlation matrix.]
{\includegraphics[width=0.99\columnwidth]{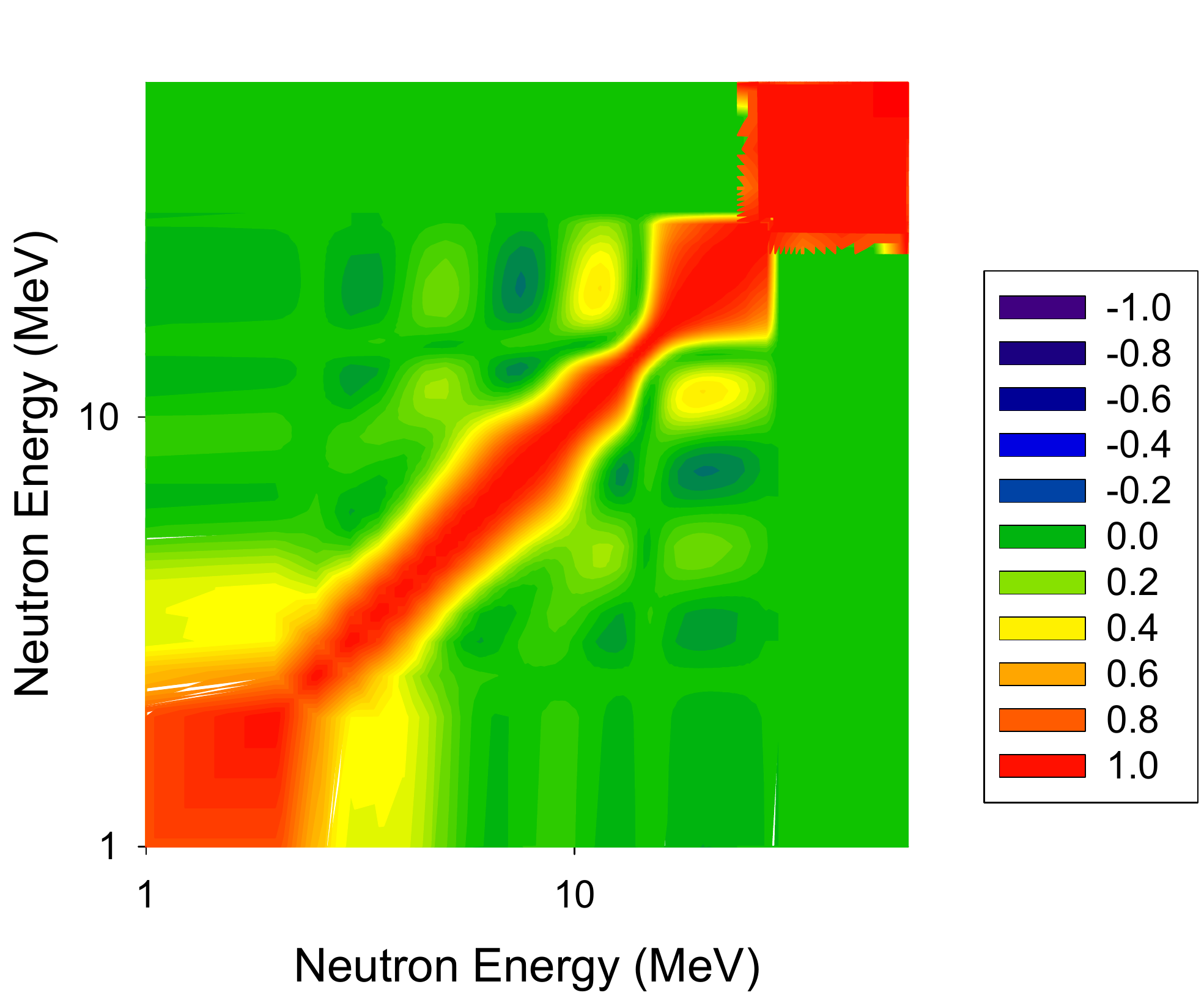}}
\subfigure[~One-sigma uncertainties in \% ($\equiv 100\times \frac{\sqrt{cov(i,i)}}{\mu_i}$), being $\mu_i$ the corresponding cross-section mean value.]
{\includegraphics[width=0.99\columnwidth]{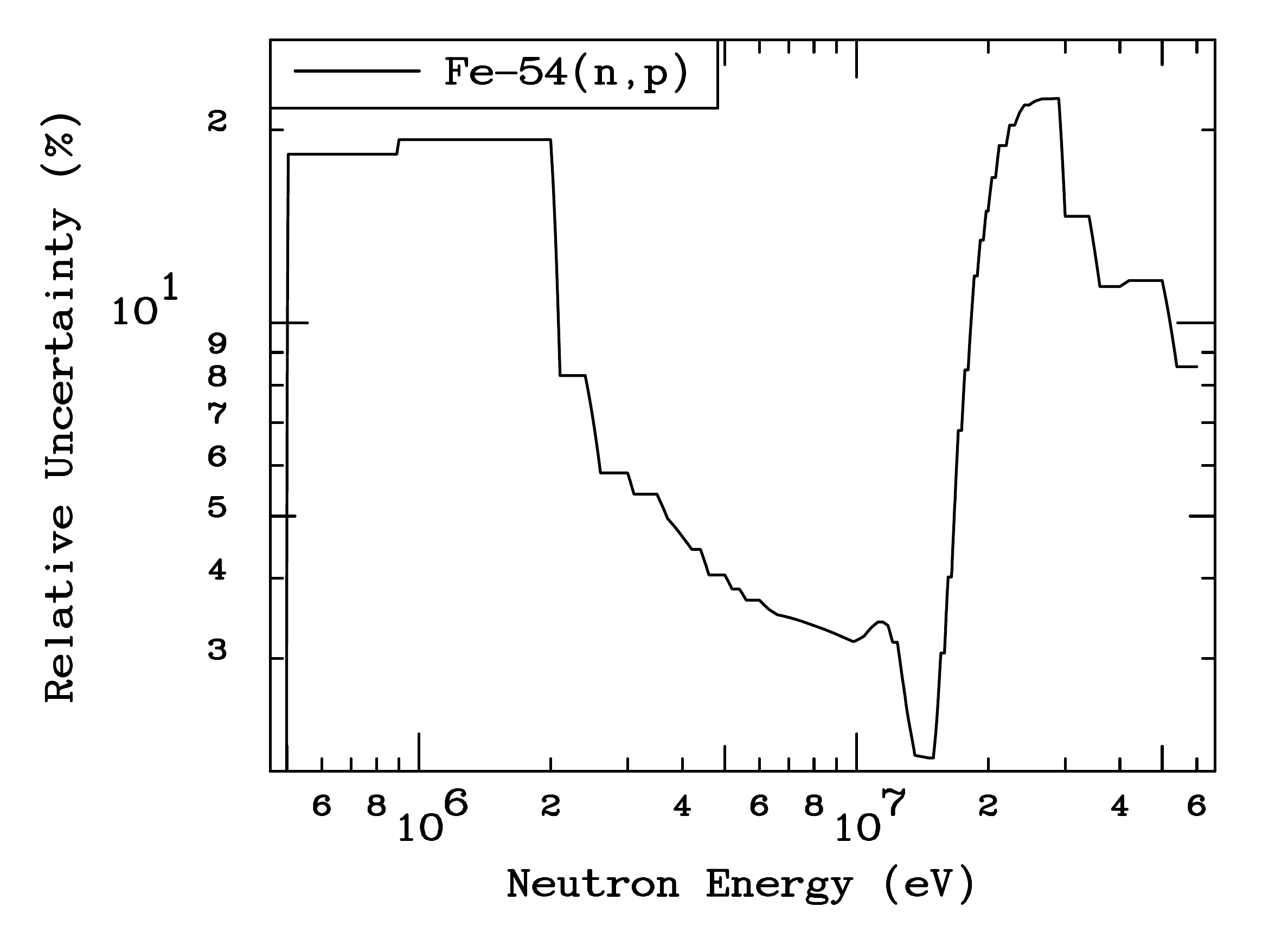}}
\vspace{-3mm}
\caption{(Color online) Uncertainties and correlations of the $^{54}$Fe(n,p)$^{54}$Mn cross-section evaluation in \mbox{IRDFF-II} library.}
\label{pFe54-unc}
\vspace{-3mm}
\end{figure}

The plot of the correlation matrix for the $^{54}$Fe(n,X)$^{54}$Mn reaction is shown in Fig.~\ref{pFe54-unc}(a). The plot of the 1-sigma uncertainty is shown in Fig.~\ref{pFe54-unc}(b).
\begin{figure}[bth]
 \includegraphics[width=\columnwidth]{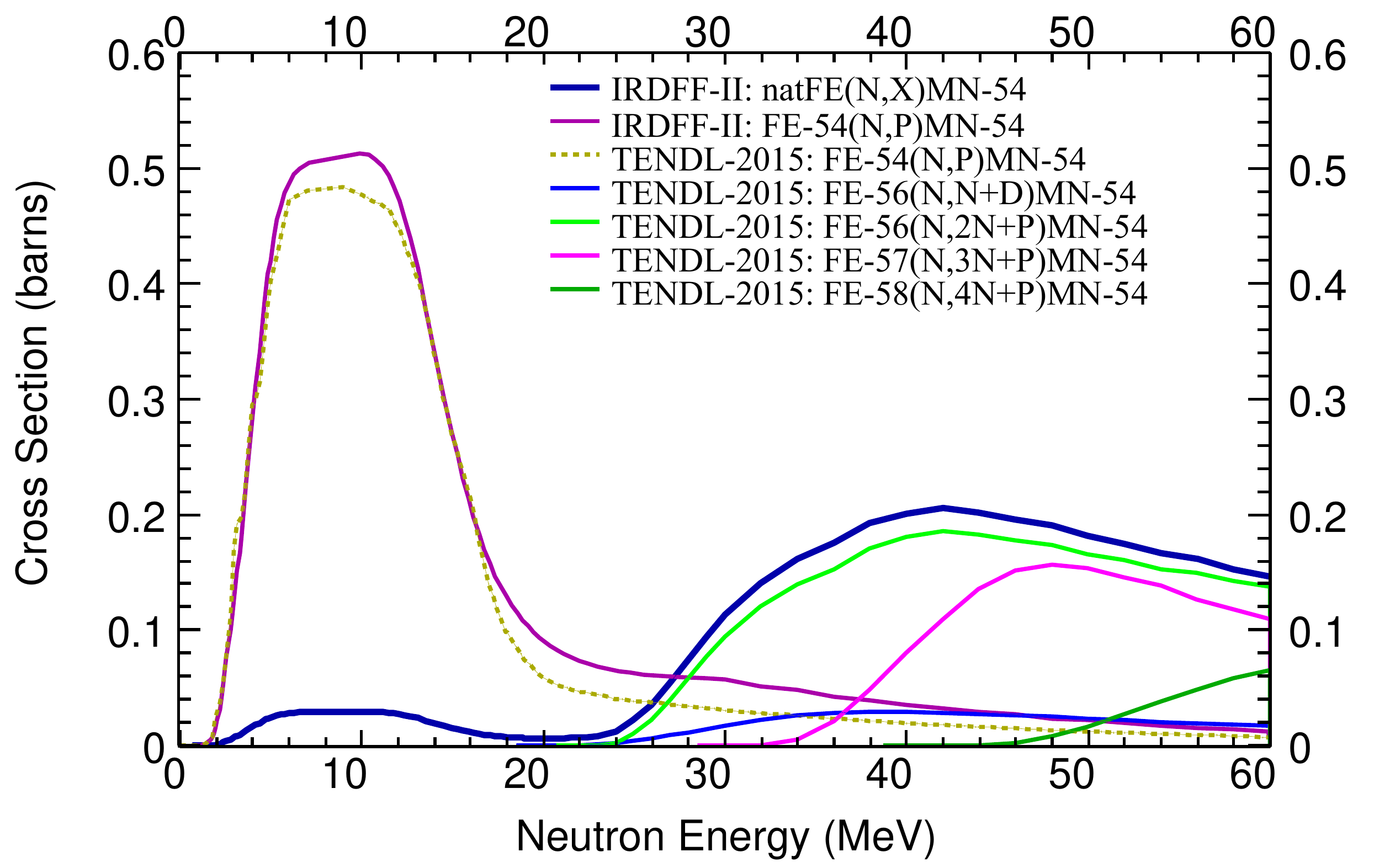}
\caption{Contributing reaction cross sections to the $^{54}$Mn production from neutrons incident on a $^{\mbox{nat}}$Fe target. Note that the cross sections are not scaled by the isotopic abundance.}
\label{xFe-Mn54}
\end{figure}

As in the case of nickel, the $^{54}$Fe(n,p)$^{54}$Mn reaction cross sections are traditionally included in dosimetry libraries, but the users should note that several neutron induced reactions from the other isotopes lead to the same product. The thresholds for the competing reactions are fairly high and they become significant only well above 20~MeV. Nevertheless, the user is advised to use the $^{\mbox{nat}}$Fe(n,X)$^{54}$Mn cross sections that are newly introduced into the \mbox{IRDFF-II} library and are described below, noting that the cross sections are not validated for dosimetry purposes in neutron fields with a significant contribution of the flux above 20~MeV.

\begin{figure}[tbh]
\vspace{-2mm}
\subfigure[~Cross-section correlation matrix.]
{\includegraphics[width=\columnwidth]{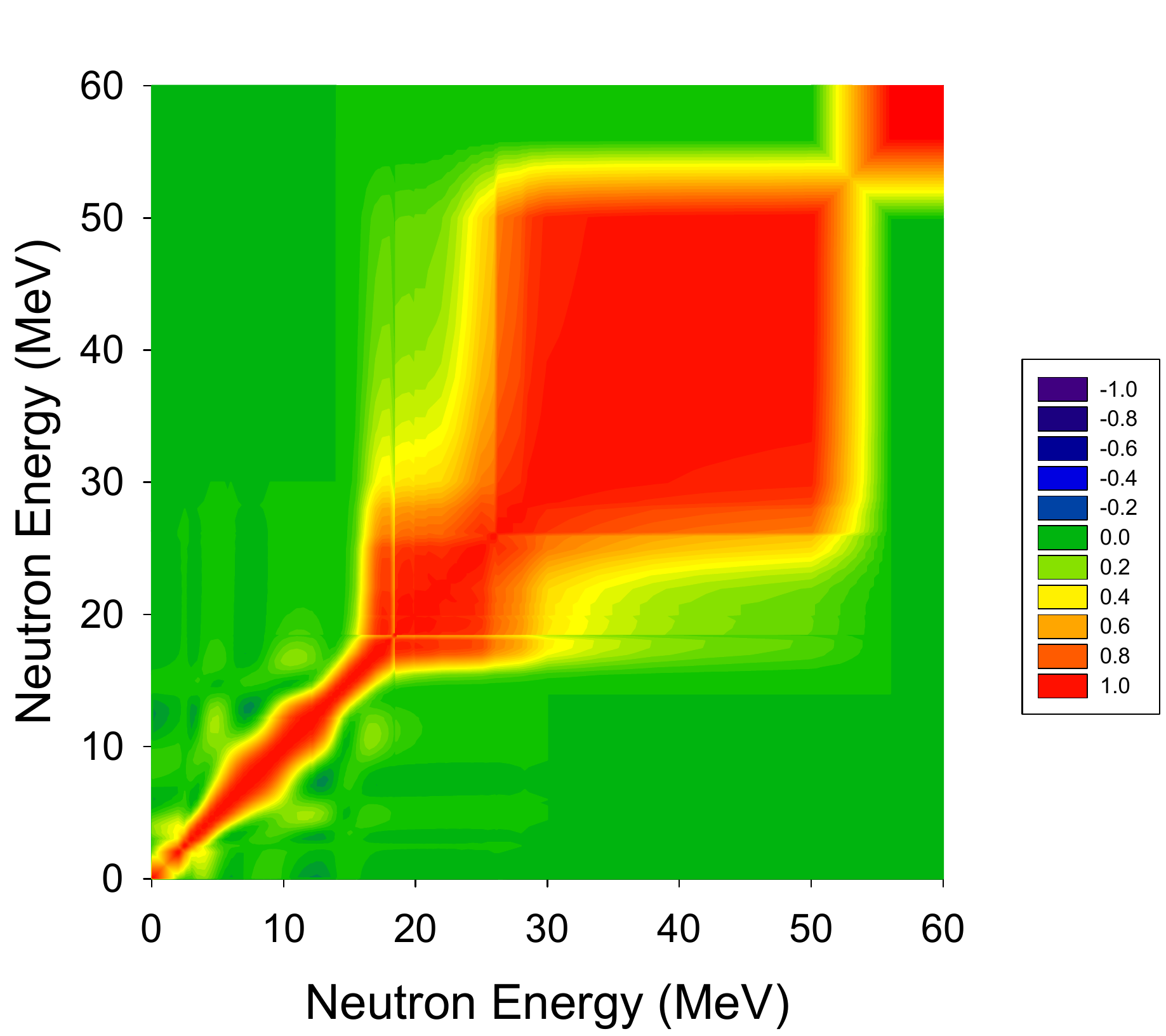}}
\subfigure[~One-sigma uncertainties in \% ($\equiv 100\times \frac{\sqrt{cov(i,i)}}{\mu_i}$), being $\mu_i$ the corresponding cross-section mean value.]
{\includegraphics[width=\columnwidth]{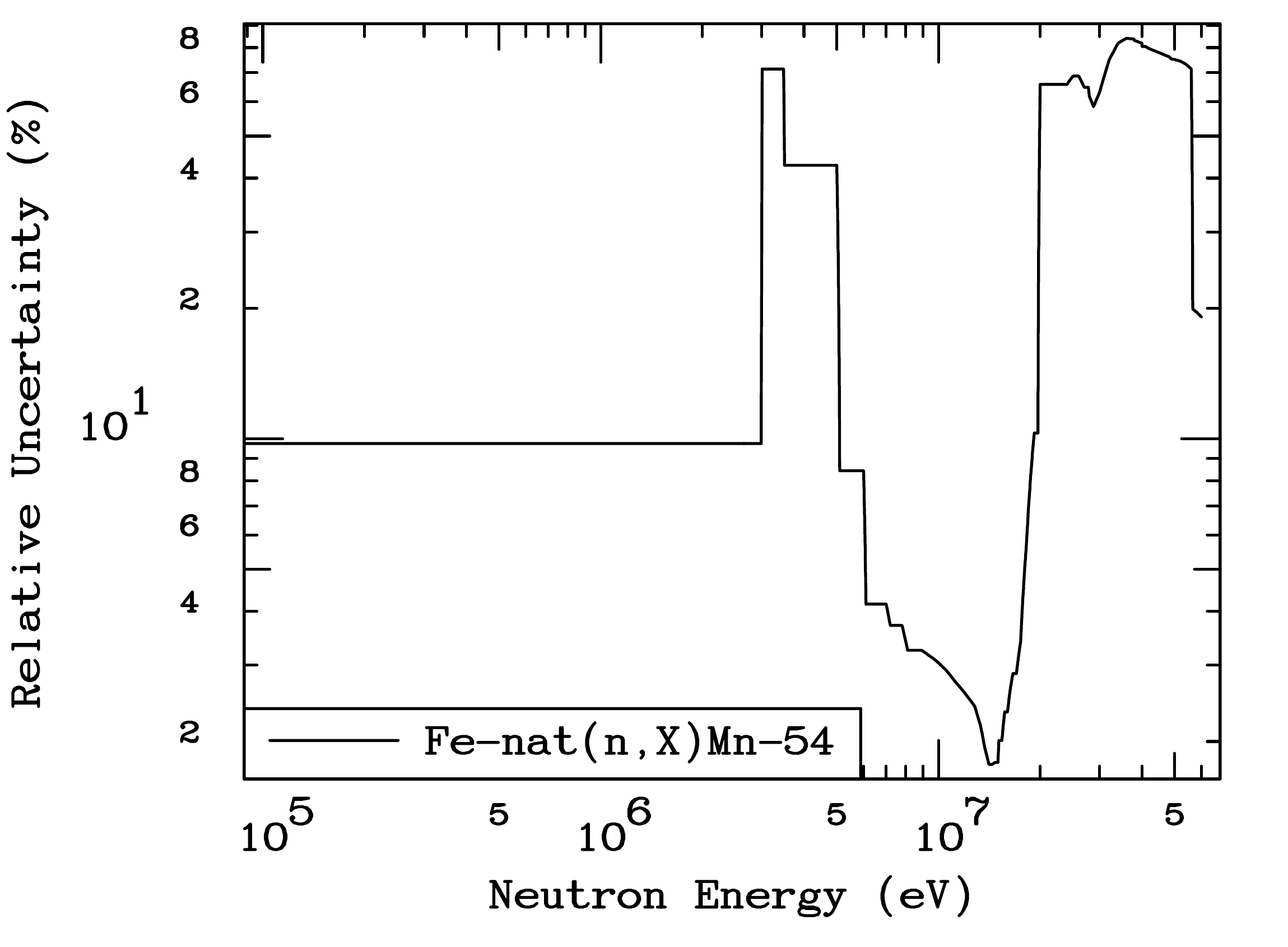}}
\vspace{-3mm}
\caption{(Color online) Uncertainties and correlations of the $^{\mbox{nat}}$Fe(n,X)$^{54}$Mn cross-section evaluation in \mbox{IRDFF-II} library.}
\label{Fe(n,X)-unc}
\vspace{-3mm}
\end{figure}

\subsubsection{$^{\mbox{nat}}$Fe(n,X)$^{54}\!$Mn}
Elemental iron consists mainly of isotope $^{56}$Fe, but the minor isotopes are sometimes important as they may lead to reactions that produce the same residual. Monitor samples are usually made of natural elements, so it is advantageous to provide the data for the residual production from the element, since production paths may include a multitude of minor channels which a user would need to consider. The cross sections and the covariance data for the $^{\mbox{nat}}$Fe(n,X)$^{54}$Mn reaction were assembled from the \mbox{IRDFF-II} evaluation for the $^{54}$Fe(n,p) reaction (described in the previous section), which is the dominant contribution at low energies. The contributing reactions at higher energies were taken from the \mbox{TENDL-2015} library, as shown in Fig.~\ref{xFe-Mn54}. Note that above 20~MeV the $^{54}$Mn production cross section is dominated by the contributions from reactions on $^{56}$Fe. For neutron dosimetry at these energies the use of the elemental radionuclide production cross sections is essential.

The plot of the correlation matrix for the $^{\mbox{nat}}$Fe(n,X)$^{54}$Mn reaction is shown in Fig.~\ref{Fe(n,X)-unc}(a). The plot of the 1-sigma uncertainty is shown in Fig.~\ref{Fe(n,X)-unc}(b).

\subsubsection{$^{58}\!$Fe(n,$\gamma$)$^{59}\!$Fe}
\begin{figure}[!hbtp]
%\vspace{-2mm}
\includegraphics[width=\columnwidth]{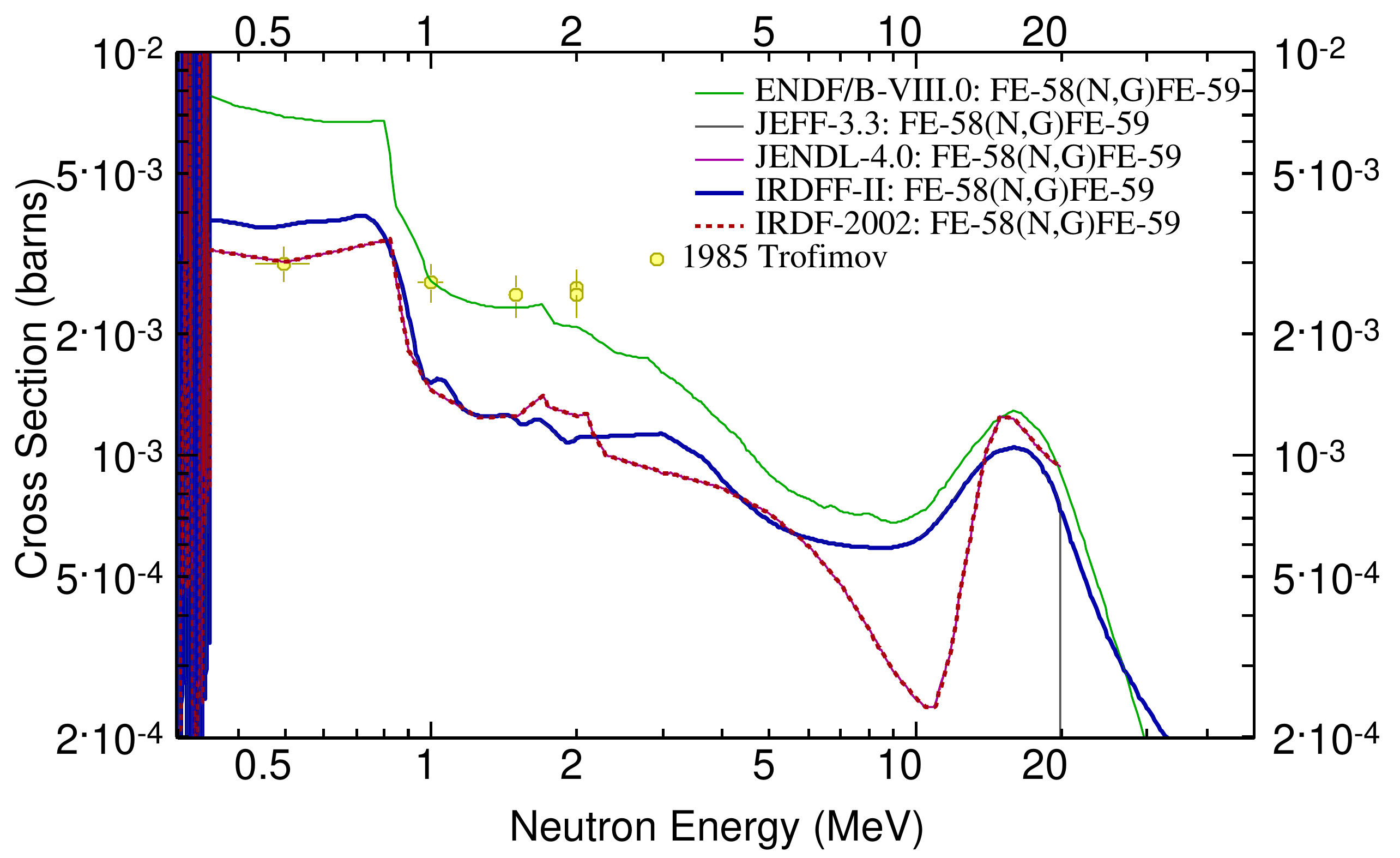}
\caption{(Color online) \textit Comparison of $^{58}$Fe(n,$\gamma$)$^{59}$Fe cross-section evaluations with experimental data from EXFOR.}
\label{gFe58-exp}
\end{figure}

The $^{58}$Fe(n,$\gamma$)$^{59}$Fe reaction belongs to a minor isotope of iron, with abundance of less than 0.3~\% and pronounced resonance structure, hence the measured cross section data are less precise. Through an IAEA-sponsored activity M. Moxon re-evaluated the resonance parameters of $^{58}$Fe in 2004~\cite{Mox04}, which were adopted in the JEFF and ENDF/B libraries. The evaluation removed some of the discrepancies with the measured thermal capture cross section and the resonance integral. The cross sections of $^{58}$Fe up to 60~MeV for \mbox{IRDFF-II} were taken from the \mbox{JEFF-3.1} library. Comparison of the cross sections with experimental data in the fast energy range is shown in Fig.~\ref{gFe58-exp}; big discrepancies are shown among different evaluated libraries. The cross section comparison in the resonance range is not shown, since many libraries adopt the Moxon evaluation. The figure also shows the comparison between IRDF-2002 and \mbox{IRDFF-II}, which remained unchanged from version IRDFF-v1.05.

\begin{figure}[!thb]
\vspace{-3mm}
\subfigure[~Cross-section correlation matrix.]
{\includegraphics[width=\columnwidth]{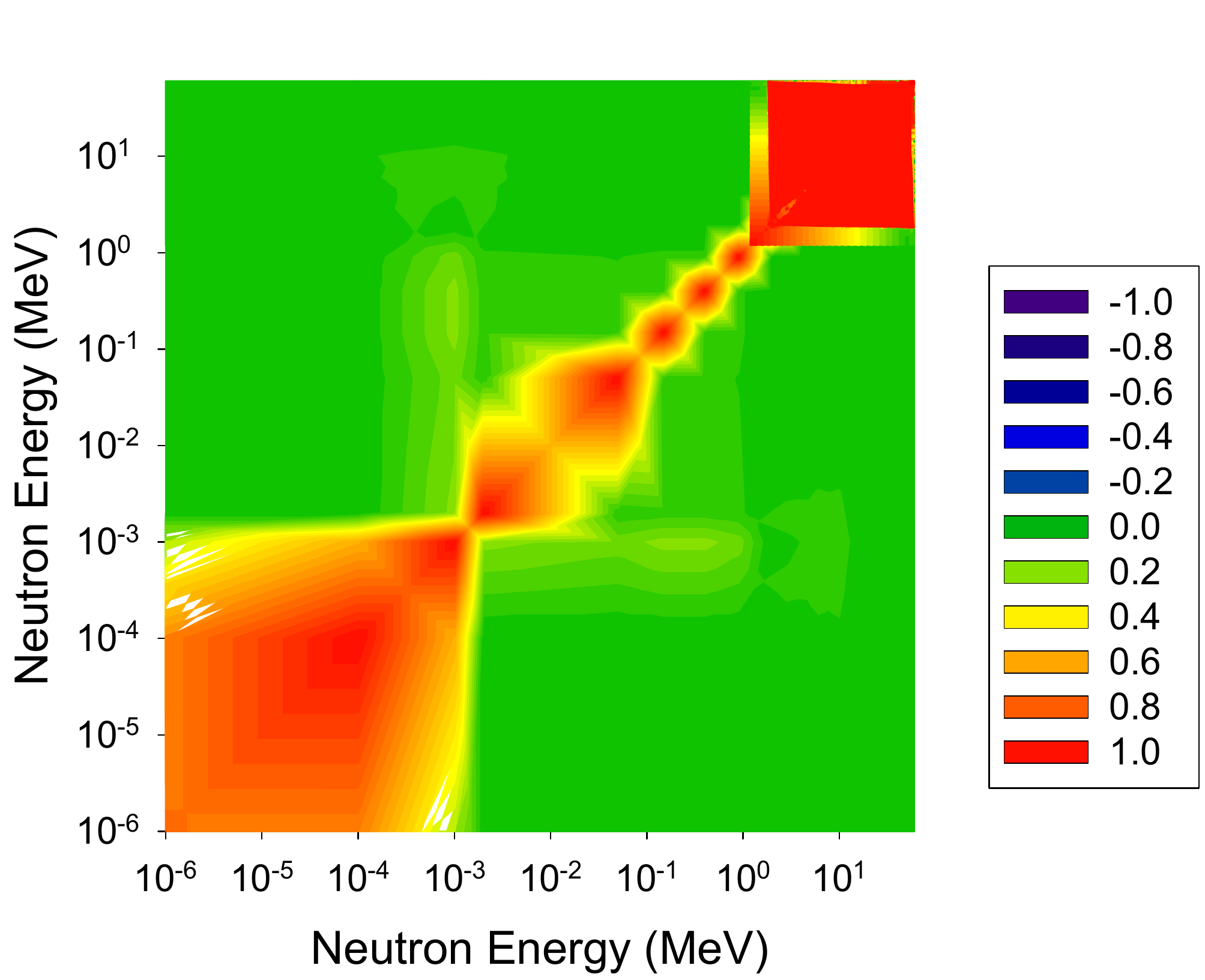}}
\subfigure[~One-sigma uncertainties in \% ($\equiv 100\times \frac{\sqrt{cov(i,i)}}{\mu_i}$), being $\mu_i$ the corresponding cross-section mean value.]
{\includegraphics[width=\columnwidth]{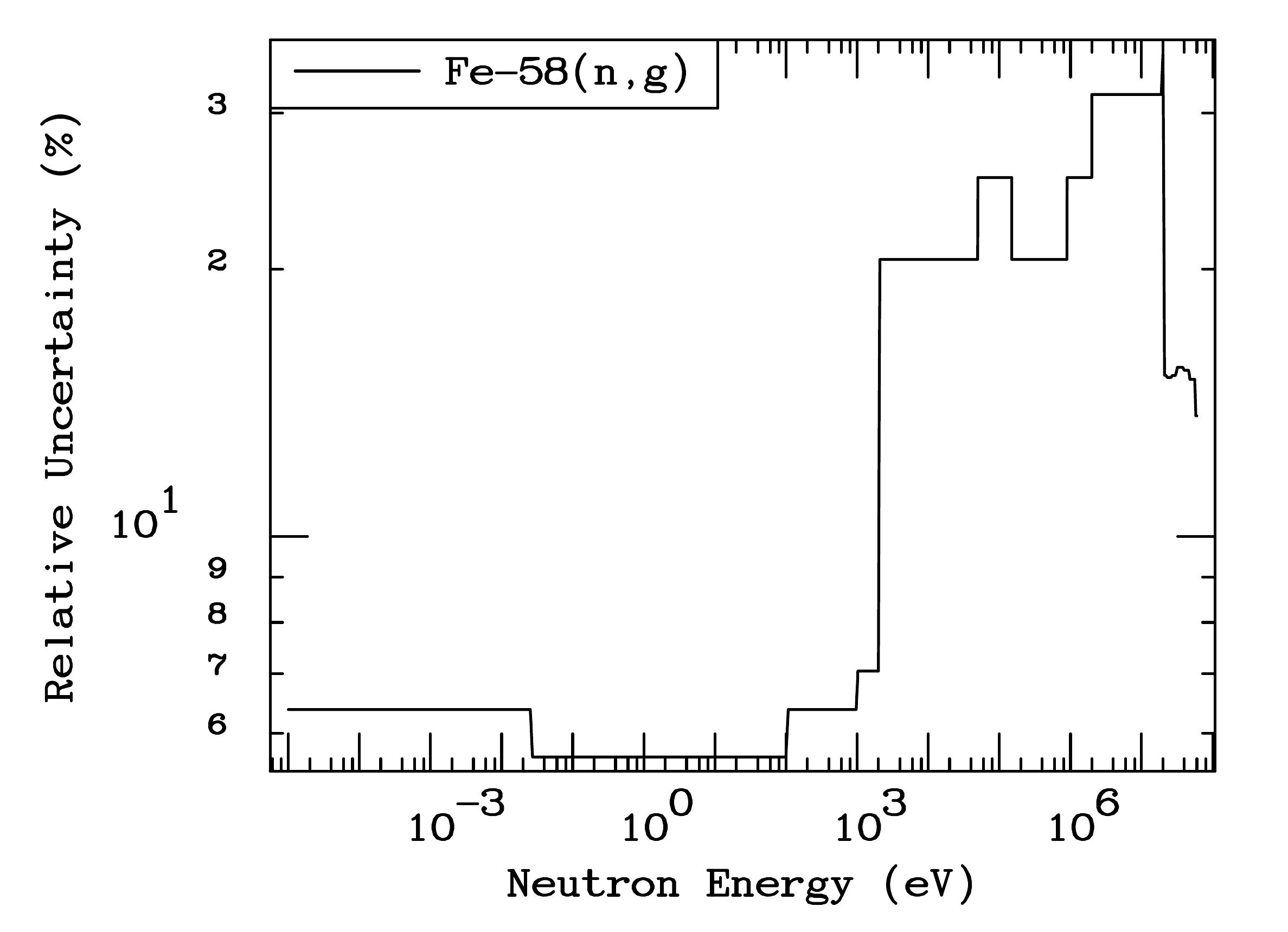}}
\vspace{-2mm}
\caption{(Color online) Uncertainties and correlations of the $^{58}$Fe(n,$\gamma$)$^{59}$Fe cross-section evaluation in \mbox{IRDFF-II} library.}
\label{gFe58-unc}
\vspace{-2mm}
\end{figure}

The plot of the correlation matrix for the $^{58}$Fe(n,$\gamma$)$^{59}$Fe reaction is shown in Fig.~\ref{gFe58-unc}(a). The plot of the 1-sigma uncertainty is shown in Fig.~\ref{gFe58-unc}(b).

The $^{58}$Fe(n,$\gamma$)$^{59}$Fe reaction is interesting for dosimetry applications because it is \ql clean\qr in the sense that it does not have any contributions from other isotopes of the elemental iron. The drawback is its low abundance, which is inherently known with a lower relative precision and this makes precise differential measurements more difficult, because they require enriched samples. The drawbacks are reflected in the lower accuracy of the evaluated cross sections.

\subsubsection{$^{32}\!$S(n,p)$^{32}\!$P}
The residual nuclide from the $^{32}$S(n,p)$^{32}$P reaction is $^{32}$P, which, as seen from Table~\ref{Table_Long_XIVa}, is a pure beta emitter. As with all beta dosimetry, care must be taken in the dosimetry measurement process to ensure that only the $^{32}$P, and no other activation products are being counted. Potential interfering reaction products may arise from activation of any of the naturally occurring isotopes in the target material, \eg, $^{32}$S, $^{33}$S, or $^{35}$S. There are many ways to addressing the potential interferents~\cite{A265}. For beta measurements performed on samples irradiated in fission reactor environments, a common practice for eliminating interference from other beta emitters is to use a cadmium thermal neutron shield to reduce the generation of capture products and to wait 24~hours between irradiation and counting so that residual reaction products, such as $^{31}$S, $^{34}$P, $^{31}$Si, and $^{37}$S, can die away before the counting is started. To increase the efficiency of the counting, the irradiated sulfur can also be burned, leaving a residue of $^{32}$P that can be counted without the need for corrections to account for absorption of the beta particles by the bulk sulfur in the irradiated material~\cite{A265}.

The $^{32}$S(n,p)$^{32}$P reaction cross sections and covariances  below 20~MeV were evaluated by Zolotarev~\cite{Zol13} in 2013. Extension to 60~MeV was done at the IAEA based on \mbox{TENDL-2013}, renormalizing the cross sections at 20~MeV for continuity.

\begin{figure}[!thb]
\vspace{-2mm}
\subfigure[~Comparison to selected experimental data from EXFOR \cite{EXF08}.]
{\includegraphics[width=\columnwidth]{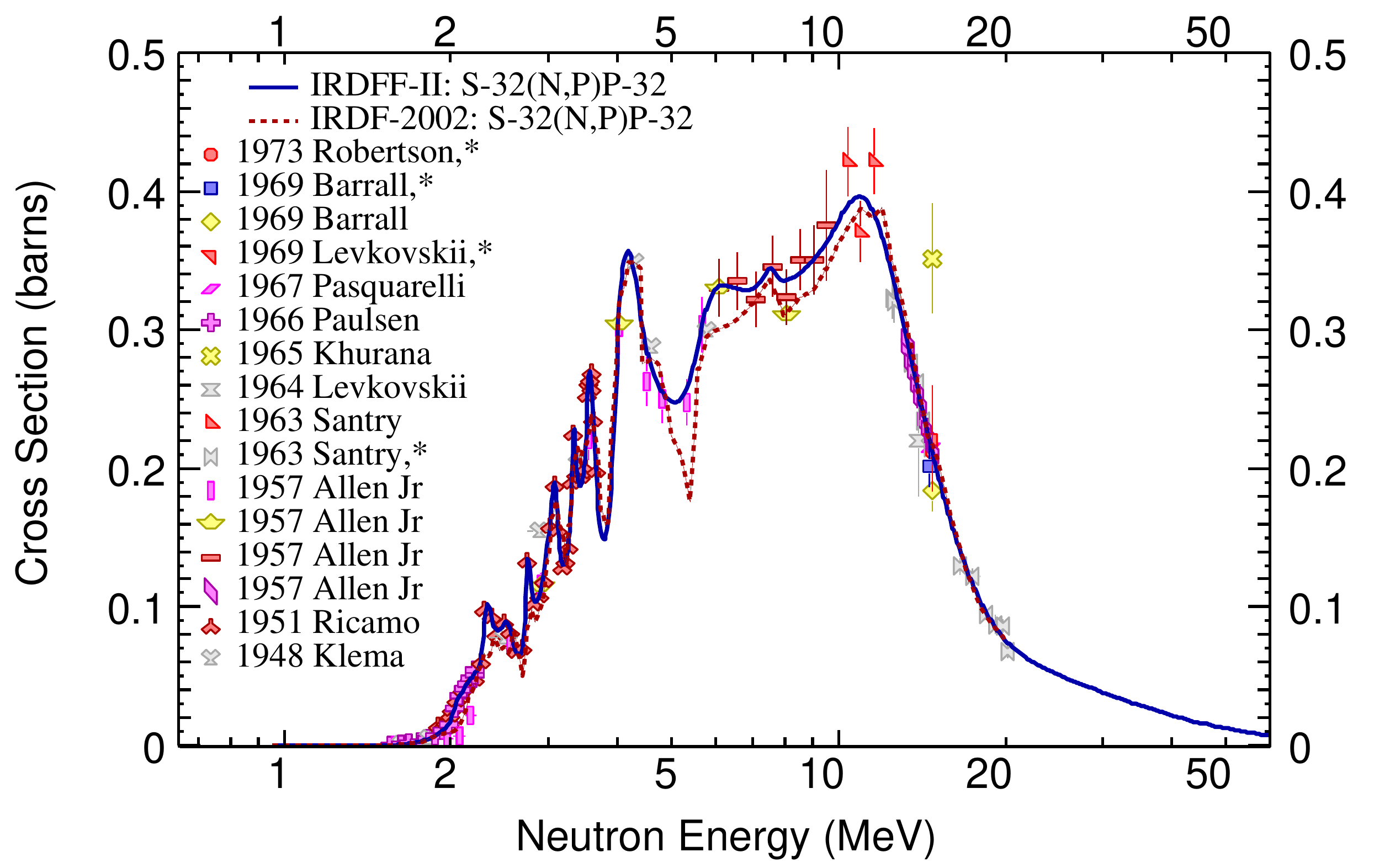}}
\subfigure[~Ratio of cross-section evaluations to \mbox{IRDFF-II} evaluation.]
{\includegraphics[width=\columnwidth]{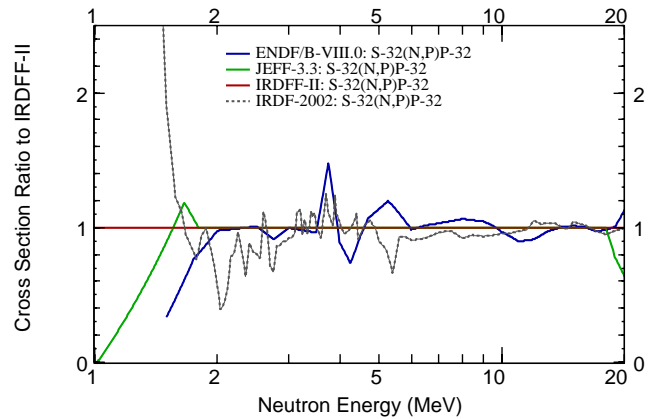}}
\vspace{-3mm}
\caption{(Color online) $^{32}$S(n,p)$^{32}$P cross-section evaluation in \mbox{IRDFF-II} library relative to other data.}
\label{pS32}
\vspace{-3mm}
\end{figure}

Comparison of the cross sections with experimental data is shown in Fig.~\ref{pS32}(a). The figure also shows the comparison between \mbox{IRDF-2002} and \mbox{IRDFF-II}, which remained unchanged from \mbox{IRDFF-v1.05}. Cross sections in different evaluated libraries differ significantly near threshold as shown in Fig.~\ref{pS32}(b), where the ratios of the various evaluations to the \mbox{IRDFF-II} recommended cross section are given.
%These differences may be important if this reaction monitor is used in reactor spectra.

\begin{figure}[!thb]
\vspace{-4mm}
\subfigure[~Cross-section correlation matrix.]
{\includegraphics[width=\columnwidth]{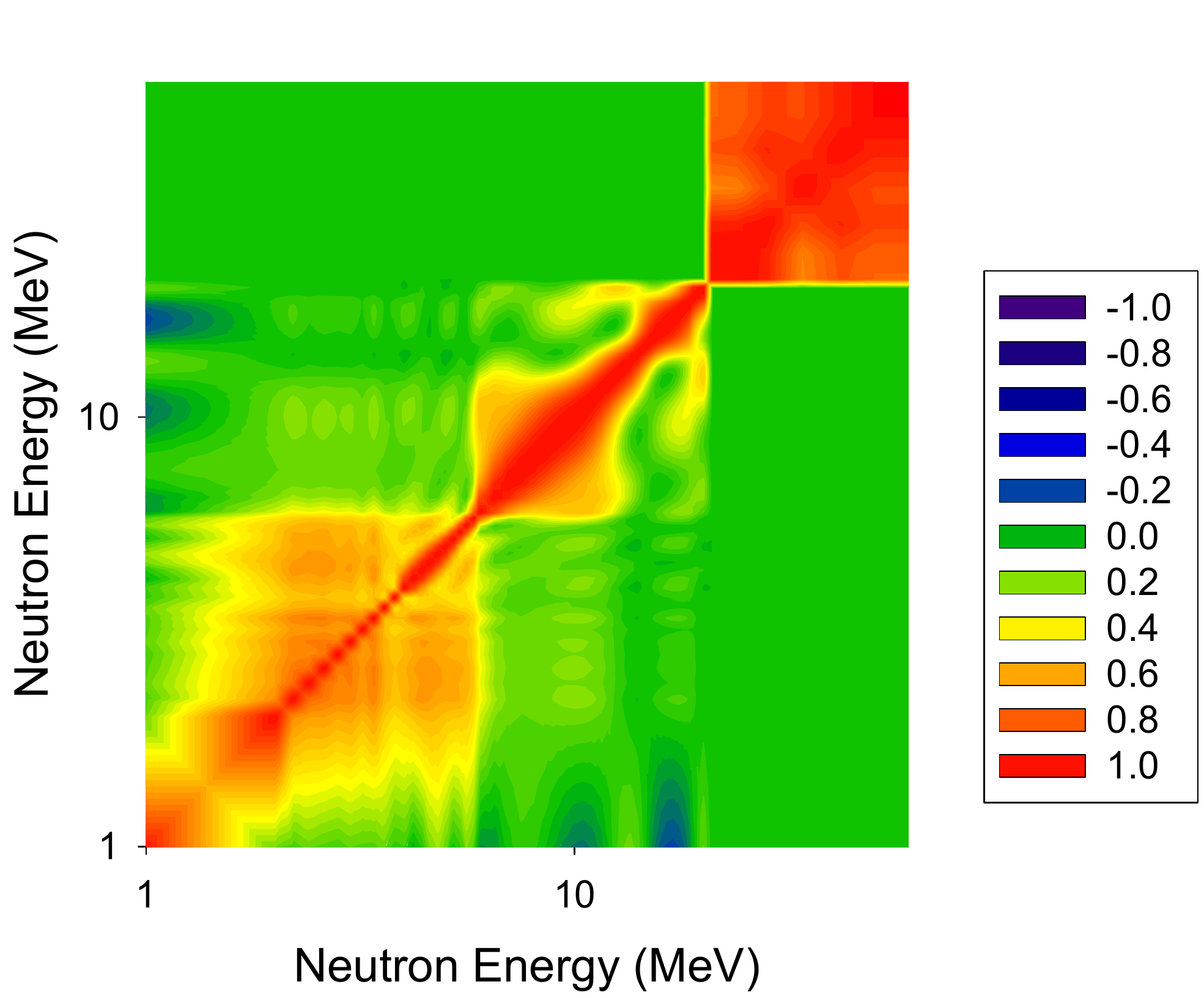}}
\subfigure[~One-sigma uncertainties in \% ($\equiv 100\times \frac{\sqrt{cov(i,i)}}{\mu_i}$), being $\mu_i$ the corresponding cross-section mean value.]
{\includegraphics[width=\columnwidth]{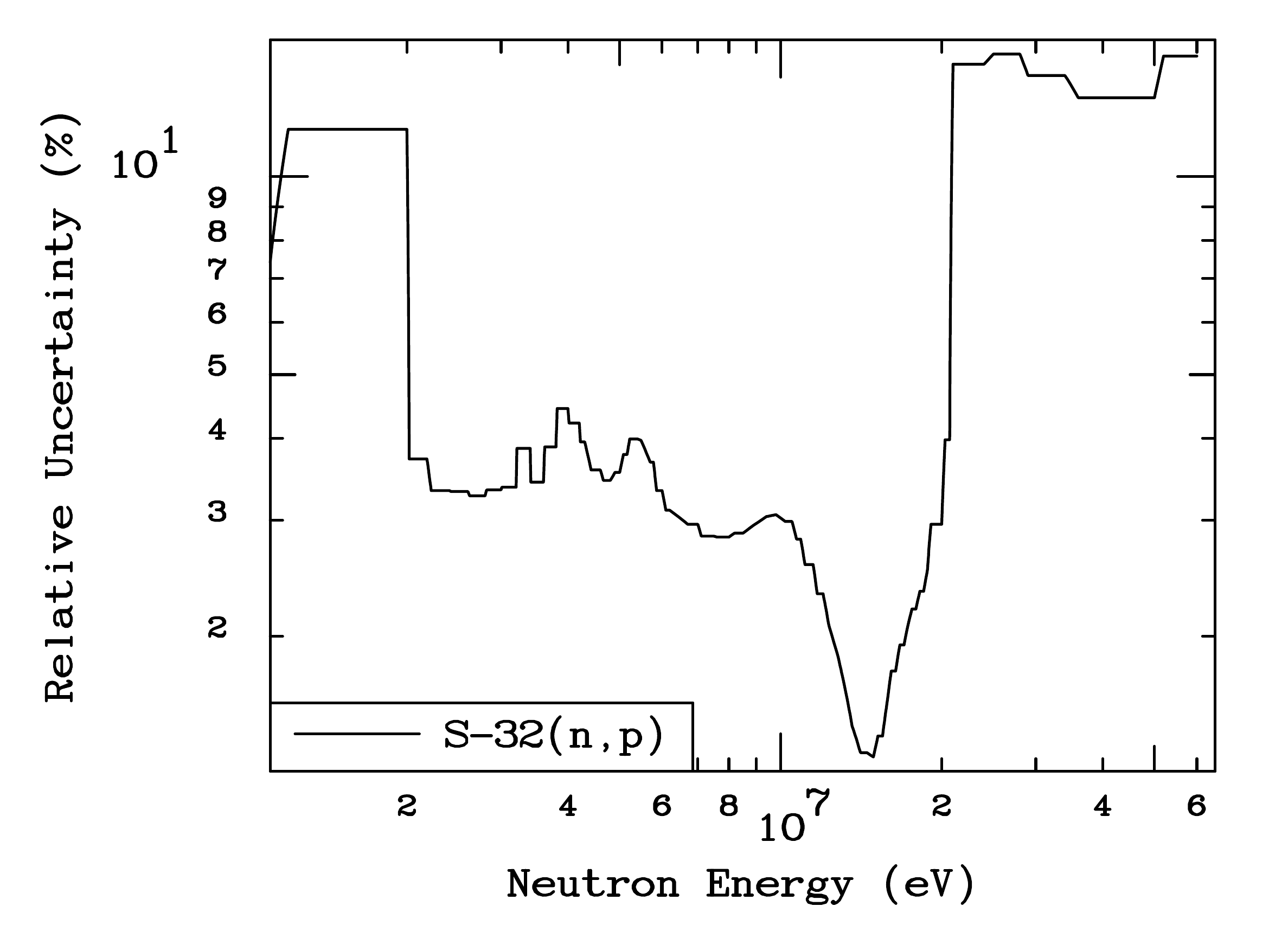}}
\vspace{-3mm}
\caption{(Color online) Uncertainties and correlations of the $^{32}$S(n,p)$^{32}$P cross-section evaluation in \mbox{IRDFF-II} library.}
\label{pS32-unc}
\vspace{-3mm}
\end{figure}

The plot of the correlation matrix for the $^{32}$S(n,p)$^{32}$P reaction is shown in Fig.~\ref{pS32-unc}(a). The plot of the 1-sigma uncertainty is shown in Fig.~\ref{pS32-unc}(b).

Reactions on the heavier isotopes of sulphur may lead to the same reaction product $^{32}$P. In particular, the cross sections for the $^{33}$S(n,np)$^{32}$P reaction rises sharply above 12~MeV, but the natural abundance of $^{33}$S is only 0.75~\%. However, at energies above 30~MeV the $^{34}$S(n,2np)$^{32}$P may contribute to the $^{32}$P production, so the $^{32}$S(n,p)$^{32}$P reaction monitor should not be used for high energy applications. The reaction $^{\mbox{nat}}$S(n,X)$^{32}$P should be used instead.

\subsubsection{$^{90}\!$Zr(n,2n)$^{89}\!$Zr}
\begin{figure}[!thb]
\vspace{-2mm}
\subfigure[~Comparison to selected experimental data from EXFOR \cite{EXF08}.]
{\includegraphics[width=\columnwidth]{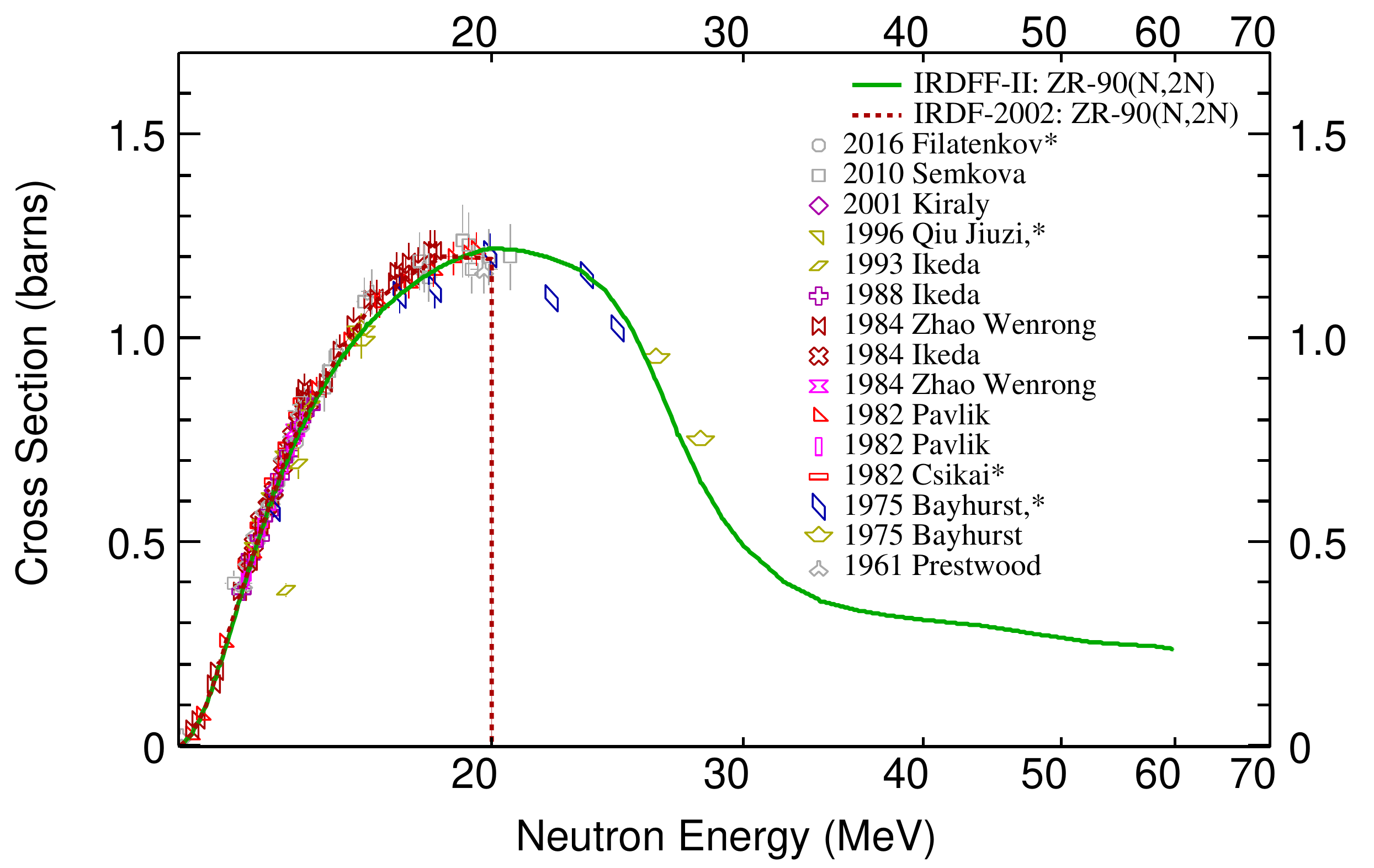}}
\subfigure[~Ratio of cross-section evaluations to \mbox{IRDFF-II} evaluation.]
{\includegraphics[width=\columnwidth]{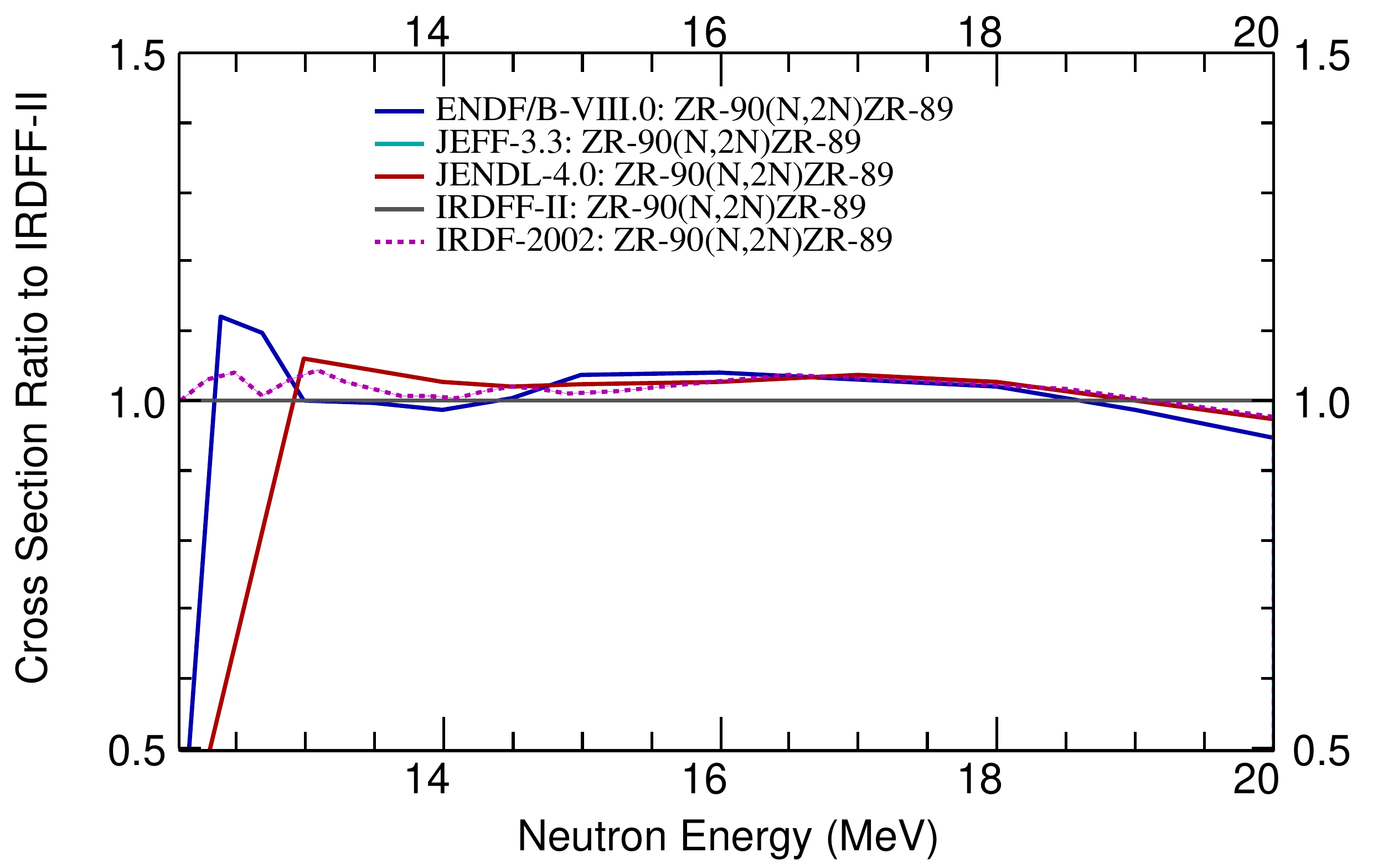}}
\vspace{-3mm}
\caption{(Color online) $^{90}$Zr(n,2n)$^{89}$Zr cross-section evaluation in \mbox{IRDFF-II} library relative to other data.}
\label{nZr90}
\vspace{-3mm}
\end{figure}

The $^{90}$Zr(n,2n)$^{89}$Zr reaction cross sections and covariances  below 20~MeV were evaluated by Zolotarev~\cite{Zol09} in 2009. Extension to 60~MeV was done at the IAEA based on \mbox{TENDL-2013}, renormalizing the cross sections at 20~MeV for continuity. Comparison of the cross sections with experimental data is shown in Fig.~\ref{nZr90}(a). The figure also shows the comparison between \mbox{IRDF-2002} and \mbox{IRDFF-II}, which remained unchanged from version \mbox{IRDFF-v1.05}. Cross sections in different evaluated libraries differ significantly near threshold as shown in Fig.~\ref{nZr90}(b), where the ratios of the various evaluations to the \mbox{IRDFF-II} recommended cross section are given. These differences may be important if this reaction monitor is used in reactor spectra.

\begin{figure}[!thb]
\vspace{-4mm}
\subfigure[~Cross-section correlation matrix.]
{\includegraphics[width=\columnwidth]{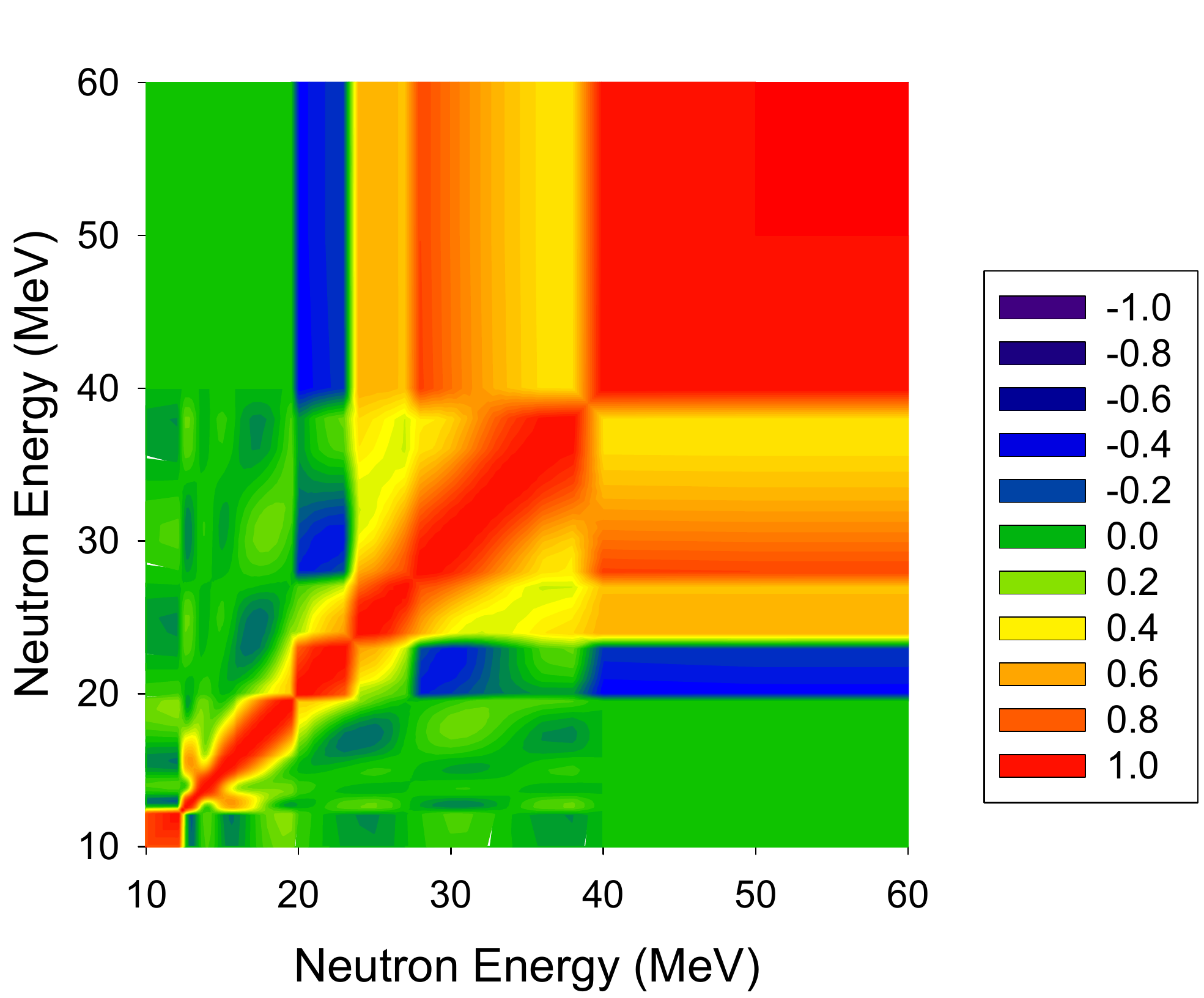}}
\subfigure[~One-sigma uncertainties in \% ($\equiv 100\times \frac{\sqrt{cov(i,i)}}{\mu_i}$), being $\mu_i$ the corresponding cross-section mean value.]
{\includegraphics[width=\columnwidth]{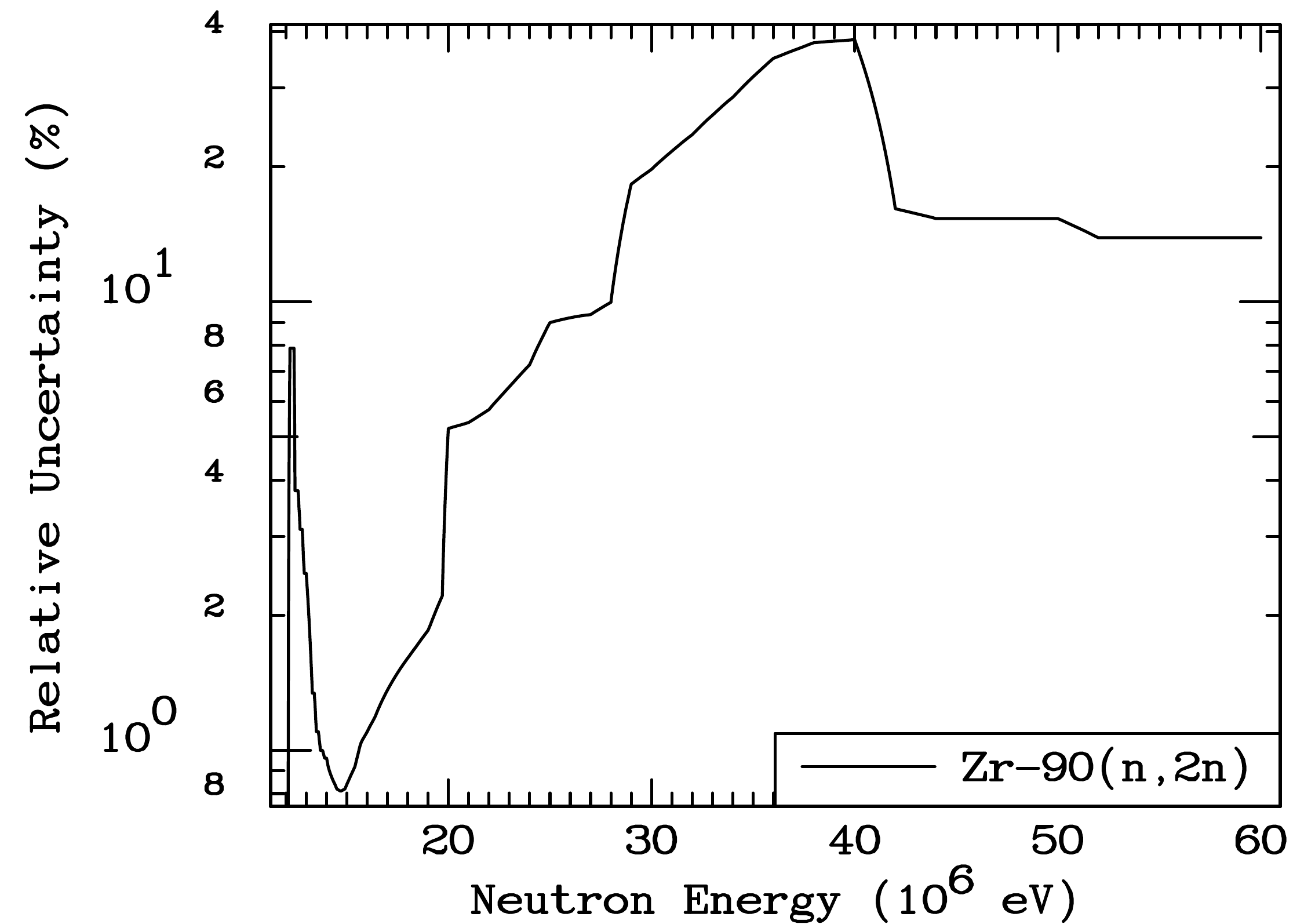}}
\vspace{-3mm}
\caption{(Color online) Uncertainties and correlations of the $^{90}$Zr(n,2n)$^{89}$Zr cross-section evaluation in \mbox{IRDFF-II} library.}
\label{nZr90-unc}
\vspace{-3mm}
\end{figure}

The plot of the correlation matrix for the $^{90}$Zr(n,2n)$^{89}$Zr reaction is shown in Fig.~\ref{nZr90-unc}(a). The plot of the 1-sigma uncertainty is shown in Fig.~\ref{nZr90-unc}(b).
Reactions on the heavier isotopes of Zr may lead to the same reaction product $^{89}$Zr. In particular, the cross sections for the $^{91}$Zr(n,3n)$^{89}$Zr reaction rises sharply above 20~MeV. The use of the $^{\mbox{nat}}$Zr(n,X)$^{89}$Zr reaction cross sections, which are present in the \mbox{IRDFF-II} library, is strongly recommended, even though the user should be aware that the minor contributing channels represent a correction and do not meet the dosimetry quality standards, so even the reaction $^{\mbox{nat}}$Zr(n,X)$^{89}$Zr should be used with caution in neutron fields with a significant contribution of neutrons with energies above 20~MeV.

\newpage
\subsubsection{$^{23}\!$Na(n,2n)$^{22}\!$Na}
Sodium is a mono-isotopic naturally occurring element and is convenient for neutron dosimetry because it does not have competing reactions from different isotopes producing the same residual. The $^{23}$Na(n,2n)$^{22}$Na has a relatively high threshold and the plateau above 20~MeV; it can be used for probing the very high-energy tail of the fission spectrum.

\begin{figure}[!thb]
\vspace{-2mm}
\subfigure[~Comparison to selected experimental data from EXFOR \cite{EXF08} and new data from Ref.~\cite{Sim19} (Simeckova'19).]
{\includegraphics[width=\columnwidth]{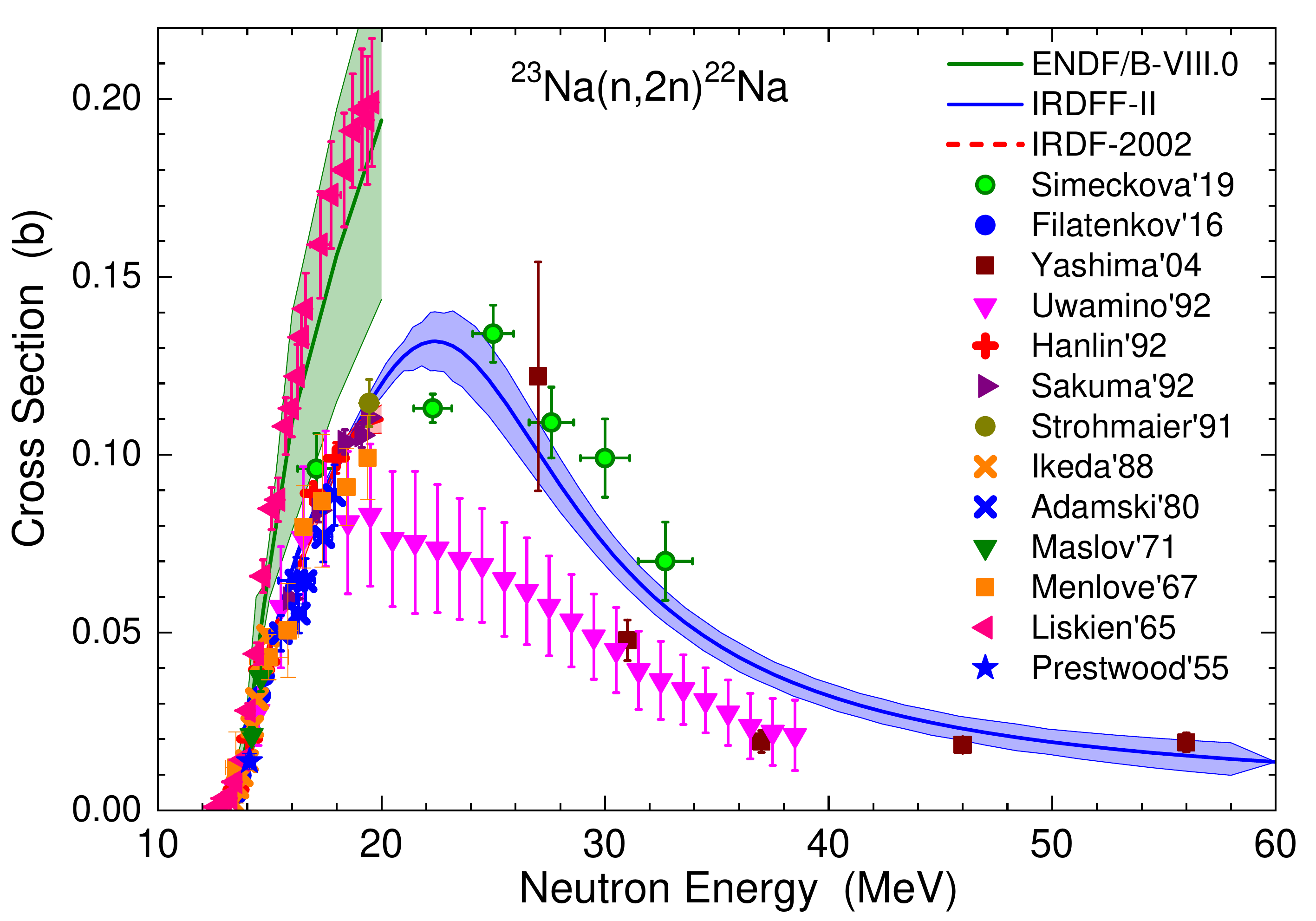}}
\subfigure[~Ratio of cross-section evaluations to \mbox{IRDFF-II} evaluation.]
{\includegraphics[width=\columnwidth]{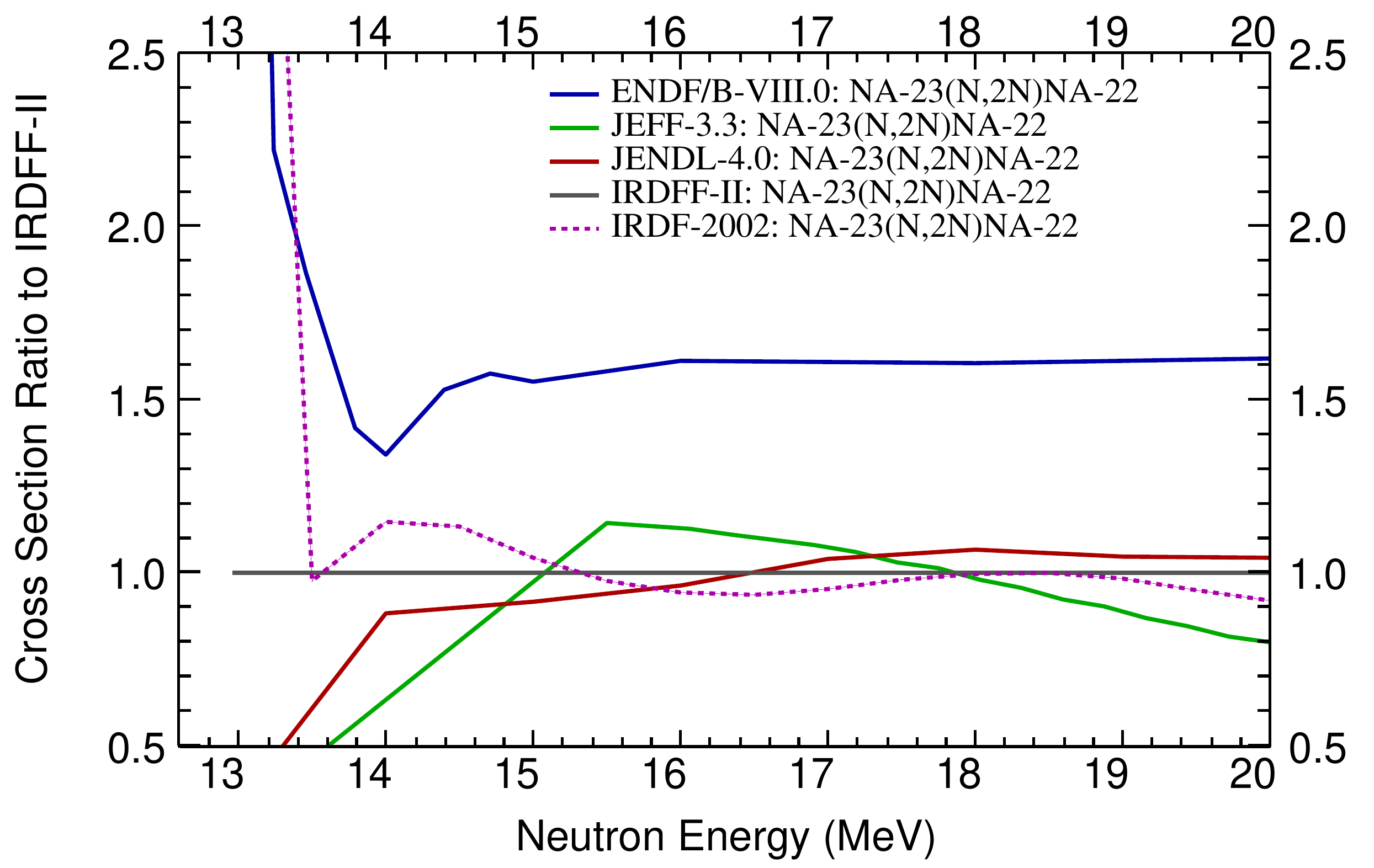}}
\vspace{-3mm}
\caption{(Color online) $^{23}$Na(n,2n)$^{22}$Na cross-section evaluation in \mbox{IRDFF-II} library relative to other data.}
\label{nNa23}
\vspace{-3mm}
\end{figure}

The $^{23}$Na(n,2n)$^{22}$Na reaction cross sections and covariances in the entire energy range up to 60~MeV were evaluated by Zolotarev~\cite{Zol17} in 2017.
In 2019 this reaction cross section was measured at Nuclear Physics Institute of the Czech Academy of Sciences employing the p-Li quasi-monoenergetic neutrons with energies between 17 and 33~MeV~\cite{Sim19}. As seen in Fig.~\ref{nNa23}(a), these data confirm the latest Zolotarev evaluation.
%Comparison of the cross sections with experimental data is shown in Fig.~\ref{nNa23}(a).
The figure also shows the comparison between \mbox{IRDF-2002} and \mbox{IRDFF-II}, which remained unchanged from version \mbox{IRDFF-v1.05}. Cross sections in different evaluated libraries differ significantly near threshold as shown in Fig.~\ref{nNa23}(b), where the ratios of the various evaluations to the \mbox{IRDFF-II} recommended cross section are given. The ENDF/B-VIII.0 evaluation followed a different set of experimental data. Therefore, differences are observed between evaluations, but also between measurements.

\begin{figure}[!thb]
\vspace{-4mm}
\subfigure[~Cross-section correlation matrix.]
{\includegraphics[width=\columnwidth]{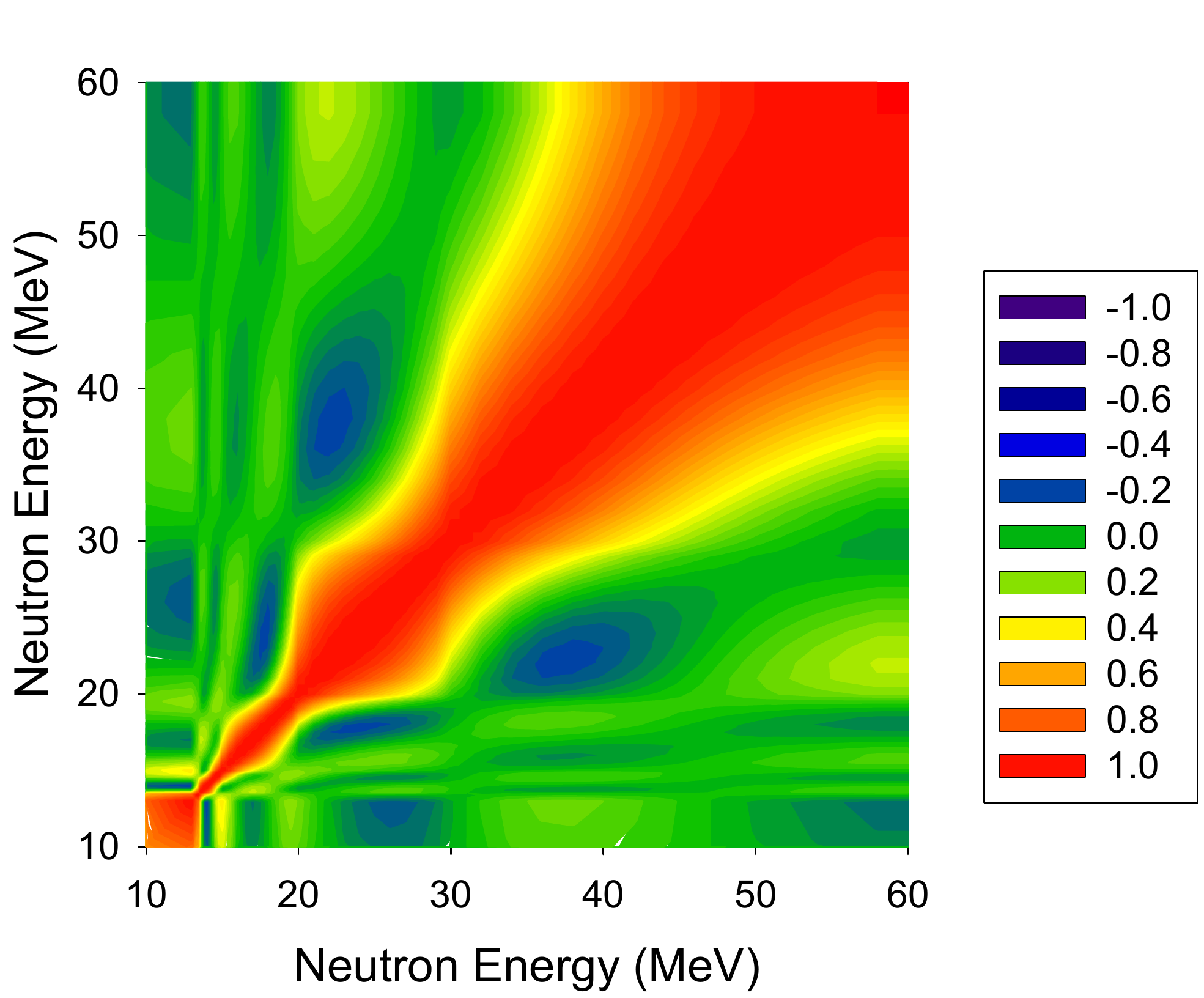}}
\subfigure[~One-sigma uncertainties in \% ($\equiv 100\times \frac{\sqrt{cov(i,i)}}{\mu_i}$), being $\mu_i$ the corresponding cross-section mean value.]
{\includegraphics[width=\columnwidth]{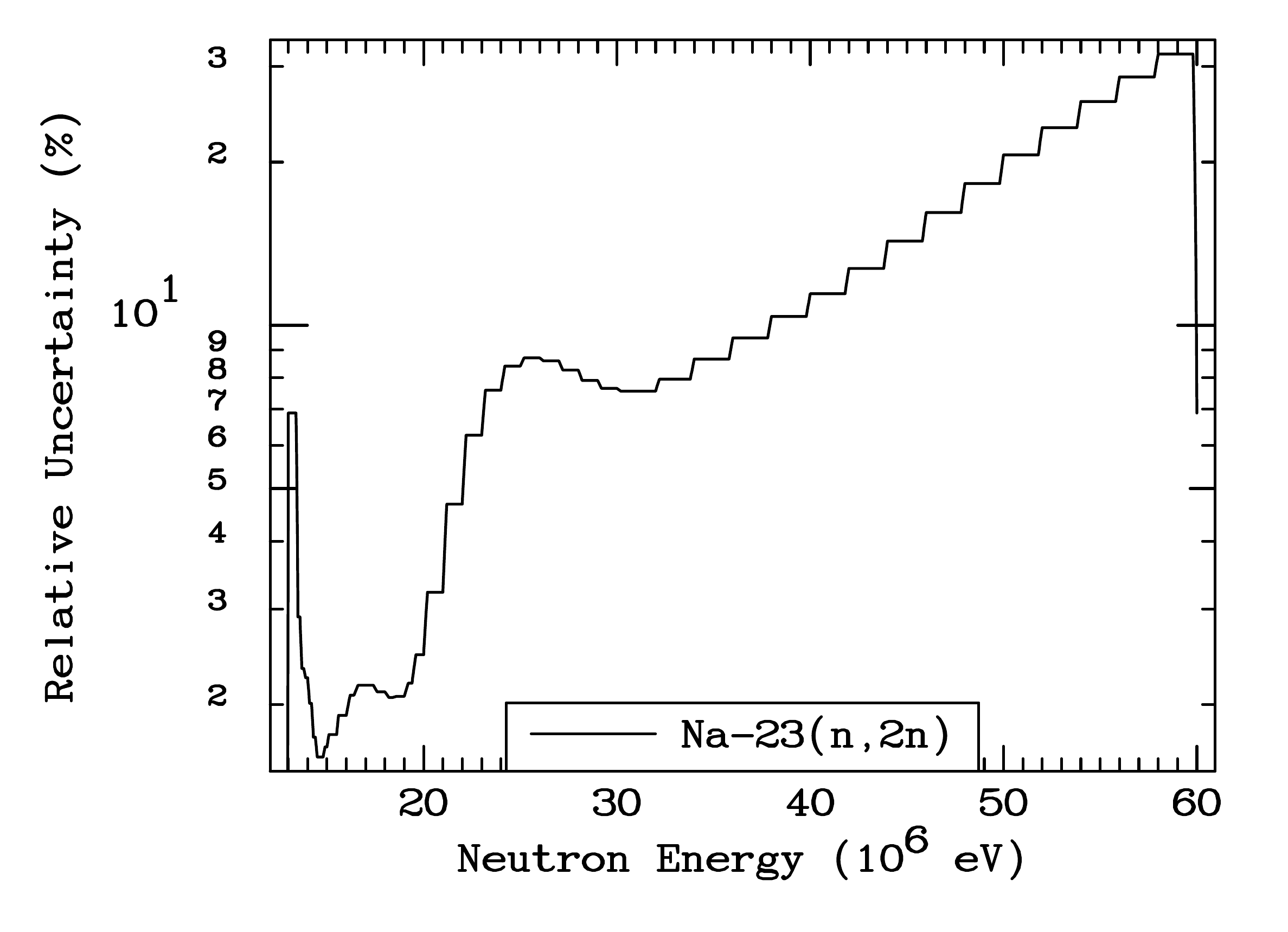}}
\vspace{-3mm}
\caption{(Color online) Uncertainties and correlations of the $^{23}$Na(n,2n)$^{22}$Na cross-section evaluation in \mbox{IRDFF-II} library.}
\label{nNa23-unc}
\vspace{-3mm}
\end{figure}

The covariance matrix of the $^{23}$Na(n,2n)$^{22}$Na reaction cross sections is shown in Fig.~\ref{nNa23-unc}(a) and the corresponding 1-sigma uncertainty in Fig.~\ref{nNa23-unc}(b).

\newpage
\subsubsection{$^{23}\!$Na(n,$\gamma$)$^{24}\!$Na}
\begin{figure}[!thb]
\vspace{-2mm}
\subfigure[~Comparison to selected experimental data from EXFOR \cite{EXF08}.]
{\includegraphics[width=\columnwidth]{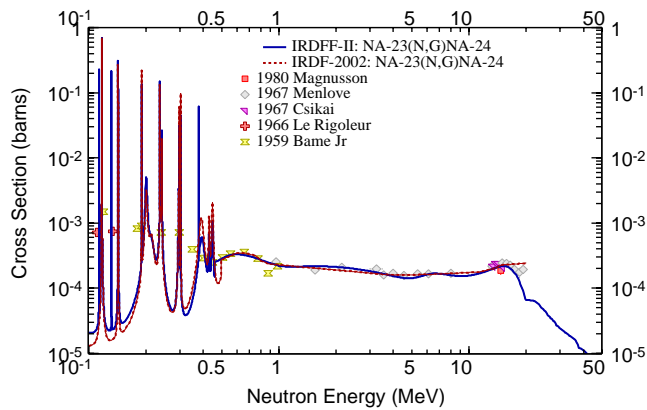}}
\subfigure[~Ratio of cross-section evaluations to \mbox{IRDFF-II} evaluation.]
{\includegraphics[width=\columnwidth]{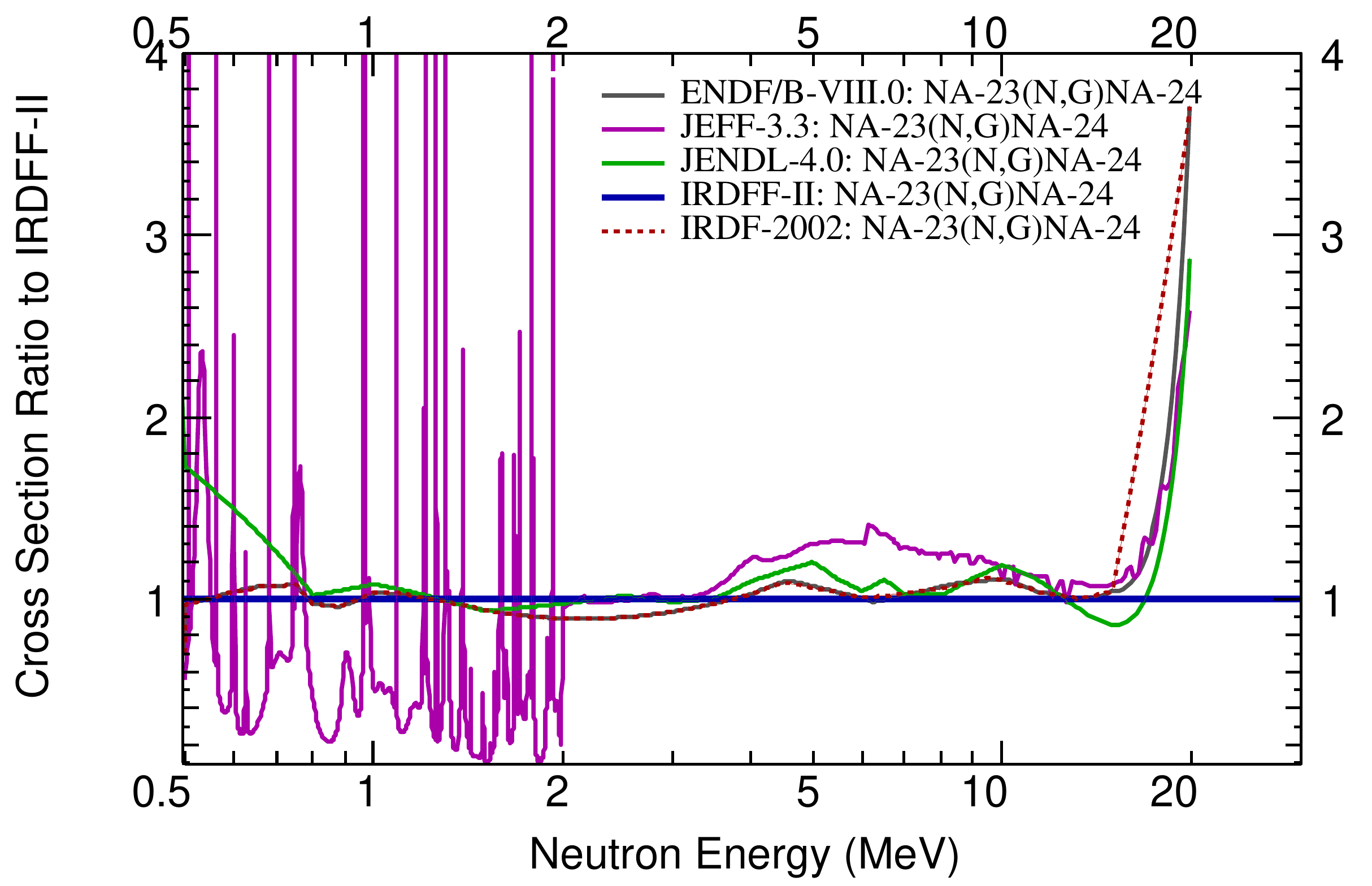}}
\vspace{-3mm}
\caption{(Color online) $^{23}$Na(n,$\gamma$)$^{24}$Na cross-section evaluation in \mbox{IRDFF-II} library relative to other data.}
\label{gNa23}
\vspace{-3mm}
\end{figure}

The capture reaction has a strong resonance near 2.8~keV and is highly sensitive to the epithermal energies in reactor spectra.
The $^{23}$Na(n,$\gamma$)$^{24}$Na reaction resonance parameters, cross sections and covariances up to 20~MeV were also evaluated by Zolotarev~\cite{Zol17}.
Comparison of the cross sections with experimental data in the fast energy range is shown in Fig.~\ref{gNa23}(a). The figure also shows the comparison between \mbox{IRDF-2002} and \mbox{IRDFF-II}, which is different from \mbox{IRDFF-v1.05}. Cross sections in different evaluated libraries differ significantly as shown in Fig.~\ref{gNa23}(b), where the ratios of the various evaluations to the \mbox{IRDFF-II} recommended cross section are given.

\begin{figure}[!thb]
\vspace{-4mm}
\subfigure[~Cross-section correlation matrix.]
{\includegraphics[width=\columnwidth]{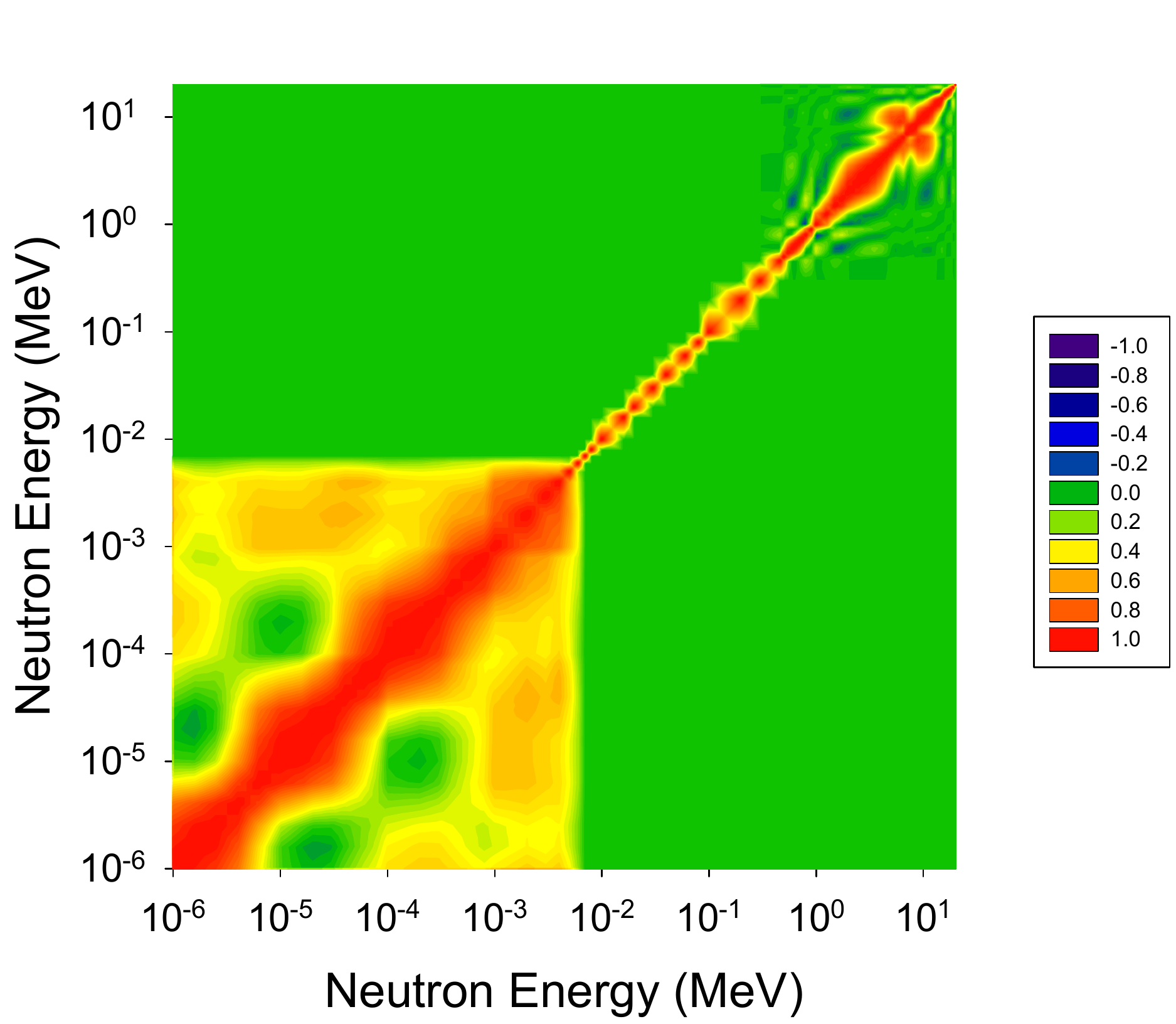}}
\subfigure[~One-sigma uncertainties in \% ($\equiv 100\times \frac{\sqrt{cov(i,i)}}{\mu_i}$), being $\mu_i$ the corresponding cross-section mean value.]
{\includegraphics[width=\columnwidth]{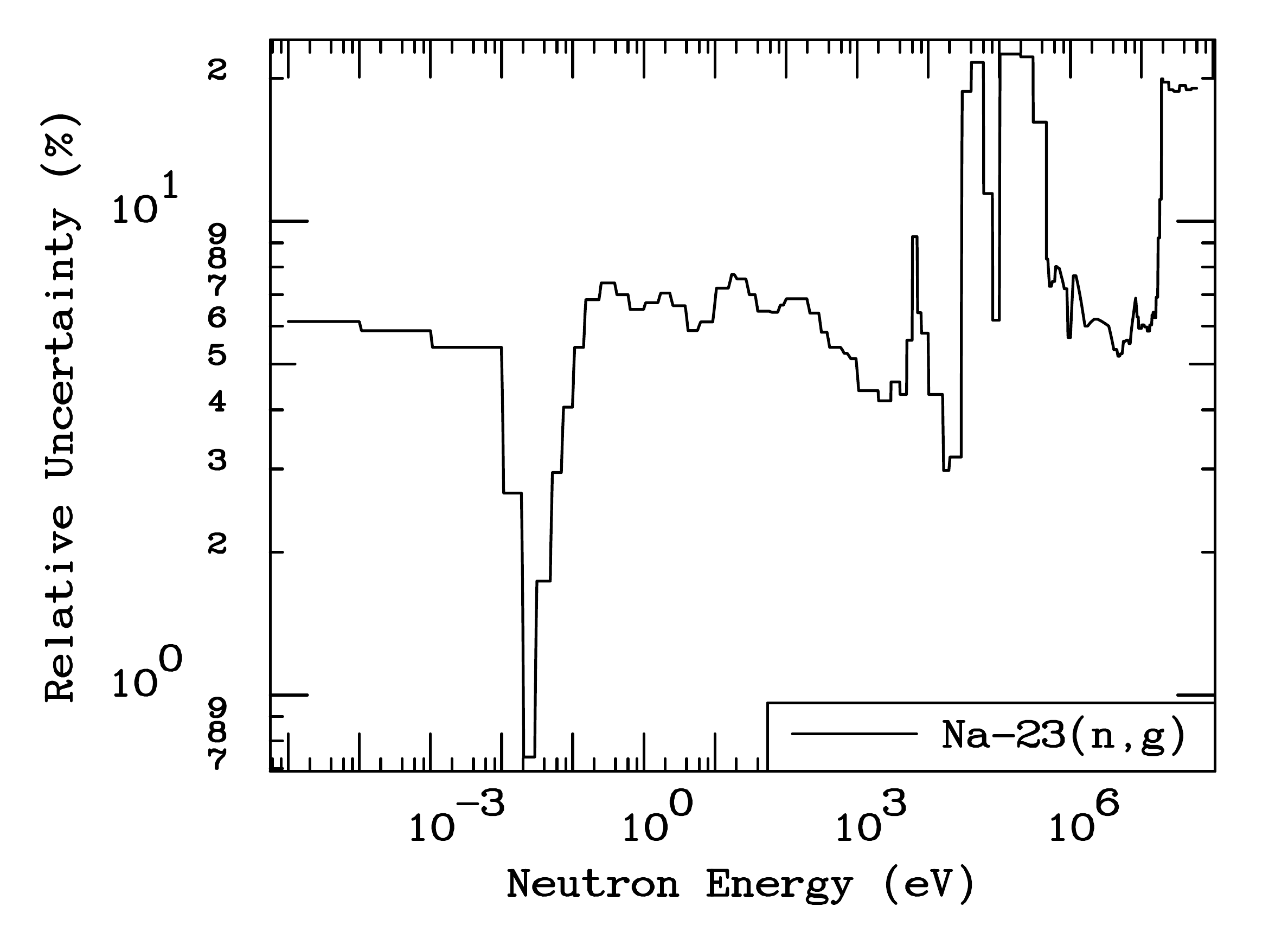}}
\vspace{-3mm}
\caption{(Color online) Uncertainties and correlations of the $^{23}$Na(n,$\gamma$)$^{24}$Na cross-section evaluation in \mbox{IRDFF-II} library.}
\label{gNa23-unc}
\vspace{-3mm}
\end{figure}

The covariance matrix of the $^{23}$Na(n,$\gamma$)$^{24}$Na reaction cross sections is shown in Fig.~\ref{gNa23-unc}(a) and the corresponding 1-sigma uncertainty in Fig.~\ref{gNa23-unc}(b).

\subsubsection{$^{24}\!$Mg(n,p)$^{24}\!$Na}
\begin{figure}[!thb]
\vspace{-2mm}
\subfigure[~Comparison to selected experimental data from EXFOR \cite{EXF08}.]
{\includegraphics[width=\columnwidth]{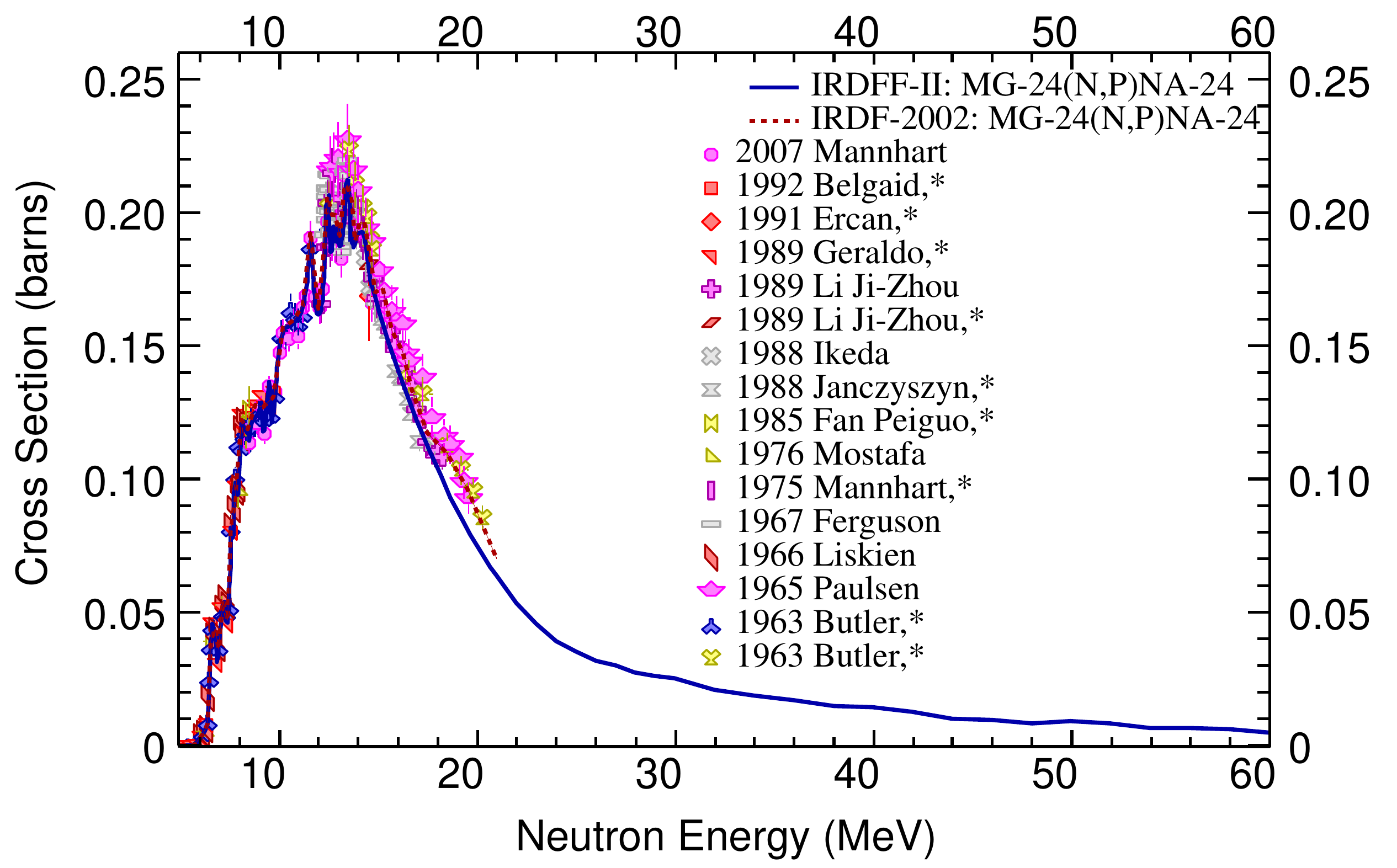}}
\subfigure[~Ratio of cross-section evaluations to \mbox{IRDFF-II} evaluation.]
{\includegraphics[width=\columnwidth]{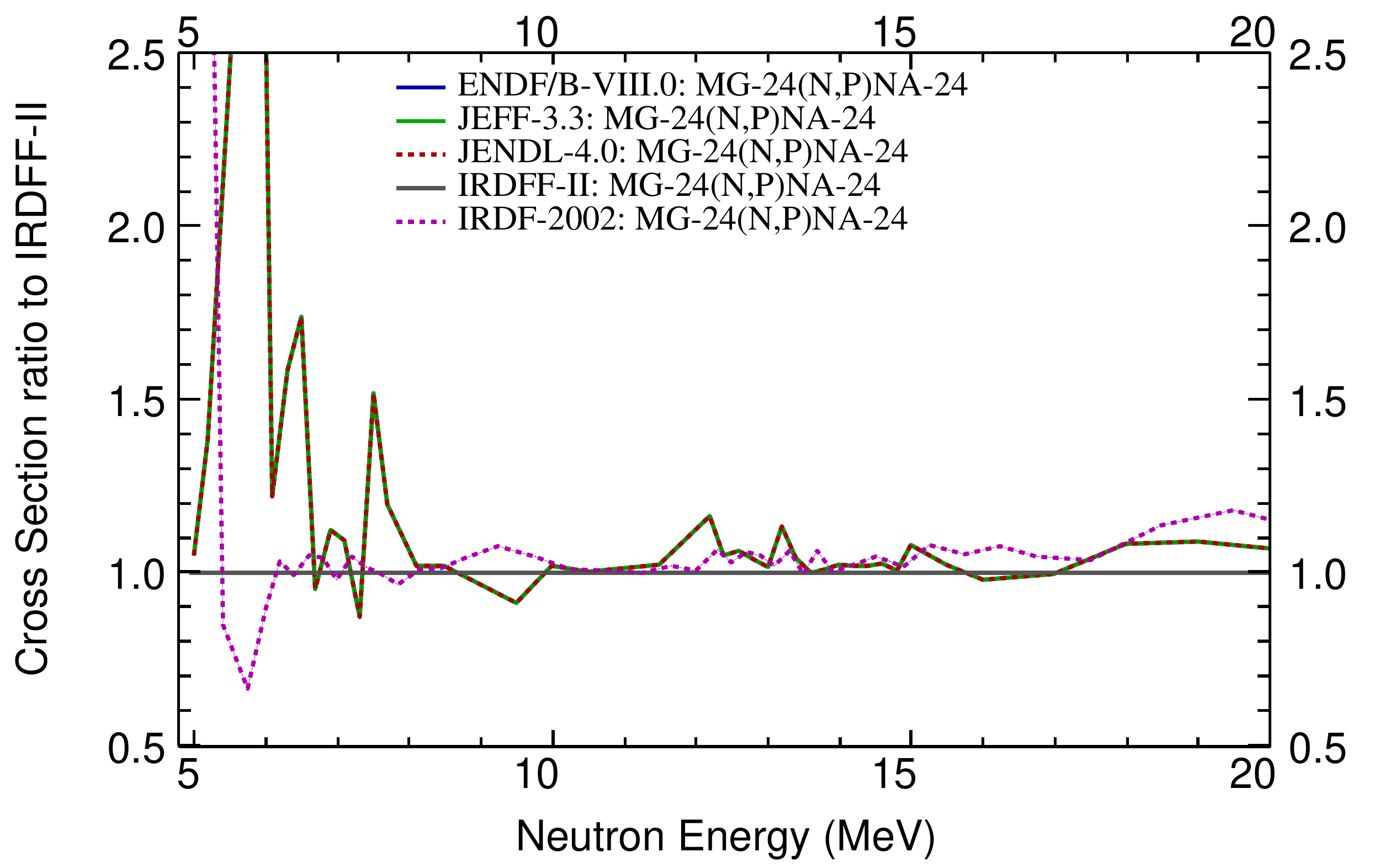}}
\vspace{-3mm}
\caption{(Color online) $^{24}$Mg(n,p)$^{24}$Na cross-section evaluation in \mbox{IRDFF-II} library relative to other data.}
\label{pMg24}
\vspace{-3mm}
\end{figure}

The $^{24}$Mg(n,p)$^{24}$Na reaction cross sections and covariances  below 21~MeV were evaluated by Zolotarev~\cite{Zol08} in 2008. Extension to 60~MeV was done at the IAEA based on \mbox{TENDL-2010}, renormalizing the cross sections at 21~MeV for continuity. Comparison of the cross sections with experimental data is shown in Fig.~\ref{pMg24}(a). The figure also shows the comparison between \mbox{IRDF-2002} and \mbox{IRDFF-II}, which remained unchanged from version \mbox{IRDFF-v1.05}. Cross sections in different evaluated libraries differ significantly near threshold as shown in Fig.~\ref{pMg24}(b), where the ratios of the various evaluations to the \mbox{IRDFF-II} recommended cross section are given. These differences may be important if this reaction monitor is used in reactor spectra.

\begin{figure}[!thb]
%\vspace{-4mm}
\subfigure[~Cross-section correlation matrix.]
{\includegraphics[width=\columnwidth]{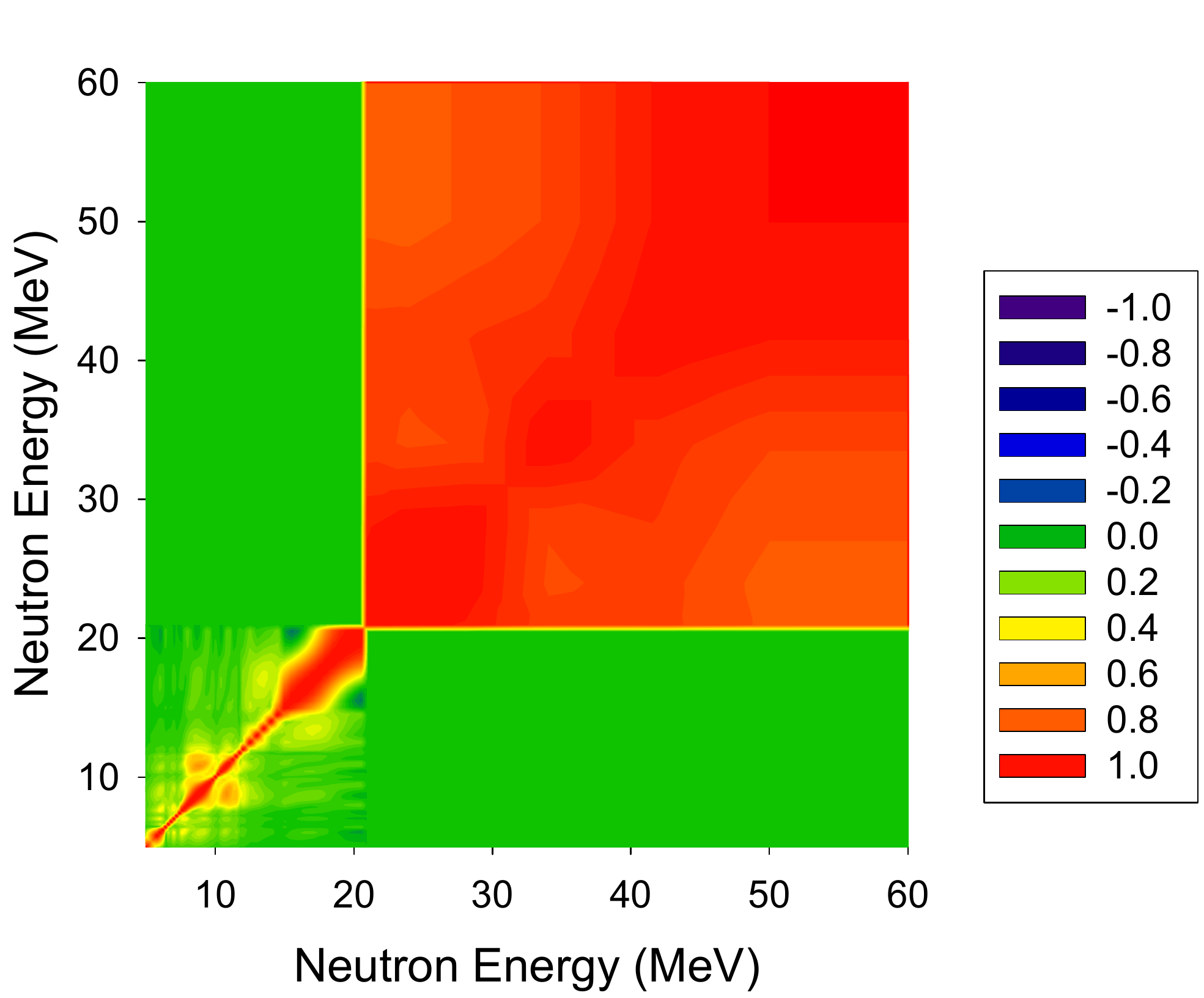}}
\subfigure[~One-sigma uncertainties in \% ($\equiv 100\times \frac{\sqrt{cov(i,i)}}{\mu_i}$), being $\mu_i$ the corresponding cross-section mean value.]
{\includegraphics[width=\columnwidth]{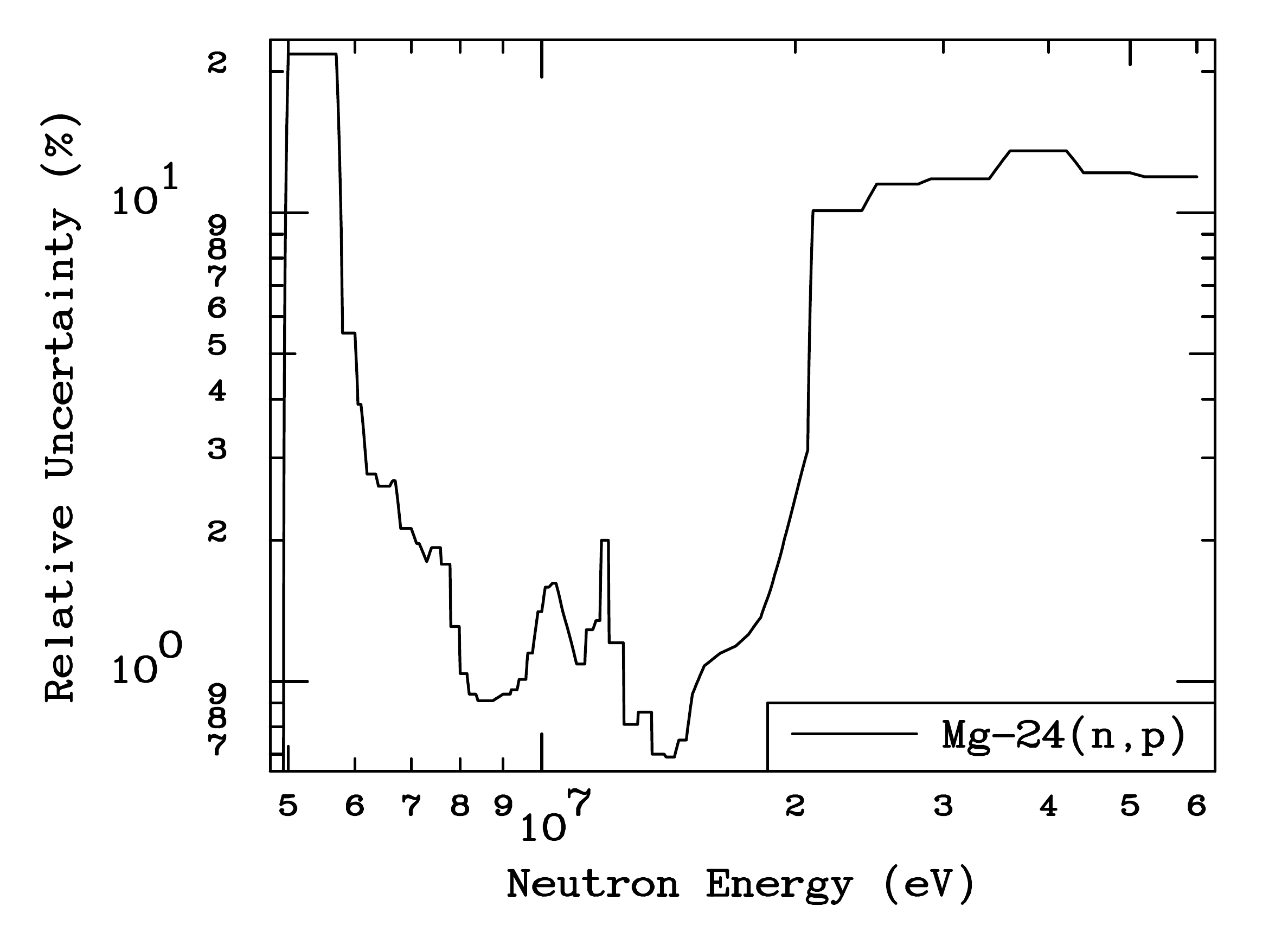}}
\vspace{-3mm}
\caption{(Color online) Uncertainties and correlations of the $^{24}$Mg(n,p)$^{24}$Na cross-section evaluation in \mbox{IRDFF-II} library.}
\label{pMg24-unc}
%\vspace{-3mm}
\end{figure}

The plot of the correlation matrix for the $^{24}$Mg(n,p)$^{24}$Na reaction is shown in Fig.~\ref{pMg24-unc}(a). The plot of the 1-sigma uncertainty is shown in Fig.~\ref{pMg24-unc}(b).

Reactions on the heavier isotopes of Mg may lead to the same reaction product $^{24}$Na. In particular, the cross sections for the $^{25}$Mg(n,np)$^{24}$Na reaction rises sharply above 18~MeV. The use of the $^{\mbox{nat}}$Mg(n,X)$^{24}$Na reaction cross sections, which are present in the \mbox{IRDFF-II} library, is strongly recommended, especially for neutron fields that have a significant fraction of neutrons with energies above 18~MeV. However, the user should be aware that the minor contributing channels represent a correction and do not meet the same dosimetry quality standards.
\vfill

\newpage
\subsubsection{$^{\mbox{nat}}$Mg(n,X)$^{24}\!$Na}
\begin{figure}[!hbtp]
\includegraphics[width=\columnwidth]{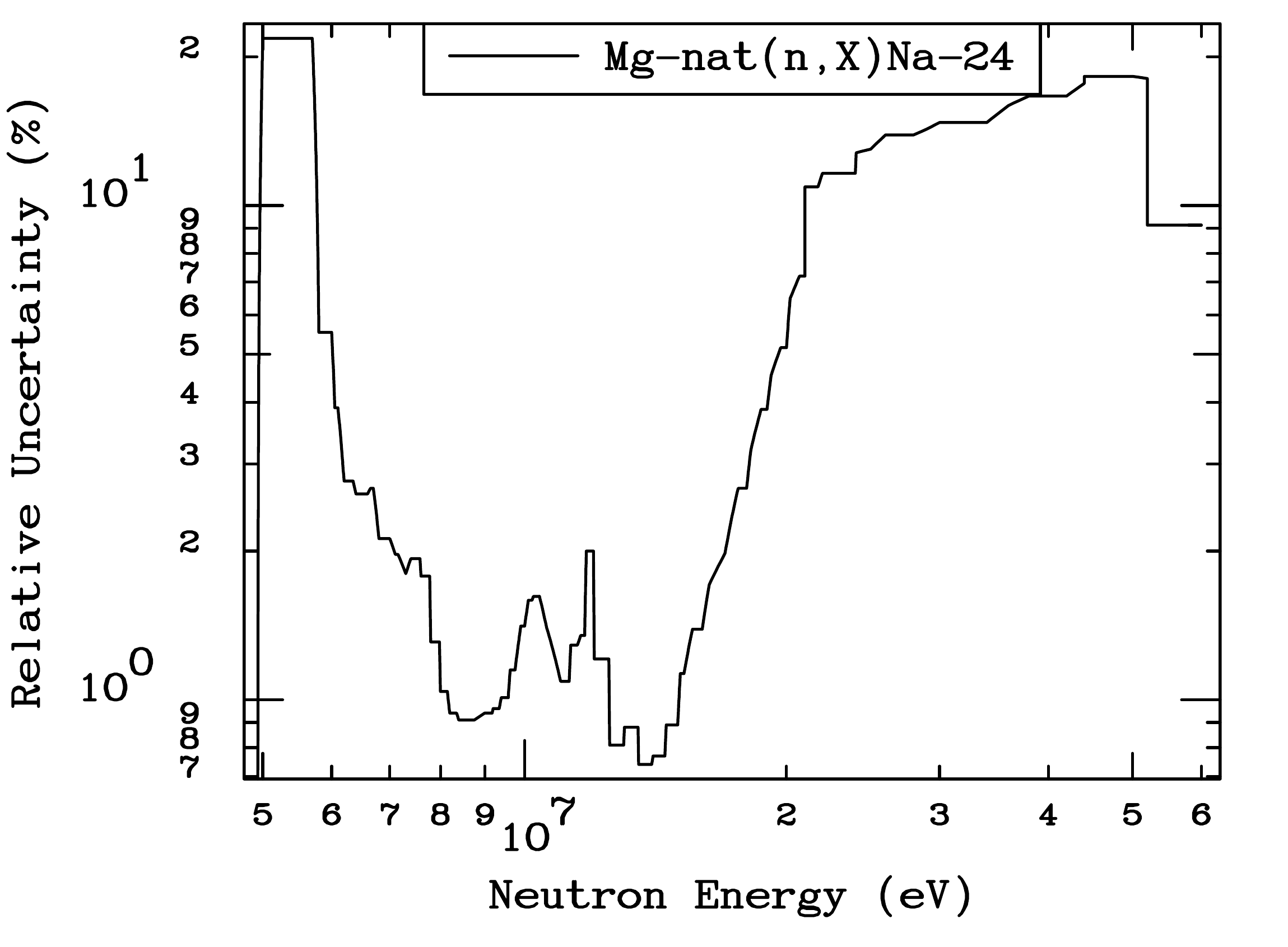}
\caption{$^{nat}$Mg(n,X)$^{24}$Na cross-section 1-sigma uncertainties.}
\label{Fig:xMg-Na24-unc}
\end{figure}

The major contribution to the $^{24}$Na production comes from the $^{24}$Mg(n,p)$^{24}$Na reaction, which is present in the \mbox{IRDFF-II} library, supplemented by the contributions of other channels, including $^{25}$Mg(n,np), $^{25}$Mg(n,d), $^{26}$Mg(n,2np), $^{25}$Mg(n,nd) and $^{25}$Mg(n,t) that are taken from the \mbox{TENDL-2015} library. The contribution from the $^{25}$Mg(n,np) rises sharply above about 18~MeV. The uncertainty and the covariance matrix below this energy are practically the same as for $^{24}$Mg(n,p)$^{24}$Na; see the uncertainties plotted in Fig.~\ref{Fig:xMg-Na24-unc}. It is strongly recommended to use $^{\mbox{nat}}$Mg(n,X)$^{24}$Na reaction cross sections, especially in neutron fields with a significant fraction of neutrons with energies above 15~MeV.
\begin{figure}[!hbtp]
\includegraphics[width=\columnwidth]{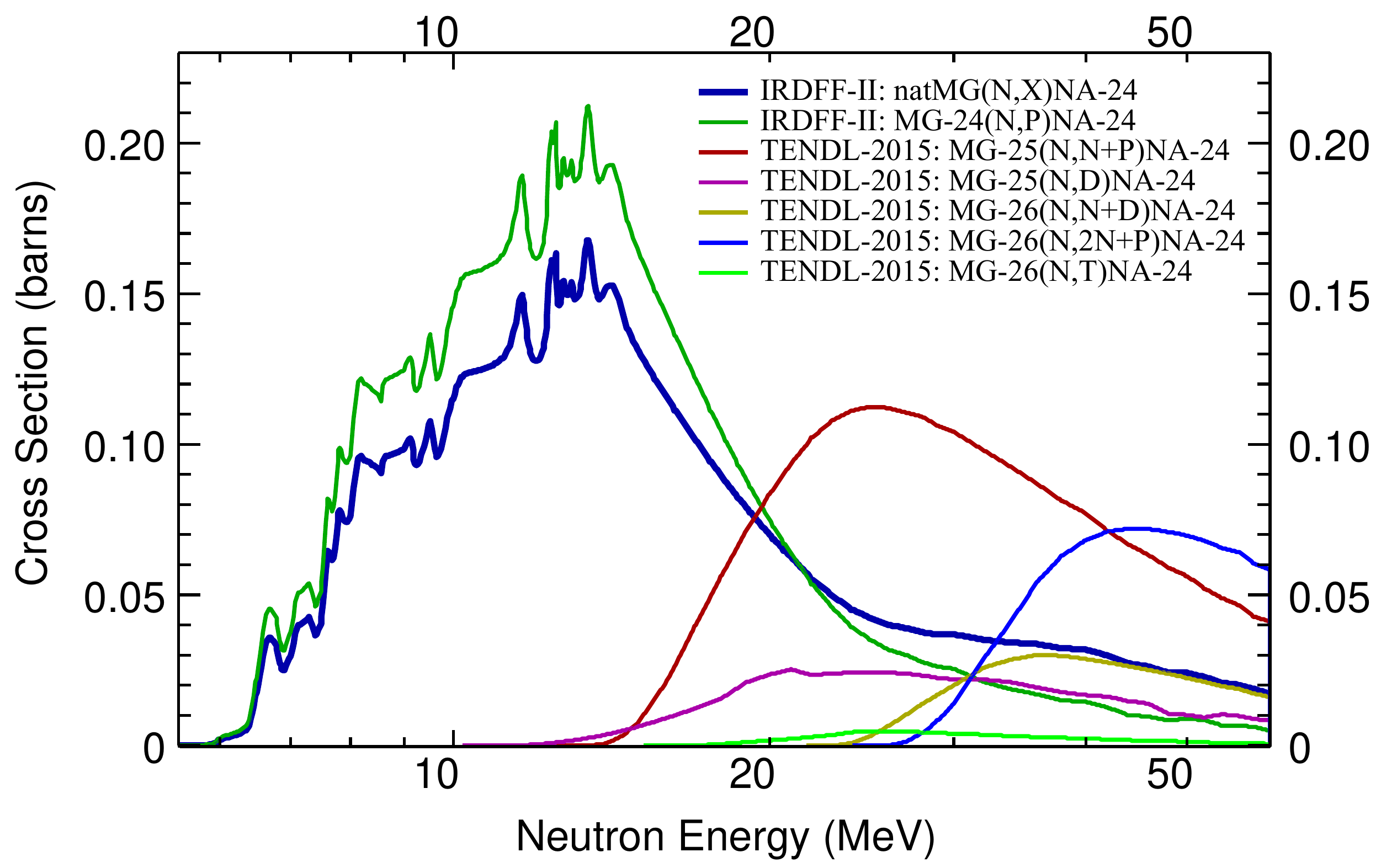}
\caption{(Color online) Contributing reactions to the $^{24}$Na production from neutrons incident on $^{\mbox{nat}}$Mg target.  Note that the cross sections are not scaled by the isotopic abundance.}
\label{Fig:xMg-Na24}
\end{figure}

\newpage
\subsubsection{$^{27}\!$Al(n,p)$^{27}\!$Mg}
\begin{figure}[!thb]
\vspace{-2mm}
\subfigure[~Comparison to selected experimental data from EXFOR \cite{EXF08}.]
{\includegraphics[width=\columnwidth]{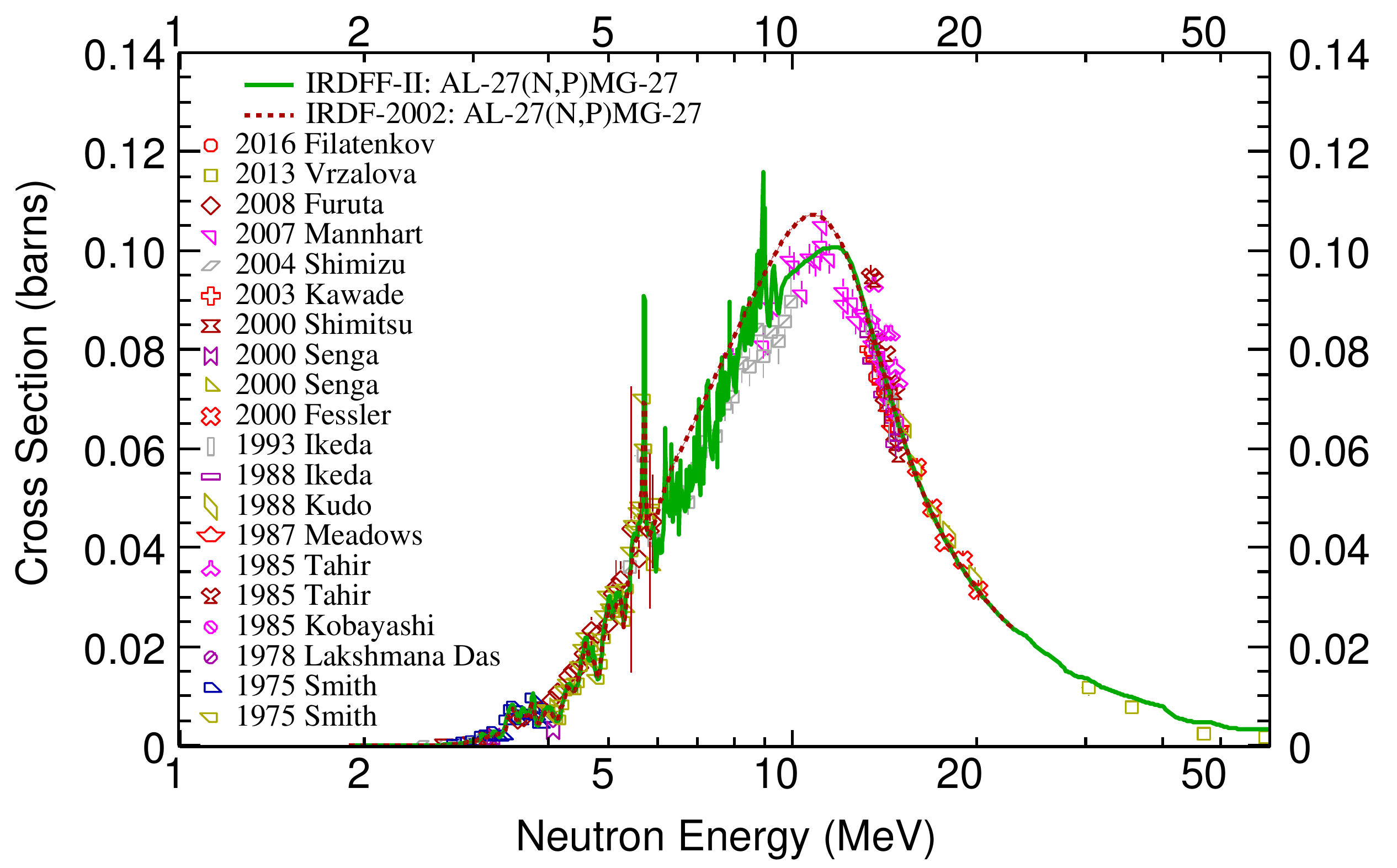}}
\subfigure[~Ratio of cross-section evaluations to \mbox{IRDFF-II} evaluation.]
{\includegraphics[width=\columnwidth]{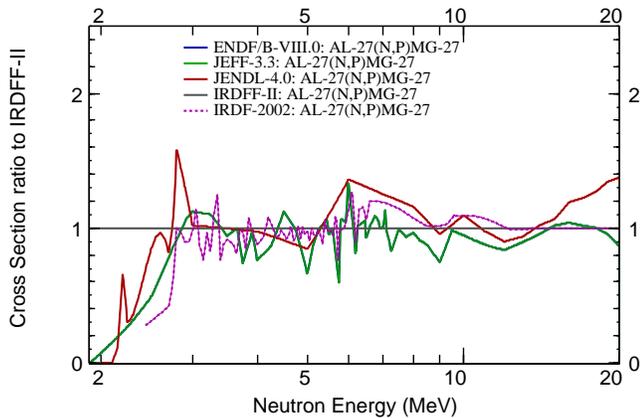}}
\vspace{-3mm}
\caption{(Color online) $^{27}$Al(n,p)$^{27}$Mg cross-section evaluation in \mbox{IRDFF-II} library relative to other data.}
\label{pAl27}
\vspace{-3mm}
\end{figure}

The $^{27}$Al(n,p)$^{27}$Mg reaction cross sections were evaluated by Zolotarev up to 22~MeV in 2004~\cite{Zol04}. Extension to 60~MeV was done at the IAEA based on \mbox{TENDL-2010}, renormalizing the cross sections at 22~MeV for continuity. Comparison of the cross sections with experimental data is shown in Fig.~\ref{pAl27}(a). The figure also shows the comparison between \mbox{IRDF-2002} and \mbox{IRDFF-II}, which remained unchanged from version \mbox{IRDFF-v1.05}.

Cross sections in different evaluated libraries are shown in Fig.~\ref{pAl27}(b) as ratios of the various evaluations to the \mbox{IRDFF-II} recommended cross section.

\begin{figure}[!thb]
\vspace{-4mm}
\subfigure[~Cross-section correlation matrix.]
{\includegraphics[width=\columnwidth]{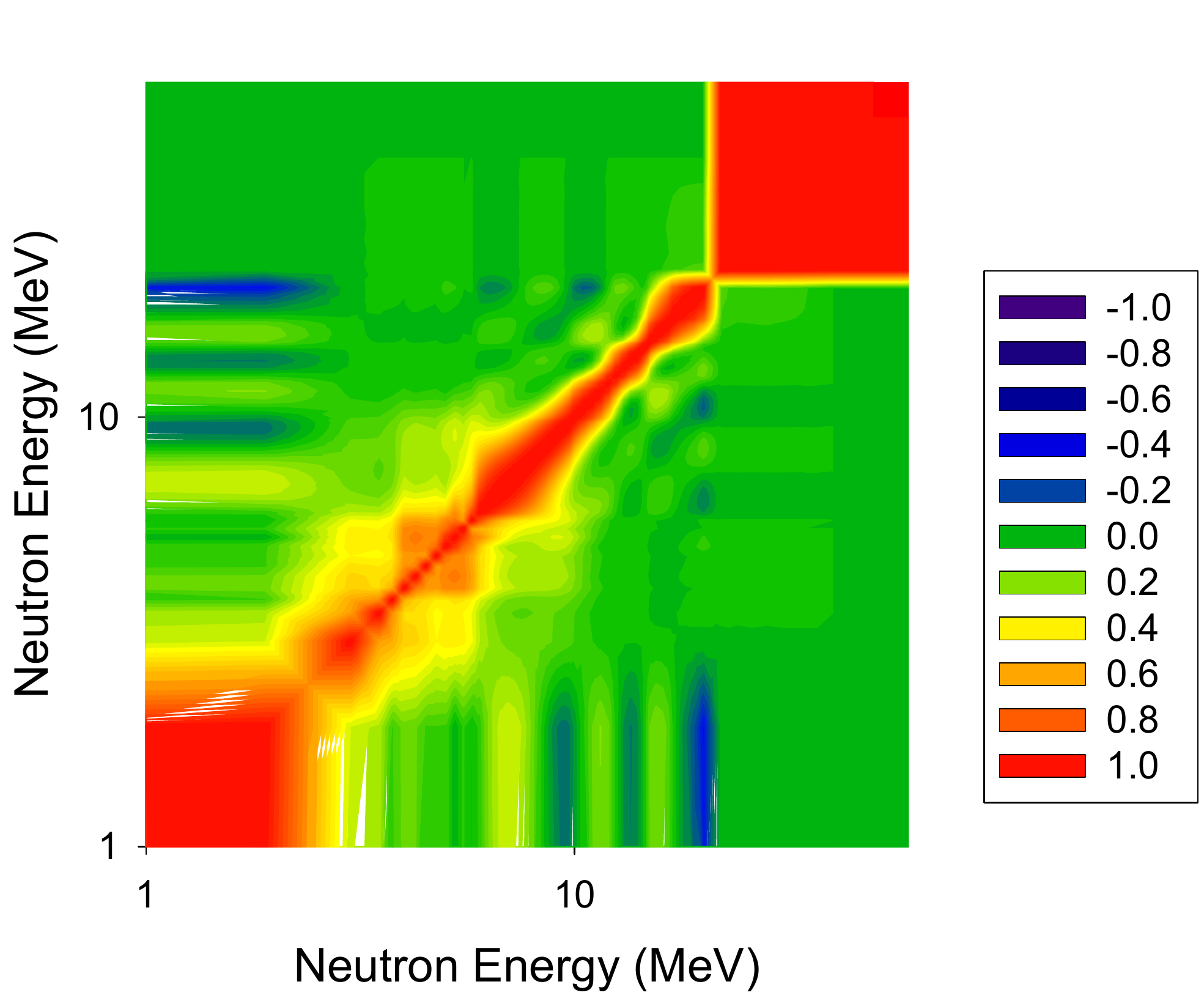}}
\subfigure[~One-sigma uncertainties in \% ($\equiv 100\times \frac{\sqrt{cov(i,i)}}{\mu_i}$), being $\mu_i$ the corresponding cross-section mean value.]
{\includegraphics[width=\columnwidth]{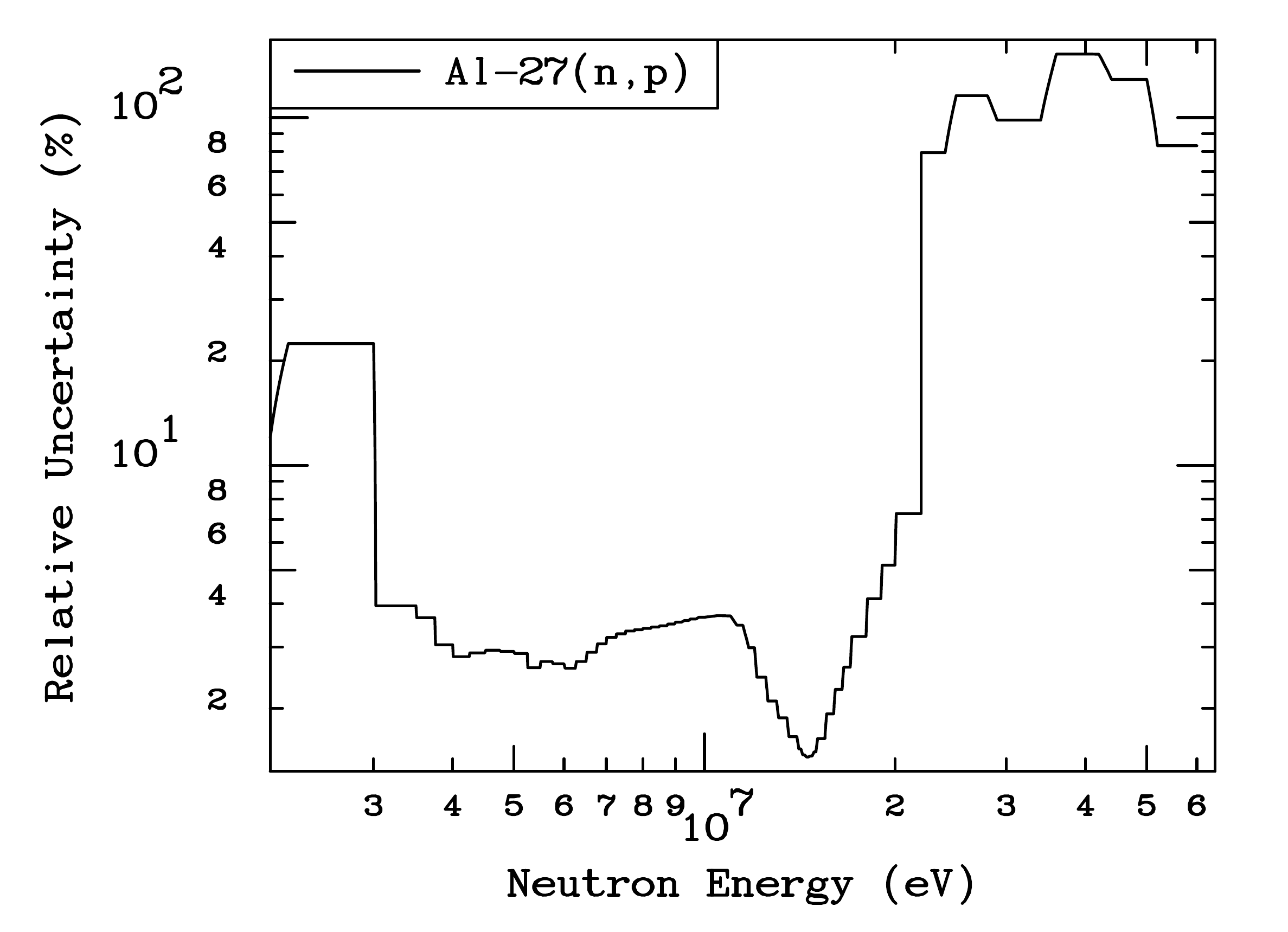}}
\vspace{-3mm}
\caption{(Color online) Uncertainties and correlations of the $^{27}$Al(n,p)$^{27}$Mg cross-section evaluation in \mbox{IRDFF-II} library.}
\label{pAl27-unc}
\vspace{-3mm}
\end{figure}

The plot of the correlation matrix for the $^{27}$Al(n,p)$^{27}$Mg reaction is shown in Fig.~\ref{pAl27-unc}(a). The plot of the 1-sigma uncertainty is shown in Fig.~\ref{pAl27-unc}(b).
\vfill

\newpage
\subsubsection{$^{27}\!$Al(n,$\alpha$)$^{24}\!$Na}
\begin{figure}[!thb]
%\vspace{-2mm}
\subfigure[~Comparison to selected experimental data from EXFOR \cite{EXF08}.]
{\includegraphics[width=\columnwidth]{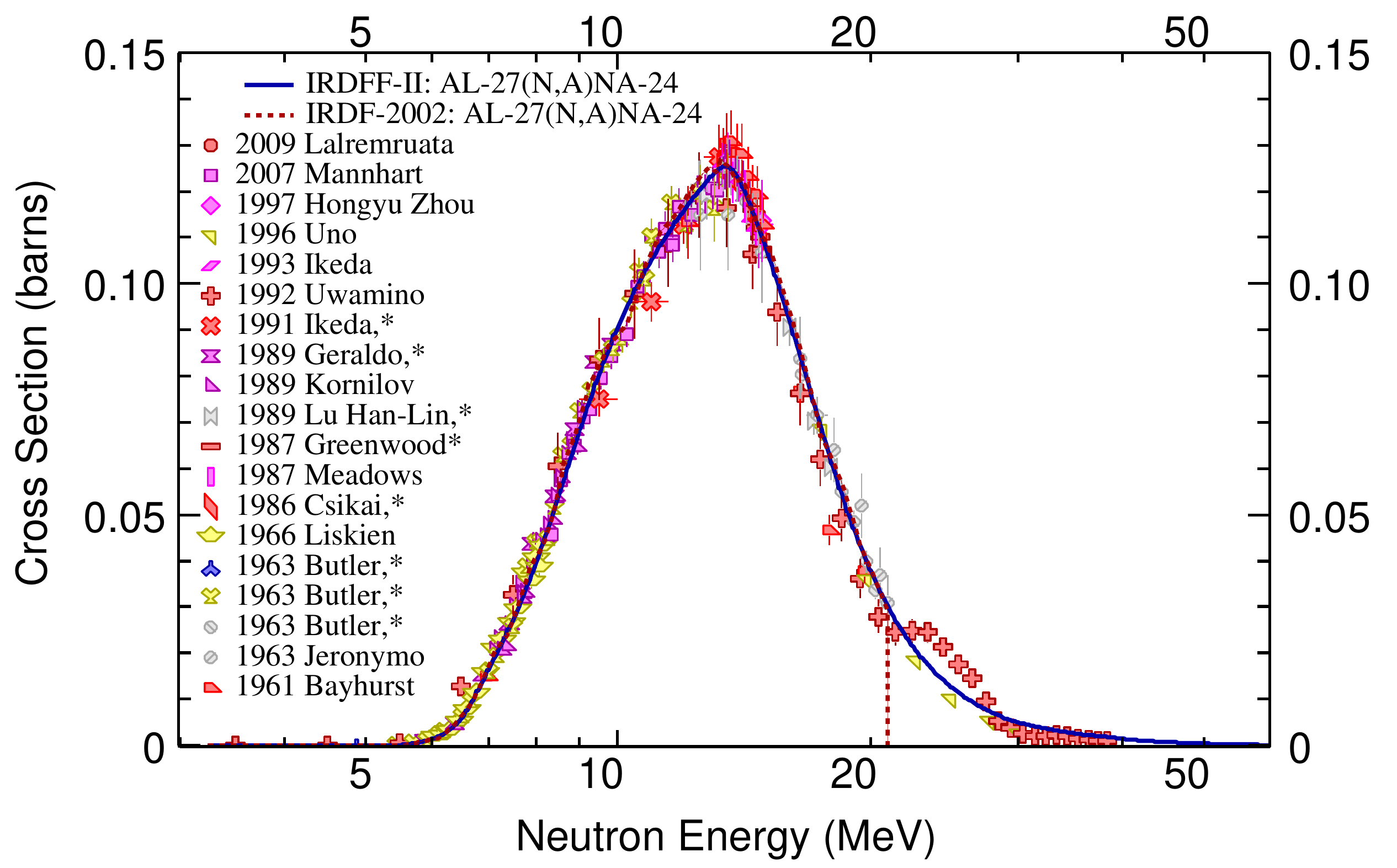}}
\subfigure[~Ratio of cross-section evaluations to \mbox{IRDFF-II} evaluation.]
{\includegraphics[width=\columnwidth]{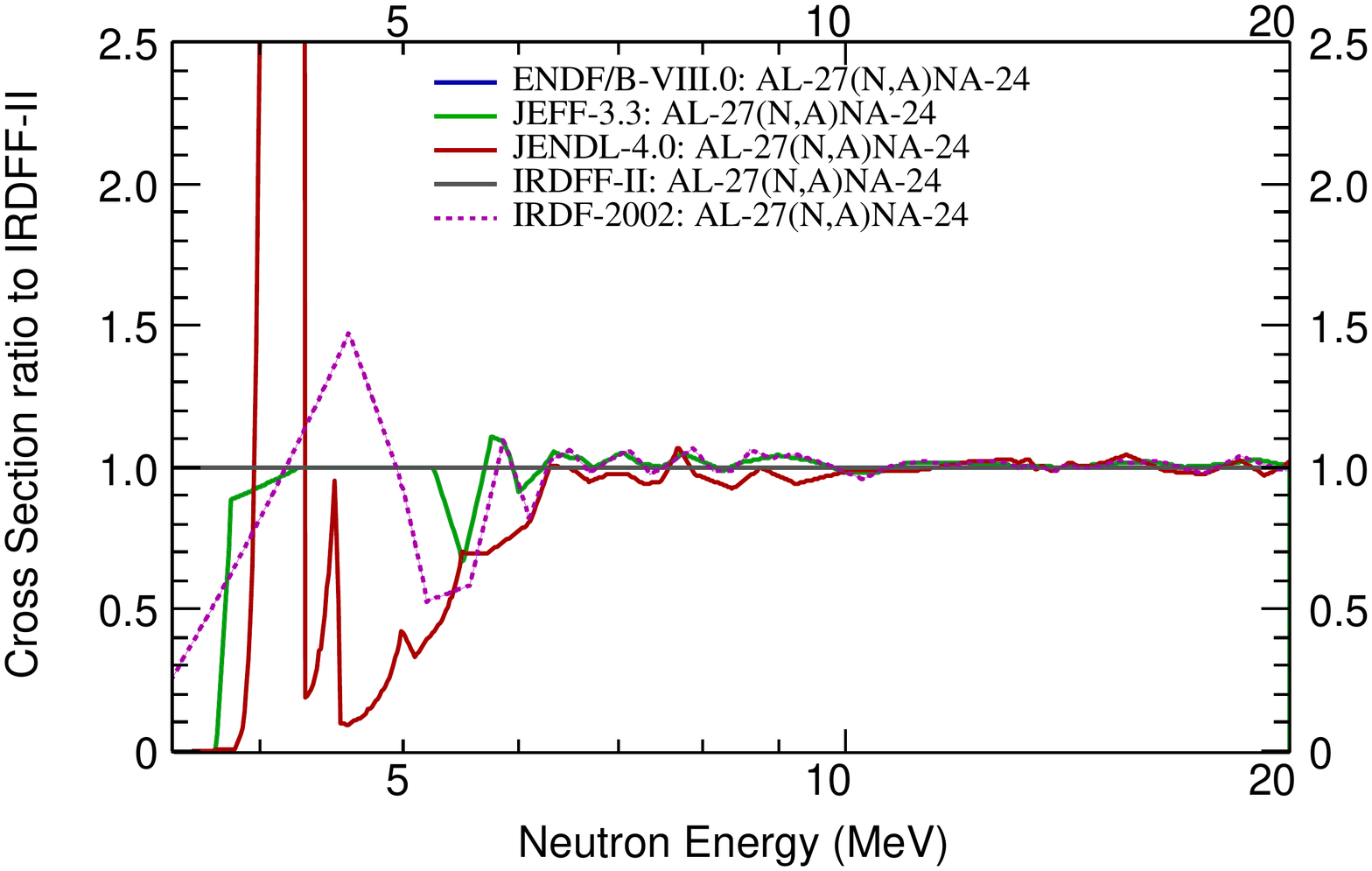}}
\vspace{-3mm}
\caption{(Color online) $^{27}$Al(n,$\alpha$)$^{24}$Na cross-section evaluation in \mbox{IRDFF-II} library relative to other data.}
\label{aAl27}
%\vspace{-3mm}
\end{figure}

In 2004 the $^{27}$Al(n,$\alpha$)$^{24}$Na reaction cross sections were evaluated by Zolotarev up to 40~MeV~\cite{Zol09}, but the cross sections had unphysical behaviour above 30~MeV. Extension between 30~MeV and 60~MeV was done at the IAEA based on \mbox{TENDL-2010}, renormalizing the cross sections at 30~MeV for continuity. Comparison of the cross sections with experimental data is shown in Fig.~\ref{aAl27}(a). The figure also shows the comparison between \mbox{IRDF-2002} and \mbox{IRDFF-II}, which remained unchanged from version \mbox{IRDFF-v1.05}. Cross sections in different evaluated libraries are shown in Fig.~\ref{aAl27}(b) as ratios of the various evaluations to the \mbox{IRDFF-II} recommended cross section.

\begin{figure}[!thb]
%\vspace{-4mm}
\subfigure[~Cross-section correlation matrix.]
{\includegraphics[width=\columnwidth]{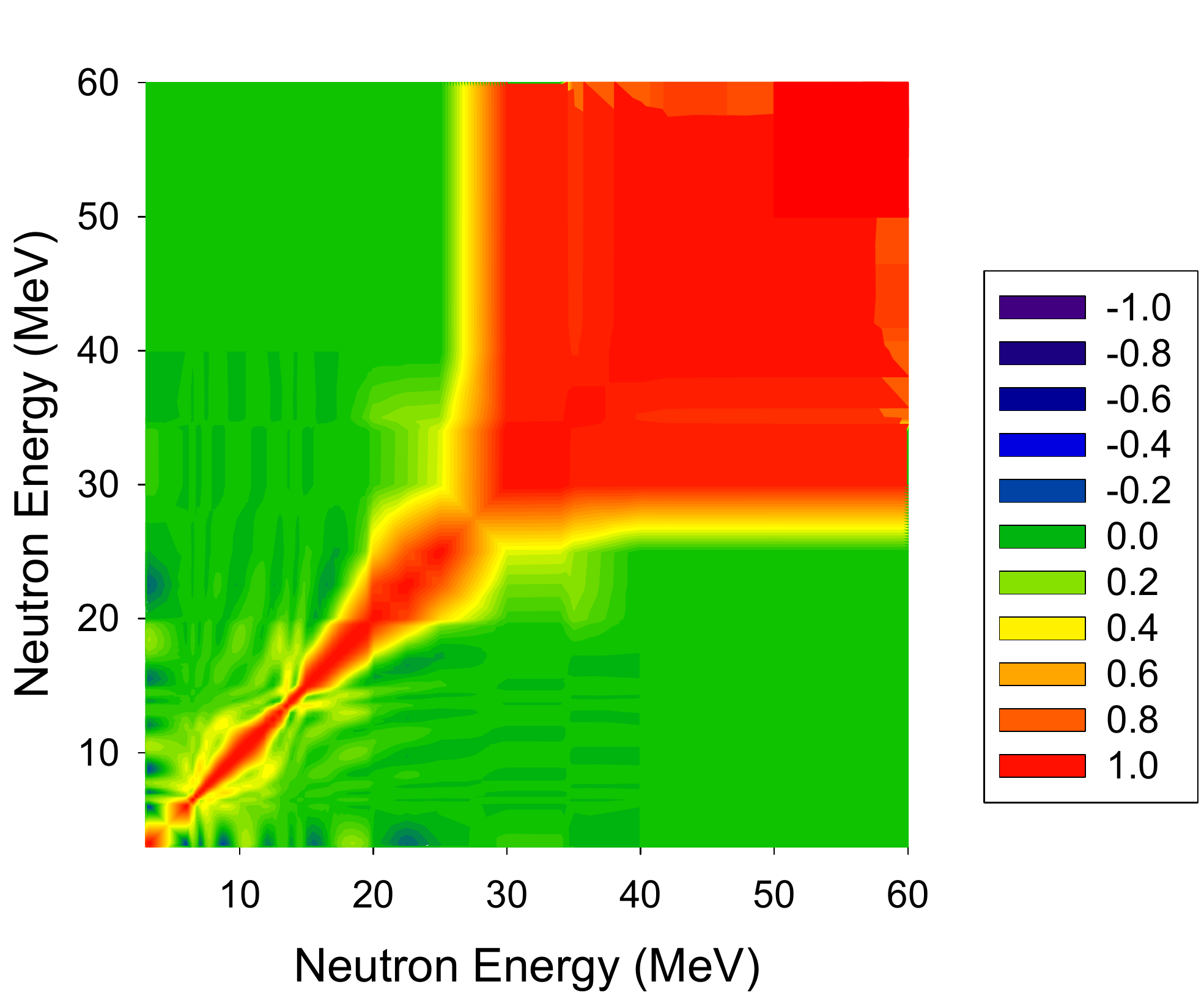}}
\subfigure[~One-sigma uncertainties in \% ($\equiv 100\times \frac{\sqrt{cov(i,i)}}{\mu_i}$), being $\mu_i$ the corresponding cross-section mean value.]
{\includegraphics[width=\columnwidth]{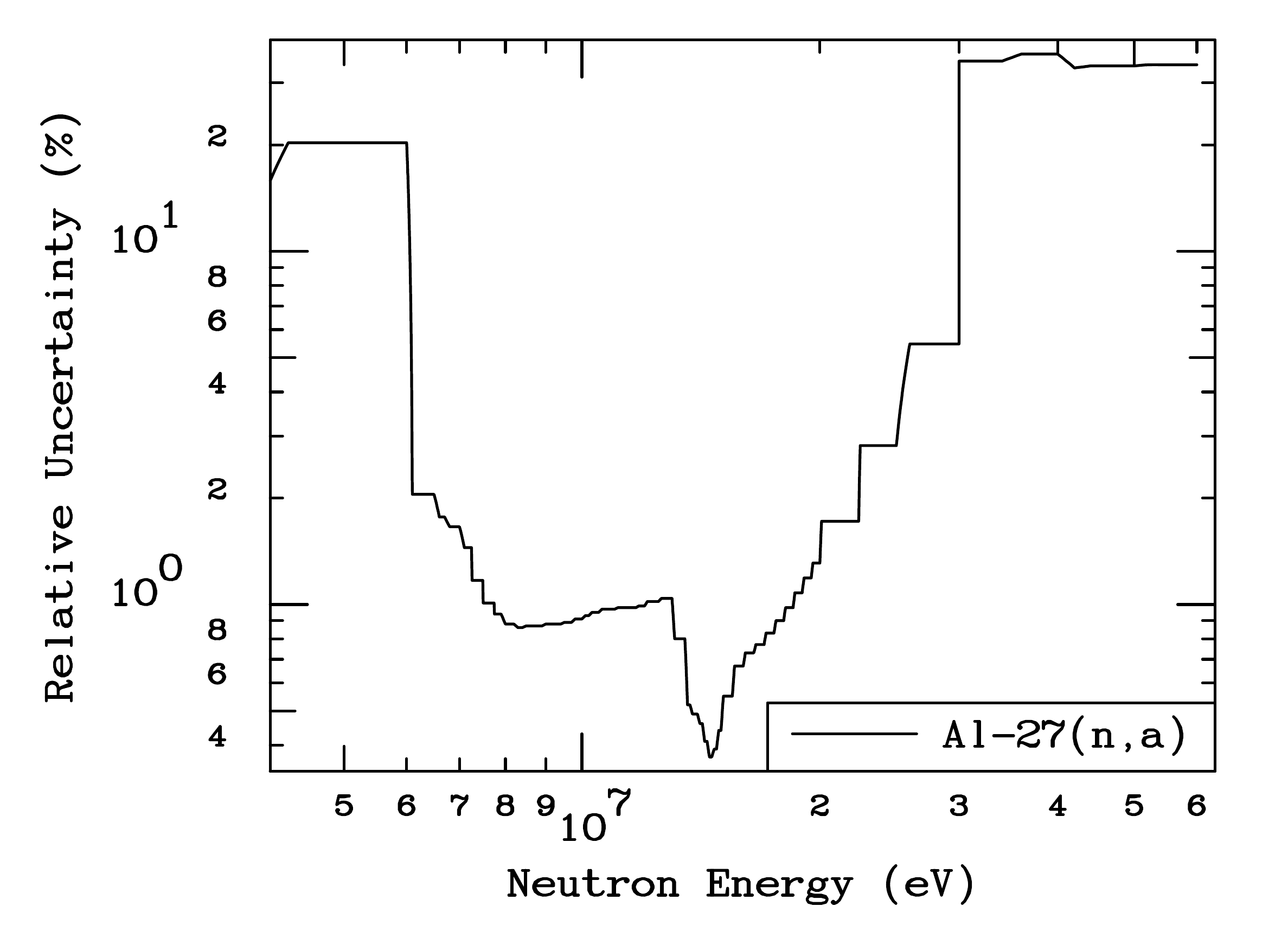}}
\vspace{-3mm}
\caption{(Color online) Uncertainties and correlations of the $^{27}$Al(n,$\alpha$)$^{24}$Na cross-section evaluation in \mbox{IRDFF-II} library.}
\label{aAl27-unc}
%\vspace{-3mm}
\end{figure}

The plot of the correlation matrix for the $^{27}$Al(n,$\alpha$)$^{24}$Na reaction is shown in Fig.~\ref{aAl27-unc}(a). The plot of the 1-sigma uncertainty is shown in Fig.~\ref{aAl27-unc}(b).

\newpage
Aluminium is a convenient monitor for neutron dosimetry because it is a mono-isotopic element. However, at energies above 30~MeV there appear competing reactions that produce the same $^{24}$Na residual. It is strongly recommended to use $^{27}$Al(n,X)$^{24}$Na reaction cross sections, which are provided in IRDFF-II, especially in neutron fields with a significant fraction of neutrons with energies above 30~MeV.

\newpage
\subsubsection{$^{47}\!$Ti(n,p)$^{47}\!$Sc}
\begin{figure}[!thb]
%\vspace{-2mm}
\subfigure[~Comparison to selected experimental data from EXFOR \cite{EXF08}.]
{\includegraphics[width=\columnwidth]{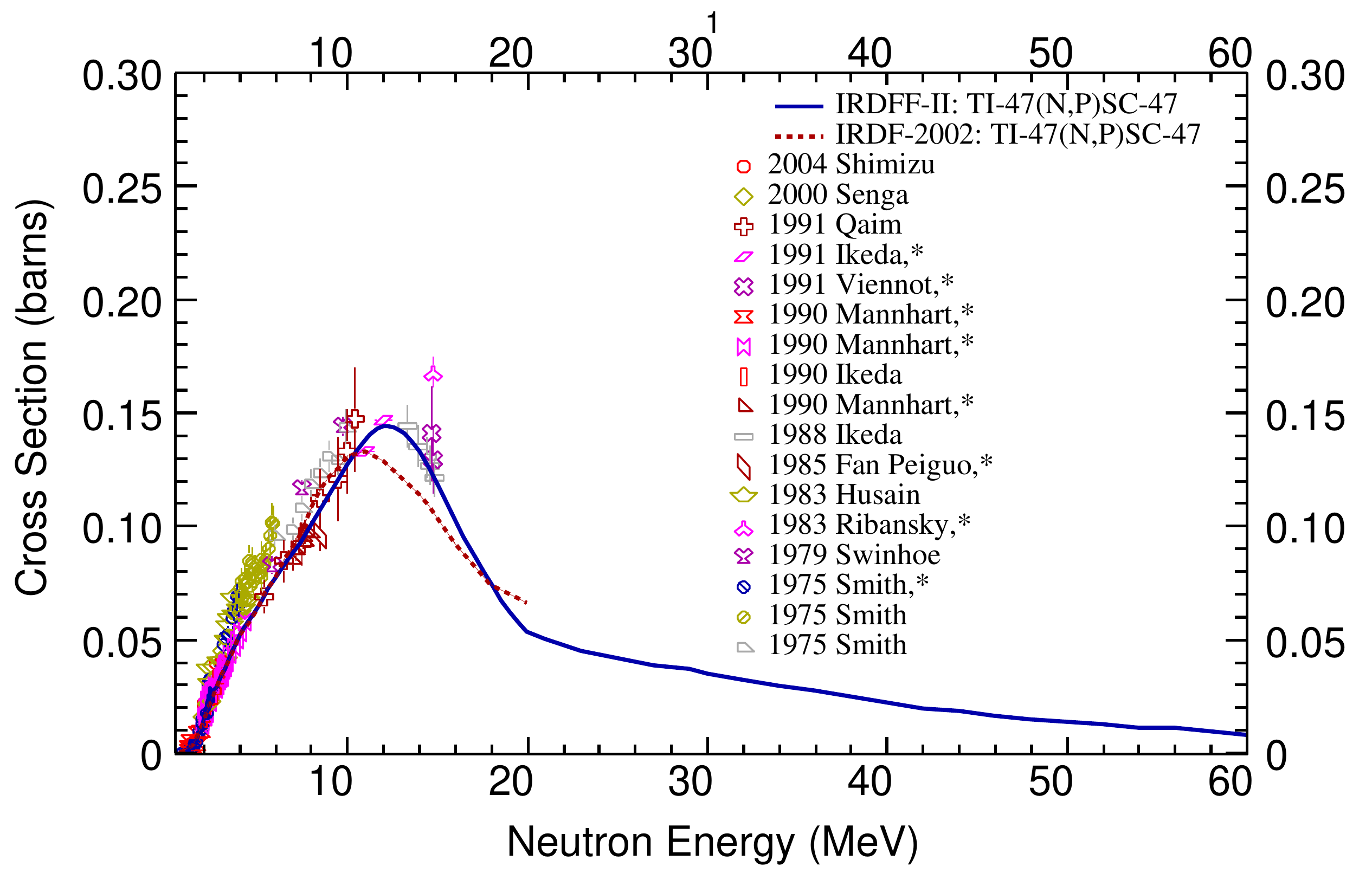}}
\subfigure[~Ratio of cross-section evaluations to \mbox{IRDFF-II} evaluation.]
{\includegraphics[width=\columnwidth]{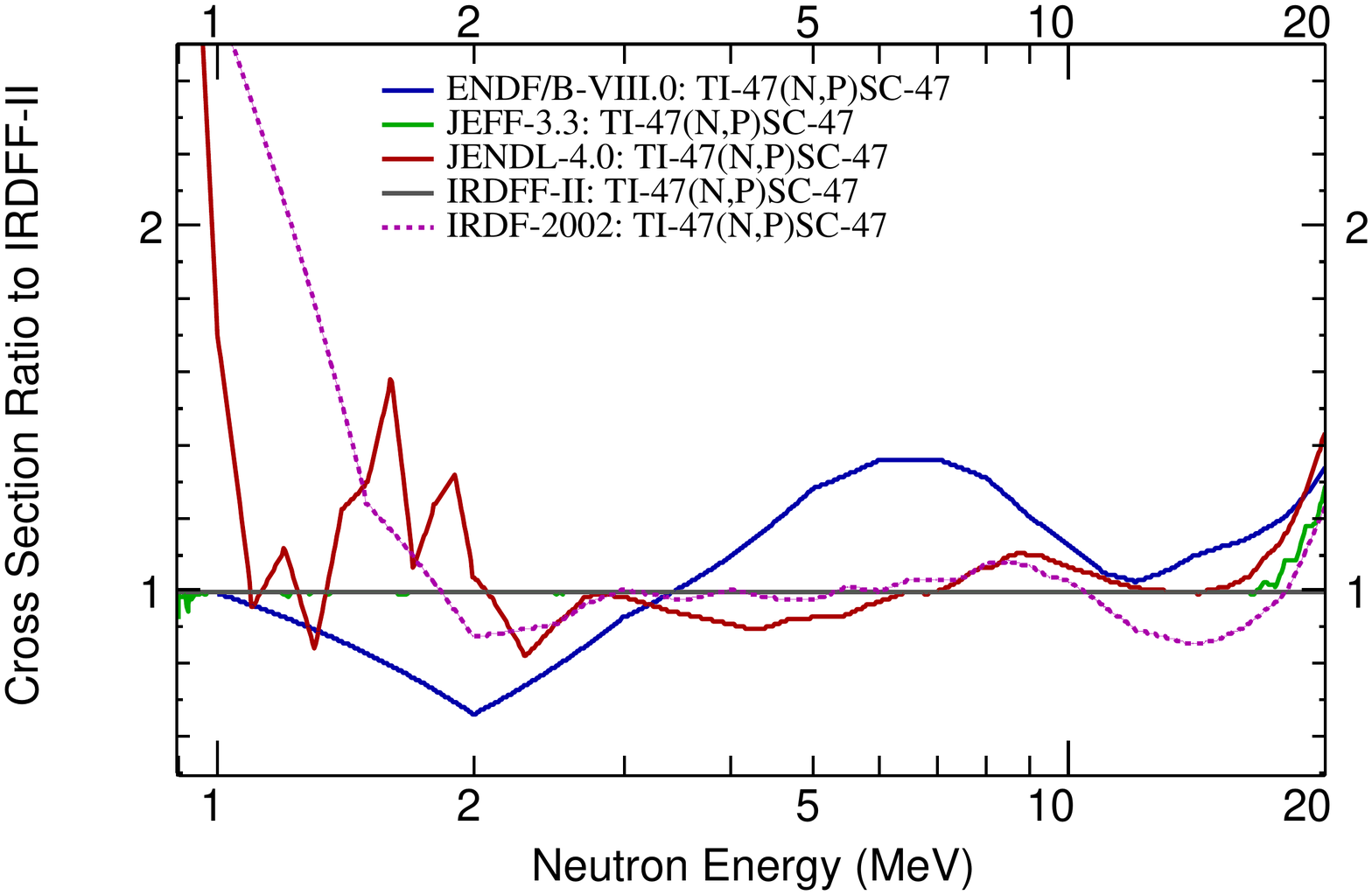}}
%\vspace{-3mm}
\caption{(Color online) $^{47}$Ti(n,p)$^{47}$Sc cross-section evaluation in \mbox{IRDFF-II} library relative to other data.}
\label{pTi47}
%\vspace{-3mm}
\end{figure}

The $^{47}$Ti(n,p)$^{47}$Sc reaction cross sections were evaluated by Zolotarev up to 20~MeV in 2002 for the Russian Dosimetry Library RRDF-2002 within the BROND-2 project. Extension to 60~MeV was done at the IAEA based on \mbox{TENDL-2011}, renormalizing the cross sections at 22~MeV for continuity. Comparison of the cross sections with experimental data is shown in Fig.~\ref{pTi47}(a). The figure also shows the comparison between \mbox{IRDF-2002} and \mbox{IRDFF-II}, which remained unchanged from version \mbox{IRDFF-v1.05}.
Cross sections in different evaluated libraries are shown in Fig.~\ref{pTi47}(b) as ratios of the various evaluations to the \mbox{IRDFF-II} recommended cross section.

\begin{figure}[!thb]
%\vspace{-4mm}
\subfigure[~Cross-section correlation matrix.]
{\includegraphics[width=\columnwidth]{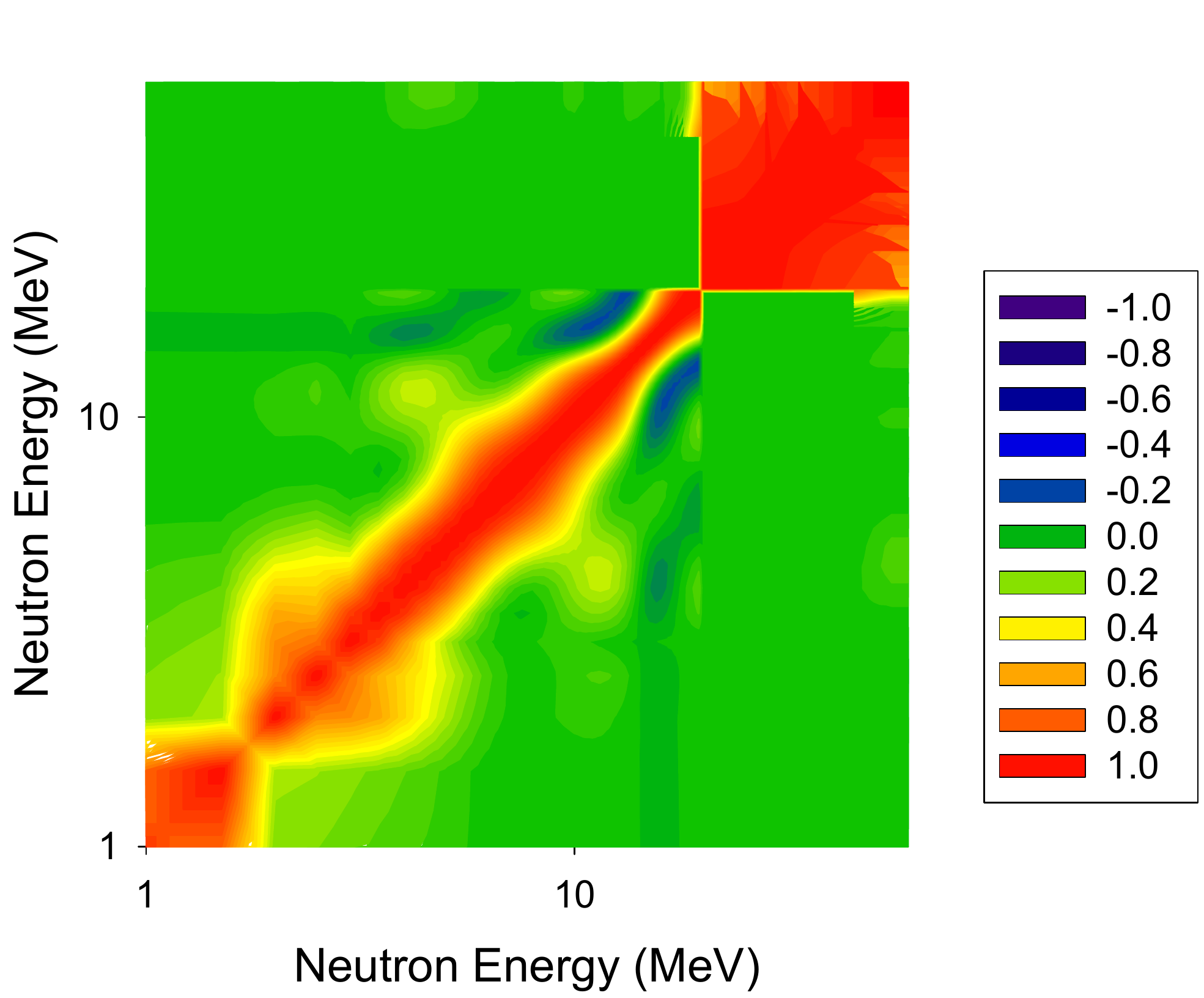}}
\subfigure[~One-sigma uncertainties in \% ($\equiv 100\times \frac{\sqrt{cov(i,i)}}{\mu_i}$), being $\mu_i$ the corresponding cross-section mean value.]
{\includegraphics[width=\columnwidth]{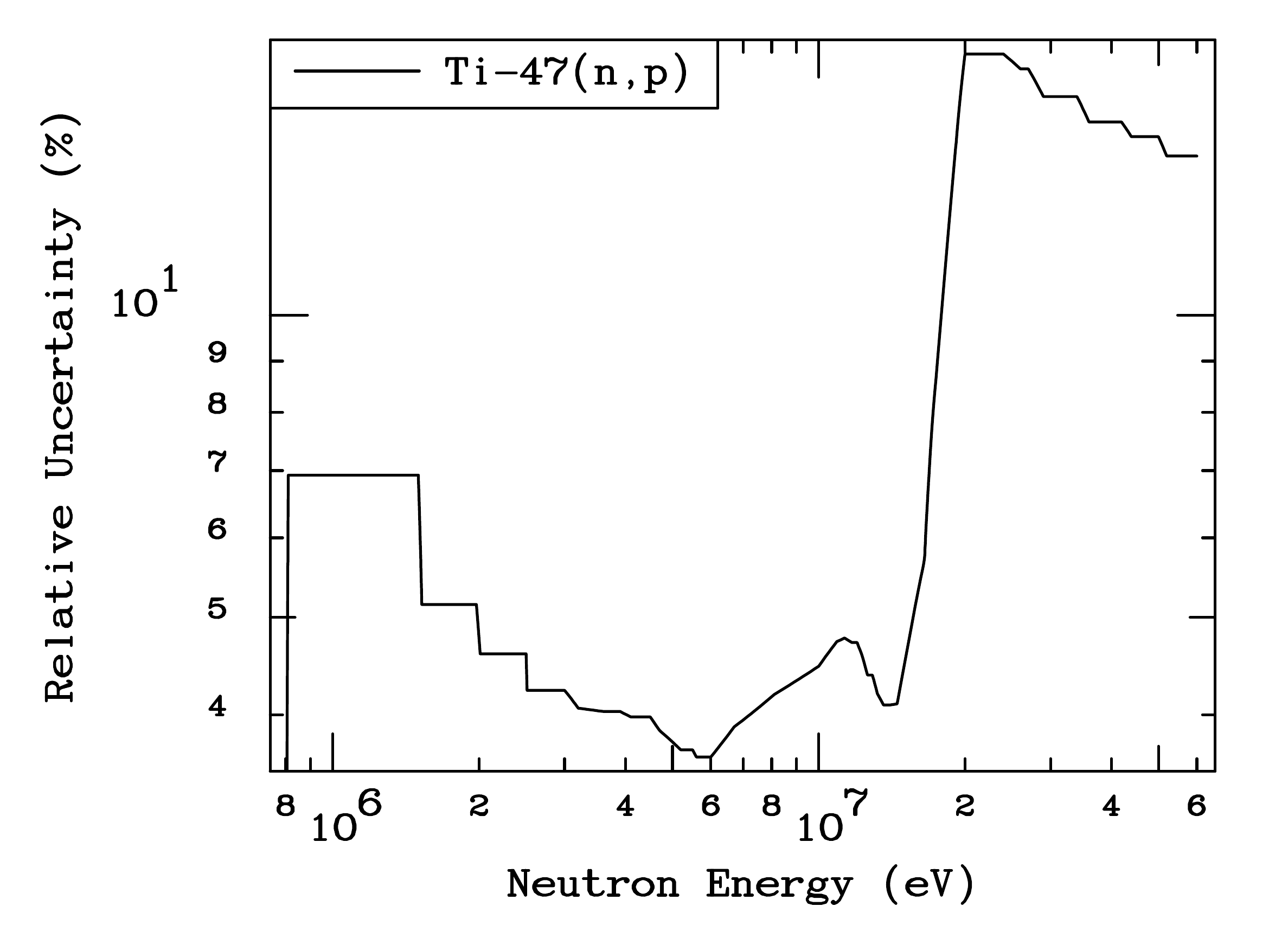}}
%\vspace{-3mm}
\caption{(Color online) Uncertainties and correlations of the $^{47}$Ti(n,p)$^{47}$Sc cross-section evaluation in \mbox{IRDFF-II} library.}
\label{pTi47-unc}
%\vspace{-3mm}
\end{figure}

The plot of the correlation matrix for the $^{47}$Ti(n,p)$^{47}$Sc reaction is shown in Fig.~\ref{pTi47-unc}(a). The plot of the 1-sigma uncertainty is shown in Fig.~\ref{pTi47-unc}(b).

Titanium is a monitor elemental material that has a large number of naturally occuring isotopes that lead to the same residual. The $^{47}$Ti(n,p)$^{47}$Sc reaction is just one example, where the residual $^{47}$Sc can be produced from the heavier isotopes in reactions like $^{48}$Ti(n,np)$^{47}$Sc, \textit{etc}. The heavier isotopes begin to contribute at energies around 14~MeV. For this reason, Reactions like $^{\mbox{nat}}$Ti(n,X)$^{47}$Sc were introduced into the \mbox{IRDFF-II} library, as illustrated in the section below.

\newpage
\begin{figure}[!hbtp]
\vspace{3mm}
\includegraphics[width=\columnwidth]{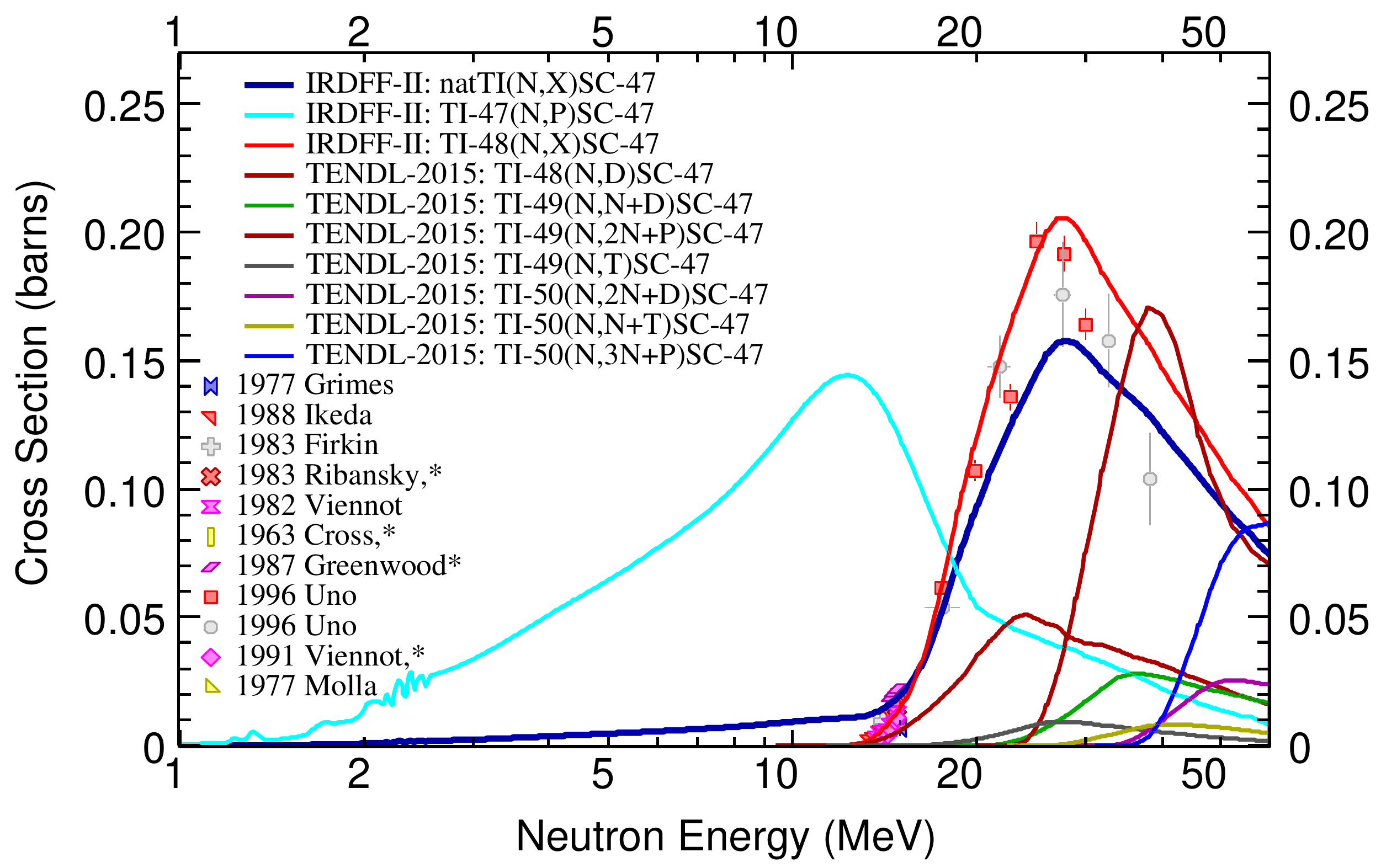}
\caption{(Color online) Contributing reactions to the $^{47}$Sc production from neutrons incident on $^{\mbox{nat}}$Ti target. Note that the cross sections are not scaled by the isotopic abundance.}
\label{Fig:xTi-Sc47}
\end{figure}

\subsubsection{$^{\mbox{nat}}$Ti(n,X)$^{47}\!$Sc}
The major contribution to the $^{47}$Sc production comes from the $^{47}$Ti(n,p)$^{47}$Sc reaction, which is present in the \mbox{IRDFF-II} library, supplemented by the contributions of other channels, including $^{48}$Ti(n,np), $^{48}$Ti(n,d), $^{49}$Ti(n,2np), $^{49}$Ti(n,nd), $^{49}$Ti(n,t), $^{50}$Ti(n,2nd), $^{50}$Ti(n,nt) and $^{50}$Ti(n,3np)  that are taken from the \mbox{TENDL-2015} library. Relevant cross sections are compared to the available experimental data in Fig.~\ref{Fig:xTi-Sc47}. The contribution from the $^{48}$Ti(n,np+pn) rises sharply above about 14~MeV. The uncertainty and the covariance matrix below this energy are practically the same as for the $^{47}$Ti(n,p)$^{47}$Sc reaction. %; , see the uncertainties plotted in Fig.~\ref{Fig:xTi-Sc47-unc}.
It is strongly recommended to use $^{\mbox{nat}}$Ti(n,X)$^{47}$Sc reaction cross section for neutron dosimetry applications, especially in neutron fields with a significant fraction of neutrons with energies above 14~MeV.
%\begin{figure}[hbtp]
%\includegraphics[width=\columnwidth]{Figures/xTi-Sc47-unc.pdf}
%\caption{$^{nat}$Ti(n,X)$^{47}$Sc cross-section 1-sigma uncertainties.}
%\label{Fig:xTi-Sc47-unc}
%\end{figure}

\subsubsection{$^{59}\!$Co(n,2n)$^{58}\!$Co and $^{59}\!$Co(n,3n)$^{57}\!$Co}
The $^{59}$Co(n,2n)$^{58}$Co reaction cross sections and covariances in energy range up to 20~MeV were evaluated by Zolotarev~\cite{Zol09} in 2009. Extension to 60~MeV was done at the IAEA based on \mbox{TENDL-2010}, renormalizing the cross sections at 20~MeV for continuity.

\begin{figure}[!thb]
%\vspace{-2mm}
\subfigure[~Comparison to selected experimental data from EXFOR \cite{EXF08}.]
{\includegraphics[width=\columnwidth]{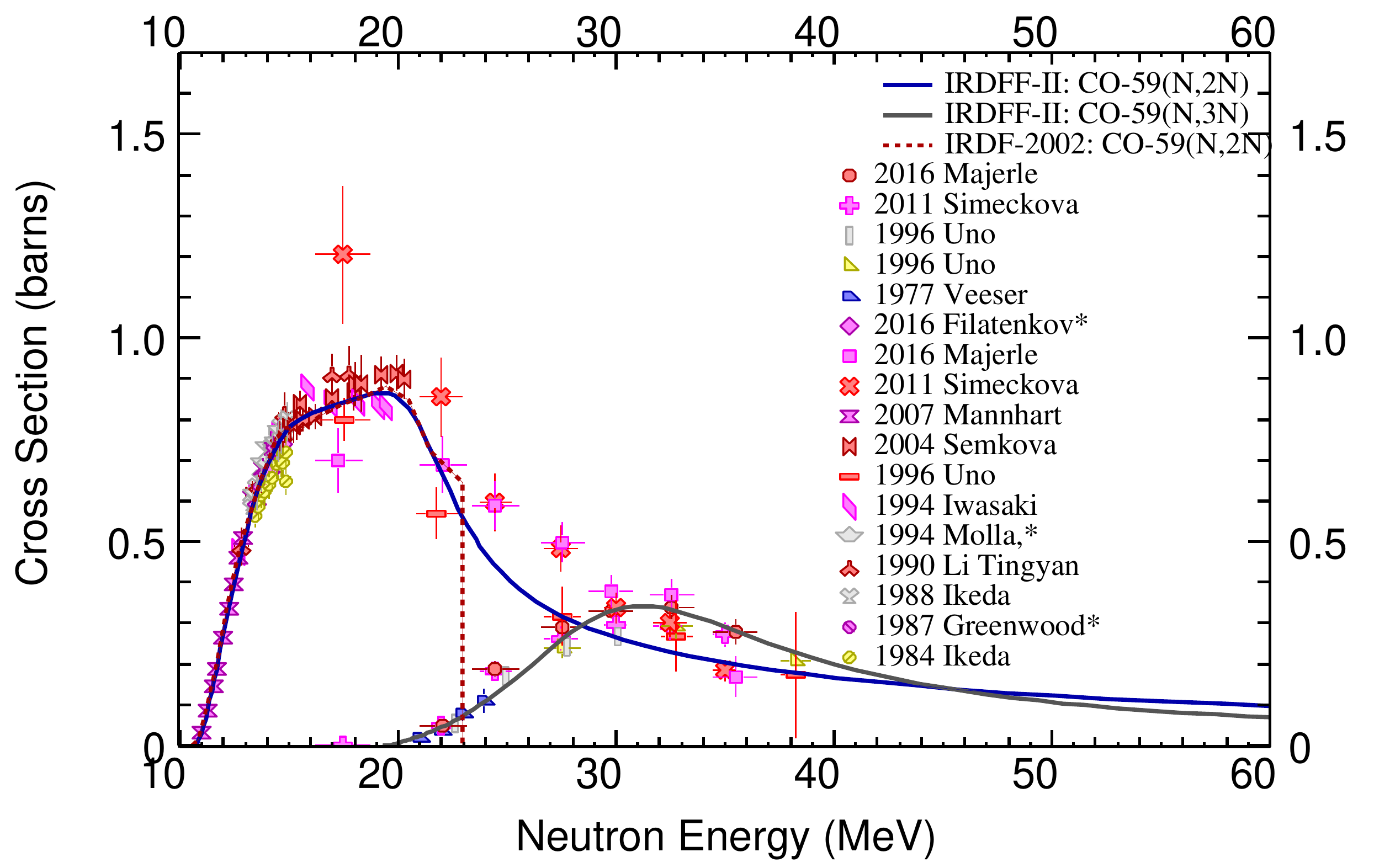}}
\subfigure[~Ratio of cross-section evaluations to \mbox{IRDFF-II} evaluation.]
{\includegraphics[width=\columnwidth]{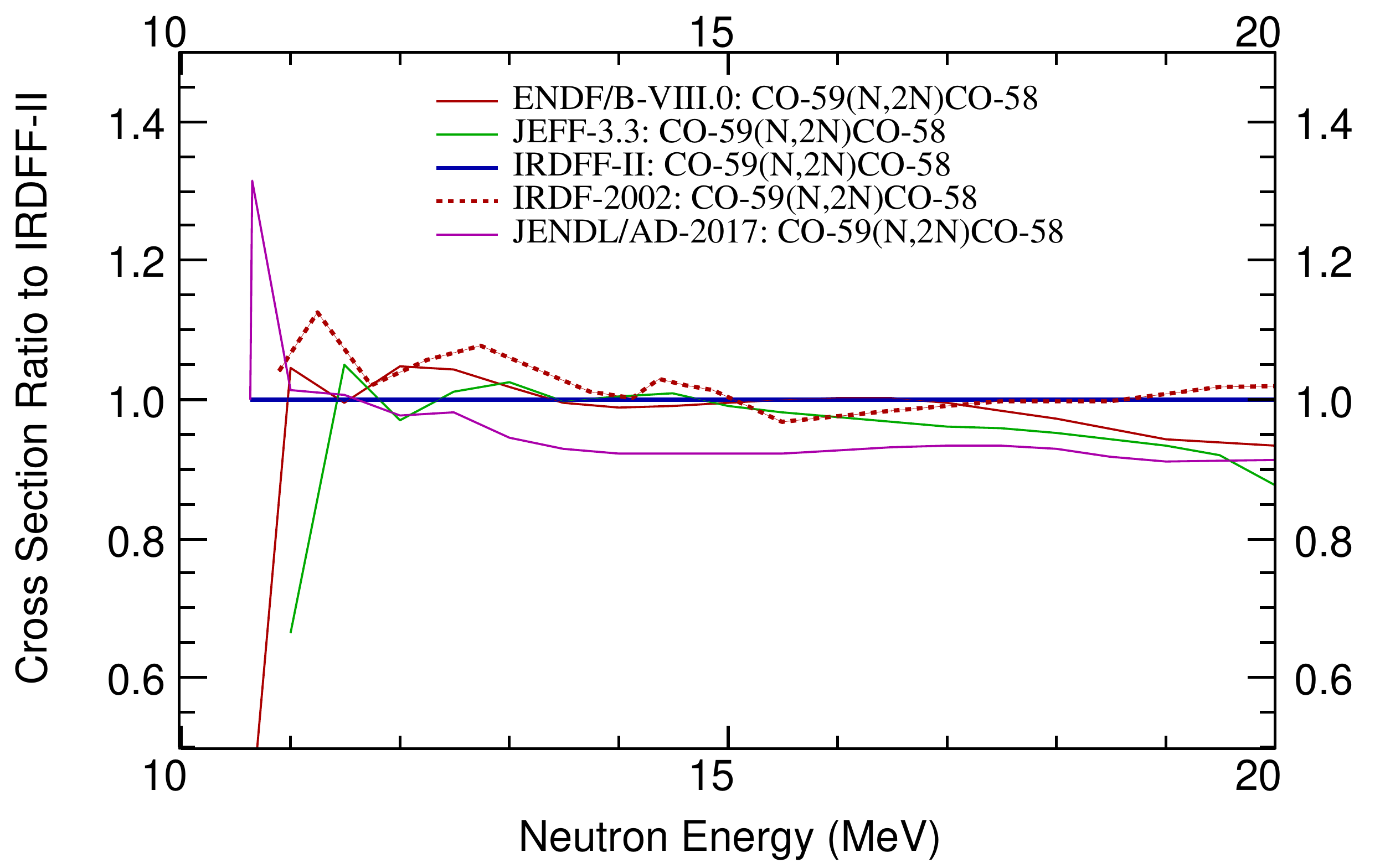}}
%\vspace{-3mm}
\caption{(Color online) $^{59}$Co(n,2n)$^{58}$Co and $^{59}$Co(n,3n)$^{57}$Co  cross-section evaluations in \mbox{IRDFF-II} library relative to other data.}
\label{nCo59}
%\vspace{-3mm}
\end{figure}

The $^{59}$Co(n,3n)$^{57}$Co reaction cross sections and covariances in energy range up to 20~MeV were evaluated by Zolotarev~\cite{Zol10} in 2010. Extension to 60~MeV was done at the IAEA based on \mbox{TENDL-2010}, renormalizing the cross sections at 20~MeV for continuity. Comparison of the cross sections with experimental data for both reactions is shown in Fig.~\ref{nCo59}(a). The reaction cross sections were not present in \mbox{IRDF-2002}. In \mbox{IRDFF-II} they remained unchanged from version \mbox{IRDFF-v1.05}.
The $^{59}$Co(n,2n)$^{58}$Co cross sections in different evaluated libraries differ significantly near threshold as shown in Fig.~\ref{nCo59}(b), where the ratios of the various evaluations to the \mbox{IRDFF-II} recommended cross section are given. The cross sections in JENDL/AD-2017 are significantly lower compared to other evaluations.

\begin{figure}[!hbtp]
\includegraphics[width=0.98\columnwidth]{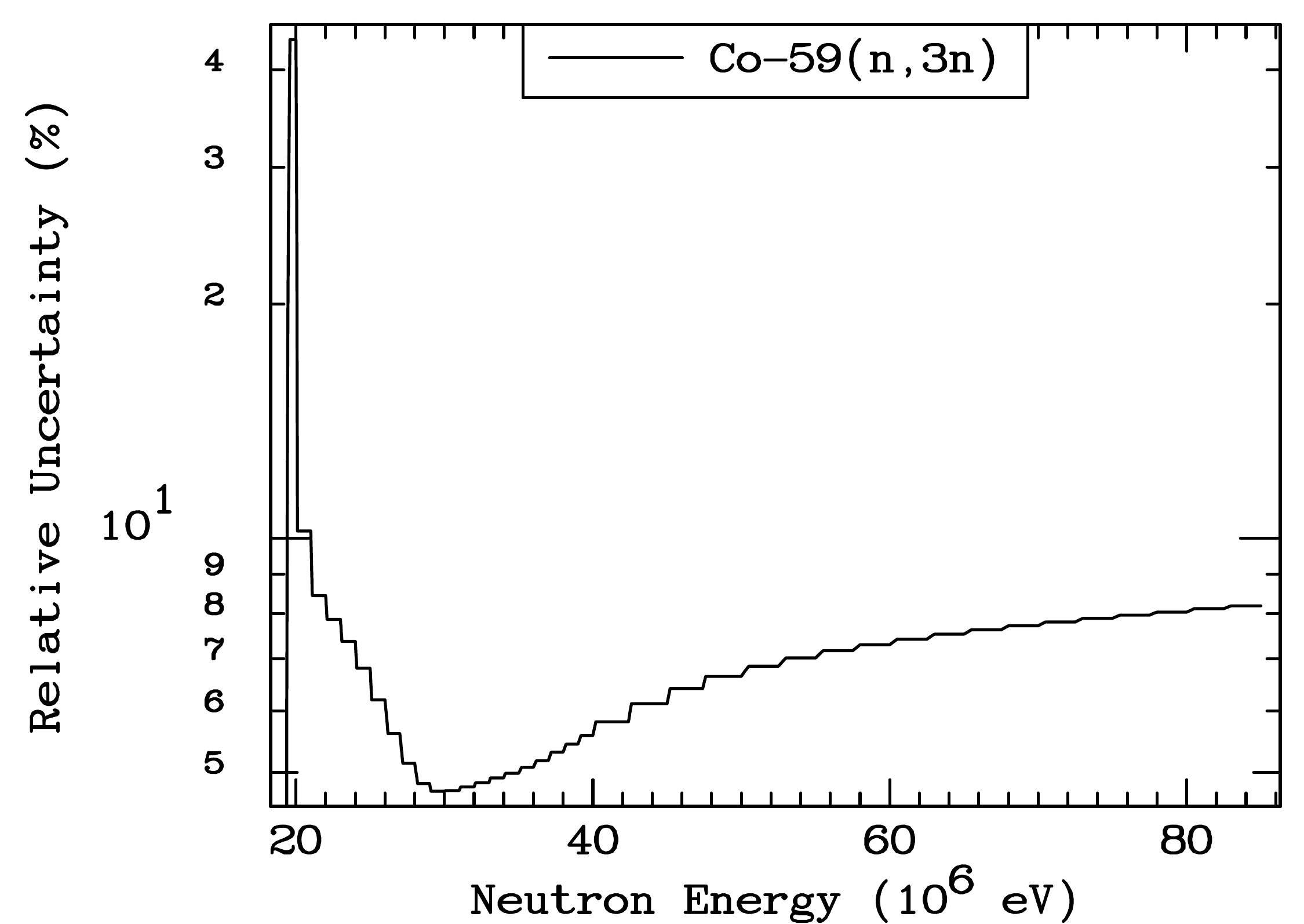}
\caption{$^{59}$Co(n,3n)$^{57}$Co cross-section 1-sigma uncertainties.}
\label{Co59_17_IRDFF_17_stddev_format}
\end{figure}
The plot of the 1-sigma uncertainty  for the $^{59}$Co(n,3n)$^{57}$Co reaction is shown in Fig.~\ref{Co59_17_IRDFF_17_stddev_format}.
%The plot of the correlation matrix for the $^{59}$Co(n,3n)$^{57}$Co reaction is shown in Fig.~\ref{Co593_IRDFF-II_cor_lsl_gen_log}.

\begin{figure}[!thb]
%\vspace{-4mm}
\subfigure[~Cross-section correlation matrix.]
{\includegraphics[width=\columnwidth]{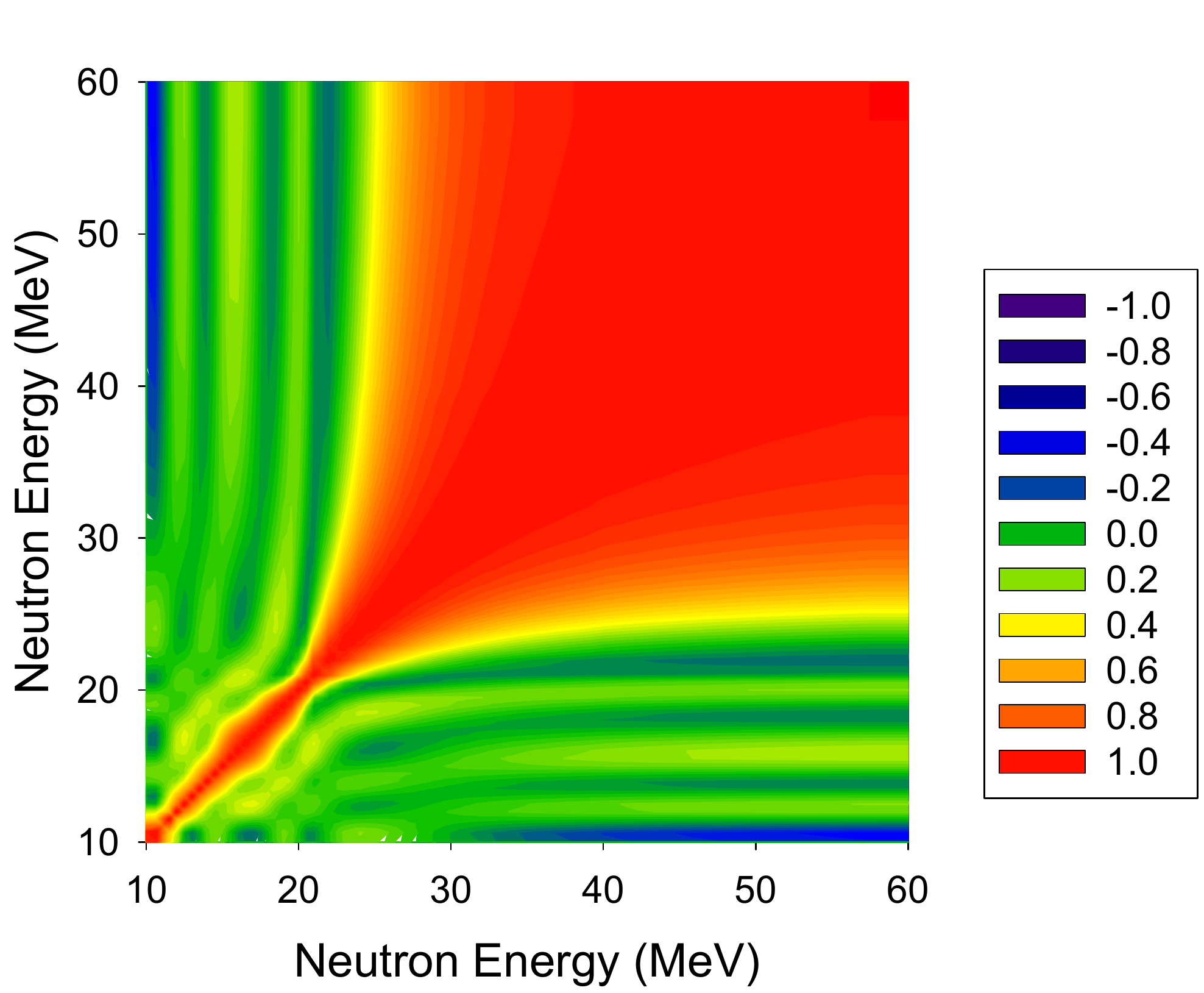}}
\subfigure[~One-sigma uncertainties in \% ($\equiv 100\times \frac{\sqrt{cov(i,i)}}{\mu_i}$), being $\mu_i$ the corresponding cross-section mean value.]
{\includegraphics[width=\columnwidth]{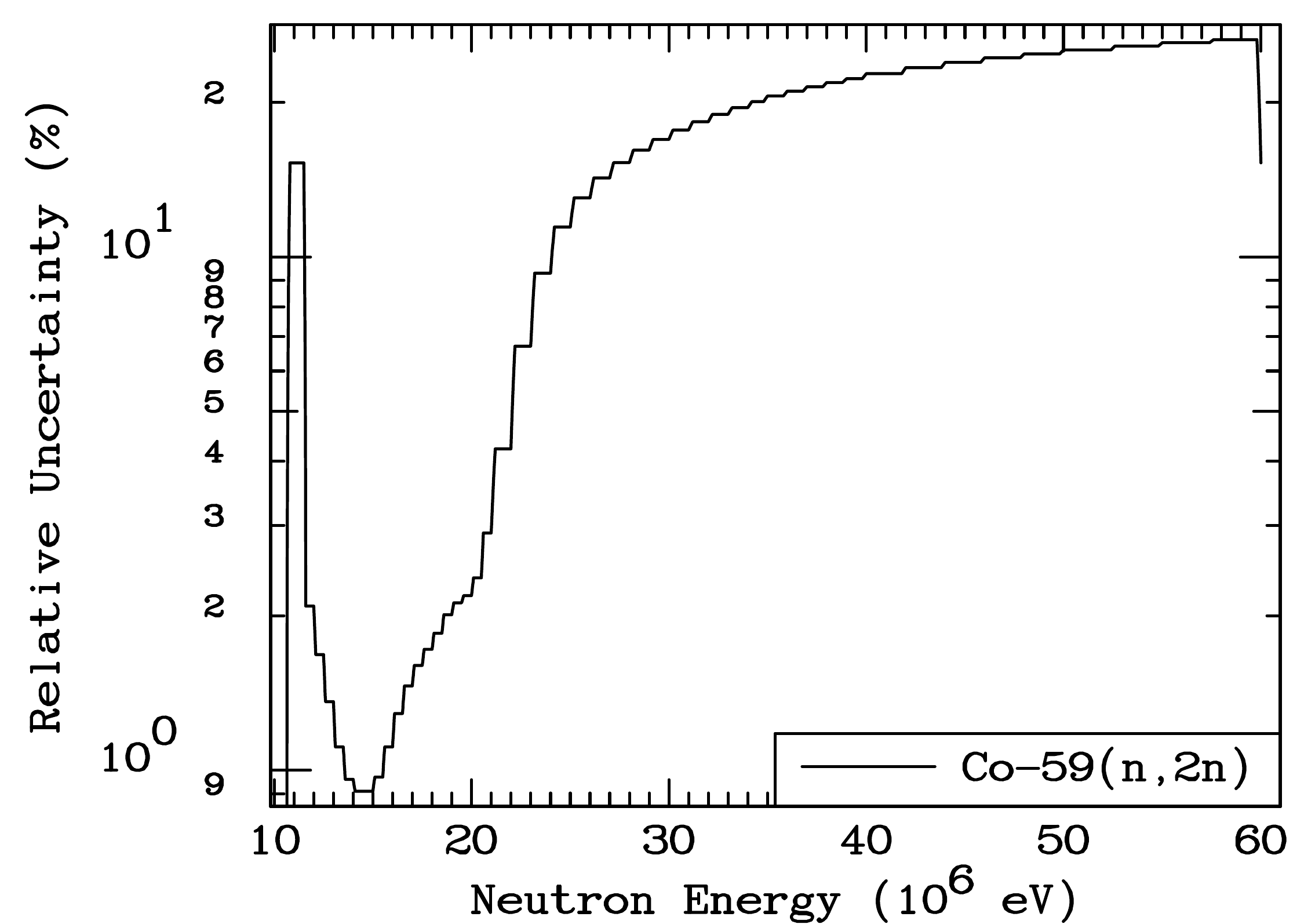}}
%\vspace{-3mm}
\caption{(Color online) Uncertainties and correlations of the $^{59}$Co(n,2n)$^{58}$Co cross-section evaluation in \mbox{IRDFF-II} library.}
\label{nCo59-unc}
%\vspace{-3mm}
\end{figure}

The plot of the correlation matrix for the $^{59}$Co(n,2n)$^{58}$Co reaction is shown in Fig.~\ref{nCo59-unc}(a). The plot of the 1-sigma uncertainty is shown in Fig.~\ref{nCo59-unc}(b).

The $^{59}$Co(n,2n)$^{58}$Co and $^{59}$Co(n,3n)$^{57}$Co reactions are examples of monitor reactions that can be used to probe neutron fields at higher energies.
\clearpage

\subsubsection{$^{55}\!$Mn(n,2n)$^{54}\!$Mn}
\begin{figure}[!thb]
%\vspace{-2mm}
\subfigure[~Comparison to selected experimental data from EXFOR \cite{EXF08}.]
{\includegraphics[width=\columnwidth]{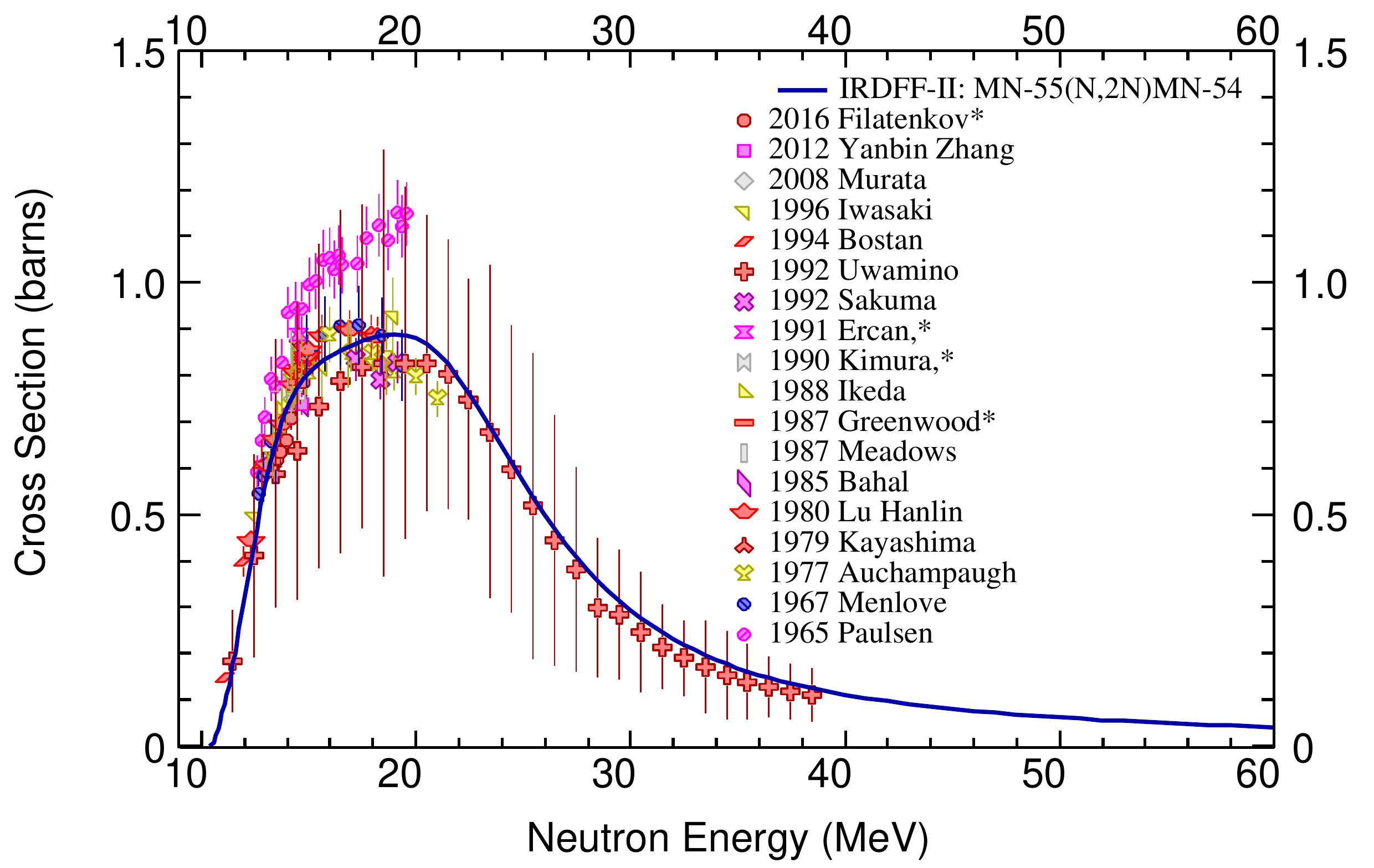}}
\subfigure[~Ratio of cross-section evaluations to \mbox{IRDFF-II} evaluation.]
{\includegraphics[width=\columnwidth]{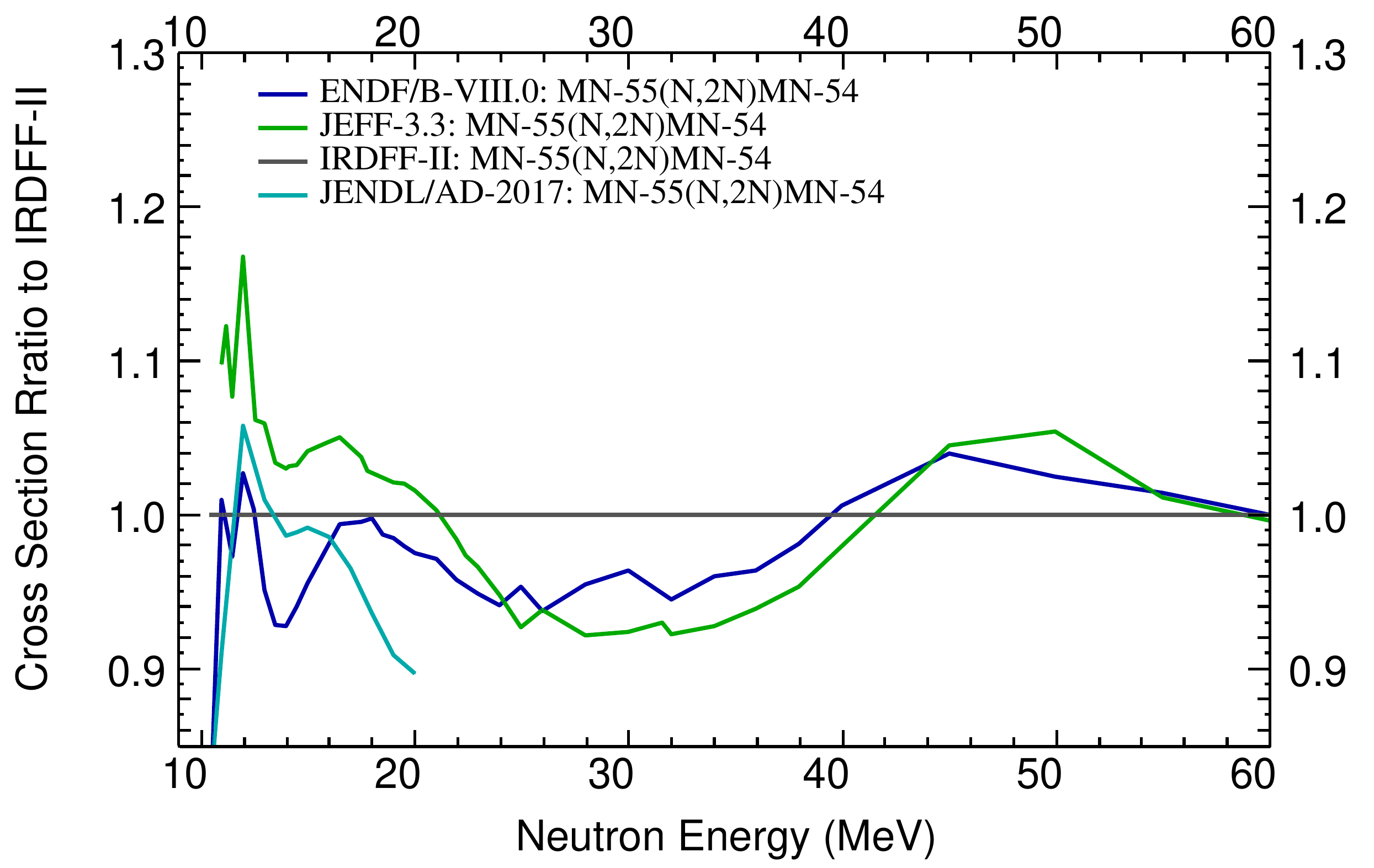}}
%\vspace{-3mm}
\caption{(Color online) $^{55}$Mn(n,2n)$^{54}$Mn  cross-section evaluation in \mbox{IRDFF-II} library relative to other data.}
\label{nMn55}
%\vspace{-3mm}
\end{figure}

The $^{55}$Mn(n,2n)$^{54}$Mn reaction cross sections and covariances in energy range up to 20~MeV were evaluated by Zolotarev~\cite{Zol09} in 2009. Extension to 60~MeV was done at the IAEA based on \mbox{TENDL-2010}, renormalizing the cross sections at 20~MeV for continuity. Comparison of the cross sections with experimental data is shown in Fig.~\ref{nMn55}(a). The reaction cross sections were not present in \mbox{IRDF-2002}. In \mbox{IRDFF-II} they were updated in version \mbox{IRDFF-v1.05}. The $^{55}$Mn(n,2n)$^{54}$Mn cross sections in different evaluated libraries differ significantly near threshold as shown in Fig.~\ref{nMn55}(b), where the ratios of the various evaluations to the \mbox{IRDFF-II} recommended cross section are given.

\begin{figure}[!thb]
%\vspace{-4mm}
\subfigure[~Cross-section correlation matrix.]
{\includegraphics[width=\columnwidth]{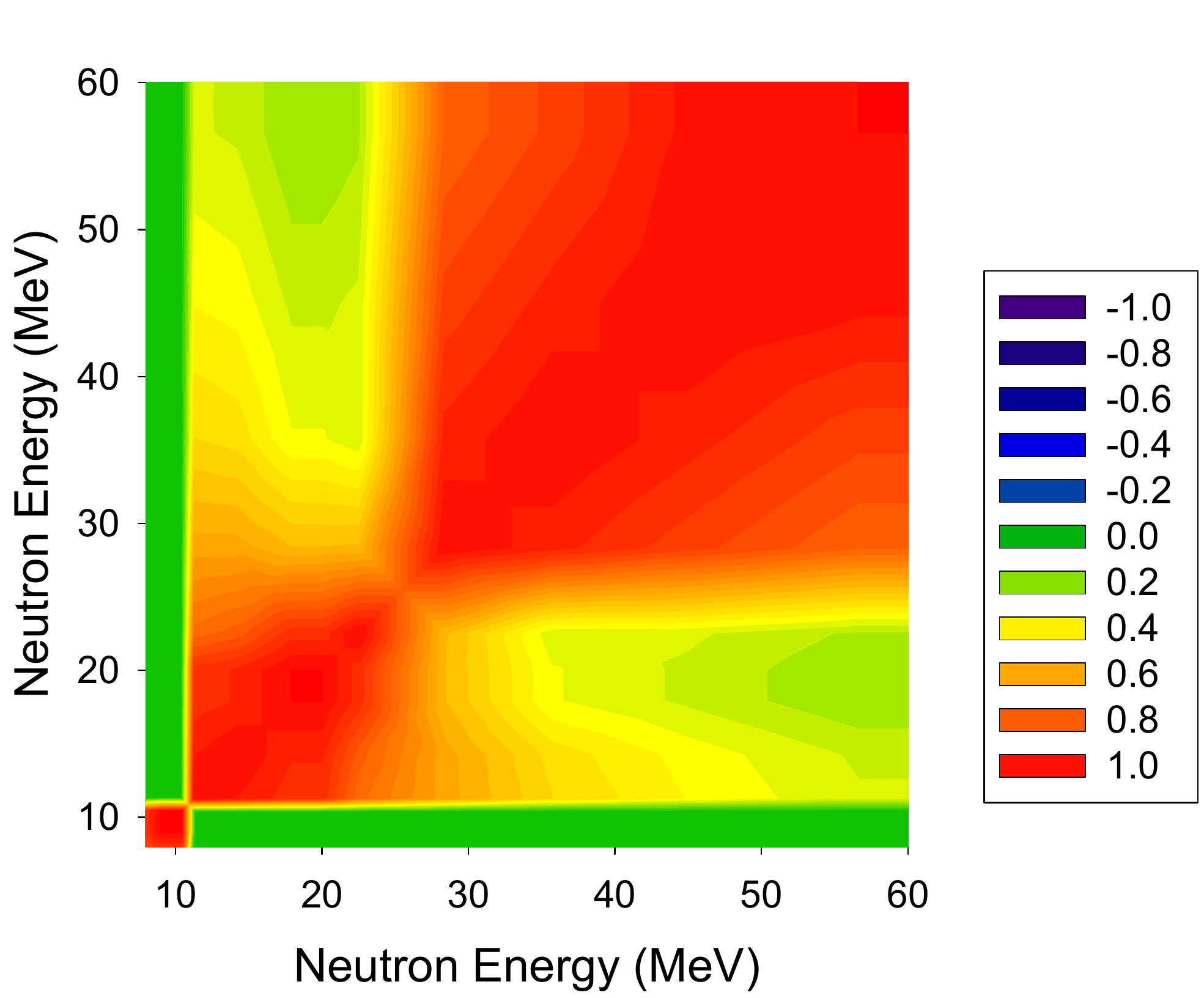}}
\subfigure[~One-sigma uncertainties in \% ($\equiv 100\times \frac{\sqrt{cov(i,i)}}{\mu_i}$), being $\mu_i$ the corresponding cross-section mean value.]
{\includegraphics[width=\columnwidth]{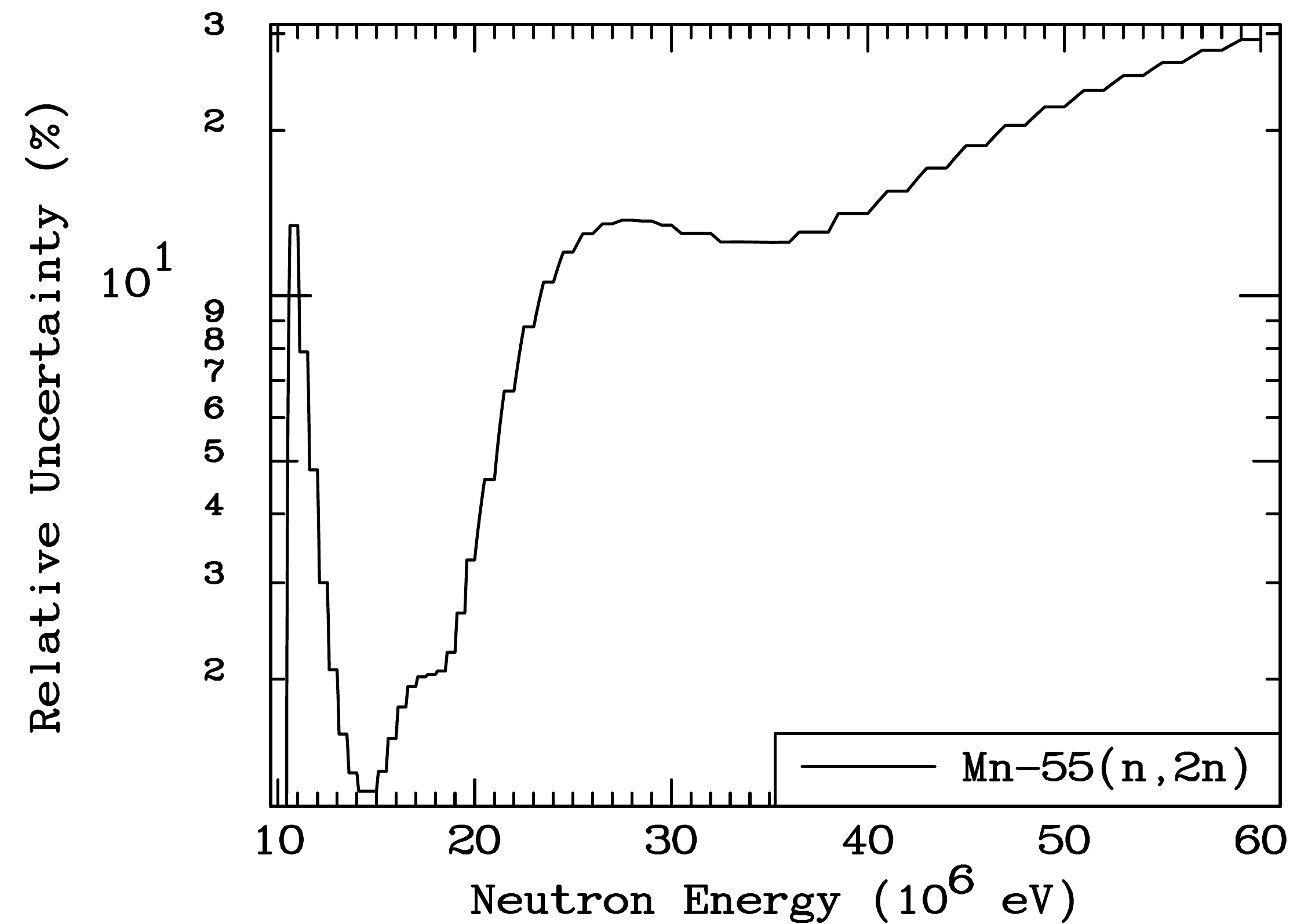}}
%\vspace{-3mm}
\caption{(Color online) Uncertainties and correlations of the $^{55}$Mn(n,2n)$^{54}$Mn cross-section evaluation in \mbox{IRDFF-II} library.}
\label{nMn55-unc}
%\vspace{-3mm}
\end{figure}

The plot of the correlation matrix for the $^{55}$Mn(n,2n)$^{54}$Mn reaction is shown in Fig.~\ref{nMn55-unc}(a). The plot of the 1-sigma uncertainty is shown in Fig.~\ref{nMn55-unc}(b).

The $^{55}$Mn(n,2n)$^{54}$Mn reaction is also one that can be used to probe neutron fields at higher energies.

\clearpage
\subsubsection{$^{67}\!$Zn(n,p)$^{67}\!$Cu}
\begin{figure}[!thb]
%\vspace{-2mm}
\subfigure[~Comparison to selected experimental data from EXFOR \cite{EXF08}.]
{\includegraphics[width=\columnwidth]{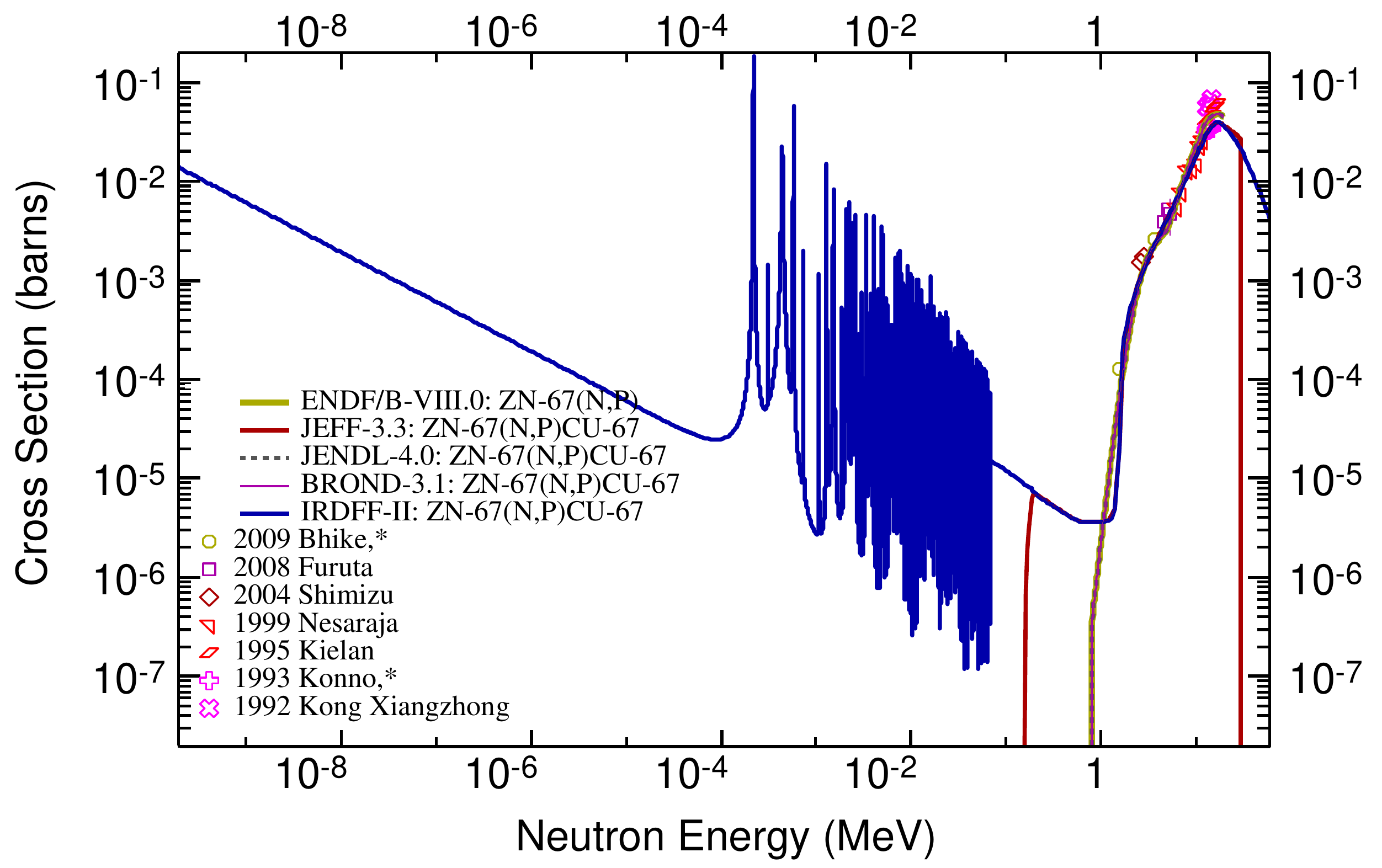}}
\subfigure[~Ratio of cross-section evaluations to \mbox{IRDFF-II} evaluation.]
{\includegraphics[width=\columnwidth]{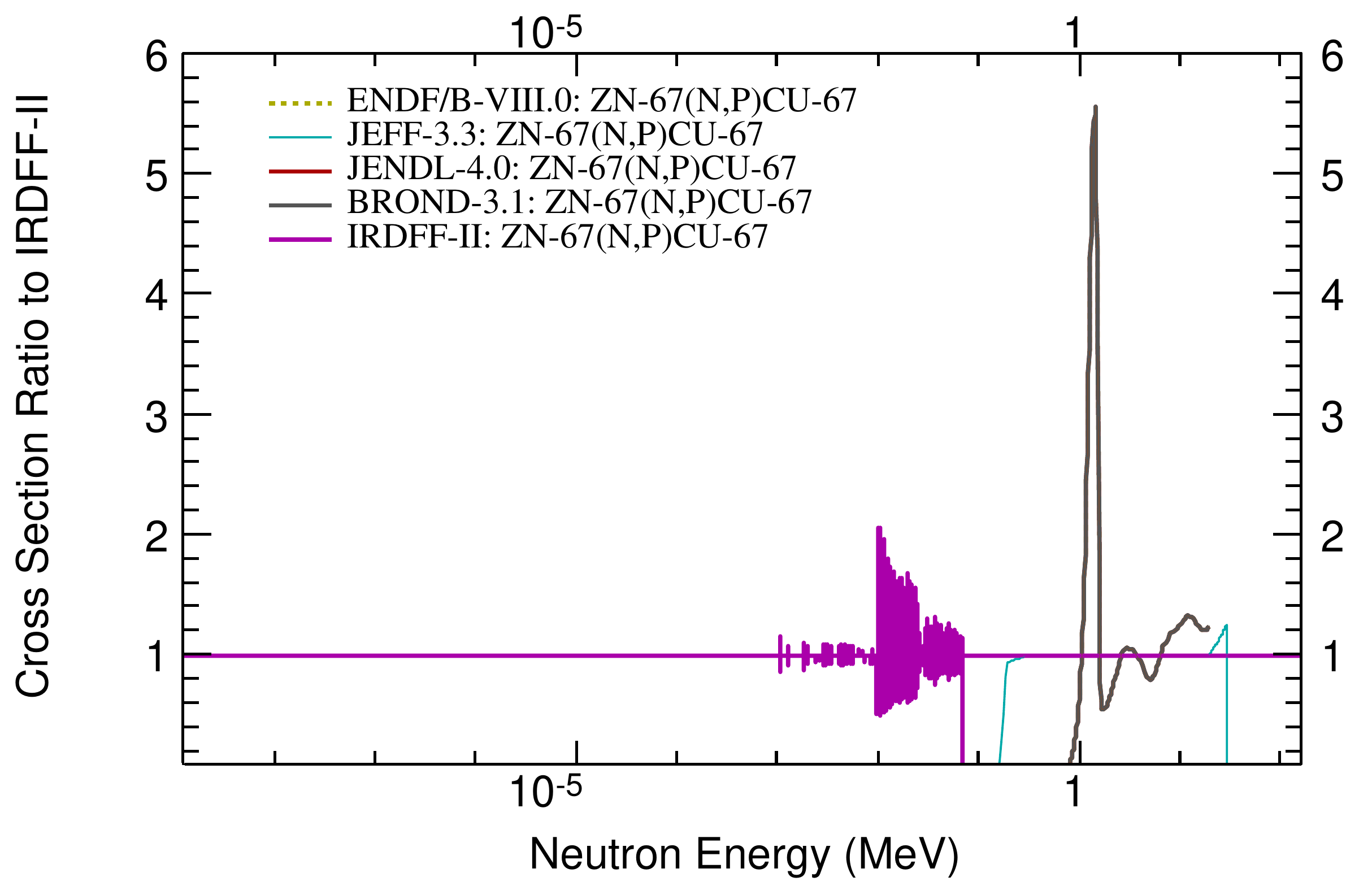}}
%\vspace{-3mm}
\caption{(Color online) $^{67}$Zn(n,p)$^{67}$Cu  cross-section evaluation in \mbox{IRDFF-II} library relative to other data.}
\label{pZn67}
%\vspace{-3mm}
\end{figure}

The $^{67}$Zn(n,p)$^{67}$Cu reaction cross sections and covariances in energy range up to 20~MeV were evaluated by Zolotarev~\cite{Zol08} in 2008. Extension to 60~MeV was done at the IAEA based on \mbox{TENDL-2010}, renormalizing the cross sections at 20~MeV for continuity. This reaction is exothermic and has a positive Q-value, therefore it has no real threshold. However, the difference is the relatively large thermal capture cross section. The capture cross sections from JENDL-4 were re-scaled to give a thermal value of 0.00123 barns and adjusted above 600 keV to join smoothly with the (n,p) cross section evaluation by Zolotarev. A fully correlated uncertainty of 25~\% was assigned arbitrarily to the cross sections below the pseudo-threshold. The plot of the cross sections is shown in Fig.~\ref{pZn67}(a). The cross sections for this reaction were not present in \mbox{IRDF-2002}. In \mbox{IRDFF-II} they remained unchanged from version \mbox{IRDFF-v1.05}. The $^{67}$Zn(n,p)$^{67}$Cu cross sections in different evaluated libraries differ significantly, especially near the apparent threshold as shown in Fig.~\ref{pZn67}(b), where the ratios of the various evaluations to the \mbox{IRDFF-II} recommended cross section are given.

\begin{figure}[!thb]
%\vspace{-4mm}
\subfigure[~Cross-section correlation matrix.]
{\includegraphics[width=\columnwidth]{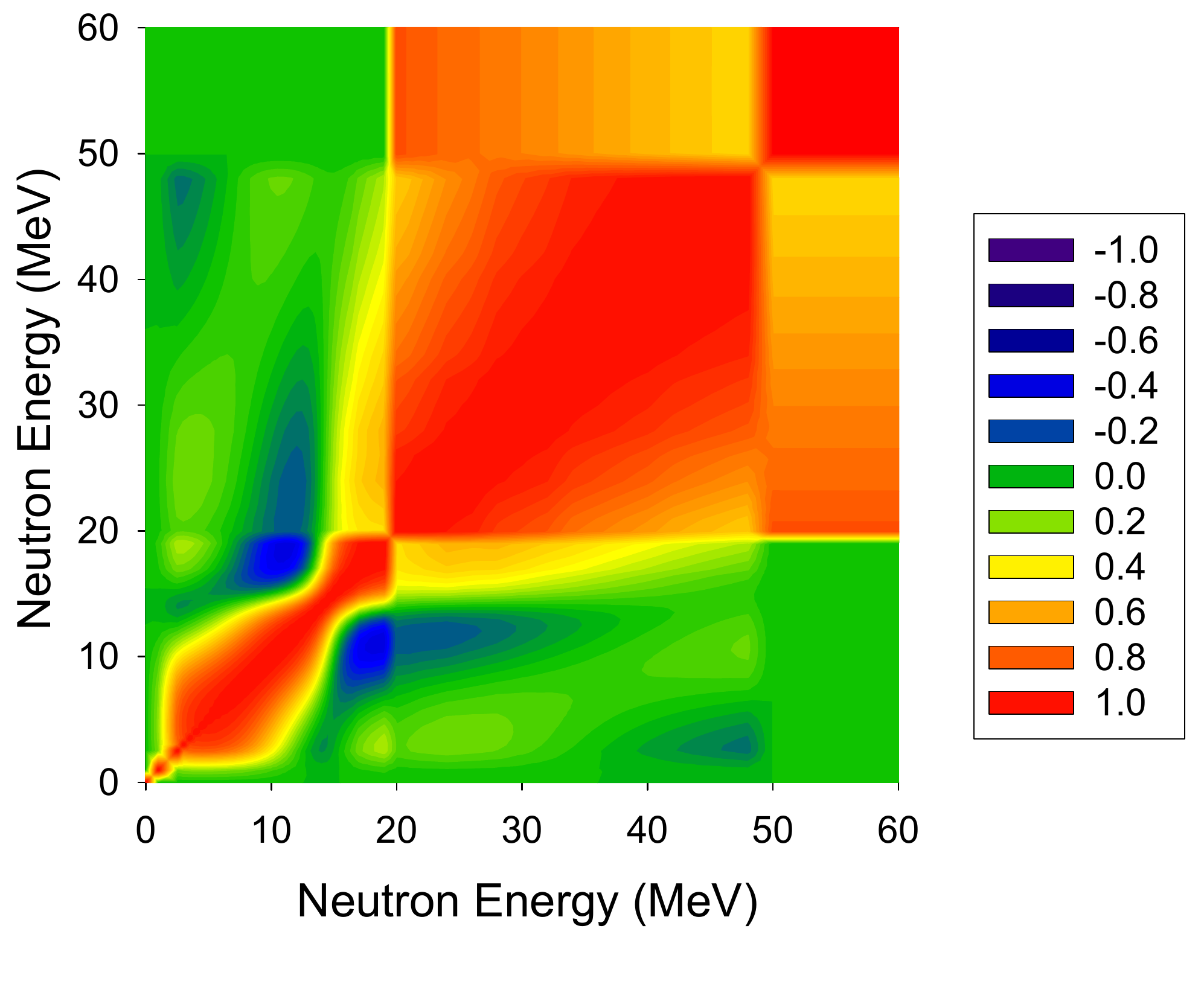}}
\subfigure[~One-sigma uncertainties in \% ($\equiv 100\times \frac{\sqrt{cov(i,i)}}{\mu_i}$), being $\mu_i$ the corresponding cross-section mean value.]
{\includegraphics[width=\columnwidth]{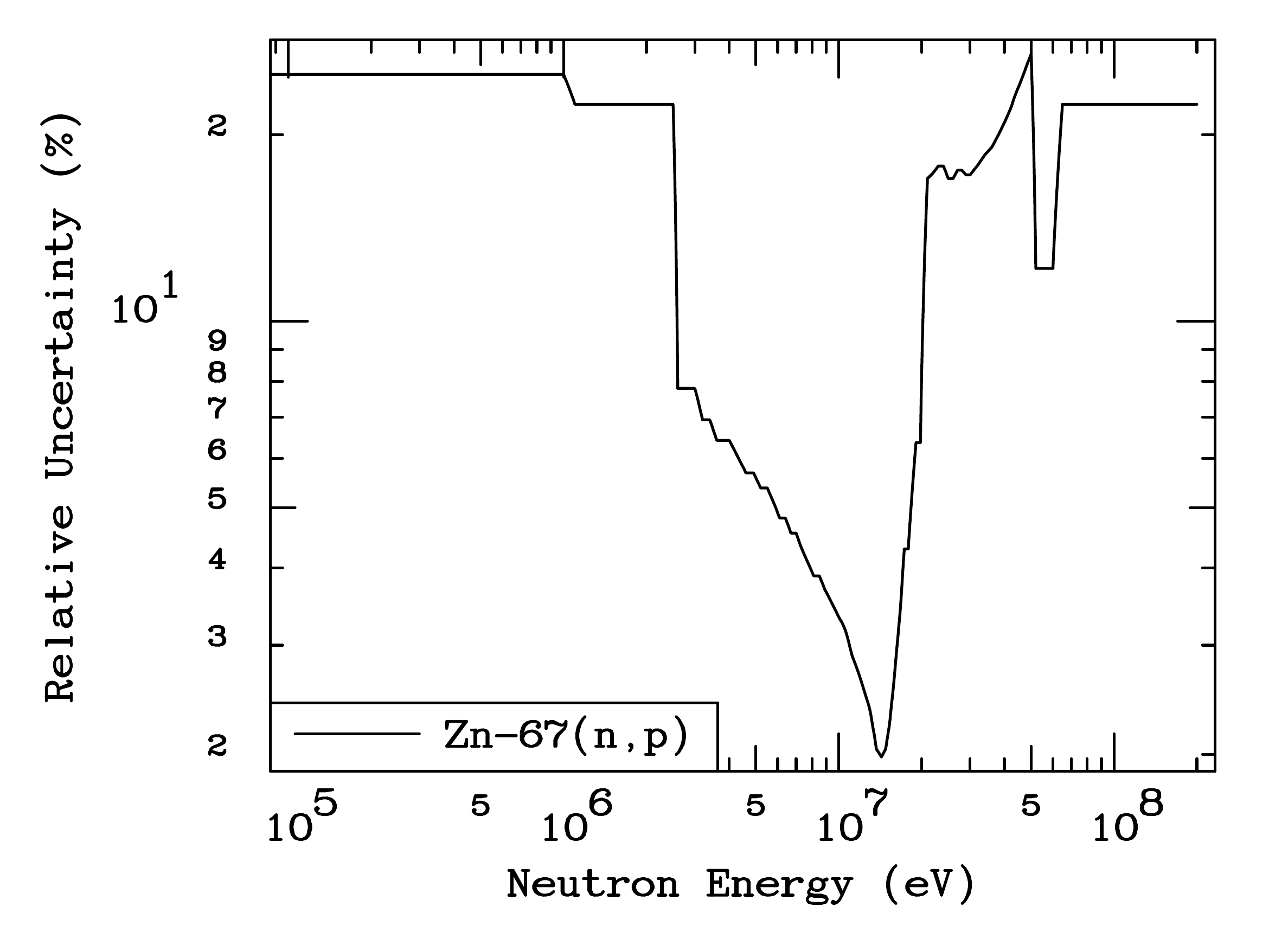}}
%\vspace{-3mm}
\caption{(Color online) Uncertainties and correlations of the $^{67}$Zn(n,p)$^{67}$Cu cross-section evaluation in \mbox{IRDFF-II} library.}
\label{pZn67-unc}
%\vspace{-3mm}
\end{figure}

The plot of the correlation matrix for the $^{67}$Zn(n,p)$^{67}$Cu reaction is shown in Fig.~\ref{pZn67-unc}(a). The plot of the 1-sigma uncertainty is shown in Fig.~\ref{pZn67-unc}(b).

The $^{67}$Zn(n,p)$^{67}$Cu reaction is an example of a reaction with a positive Q-value, where the cross sections were extended down to thermal energies and a resonance evaluation was included.

\clearpage
\subsubsection{$^{199}\!$Hg(n,n')$^{199m}\!$Hg}
\begin{figure}[!thb]
%\vspace{-2mm}
\subfigure[~Comparison to selected experimental data from EXFOR \cite{EXF08}.]
{5\includegraphics[width=\columnwidth]{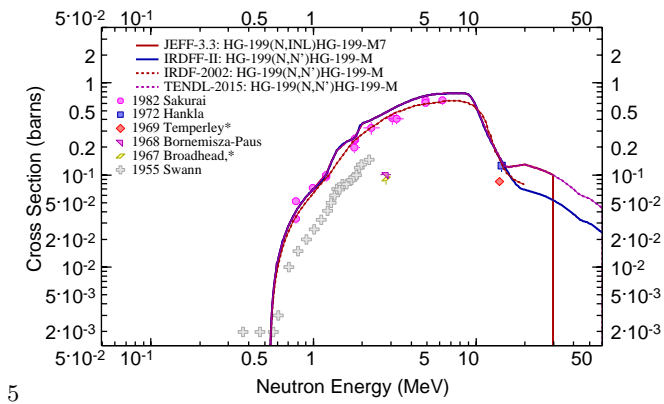}}
\subfigure[~Ratio of cross-section evaluations to \mbox{IRDFF-II} evaluation.]
{\includegraphics[width=\columnwidth]{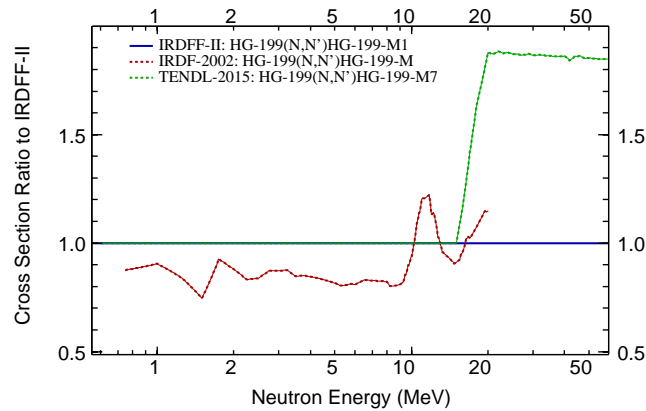}}
%\vspace{-3mm}
\caption{(Color online) $^{199}$Hg(n,n')$^{199m}$Hg cross-section evaluation in \mbox{IRDFF-II} library relative to other data.}
\label{iHg199}
%\vspace{-3mm}
\end{figure}

The $^{199}$Hg(n,n')$^{199m}$Hg reaction cross sections and covariances in energy range up to 20~MeV were evaluated by Zolotarev~\cite{Zol08} in 2008. Extension to 60~MeV was done at the IAEA based on \mbox{TENDL-2010}, renormalizing the cross sections at 20~MeV for continuity. Comparison with experimental data is shown in Fig.~\ref{iHg199}(a). The figure also shows the comparison between \mbox{IRDF-2002} and \mbox{IRDFF-II}, which remained unchanged from version \mbox{IRDFF-v1.05}. The $^{199}$Hg(n,n')$^{199m}$Hg cross sections represent the excitation of a metastable state and are not present in other major libraries, hence only the comparison with IRDF-2002 is shown in Fig.~\ref{iHg199}(b), where the ratios to the \mbox{IRDFF-II} recommended cross section are given.

\begin{figure}[!thb]
\vspace{-4mm}
\subfigure[~Cross-section correlation matrix.]
{\includegraphics[width=\columnwidth]{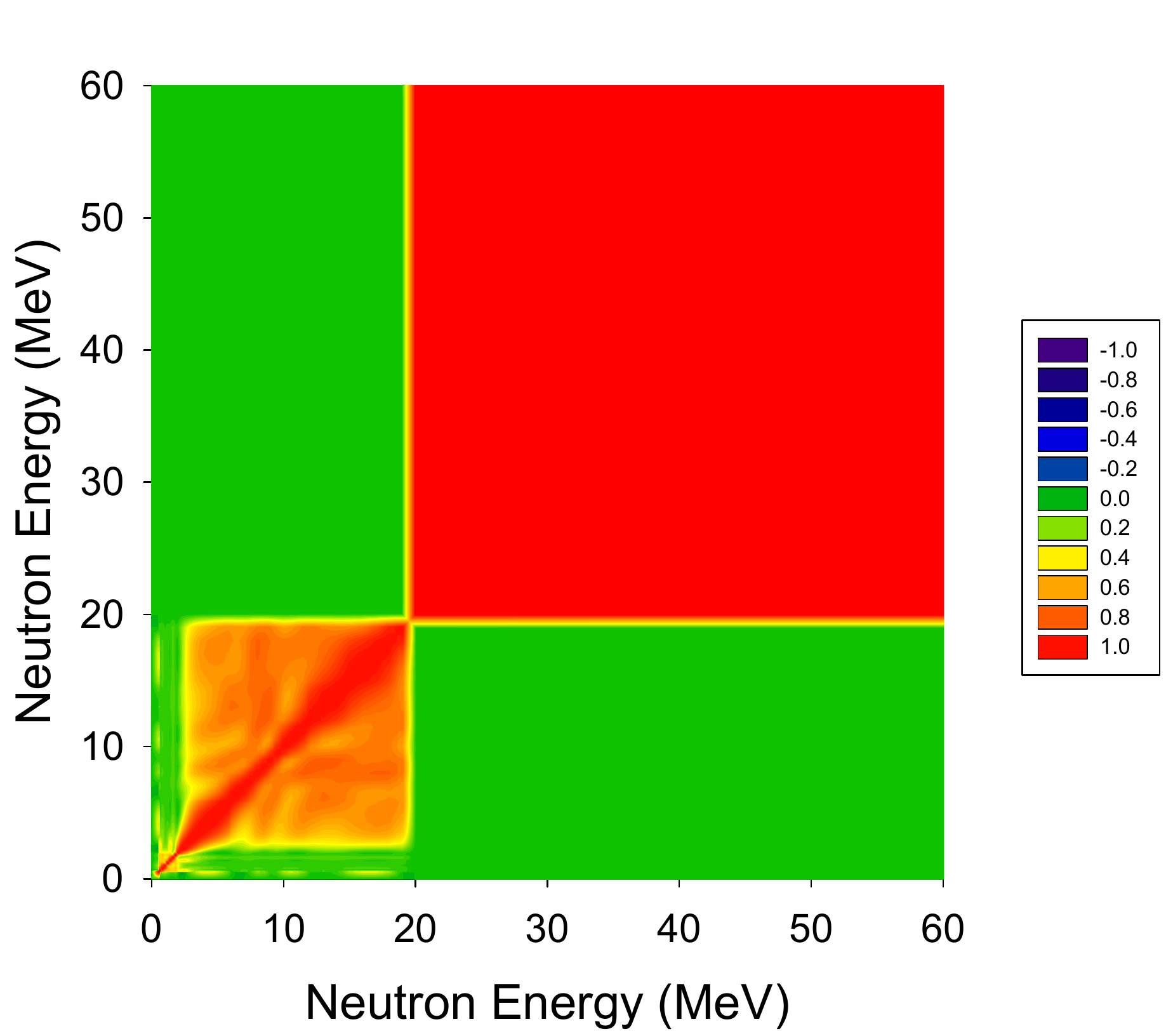}}
\subfigure[~One-sigma uncertainties in \% ($\equiv 100\times \frac{\sqrt{cov(i,i)}}{\mu_i}$), being $\mu_i$ the corresponding cross-section mean value.]
{\includegraphics[width=\columnwidth]{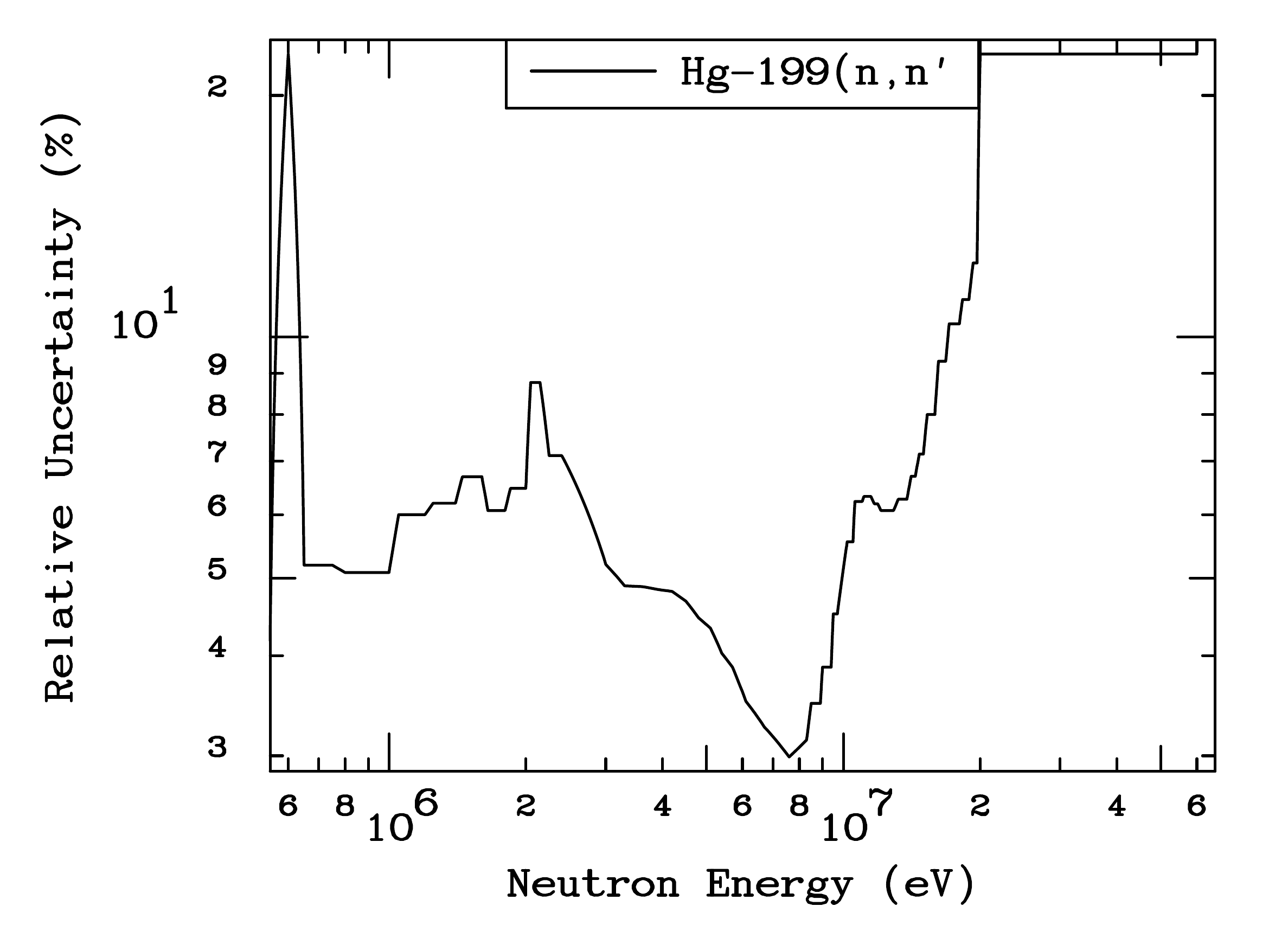}}
\vspace{-3mm}
\caption{(Color online) Uncertainties and correlations of the $^{199}$Hg(n,n')$^{199m}$Hg cross-section evaluation in \mbox{IRDFF-II} library.}
\label{nHg199-unc}
\vspace{-3mm}
\end{figure}

The plot of the correlation matrix for the $^{199}$Hg(n,n')$^{199m}$Hg reaction is shown in Fig.~\ref{nHg199-unc}(a). The plot of the 1-sigma uncertainty is shown in Fig.~\ref{nHg199-unc}(b).

\clearpage
\subsubsection{$^{89}\!$Y(n,2n)$^{88}\!$Y}
\begin{figure}[!thb]
%\vspace{-2mm}
\subfigure[~Comparison to selected experimental data from EXFOR \cite{EXF08}.]
{\includegraphics[width=\columnwidth]{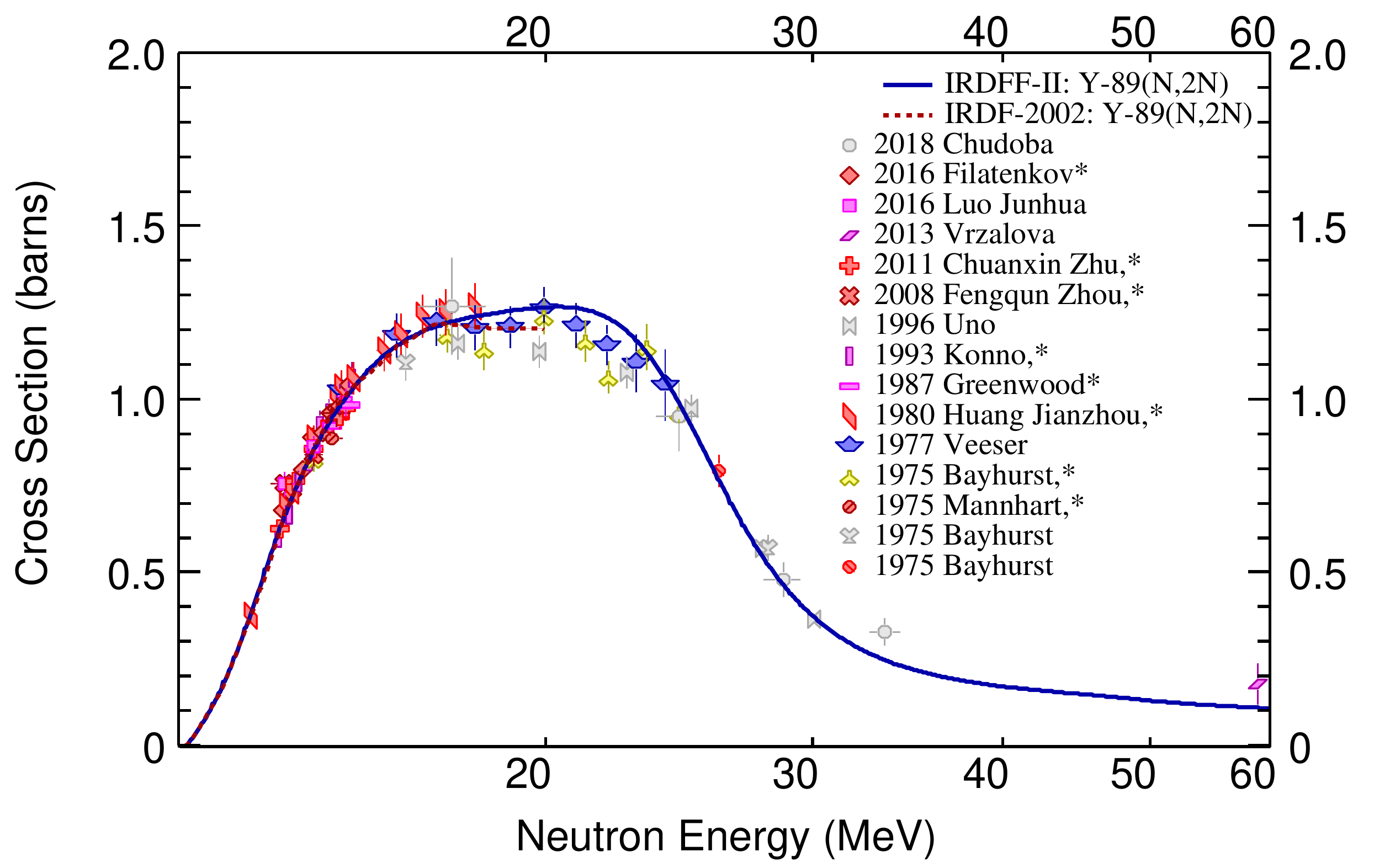}}
\subfigure[~Ratio of cross-section evaluations to \mbox{IRDFF-II} evaluation.]
{\includegraphics[width=\columnwidth]{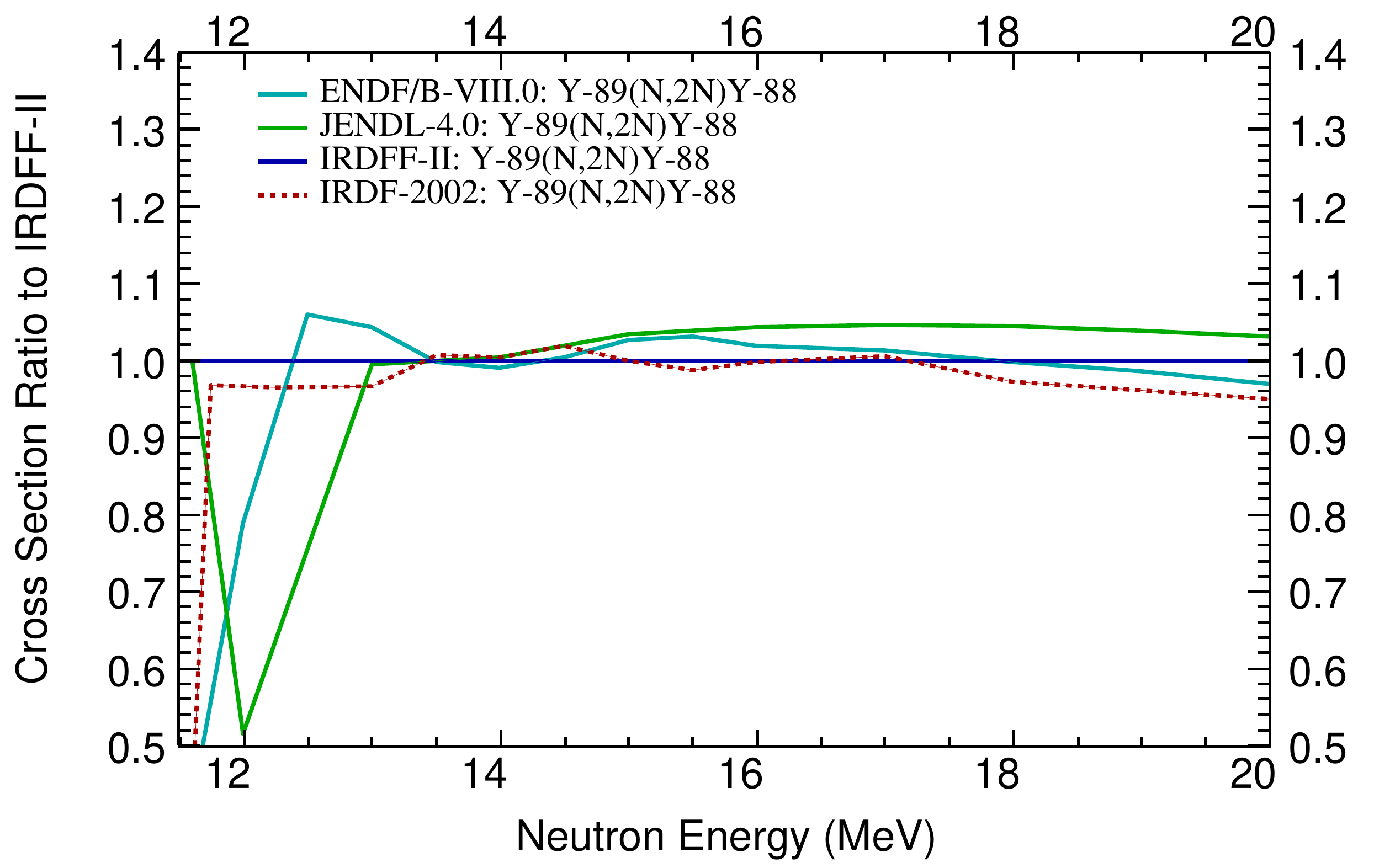}}
%\vspace{-3mm}
\caption{(Color online) $^{89}$Y(n,2n)$^{88}$Y cross-section evaluation in \mbox{IRDFF-II} library relative to other data.}
\label{nY89}
%\vspace{-3mm}
\end{figure}

The $^{89}$Y(n,2n)$^{88}$Y reaction cross sections and covariances in energy range up to 20~MeV were evaluated by Zolotarev~\cite{Zol09} in 2009. Extension to 60~MeV was done at the IAEA based on \mbox{TENDL-2010}, renormalizing the cross sections at 20~MeV for continuity. Comparison of the cross sections with experimental data is shown in Fig.~\ref{nY89}(a). The figure also shows the comparison between \mbox{IRDF-2002} and \mbox{IRDFF-II}, which remained unchanged from version \mbox{IRDFF-v1.05}.

The $^{89}$Y(n,2n)$^{88}$Y cross sections in different evaluated libraries differ significantly near threshold as shown in Fig.~\ref{nY89}(b), where the ratios of the various evaluations to the \mbox{IRDFF-II} recommended cross section are given.

\begin{figure}[!thb]
%\vspace{-4mm}
\subfigure[~Cross-section correlation matrix.]
{\includegraphics[width=\columnwidth]{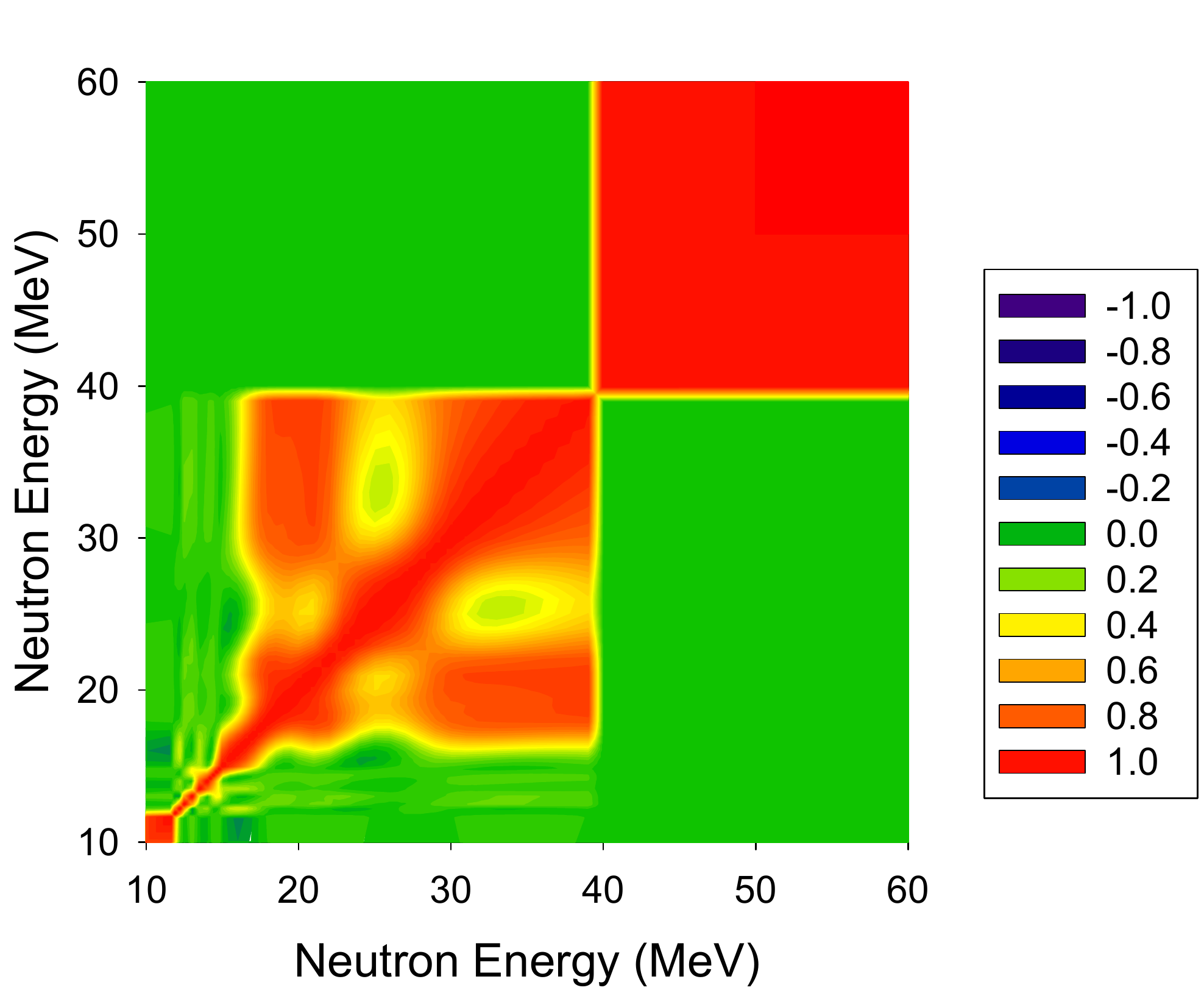}}
\subfigure[~One-sigma uncertainties in \% ($\equiv 100\times \frac{\sqrt{cov(i,i)}}{\mu_i}$), being $\mu_i$ the corresponding cross-section mean value.]
{\includegraphics[width=\columnwidth]{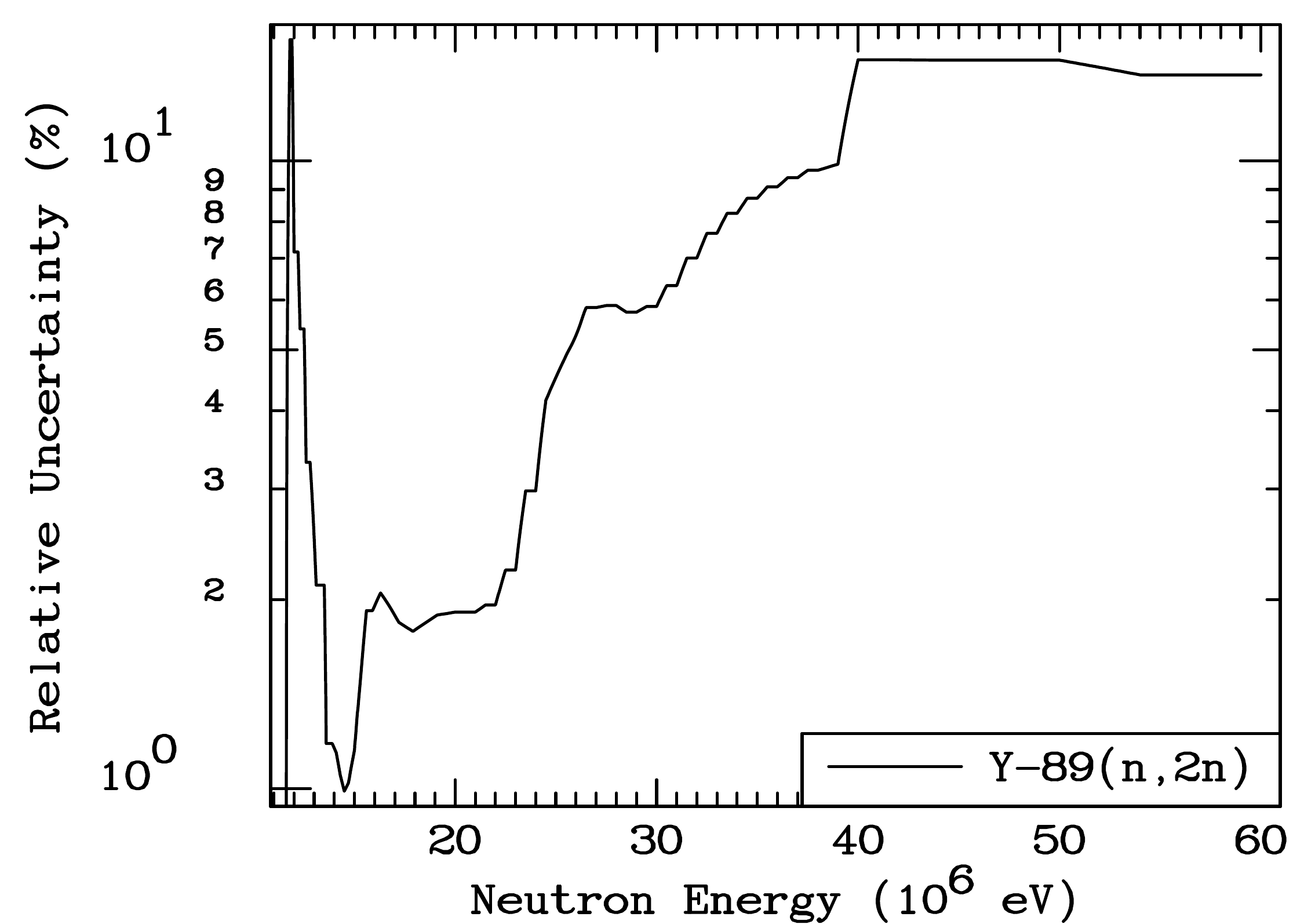}}
%\vspace{-3mm}
\caption{(Color online) Uncertainties and correlations of the $^{89}$Y(n,2n)$^{88}$Y cross-section evaluation in \mbox{IRDFF-II} library.}
\label{nY89-unc}
%\vspace{-3mm}
\end{figure}

\newpage
The plot of the correlation matrix for the $^{89}$Y(n,2n)$^{88}$Y reaction is shown in Fig.~\ref{nY89-unc}(a). The plot of the 1-sigma uncertainty is shown in Fig.~\ref{nY89-unc}(b).
The $^{89}$Y(n,2n)$^{88}$Y reaction is another one that can be used to probe neutron fields at higher energies.

\clearpage
\subsubsection{$^{169}\!$Tm(n,2n)$^{168}\!$Tm and $^{169}\!$Tm(n,3n)$^{167}\!$Tm}
\begin{figure}[!thb]
%\vspace{-2mm}
\subfigure[~Comparison to selected experimental data from EXFOR \cite{EXF08}.]
{\includegraphics[width=\columnwidth]{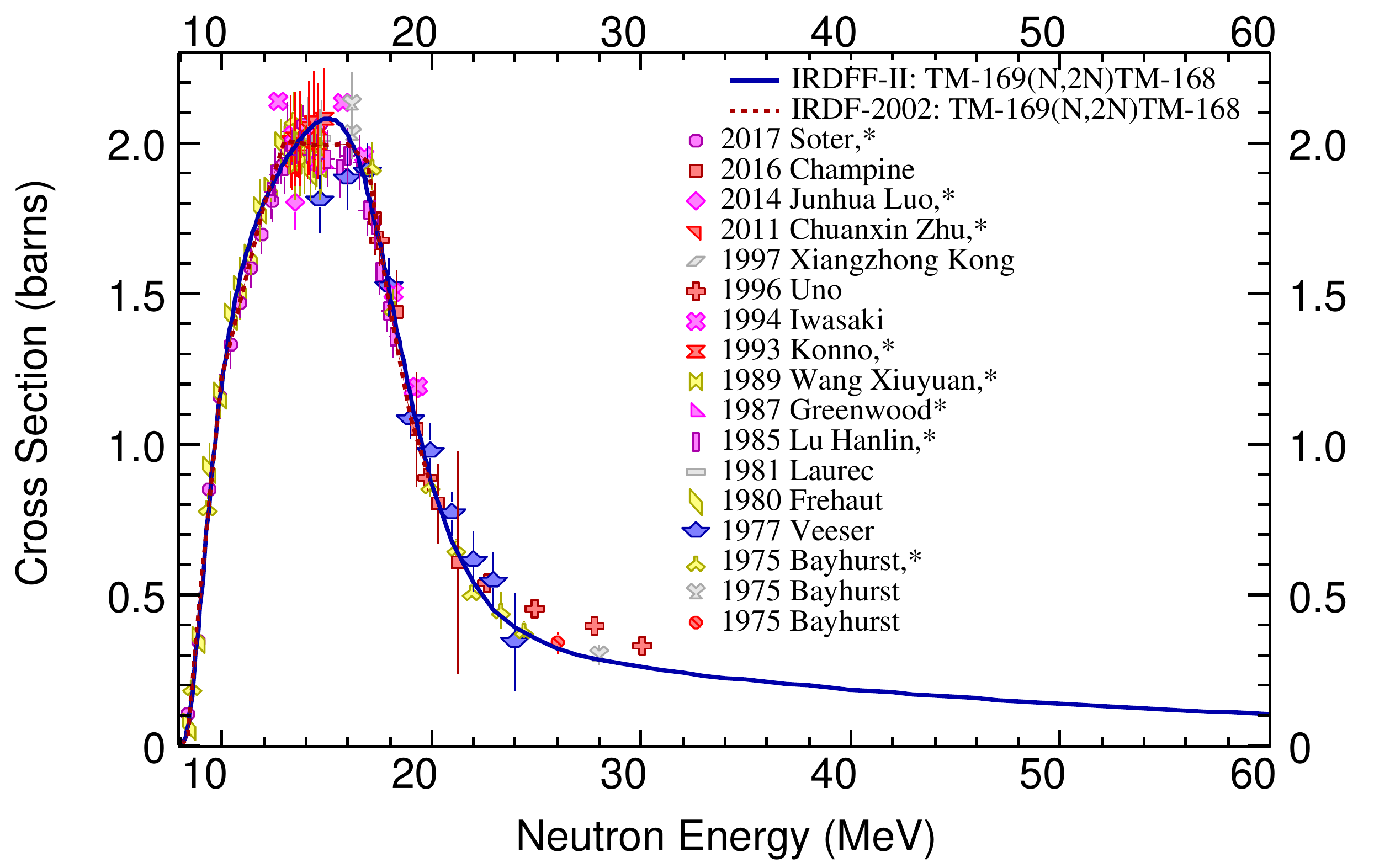}}
\subfigure[~Ratio of cross-section evaluations to \mbox{IRDFF-II} evaluation.]
{\includegraphics[width=\columnwidth]{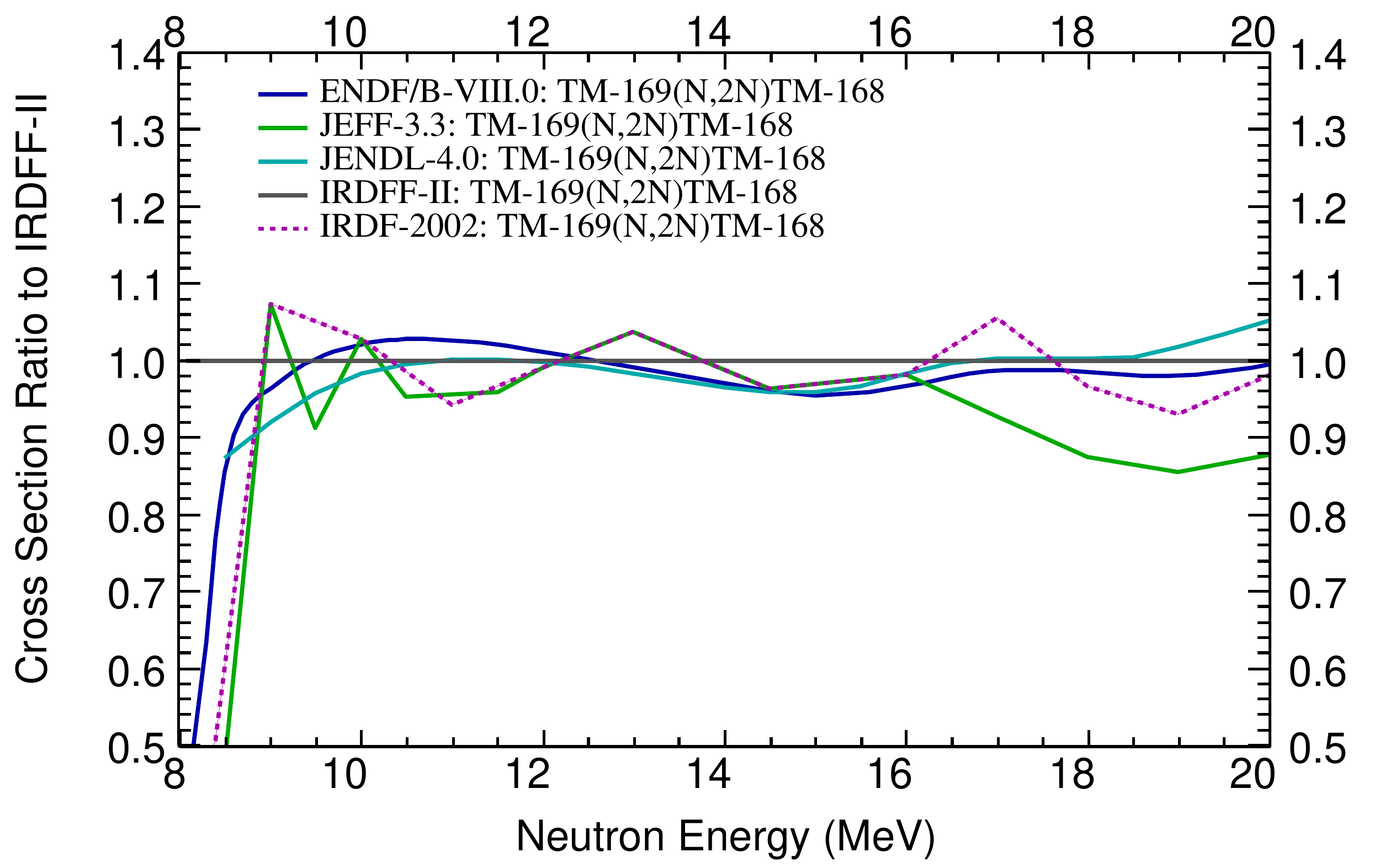}}
%\vspace{-3mm}
\caption{(Color online) $^{169}$Tm(n,2n)$^{168}$Tm cross-section evaluation in \mbox{IRDFF-II} library relative to other data.}
\label{nTm169}
%\vspace{-3mm}
\end{figure}

\begin{figure}[!thb]
%\vspace{-2mm}
\subfigure[~Comparison to selected experimental data from EXFOR \cite{EXF08}.]
{\includegraphics[width=\columnwidth]{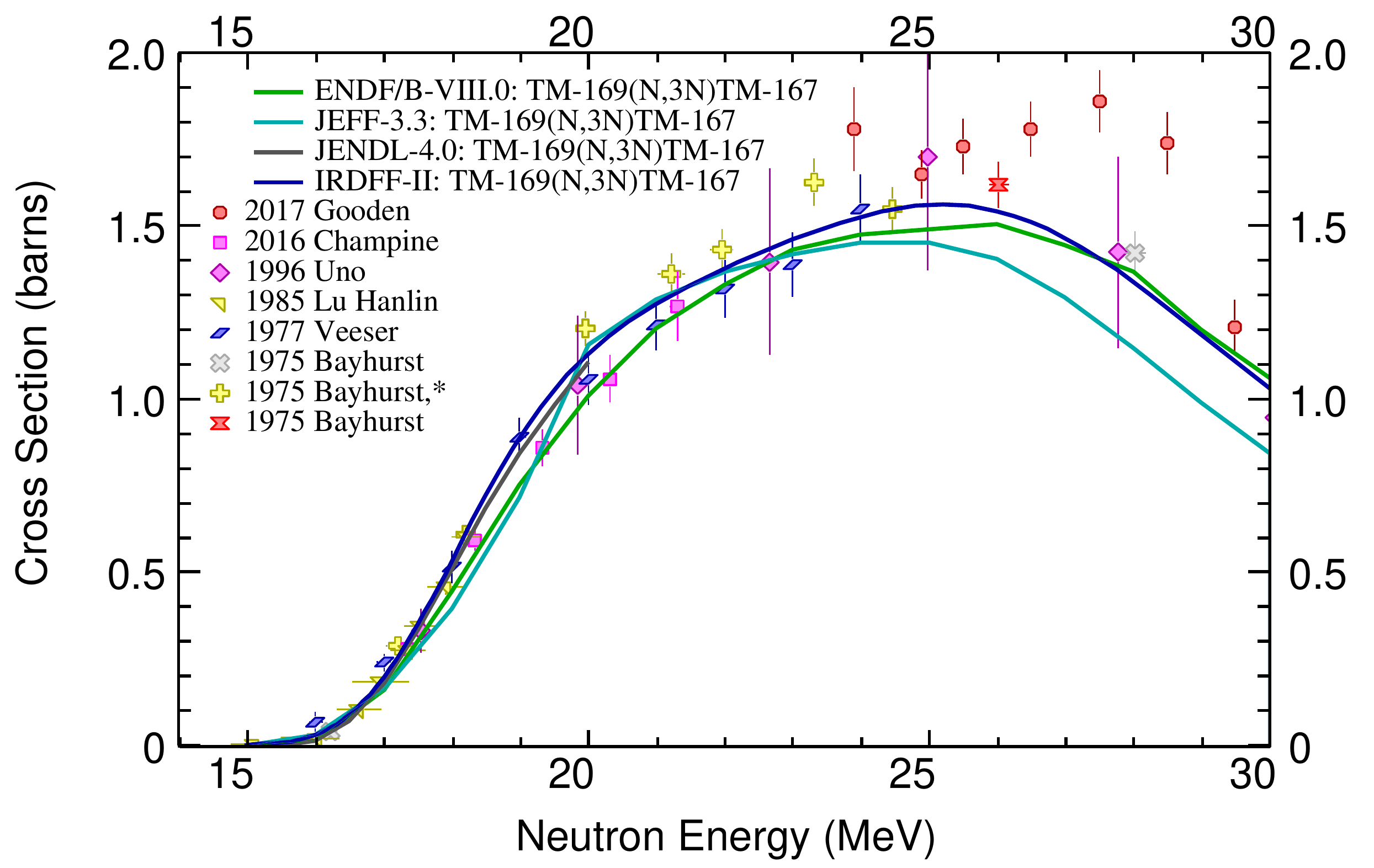}}
\subfigure[~Ratio of cross-section evaluations to \mbox{IRDFF-II} evaluation.]
{\includegraphics[width=\columnwidth]{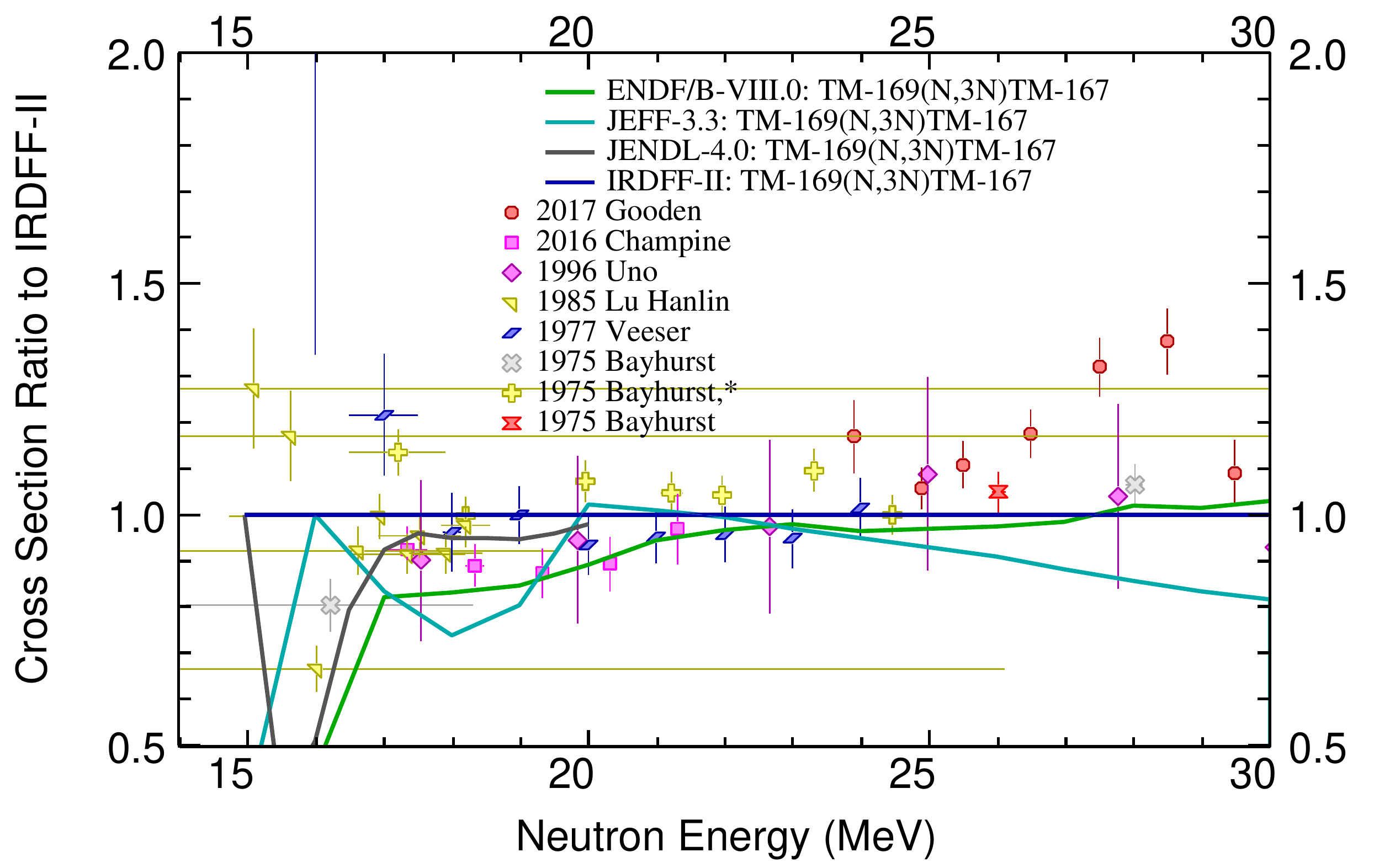}}
%\vspace{-3mm}
\caption{(Color online) $^{169}$Tm(n,3n)$^{167}$Tm cross-section evaluation in \mbox{IRDFF-II} library relative to other data.}
\label{n3Tm169}
%\vspace{-3mm}
\end{figure}

The $^{169}$Tm(n,2n)$^{168}$Tm reaction cross sections and covariances in energy range up to 40~MeV were evaluated by Zolotarev~\cite{Zol10}. Extension to 60~MeV was done at the IAEA based on \mbox{TENDL-2010}, renormalizing the cross sections at 40~MeV for continuity. The $^{169}$Tm(n,3n)$^{167}$Tm reaction cross sections and covariances in energy range up to 60~MeV were evaluated by Zolotarev~\cite{Zol13} in 2003; no extension in energy was necessary. Comparison of the cross sections with experimental data for $^{169}$Tm(n,2n)$^{168}$Tm and $^{169}$Tm(n,3n)$^{167}$Tm  reactions is shown in Figs.~\ref{nTm169}(a) and~\ref{n3Tm169}(a), respectively. The differences in reaction cross sections for $^{169}$Tm(n,2n)$^{168}$Tm from \mbox{IRDF-2002} are shown, which were unchanged in \mbox{IRDFF-II}. The cross sections for $^{169}$Tm(n,3n)$^{167}$Tm were not present in \mbox{IRDF-2002}; in \mbox{IRDFF-II} they remained unchanged from version \mbox{IRDFF-v1.05}. The $^{169}$Tm(n,2n)$^{168}$Tm cross sections in different evaluated libraries differ near threshold are shown in Fig.~\ref{nTm169}(b), where the ratios of the various evaluations to the \mbox{IRDFF-II} recommended cross section are given. Differences in the $^{169}$Tm(n,3n)$^{167}$Tm cross sections are even larger. Fig.~\ref{n3Tm169}(b) includes experimental data, which tend to support the \mbox{IRDFF-II} evaluation.

\begin{figure}[!thb]
%\vspace{-4mm}
\subfigure[~Cross-section correlation matrix.]
{\includegraphics[width=\columnwidth]{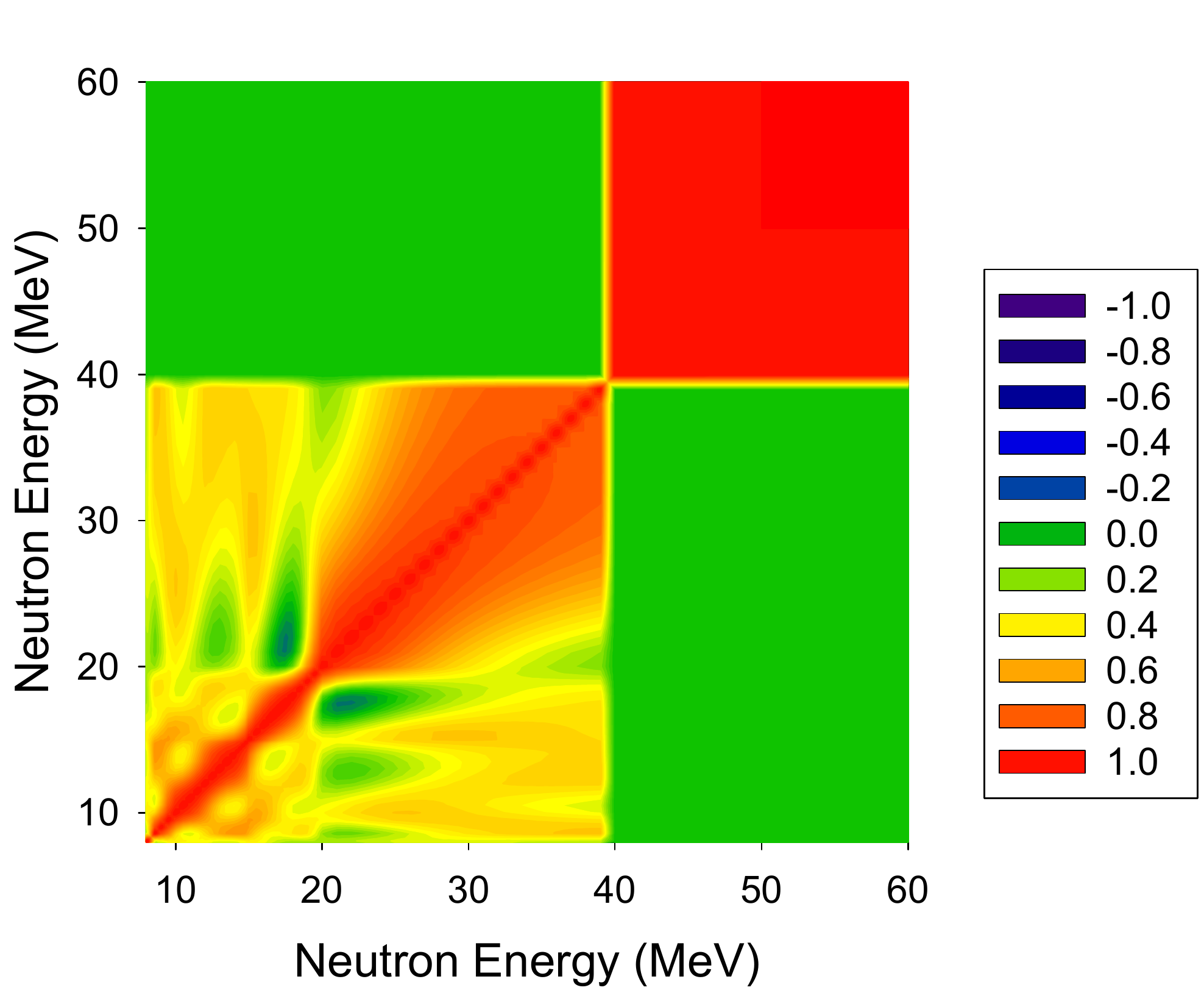}}
\subfigure[~One-sigma uncertainties in \% ($\equiv 100\times \frac{\sqrt{cov(i,i)}}{\mu_i}$), being $\mu_i$ the corresponding cross-section mean value.]
{\includegraphics[width=\columnwidth]{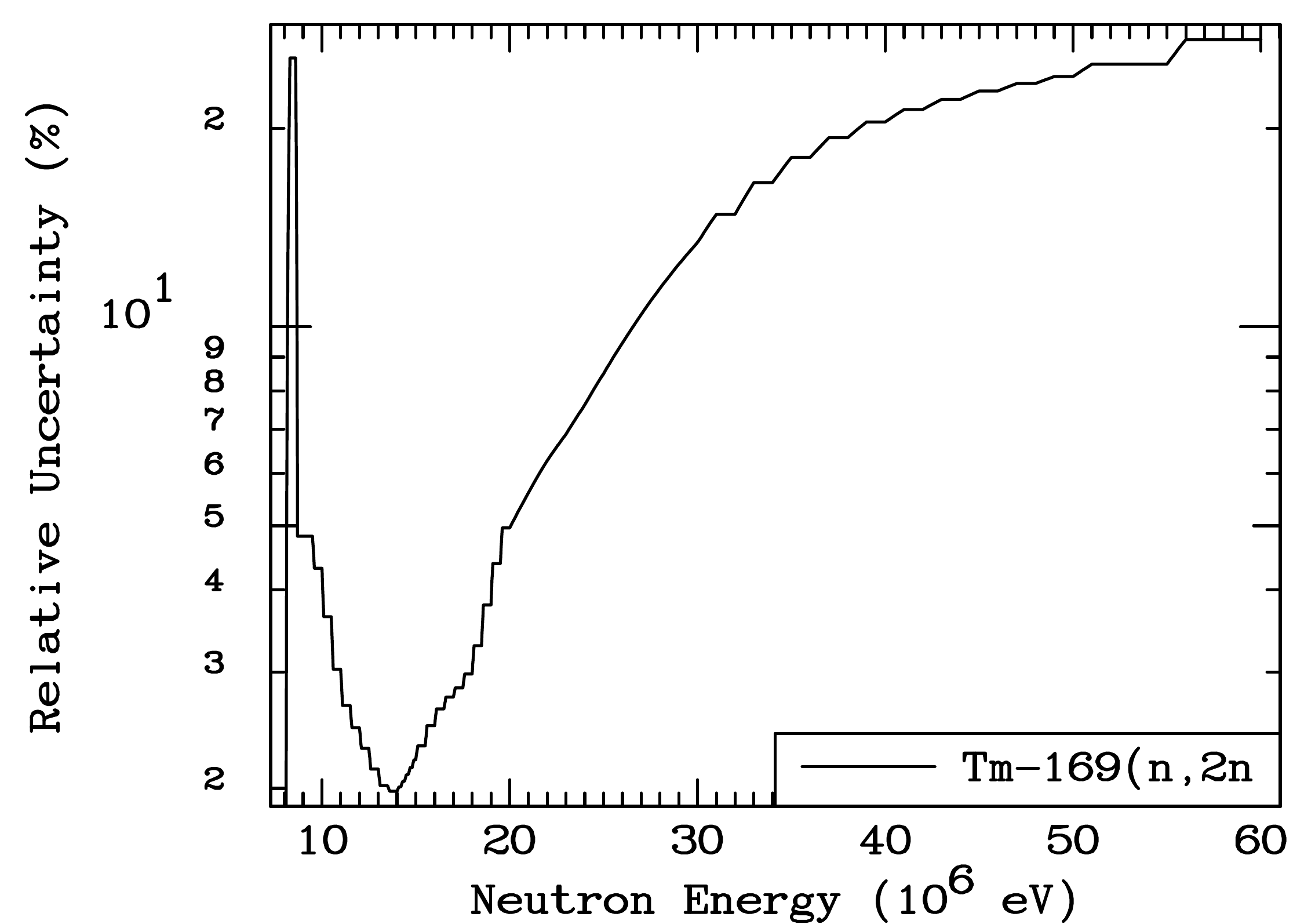}}
%\vspace{-3mm}
\caption{(Color online) Uncertainties and correlations of the $^{169}$Tm(n,2n)$^{168}$Tm cross-section evaluation in \mbox{IRDFF-II} library.}
\label{nTm169-unc}
%\vspace{-3mm}
\end{figure}

\newpage
The plot of the correlation matrix for the $^{169}$Tm(n,2n)$^{168}$Tm reaction is shown in Fig.~\ref{nTm169-unc}(a). The plot of the 1-sigma uncertainty is shown in Fig.~\ref{nTm169-unc}(b).

\begin{figure}[!thb]
%\vspace{-4mm}
\subfigure[~Cross-section correlation matrix.]
{\includegraphics[width=\columnwidth]{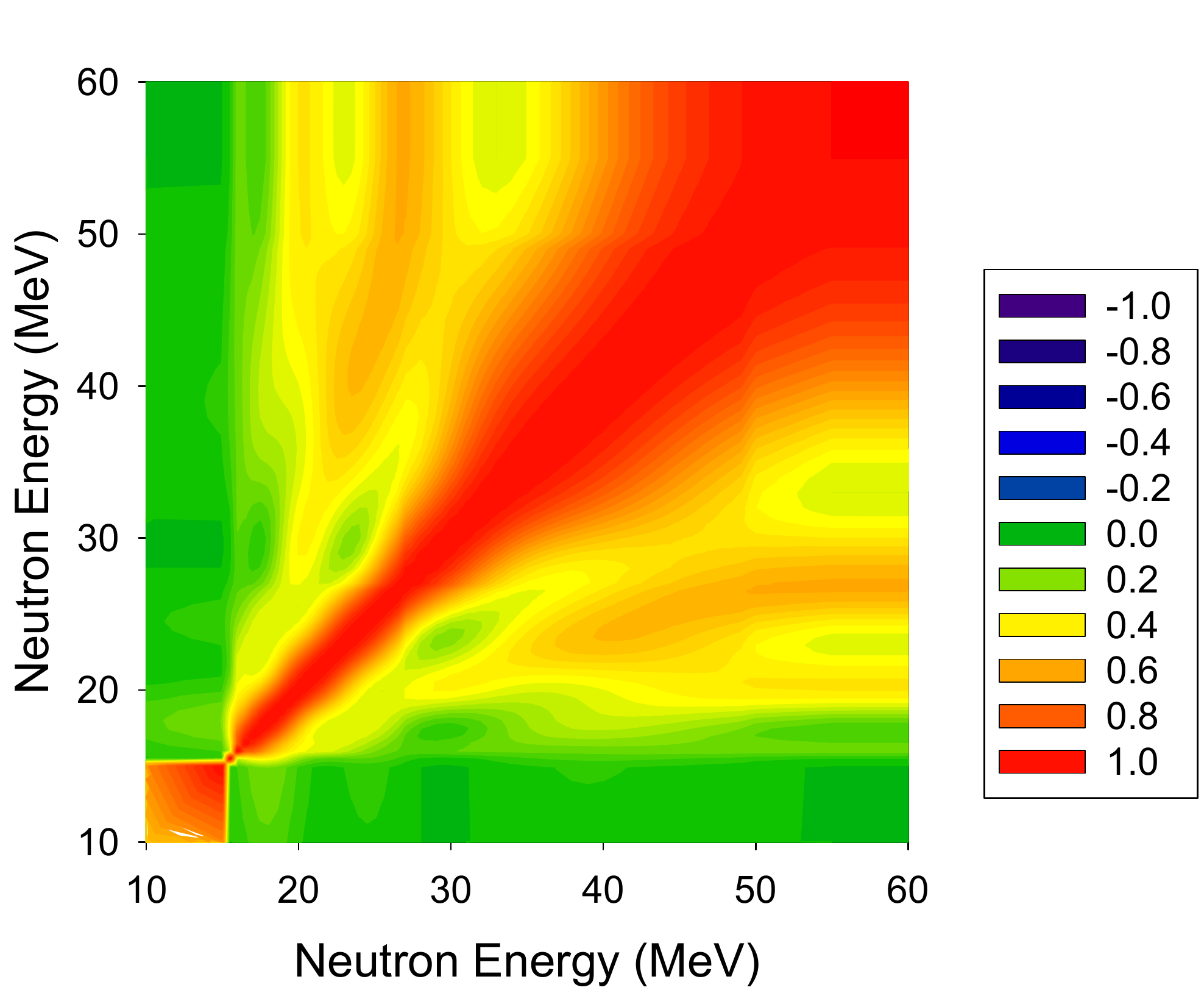}}
\subfigure[~One-sigma uncertainties in \% ($\equiv 100\times \frac{\sqrt{cov(i,i)}}{\mu_i}$), being $\mu_i$ the corresponding cross-section mean value.]
{\includegraphics[width=\columnwidth]{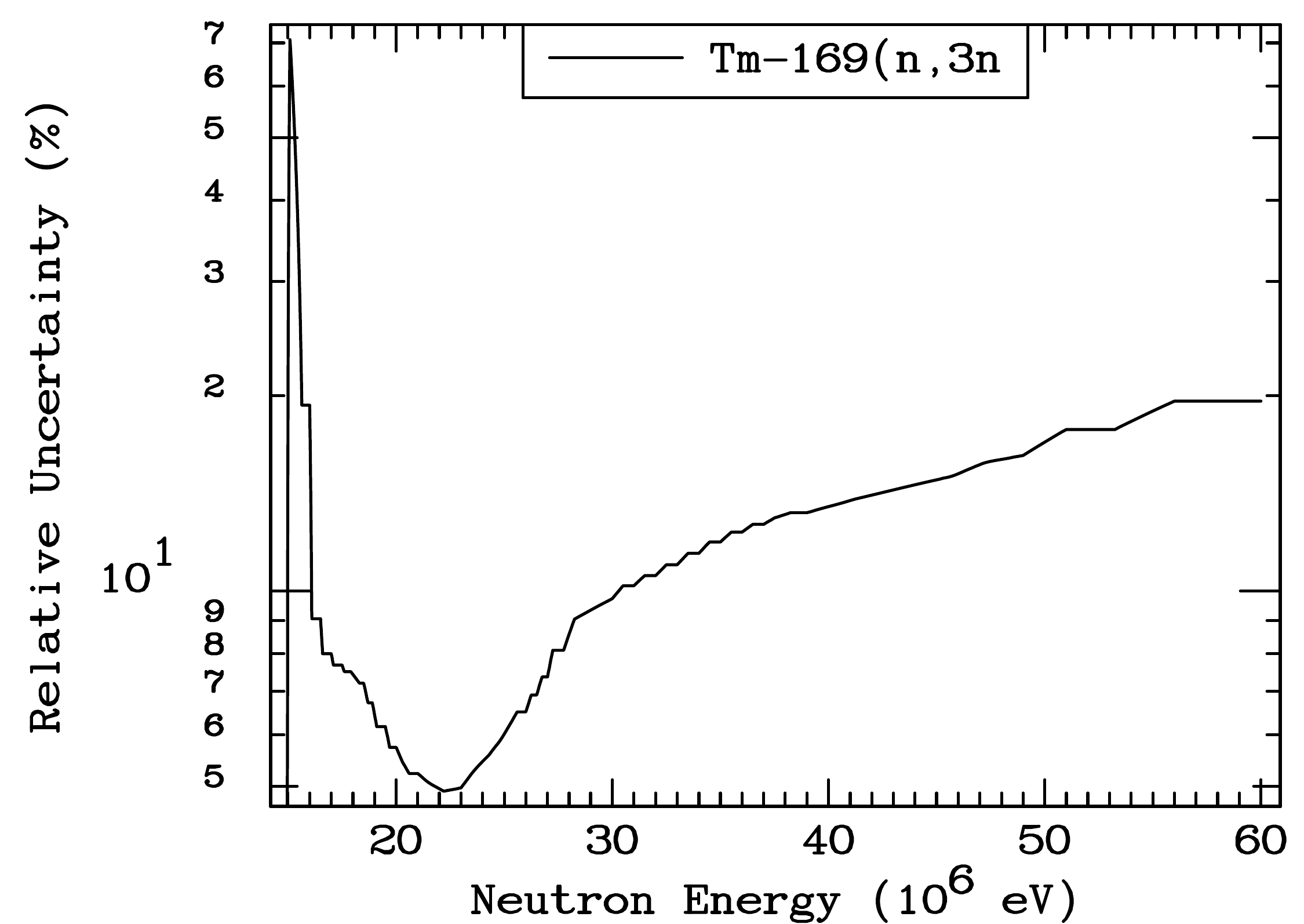}}
%\vspace{-3mm}
\caption{(Color online) Uncertainties and correlations of the $^{169}$Tm(n,3n)$^{167}$Tm cross-section evaluation in \mbox{IRDFF-II} library.}
\label{n3Tm169-unc}
%\vspace{-3mm}
\end{figure}

The plot of the correlation matrix for the $^{169}$Tm(n,3n)$^{167}$Tm reaction is shown in Fig.~\ref{n3Tm169-unc}(a). The plot of the 1-sigma uncertainty is shown in Fig.~\ref{n3Tm169-unc}(b).

The $^{169}$Tm(n,2n)$^{168}$Tm and $^{169}$Tm(n,3n)$^{167}$Tm reactions are examples of monitor reactions that can be used to probe neutron fields at higher energies.

\clearpage
\subsubsection{$^{209}\!$Bi(n,3n)$^{207}\!$Bi, $^{209}\!$Bi(n,4n)$^{206}\!$Bi, and $^{209}\!$Bi(n,5n)$^{205}\!$Bi}
\begin{figure}[hbtp]
  \includegraphics[width=\columnwidth]{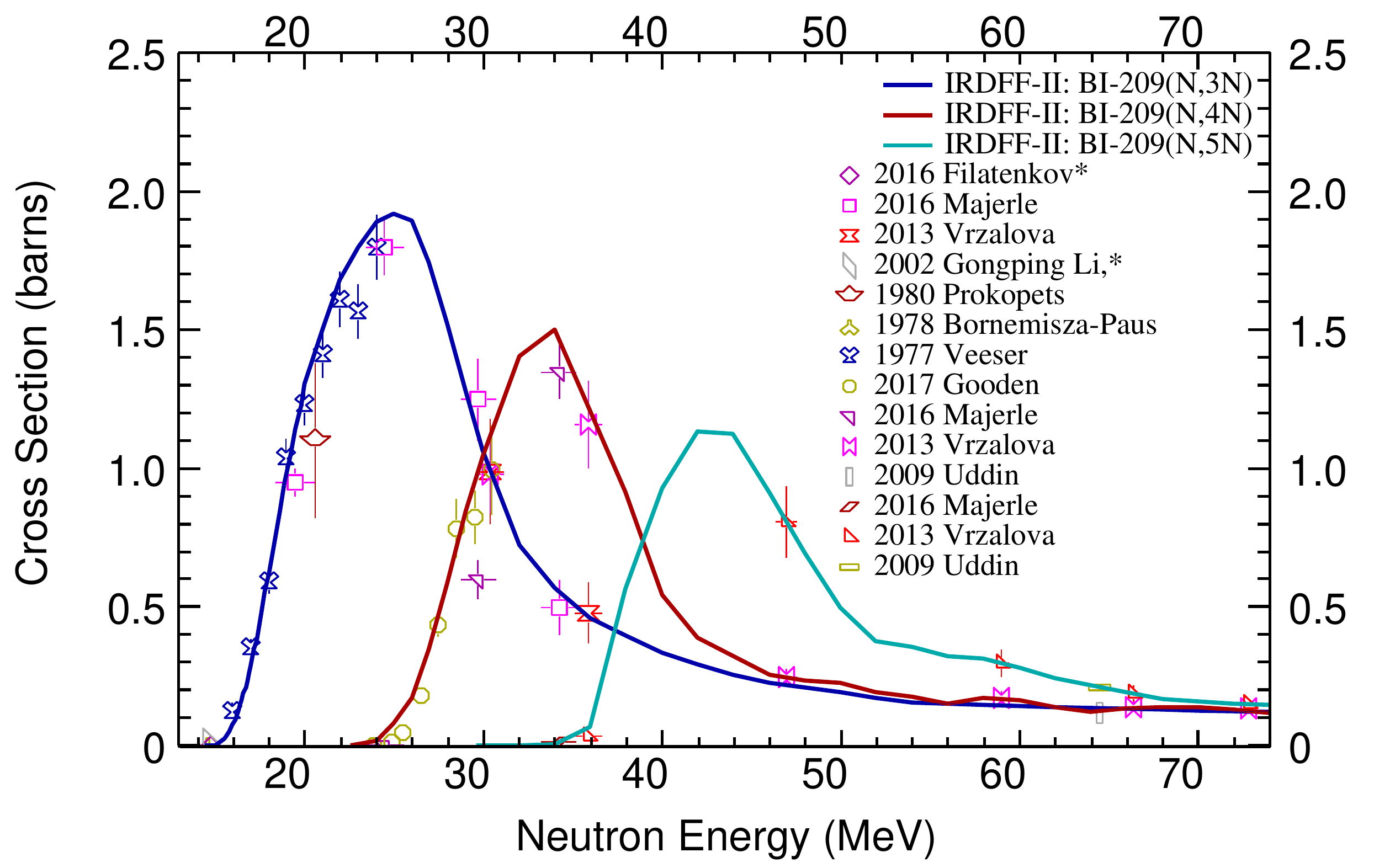}
  \caption{(Color online) \textit Comparison of $^{209}$Bi(n,xn) cross-section evaluations with experimental data from EXFOR.}
  \label{Bi209-n3n-n4n-n5n-exp}
\end{figure}

The $^{209}$Bi reaction cross sections for multiple neutron emission with n=2--6 were evaluated up to 100~MeV by Pronyaev in 2013 using a GMA code and smoothed and extended up to 400~MeV by Pronyaev in 2019. Comparison of the cross sections for $^{209}$Bi(n,3n), $^{209}$Bi(n,4n) and $^{209}$Bi(n,5n) reactions with experimental data is shown in Fig.~\ref{Bi209-n3n-n4n-n5n-exp}. No comparison with other libraries is given, since the evaluated cross sections from other libraries do not extend to energies above 20 MeV which are supported by the new \mbox{IRDFF-II} library. The cross sections for these reactions were not present in \mbox{IRDF-2002}.

\begin{figure}[!thb]
\vspace{-4mm}
\subfigure[~Cross-section correlation matrix.]
{\includegraphics[width=\columnwidth]{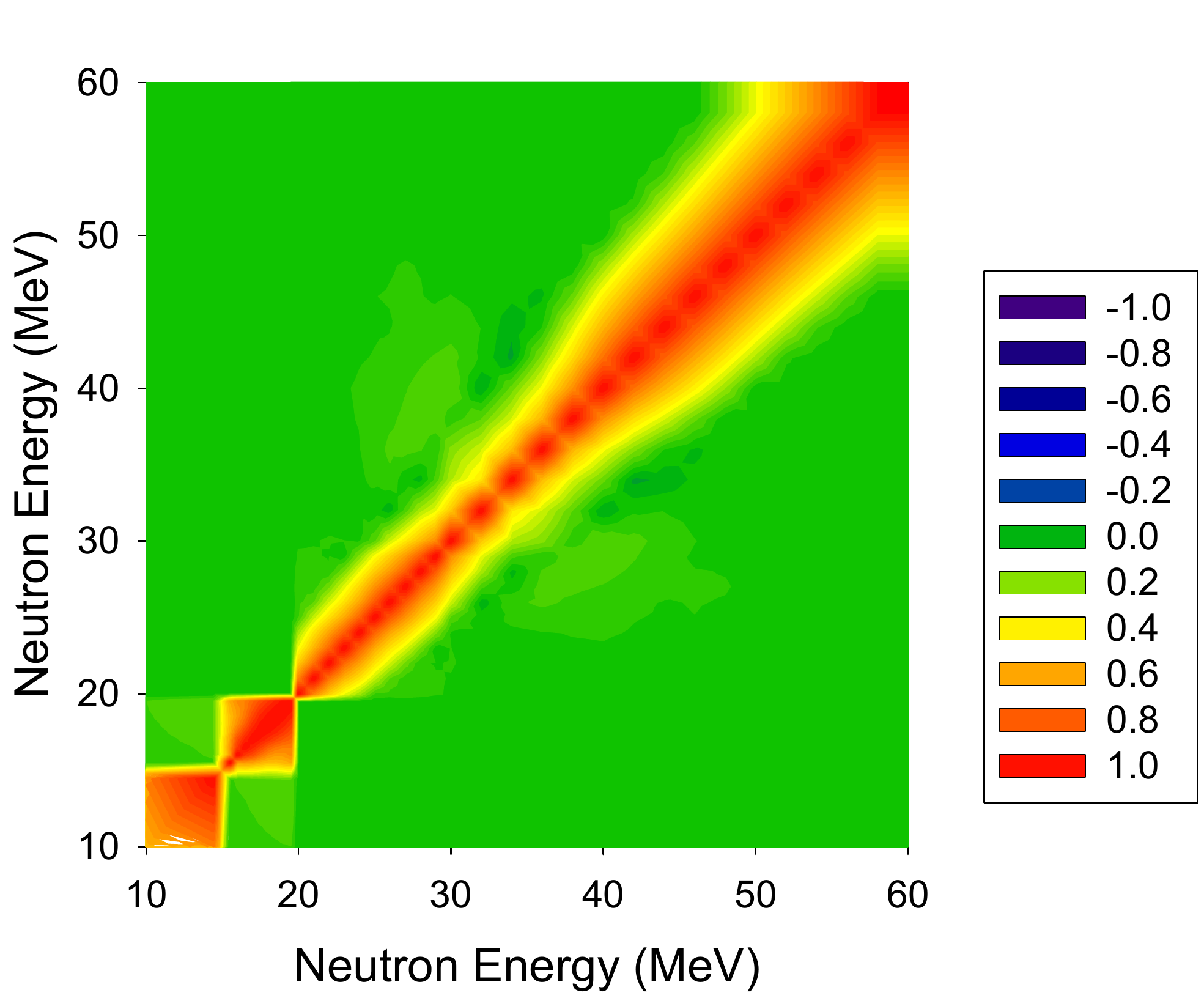}}
\subfigure[~One-sigma uncertainties in \% ($\equiv 100\times \frac{\sqrt{cov(i,i)}}{\mu_i}$), being $\mu_i$ the corresponding cross-section mean value.]
{\includegraphics[width=\columnwidth]{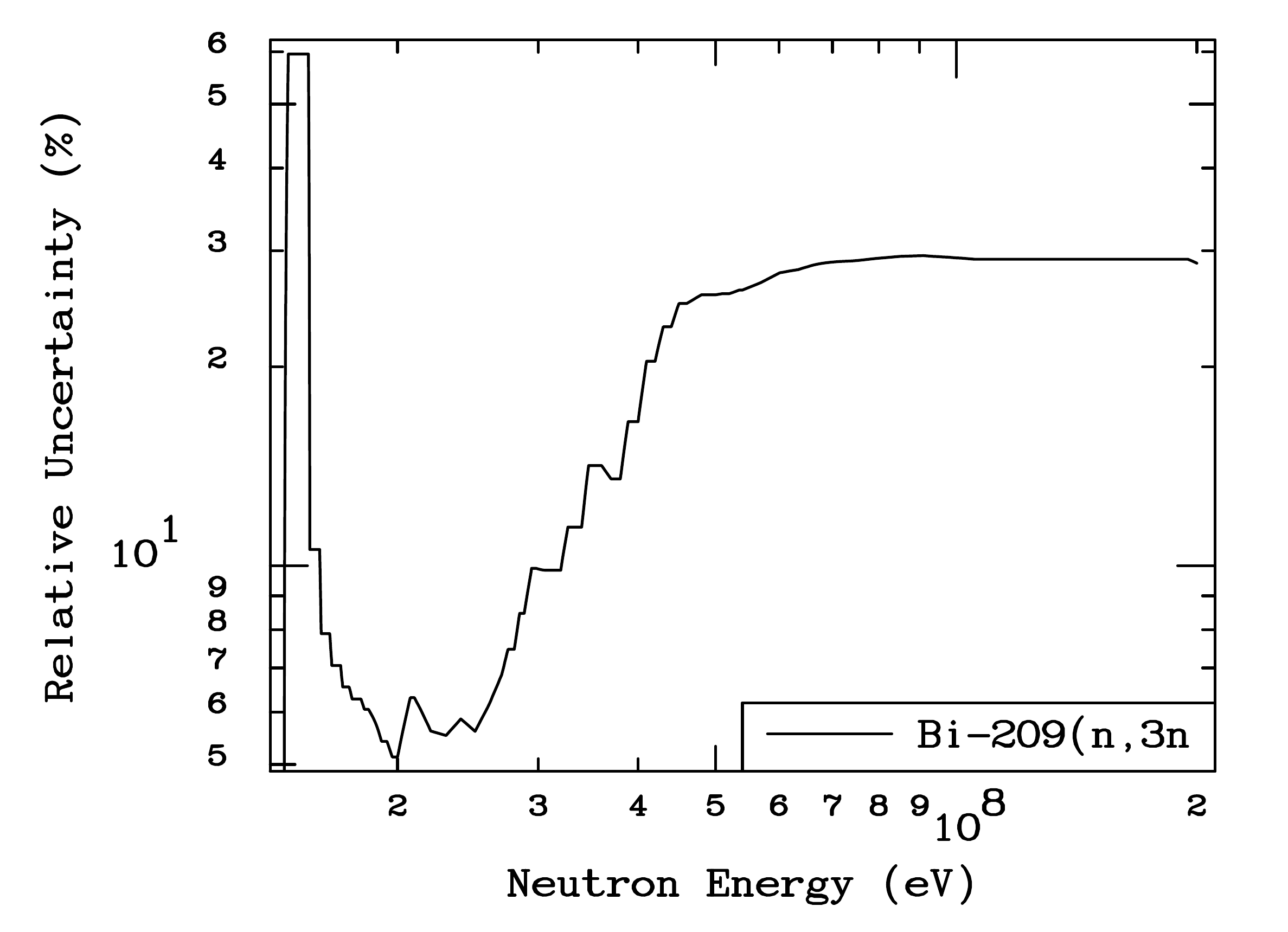}}
\vspace{-3mm}
\caption{(Color online) Uncertainties and correlations of the $^{209}$Bi(n,3n)$^{207}$Bi cross-section evaluation in \mbox{IRDFF-II} library.}
\label{n3Bi209-unc}
\vspace{-3mm}
\end{figure}

As an example for these reactions, the plot of the correlation matrix for the $^{209}$Bi(n,3n)$^{207}$Bi reaction is shown in Fig.~\ref{n3Bi209-unc}(a). The plot of the 1-sigma uncertainty is shown in Fig.~\ref{n3Bi209-unc}(b).

The $^{209}$Bi(n,xn) reactions in the \mbox{IRDFF-II} library are examples of monitor reactions that are intended for probing high-energy neutron fields that extend well above 10 MeV up to the maximum energy of 60~MeV.

\newpage
\subsubsection{$^{237}\!$Np(n,f)}
\begin{figure}[!thb]
\vspace{-2mm}
\subfigure[~Comparison to selected experimental data from EXFOR \cite{EXF08}.]
{\includegraphics[width=\columnwidth]{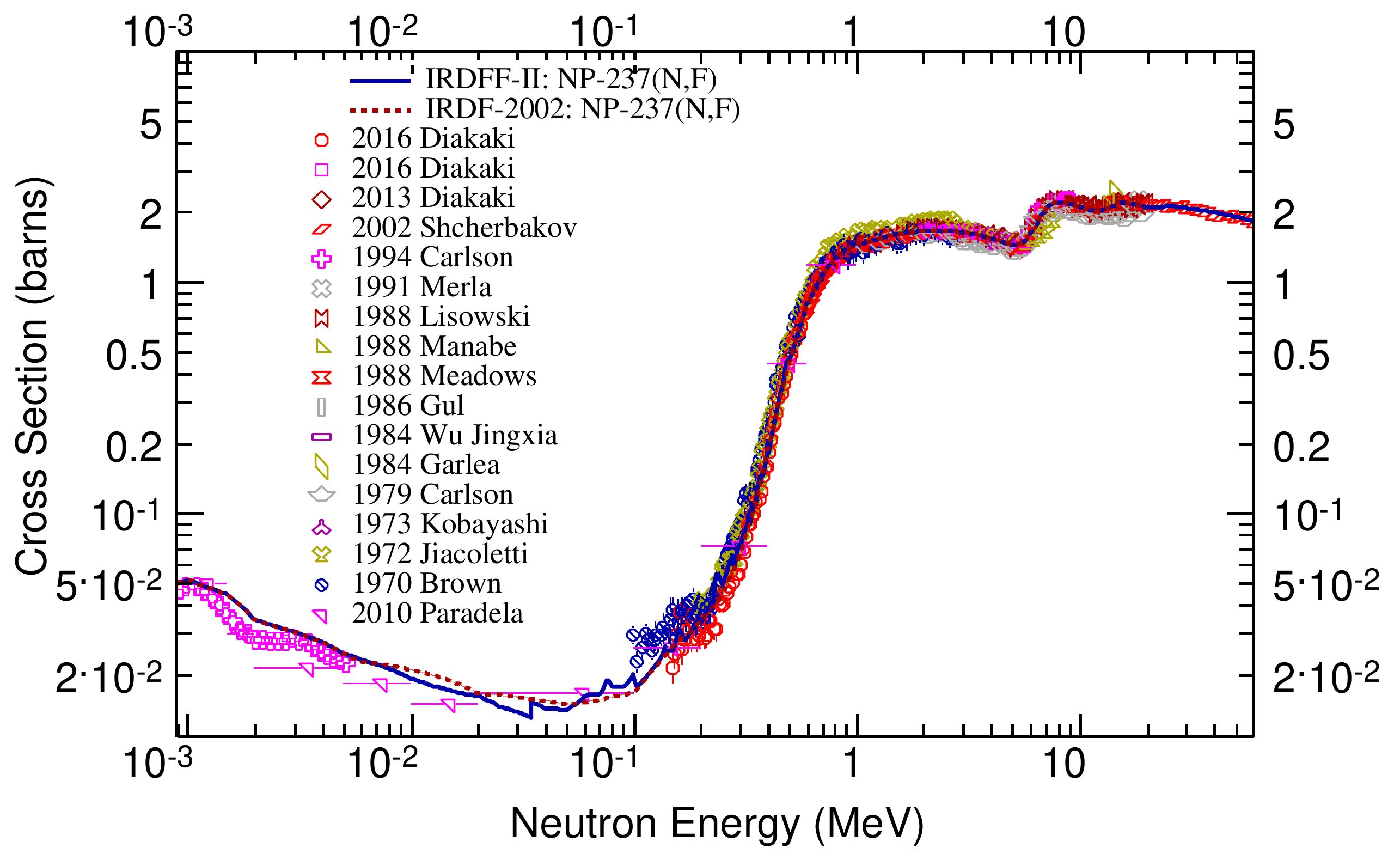}}
\subfigure[~Ratio of cross-section evaluations to \mbox{IRDFF-II} evaluation.]
{\includegraphics[width=\columnwidth]{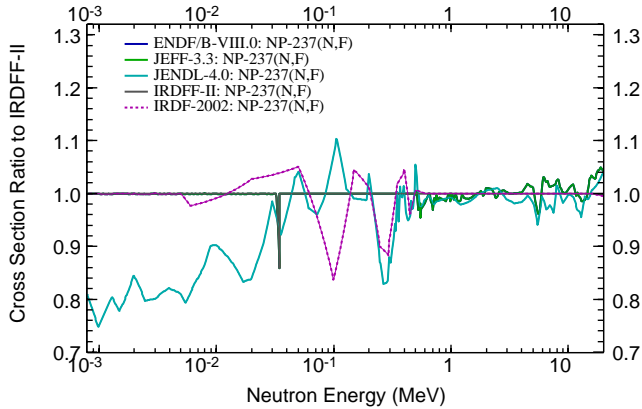}}
\vspace{-3mm}
\caption{(Color online) $^{237}$Np(n,f) cross-section evaluation in \mbox{IRDFF-II} library relative to other data.}
\label{fNp237}
\vspace{-3mm}
\end{figure}

The cross sections for the $^{237}$Np(n,f) reaction below 0.5~MeV were taken from the \mbox{JEFF-3.3} evaluation. In the energy interval 0.5--20~MeV the data were taken from \mbox{RRDF-2002}. Above this energy the JENDL-4/HE evaluation was used, but it does not include the covariances. The difference in the cross sections from \mbox{TENDL-2015} in 11 subintervals was calculated and the relative difference was added to the diagonal term of the TENDL-2015 covariance matrix in the 20~MeV to 60~MeV energy interval to account for the fact that the original covariances do not correspond to the same cross section evaluation. Comparison of the cross sections with experimental data is shown in Fig.~\ref{fNp237}(a). The $^{237}$Np(n,f) cross sections in different evaluated libraries are shown in Fig.~\ref{fNp237}(b), where the ratios of the various evaluations to the \mbox{IRDFF-II} recommended cross section are given. Eigenvalues of covariances were checked to be positive, but it is worth noting that the quality of the evaluation above 20 MeV should be improved if this reaction will be used at those energies.

\begin{figure}[!thb]
\vspace{-4.5mm}
\subfigure[~Cross-section correlation matrix.]
{\includegraphics[width=0.98\columnwidth]{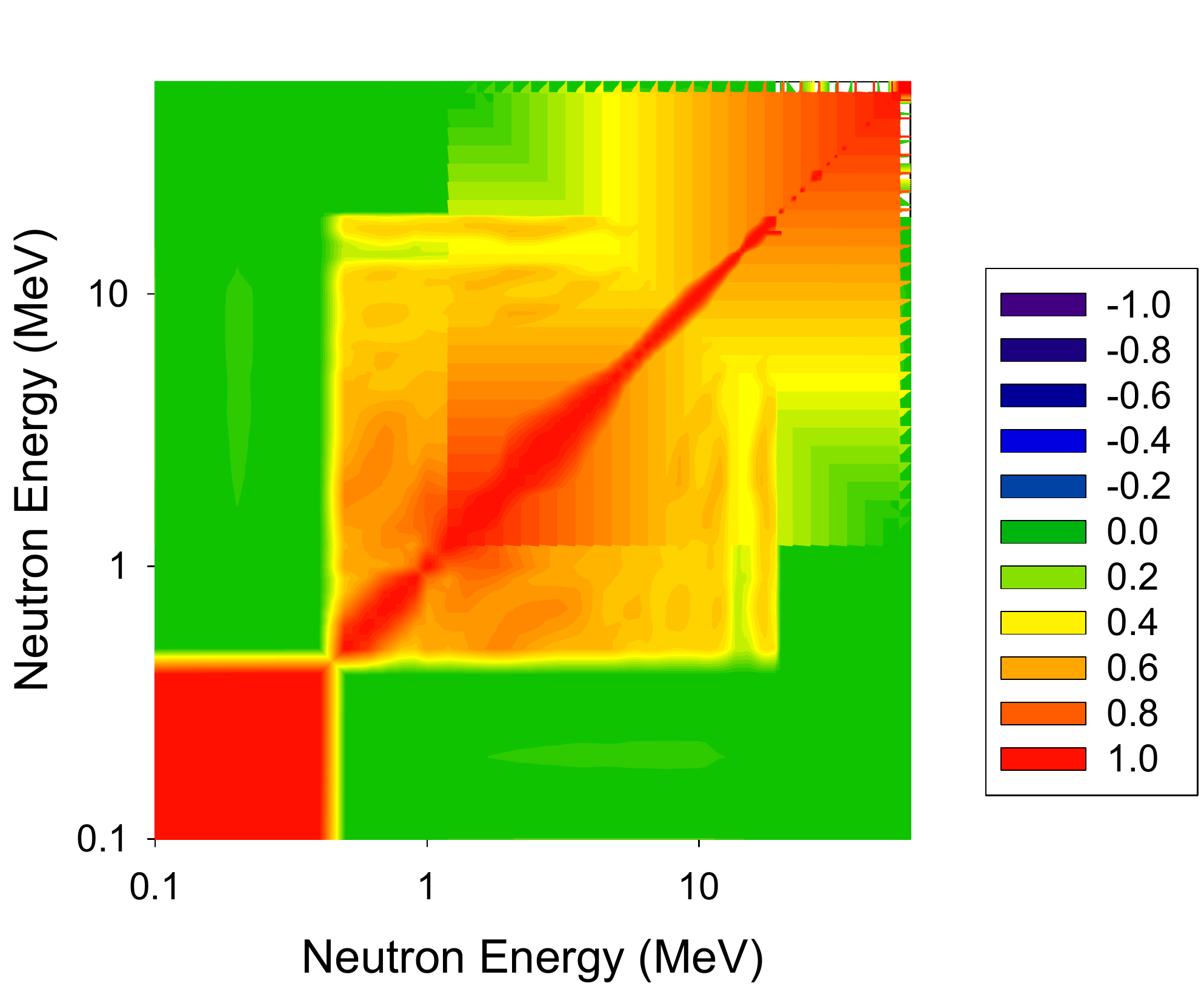}}
\subfigure[~One-sigma uncertainties in \% ($\equiv 100\times \frac{\sqrt{cov(i,i)}}{\mu_i}$), being $\mu_i$ the corresponding cross-section mean value.]
{\includegraphics[width=0.98\columnwidth]{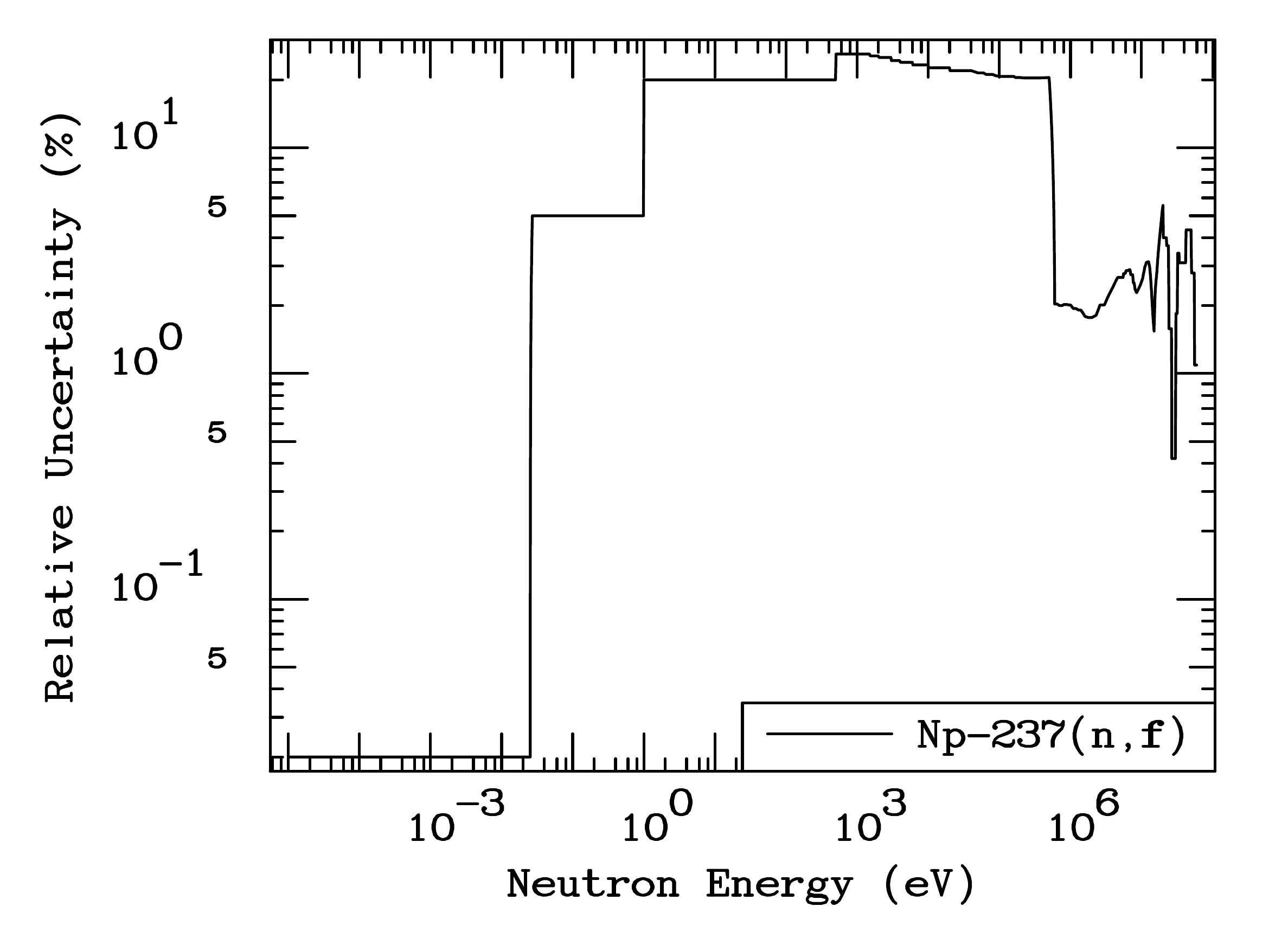}}
\vspace{-3mm}
\caption{(Color online) Uncertainties and correlations of the $^{237}$Np(n,f) cross-section evaluation in \mbox{IRDFF-II} library.}
\label{fNp237-unc}
\vspace{-3mm}
\end{figure}

The plot of the correlation matrix for the $^{237}$Np(n,f) reaction is shown in Fig.~\ref{fNp237-unc}(a). The plot of the 1-sigma uncertainty is shown in Fig.~\ref{fNp237-unc}(b).

\newpage
%\subsubsection{$^{197}\!$Au(n,$\gamma$)$^{198}\!$Au, $^{235}\!$U(n,f), $^{238}\!$U(n,f), $^{238}\!$U(n,$\gamma$)$^{239}\!$U, and $^{6}\!$Li(n,t)$^{4}\!$He}
\subsubsection{$^{197}\!$Au(n,$\gamma$)$^{198}\!$Au, $^{235}\!$U(n,f), $^{238}\!$U(n,f), $^{238}\!$U(n,$\gamma$)$^{239}\!$U, $^{10}\!$B(n,$\alpha$)$^{7}\!$Li, and $^{6}\!$Li(n,t)$^{4}\!$He} \label{Sec_V_Y}
The reactions in this section are Neutron Standards (or reaction cross section evaluated together with the Standards) in certain limited energy ranges and are described in detail in a previous publication~\cite{Car18}. What is shown here are examples, where the content of the \mbox{IRDFF-II} library significantly exceeds the energy range of the IAEA Standards~2017.

\begin{figure}[!thb]
\vspace{-3mm}
\subfigure[~Comparison to selected experimental data from EXFOR \cite{EXF08}.]
{\includegraphics[width=\columnwidth]{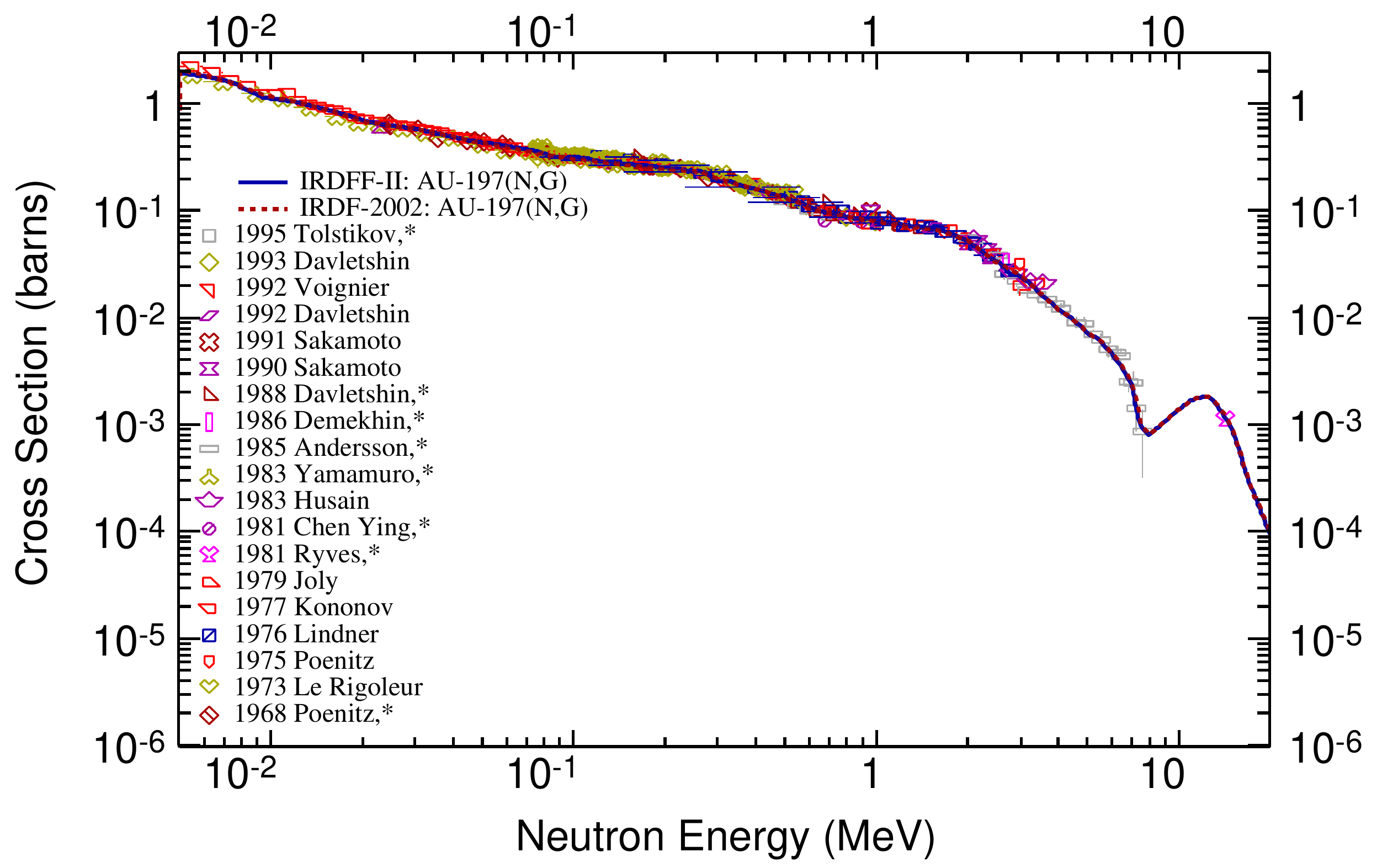}}
\subfigure[~Ratio of cross-section evaluations to \mbox{IRDFF-II} evaluation.]
{\includegraphics[width=\columnwidth]{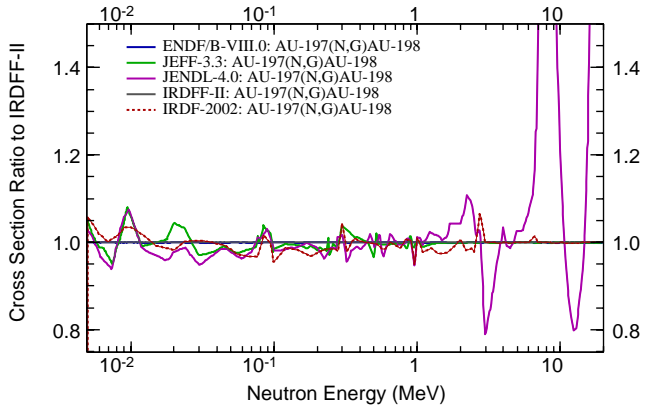}}
\vspace{-3mm}
\caption{(Color online) $^{197}$Au(n,$\gamma$)$^{198}$Au cross-section evaluation in \mbox{IRDFF-II} library relative to other data.}
\label{gAu197}
\vspace{-3mm}
\end{figure}

The $^{197}$Au(n,$\gamma$)$^{198}$Au reaction cross sections are compared to experimental data in Fig.~\ref{gAu197}(a). The capture cross sections of gold are probably the best known capture cross sections over a broad energy range; this is why they are declared a standard at thermal and in the energy range 0.2~MeV to 2.5~MeV~\cite{Sch14}. The \mbox{ENDF/B-VIII.0} evaluation is consistent with the IAEA Standards~2017 for this reaction, hence it was adopted for \mbox{IRDFF-II}. Comparison with other evaluations is shown as ratio to \mbox{IRDFF-II} in Fig.~\ref{gAu197}(b) for the fast energy range, since the comparison in the resonance range is not meaningful due to the resonances.

\begin{figure}[!thb]
\vspace{-5mm}
\subfigure[~Cross-section correlation matrix.]
{\includegraphics[width=\columnwidth]{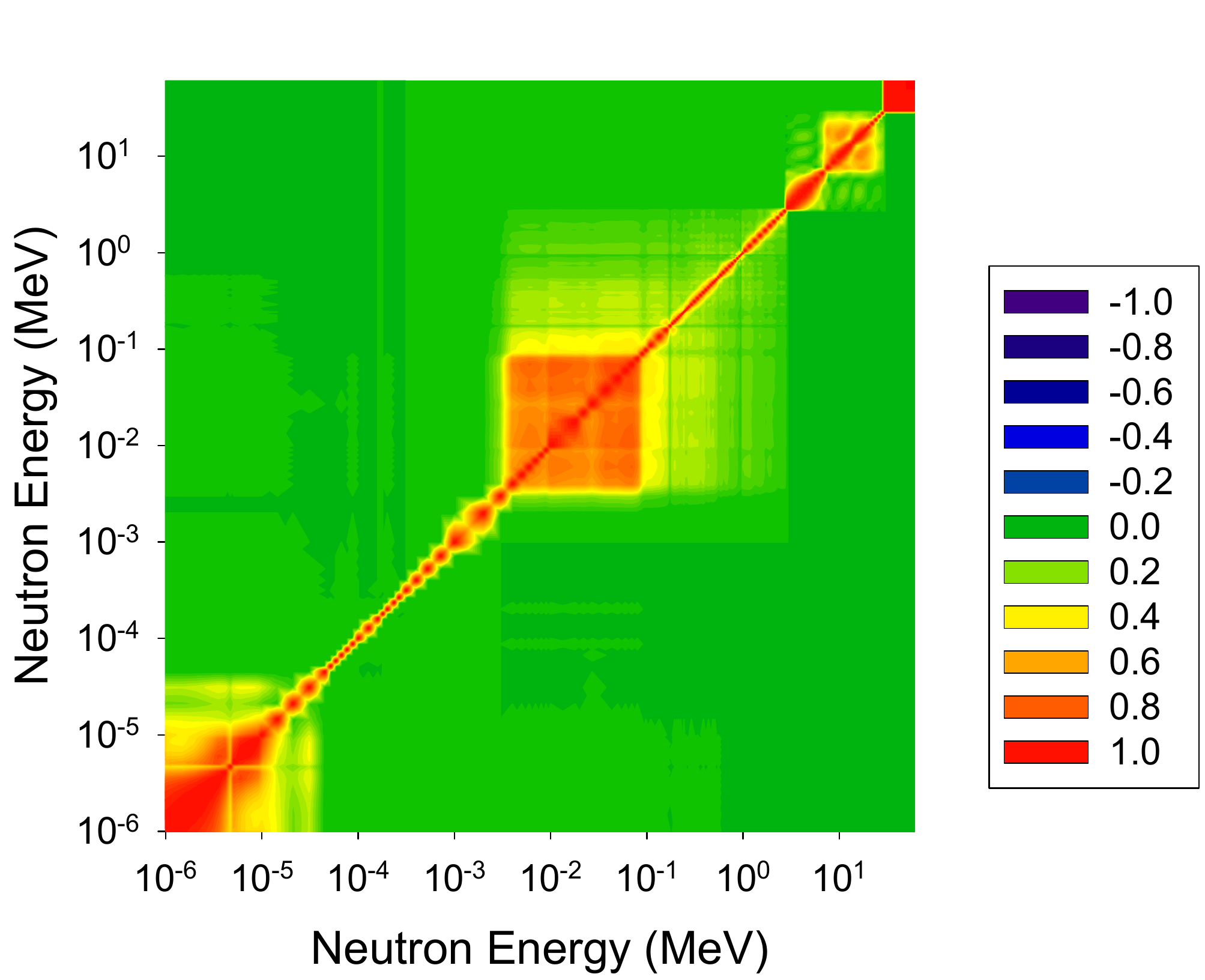}}
\subfigure[~One-sigma uncertainties in \% ($\equiv 100\times \frac{\sqrt{cov(i,i)}}{\mu_i}$), being $\mu_i$ the corresponding cross-section mean value.]
{\includegraphics[width=\columnwidth]{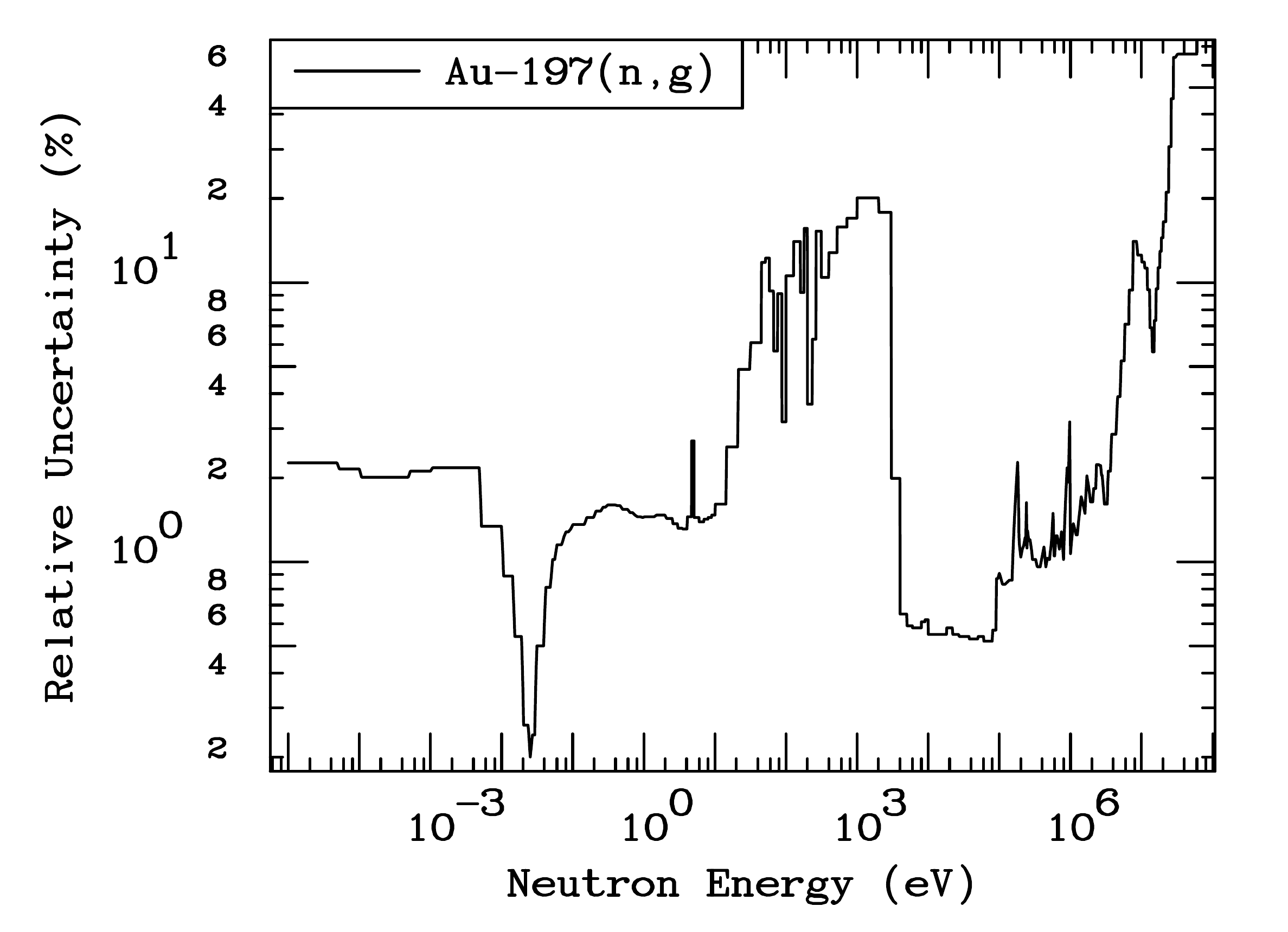}}
\vspace{-3mm}
\caption{(Color online) Uncertainties and correlations of the $^{197}$Au(n,$\gamma$)$^{198}$Au cross-section evaluation in \mbox{IRDFF-II} library.}
\label{gAu197-unc}
\vspace{-4mm}
\end{figure}

The correlation matrix is given in Fig.~\ref{gAu197-unc}(a) and the corresponding uncertainty is shown in Fig.~\ref{gAu197-unc}(b).

\begin{figure}[!hbtp]
  \includegraphics[width=\columnwidth]{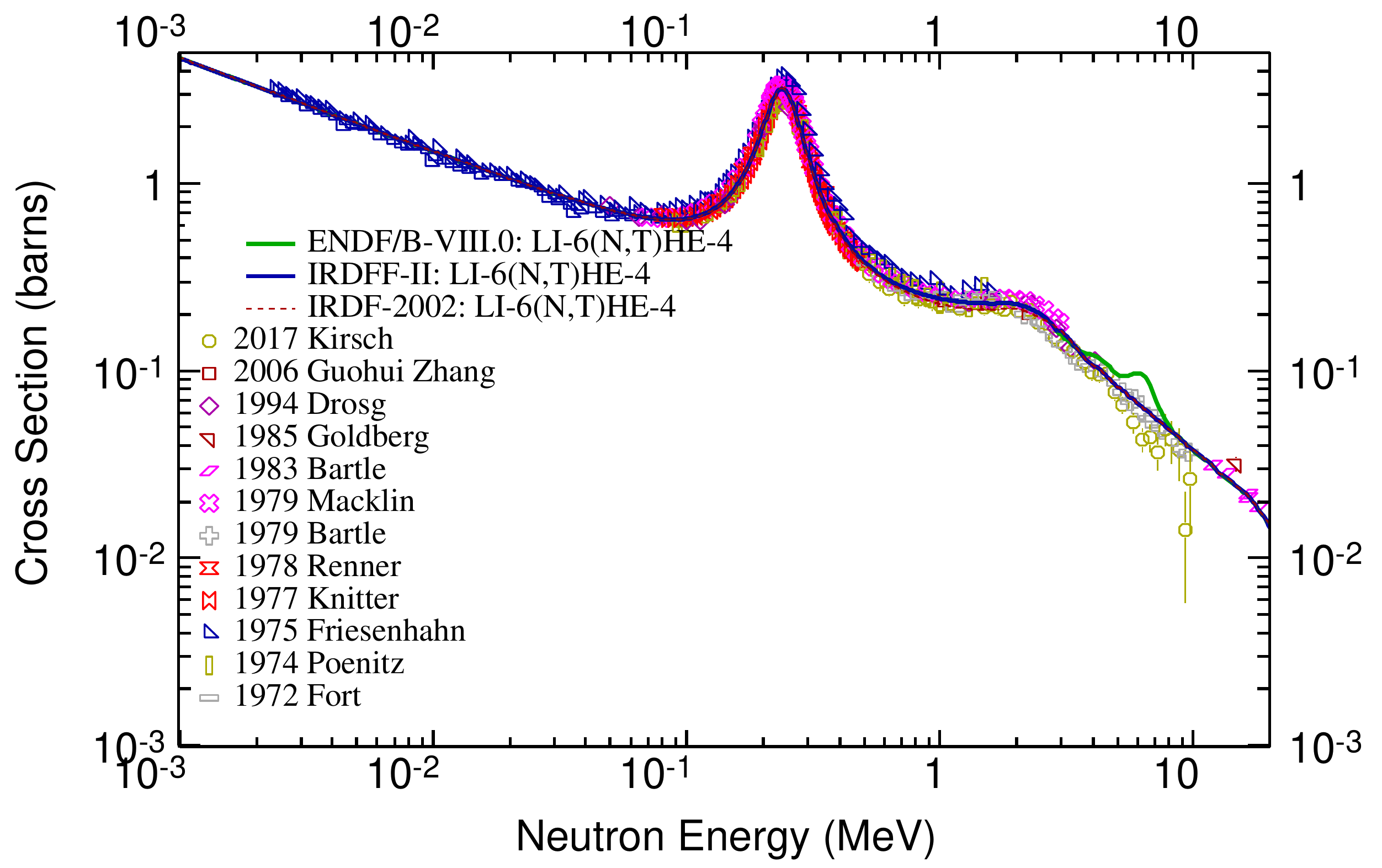}
  \caption{(Color online) \textit Comparison of $^{6}$Li(n,t)$^{4}$He cross sections with experimental data from EXFOR.}
  \label{tLi6-exp}
\end{figure}

The tritium and $^4$He production reaction $^{6}$Li(n,t)$^{4}$He has also been evaluated for the IAEA Neutron Standards~2017 in the energy range from 0.0253~eV to 1.4~MeV, being the standard energy region from the thermal point up to 1~MeV. The actual evaluation was performed over a broader energy range and is consistent with the contents of \mbox{ENDF/B-VIII.0} up to 20~MeV. However, by a careful consideration of available experimental evidence it was concluded that the structure in the cross section near 5~MeV that appears in the ENDF/B-VIII.0 is questionable. Note that this regions is well outside the Neutron Standards range.
At 2.75~MeV and between 8~MeV and 20~MeV the \mbox{ENDF/B-VIII.0} cross sections for this reaction are equal to those in \mbox{EAF-2010}. To smooth out the structure in the \mbox{ENDF/B-VIII.0} cross sections the data form \mbox{EAF-2010} were adopted above 2.75~MeV up to 60~MeV. In this respect \mbox{IRDFF-II} differs from \mbox{ENDF/B-VIII.0}. Comparison of the cross sections for the $^{6}$Li(n,X)$^{3}$H reaction with experimental data and with \mbox{ENDF/B-VIII.0} and \mbox{IRDFF-II} is shown in Fig.~\ref{tLi6-exp}, from which the anomalous behaviour of the \mbox{ENDF/B-VIII.0} cross sections can be seen.

\begin{figure}[!thb]
\vspace{-4mm}
\subfigure[~Cross-section correlation matrix.]
{\includegraphics[width=0.98\columnwidth]{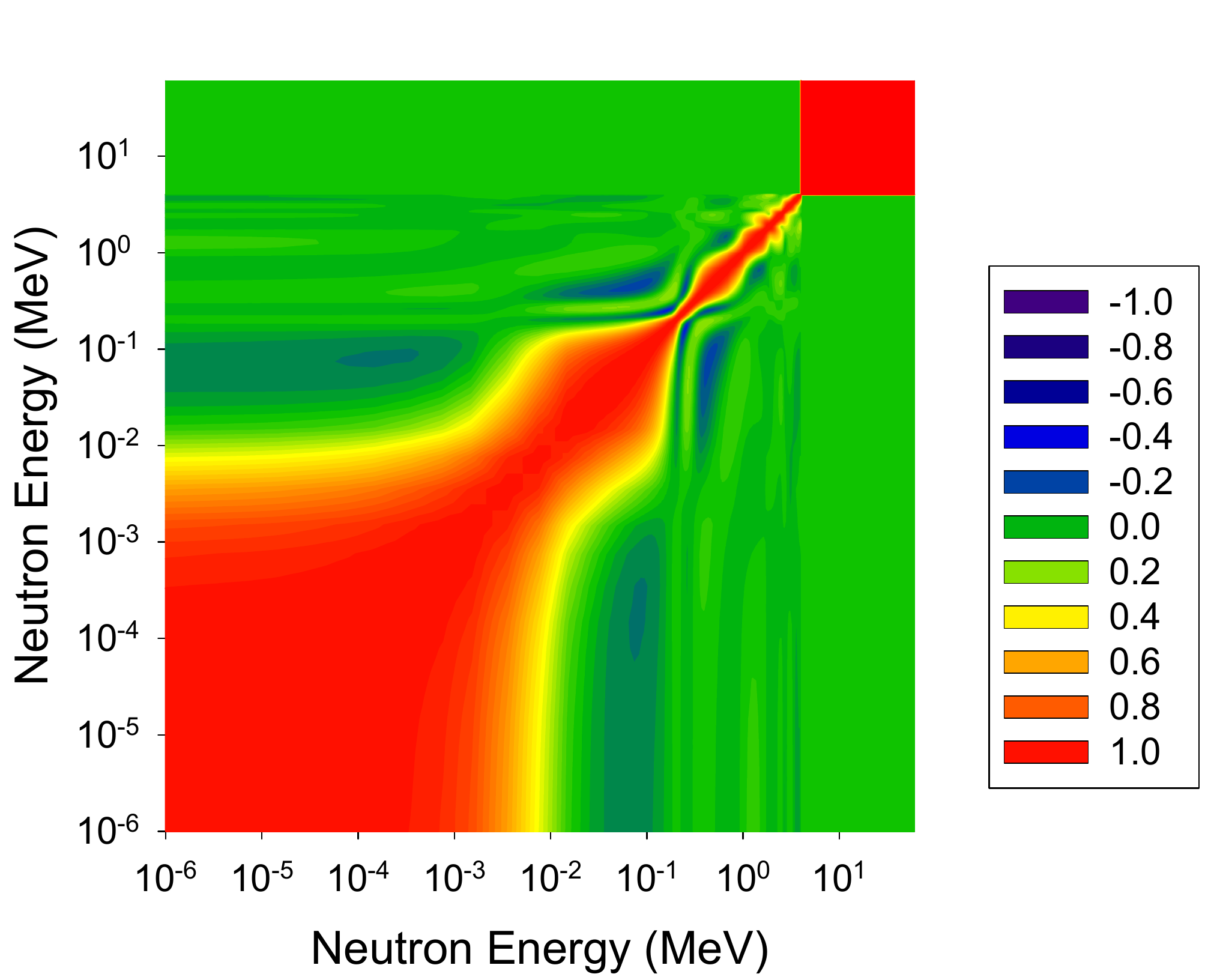}}
\subfigure[~One-sigma uncertainties in \% ($\equiv 100\times \frac{\sqrt{cov(i,i)}}{\mu_i}$), being $\mu_i$ the corresponding cross-section mean value.]
{\includegraphics[width=0.98\columnwidth]{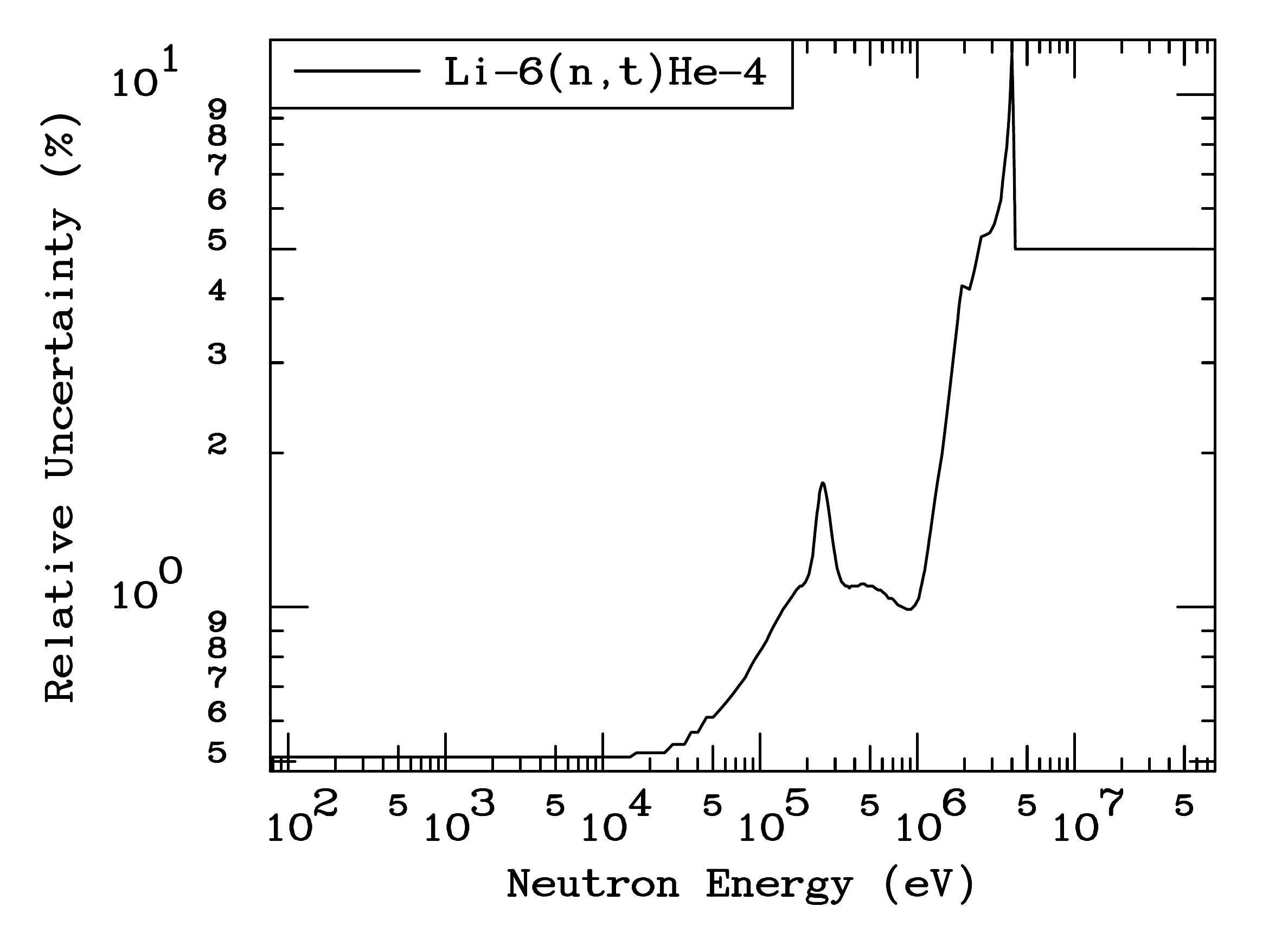}}
\vspace{-3mm}
\caption{(Color online) Uncertainties and correlations of the $^{6}$Li(n,t)$^{4}$He cross-section evaluation in \mbox{IRDFF-II} library.}
\label{tLi6-unc}
\vspace{-4mm}
\end{figure}

The correlation matrix is given in Fig.~\ref{tLi6-unc}(a) and the corresponding uncertainty is shown in Fig.~\ref{tLi6-unc}(b).

The $^{10}$B(n,$\alpha$) reaction cross section was also evaluated as part of the IAEA Neutron Standards~2017 from thermal to 1~MeV~\cite{Car18}. Both (n,$\alpha$) reactions on the ground and the first excited state of the $^{10}$B target were evaluated and coded as MT=800 and MT=801 in ENDF-6 format. The \mbox{ENDF/B-VIII.0} evaluation is consistent with these Standards.

\subsubsection{$^{6}\!$Li(n,X)$^{3}\!$H and $^{6}\!$Li(n,X)$^{4}\!$He} \label{Sec_V_AA}

\begin{table*}[!bhtp]
\vspace{-5mm}
\caption{Gas production reactions in IRDFF-II library from neutrons interacting with $^{6}$Li target. The MT numbers (reactions) used in IRDFF-II assembly following the ENDF-6 format, and the relevant reaction products are highlighted in bold.} \label{Li6_gas}
\begin{tabular}{ c | l | l | l c l | l | l }
\hline \hline
 MT            & LR & Gas      & Reaction               &             & Products                                                     & \multicolumn{2}{c}{Source evaluations of IRDFF-II} \\
\hline
\T
\textbf{105}   &    & $^{3}$H  & $^{6}$Li(n,t)$^{4}$He  &$\rightarrow$& \textbf{$^{3}$H}  +  $^{4}$He                                & EAF-2010(r) ($E>20$) & ENDF/B-VIII.0 ($E<20$)      \\
\hline
\T
\textbf{24}    &    & $^{4}$He & $^{6}$Li(n,2n$\alpha$) &$\rightarrow$& 2n + \textbf{$^{4}$He} + $^{1}$H                             & EAF-2010(r) ($E>20$) & ENDF/B-VIII.0 ($E<20$)      \\
\textbf{32}    &    &          & $^{6}$Li(n,nd)         &$\rightarrow$& n + $^{2}$H + \textbf{$^{4}$He}                              & EAF-2010(r) ($E>20$) & ENDF/B-VIII.0 MT51,52,...   \\
51,52,...      & 32 &          & $^{6}$Li(n,n')         &$\rightarrow$& n + ($^{6}$Li $\rightarrow$ $^{2}$H + \textbf{$^{4}$He})     &                      & ENDF/B-VIII.0 ($E<20$)      \\
\textbf{105}   &    &          & $^{6}$Li(n,t)$^{4}$He  &$\rightarrow$& $^{3}$H + \textbf{$^{4}$He}                                  & EAF-2010(r) ($E>20$) & ENDF/B-VIII.0 ($E<20$)      \\
\hline \hline
\end{tabular}
\vspace{-3mm}
\end{table*}

The gas-production reactions from neutrons incident on light targets were not evaluated separately. Direct tritium production in $^{6}$Li, with $^{4}$He as the residual was evaluated within \mbox{IAEA Standards 2017}~\cite{Car18} and is described in Sec.~\ref{Sec_V_Y}. Other gas-production cross sections were assembled by reviewing existing evaluated data libraries and choosing the most appropriate combination of data, as described below. In particular, the libraries that were considered include \mbox{ENDF/B-VIII.0}, \mbox{EAF-2010} European Activation File and  \mbox{TENDL-2015} sublibrary "s60", which contains explicit cross section representation with covariances up to 60~MeV. The \mbox{TENDL-2015}.s60 library comes from TALYS nuclear model calculations, which are not best suited for light nuclei, but the library was considered because of its completeness in terms of reactions. However, \mbox{TENDL-2015} library was found inappropriate for neutron induced reactions on boron and lithium isotopes by extensive comparison with available experimental data and other available evaluations, and will not be discussed in this subsection.

\begin{figure}[!thb]
\vspace{-1mm}
\subfigure[~Comparison of data from selected evaluated libraries relative to the \mbox{IRDFF-II} evaluation.]
{\includegraphics[width=\columnwidth]{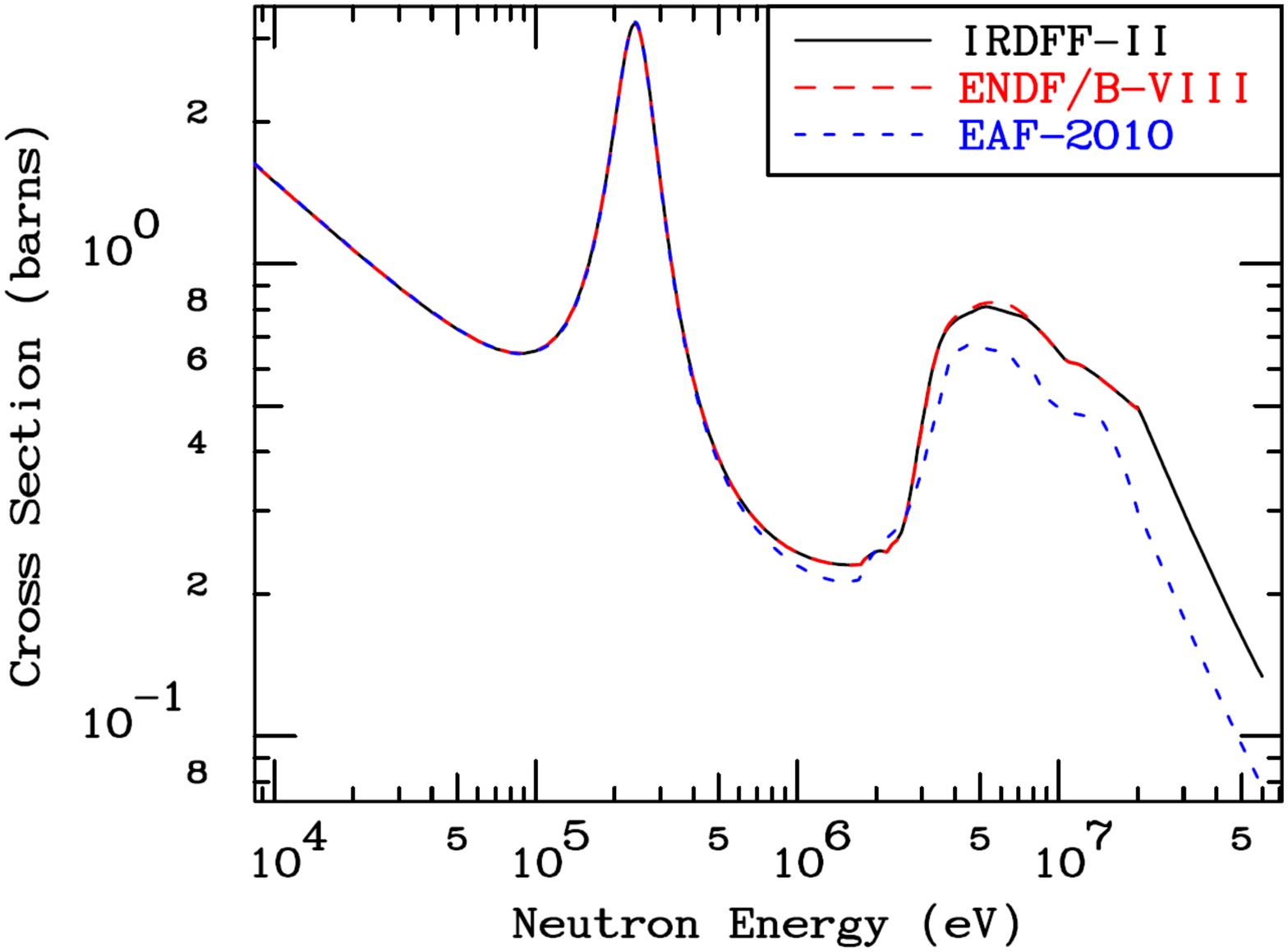}}
\subfigure[~Contribution of different reaction channels to the total $^{4}$He gas-production cross-sections in the \mbox{IRDFF-II} evaluation.]
{\includegraphics[width=\columnwidth]{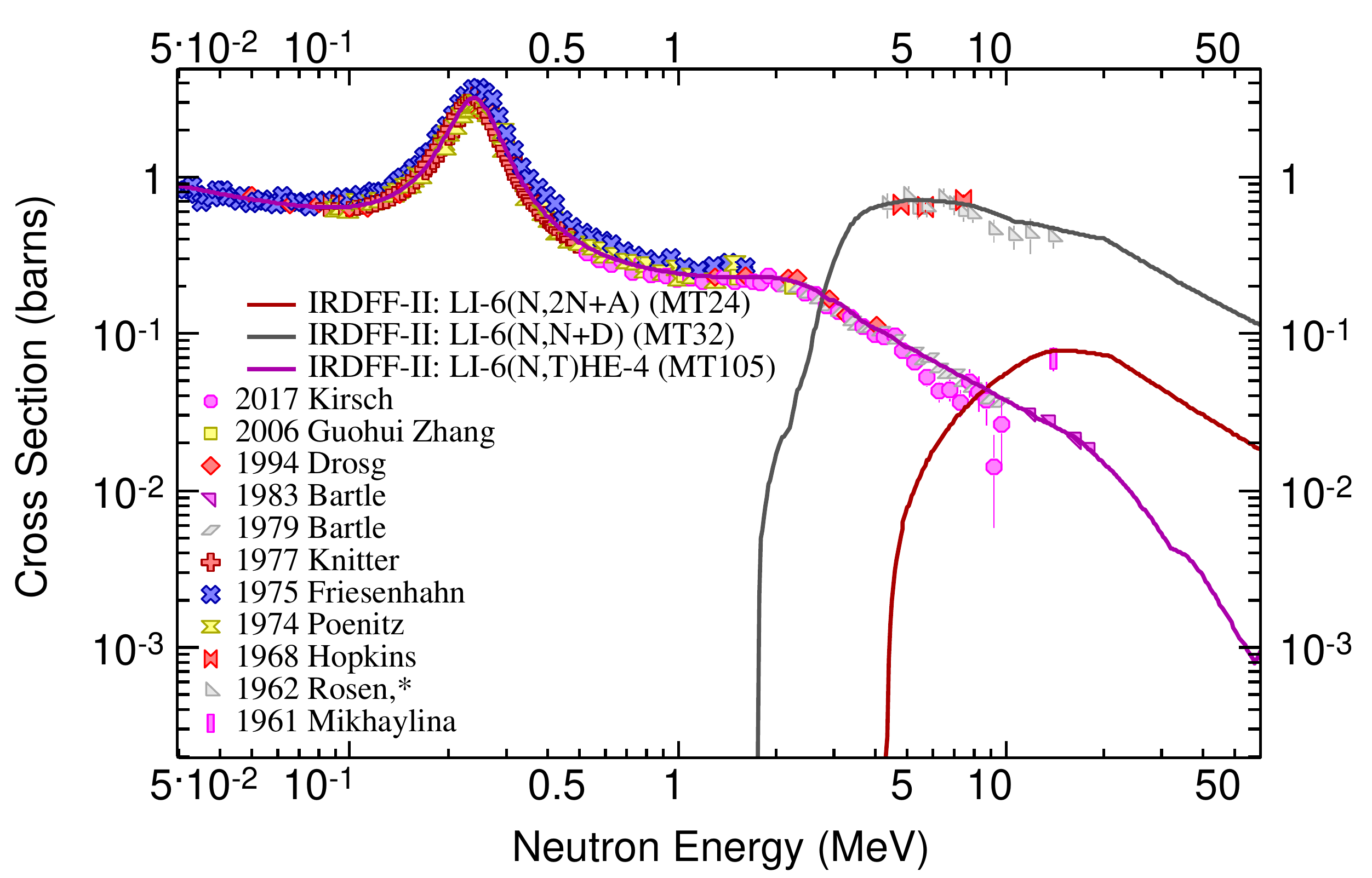}}
\vspace{-3mm}
\caption{(Color online) Comparison of the $^{4}$He gas-production cross sections for the n+$^{6}$Li reaction in the \mbox{IRDFF-II} library.}
\label{Li6_He4}
\vspace{-3mm}
\end{figure}

Tritium production from $^{6}$Li in \mbox{IRDFF-II} corresponds uniquely to the $^{6}$Li(n,t)$^4$He reaction (ENDF-6 MT~105), which is a Neutron Standard up to 1~MeV and was discussed in the previous section; a comparison of the cross section with experimental data and with \mbox{ENDF/B-VIII.0} is shown in Fig.~\ref{tLi6-exp}.
%The tritium-production cross sections from different libraries are shown in Fig.~\ref{Li6_H3}.

Cross sections for the $^{3}$He production are only available in the \mbox{TENDL-2015} library, and were neglected for the assembly of the \mbox{IRDFF-II} library.

Since the neutron metrology community also uses mass spectrometry methods to measure $^{4}$He production in materials such as $^{6}$Li, the new \mbox{IRDFF-II} library also includes this reaction. The $^{4}$He nucleus is the residual of the $^{6}$Li(n,t) reaction that is discussed in Sec.~\ref{Sec_V_AA}. There are other reaction channels that can produce $^{4}$He, as can be seen in Table~\ref{Li6_gas}.
As a contributor to the $^{4}$He production, all libraries contain the reaction $^{6}$Li(n,2n$\alpha$) (MT~24). The \mbox{ENDF/B-VIII.0} evaluation was adopted up to 20~MeV, and the \mbox{EAF-2010} library from 20 up to 60~MeV, renormalized for continuity at 20 MeV. The \mbox{EAF-2010} library contains reaction $^{6}$Li(n,nd) MT~32 explicitly, while in the \mbox{ENDF/B-VIII.0} library this reaction is represented by discrete level inelastic cross sections $^{6}$Li(n,n’) of the MT~50 series, with a consecutive break-up of the residual. Note that not all discrete levels contribute to the $^{4}$He production, governed by the LR flag in the ENDF file. The MT~32 contribution reconstructed from the discrete inelastic levels in the \mbox{ENDF/B VIII.0} library was adopted for \mbox{IRDFF-II}, while renormalized EAF-2010 cross sections were adopted from 20 up to 60~MeV. The total $^{4}$He production in \mbox{IRDFF-II} is higher than in EAF-2010 library due to the larger cross section from the $^{6}$Li(n,n’) channels.
Fig.~\ref{Li6_He4}(a) compares the $^{4}$He gas production cross sections in \mbox{IRDFF-II}, \mbox{ENDF/B-VIII.0} and \mbox{EAF-2010} libraries. Fig.~\ref{Li6_He4}(b) shows the relative contribution to the $^{4}$He production cross section from the various reaction channels in \mbox{IRDFF-II}.

\subsubsection{$^{7}\!$Li(n,X)$^{3}\!$H and $^{7}\!$Li(n,X)$^{4}\!$He}

Since elemental Li is composed of both $^{6}$Li and $^{7}$Li, different enrichments of Li are often used in neutron metrology measurements. Thus, the \mbox{IRDFF-II} library also provides the tritium and $^{4}$He gas-production cross sections in $^{7}$Li. This enables users to combine the cross sections in a manner that represents the composition of their dosimeters.

\begin{figure}[!hbtp]
 \includegraphics[width=\columnwidth]{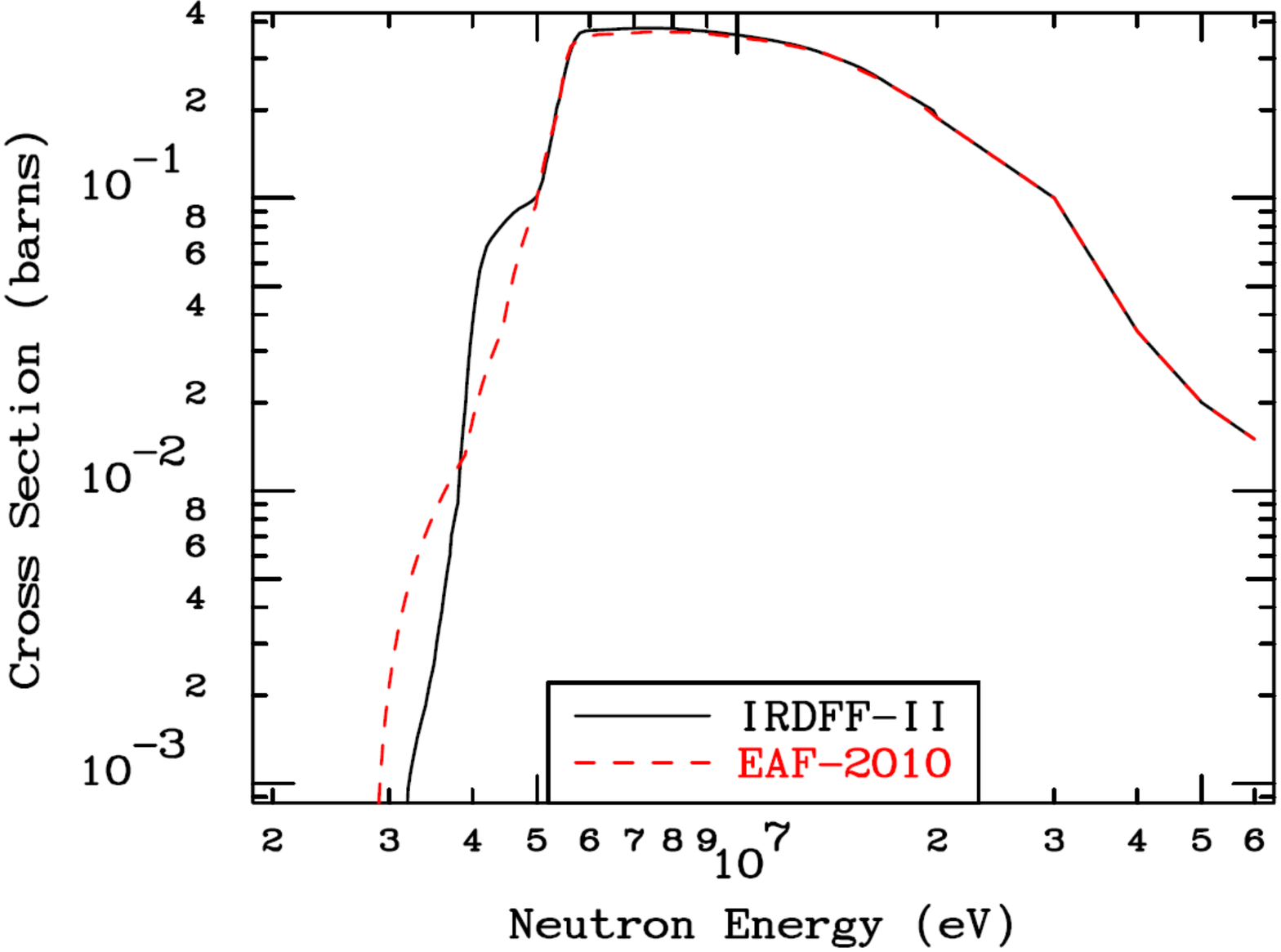}
\caption{(Color online) Comparison of the tritium production cross sections for the n+$^{7}$Li reaction for different evaluated data libraries.}
\label{Li7_H3}
\end{figure}

The list of reactions contributing to tritium production is shown in Table~\ref{Li7_gas}. In the \mbox{EAF-2010} library the tritium production is represented by reaction $^{7}$Li(n,n$\alpha$) MT~22, while in the \mbox{ENDF/B VIII.0} library this reaction is represented by discrete level inelastic cross sections $^{7}$Li(n,n’) of the MT 50 series, with a consecutive break-up of the residual from most of the discrete levels into $^{3}$H and $^{4}$He. Reaction MT~33 reconstructed from the discrete inelastic levels from \mbox{ENDF/B VIII.0} was adopted for \mbox{IRDFF-II}, extrapolated above 20~MeV with MT~22 from \mbox{EAF-2010} since the two reactions produce the same set of emitted particles. The tritium production cross sections are plotted in Fig.~\ref{Li7_H3}. The bumps in the cross section near threshold come from the contributions of different discrete levels. The \mbox{ENDF/B-VIII.0} cross sections are equal to \mbox{IRDFF-II} and are not included in the plot.

\begin{figure}[!thb]
\vspace{-2mm}
\subfigure[~Comparison of data from selected evaluated libraries relative to the \mbox{IRDFF-II} evaluation]
{\includegraphics[width=\columnwidth]{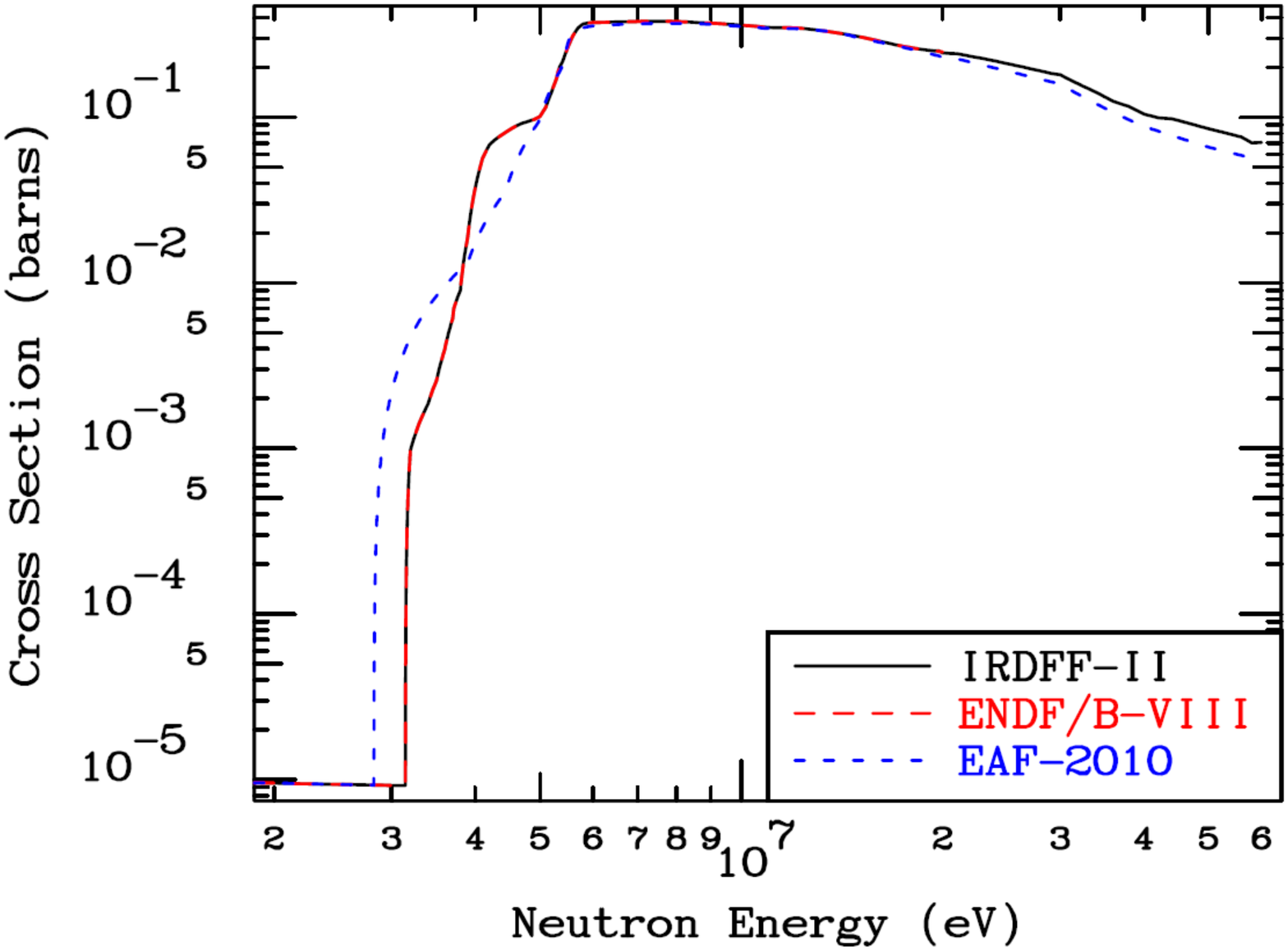}}
\subfigure[~Contribution of different reaction channels to the total $^{4}$He gas-production cross-sections in the \mbox{IRDFF-II} evaluation.]
{\includegraphics[width=\columnwidth]{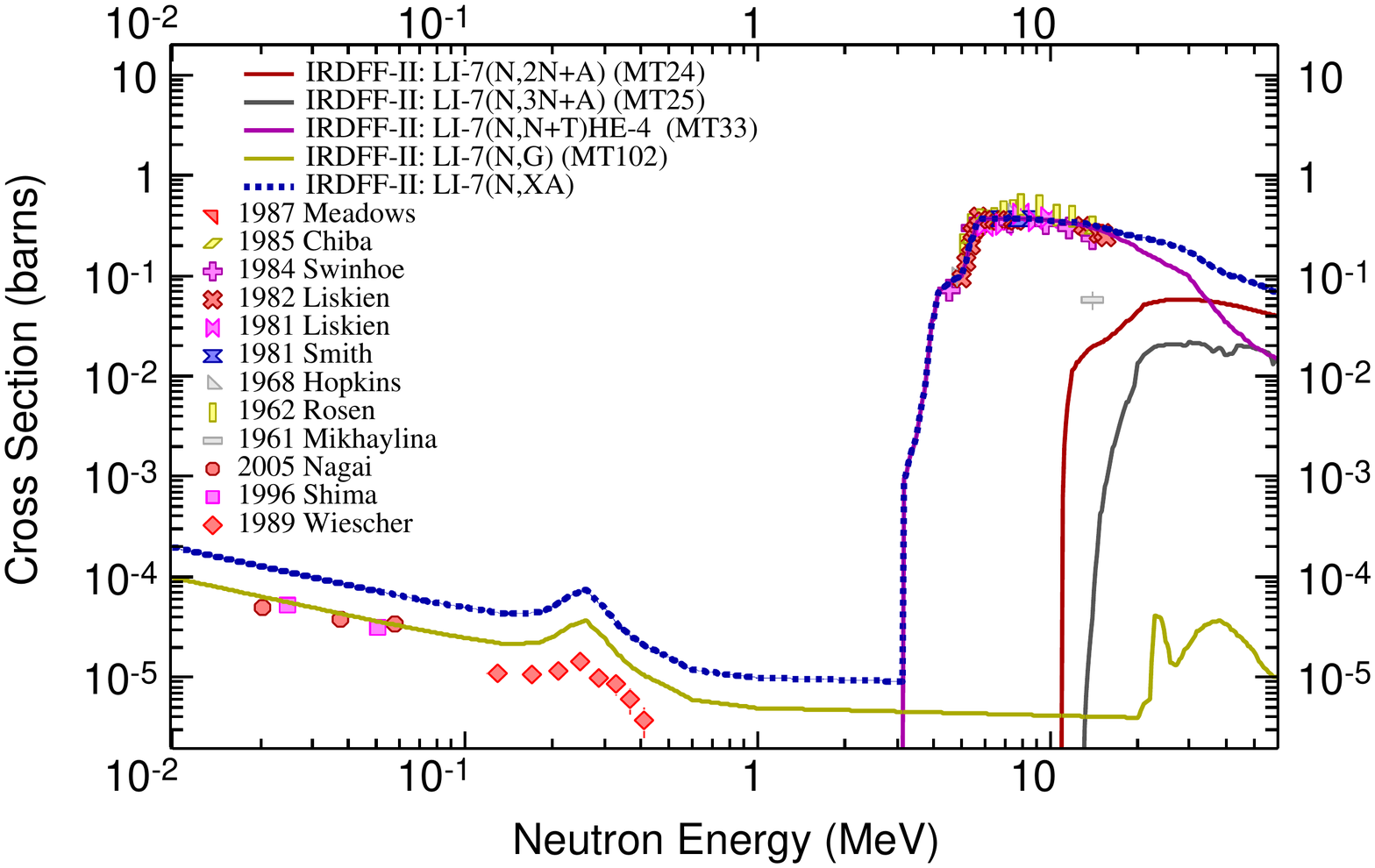}}
\vspace{-3mm}
\caption{(Color online) Comparison of the $^{4}$He gas-production cross sections for the n+$^{7}$Li reaction in the \mbox{IRDFF-II} library.}
\label{Li7_He4}
\vspace{-3mm}
\end{figure}

\begin{table*}[!htb]
\vspace{-5mm}
\caption{Gas production reactions in IRDFF-II library from neutrons interacting with $^{7}$Li target. The MT numbers (reactions) used in IRDFF-II assembly following the ENDF-6 format, and the relevant reaction products are highlighted in bold.} \label{Li7_gas}
\begin{tabular}{ c | l | l | l c l | l | l }
\hline \hline
 MT         & LR & Gas      & Reaction              &             & Products                                                                   & \multicolumn{2}{c}{Source evaluations of IRDFF-II}\\
\hline
\T
 22         &    & $^{3}$H  & $^{7}$Li(n,n$\alpha$) &$\rightarrow$& n + $^{4}$He + \textbf{$^{3}$H}                                            & EAF-2010(r) ($E>20$) &                            \\
 52,53,...  & 33 &          & $^{7}$Li(n,n')        &$\rightarrow$& n + ($^{7}$Li$^*$ $\rightarrow$ \textbf{$^{3}$H} + $^{4}$He)               &                      & ENDF/B-VIII.0 ($E<20$)     \\
\textbf{33} &    &          & $^{7}$Li(n,nt)        &$\rightarrow$& n + \textbf{$^{3}$H} + $^{4}$He                                            & EAF-2010(r) MT22     & ENDF/B-VIII.0 MT 52,53,... \\
\hline
\T

\textbf{24} &    &          & $^{7}$Li(n,2n$\alpha$)&$\rightarrow$& 2n + \textbf{$^{4}$He} + $^{2}$H                                           & EAF-2010(r) ($E>20$) & ENDF/B-VIII.0 ($E<20$)     \\
\textbf{25} &    &          & $^{7}$Li(n,3n$\alpha$)&$\rightarrow$& 3n + \textbf{$^{4}$He} + $^{1}$H                                           & EAF-2010(r) MT28     & ENDF/B-VIII.0 ($E<20$)     \\
 28         &    &          & $^{7}$Li(n,np)        &$\rightarrow$& 3n + \textbf{$^{4}$He} + $^{1}$H                                           & EAF-2010(r) ($E>20$) &                            \\
 22         &    & $^{4}$He & $^{7}$Li(n,n$\alpha$) &$\rightarrow$& n + \textbf{$^{4}$He} + $^{3}$H                                            & EAF-2010(r) ($E>20$) &                            \\
 52,53,...  & 33 &          & $^{7}$Li(n,n')        &$\rightarrow$& n + ($^{7}$Li$^*$ $\rightarrow$ $^{3}$H + \textbf{$^{4}$He})               &                      & ENDF/B-VIII.0 ($E<20$)     \\
\textbf{33} &    &          & $^{7}$Li(n,nt)        &$\rightarrow$& n + $^{3}$H + \textbf{$^{4}$He}                                            & EAF-2010(r) MT22     & ENDF/B-VIII.0 MT 52,53,... \\
\textbf{102}&    &          & $^{7}$Li(n,$\gamma$)  &$\rightarrow$& $^{8}$Li $\rightarrow$(0.8 s) ($^{8}$Be $\rightarrow$ \textbf{2 $^{4}$He}) & EAF-2010(r) ($E>20$) & ENDF/B-VIII.0 ($E<20$)     \\
\hline\hline
\end{tabular}
\vspace{-3mm}
\end{table*}

Cross sections for the $^{3}$He production are only available in the \mbox{TENDL-2015} library.
The reaction threshold is near 30~MeV. The absolute magnitude of the cross sections is about 2~mb at the peak, according to \mbox{TENDL-2015}. Considering that the data from TALYS calculations for the gas production data in light elements in \mbox{TENDL-2015}.s60 are not reliable and there is no alternative source for the $^{3}$He production reaction cross sections, these cross sections were not included in the \mbox{IRDFF-II} library.

The list of reactions contributing to $^{4}$He production is also shown in Table~\ref{Li7_gas}. The libraries \mbox{EAF-2010} and \mbox{ENDF/B-VIII.0} contain reactions $^{7}$Li(n,2n$\alpha$) MT~24 and $^{7}$Li(n,$\gamma$) MT~102; the latter is adopted for \mbox{IRDFF-II}. The \mbox{ENDF/B-VIII.0} library contains discrete inelastic levels where the residual is flagged to break up as $^{7}$Li(n,nt) MT~33. The summed contributions from the respective levels are stored in \mbox{IRDFF-II} as MT~33. The \mbox{EAF-2010} library takes a different approach, it contains the $^{7}$Li(n,n$\alpha$) MT~22 reaction, which involves exactly the same outgoing particles. The \mbox{ENDF/B-VIII.0} library also contains the MT~25 reaction, which has a higher threshold and is not present in \mbox{EAF-2010}. This reaction was adopted from \mbox{ENDF/B-VIII.0} and extrapolated to 60~MeV with the shape of the $^{7}$Li(n,np) reaction, which is mostly equivalent. The contribution of this reaction is mainly responsible for the higher total alpha-production cross section in \mbox{IRDFF-II} compared to \mbox{EAF-2010}. The $^{4}$He production cross sections from different libraries are plotted in Fig.~\ref{Li7_He4}(a) and the contribution of different reaction channels in \mbox{IRDFF-II} are shown in Fig.~\ref{Li7_He4}(b). Note the capture cross section as a contributor to the $^{4}$He production at low energies. Two alpha particles are produced after the disintegration of the capture product, hence the  $^{4}$He production cross section is by a factor of two higher than the capture cross section itself.

There is an issue regarding the $^{7}$Li(n,$\gamma$) MT~102 reaction, which is found in all libraries and is also included in \mbox{IRDFF-II}. The residual is $^{8}$Li, and this nuclide decays into $^{8}$Be with a half-life of 0.5~s, which, in turn, breaks up almost instantaneously into two alpha particles. In general, the contribution of this reaction is small, but it is the only contribution to alpha production at thermal energies –- a region where the cross section is about 0.1~b -- and its multiplicity is two. Beware that the contribution of this reaction is not included in the gas production cross sections generated from the \mbox{ENDF/B-VIII.0} data with the current version of the \mbox{NJOY-2016} code because the reaction is not flagged for break-up. Due to the very short half-life involved, the radiative capture reaction was included as a contributor to alpha production in the present work for \mbox{IRDFF-II}.

\subsubsection{$^{\mbox{nat}}$Li(n,X)$^{4}\!$He} \label{Sec_V_CC}
\begin{figure}[!hbtp]
\includegraphics[width=\columnwidth]{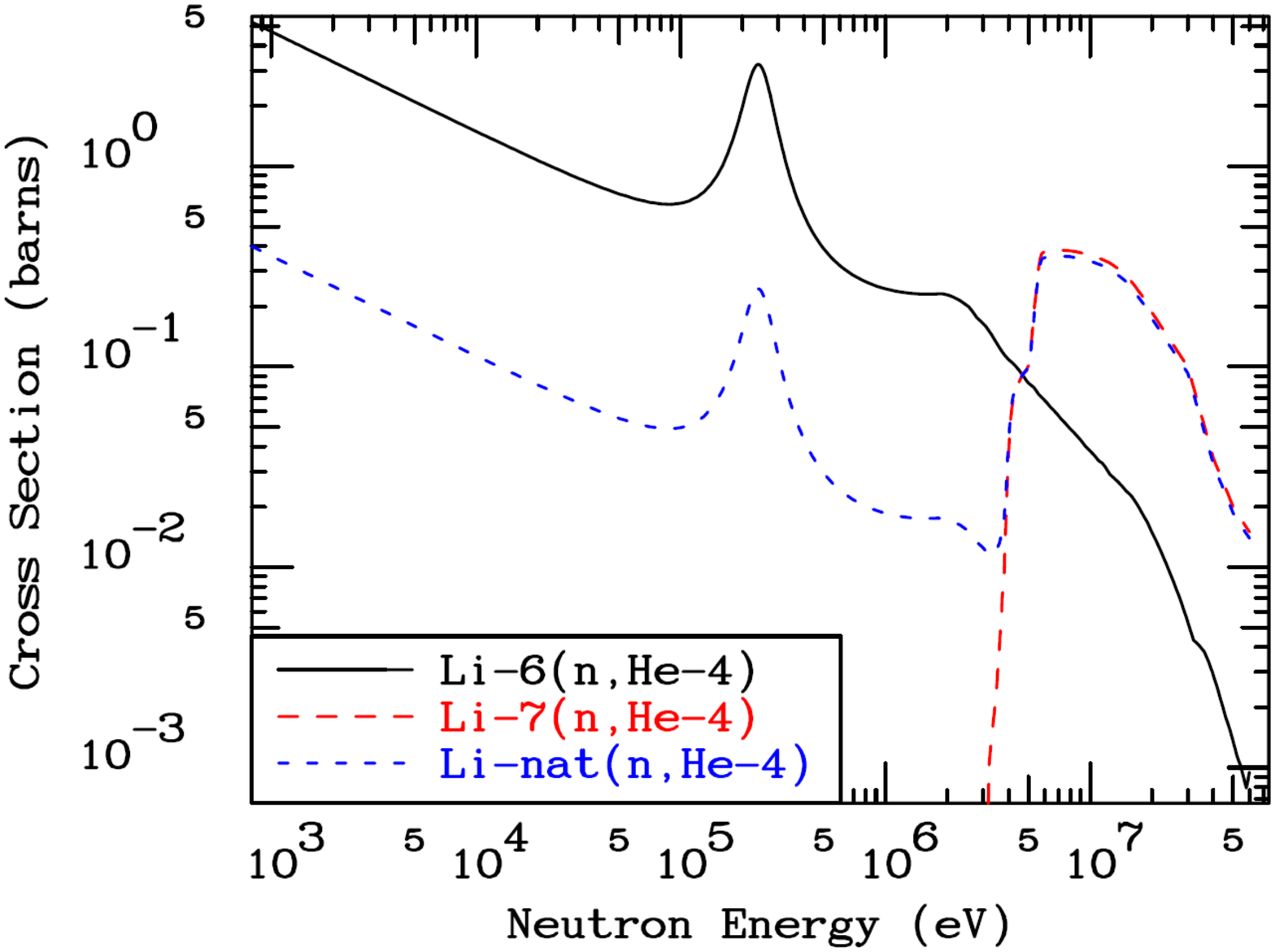}
\vspace{-2mm}
\caption{(Color online) Comparison of the \mbox{IRDFF-II} $^{4}$He gas-production cross sections in neutron induced reactions on $^{6}$Li, $^{7}$Li, and $^{\mbox{nat}}$Li targets.}
\label{LiNat_He4}
\vspace{-2mm}
\end{figure}

In order to support application of the \mbox{IRDFF-II} library to irradiations that use a lithium dosimeter having naturally occurring lithium with nominal abundances for $^{6}$Li and $^{7}$Li, the \mbox{IRDFF-II} library also provides an entry for the $^{4}$He gas production on elemental lithium. The cross sections for the gas production in natural element lithium were constructed from the isotopic data in \mbox{IRDFF-II} using the recommended natural abundance data. Fig.~\ref{LiNat_He4} compares the $^{4}$He gas production on $^{6}$Li, $^{7}$Li, and $^{\mbox{nat}}$Li targets. The $^{4}$He gas production cross sections of $^{\mbox{nat}}$Li are normalized per Li atom of the element with an average mass.

\clearpage
\subsubsection{$^{10}\!$B(n,X)$^{4}\!$He}
Gas production in the boron isotopes is very complex because many reactions contribute and very often the residual nuclei undergo further break-up. The complexity of the gas production cross sections in terms of contributing reactions is captured in Table~\ref{B10_gas}. As in the case of Li, the cross sections from the \mbox{TENDL-2015} library are not considered in the analysis because boron is a light element and TALYS calculations underlying \mbox{TENDL-2015}.s60 are not appropriate in such cases. The discussion in this section is limited to the $^{4}$He production from neutrons incident on $^{10}$B, for tritium production relevant contributing reactions are listed in Table~\ref{B10_gas}.
\begin{figure}[!thb]
%\vspace{-2mm}
\subfigure[~Comparison of data from evaluated libraries relative to Kneff \etal measurement~\cite{Kne86}.]
{\includegraphics[width=\columnwidth]{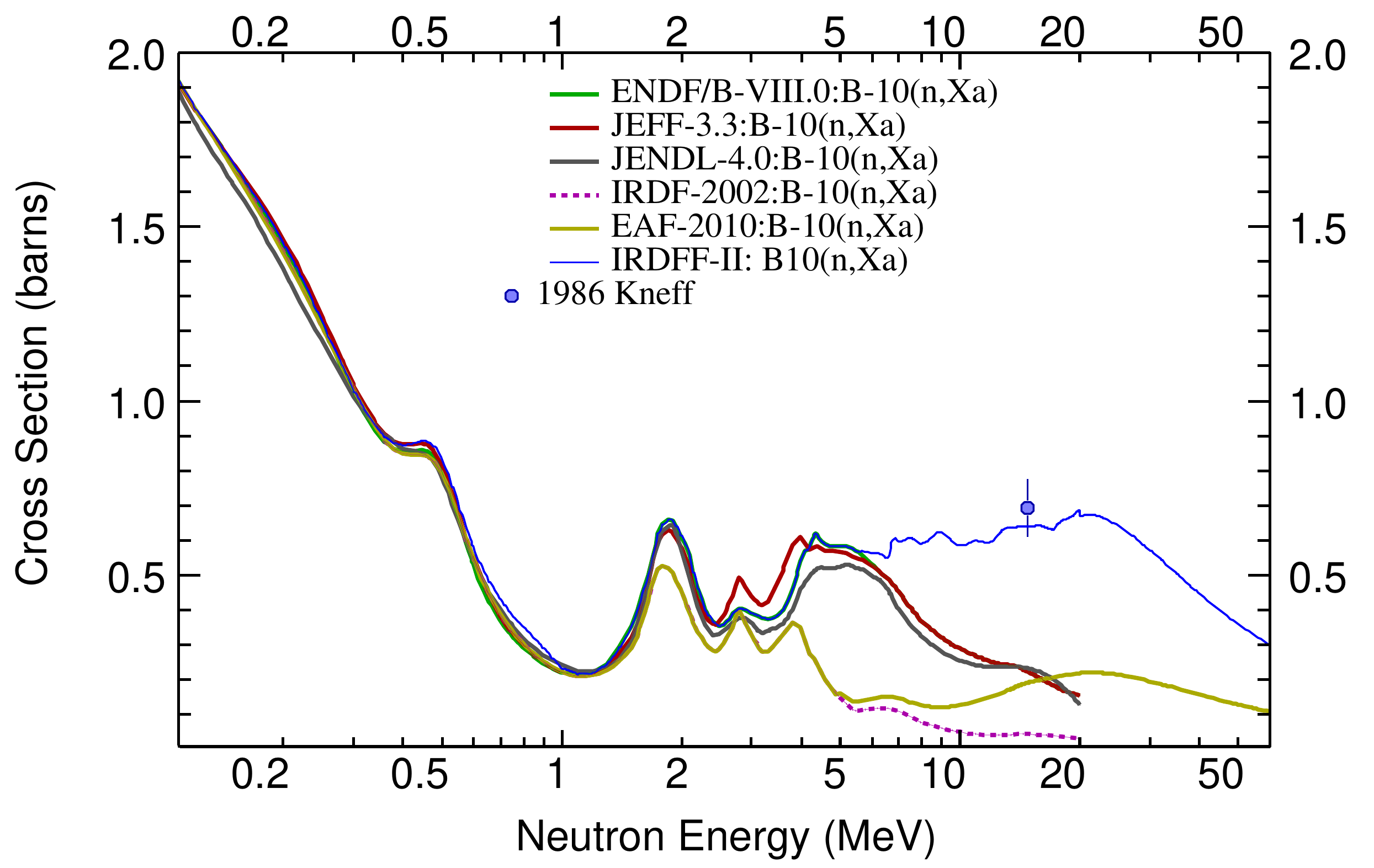}}
\subfigure[~Contribution of different reaction channels to the total $^{4}$He gas-production cross-sections in the \mbox{IRDFF-II} evaluation.]
{\includegraphics[width=\columnwidth]{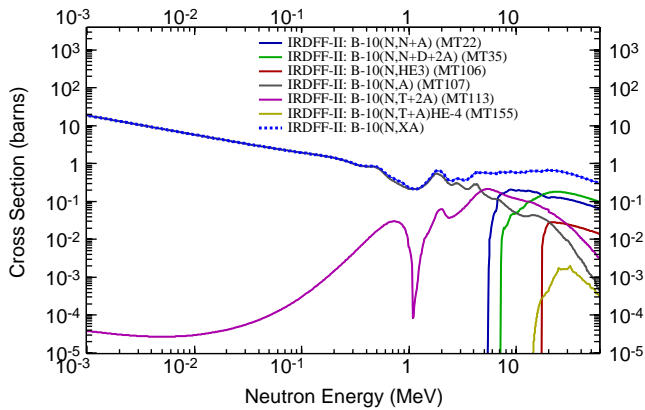}}
\vspace{-3mm}
\caption{(Color online) Comparison of the $^{4}$He gas-production cross sections for the n+$^{10}$B reaction in the \mbox{IRDFF-II} library.}
\label{B10_He4}
\vspace{-3mm}
\end{figure}

\begin{table*}[t]
\vspace{-6mm}
\caption{Gas production reactions from neutrons interacting with $^{10}$B target. The MT numbers (reactions) used in IRDFF-II assembly following the ENDF-6 format, and the relevant reaction products are highlighted in bold.}
\label{B10_gas}
\begin{tabular}{ c | l | l | l c l | l | l }
\hline \hline
 MT          & LR & Gas      & Reaction               &             & Products                                                                 & \multicolumn{2}{c}{Source evaluations of IRDFF-II} \\
 \hline
\T
 105         &    & $^3$H    & $^{10}$B(n,t)          &$\rightarrow$& \textbf{$^{3}$H} + ($^{8}$Be$^*$ $\rightarrow$ 2 $^4$He)                 &EAF-2010(r) ($E>20$) &                       \\
 700         &    &          & $^{10}$B(n,t)          &$\rightarrow$& \textbf{$^{3}$H} + ($^{8}$Be$^*$ $\rightarrow$ 2 $^4$He)                 &                     &ENDF/B-VIII.0 ($E<20$) \\
 \textbf{113}&    &          & $^{10}$B(n,t2$\alpha$) &$\rightarrow$& \textbf{$^{3}$H} + 2 $^{4}$He                                            &EAF-2010(r) MT105    &ENDF/B-VIII.0 MT700    \\
 \textbf{155}&    &          & $^{10}$B(n,t$\alpha$)  &$\rightarrow$& \textbf{$^{3}$H} + 2 $^{4}$He                                            &EAF-2010             &                       \\
\hline
\T
 \textbf{106}&    & $^{3}$He & $^{10}$B(n,$^{3}$He)   &$\rightarrow$& \textbf{$^{3}$He} + ($^{8}$Li $\rightarrow$ $^{8}$Be$^*$ $\rightarrow$ 2 $^{4}$He) &EAF-2010   &                       \\
\hline
\T
 \textbf{22} &    & $^{4}$He & $^{10}$B(n,n$\alpha$)  &$\rightarrow$& n + \textbf{$^{4}$He} + $^{6}$Li                                         &EAF-2010(r) ($E>20$) & ENDF/B-VIII.0 MT55,56,...\\
 32          &    &          & $^{10}$B(n,nd)         &$\rightarrow$& n + $^{2}$H + ($^{8}$Be$^*$ $\rightarrow$ \textbf{2 $^{4}$He})           &EAF-2010(r) ($E>20$) &                          \\
 \textbf{35} &    &          & $^{10}$B(n,nd2$\alpha$)&$\rightarrow$& n + $^{2}$H + \textbf{2 $^{4}$He}                                        &EAF-2010(r) MT32     & ENDF/B-VIII.0 MT55,56,...\\
% 41         &    &          & $^{10}$B(n,2np)        &$\rightarrow$& 2 n + $^{1}$H + ($^{8}$Be$^*$ $\rightarrow$ 2 $^{4}$He)                  &EAF-2010(r)          &                          \\
 55,56,...   & 22 &          & $^{10}$B(n,n’)         &$\rightarrow$& n + \textbf{$^{4}$He} + $^{6}$Li                                       &                     &ENDF/B-VIII.0 ($E<20$)    \\
 55,56,...   & 35 &          & $^{10}$B(n,n’)         &$\rightarrow$& n + $^{2}$H + \textbf{2 $^{4}$He}                                      &                     &ENDF/B-VIII.0 ($E<20$)    \\
 105         &    &          & $^{10}$B(n,t)          &$\rightarrow$& $^{3}$H + ($^{8}$Be$^*$ $\rightarrow$ \textbf{2 $^{4}$He})               &EAF-2010(r) ($E>20$) &                          \\
 700         &    &          & $^{10}$B(n,t)          &$\rightarrow$& $^{3}$H + ($^{8}$Be $\rightarrow$ \textbf{2 $^{4}$He})                   &                     &ENDF/B-VIII.0 ($E<20$)    \\
 \textbf{106}&    &          & $^{10}$B(n,$^{3}$He)   &$\rightarrow$& $^{3}$He + ($^{8}$Li$^*$ $\rightarrow ^{8}$Be$^*$ $\rightarrow$ \textbf{2 $^{4}$He}) &EAF-2010 &                          \\
 \textbf{107}&    &          & $^{10}$B(n,$\alpha$)   &$\rightarrow$& \textbf{$^{4}$He} + $^{7}$Li                                             &EAF-2010(r) ($E>20$) &ENDF/B-VIII.0 MT800,801   \\
 800         &    &          & $^{10}$B(n,$\alpha$)   &$\rightarrow$& \textbf{$^{4}$He} + $^{7}$Li                                             &                     &ENDF/B-VIII.0 ($E<20$)    \\
 801         &    &          & $^{10}$B(n,$\alpha$)   &$\rightarrow$& \textbf{$^{4}$He} + $^{7}$Li$^*$                                         &                     &ENDF/B-VIII.0 ($E<20$)    \\
 \textbf{113}&    &          & $^{10}$B(n,t2$\alpha$) &$\rightarrow$& $^{3}$H + \textbf{2 $^{4}$He}                                            &EAF-2010(r) MT105    &ENDF/B-VIII.0 MT700       \\
 \textbf{155}&    &          & $^{10}$B(n,t$\alpha$)  &$\rightarrow$& $^{3}$H + \textbf{2 $^{4}$He}                                            &EAF-2010             &                          \\
\hline\hline
\end{tabular}
\vspace{-3mm}
\end{table*}

The $^{10}$B(n,$\alpha$) reaction cross sections are part of the IAEA Neutron Standards~2017 from thermal to 1~MeV. The \mbox{ENDF/B-VIII.0} evaluation is consistent with these Standards. The library also contains discrete level inelastic cross sections of the MT~50 series, which are flagged for break-up of the residual, equivalent to the reactions under MT~22 and~35. In addition there are also discrete level cross sections for the $^{10}$B(n,t) MT~700 and $^{10}$B(n,$\alpha$) MT~800 series, that can be summed into MT~105 and MT~107, respectively. For \mbox{IRDFF-II} the $^{10}$B(n,$\alpha$) MT~107 cross sections are taken from the \mbox{ENDF/B-VIII.0} library up to 1~MeV because these data agree with the IAEA Neutron Standards~2017. Extrapolation to 60~MeV is done using the shape of the $^{10}$B(n,$\alpha$) MT~107 cross sections from the \mbox{EAF-2010} library.
\begin{figure}[!hbtp]
\vspace{-2mm}
\includegraphics[width=0.98\columnwidth]{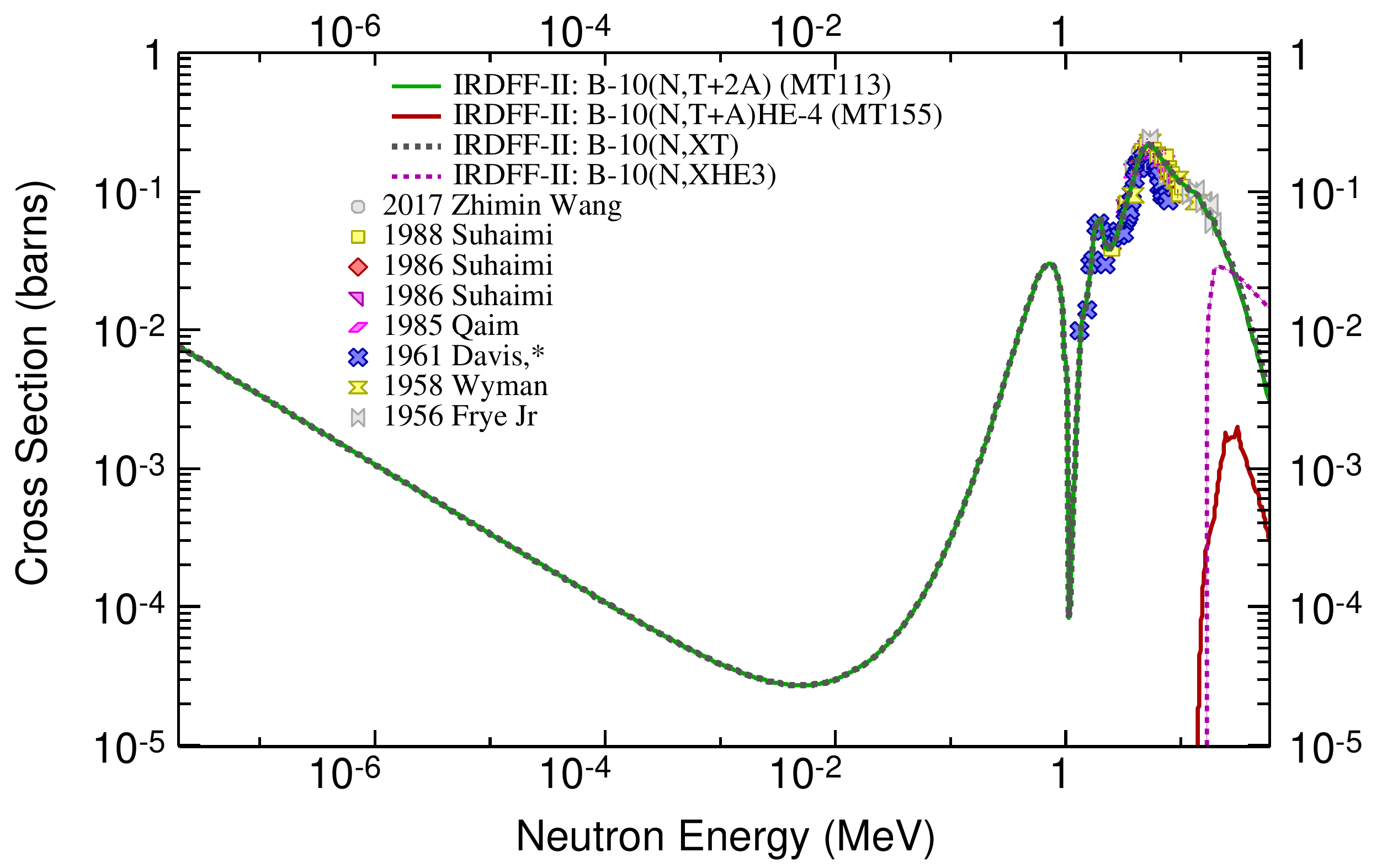}
\caption{(Color online) Comparison of tritium-gas production cross sections for the n+$^{10}$B reaction for different reaction channels in the \mbox{IRDFF-II} library.}
\label{B10_H3}
\vspace{-3mm}
\end{figure}

In the \mbox{IRDFF-II} library the $^{10}$B(n,n$\alpha$) MT~22 cross sections are taken from the summed contributions of the discrete level inelastic cross sections in \mbox{ENDF/B-VIII.0} up to 20~MeV. Extrapolation to 60~MeV is done using the shape of the $^{10}$B(n,n$\alpha$) MT~22 cross sections from the \mbox{EAF-2010} library, normalized for continuity at~20 MeV. The data from the \mbox{ENDF/B-VIII.0} library do not have a dip near 10~MeV, which is observed in the \mbox{EAF-2010} data.

The cross section curves from the various libraries are shown in Fig.~\ref{B10_He4}(a); plots of selected contributing reactions to $^4$He gas production are shown in Fig.~\ref{B10_He4}(b).
The plots of contributing reactions to tritium gas production are shown in Fig.~\ref{B10_H3}.

\subsubsection{$^{11}\!$B(n,X)$^{4}\!$He}
Many reaction channels from neutrons incident on $^{11}$B result in $^{4}$He production and/or tritium production, as seen from Table~\ref{B11_gas}.
There are 7 such reactions in the \mbox{EAF-2010} library.
\begin{table*}[t]
\vspace{-2mm}
\caption{Gas production reactions from neutrons interacting with $^{11}$B target. The MT numbers (reactions) used in IRDFF-II assembly following the ENDF-6 format, and the relevant reaction products are highlighted in bold.}
\label{B11_gas}
\begin{tabular}{ c | l | l | l c l | l | l }
\hline \hline
 MT          & LR & Gas      & Reaction               &             & Products                                                                        & \multicolumn{2}{c}{Source evaluations of IRDFF-II}           \\
 \hline
\T
\textbf{33}  &    & $^{3}$H  & $^{11}$B(n,nt)              &$\rightarrow$& n + $^{3}$H + ($^{8}$Be$\rightarrow$2 $^{4}$He)                            &EAF-2010             &                          \\
\textbf{105} &    &          & $^{11}$B(n,t)               &$\rightarrow$& $^{3}$H + $^{9}$Be                                                         &EAF-2010(r) ($E>20$) & ENDF/B-VIII.0 ($E<20$)   \\
\hline
\T
\textbf{11}  &    & $^{4}$He & $^{11}$B(n,2nd)             &$\rightarrow$& 2 n + $^{2}$H + ($^{8}$Be $\rightarrow$ \textbf{2 $^{4}$He})               &EAF-2010             &                          \\
\textbf{22}  &    &          & $^{11}$B(n,n$\alpha$)       &$\rightarrow$& n + \textbf{$^{4}$He} + $^{7}$Li                                           &EAF-2010(r) ($E>20$) & ENDF/B-VIII.0(r) ($E<20$)\\
\textbf{24}  &    &          & $^{11}$B(n,2n$\alpha$)      &$\rightarrow$& 2 n + \textbf{$^{4}$He} +$^{6}$Li                                          &EAF-2010             &                          \\
\textbf{33}  &    &          & $^{11}$B(n,nt)              &$\rightarrow$& n + $^{3}$H + ($^{8}$Be $\rightarrow$ \textbf{2 $^{4}$He})                 &EAF-2010             &                          \\
\textbf{107} &    &          & $^{11}$B(n,$\alpha$)        &$\rightarrow$& \textbf{$^{4}$He} + ($^{8}$Li$\rightarrow$$^{8}$Be$\rightarrow$2 $^{4}$He) &EAF-2010(r) ($E>20$) & ENDF/B-VIII.0 ($E<20$)   \\
\textbf{117} &    &          & $^{11}$B(n,d$\alpha$)       &$\rightarrow$& $^{2}$H + \textbf{$^{4}$He} + ($^{6}$He $\rightarrow$ $^{6}$Li)            &EAF-2010             &                          \\
\hline\hline
\end{tabular}
\vspace{-3mm}
\end{table*}

\begin{figure}[!hbtp]
\vspace{-1mm}
\includegraphics[width=\columnwidth]{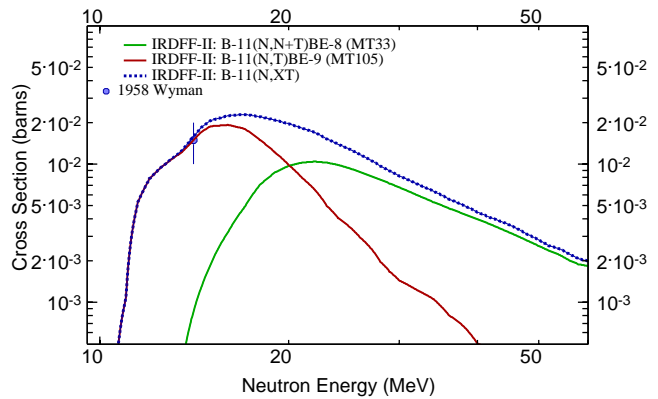}
\caption{(Color online) Comparison of tritium gas-production cross sections for the n+$^{11}$B reaction for different reaction channels in the \mbox{IRDFF-II} library.}
\label{B11_H3}
\vspace{-1mm}
\end{figure}

The \mbox{ENDF/B-VIII.0} library contains only $^{11}$B(n,n$\alpha$) MT~22 and $^{11}$B(n,$\alpha$) MT~107 reactions; the first one is much higher than the corresponding one in the \mbox{EAF-2010} library because it was normalized to the measurement by Kneff (EXFOR \#10933019); It can be understood from the original article that the measured quantity was actually the total helium production, therefore the normalisation in \mbox{ENDF/B-VIII.0} is not correct.
\begin{figure}[!hbtp]
\vspace{-1mm}
\includegraphics[width=\columnwidth]{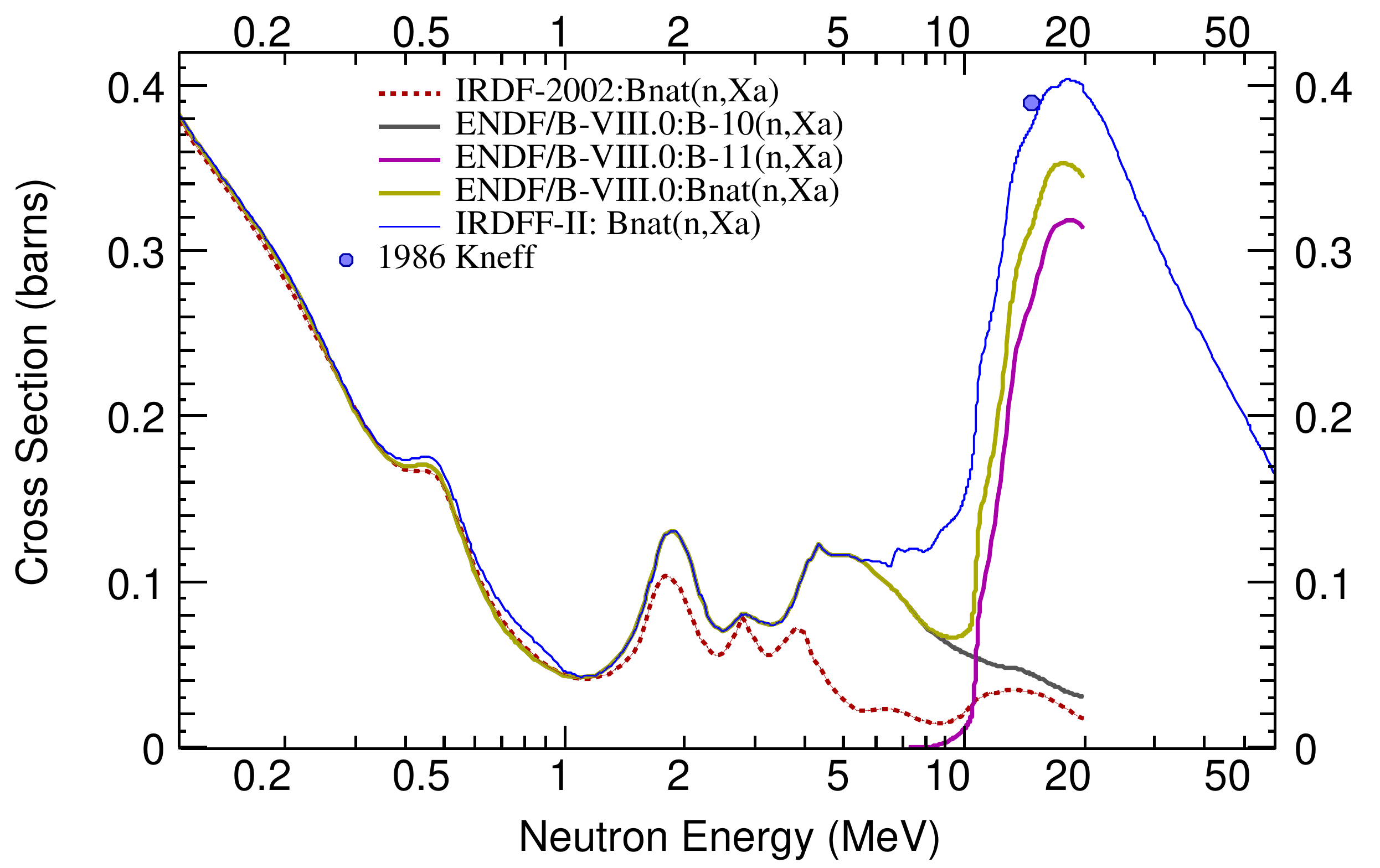}
\caption{(Color online) Comparison of the \mbox{IRDFF-II} $^{4}$He gas-production cross sections in neutron induced reactions on  $^{\mbox{nat}}$B target relative to Kneff \etal measurement~\cite{Kne86}.}
\label{Bnat_He4}
\vspace{-1mm}
\end{figure}

\begin{figure}[!thb]
\vspace{-2mm}
\subfigure[~Comparison of data from evaluated libraries relative to Kneff \etal measurement~\cite{Kne86}.]
{\includegraphics[width=\columnwidth]{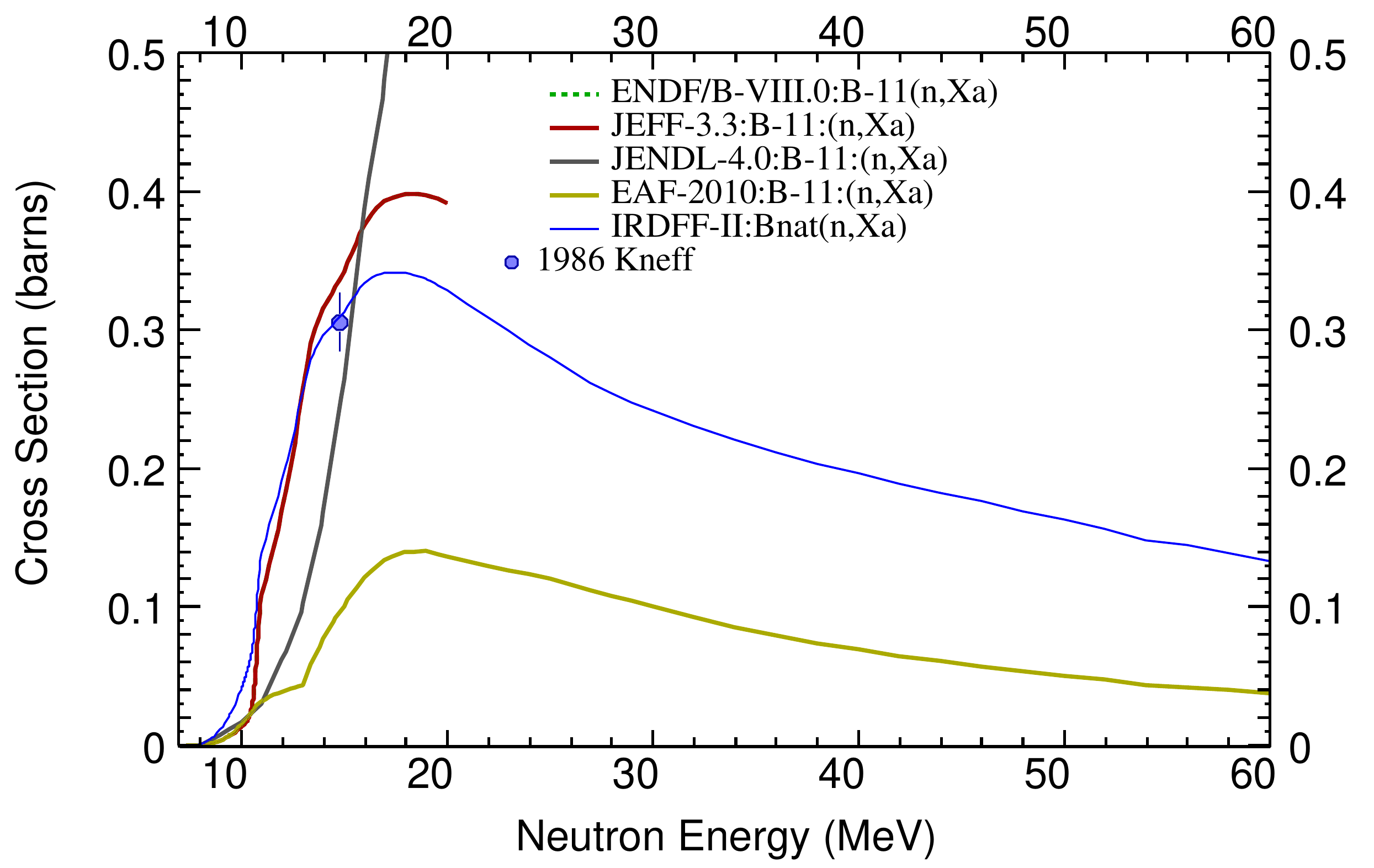}}
\subfigure[~Contribution of different reaction channels to the total $^{4}$He gas-production cross-sections in the \mbox{IRDFF-II} evaluation.]
{\includegraphics[width=\columnwidth]{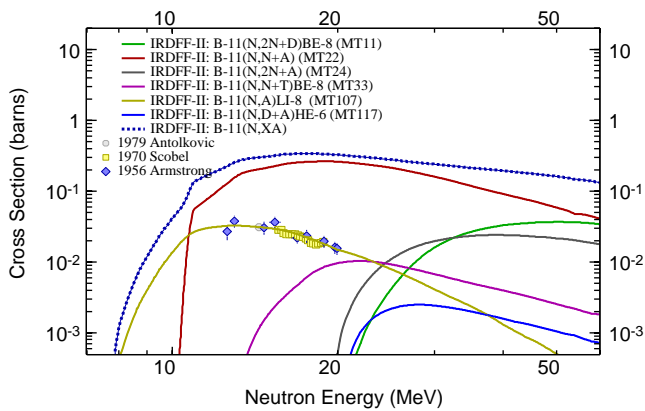}}
\vspace{-3mm}
\caption{(Color online) Comparison of the $^{4}$He gas-production cross sections for the n+$^{11}$B reaction in the \mbox{IRDFF-II} library.}
\label{B11_He4}
\vspace{-3mm}
\end{figure}

Reactions in the \mbox{IRDFF-II} library are adopted from the \mbox{EAF-2010} library, except for those that are also present in \mbox{ENDF/B-VIII.0}. The $^{11}$B(n,n$\alpha$) MT~22 from \mbox{ENDF/B-VIII.0} was scaled down by 30~\% to agree with the total helium-production measurement by Kneff. The cross sections above 20~MeV are taken from the \mbox{EAF-2010} library, scaled for continuity at 20~MeV.

The cross-section curves are shown in Fig.~\ref{B11_He4}(a). Examples of contributing reactions to $^{4}$He and tritium gas production from neutrons interacting with $^{11}$B are shown in Figs.~\ref{B11_He4}(b) and ~\ref{B11_H3}.

\subsubsection{$^{\mbox{nat}}$B(n,X)$^{4}\!$He}
Since various $^{10}$B enrichments can be used for the dosimeter, users of the library can combine $^{4}$He gas generation contributions from $^{10}$B and $^{11}$B to get the dosimeter response. Some dosimeters use the boron in its natural abundance. To serve as a check for how the individual isotopic contributions should be combined, the \mbox{IRDFF-II} library also includes the $^{4}$He gas generation cross section from elemental boron with the naturally occurring isotopic abundances. Fig.~\ref{Bnat_He4} compares the $^{4}$He gas production from the separate isotopes and from the elemental boron file and experimental data at 14~MeV measured by Kneff in 1986~\cite{Kne86}. The \mbox{IRDFF-II} is in excellent agreement with measured data.

\newpage
\section{NEUTRON BENCHMARK FIELDS} \label{Sec_VI}
The following subsections describe the source for the neutron spectral data used to characterize the benchmark neutron fields.

\subsection{Standard Field}\label{Sec_VI_A}

A standard benchmark neutron field is defined \cite{A170, Field} as a permanent and reproducible neutron field that is characterized to state-of-the-art accuracy in terms of neutron fluence rate and energy spectra, and one where the associated spatial and angular distributions, and other important field quantities, have been verified by interlaboratory measurements. A given neutron field is considered to be a ``standard'' only over a specified neutron energy range and there is only one type of ``standard neutron field'' for a given energy range. The recognized ``standard neutron field'' is the $^{252}$Cf spontaneous fission field. Secs.~\ref{Sec_VII_A} and ~\ref{Sec_VIII_A} address how this field is used to provide supporting \mbox{IRDFF-II} validation evidence and how consistent are those evidences.

\begin{figure}[!htb]
\vspace{-3mm}
   \includegraphics[width=0.99\columnwidth]{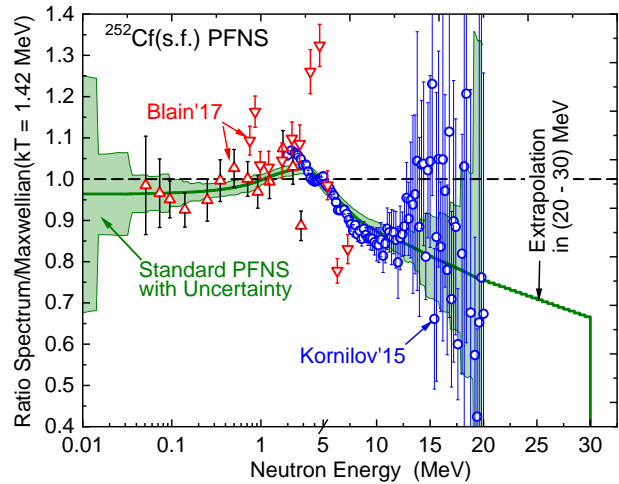}
   \caption{(Color online) Standard $^{252}$Cf(s.f.) prompt fission neutron spectrum (PFNS) and its uncertainties \cite{Car18} as a ratio to the Maxwellian distribution with kT =1.42 MeV (both spectra are normalised to unity). Spectrum extrapolation from 20~MeV to 30 MeV is shown by dash line. The recent new measurements \cite{Kor15} and \cite{Bla17}  are plotted as symbols. Note that the scale changes from log to linear above 5~MeV.}
   \label{fig:Cf252_spect}
\vspace{-3mm}
\end{figure}

\subsubsection{$^{252}$Cf(s.f.)}
The $^{252}$Cf spontaneous fission neutron field is defined to be a standard neutron field over the energy range from 0.1~MeV to 20~MeV. %(0.5 up to 8??)

The standard reference prompt fission neutron spectrum from spontaneous fission of $^{252}$Cf(s.f.) \cite{Man89} is characterized based on neutron time-of-flight measurements and subsequent model-independent evaluation. This work provided the spectrum definition incorporated into the \mbox{IRDFF-II} library up to 20 MeV. The spectrum was then extrapolated up to 30~MeV of neutron outgoing energies by adding a point at 30~MeV, preserving the energy slope as below 20~MeV. All $^{252}$Cf(s.f.) spectrum evaluations are based on original Mannhart work~\cite{Man87}. There are some minor differences above 10 MeV outgoing neutron energy between IDRFF-II spectrum and the one included in the \mbox{ENDF/B-VIII.0} library.

\begin{figure}[!thb]
\vspace{-4mm}
\subfigure[~$^{252}$Cf(s.f.) spectrum correlation matrix.]
{\includegraphics[width=0.975\columnwidth]{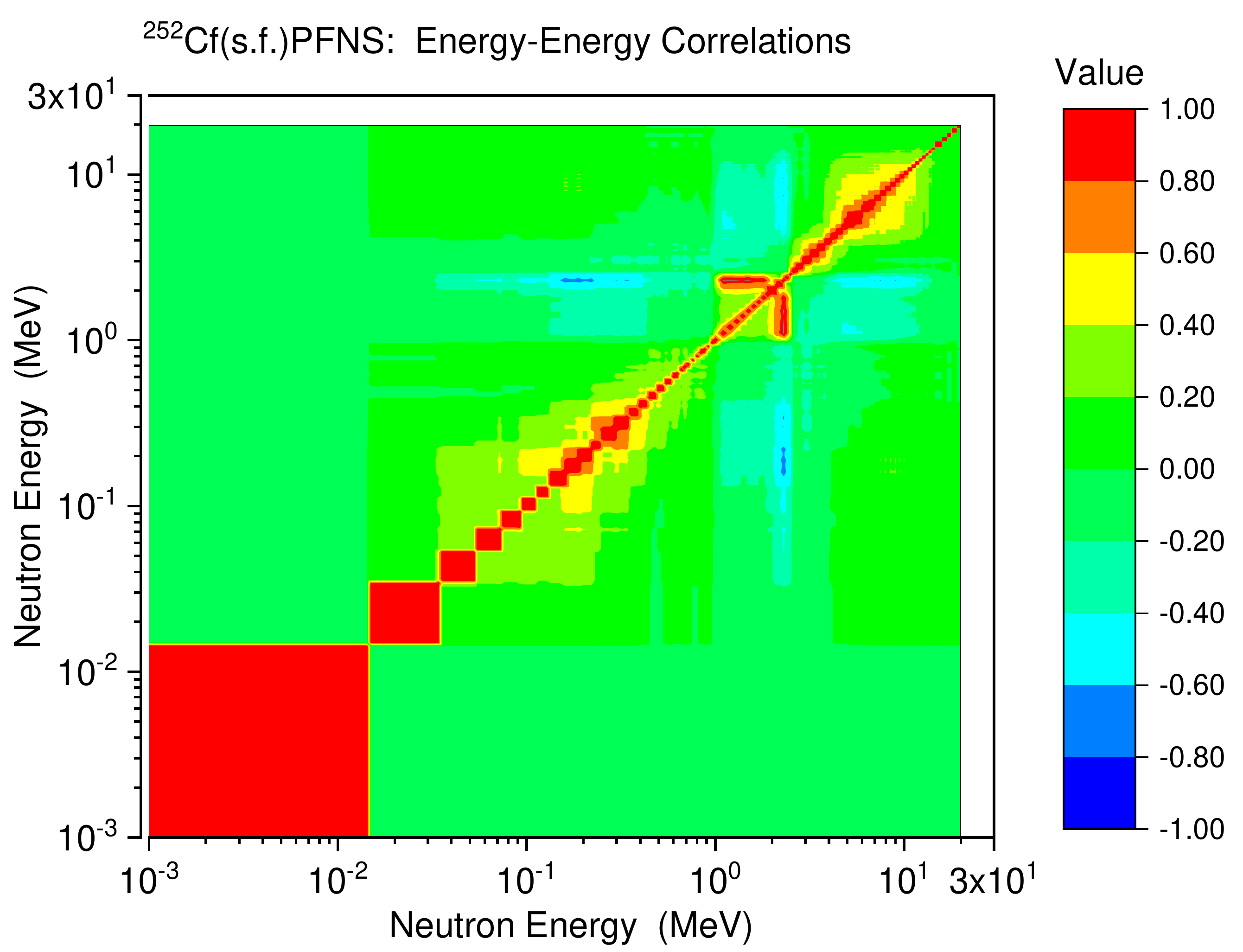}}
\subfigure[~One-sigma uncertainties in \% ($\equiv 100\times \frac{\sqrt{cov(i,i)}}{\mu_i}$), being $\mu_i$ the corresponding spectrum mean value.]
{\includegraphics[width=0.975\columnwidth]{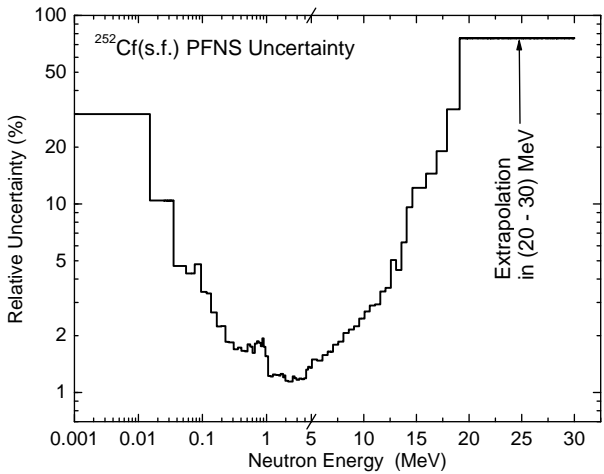}}
\vspace{-3mm}
\caption{(Color online) Uncertainties and correlations of the $^{252}$Cf(s.f.) standard neutron benchmark spectrum in \mbox{IRDFF-II} library.}
\label{Cf252-sp}
\vspace{-4mm}
\end{figure}

Figs.~\ref{Cf252-sp}(a) ~\ref{Cf252-sp}(b) shows the correlation matrix and  the energy-dependent standard deviation for the $^{252}$Cf(s.f.) spectrum, correspondingly.

The spectrum is shown in Fig.~\ref{fig:Cf252_spect}. In the figure the y-axis is shown as a ratio to the Maxwellian distribution with kT = 1.42 MeV to better capture the large variation in neutron population with increasing energy. Since the time-of-flight (TOF) spectral characterization data only addressed the region up to 20 MeV, the higher energy portion of the spectrum is obtained from the extrapolation process described above. Significant increases in the spectral uncertainties are seen above 10 MeV and continuing up to the TOF cut-off energy at 20 MeV. Above this energy, the uncertainty in the extrapolated spectrum is assumed to be the same as that at the 20-MeV last measured value, which is 75~\%.

Since the evaluation of the $^{252}$Cf(s.f.) prompt fission spectrum performed by W. Mannhart in 1987~\cite{Man87}, several new measurements have been carried out by the time-of-flight technique. The two newest measurements are discussed below. In 2015, N. Kornilov has measured the PFNS in the neutron range 2~MeV to 20~MeV~\cite{Kor15}. In 2017, E. Blain and co-workers have reported two data sets measured by different detectors: from 0.05 to 2.8 MeV and from 0.75 to 7.3 MeV, however both - in arbitrary units~\cite{Bla17}. Regrettably these new experiments, as seen in Fig.~\ref{fig:Cf252_spect}, have rather large uncertainties or fluctuations, thus contributing little to the improvement of the existing $^{252}$Cf(s.f.) standard.

\subsection{Experimentally Measured Potential Reference Neutron Fields}\label{Sec_VI_B}

Reference neutron benchmark fields are defined \cite{A170, Field} as permanent and reproducible neutron fields, less-well-characterized than a standard field, but acceptable as a measurement reference by a community of users. As part of the validation effort, an IAEA-sponsored Coordinated Research Project has focused on addressing improving the fidelity of the spectral characterization at candidate reference benchmark fields and using data gathered in these fields as part of the validation evidence for the \mbox{IRDFF-II} library. The following subsections address the spectral characterization data that have been used in the supporting analysis reported here.

\begin{figure}[!htbp]
\vspace{-3mm}
   \includegraphics[width=\columnwidth]{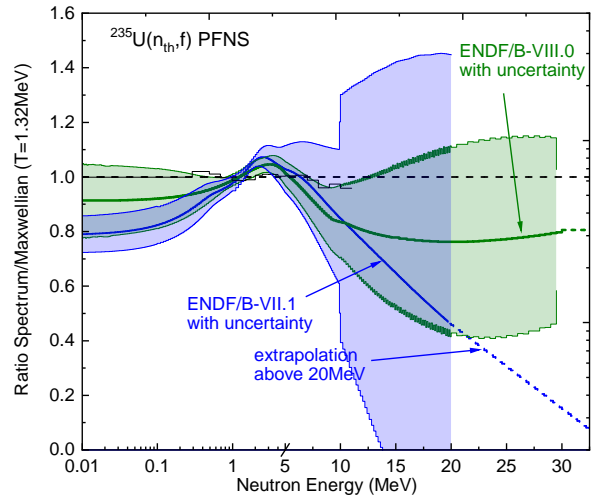}
   \caption{(Color online) Reference $^{235}$U(n$_{th}$,f) prompt fission neutron spectrum (PFNS)~\cite{PFNS1,PFNS2,PFNS} and its uncertainties as a ratio to the Maxwellian distribution with kT = 1.32 MeV (the spectra are normalised to unity). Red curve - from ENDF/B-VIII.0 \cite{ENDF8}, black curve -- from ENDF/B-VII.1 \cite{ENDFB71}. Extrapolations beyond 20~MeV or 30 MeV are shown by dashed  lines. Note that the scale changes from log to linear above 5~MeV.}
   \label{fig:U235_spect}
\vspace{-3mm}
\end{figure}

\subsubsection{$^{235}$U(n$_{th}$,f)}\label{Sec_VI_B1}

A new $^{235}$U thermal fission spectrum evaluation was developed within an IAEA-sponsored project using a least-square code GMAP in conjunction with shape data relative to the very-well-characterized $^{252}$Cf(s.f.) prompt fission neutron spectrum (PFNS). This new spectrum was the result of contributions from the IAEA CIELO collaboration~\cite{OECD,CIELO:2014,CIELO:2018,CIELO-IAEA}, the IAEA Neutron Standard committee~\cite{Car18}, and the IAEA Prompt Fission Neutron Spectra project~\cite{PFNS1,PFNS2,PFNS}. The spectrum incorporated into the \mbox{IRDFF-II} library was converted from the ENDF/B-VIII.0 MF5/MT18 entry~\cite{ENDF8}. The associated covariance matrix was taken from the MF35/MT18 entry. Fig.~\ref{fig:U235_spect} shows this $^{235}$U thermal neutron fission spectrum -- represented as a ratio to a Maxwellian distribution at a temperature of kT=1.32 MeV.

\begin{figure}[!thb]
\vspace{-4mm}
\subfigure[~$^{235}$U(n$_{th}$,f) spectrum correlation matrix.]
{\includegraphics[width=0.98\columnwidth]{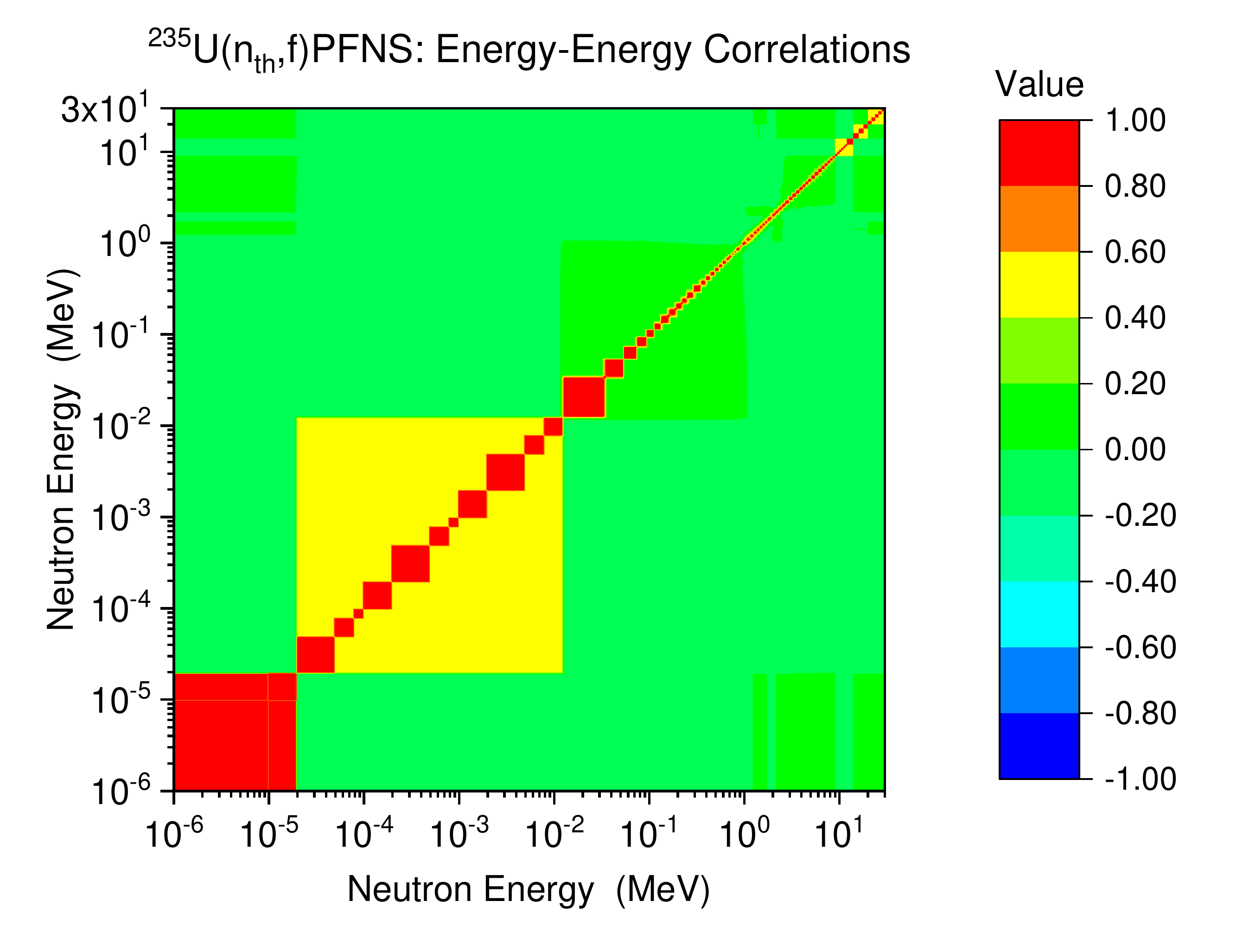}}
\subfigure[~One-sigma uncertainties in \% ($\equiv 100\times \frac{\sqrt{cov(i,i)}}{\mu_i}$), being $\mu_i$ the corresponding spectrum mean value.]
{\includegraphics[width=0.98\columnwidth]{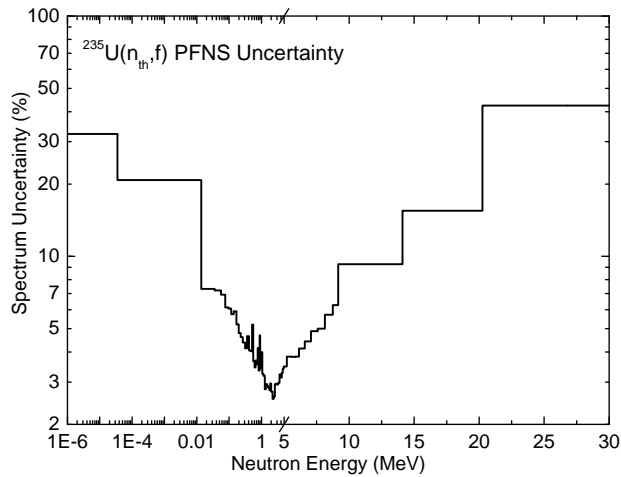}}
\vspace{-3mm}
\caption{(Color online) Uncertainties and correlations of the $^{235}$U(n$_{th}$,f) standard neutron benchmark spectrum in \mbox{IRDFF-II} library.}
\label{U235-sp}
\vspace{-4mm}
\end{figure}

Figs.~\ref{U235-sp}(a) ~\ref{U235-sp}(b) shows the correlation matrix and  the energy-dependent standard deviation for the $^{235}$U(n$_{th}$,f) reference spectrum, correspondingly.

\subsubsection{$^{233}$U(n$_{th}$,f) and $^{239}$Pu(n$_{th}$,f)}\label{Sec_VI_B2}
\begin{figure}[htbp]
\vspace{-3mm}
   \includegraphics[width=\columnwidth]{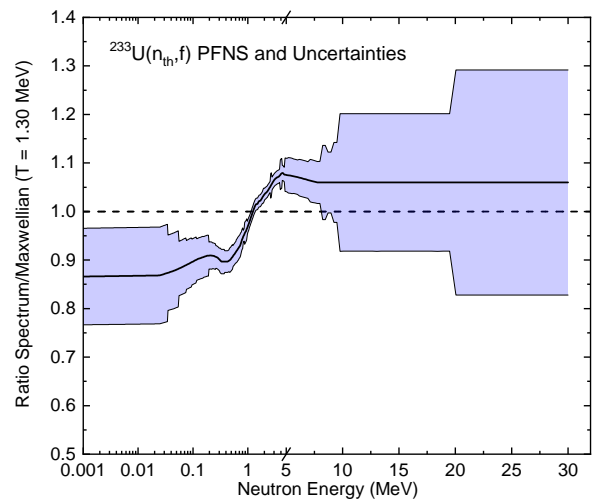}
   \caption{(Color online) The $^{233}$U(n$_{th}$,f) prompt fission neutron spectrum (PFNS) ratios to the Maxwellian distribution with $kT$=1.30~MeV. Note that the scale changes from log to linear above 5~MeV.}
   \label{U233_spect}
\end{figure}

\begin{figure}[htbp]
\vspace{-3mm}
   \includegraphics[width=\columnwidth]{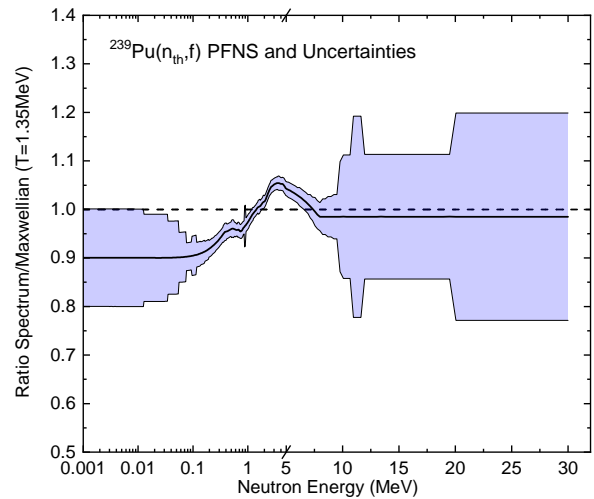}
   \caption{(Color online) The $^{239}$Pu prompt fission neutron spectra (PFNS) ratios to the Maxwellian distribution with $kT$=1.35~MeV. Note that the scale changes from log to linear above 5~MeV.}
   \label{Pu239_spect}
\end{figure}

The PFNS of thermal-neutron induced fission on $^{233}$U and $^{239}$Pu targets were evaluated together with the PFNS of thermal-neutron induced fission on $^{235}$U target within the IAEA CRP on the Prompt Fission Neutron Spectra of Major Actinides~\cite{PFNS}. A very few measured reaction rates are available, one set for the $^{233}$U(n,f) PFNS, none for $^{239}$Pu(n,f), but some measurements were reported as double ratios of spectral indices to the ratio in PFNS of Pu. Since the spectra are not readily available elsewhere, they are included in the \mbox{IRDFF-II} spectra file. The plot of the $^{233}$U spectrum is shown in Fig.~\ref{U233_spect} as a ratio to a Maxwellian spectrum with temperature $kT$=1.30~MeV. The plot of the $^{239}$Pu spectrum is shown in Fig.~\ref{Pu239_spect} as a ratio to a Maxwellian spectrum with temperature $kT$=1.35~MeV. These spectra were not modified above 10 MeV, but the proposed modification of the high-energy tail was described in detail in Ref.~\cite{PFNS}.

\subsubsection{$^{9}$Be(d,n) Reaction Neutron Field}\label{Sec_VI_B3}

%A cyclotron can be used to accelerate a deuteron into a beryllium target to get a high energy neutron spectrum. The reference spectra incorporated into the \mbox{IRDFF-II} library correspond to a 16~MeV and 40~MeV deuteron beam incident on a beryllium cylinder (38~mm in diameter and 38~mm long), respectively. A 12.7~mm diameter hole is bored along the axis of the cylinder leaving just enough material to stop the 40~MeV deuteron beam. One implementation of this field is at the Oak Ridge Isochronous Cyclotron (ORIC)~\cite{Sal77}. A time-of-flight spectral characterization for this facility exists and the characterization data appears in the EXFOR entry No.~\href{http://www-nds.iaea.org/EXFOR/C1832.002}{C1832.002}.

%The $^{9}$Be(d,n) neutron time-of-ight spectrum was also measured with deuteron energy of 16~MeV at the Argonne National Laboratory tandem accelerator (ref) and the spectrum is also shown in Fig.~\ref{Fig:XI_E}.

The development and qualification of fusion reactor materials requires irradiation in a neutron spectrum that includes very high flux at 14~MeV along with lower energy neutrons to simulate the first wall spectrum at proposed fusion power reactors.  Since fusion reactors do not exist for the testing and fission reactors lack any significant flux at 14~MeV, it has been proposed to produce suitable neutron fields by stopping intense deuteron beams produced by particle accelerators in thick Be or Li targets.  This produces a very high flux of neutrons over a continuous energy spectrum that can be designed to peak around 14~MeV.  The \href{https://www.ifmif.org/}{IFMIF} (International Fusion Materials Irradiation Facility)~\cite{IFMIF} is being proposed to study radiation damage in candidate fusion reactor materials.

\begin{figure}[htbp]
\includegraphics[width=\columnwidth]{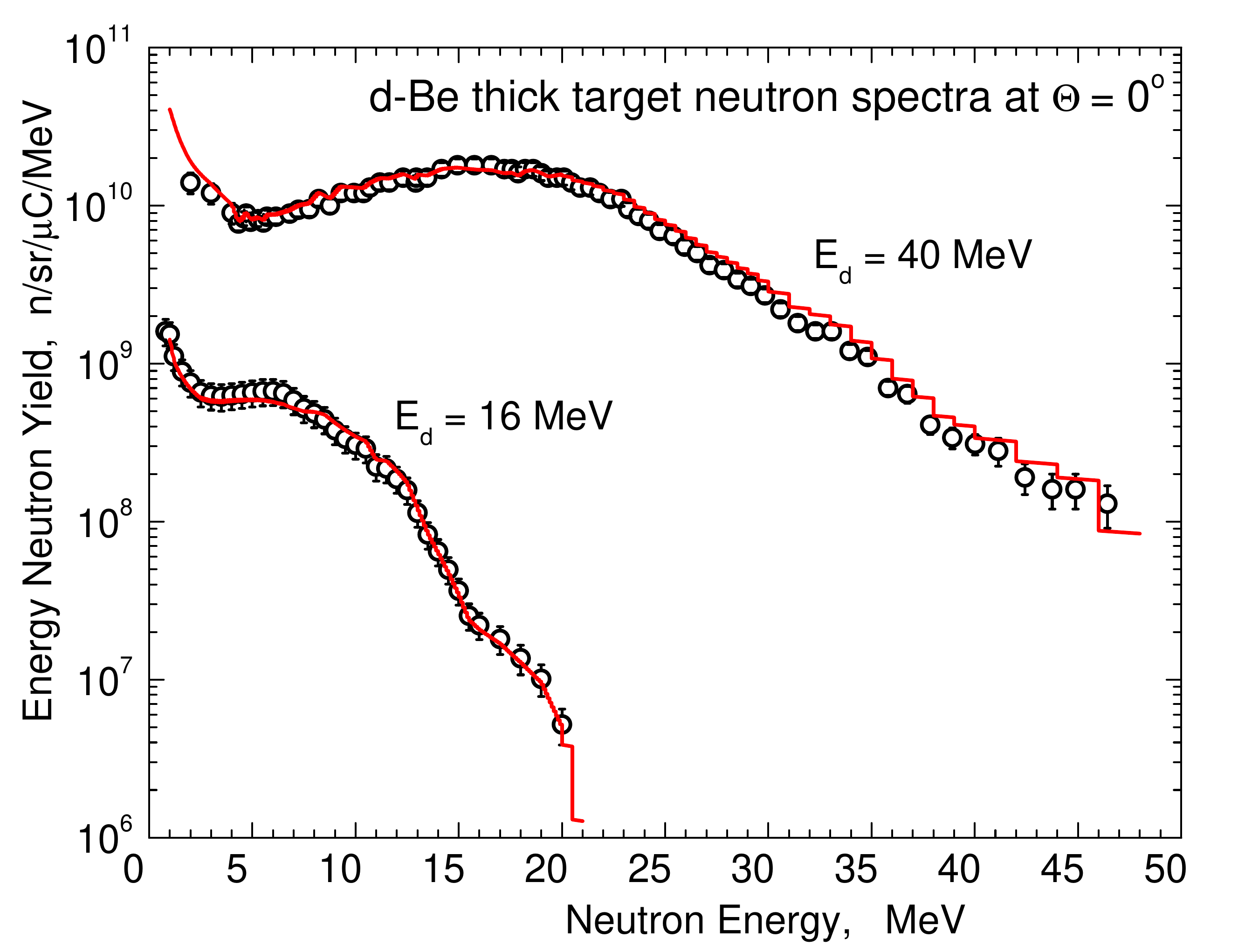}
\caption{(Color online) Neutron time-of-flight spectra are shown for thick Be targets at deuteron energies of 16 and 40~MeV at zero degrees.  The red lines show neutron spectral adjustment in 725 groups using the new \mbox{IRDFF-II} natural element cross sections with STAYSL-PNNL.}
\label{Fig:XI_E}
\end{figure}

Experiments were performed at deuteron energies of 16~MeV and 40~MeV producing neutron spectra as shown in Fig.~\ref{Fig:XI_E}. The neutron spectrum produced with deuteron energy of 16~MeV was measured at the Argonne National Laboratory tandem accelerator~\cite{Gre78}. The neutron spectrum produced with deuteron energy of 40~MeV was measured at the Oak Ridge Isochronous Cyclotron (ORIC)~\cite{Sal77}. The data from this facility can be found in the EXFOR entry No.~\href{http://www-nds.iaea.org/EXFOR/C1832.002}{C1832.002}. The neutron spectra were very well characterized by using neutron time-of-flight technique.  This method accurately determines the energy of neutrons interacting with a neutron detector at a fixed distance from the beryllium target by measuring the time required for a neutron to be created in the target and then travel to the neutron detector. The deuteron beam is pulsed to provide the start time and the stop time is determined when a neutron is detected.   The absolute accuracy of the neutron flux is determined by the efficiency of the neutron detector, which is about 10~\%.  Dosimetry materials were then irradiated in prepared geometries resulting in accurately measured reaction rates in well-characterized neutron spectra.

\subsection{Reference Neutron Fields from Detailed Computational Models}\label{Sec_VI_C}

\subsubsection{SPR-III Fast Burst Reactor Central Cavity}

SPR-III is an advanced fast-burst Godiva-type reactor with a large 16.5-cm central cavity, commissioned in 1975 at Sandia National Laboratory, NM, U.S.A.~\cite{Gri15a}. This is an unmoderated cylindrical assembly of uranium enriched to 93~\% $^{235}$U and alloyed with 10-wgt-\% molybdenum to ensure phase stability. The core consists of eighteen fuel plates mechanically fastened into two halves of nine plates each. The mass of the individual plates varies between 6.8 and 15.4~kg and the total mass of the core is about 258~kg. The nine upper plates are held stationary by the core support structure; the nine lower plates are attached to an electromechanical drive mechanism. Four reflector-type control devices are used; three for control and the fourth is the burst element. The control elements are used to establish a critical configuration of the core and to adjust the pulse yields. Although it was developed primarily for the radiation testing of electronic parts and systems, it has been used in a wide variety of research activities.  It is positioned in the centre of an air-filled shield building called a Kiva.  SPR-III has a $^{10}$B shroud that serves to decouple it from room return. The reactor can be operated in steady-state (up to 10 kW power) or pulsed mode (10 MJ in an 76~$\mu$s FWHM pulse that yields approximately 5x10$^{14}$~n/cm$^{2}$ in the reactor central cavity).

The reference SPR-III spectrum adopted within the \mbox{IRDFF-II} library characterizes the fuel centreline of the central cavity. High fidelity calculations using the MCNP, version 6.2.0, point cross section Monte Carlo code with the ENDF/B-VIII.0 cross sections and a detailed 3D geometry were performed to provide the best possible calculated neutron spectrum. The reactor, support structures, Kiva, and dosimetry test fixture were modelled in detail. The coupled neutron/photon calculation was performed using the MPI version of the code with 40 processors and using the L'Ecuyer 63-bit random number generator. The calculation used 5015 cycles with 2000000 source neutrons per cycle, of which 15 cycles were used to establish the normal fission distribution before beginning tally accumulation, for a total neutron source tally sample size of 100 billion neutrons. The neutron spectrum was scored in the 640-group SAND-II energy structure.

\begin{figure}[htbp]
\includegraphics[width=\columnwidth]{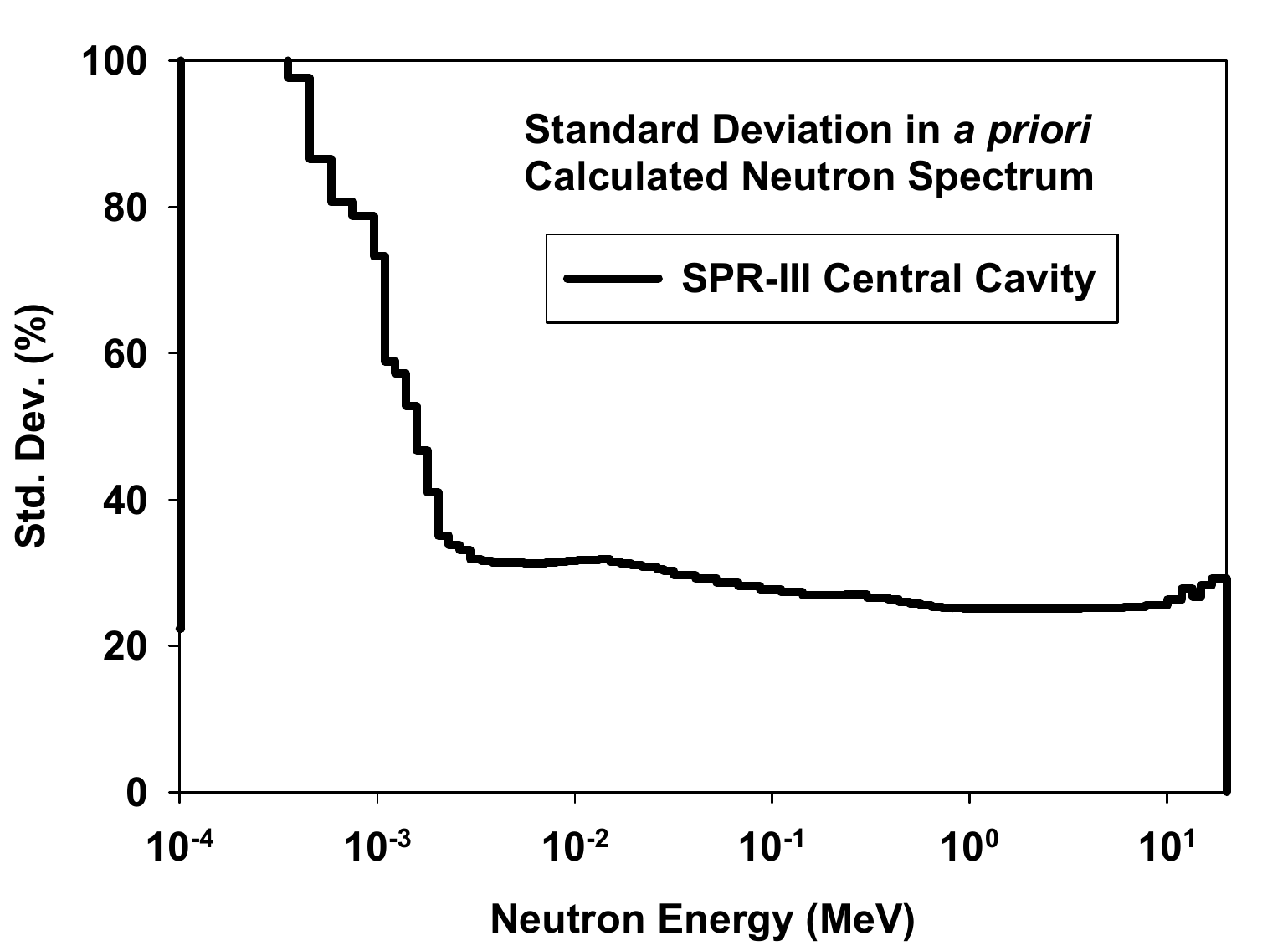}
\caption{Standard deviation for the calculated SPR-III fast burst reactor central cavity neutron spectrum.}
\label{fig:image21}
\end{figure}

\begin{figure}[htbp]
\includegraphics[width=1.02\columnwidth]{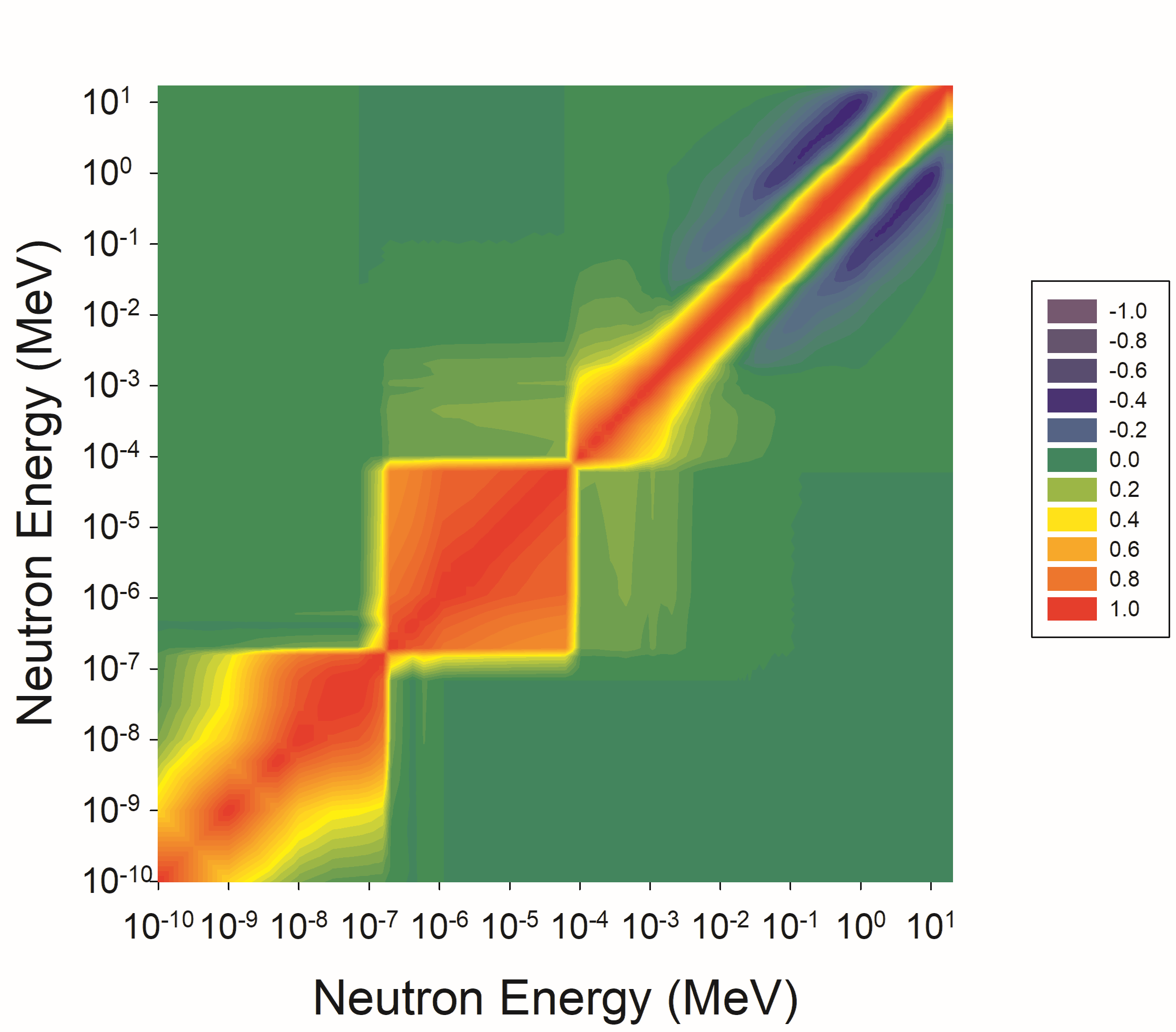}
\caption{(Color online) Correlation matrix for the Calculated SPR-III fast burst reactor central cavity neutron spectrum.}
\label{fig:image22}
\end{figure}

For this benchmark neutron field to be used to support the validation of the \mbox{IRDFF-II} cross sections, it is important that there be an uncertainty estimate of the calculated ``\textit{a priori}'' spectrum, in the form of an energy-dependent covariance matrix, and that this uncertainty estimate be derived independent of the activation measurements that are being used in the intended validation activity. This derivation of an energy-dependent uncertainty for a fine-group calculated spectrum is a very challenging task and a task where existing uncertainty quantification (UQ) methodologies are deficient and active work is still needed by the community. In addition to consideration of the numerical uncertainty, great care must be taken to not only address the uncertainty in the underlying transport cross sections and material geometry used in the radiation transport calculation, but to also investigate existing sources of ``model defect'', \ie, potentially correlated sources of uncertainty in the calculational models not directly captured in the user input parameters. A covariance matrix for this calculated spectrum was derived using the methods detailed in Reference \cite{Wil98} and subject-matter expertise based on previous modelling activities and subsequent comparison with the results obtained through spectrum adjustment methods. This ``\textit{a priori}'' covariance matrix for the reference neutron, derived independent of the reaction activities addressed later in this paper as part of the \mbox{IRDFF-II} validation evidence, is included as part of the \mbox{IRDFF-II} file that was outlined in Sec.~II.H. This content from the covariance matrix is seen in Figs.~\ref{fig:image21} and \ref{fig:image22}. Fig.~\ref{fig:image21} shows the energy-dependent uncertainty in the neutron spectrum. Fig.~\ref{fig:image22}  shows the correlation matrix.

% \subsubsection{Fast Burst Reactor 6-inch Leakage.}
%
% In addition to the SPR-III fast burst central cavity addressed in Sec. VIII\_B.2, the set of reference neutron fields also includes a leakage spectrum from a 93~\%-enriched $^{235}$U assembly. This spectrum represents a 640-group tally MCNP calculation of the leakage just outside of a Godiva-like assemble, 6-inches from the centre of the core. The effects of the experiment platform/table and room return have also been included in the computational model. Fig.~\ref{fig:image30} compares this calculated spectrum to that for other reference fields. This FBR leakage spectra is seen to show a more degraded fission spectra than that SPR-III central cavity due to neutron scattering within the fuel and room return. Both the SPR-III central cavity and this 6-inch leakage spectra show a very significant epithermal and thermal neutron component when compared to the $^{252}$Cf(s.f.) fission standard benchmark spectrum.

 \begin{figure}[htbp]
 \vspace{-3mm}
    \includegraphics[width=\columnwidth]{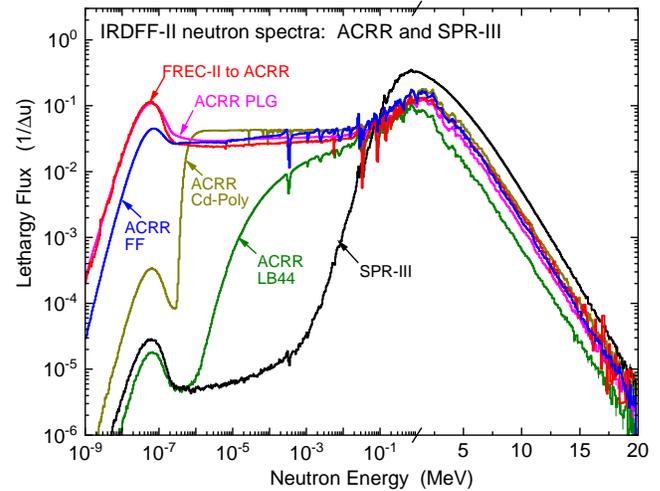}
 \vspace{-2mm}
    \caption{(Color online) Neutron reference reactor spectra for ACRR fission fields with various \ql Bucket\qr modifying environments. Note that the scale changes from log to linear above 1~MeV.}
    \label{fig:image30}
 \vspace{-3mm}
 \end{figure}

\subsubsection{ACRR Pool-type Reactor Central Cavity}

ACRR is a pulse and steady-state, pool-type research reactor located at Sandia National Laboratory, Albuquerque, U.S.A. The ACRR maintains a large, dry irradiation cavity at the centre of its core.  ACRR is typically used to perform irradiation testing where a high neutron fluence is required for a short period of time. Historically, ACRR has been used for a wide variety of experiment campaigns including electronics damage testing, nuclear fuels testing, space nuclear thermal propulsion testing, and medical isotope production. ACRR is currently fully operational and is used to add validation evidence for new dosimetry-quality reactions with a response in the epithermal and fast fission energy regions. ACRR's main attributes include a large, dry central irradiation cavity, epithermal neutron fluence, and large pulsing capabilities that permit a high irradiation and rapid recovery of the dosimeter.

ACRR can operate in a steady-state, transient, or pulse mode. In the steady-state mode, the operating power level is limited to 4 MW. In the pulse mode, a maximum pulse size of 250 MJ with a full-width half-maximum (FWHM) of 6 ms. In the transient mode, the reactor power shape can be tailored to the desired requirements for a total reactor energy deposition of 300 MJ. The transient mode can be used to increase the reactor power quickly; for example, a ramp increase in power level linear with time from low power to high power. ACRR is fueled by a 236 element array of UO$_{2}$-BeO fuel elements. The fuel is uranium enriched to 35 weight percent $^{235}$U, with 21.5 weight percent UO$_{2}$ and 78.5 weight percent BeO. The ACRR fuel elements are stainless steel clad, 3.747 cm in outer diameter and 52.25 cm in fuel length. Within the elements are niobium cups that hold the UO$_{2}$BeO fuel pieces. ACRR is controlled by two fuel-followed safety rods, three poison (void-followed) transient rods, and six fuel-followed control rods. The control rods make up part of the 236 elements for the normal core configuration. Further information regarding the ACRR irradiation environment can be found in internal laboratory documents \cite{DeP06}.

\begin{figure}[htbp]
\includegraphics[width=1.0\columnwidth]{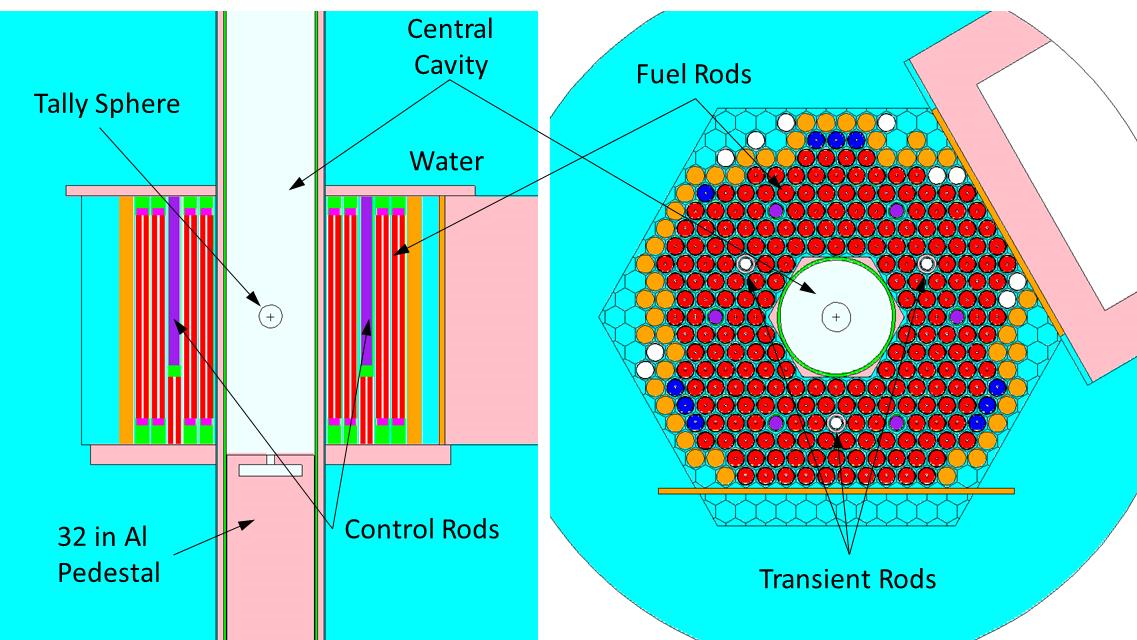}
\caption{(Color online) Computational model for the ACRR pool-type reactor central neutron spectrum.}
\label{ACRR_core_layout}
\end{figure}

The ACRR reference spectrum \cite{Par15} adopted within the \mbox{IRDFF-II} library characterizes the core centreline of the central cavity with the 32-inch pedestal. High fidelity calculations were performed to provide the best possible calculated neutron spectrum. This is the spectrum that should be used as input to various spectrum adjustment codes, or as the trial spectrum in iterative unfolding codes. The radiation transport was performed with the continuous energy, three-dimensional Monte Carlo particle transport code MCNP \cite{MCNP}. The MCNP model of the ACRR central cavity with the 32-inch pedestal is shown in Fig.~\ref{ACRR_core_layout}. The coupled neutron/photon calculation was performed using the MPI version of the code with 40 processors and using the L'Ecuyer 63-bit random number generator. The calculation used 5015 cycles with 2000000 source neutrons per cycle, of which 15 cycles were used to establish the normal fission distribution before beginning tally accumulation, for a total neutron source tally sample size of 100 billion neutrons. The neutron spectrum was scored in the 640-bin SAND-II energy group structure.

\begin{figure}[htbp]
\vspace{-3mm}
\includegraphics[width=1.0\columnwidth]{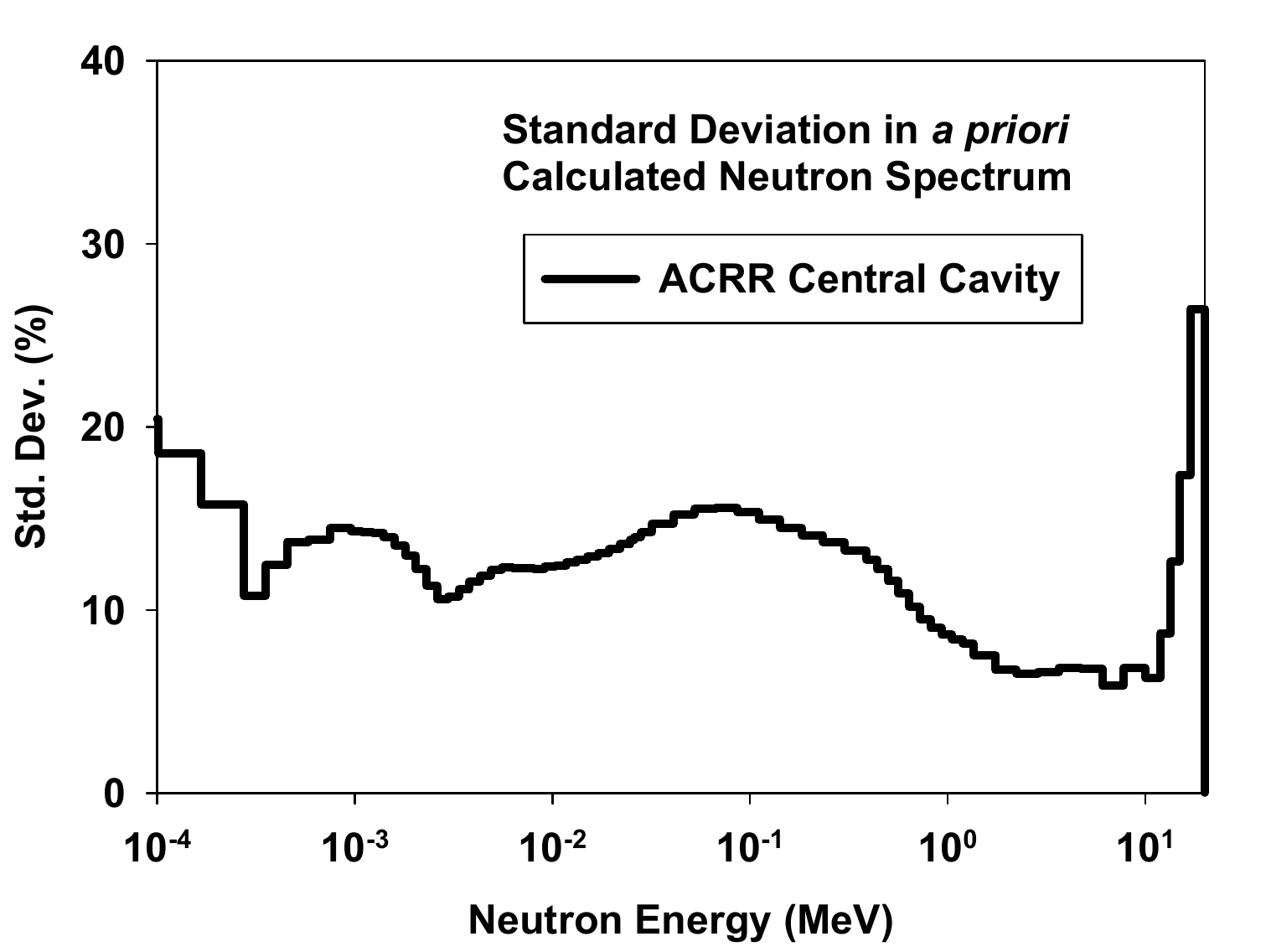}
\caption{Standard deviation for the calculated ACRR pool-type reactor central neutron spectrum.}
\label{fig:image24}
\vspace{-2mm}
\end{figure}

\begin{figure}[htbp]
\vspace{-2mm}
\includegraphics[width=1.0\columnwidth]{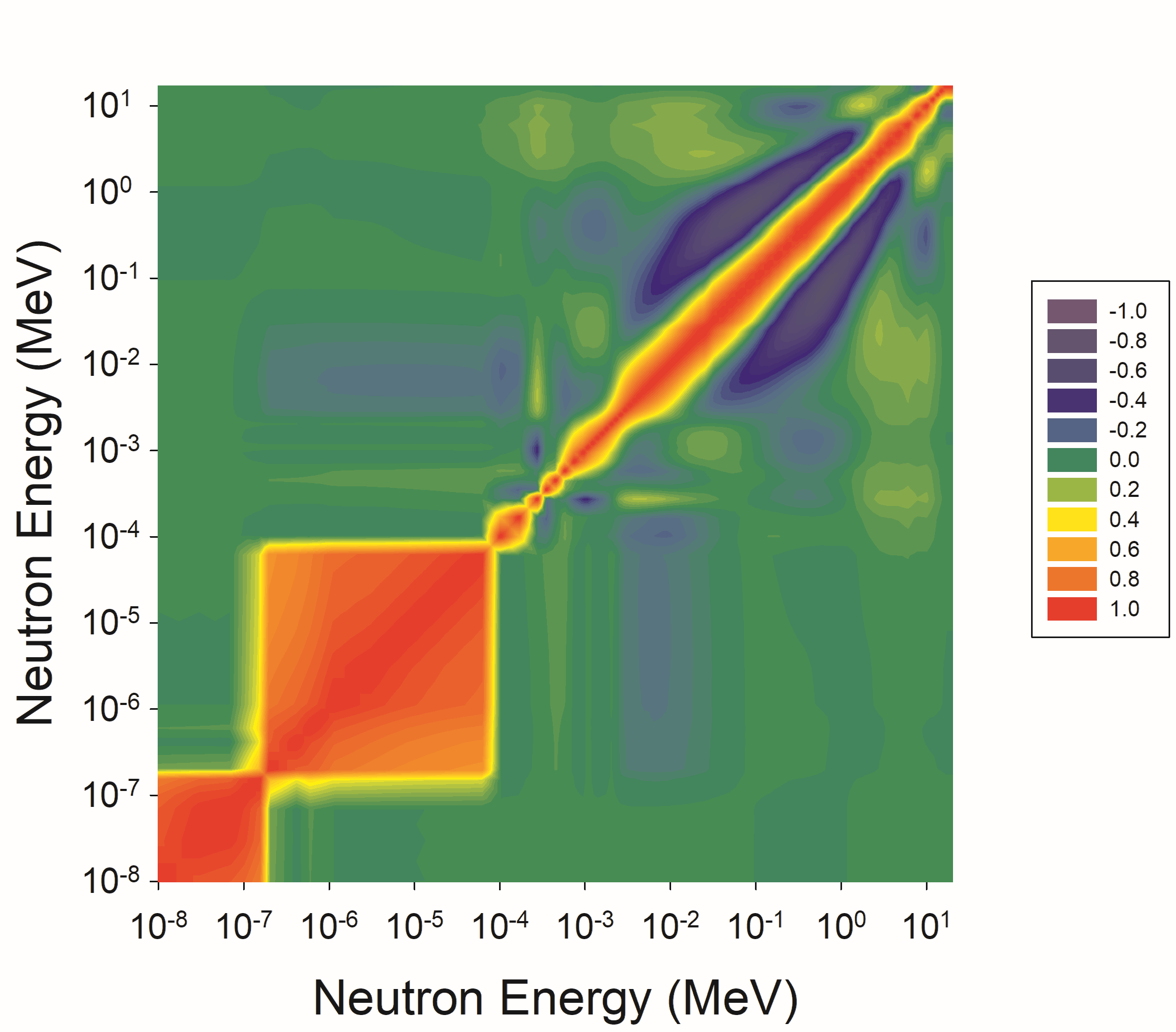}
\caption{(Color online) Correlation matrix for the calculated ACRR pool-type reactor central neutron spectrum.}
\label{fig:image25}
\vspace{-2mm}
\end{figure}

For this benchmark neutron field to be used in support of the validation of the \mbox{IRDFF-II} cross sections, a calculated "\textit{a priori}" spectrum, in the form of an energy-dependent covariance matrix, and one that has been developed independent of the activation measurements that are being used in the intended validation activity, is required. The covariance matrix for this calculated spectrum was derived using the methods detailed in Reference \cite{Wil98} and subject-matter expertise based on previous modelling activities and subsequent comparison with the results obtained through spectrum adjustment methods. This "\textit{a priori}" covariance matrix for the reference neutron, derived independent of the reaction activities is included as part of the \mbox{IRDFF-II} file that was outlined in Sec.~II.H. This content from the covariance matrix is seen in Figs.~\ref{fig:image24} and \ref{fig:image25}. Fig.~\ref{fig:image24}  shows the energy-dependent uncertainty in the neutron spectrum. Fig.~\ref{fig:image25}  shows the correlation matrix.

\subsubsection{ACRR Pool-type Reactor Pb-B$_{4}$C (LB44) Bucket}

\begin{figure}[htbp]
\vspace{-3mm}
 \includegraphics[width=\columnwidth]{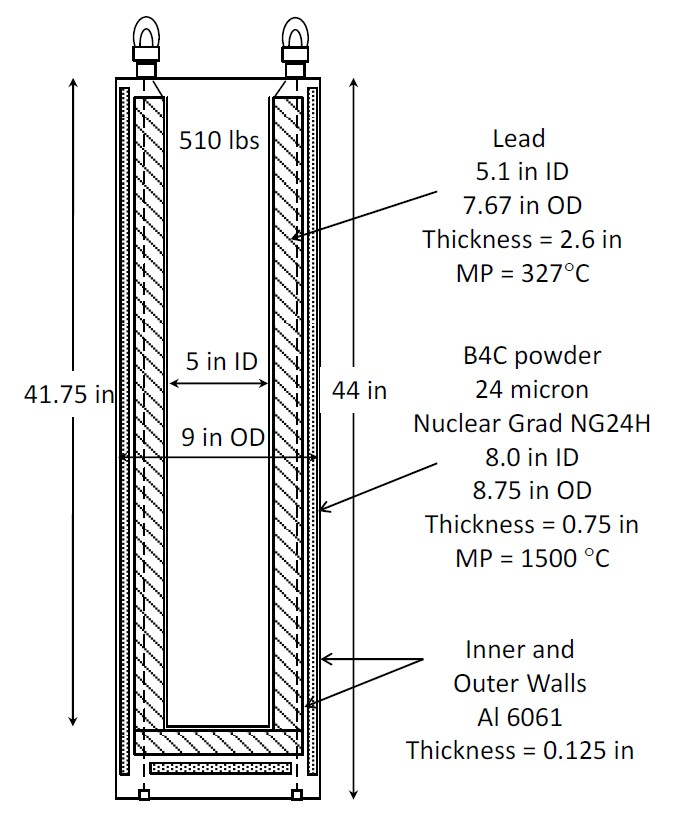}
\vspace{-2mm}
   \caption{\ql Bucket\qr used to tailor the ACRR-LB44 neutron spectrum.}
   \label{fig:image31}
\vspace{-3mm}
\end{figure}

The ACRR maintains an epithermal neutron fluence spectrum in the core and central cavity. This allows for the neutron energy fluence spectrum to be tailored to the desired specifications of an experiment by introducing various spectrum modifying buckets. Moderators can be used within the cavity to thermalize the neutron spectrum. Boron and lead can be used to increase the fast neutron fluence ratio and decrease the gamma-ray fluence, respectively. The ACRR-LB44 neutron field uses a 44-inch tall lead and B$_{4}$C ``bucket'' to moderate neutron field seen at a 5-inch inner diameter experiment location. The primary purpose of this ``bucket'', depicted in Fig.~\ref{fig:image31}, is to reduce the thermal component of the neutron spectrum as well as the associated gamma exposure. This neutron spectrum is calculated using the MCNP code in a manner similar to that described in Sec.~VI.B.3. The calculated neutron spectrum is shown in Fig.~\ref{fig:image30}.

\subsubsection{ACRR Pool-type Reactor Cd-Polyethylene Bucket}

\begin{figure}[!htbp]
\vspace{-2mm}
  \includegraphics[width=\columnwidth]{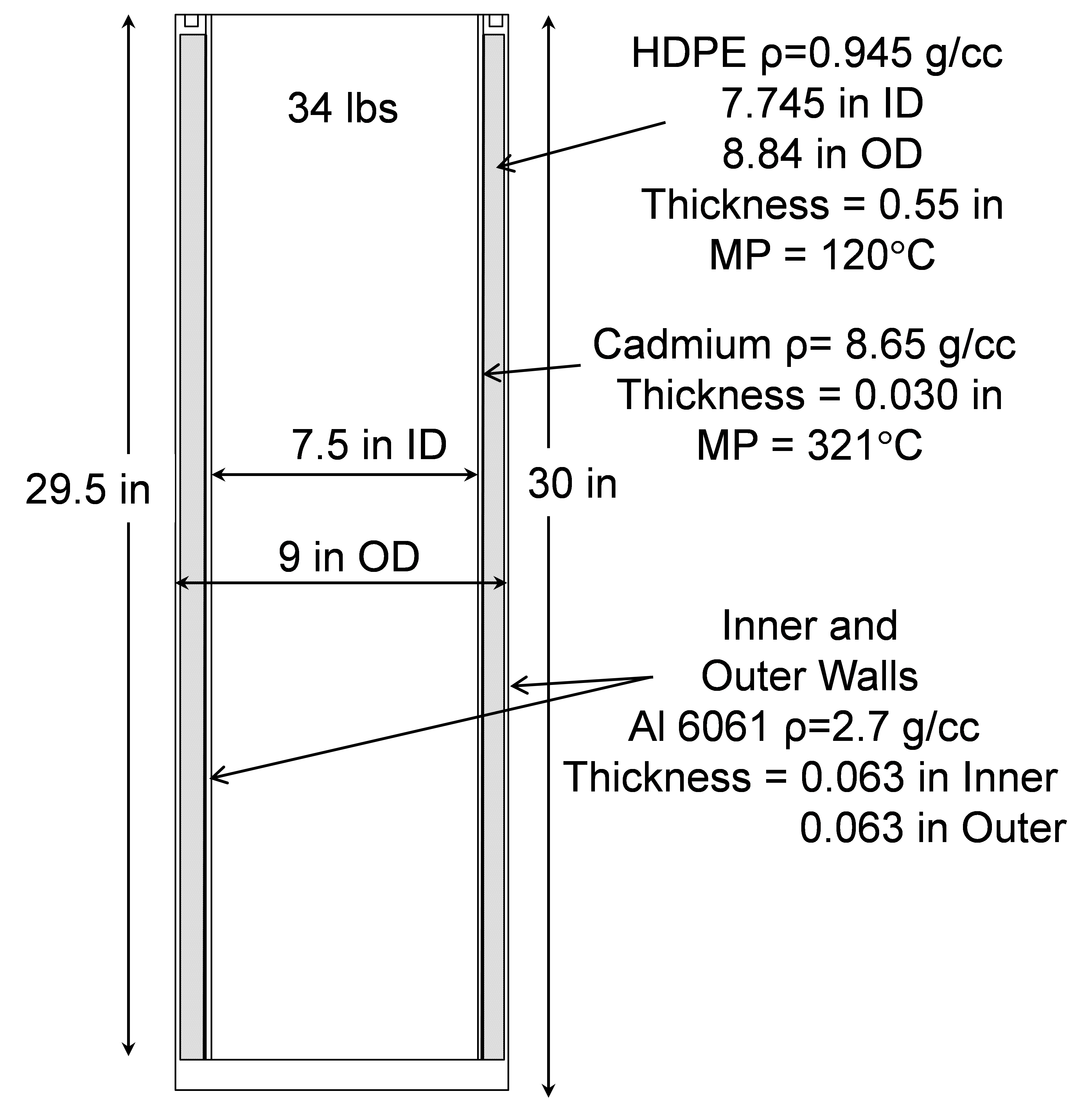}
%\vspace{-2mm}
   \caption{\ql Bucket\qr used to tailor the ACRR-Cd-poly neutron spectrum.}
   \label{fig:image33}
\vspace{-2mm}
\end{figure}

A cadmium-polyethylene insert is used to tailor the neutron spectrum for this field. This Cd-poly ``bucket'' is designed to enhance the gamma ionization component in the mixed neutron/gamma reactor field by a factor of two. It also significantly reduces the thermal neutron component while leaving the epithermal component of the cavity spectrum. The ``bucket'' provides an inner experiment diameter of about 7.5-inches and accommodates packages with a height of 30-inches. The outer portion of the bucket is fabricated from high density polyethylene with a thickness of 0.55-inches and the inner cadmium sheet is 0.02-inches. The purpose of the bucket, depicted in Fig.~\ref{fig:image33}, is to provide variation in the neutron-to-gamma radiation field when testing electronic components.  The calculated neutron spectrum is shown in Fig.~\ref{fig:image30}.

\subsubsection{ACRR Pool-type Reactor Polyethylene-lead-graphite (PLG) Bucket}

\begin{figure}[!htbp]
%\vspace{-2mm}
  \includegraphics[width=\columnwidth]{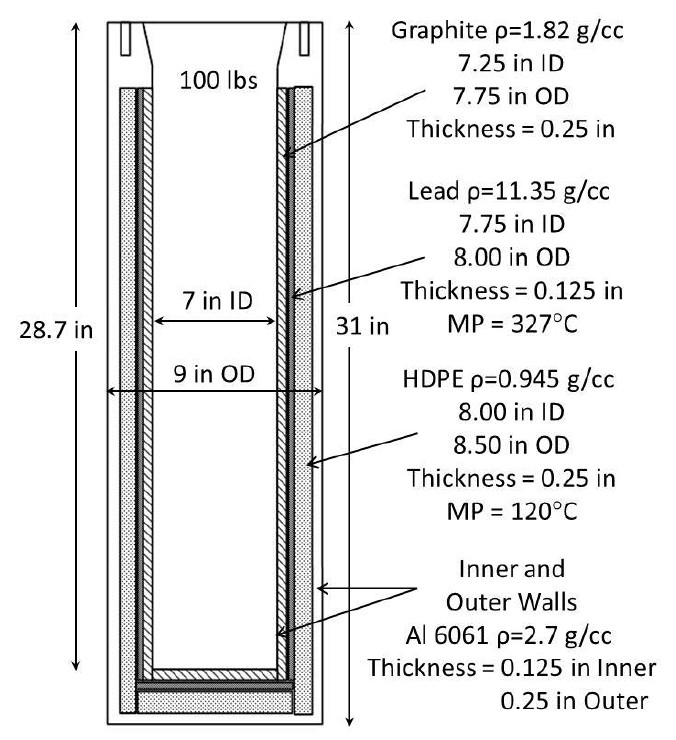}
\vspace{-2mm}
   \caption{\ql Bucket\qr used to tailor the ACRR-PLG neutron spectrum.}
   \label{fig:image32}
%\vspace{-2mm}
\end{figure}

A polyethylene-lead-graphite (PLG) insert is used to tailor the neutron spectrum for this field. This PLG bucket is designed to fit within the 23.4-cm diameter ACRR central cavity. The purpose of the bucket, depicted in Fig.~\ref{fig:image32}, is to provide some neutron moderation, reducing the fast neutron component relative to the total neutron fluence. The calculated neutron spectrum is shown in Fig.~\ref{fig:image30}.

\begin{figure}[!htbp]
%\vspace{-2mm}
    \includegraphics[width=\columnwidth]{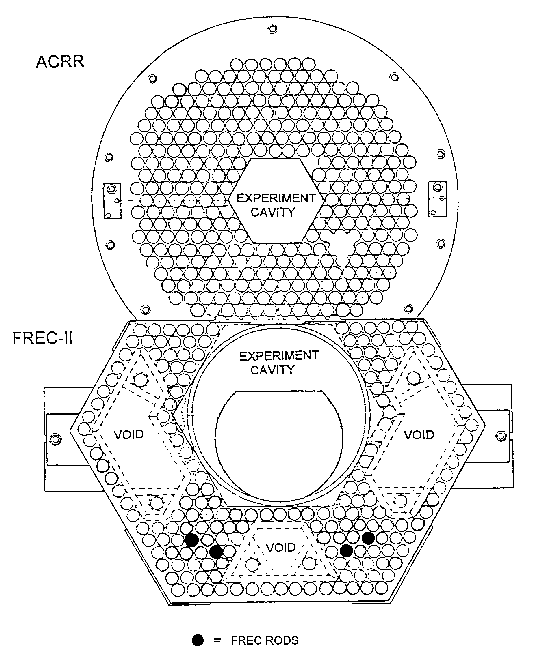}
\vspace{-2mm}
   \caption{ACRR fueled external cavity neutron spectrum.}
   \label{fig:image34}
%\vspace{-2mm}
\end{figure}

\subsubsection{ACRR Fueled Ring External Cavity (FREC-II)}

FREC-II is an external cavity attached to the ACRR and it can be ``coupled'' or ``de-coupled'' to the ACRR core depending on the experimenter's desires.  In the ``de-coupled'' mode, FREC-II is tilted away from the core several degrees and the Ni plate is attached to the ACRR.  In this configuration, the FREC-II has minimal effect on the neutronic behavior of the reactor.  With the FREC-II ``coupled,'' a significant neutron and gamma-ray flux exists in the 20-inch, dry, FREC-II cavity.

Fig.~\ref{fig:image34} depicts the FREC-II cavity as it is coupled to the ACRR. The FREC-II uses uranium/zirconium-hydride (U-ZrH) TRIGA fuel. FREC-II is fueled with a 186-element array of UZrH$_{1.625\ }$fuel elements. The fuel is 12 weight percent uranium enriched to 20 weight percent U-235.  The hydrogen atomic ratio is 1.625 hydrogen atoms per Zr atom. The FREC-II fuel elements are stainless steel clad, 1.5 inches in diameter and 15 inches in fuel length. The UZrH fuel in each element is a single solid cylinder with a 0.25-inch diameter hole at its centre. The centre region maintains a Zr rod with the same diameter. FREC-II has four neutron poison elements that are fuel followed and can be positioned full-in (down position-poison in), full-out (up position-poison out), or any position in between. This feature allows for radial flux tilting in the 20-inch cavity. The FREC-II cavity allows for some advantages compared to the ACRR central cavity for consideration by experimenters: 1) the FREC-II can accommodate larger experiments, since its cavity is 20 inches in diameter compared to the 9-inch diameter central cavity; 2) the neutron and gamma-ray flux can be radially tilted by using the FREC rods and positioning the experiment within the cavity; and 3) the smaller neutron flux allows for a maximum pulse with a short pulse width ($\sim$8 ms) and smaller total neutron fluence. This same effect could be achieved in the central cavity by positioning the experiment package higher in the cavity.

The calculated neutron spectrum is shown in Fig.~\ref{fig:image30}.

\begin{figure*}[!thbp]
\vspace{-18mm}
\centering
   \includegraphics[width=1.1\textwidth]{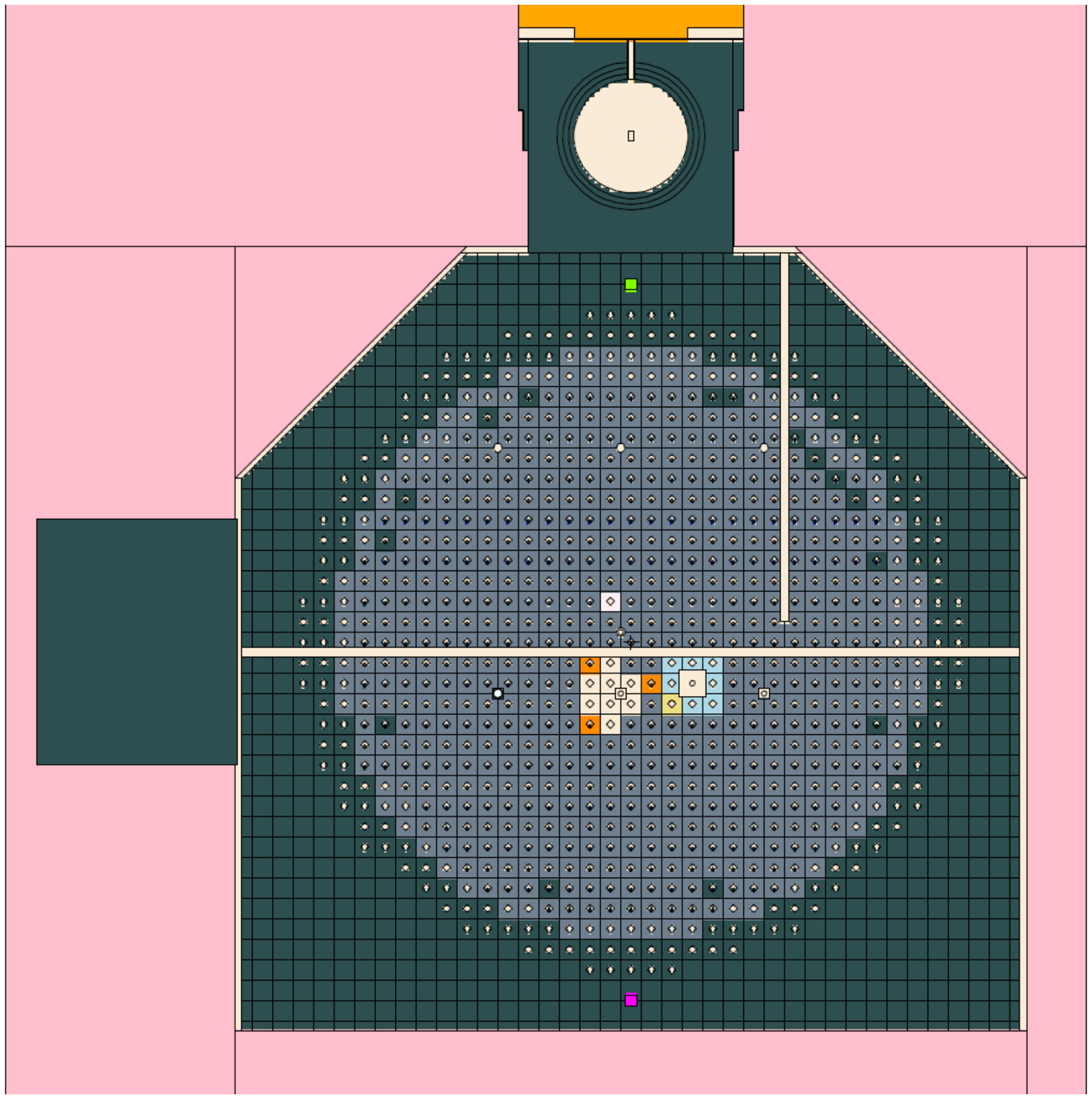}
   \caption{(Color online) Computational model of the Mol-BR1 irradiation facility.}
   \label{Mol-BR1_Core_Layout}
\vspace{-3mm}
\end{figure*}

\subsubsection{Mol-BR1 Cavity $^{235}$U Fission Spectrum}
A number of standard neutron fields were designed in the BR1 reactor at the Belgian Nuclear Research Centre SCK$\bullet$CEN in Mol. They are designated as Mark-I, Mark-II, Mark-IIA or Mark-III. Many measurements of cross sections in these standard fields were reported~\cite{Fab82,Gil84,Man85_2}. In particular, most of the measurements of the $^{235}$U(n$_{th}$,f) PFNS spectrum-averaged cross sections were measured in the Mark-IIA neutron field, which is very similar to the current Mark-III. The standard fields are differenced by enrichment, height and thickness of the $^{235}$U cylinder and by the thickness of the Al claddings inside and outside of the uranium cylinder. The Mark-III standard neutron field was generated by means of a flexible $^{235}$U source arrangement located at the centre of a spherical cavity in the graphite thermal column \cite{Fab82}. The large 100~cm diameter cavity has minimized the wall return corrections while permitting a large volume for active instruments. The access to the centre of the field is enabled through a long 1~mm thick Cd tube, which is loaded with the irradiation samples. The source of fission neutrons is a $^{235}$U cylinder wrapped around the cadmium tube. The Mark-III configuration was re-analysed recently~\cite{Wag16} and the computational model of the core layout is shown in Fig.~\ref{Mol-BR1_Core_Layout}, with the irradiation facility at the top of the figure.

\begin{figure}[htbp]
\includegraphics[width=\columnwidth]{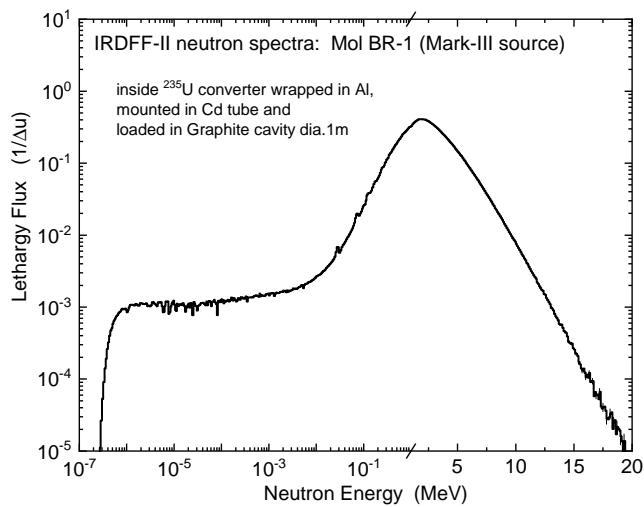}
\caption{The calculated fission neutron spectrum inside Mark-III U-235 converter in the Mol BR1 reactor graphite cavity. Note that the scale changes from log to linear above 1~MeV.}
\label{fig: Mol_BR1_spect}
\end{figure}

To refer the dosimetry reaction rates measured in the cavity centre to the ones in the $^{235}$U(n$_{th}$,f) PFNS, the neutron field perturbation corrections for the used facility configurations were calculated and applied by the authors of the measurements. The absolute neutron fluence was established relative to the $^{252}$Cf source, or reactions of interest were measured as chosen ratios.

The neutron spectrum inside the MARK-III converter loaded in the BR1 cavity, computed by MCNP-6 with ENDF/B-VIII.0 neutron data, is depicted in Fig.~\ref{fig: Mol_BR1_spect}. The standard fission neutron field at the BR-1 reactor has been made available for international cooperative efforts for neutron metrology. Thus the large fraction of the existing cross sections averaged over the reference $^{235}$U(n$_{th}$,f) prompt spectrum was measured there.

\subsubsection{LR-0 Reactor at \v{R}e\v{z}}
The LR-0 reactor of the Research Centre \v{R}e\v{z} near Prague, Czech Republic is an experimental zero-power light-water-moderated reactor. The core is placed in a 6.5 m high Al tank with diameter of 3.5~m. The large reactor vessel is a legacy of the former heavy water reactor project which was carried out back in the 60’s. At the end of the 70’s, the refurbishment was accomplished and the reactor core was converted to light-water-moderated to support research in this field. The first criticality of the light water version was reached in December 1982.

\begin{figure*}[hbtp]
\includegraphics[width=\textwidth]{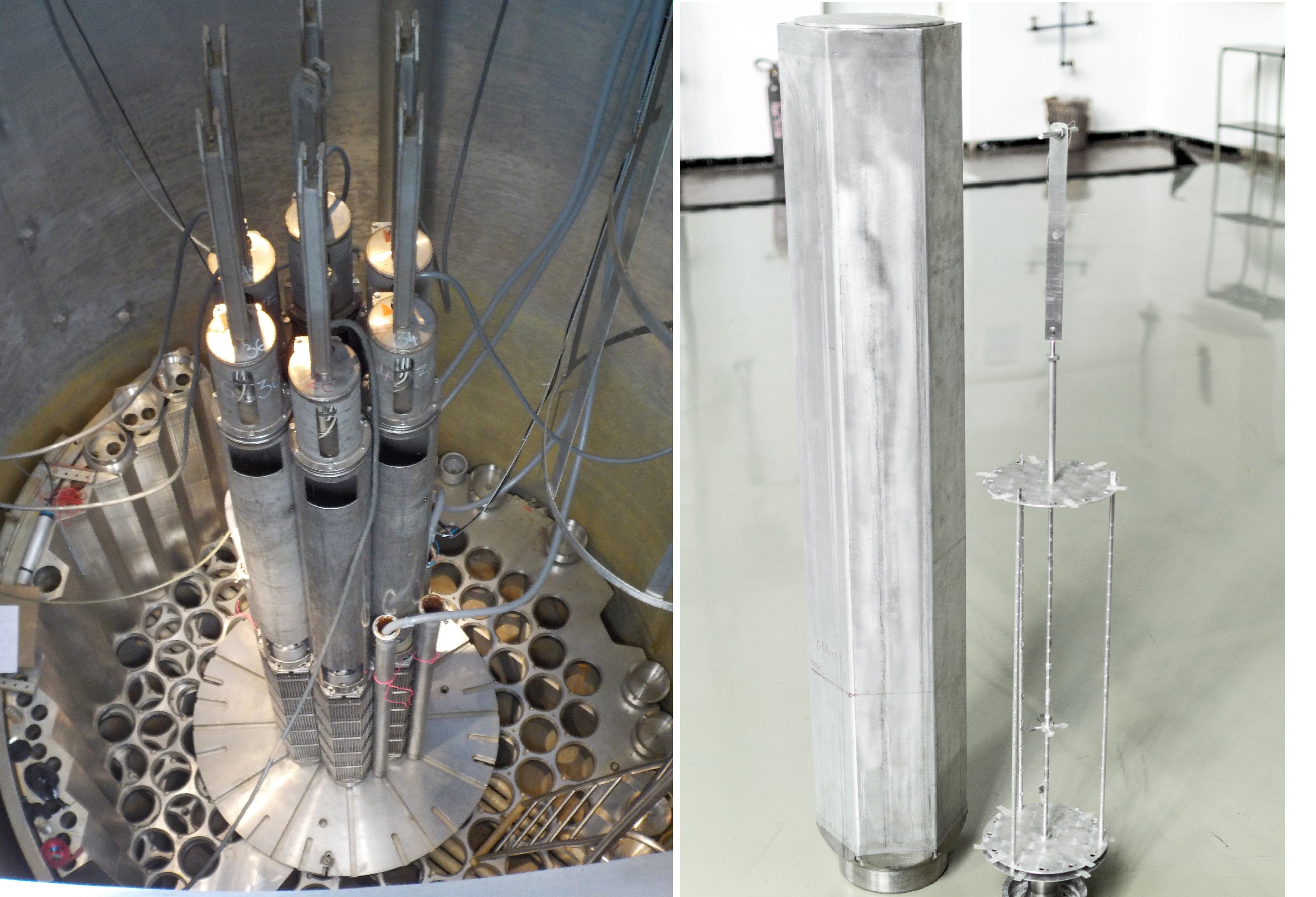}
  \caption{(Color online) Core(left), special Al assembly (right) (from \textsc{Ann. Nucl. En. }\textbf{112}, pp. 759--768 (2018).).}
  \label{LR0_reactor_layout}
\end{figure*}
\begin{figure}[htbp]
  \centering
  \includegraphics[width=\columnwidth]{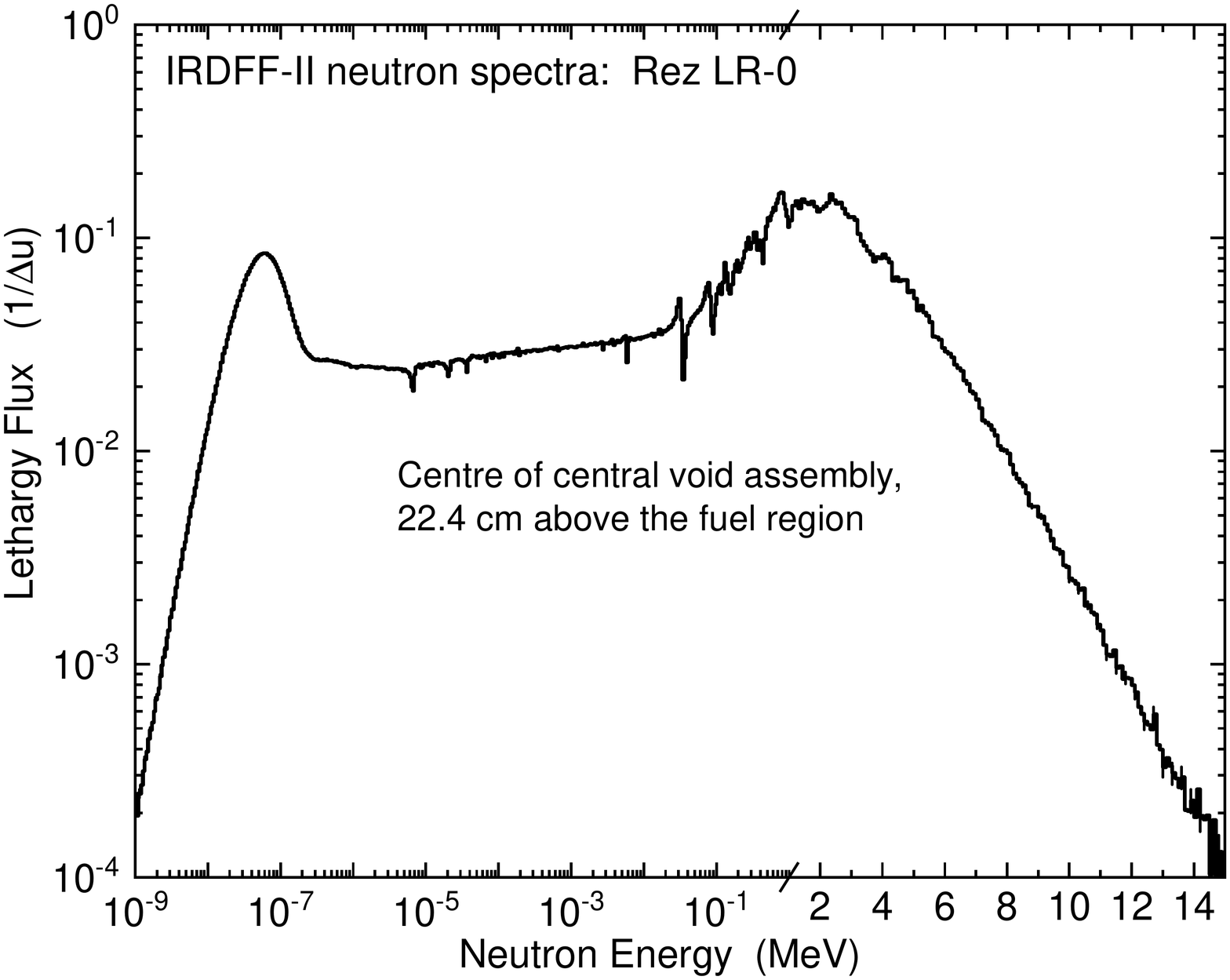}
  \caption{Neutron field in the centre of the dry assembly of the \v{R}e\v{z}-LR0 reactor. Note that the scale changes from log to linear above 1~MeV.}
  \label{Rez-LR0_Spectrum}
\end{figure}
The main scope of the LR-0 research were some lattice studies of the VVER-440 and VVER-1000 reactor. Nominal thermal neutron flux density is 10$^{9}$~n.cm$^{-2}$.s$^{-1}$, which corresponds to the thermal power of 1~kW. Thanks to the low power, the fuel is considered to be physically fresh during all experiments. Later, the full-scale reflector segment of the VVER-1000 reflector was installed to the LR-0 vessel. Thus, the so called "VVER-1000 Mock-up" experiments were enabled. The only significant difference between a real \mbox{VVER-1000} reactor core and the \mbox{VVER-1000} Mock-up is the height, because the LR-0 does not need so much fuel as an energy-producing reactor. Despite the shorter fuel assemblies, the LR-0 reactor can be used for \mbox{VVER-1000} and \mbox{VVER-440} reactor experiments focused on criticality, neutron transport and dosimetry to support services for utilities. For reproducibility, the full-scale simulators of the \mbox{VVER-1000} internals are permanently installed in the LR-0 tank (see Fig.~\ref{LR0_reactor_layout}, left).

The large reactor vessel offers space for the variety of experiments with cores physically and neutronically decoupled from the \mbox{VVER-1000} Mock-up. Based on this, smaller cores consisting of six hexagonal \mbox{VVER-1000} fuel assemblies were designed to extend the experimental portfolio. The small core occupying only seven lattice positions is the simplest assembly which can be formed in the LR-0 reactor and it was also benchmarked. This benchmark core contains fuel assemblies with enrichment of 3.3~\% surrounding one void hexagonal channel exactly fitting to the fuel lattice. As mentioned, the LR-0 reaches only negligible burn-up, thus redistribution of the fission source does not occur due to this effect and high reproducibility of experiments can be guaranteed.

In the past, neutron spectra~\cite{Kos18_3} and flux distribution~\cite{Kos18_4} in the centre of the dry assembly have been validated as well as the criticality~\cite{Kos16_1} and neutron emission distribution~\cite{Kos16_2} of the driver core. Therefore, the field in the void assembly can be understood as the reference neutron field. Low flux experiments at 2$\times 10^{7}$~n.cm$^{-2}$.s$^{-1}$ in flux over 1~MeV with some larger samples have been performed. Only low activities of threshold reactions can be induced in this field, hence closer measuring geometry during gamma spectrometry was selected. This is possible thanks to a well-defined efficiency curve in the used HPGe semiconductor measurement arrangement. More details concerning the core description can be found in the IRPhE Project Handbook~\cite{Kos18_5}. The spectrum in the centre of the dry assembly of the \v{R}e\v{z}-LR0 reactor is shown in Fig.~\ref{Rez-LR0_Spectrum}.

    \begin{figure*}[!thbp]
        \centering
        \includegraphics[width=\textwidth]{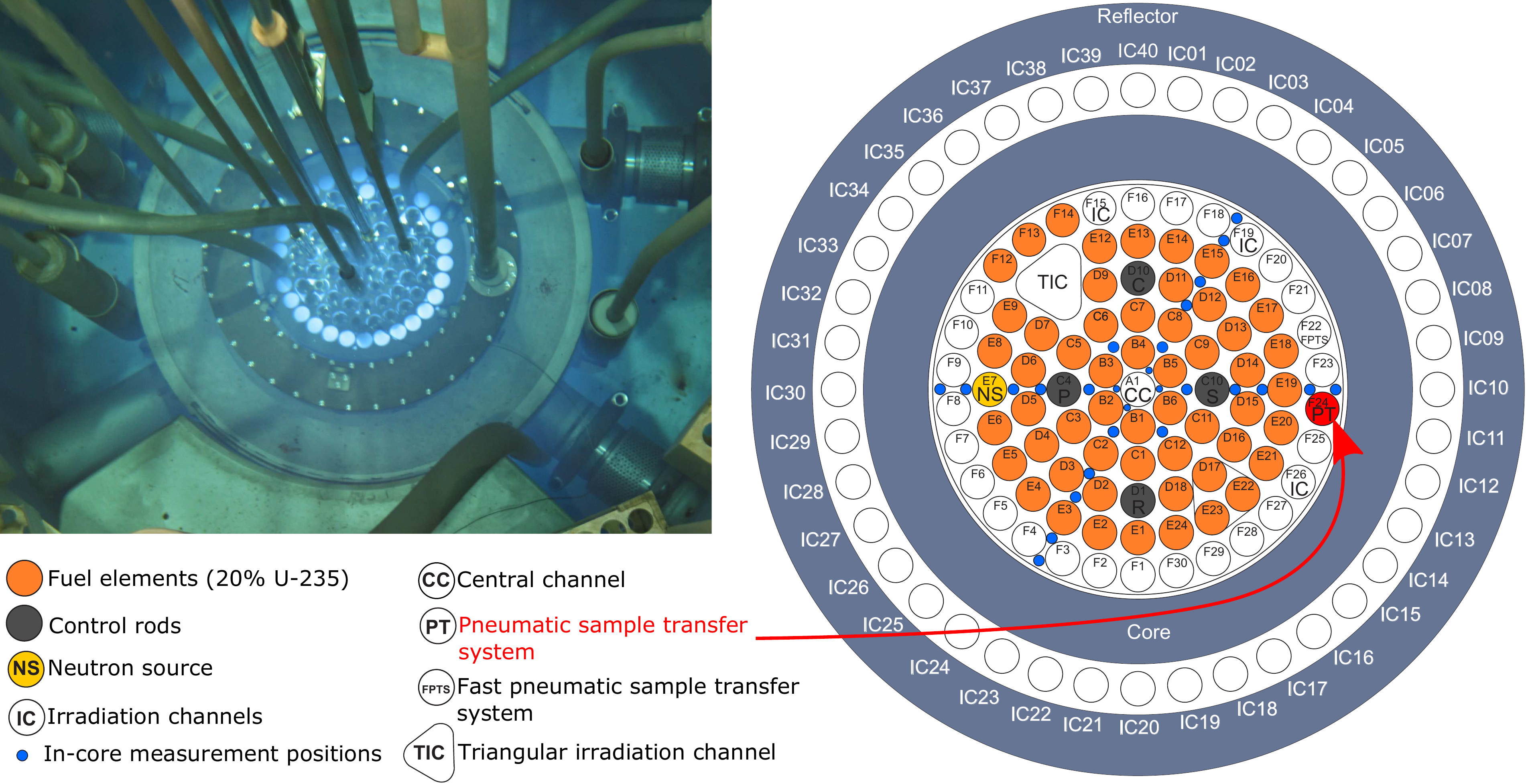}
        \caption{(Color online) Left: Photograph of the JSI TRIGA reactor core at 250 kW (full power). Right: schematic diagram of the reactor core configuration.}
        \label{JSI_TRIGA}
    \end{figure*}

\subsubsection{TRIGA-JSI Pneumatic Tube Irradiation Channel}
The Jo\v{z}ef Stefan Institute (JSI) TRIGA reactor in Ljubljana, Slovenia is a 250~kW light-water pool-type TRIGA Mark II reactor with uranium zirconium hydride fuel elements and an annular graphite reflector, cooled by natural convection. The reactor achieved first criticality in 1966. Due to its very well characterized irradiation fields its main applications, at present, are Neutron Activation Analysis (NAA), radiation hardness testing of detectors and electronic components (\ie, reference neutron radiation hardness irradiation facility for the CERN LHC accelerator and other accelerators within the EC H2020 project AIDA-II - Advanced European Infrastructures for Detectors at Accelerators), support for testing of neutron instrumentation detectors (fission chambers, self-powered neutron detectors) support for development of radiation-hard products, code testing and validation.

The JSI TRIGA reactor facility is described in detail in the publications~\cite{Jeraj_ICSBEP,Stancar_IRPHE}. The reactor core is located at the bottom of a 6.25~m high open tank, 2~m in diameter. The core has a cylindrical configuration with 91 core locations which can be filled either by fuel elements or other components like control rods, a neutron source, irradiation channels, \textit{etc}. The core lattice has an annular but not periodic structure. The fuel elements are arranged in six concentric rings: A, B, C, D, E and F, with 1, 6, 12, 18, 24 and 30 fuel locations per ring, respectively. The graphite reflector surrounding the reactor core is enclosed in aluminum casing and also houses a special irradiation facility, the rotary specimen rack, with 40 irradiation positions.

The reactor core is supported by a two 3/4-inch aluminium grid plates, which provide accurate axial and lateral positioning of the core components. The fuel elements are cylindrical with stainless steel (SS-304) cladding. Their total length is approximately 28 inches, their diameter is 1.478 inches. The fuel material in each element is 15 inches in length. The fuel elements are equipped with cylindrical graphite inserts at the top and bottom ends, 2.6 inches and  3.7 inches in length, respectively, which act as axial reflectors. In the centre of the fuel material there is a 0.25-inch-diameter hole which is filled by a zirconium rod. Between the fuel meat and the bottom graphite insert there is a 1/32-inch-thick molybdenum disc. The fuel itself is a homogeneous mixture of uranium and zirconium hydride with 12 wt.\% uranium, enriched to 20~\%.

Three control rods of the fueled-follower type are used in the reactor, \ie, the Regulating (R), Shim (C), and Safety (S) control rods, identical in geometry and composition. A fourth control rod, the Transient Rod (T), is equipped with a pneumatic ejection system and enables reactor pulse operation. It is similar to the fueled-follower control rods, but with a so-called air follower, which replaces the fuel part in the fueled-follower control rods. The purpose of the air-follower is to reduce power peaking that could appear when the transient rod is in its fully withdrawn position during steady state operation.

During normal operation the reactor core is equipped with 7 irradiation channels. Six of the irradiation channels occupy one core position: the Central Channel (CC), located in the reactor core centre (A1), and five irradiation channels located in the outer F core ring (F15, F19, F22, F24 and F26). Two of these irradiation channels are equipped with pneumatic sample transfer systems, \ie, the channel in position F22, also denoted as Fast Pneumatic Transfer System (FPTS), and in position F24, also denoted as Pneumatic Tube (PT). One channel, the Triangular Irradiation Channel (TIC) is larger and occupies three core positions (D8, E10 and E11), arranged in a triangle, and enables irradiations of objects with maximal lateral dimensions of approximately 50~mm. Fig.~\ref{JSI_TRIGA} displays a photograph of the TRIGA-JSI reactor core and a schematic diagram of the core configuration.

\begin{table}[!hbtp]
\vspace{-4mm}
    \caption{Dimensions of the Cd, BN, B$_4$C and $^{10}$B$_4$C neutron filters used in the experiments at the JSI TRIGA reactor.}
    \label{Filter_dimensions}
    \centering
        \begin{tabular}{c|c|c|c|c}
            \hline  \hline
            Filter & Outer    & Height & Inner    & Inner  \\
                   & diameter &        & diameter & height \\
                   & [mm]     & [mm]   & [mm]     & [mm]   \\           \hline
  \T
            Cd (1st campaign) & 10 & 6  & 8 & 4 \\
            BN (1st campaign) & 13 & 14 & 5 & 6 \\
            BN (2nd campaign) & 13 & 14 & 5 & 5 \\
            B$_4$C (2nd campaign) & 13 & 15 & 5 & 5 \\
\T          $^{10}$B$_4$C (2nd campaign) & \multirow{2}{*}{13} & \multirow{2}{*}{13} & \multirow{2}{*}{5} & \multirow{2}{*}{5} \\
\T          ($^{10}$B enrichment $>$ 96~\%) & & & \\
            \hline \hline
        \end{tabular}
\end{table}

\begin{figure}[!htbp]
\vspace{-2mm}
  \centering
\includegraphics[width=\columnwidth]{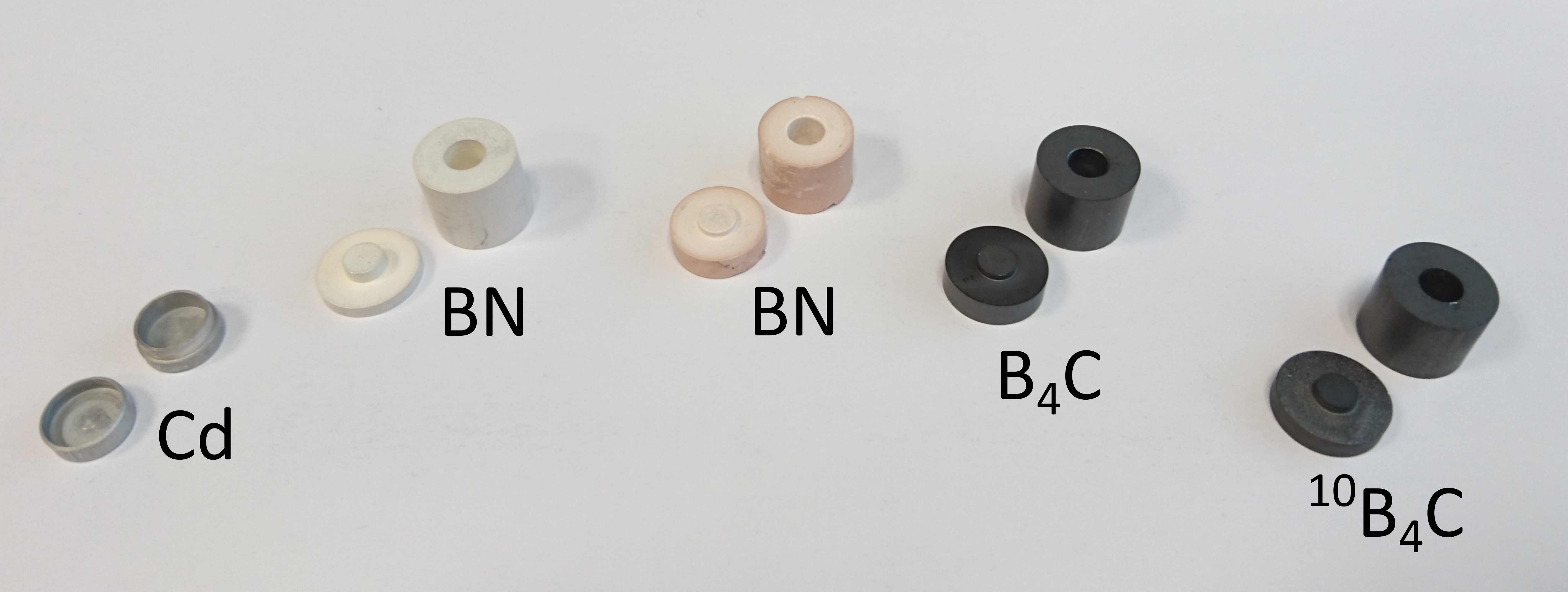}
    \caption{(Color online) Neutron filters used in the experiments at the JSI TRIGA reactor. From left to right: Cd (1st campaign), BN (1st campaign), BN (2nd campaign), B$_4$C (2nd campaign), $^{10}$B$_4$C (2nd campaign).}
    \label{Filter_photo}
\vspace{-2mm}
\end{figure}
\begin{figure}[!htbp]
\vspace{-2mm}
  \centering
  \includegraphics[width=\columnwidth]{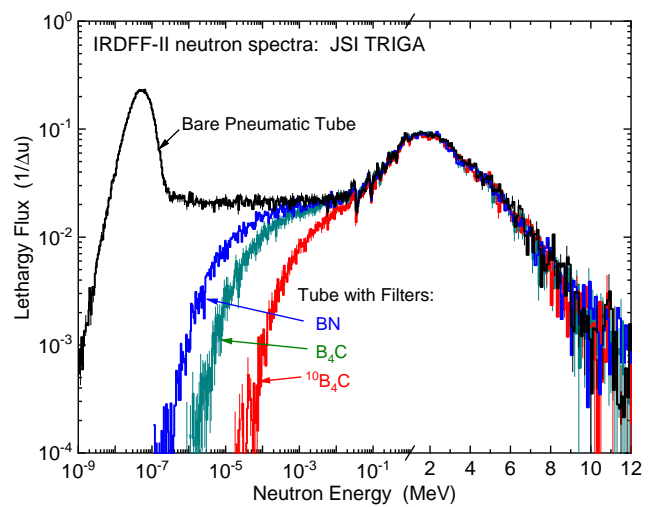}
\vspace{-2mm}
    \caption{(Color online) Unfiltered and filtered neutron fields in the TRIGA-JSI reactor. Note that the scale changes from log to linear above 1~MeV.}
    \label{TRIGA-JSI_Spectra}
\vspace{-4mm}
\end{figure}

The measured quantities in the experiments were reaction rate ratios, either unfiltered relative to filtered, or one reaction rate relative to another, derived from activation measurements with gamma spectrometry. Irradiations were performed in two experimental campaigns in the TRIGA-JSI reactor, both in the Pneumatic Tube (PT) irradiation channel. In the first campaign, cadmium (Cd) and boron nitride (BN) filters were employed. The objectives of the campaign were two: firstly the characterization of the neutron spectrum in the PT irradiation channel, and secondly, to investigate the suitability of boron nitride filters to shift the energy sensitivity of radiative capture reactions to the epithermal energy range, for cross-section validation \cite{Radulovic_NIMA-2016}. In the second campaign, boron nitride (BN), boron carbide (B$_4$C) and enriched boron carbide ($^{10}$B$_4$C) filters were employed. The objective of the campaign was to investigate new candidate radiative capture reactions, which are sensitive to the epithermal energy range (broadly between 1 keV and 1 MeV) under boron filters~\cite{Radulovic_NENE2018}, which could be applicable for epithermal neutron dosimetry. The filter dimensions are reported in Table~\ref{Filter_dimensions}. Fig.~\ref{Filter_photo} displays the filters. Fig.~\ref{TRIGA-JSI_Spectra} shows the spectra with and without covers.

\subsubsection{ISNF}\label{sssect:ISNF}
The ISNF facility is set up in the spherical cavity inside the graphite column of the reactor. Eight enriched $^{235}$U fission discs, symmetrically mounted in the cavity, convert the thermal neutron flux with an average energy $\mathrm{<}$E$\mathrm{>}$ = 0.0253 eV to the fast fission spectrum. The summary flux is transmitted by the enriched $^{10}$B shell and it produces the standard' neutron field at the centre of the cavity.

The effective spectrum range E$_{5~\%}$ - E$_{95~\%}$, where the fraction of spectrum equals 5~\% and 95~\% of the total neutron fluence, lies between 8 keV and 4.1 MeV, an averaged energy $\mathrm{<}$E$\mathrm{>}$ equals to 970 keV. The spectrum-averaged cross sections were measured at ISFN assembly operated at the research reactor of the U.S. National Bureau of Standards (Gaithersburg) \cite{Gru78, Oli84, Lam88}.

\begin{figure}[!htbp]
\vspace{-2mm}
   \includegraphics[width=\columnwidth]{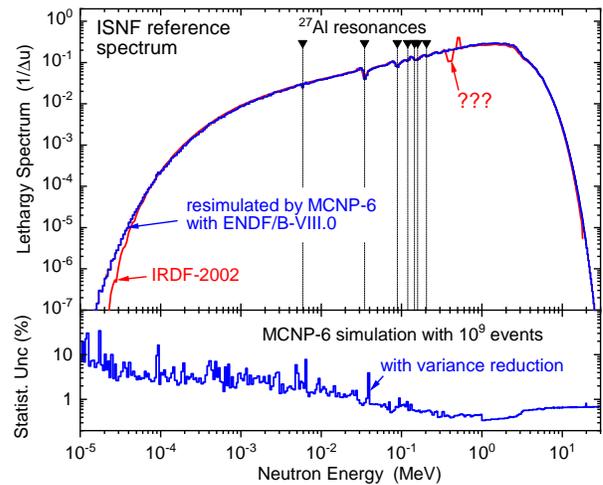}
\vspace{-2mm}
   \caption{(Color online) The ISNF reference spectrum: original from IRDF-2002 (red curve, the question marks point to unphysical oscillations) and recalculated in this work (blue). Arrows with dashed lines indicate the resonance energies of the Al total cross sections. The bottom part shows the statistical uncertainties obtained during the Monte Carlo simulation with variance reduction techniques.}
   \label{fig:ISNF_spect}
\vspace{-2mm}
\end{figure}

\begin{figure}[!htbp]
\vspace{-2mm}
   \includegraphics[width=\columnwidth]{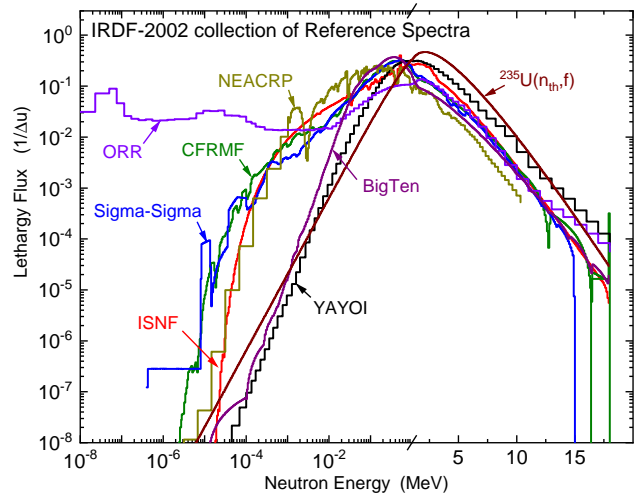}
   \caption{(Color online) Neutron reference reactor spectra available in the IRDF-2002 collection~\cite{Spectra02}. Note that the scale changes from log to linear above 1~MeV.}
   \label{fig:IRDF2002_spect}
\vspace{-2mm}
\end{figure}

The ISNF spectrum available in IRDF-2002, as seen in Figs.~\ref{fig:ISNF_spect} and \ref{fig:IRDF2002_spect}, has obviously unphysical irregularities in the vicinity of 0.4 MeV marked with question marks in Fig.~\ref{fig:ISNF_spect}. Failing to find the original data, we went back to the available description of the facility \cite{Gru78}. The detailed information documented there gave us an opportunity to conduct a re-analysis of the spectrum using the MCNP-6 code with neutron transport cross sections including $^{235}$U(n$_{th}$,f) PFNS from the ENDF/B-VIII.0 library. The calculated neutron energy distribution within the irradiation volume is shown in Fig.~\ref{fig:ISNF_spect}. This figure shows that the recalculated energy distribution is very close to the ISNF spectrum representation given in the IRDF-2002 library. It indeed exhibits resonances caused by the Al cap of the boron shell, but no longer exhibits the oscillations in the vicinity of 0.4~MeV.

This re-simulated ISFN spectrum, calculated in the 725-groups presentation, is used to compute the SACS and to support the validation of the \mbox{IRDFF-II} cross sections (see Sec.~\ref{Sec_VII_L} for details). The indicated numerical uncertainties only capture the statistical uncertainty of the Monte Carlo simulation and do not represent the complete spectrum uncertainty. To reach a similar numerical uncertainty over the whole neutron energy range, at the 1~\% level, variance reduction techniques were used. The impact on statistical uncertainty of the variance reduction is shown in the bottom panel of Fig.~\ref{fig:ISNF_spect}.

\subsubsection{Sigma--Sigma}\label{sssect:Sigma}
Sigma-Sigma is a thermal-fast coupled spherical source assembly located within a conventional graphite thermal column of fission reactor. The column is a source of thermal neutrons that produce the fast fission spectrum in the spherical natural uranium shell (thermal-fast convertor). The inner boron carbide shell shapes the neutron low-energy tail in such a way to increase the reaction rates within an appropriate energy range.

The effective spectrum range, E$_{5~\%}$ - E$_{95~\%}$ fluence metric, is 20 keV -- 3.7 MeV, with an averaged energy $\mathrm{<}$E$\mathrm{>}$ = 730 keV. The Sigma-Sigma assemblies were operated at the Imperial College of Science and Technology (London) \cite{Bes73, Han78}, at the Belgian Nuclear Research Centre SCK$\bullet$CEN, in Mol (Belgium) \cite{Fab75, Fab78}, and at the Institute for Nuclear Technology (Bucharest) in cooperation with SCK$\bullet$CEN \cite{Gar78a, Gar81}.

\begin{figure}[htbp]
   \includegraphics[width=\columnwidth]{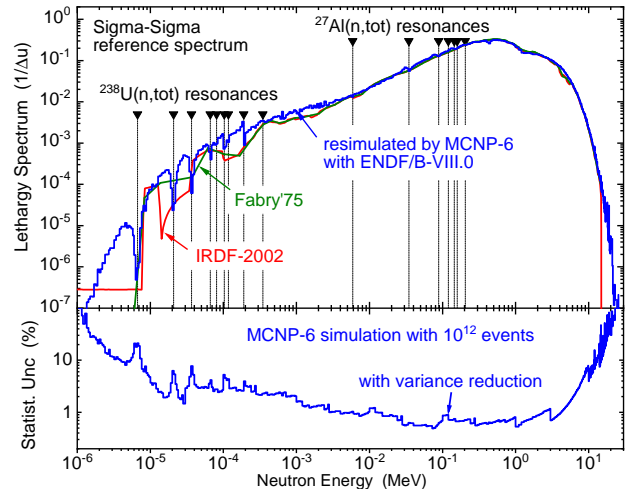}
   \caption{(Color online) The Sigma-Sigma reference spectrum: original from IRDF-2002 (red curve), published by Fabry~\cite{Fab75} (green) and recalculated in this work (blue). The bottom part shows the statistical uncertainties obtained in the Monte-Carlo simulations with variance reduction techniques.}
   \label{fig:SigSig_spect}
\end{figure}

The Sigma-Sigma reference spectrum available in \mbox{IRDF-2002} exhibits a suspicious spectrum shape below 10 keV, as seen in Fig.~\ref{fig:IRDF2002_spect}. The paper of A.~Fabry and co-workers in 1975~\cite{Fab75} provided a tabulated spectrum which was based on the measurement and discrete ordinate calculation with nuclear data available at that time. Comparison of these two spectra shown in Fig.~\ref{fig:SigSig_spect} confirms the unreliability of the \mbox{IRDF-2002} data below $\mathrm{\sim}$10 keV.

Fortunately, several papers \cite{Fab75, Fab71, Oli82} have documented detailed specifications of the Sigma-Sigma assemblies. Using this information, we have constructed an MCNP model and re-simulated the neutron spectrum for this facility while employing the modern neutron fission and transport data from the \mbox{ENDF/B-VIII.0} library. The computed spectrum  is close to the original one by Fabry and the \mbox{IRDF-2002} spectrum over most of the energy range. However, substantial differences are observed below 1~keV, the region where the resonances of the $^{238}$U(n,tot) cross sections produce a fine structure in the spectrum. The newly simulated Sigma-Sigma spectrum was used for the validation of \mbox{IRDFF-II} cross sections, as reported in Sec.~\ref{Sec_VII_L}. The indicated uncertainties only represent the numerical statistical contributions and have been reduced to a few percent by applying variance reduction techniques.

\subsubsection{FNS--Graphite Neutron Field with a D-T Source}
\begin{figure}[htbp]
\vspace{-2mm}
   \includegraphics[width=\columnwidth]{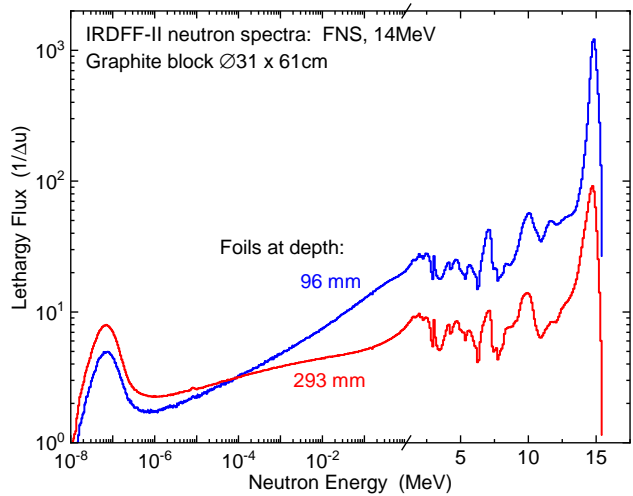}
   \caption{(Color online) The FNS--Graphite reference spectrum from a D-T neutron source at 96~mm and 293~mm depth in the graphite block. Note that the scale changes from log to linear above 1~MeV.}
   \label{FNS-Graphite}
\vspace{-2mm}
\end{figure}

An experiment was performed at the JAEA-FNS facility in Tokai-mura, Ibaraki-ken, Japan to measure spectral indices in a graphite block irradiated with a D-T source. Reactor-grade graphite blocks were stacked to form a pseudo-cylindrical assembly of 31.4 cm in equivalent radius and 61.0 cm in thickness. The average density of the graphite blocks was 1.64 g/cm3. The graphite assembly was placed at a distance of 20.3 cm from the DT neutron source. Several foils for the reaction rate measurement of forty dosimetry reactions were inserted into small spaces between the graphite blocks along the central axis of the assembly; the reaction rates of the $^{93}$Nb(n,2n)$^{92m}$Nb, $^{115}$In(n,n')$^{115m}$In and $^{197}$Au(n,$\gamma$)$^{198}$Au reactions were measured at four points (9.6, 19.3, 29.3 and 39.5 cm in depth) in the graphite assembly in order to check the reproducibility of the previous in situ experiment on graphite, while those of the other reactions were measured at two points (9.6 and 29.3 cm in depth). The plot of the spectra corresponding to the two depths in graphite are shown in Fig.~\ref{FNS-Graphite}.

\subsubsection{Godiva Neutron Field from the ICSBEP Handbook}
\begin{figure}[!htbp]
        \centering
        \includegraphics[width=\columnwidth]{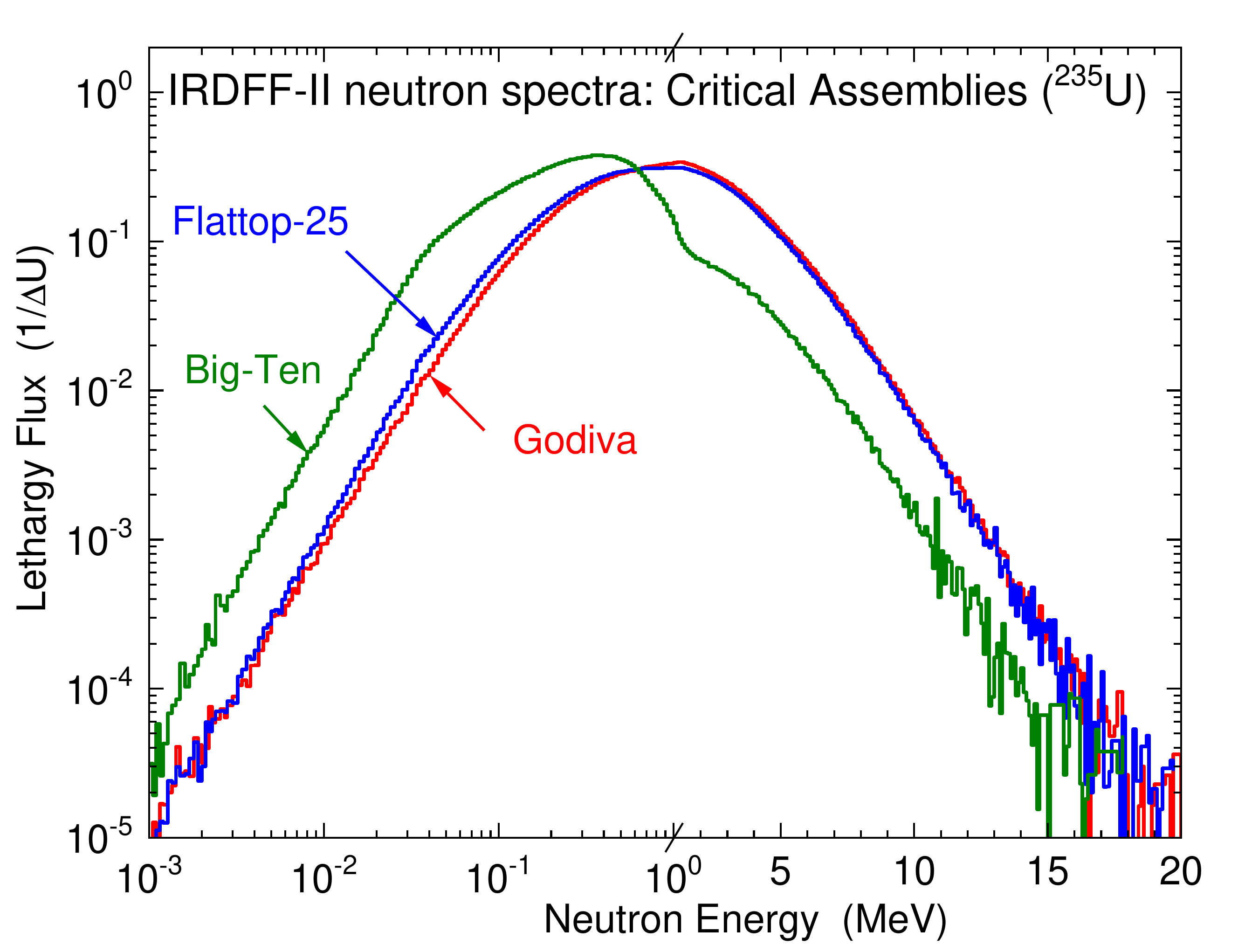}
        \caption{(Color online) Neutron spectra for  several critical assemblies containing  $^{235}$U from the ICSBEP Handbook~\cite{ICSBEP:2016}: Godiva, Flattop-25 and Big-Ten. The spectra are presented in the 725 energy groups and normalized to unity. Note that the scale changes from log to linear above 1~MeV.}
        \label{Crits_Spect_U}
    \end{figure}

The Los Alamos National Laboratory HEU-MET-FAST-001 (\ie, Godiva) critical assembly~\cite{LaB02} consists of a number of discrete highly-enriched uranium metal pieces that produce a near spherical assembly when joined into a critical configuration.  This assembly is modelled as a bare sphere with a 8.7404~cm radius.  Reaction rate measurements were made at the assembly centre using a combination of fission counters and foils.  Flux and reaction rate tallies were extracted from a detailed MCNP~6.2~\cite{MCNP} calculation.  The tally region consists of a 0.5~cm diameter sphere centered within the critical assembly and the calculated flux and reaction rates result from a 10 billion neutron history job (50 warm-up cycles followed by 4~000 active cycles with 2~500~000 neutrons per cycle).  Cross sections for the criticality transport calculation are from \mbox{ENDF/B-VIII.0}~\cite{ENDF8} using 293.6~K ACE files generated by the MCNP Data Team~\cite{Con18}.  Godiva's calculated flux spectrum for this central region is shown in Fig.~\ref{Crits_Spect_U}.

\subsubsection{Flattop-25 Neutron Field from the ICSBEP Handbook}
The Los Alamos National Laboratory HEU-MET-FAST-028 (\ie, Flattop-25) critical assembly~\cite{Bre99} consists of a highly-enriched uranium metal central core surrounded by a large natural uranium metal reflector.  When assembled in a critical configuration the core and reflector are nearly spherical in shape.  Hence this assembly is modelled as a bare sphere of highly-enriched uranium surrounded by a large, spherical natural uranium reflector.  The central core radius is 6.1156 cm while the spherical reflector radius is 18.0086 cm.  There is no gap between the central core and the reflector in the computer model.  Reaction rate measurements were made using fission counters and foils at or near the core centre with selected measurements performed at various radial distances from the core centre extending into the reflector.  In this report we focus on the central core.  The tally region consists of a 0.5 cm diameter sphere centered within the critical assembly and the calculated flux and reaction rates result from a 5 billion neutron history MCNP~6.2 job (50 warm-up cycles followed by 2 000 active cycles with 2 500 000 neutrons per cycle).  Cross sections for the kcode transport job are from \mbox{ENDF/B-VIII.0}~\cite{ENDF8} using 293.6~K ACE files generated by the MCNP Data Team~\cite{Con18}.  The Flattop-25 central region multigroup calculated flux spectrum is shown in Fig.~\ref{Crits_Spect_U}.

\subsubsection{Big-Ten Neutron Field from the ICSBEP Handbook}\label{sssect:BigTen}
The Los Alamos National Laboratory IEU-MET-FAST-007 (\ie, Big-10) critical assembly~\cite{Sap99} is a large cylindrical assembly consisting of a heterogeneous mix of solid and annular uranium metal plates with enrichments vary from \ql depleted\qr to highly-enriched.  The assembly has a central region consisting of 10~\% enriched uranium where reaction rate and flux measurements were made.  The calculational model incorporates the uranium metal plates so that the various \ql depleted"\qrs, normal uranium, 10~\% enriched uranium and highly-enriched uranium regions are accurately portrayed.  For calculated flux and reaction rate tallies a small, 0.5 cm diameter sphere is situated within the 10~\% enriched region where the fission counters and foils were positioned.  The MCNP~6.2 calculation was run for 5 billion active histories (50 warm-up cycles followed by 2 000 active cycles with 2 500 000 neutrons per cycle).  Cross sections for the kcode transport job are from \mbox{ENDF/B-VIII.0}~\cite{ENDF8} using 293.6~K ACE files generated by the MCNP Data Team~\cite{Con18}.  The calculated multigroup flux spectrum within this tally sphere is shown in Fig.~\ref{Crits_Spect_U}.

\subsubsection{CFRMF}\label{sssect:CFRMF}
The Coupled Fast Reactivity Measurements Facility (CFRMF) is located at Idaho National Laboratory (INL) in Idaho Falls, Idaho, U.S.A. The CFRMF is a zoned core critical assembly with a fast-neutron spectrum zone in the centre of an enriched $^{235}$U and $^{10}$B annulus inside the uranium block \cite{Rog75, Rog77, Rog78}. The water moderated thermal driver zone are conventional plate-type fuel elements of fully enriched uranium clad in aluminum.

\begin{figure}[htbp]
   \includegraphics[width=\columnwidth]{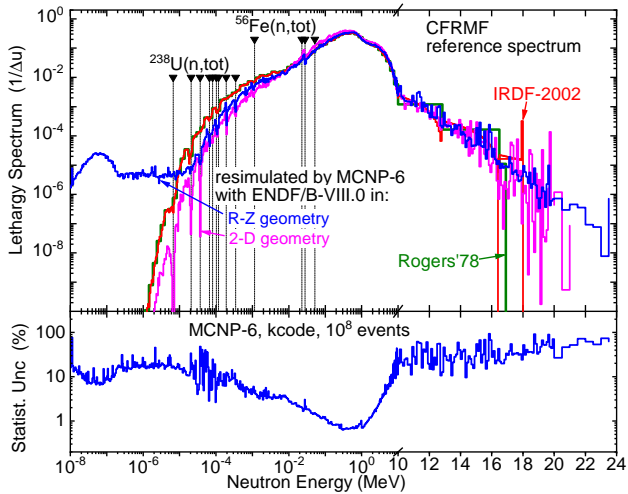}
   \caption{(Color online) The CFRMF reference spectrum: original from IRDF-2002 (red curve), published by Rogers~\cite{Rog78} (green) and recalculated in this work using 2-D (pink) and R-Z (blue) models. The bottom part shows the statistical uncertainties obtained in the Monte-Carlo criticality simulations with variance reduction. Note that the scale changes from log to linear above 1~MeV.}
   \label{fig:CFRMF_spect}
\end{figure}

In the past there was a continuing effort to improve the characterization of the CFRMF spectrum both through calculation and spectrometry measurements. Different deterministic and Monte-Carlo radiation transport computer codes have been employed in the calculations. J.W. Rogers and co-workers \cite{Rog78} have published the results of a one-dimensional calculation with S6/P1 approximation and 71-group average cross sections from ENDF/B-IV covering the energy range 16 MeV to epithermal. This spectrum, as seen in Fig.~\ref{fig:CFRMF_spect}, agrees with the spectrum assigned for CFRMF in the IRDF-2002 which is represented there at 461 energy points. Exceptions include three points that deviate from the smooth shape above 10 MeV as seen in Fig.~\ref{fig:IRDF2002_spect}.

The one-dimensional model of the CRFMF facility is documented in a published paper \cite{Rog75}. We used this model to re-calculate the CFRMF critical assembly by the MCNP-6 code and employing up-to-date data from ENDF/B-VIII.0. As seen in  Fig.~\ref{fig:CFRMF_spect}, the newly calculated spectrum demonstrates more fine structure resulting from the strongest resonances in the neutron total cross section for $^{235}$U and $^{56}$Fe (component of stainless steel claddings). However, our calculation is lower than the IRDF-2002 and Roger \cite{Rog78} spectra for energies below $\mathrm{\sim}$10 keV. Another and more realistic model in the R-Z geometry was published in \cite{Rog78}. It resulted in a representation of the CFRMF spectrum that is very similar to the cylindrical representation, shown in Fig.~\ref{fig:CFRMF_spect}, but additionally brings in the thermal and epithermal neutrons, which penetrate through the bottom and top ends of the access tube that accommodates the activation samples.

Since the reason for this systematic difference is not clear at this time, we are investigating refinements in the model geometry and have continued to use use the IRDF-2002 spectrum (after correction of values at energies 12.7, 16.3 and 17.9 MeV) for the calculation of the spectrum-averaged cross sections. Calculation of the spectrum with this model predicts that the 5~\% and 95~\% limits are between 15~keV and 2.8~MeV and its average energy $\mathrm{<} E \mathrm{>}$ equals 741~keV.

\subsubsection{IPPE--BR1 Neutron Field from the ICSBEP Handbook}
The Institute of Physics and Power Engineering FUND-IPPE-FR-MULT-RRR-001 is a Pu metal fuelled critical assembly~\cite{Kho09} located in Obninsk, Russia. The core consists of (i) an empty centre tube where foils to be irradiated are located, (ii) 77 Pu metal rods are set on a hexagonal pitch, (iii) 31 peripheral copper metal rods to fill out the inner core region, (iv) annular uranium metal cylinders used for reactivity control and as radial reflectors, and (v) axial reflectors of uranium metal (lower) and copper metal (upper).  The active fuel region is 13.1~cm tall.  The MCNP~6.2 model includes a tally region at the core mid-plane in the voided centre rod position.  This tally region is 1~cm tall.  The MCNP~6.2 model was run for 5 billion active histories (50 warm-up cycles followed by 2~000 active cycles with 2~500~000 neutrons per cycle).  Cross sections for the kcode transport job are from \mbox{ENDF/B-VIII.0}~\cite{ENDF8} using 293.6~K ACE files generated by the MCNP Data Team~\cite{Con18}.  The calculated multigroup flux spectrum within this tally sphere is shown in Fig.~\ref{Crits_Spect_Pu}.

\subsection{Potential Reference Neutron Fields from Detailed Computational Models}\label{Sec_VI_D}
The neutron fields for several well-known Pu-fuelled critical assemblies are shown in this section; namely the LANL \ql Jezebel\qr (a bare, mostly $^{239}$Pu, sphere), \ql Jezebel-240\qr (a bare Pu sphere with higher $^{240}$Pu content), \ql Flattop-Pu\qr (a Jezebel-like Pu spherical core reflected by natural Uranium) and \ql Thor\qr (a Jezebel-like Pu sphere reflected by Thorium).  Criticality calculations yield accurate $k_{\mbox{eff}}$ predictions but as will be shown in Sec.~\ref{Sec_VII_K} the reaction rate C/E results exhibit trends not seen in the Uranium fuelled assemblies.  This suggests that some aspect of these Pu-fuelled assemblies, perhaps the $^{239}$Pu PFNS, has unrecognized deficiencies.  Hence while we present our calculations for these assemblies, the reader is cautioned that these neutron fields are not judged to be of benchmark quality at this time.

\subsubsection{Jezebel Neutron Field from the ICSBEP Handbook}
The Los Alamos National Laboratory PU-MET-FAST-001 (\ie, Jezebel) critical assembly~\cite{Fav16} consists of a number of discrete Pu metal pieces that, when assembled into a critical configuration, is nearly spherical in shape.  The Pu metal is greater than 95~\% $^{239}$Pu with an approximate 1~\% Ga alloy for phase stabilization.  This benchmark evaluation has undergone significant revisions in recent years which allow for the development of complex computer models representing the individual Pu metal pieces.  In addition, the simpler legacy model, a simple sphere with one material has been retained.  This simple model, with its updated core radius and material data, is used here.  Although the core model is 6.39157~cm in radius and the flux and reaction rate tallies are obtained for a 0.5~cm diameter spherical tally region located at the core centre.  The MCNP~6.2 calculation was run for 25 billion active histories (50 warm-up cycles followed by 10 000 active cycles with 2 500 000 neutrons per cycle).  Cross sections for the kcode transport job are from \mbox{ENDF/B-VIII.0}~\cite{ENDF8} using 293.6~K ACE files generated by the MCNP Data Team~\cite{Con18}.  The calculated multigroup flux spectrum within this tally sphere is shown in Fig.~\ref{Crits_Spect_Pu}.

\subsubsection{Jezebel-240 Neutron Field from the ICSBEP Handbook}
\begin{figure}[htbp]
        \centering
        \includegraphics[width=\columnwidth]{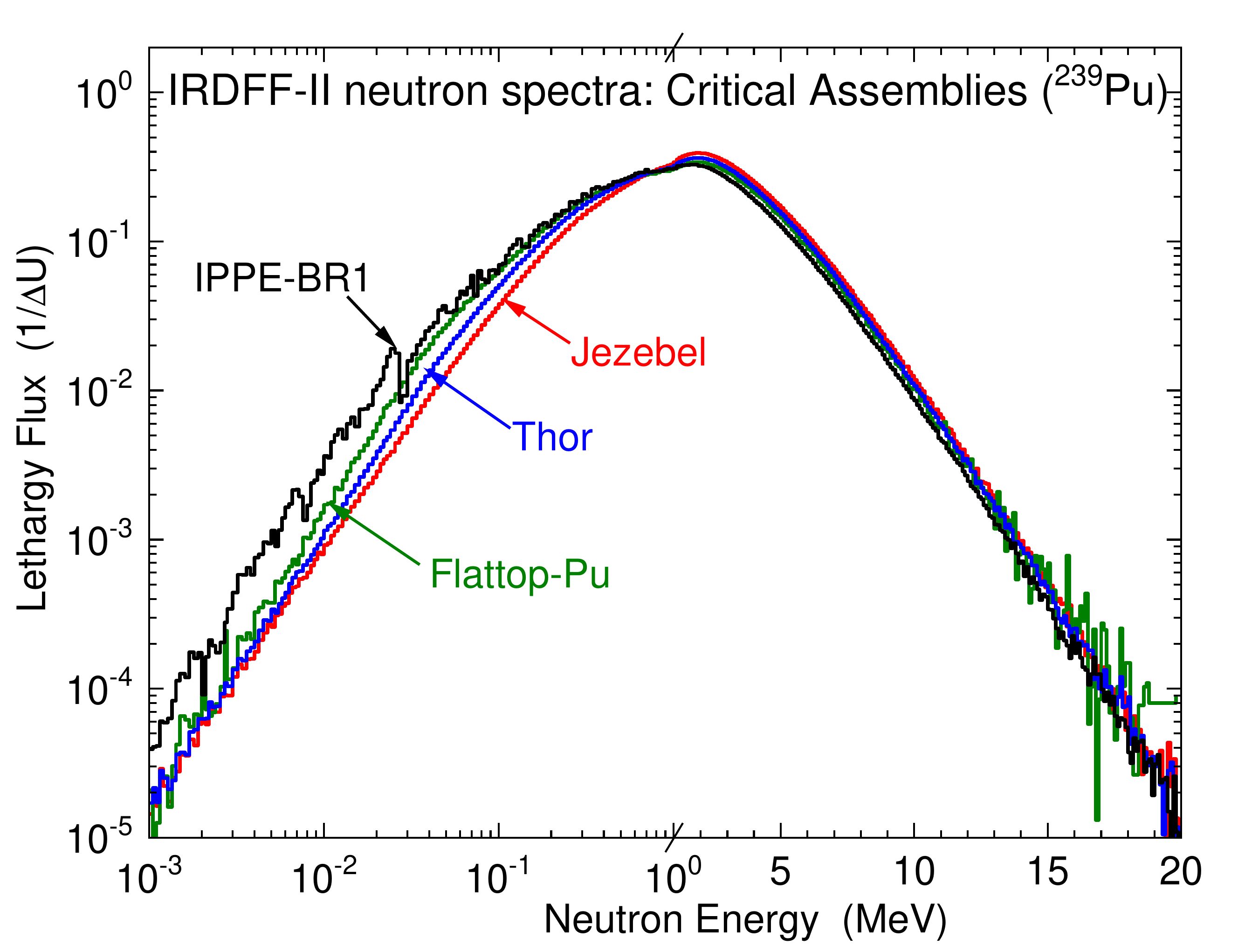}
        \caption{(Color online) Neutron spectra for  several critical assemblies containing  $^{239}Pu$ from the ICSBEP Handbook~\cite{ICSBEP:2016}: Jezebel, Flattop-Pu, Thor and IPPE-BR-1. The spectra are presented in the 725 energy groups and normalized to unity. Note that the scale changes from log to linear above 1~MeV.}
        \label{Crits_Spect_Pu}
    \end{figure}

The Los Alamos National Laboratory PU-MET-FAST-002 (\ie, Jezebel-240) critical assembly~\cite{Del99} is similar to the PMF001 assembly described above; the primary difference being in the Pu metal isotopic content.  Whereas Jezebel is $>$95~\% $^{239}$Pu, the Jezebel-240 assembly is a little less than 80~\% $^{239}$Pu, approximately 20~\% $^{240}$Pu and includes smaller amounts of $^{241}$Pu and $^{242}$Pu.  This assembly is modelled as a bare sphere 6.6595~cm in radius.  Once again a small, 0.5~cm diameter, tally region is defined at the core centre where the various measurements were made.  The MCNP~6.2 calculation was run for 10 billion active histories (50 warm-up cycles followed by 4~000 active cycles with 2~500~000 neutrons per cycle).  Cross sections for the kcode transport job are from \mbox{ENDF/B-VIII.0}~\cite{ENDF8} using 293.6~K ACE files generated by the MCNP Data Team~\cite{Con18}.  The calculated multigroup flux spectrum within this tally sphere is very similar to the one of Jezebel in Fig.~\ref{Crits_Spect_Pu} and is not shown explicitly.

\subsubsection{Flattop-Pu Neutron Field from the ICSBEP Handbook}
The Los Alamos National Laboratory PU-MET-FAST-006 (\ie, Flattop-Pu) critical assembly~\cite{Bre99_1} is similar to the Flattop-25 assembly described previously with the exception that the central core is Pu metal rather than highly enriched uranium.  The spherical Pu core radius is 4.5332~cm and the outer reflector radius is 24.1420~cm.  As with Flattop-25 there is no gap between the central core and the reflector in the computer model.  The central core flux and reaction rate tallies are obtained for a 0.5~cm diameter sphere at the located at the centre of the core.  An MCNP~6.2 calculation was run for 1 billion histories (50 warm-up cycles followed by 400 active cycles with 2~500~000 neutrons per cycle).  Cross sections for the kcode transport job are from \mbox{ENDF/B-VIII.0}~\cite{ENDF8} using 293.6~K ACE files generated by the MCNP Data Team~\cite{Con18}.  The calculated multigroup flux spectrum within this tally sphere is shown in Fig.~\ref{Crits_Spect_Pu}.

\subsubsection{THOR Neutron Field from the ICSBEP Handbook}
The Los Alamos National Laboratory PU-MET-FAST-008 (\ie, THOR) critical assembly~\cite{Bre99_2} is similar to the Flattop-Pu assembly described previously with the exception that the heavy metal reflector is a cylinder of Thorium metal.  The spherical Pu core radius is 5.310~cm and centered in the Thorium cylinder whose diameter and height are both 53.34~cm.  Once again there is no gap between the core and reflector in the computer model.  The MCNP~6.2 job includes a small, 0.5~cm diameter sphere in the centre of the core for flux and reaction rate tallies and was run for 5 billion active histories (50 warm-up cycles followed by 2~000 active cycles with 2~500~000 neutrons per cycle).  Cross sections for the kcode transport job are from \mbox{ENDF/B-VIII.0}~\cite{ENDF8} using 293.6~K ACE files generated by the MCNP Data Team~\cite{Con18}.  The calculated multigroup flux spectrum within this tally sphere is very similar to the one of Jezebel in Fig.~\ref{Crits_Spect_Pu}.

\subsection{Legacy Reference Neutron Fields}\label{Sec_VI_E}

The reference spectra incorporated within the \mbox{IRDFF-II} library build upon the spectra collected in earlier dosimetry libraries. The reference neutrons fields generated at the research reactors, and frequently used for measurement of the spectrum-averaged dosimetry cross sections, were collected and documented in the IRDF-2002 database ~\cite{Spectra02}. Some of these reference spectra were transferred over from the IRDF-90 ~\cite{IRDF90}. The IRDF-2002 library included 9 spectra. The spectra, and associated references are:
\begin{itemize}
 \item two $^{\ 235}$U thermal induced fission spectra, one from NBS (J. Grundl, private communication), second -- from the ENDF/B-V evaluation;
 \item Intermediate-energy Standard Neutron Field (ISNF) \cite{Gru78};
 \item Coupled Fast Reactivity Measurement Facility (CFRMF) \cite{Gru78};
 \item 10~\% enriched Uranium critical assembly (Big-Ten) \cite{Gru78};
 \item Coupled thermal/fast uranium and boron carbide spherical assembly (Sigma-Sigma) \cite{Gru78};
 \item Oak Ridge Research Reactor (ORR) [L. Greenwood, private communication];
 \item YAYOI reactor [L. Greenwood, private communication];
 \item Central zone flux of the NEACRP Benchmarks [B. Goal, private communication].
\end{itemize}

No uncertainties or associated correlation matrices for these spectra were provided in IRDF-2002. In 2005, the $^{252}$Cf spontaneous fission neutron spectrum and its covariance matrix, evaluated by H. Mannhart~\cite{Man08}, were added to IRDF-2002.

All compiled IRDF-2002 spectra were displayed in Fig.~\ref{fig:IRDF2002_spect} above.  It is worthwhile to stress that systematic measurements of SACS were carried out and consequently the corresponding data were compiled in the EXFOR database only for the following reference spectra: ISNF, CFRMF, Sigma-Sigma.

Only a subset of these spectra was used for the verification of the \mbox{IRDFF-II} cross sections. Big-Ten critical assembly, ISNF and Sigma--Sigma reactor facilities were re-simulated using the latest ENDF/B-VIII.0 evaluation as described in Sects.~\ref{sssect:ISNF}, \ref{sssect:Sigma}, and \ref{sssect:BigTen}, respectively. The CFRMF spectrum could not be reproduced in full, but because of the large number of measurements in this field the original spectrum provided by the authors for IRDF-2002 was adopted as described in Sec.~\ref{sssect:CFRMF}.

In ORR and YAYOI spectra unexplained structures are observed and the information describing the irradiation set-up is insufficient to develop new models. Therefore those IRDF-2002 spectra were not considered for validation. The Central zone flux of the NEACRP Benchmarks is not a real benchmark field and was discarded.

\subsection{Analytical Reference Neutron Fields}\label{Sec_VI_F}

\subsubsection{Thermal Maxwellian}
The thermal Maxwellian spectrum can be used to calculate spectrum-averaged cross sections $\overline{\sigma}$ in a well-thermalised spectrum. The spectrum corresponds to a temperature of 0.0253~eV and is normalised to $2/\sqrt{\pi}$ so that the averaged cross section can be compared to the $\sigma_{0}$ cross section value at 0.0253~eV, the ratio $\overline{\sigma}/\sigma_{0}$ being the Westcott $g$-factor.

\subsubsection{1/$E$ Spectrum in the Energy Range 0.55~eV to 2~MeV}
From neutron slowing-down theory it can be shown that high-energy neutrons in an ideal non-absorbing moderator with a constant cross section follow a 1/$E$ distribution down to thermal energies. This is a reasonable approximation for a reactor spectrum. Cadmium has a very strong resonance near 0.2~eV. A cadmium cover of thickness 1~mm effectively filters thermal neutrons below 0.55~eV. At high energies the fission spectrum in a reactor falls off rapidly above 2~MeV. The 1/$E$ spectrum defined in the energy range 0.55~eV to 2~MeV is a reasonable approximation for resonance integral calculations in a reactor spectrum under a cadmium cover, which can be compared to measured values by the cadmium ratio method, for example.

\subsubsection{1/$E$ Spectrum in the Energy Range 0.5~eV to 20~MeV}
Some authors prefer different energy boundaries to define the resonance integral. In the Atlas of Resonance Parameters by Mughabghab~\cite{Mug16}, the integration boundaries are from 0.5~eV to 20~MeV. For convenience, both forms of the 1/$E$ neutron spectrum are included in the spectra file of \mbox{IRDFF-II}.

\begin{figure}[!thbp]
\vspace{-3mm}
   \includegraphics[width=\columnwidth]{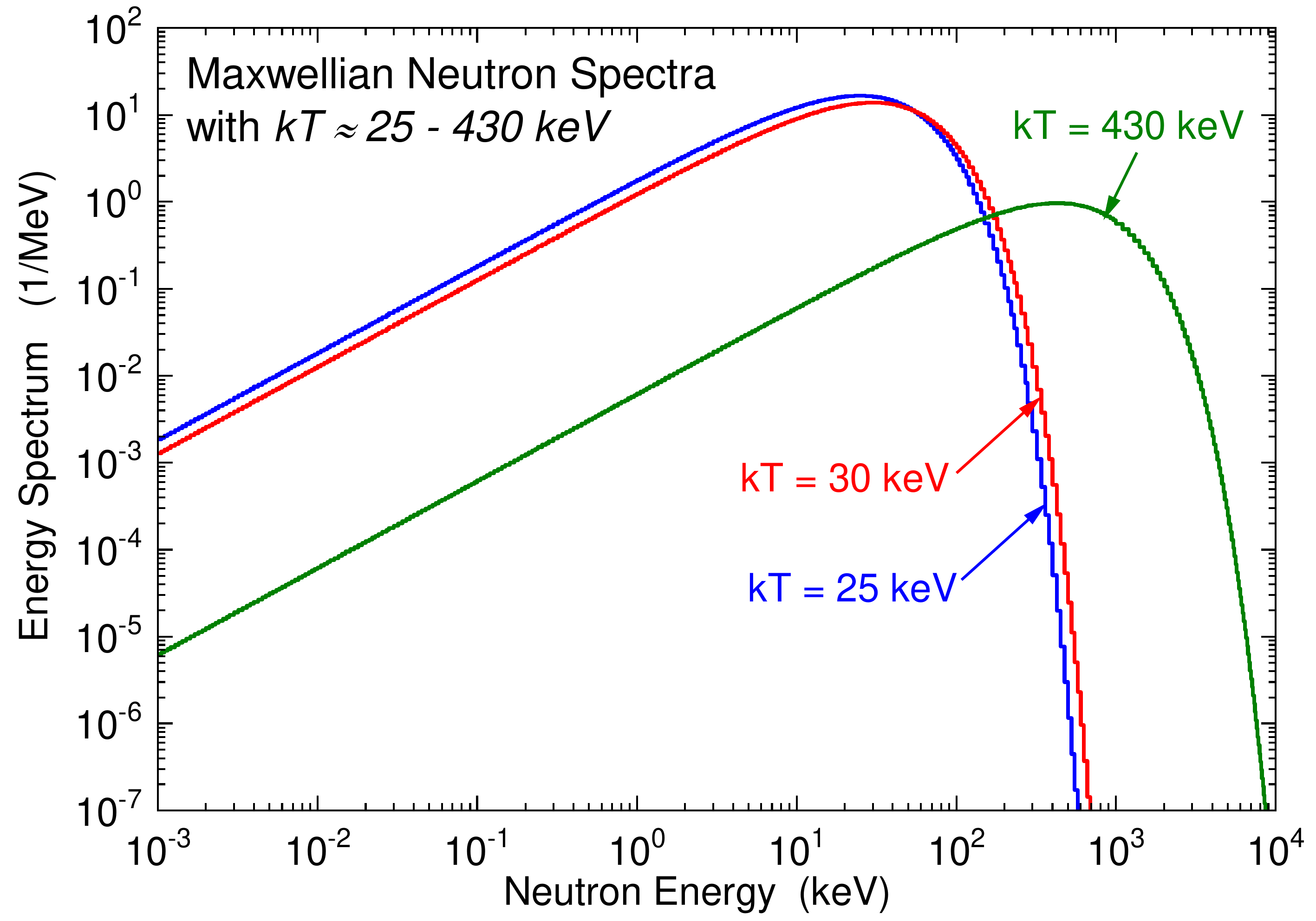}
   \caption{(Color online) Maxwellian spectra with $E_{\mbox{Lab}}$ ranging from 25~keV to 430~keV.}
   \label{fig:iMACS_spect}
\vspace{-3mm}
\end{figure}

\subsubsection{High Temperature Maxwellian}
A set of high temperature quasi-Maxwellian neutron spectra, $kT = E \approx$ 25--60~keV, are frequently used for studying the neutron capture reactions along the astrophysical s-process path \cite{Pau19}. While the ideal spectra are defined by the analytic form, these neutron spectra can be approximately realized using the thick target $^{7}$Li(p,n) reaction for incident proton energies near 2 MeV with modulation of the proton beam. The Maxwellian Averaged Cross Sections (MACS) for activation reactions are typically reported by the astrophysical community, but it must be remembered that the spectrum temperature is expressed in the centre-of-mass (CM) coordinate system, while the incident particle energy is normally given in the Lab system. Fortunately, the conversion is fairly straightforward. It can be shown that:
\begin{equation}
 \overline{\sigma}(E_{\mbox{CM}}) = \overline{\sigma}(E_{\mbox{Lab}} \times (A+1)/A ),
\end{equation}
\noindent where $A$ is the ratio of the target mass to that of a neutron. For example, for hydrogen:
\begin{equation}
 \overline{\sigma}(E_{\mbox{CM}}=30 \mbox{keV}) = \overline{\sigma}(E_{\mbox{Lab}} \approx 60 \mbox{keV} )
\end{equation}

A range of Maxwellian neutron spectra, with $E_{\mbox{Lab}}$ ranging from thermal to 60 keV, are defined within the \mbox{IRDFF-II} library and can be used to support validation activities. Fig.~\ref{fig:iMACS_spect} shows a representative sample of these spectra.

\subsubsection{Constant}
The constant $\phi=1$ is a trivial field that allows simple averages to be calculated with the same tool. Namely, it allows the calculation of flat-spectrum average cross sections or the spectrum integrals when applied to the \mbox{IRDFF-II} library of neutron spectrum fields.

\subsubsection{Linear}
The linear spectrum $\phi=E$ is a trivial field that allows the calculation of the average spectrum energy when applied to a spectrum in the library of spectrum fields in \mbox{ENDF-6} format.

\subsection{Thermal Cross Sections and Resonance Integrals} \label{Sec_VI_G}
For all of the non-threshold reactions, the thermal cross sections and resonance integrals in the \mbox{IRDFF-II} library entries have been compared to the data from the 2016 Atlas of Thermal Neutron Capture Cross Sections Resonance Integrals and Thermal Cross Sections \cite{Mug16}. Since in many cases these data were considered in the development of the underlying nuclear data evaluations, this data comparison constitutes a verification step in the library development.

An independent comparison can be made to the values derived from the Kayzero data-base for neutron activation analysis (NAA) by the k$_{0}$ standardization method.

The Kayzero data-base is a library of nuclear constants that has been developed over many  years and has been extensively tested in practical application for analytical work in certified laboratories. Most of the $k_{0}$ and $Q_{0}$ factors in the library have been measured independently of any differential cross section, so they provide a convenient source of information for the validation of thermal capture cross sections and resonance integrals. The $k_{0}$ factors are related to the cross sections:
\begin{equation}
 k_{0} = \frac{M_{s} \Theta_{a} \gamma_{a} \sigma_{0,a}}
              {M_{a} \Theta_{s} \gamma_{s} \sigma_{0,s}}
\end{equation}
where $M$ is the molar mass, $\Theta$ is the isotopic abundance, $\gamma$ is the gamma-emission probability and $\sigma$ is the cross section. Subscript $a$ stands for the measured nuclide and $s$ for the gold standard.

The $Q_{0}$ factors are simply the ratios of the resonance integral $I_{0}$ in a pure $1/E$ spectrum and the thermal cross section:
\begin{equation}
 Q_{0} = \frac{I_{0}}{\sigma_{0}}
\end{equation}

Comparison of the cross sections and resonance integrals in the \mbox{IRDFF-II} library to the values derived from the Kayzero data-base~\cite{Kayzero} provides an alternative source of validation for the thermal constants and the resonance integrals and augments the comparison with information provided in the Atlas of Resonance Parameters by Mughabghab~\cite{Mug16}.

In this implementation of k$_{0}$-based metrics, the $k_{0}$ factors from the Kayzero library, the natural abundances and the gamma emission probabilities from the \mbox{IRDFF-II} decay data library are combined and used to reconstruct the thermal cross sections and resonance integrals. In principle, the derived values correspond to idealized thermal reactor spectrum with a Maxwellian thermal component and a 1/$E$ epithermal component. The deviation from a 1/$v$ behaviour of the cross sections in the thermal range, deviation from the 1/$E$ spectrum shape in the epithermal range and the non-ideal cadmium filter are taken into account (although not rigorously in all cases), but these are corrections that are usually smaller than the experimental uncertainties. The thermal cross sections and resonance integrals derived from the Kayzero library were taken as the reference for comparison with the Atlas values
%\footnote{Note that in the Atlas the resonance integral $I$ is defined from 0.5~eV up to 20~Mev, while the measured $I$ is usually simulated by the integral from 0.55~eV up to 2~MeV. The difference between these two definitions is expected to be very small for capture reactions.}
and the quantities as extracted from the \mbox{IRDFF-II} library. Results from this validation step are presented in Sec.~\ref{Sec_VII_M}.

\begin{table*}[tb]
\vspace{-3mm}
\caption{Criteria used to evaluate consistency validation evidence based on a least-square adjustment analysis for various reactions.}
\label{tab:X_A_3}
\begin{tabular} { l | l | c c c } \hline\hline
 &  & \multicolumn{3}{c}{Status} \\ \cline{3-5}
\addlinespace[0.4mm]
\# & Metric & \cellcolor{green} Good & \cellcolor{yellow} Acceptable & \cellcolor{lred} Poor \\ \hline
\addlinespace[0.4mm]
1  & C/E & $\leq 2$ std. dev. \cellcolor{green} & $\leq$ 3 std. dev. \cellcolor{yellow} & $>$ 3 std. dev. \cellcolor{lred}\\
\addlinespace[0.4mm]
2  & C/E Interval & within [0.9,1.1]\cellcolor{green} & within [0.8,1.25] \cellcolor{yellow} & outside [0.8,1.25] \cellcolor{lred}\\
\addlinespace[0.4mm]
3  & Experimental Uncertainty & $\leq$ 15~\% \cellcolor{green} & $\leq$ 30~\% \cellcolor{yellow} & $>$ 30~\% \cellcolor{lred}\\
\addlinespace[0.4mm]
4  & Total Uncertainty & $\leq$ 25~\% \cellcolor{green} & $\leq$ 40~\% \cellcolor{yellow} & $>$ 40~\% \cellcolor{lred}\\ \hline\hline
\end{tabular}
\vspace{-2mm}
\end{table*}

\section{VALIDATION DATA -- COMPARISON OF INTEGRAL METRICS IN STANDARD AND REFERENCE NEUTRON BENCHMARK FIELDS} \label{Sec_VII}

One of the most powerful metrics employed in the validation of the \mbox{IRDFF-II} cross sections is the comparison of calculated-to-experimental (C/E) spectrum-averaged cross sections in various standard and reference benchmark neutron fields. Since an absolute determination of the neutron fluence can be difficult to assess for some of the reactor fields, a spectral index (SI), or the ratio of the activity for a given reaction to that for a common monitor reaction, is often used. The C/E ratio for the spectral indices is then the relevant metric used for validation. Another metric that supports validation is a comparison of the consistency of the relative activity for a set of measurements gathered in a given well-characterized neutron field. The following sections address the consideration of these integral metrics in the benchmark neutron fields.

The criteria that were adopted to characterize the quality of the validation evidence based on the comparison of the measured and calculated of the spectrum-averaged cross section are shown in Table~\ref{tab:X_A_3}. A ``good'' validation status requires, as a baseline criterion, that the calculated-to-experimental (C/E) ratio of the spectrum-averaged cross section be within two standard deviations of unity and, in addition, that it lies within the interval of [0.9: 1.1]. In addition to this characteristic, in order to ensure the experimental measurement had adequate fidelity we required that the experimental uncertainty be less than 15~\%. The total uncertainty in the C/E must reflect the root-mean-squared sum of the measurement uncertainty, the spectrum uncertainty, and the cross-section uncertainty. To be sure that the benchmark neutron field was adequately characterized, we also required that the total uncertainty be less than 25~\%.

A ``good'' validation status is color-coded green. An ``acceptable'' validation status is color-coded yellow and reflects a C/E ratio within three standard deviations of unity, a value within the interval [0.8, 1.25], a measurement uncertainty less than 30~\%, and a total uncertainty less than 40~\%. A ``poor'' validation status is color-coded as red and corresponds to a C/E that is more than three standard deviations from unity, or a value that is outside the interval [0.8, 1.25]. A measurement uncertainty greater than 30~\% or a total uncertainty greater than 40~\% does not necessarily reflect a ``poor'' validation state, but it does reflect an absence of a meaningful validation. These fields are left uncolored to reflect unknown status due to the poor quality of available information -- either the measurement uncertainty or the neutron field characterization.

\subsection{$^{252}$Cf SFNS} \label{Sec_VII_A}

\begin{figure}[htbp]
\vspace{-2mm}
   \includegraphics[width=0.99\columnwidth]{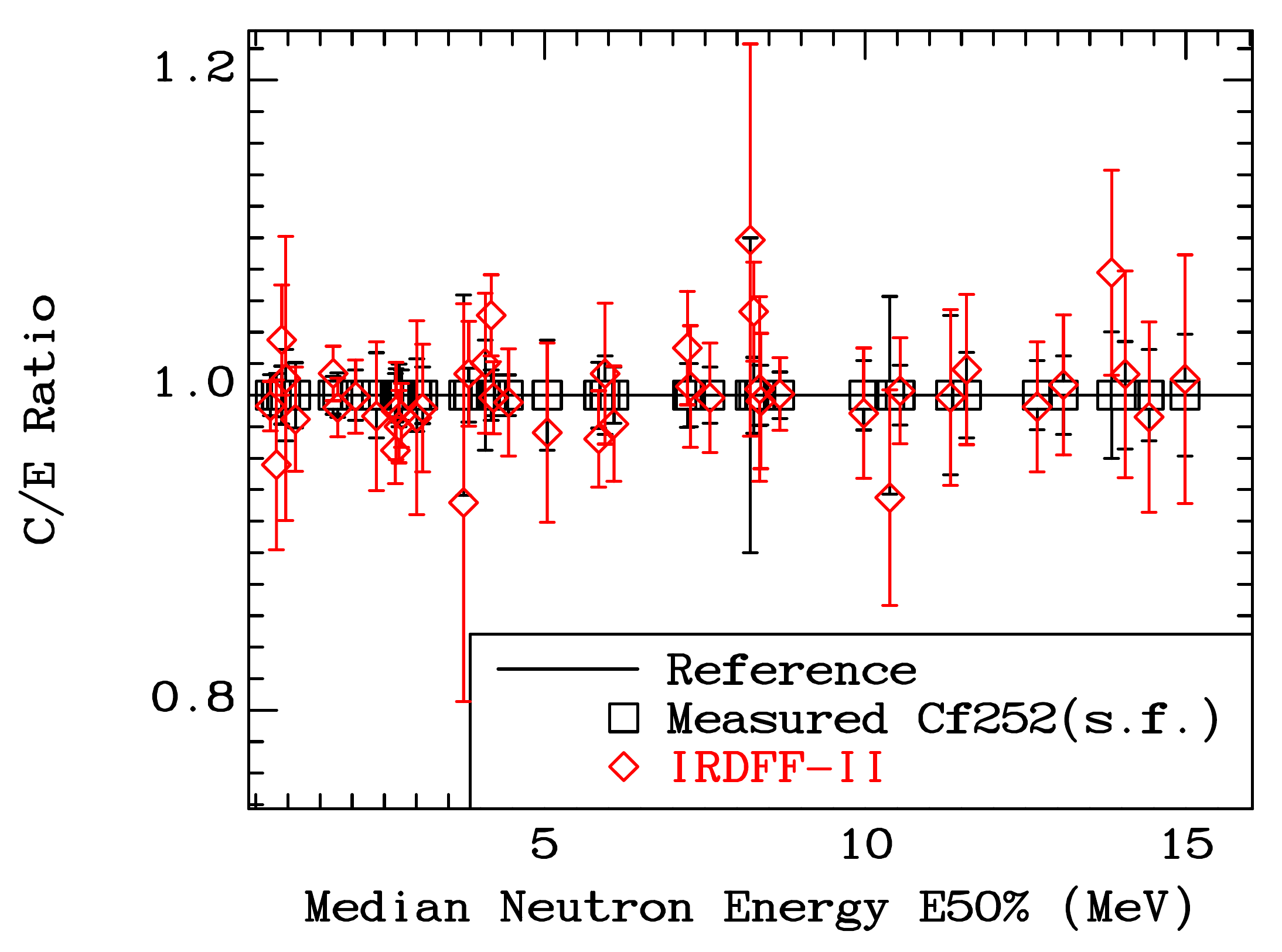}
   \caption{(Color online) C/E relative to median energy  $E_{50}$~\% for the \mbox{IRDFF-II} cross sections averaged in the $^{252}$Cf(s.f.) neutron field.
    %The C/E uncertainty bars reflect a r.m.s. combination of uncertainty contributions from the measurement, \mbox{IRDFF-II} cross sections and the spectrum.
    Plotted values are listed in Table~\ref{tab:TableLongSACScf252}.}
   \label{fig:SACS_Cf252_Mannhart}
\end{figure}

The standard $^{252}$Cf SFNS spectrum was evaluated purely from differential time-of-flight data and is independent of SACS. Therefore, the spectrum can be used to validate SACS (and vice versa).
The SACS in the standard $^{252}$Cf SFNS spectrum were evaluated mostly by Mannhart~\cite{Man06}. It is assumed that all acceptable SACS measurements available at the time were included in the evaluation. The measured values and the C/E values as a function of representative neutron energy $E_{50\%}$ (energy at which the reaction rate integral reaches 50~\% of the integral over the full energy range) are shown in Table~\ref{tab:TableLongSACScf252} and Fig.~\ref{fig:SACS_Cf252_Mannhart}.
\begin{figure}[htbp]
\vspace{-2mm}
   \includegraphics[width=0.99\columnwidth]{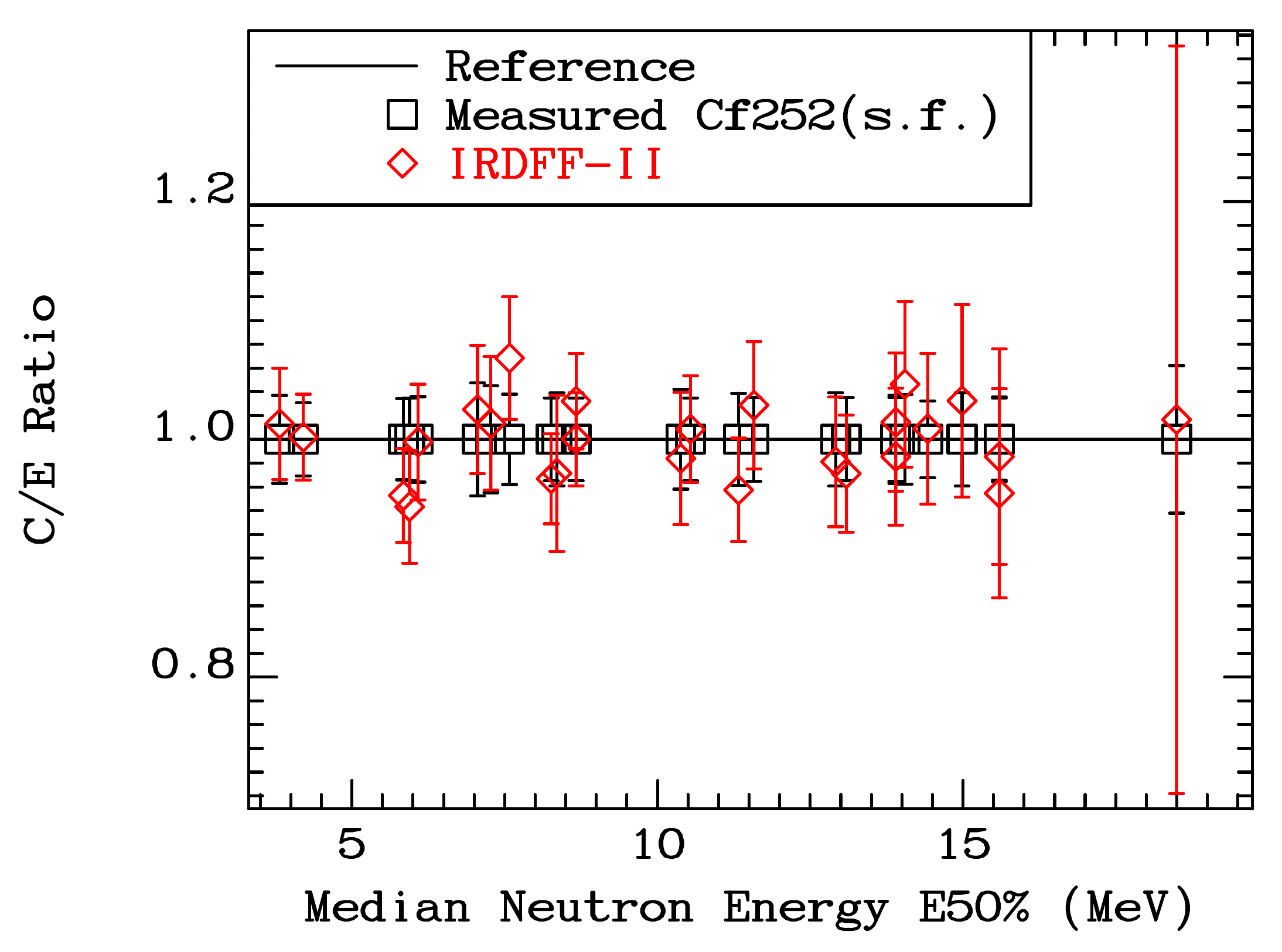}
   \caption{(Color online) C/E relative to median energy  $E_{50\%}$ for the \mbox{IRDFF-II} cross sections averaged in the $^{252}$Cf(s.f.) neutron field measured at \v{R}e\v{z}.
   Plotted values are listed in Table~\ref{tab:TableLongSACScf252_Rez}.}
   \label{fig:SACS_Cf252_Rez}
\end{figure}

More recently, new measurements were performed at \v{R}e\v{z} to confirm Manhhart evaluations. The intense $^{252}$Cf source (about $10^9$ n/cm$^2$/s) at \v{R}e\v{z} is well characterized, and the activation measurements achieved low uncertainties~\cite{Sch19_2}. The results are shown in Table~\ref{tab:TableLongSACScf252_Rez} and Fig.~\ref{fig:SACS_Cf252_Rez}. The C/E uncertainties are colour-coded as defined in Table~\ref{tab:X_A_3}. The SACS in the standard $^{252}$Cf SFNS spectrum have practically no outliers and no trend in energy is observed. A special case is the $^{169}$Tm(n,3n) reaction (Tm1693) , which has a very high threshold, with median energy $>$18~MeV, where the assigned uncertainty of the $^{252}$Cf(s.f.) spectrum is very large due to the lack of measured data, and represents the dominant contribution to the large SACS uncertainty. Still, the C/E for that very special reaction is excellent. %It is possible that the $^{252}$Cf SFNS uncertainty is over-estimated.

\begin{table*} [!htbp]
\vspace{-4mm}
\caption{Measured and calculated spectrum averaged cross sections for $^{252}$Cf(s.f.) neutron standard field.}
\label{tab:TableLongSACScf252}
% \begin{tabular}{|p{0.3in}|p{0.6in}|p{0.6in}|p{0.6in}|p{0.7in}|p{0.7in}|p{0.5in}|p{0.5in}|p{0.7in}|}
\begin{tabular}{l r r r c r r r r r r}
\hline\hline
 Reaction & $E_{50\%}$ & Measured  & Unc.  &              & Calculated& Tot.Unc.& X.S.Unc.& C/E    & Uncert.& Diff. \\
 Notation & [MeV]      & SACS [mb] & [\%]  & Reference    & SACS [mb] & [\%]    & [\%]    &        & [\%]   & [\%]  \\ \hline
\T
 Au197g   &   0.724    &   75.5000 &  1.3   & \cite{Man11} &   74.9780 &  0.90   &  0.52   &\cellcolor{green} 0.9931 &\cellcolor{green}  1.85  &\cellcolor{green} -0.69 \\
 Ta181g   &   0.818    &   87.3000 &  1.4   & \cite{Man11} &   83.4430 &  5.47   &  5.41   &\cellcolor{green} 0.9558 &\cellcolor{green}  5.64  &\cellcolor{green} -4.42 \\
 Th232g   &   0.930    &   87.0000 &  1.8   & \cite{Man11} &   93.4500 &  3.80   &  3.77   &\cellcolor{green} 1.0741 &\cellcolor{green}  4.22  &\cellcolor{green}  7.41 \\
 Cu63g    &   0.964    &   10.3000 &  2.9   & \cite{Man11} &   10.4090 &  8.44   &  8.41   &\cellcolor{green} 1.0106 &\cellcolor{green}  8.93  &\cellcolor{green}  1.06 \\
 In115gm  &   1.116    &  125.6000 &  2.1   & \cite{Man06} &  123.6800 &  2.64   &  2.61   &\cellcolor{green} 0.9847 &\cellcolor{green}  3.46  &\cellcolor{green} -1.53 \\
 U235f    &   1.703    & 1210.0000 &  1.2   & \cite{Man06} & 1226.7000 &  1.21   &  1.21   &\cellcolor{green} 1.0138 &\cellcolor{green}  1.70  &\cellcolor{green}  1.38 \\
 Pu239f   &   1.774    & 1812.0000 &  1.4   & \cite{Man06} & 1797.8000 &  1.25   &  1.25   &\cellcolor{green} 0.9922 &\cellcolor{green}  1.85  &\cellcolor{green} -0.78 \\
 Np237f   &   2.053    & 1361.0000 &  1.6   & \cite{Man06} & 1359.8000 &  1.70   &  1.69   &\cellcolor{green} 0.9991 &\cellcolor{green}  2.33  &\cellcolor{green} -0.09 \\
 Rh103m   &   2.380    &  734.3000 &  2.7   & \cite{Man08} &  724.9200 &  3.95   &  3.94   &\cellcolor{green} 0.9801 &\cellcolor{green}  2.39  &\cellcolor{green} -1.99 \\
 In115nm  &   2.673    &  197.4000 &  1.4   & \cite{Man06} &  190.4800 &  1.70   &  1.66   &\cellcolor{green} 0.9649 &\cellcolor{green}  2.18  &\cellcolor{green} -3.51 \\
 Nb93nm   &   2.685    &  147.5000 &  1.7   & \cite{Zol10} &  146.0300 &  2.61   &  2.59   &\cellcolor{green} 1.0002 &\cellcolor{green}  4.31  &\cellcolor{green}  0.02 \\
 In113nm  &   2.730    &  161.2000 &  2.0   & \cite{Man08} &  157.9900 &  1.24   &  1.18   &\cellcolor{green} 0.9801 &\cellcolor{green}  2.39  &\cellcolor{green} -1.99 \\
 U238f    &   2.768    &  325.7000 &  1.6   & \cite{Man06} &  321.5400 &  1.29   &  1.22   &\cellcolor{green} 0.9872 &\cellcolor{green}  2.09  &\cellcolor{green} -1.28 \\
 Th232f   &   3.006    &   84.5500 &  2.3   & \cite{Csi78} &   83.3500 &  5.80   &  5.78   &\cellcolor{green} 0.9864 &\cellcolor{green}  8.20  &\cellcolor{green} -1.36 \\
 Hg199nm  &   3.098    &  298.4000 &  1.8   & \cite{Man06} &  295.9400 &  3.66   &  3.63   &\cellcolor{green} 0.9918 &\cellcolor{green}  4.08  &\cellcolor{green} -0.82 \\
 BH3      &   3.737    &   11.0000 &  6.4   & \cite{Cse78} &   10.2500 & 11.97   & 11.96   &\cellcolor{green} 0.9318 &\cellcolor{green} 13.56  &\cellcolor{green} -6.82 \\
 Ti47p    &   3.817    &   19.2700 &  1.7   & \cite{Man06} &   19.5330 &  2.80   &  2.73   &\cellcolor{green} 1.0136 &\cellcolor{green}  3.26  &\cellcolor{green}  1.36 \\
 S32p     &   4.074    &   72.5400 &  3.5   & \cite{Man06} &   74.0150 &  2.59   &  2.48   &\cellcolor{green} 1.0203 &\cellcolor{green}  4.35  &\cellcolor{green}  2.03 \\
 Zn64p    &   4.167    &   40.5900 &  1.6   & \cite{Man06} &   42.6470 &  1.86   &  1.69   &\cellcolor{green} 1.0376 &\cellcolor{green}  3.67  &\cellcolor{green}  3.76 \\
 Ni58p    &   4.203    &  117.5000 &  1.3   & \cite{Man06} &  117.3100 &  1.89   &  1.74   &\cellcolor{green} 0.9984 &\cellcolor{green}  2.29  &\cellcolor{green} -0.16 \\
 Fe54p    &   4.438    &   86.8400 &  1.3   & \cite{Man06} &   86.4490 &  3.16   &  3.06   &\cellcolor{green} 0.9955 &\cellcolor{green}  3.43  &\cellcolor{green} -0.45 \\
 Pb204nm  &   5.041    &   20.8700 &  3.5   & \cite{Man06} &   20.3740 &  4.67   &  4.57   &\cellcolor{green} 0.9772 &\cellcolor{green}  5.07  &\cellcolor{green} -2.28 \\
 Al27p    &   5.843    &    4.8800 &  2.1   & \cite{Man06} &    4.7445 &  2.35   &  2.05   &\cellcolor{green} 0.9722 &\cellcolor{green}  3.18  &\cellcolor{green} -2.78 \\
 Co59p    &   5.943    &    1.6900 &  2.5   & \cite{Man06} &    1.7132 &  3.65   &  3.46   &\cellcolor{green} 1.0137 &\cellcolor{green}  4.41  &\cellcolor{green}  1.37 \\
 Ti46p    &   6.081    &   14.0700 &  1.8   & \cite{Man06} &   13.8140 &  3.28   &  3.05   &\cellcolor{green} 0.9818 &\cellcolor{green}  3.73  &\cellcolor{green} -1.82 \\
 Si28p    &   7.226    &    6.9000 &  2.0   & \cite{Zol14} &    7.1069 &  2.85   &  2.47   &\cellcolor{green} 1.0300 &\cellcolor{green}  3.46  &\cellcolor{green}  3.00 \\
 Cu63a    &   7.274    &    0.6887 &  2.0   & \cite{Man06} &    0.6925 &  3.28   &  2.97   &\cellcolor{green} 1.0055 &\cellcolor{green}  3.82  &\cellcolor{green}  0.55 \\
 Fe56p    &   7.579    &    1.4650 &  1.8   & \cite{Man06} &    1.4626 &  2.98   &  2.60   &\cellcolor{green} 0.9984 &\cellcolor{green}  3.47  &\cellcolor{green} -0.16 \\
 U2382    &   8.208    &   19.2000 &  10.   & \cite{Bli87} &   21.0920 &  5.34   &  5.10   &\cellcolor{green} 1.0442 &\cellcolor{green} 12.13  &\cellcolor{green}  4.42 \\
 Mg24p    &   8.260    &    1.9960 &  2.4   & \cite{Man06} &    2.1020 &  1.78   &  0.80   &\cellcolor{green} 1.0531 &\cellcolor{green}  3.02  &\cellcolor{green}  5.31 \\
 Ti48p    &   8.354    &    0.4247 &  1.9   & \cite{Man06} &    0.4264 &  5.53   &  5.30   &\cellcolor{green} 1.0039 &\cellcolor{green}  5.84  &\cellcolor{green}  0.39 \\
 Co59a    &   8.372    &    0.2218 &  1.9   & \cite{Man06} &    0.2210 &  3.87   &  3.54   &\cellcolor{green} 0.9963 &\cellcolor{green}  4.30  &\cellcolor{green} -0.37 \\
 Al27a    &   8.668    &    1.0160 &  1.5   & \cite{Man06} &    1.0167 &  1.77   &  0.71   &\cellcolor{green} 1.0007 &\cellcolor{green}  2.18  &\cellcolor{green}  0.07 \\
 V51a     &   9.975    &    0.0390 &  2.2   & \cite{Man06} &    0.0386 &  3.56   &  3.02   &\cellcolor{green} 0.9885 &\cellcolor{green}  4.19  &\cellcolor{green} -1.15 \\
 Tm1692   &  10.400    &    6.6900 &  6.3   &\cite{Man02}  &    6.2551 & 3.77    & 3.21    &\cellcolor{green} 1.069 &\cellcolor{green}  5.64  &\cellcolor{green}  +6.90 \\
 Au1972   &  10.542    &    5.5060 &  1.9   & \cite{Man06} &    5.5213 &  2.75   &  1.87   &\cellcolor{green} 1.0028 &\cellcolor{green}  3.30  &\cellcolor{green}  0.28 \\
 Nb932m   &  11.328    &    0.7910 &  5.1   & \cite{Man06} &    0.7899 &  2.38   &  0.84   &\cellcolor{green} 0.9986 &\cellcolor{green}  5.02  &\cellcolor{green} -0.14 \\
 I1272    &  11.580    &    2.0690 &  2.7   & \cite{Man06} &    2.1027 &  3.83   &  3.03   &\cellcolor{green} 1.0163 &\cellcolor{green}  4.70  &\cellcolor{green}  1.63 \\
 Cu652    &  12.680    &    0.6582 &  2.2   & \cite{Man06} &    0.6533 &  3.52   &  1.89   &\cellcolor{green} 0.9926 &\cellcolor{green}  4.16  &\cellcolor{green} -0.74 \\
 Co592    &  13.090    &    0.4051 &  2.5   & \cite{Man06} &    0.4078 &  3.66   &  1.52   &\cellcolor{green} 1.0066 &\cellcolor{green}  4.44  &\cellcolor{green}  0.66 \\
 Cu632    &  13.840    &    0.1844 &  4.0   & \cite{Man06} &    0.1987 &  4.51   &  1.38   &\cellcolor{green} 1.0778 &\cellcolor{green}  6.02  &\cellcolor{green}  7.78 \\
 F192     &  14.052    &    0.0161 &  3.4   & \cite{Man06} &    0.0163 &  5.50   &  2.90   &\cellcolor{green} 1.0133 &\cellcolor{green}  6.45  &\cellcolor{green}  1.33 \\
 Zr902    &  14.424    &    0.2210 &  2.9   & \cite{Man06} &    0.2179 &  5.39   &  0.91   &\cellcolor{green} 0.9861 &\cellcolor{green}  6.12  &\cellcolor{green} -1.39 \\
 Ni582    &  14.986    &   8.56E-3 &  3.6   & \cite{Man08} &   8.65E-3 &  6.79   &  1.29   &\cellcolor{green} 1.0102 &\cellcolor{green}  7.81  &\cellcolor{green}  1.02 \\ \hline
\end{tabular}
\vspace{-3mm}
\end{table*}

\begin{table*} [!htbp]
\vspace{-4mm}
\caption{New SACS for $^{252}$Cf-252(s.f.) neutron benchmark field measured at \v{R}e\v{z}.}
\label{tab:TableLongSACScf252_Rez}
% \begin{tabular}{|p{0.3in}|p{0.6in}|p{0.6in}|p{0.6in}|p{0.7in}|p{0.7in}|p{0.5in}|p{0.5in}|p{0.7in}|}
\begin{tabular}{l r r r c r r r r r r}
\hline\hline
 Reaction & $E_{50\%}$ & Measured  & Unc.  &              & Calculated& Tot.Unc.& X.S.Unc.& C/E                     & Uncert.                 & Diff. \\
 Notation & [MeV]      & SACS [mb] & [\%]  & Reference    & SACS [mb] & [\%]    & [\%]    &                         & [\%]                    & [\%]  \\ \hline
\T
 Ti47p    &  3.817     &   19.2800 & 3.70  &\cite{Sch19a} &   19.5330 & 2.80    & 2.73    &\cellcolor{green} 1.0131 &\cellcolor{green}  4.64  &\cellcolor{green}   1.31 \\
 Ni58p    &  4.203     &  117.1000 & 3.07  &\cite{Sch19a} &  117.3100 & 1.89    & 1.74    &\cellcolor{green} 1.0018 &\cellcolor{green}  3.61  &\cellcolor{green}   0.18 \\
 Al27p    &  5.843     &    4.9800 & 3.41  &\cite{Sch18_1}&    4.7445 & 2.35    & 2.05    &\cellcolor{green} 0.9527 &\cellcolor{green}  4.14  &\cellcolor{green}  -4.73 \\
 Co59p    &  5.943     &    1.8160 & 3.47  &\cite{Sch19}  &    1.7132 & 3.65    & 3.46    &\cellcolor{green} 0.9434 &\cellcolor{green}  5.04  &\cellcolor{green}  -5.66 \\
 Ti46p    &  6.081     &   13.8450 & 3.60  &\cite{Sch19a} &   13.8140 & 3.28    & 3.05    &\cellcolor{green} 0.9978 &\cellcolor{green}  4.87  &\cellcolor{green}  -0.22 \\
 Ni60p    &  7.054     &    2.7300 & 4.76  &\cite{Sch18_1}&    2.7985 & 2.27    & 1.81    &\cellcolor{green} 1.0251 &\cellcolor{green}  5.28  &\cellcolor{green}   2.51 \\
 Cu63a    &  7.274     &    0.6833 & 4.49  &\cite{Sch19a} &    0.6925 & 3.28    & 2.97    &\cellcolor{green} 1.0135 &\cellcolor{green}  5.56  &\cellcolor{green}   1.35 \\
 Fe56p    &  7.579     &    1.3690 & 3.80  &\cite{Sch19a} &    1.4626 & 2.98    & 2.60    &\cellcolor{green} 1.0684 &\cellcolor{green}  4.83  &\cellcolor{green}   6.84 \\
 Mg24p    &  8.260     &    2.1740 & 3.50  &\cite{Sch19a} &    2.1020 & 1.78    & 0.80    &\cellcolor{green} 0.9669 &\cellcolor{green}  3.92  &\cellcolor{green}  -3.31 \\
 Ti48p    &  8.354     &    0.4389 & 3.90  &\cite{Sch19a} &    0.4264 & 5.53    & 5.30    &\cellcolor{green} 0.9715 &\cellcolor{green}  6.76  &\cellcolor{green}  -2.85 \\
 Al27a    &  8.668     &    0.9851 & 3.45  &\cite{Sch18_1}&    1.0167 & 1.77    & 0.71    &\cellcolor{green} 1.0321 &\cellcolor{green}  3.88  &\cellcolor{green}   3.21 \\
 Al27a    &  8.668     &    1.0165 & 3.49  &\cite{Sch19a} &    1.0167 & 1.77    & 0.71    &\cellcolor{green} 1.0002 &\cellcolor{green}  3.92  &\cellcolor{green}   0.02 \\
 Tm1692   & 10.400     &    6.3570 & 4.20  &\cite{Sch19}  &    6.2551 & 3.77    & 3.21    &\cellcolor{green} 0.9840 &\cellcolor{green}  5.64  &\cellcolor{green}  -1.60 \\
 Au1972   & 10.542     &    5.4740 & 3.49  &\cite{Sch19}  &    5.5213 & 2.75    & 1.87    &\cellcolor{green} 1.0086 &\cellcolor{green}  4.44  &\cellcolor{green}   0.86 \\
 Nb932m   & 11.328     &    0.8250 & 3.88  &\cite{Sch19}  &    0.7899 & 2.38    & 0.84    &\cellcolor{green} 0.9574 &\cellcolor{green}  4.55  &\cellcolor{green}  -4.26 \\
 I1272    & 11.580     &    2.0440 & 3.52  &\cite{Sch18_2}&    2.1027 & 3.83    & 3.03    &\cellcolor{green} 1.0287 &\cellcolor{green}  5.20  &\cellcolor{green}   2.87 \\
 Mn552    & 12.900     &    0.4820 & 3.94  &\cite{Sch19}  &    0.4729 & 3.92    & 2.32    &\cellcolor{green} 0.9811 &\cellcolor{green}  5.56  &\cellcolor{green}  -1.89 \\
 Co592    & 13.090     &    0.4199 & 3.50  &\cite{Sch19}  &    0.4078 & 3.66    & 1.52    &\cellcolor{green} 0.9712 &\cellcolor{green}  5.06  &\cellcolor{green}  -2.88 \\
 Y892     & 13.902     &    0.3410 & 3.52  &\cite{Sch18_1}&    0.3459 & 4.53    & 1.24    &\cellcolor{green} 1.0144 &\cellcolor{green}  5.74  &\cellcolor{green}   1.44 \\
 Y892     & 13.902     &    0.3510 & 3.70  &\cite{Sch19}  &    0.3459 & 4.53    & 1.24    &\cellcolor{green} 0.9855 &\cellcolor{green}  5.85  &\cellcolor{green}  -1.45 \\
 F192     & 14.053     &    0.0156 & 3.78  &\cite{Sch18_1}&    0.0163 & 5.50    & 2.90    &\cellcolor{green} 1.0464 &\cellcolor{green}  6.67  &\cellcolor{green}   4.64 \\
 Zr902    & 14.424     &    0.2160 & 3.24  &\cite{Sch18_1}&    0.2179 & 5.39    & 0.91    &\cellcolor{green} 1.0089 &\cellcolor{green}  6.29  &\cellcolor{green}   0.89 \\
 Ni582    & 14.987     &   8.37E-3 & 3.94  &\cite{Sch19a} &   8.65E-3 & 6.79    & 1.29    &\cellcolor{green} 1.0324 &\cellcolor{green}  7.85  &\cellcolor{green}   3.24 \\
 Na232    & 15.597     &   8.70E-3 & 3.45  &\cite{Sch18_1}&   8.57E-3 & 8.52    & 1.28    &\cellcolor{green} 0.9854 &\cellcolor{green}  9.19  &\cellcolor{green}  -1.46 \\
 Na232    & 15.597     &   8.98E-3 & 3.56  &\cite{Sch18_2}&   8.57E-3 & 8.52    & 1.28    &\cellcolor{green} 0.9547 &\cellcolor{green}  9.24  &\cellcolor{green}  -4.53 \\
 Tm1693   & 18.506     &    0.0145 & 6.21  &\cite{Sch19a} &    0.0147 &30.32    & 5.76    &\cellcolor{green} 1.0166 &\cellcolor{yellow}30.95  &\cellcolor{green}   1.66 \\ \hline
\end{tabular}
\vspace{-3mm}
\end{table*}

% Table~\ref{tab:TableLongSACScf252} shows validation data for 54 reactions in the $^{252}$Cf SFNS benchmark neutron field. Column 9 of this table shows that, of these measurements, 44 were found to constitute ``good'' validation evidence in the applicable response region provided by this neutron sensitivity. An additional 6 reactions had an ``acceptable'' status, while 4 reactions showed a ``poor'' validation status based on the C/E ratio.

\newpage
\subsection{$^{235}$U(n$_{th}$,f) PFNS}  \label{Sec_VII_B}

\begin{figure}[htbp]
\vspace{-2mm}
   \includegraphics[width=\columnwidth]{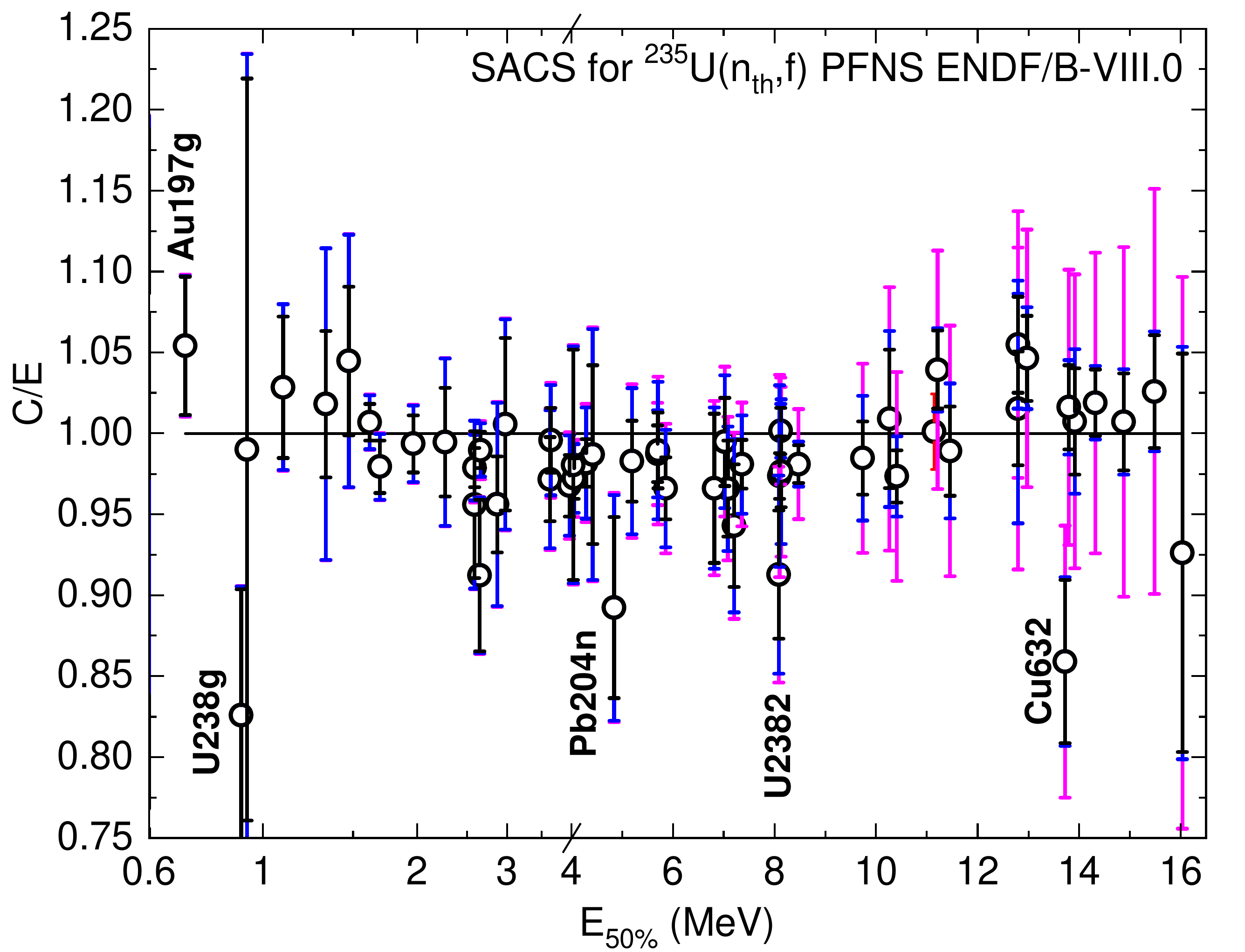}
\vspace{-1mm}
   \caption{(Color online) C/E as a function of median energy $E_{50\%}$ for the \mbox{IRDFF-II} cross sections averaged in the $^{235}$U(n$_{th}$,f) PFNS field. Uncertainty bars reflect a r.m.s. combination of uncertainty contributions from the measurement (black), \mbox{IRDFF-II} cross sections (blue) and spectra (pink). Plotted values are listed in Table~\ref{TableLongSACSu235}.}
   \label{fig:SACS_U235}
\vspace{-2mm}
\end{figure}

The reference prompt fission neutron spectrum (PFNS) for $^{235}$U(n$_{th}$,f)~\cite{PFNS,PFNS1,PFNS2} was shown in Fig.~\ref{fig:U235_spect} as ratio to the Maxwellian distribution with $kT = 1.32$~MeV. The uncertainties in that evaluated spectrum were constrained at high energy by the use of $^{90}$Zr(n,2n) SACS point.
%The general tendency of increasing spectral uncertainties above 10 MeV is again noted. The large spectrum uncertainty in the high energy region results in very large uncertainty in the C/E's in this benchmark field.
Recommended SACS in the $^{235}$U(n$_{th}$,f) PFNS are listed in Table~\ref{TableLongSACSu235}. The C/E ratios of SACS in the reference $^{235}$U(n$_{th}$,f) PFNS as a function of representative neutron energy $E_{50\%}$ are shown in Fig.~\ref{fig:SACS_U235}. For the $^{235}$U(n$_{th}$,f) PFNS field no trend in energy is observed. An outlier is the tritium production reaction on $^{10}$B (B10H3) measured by Wyman in 1958.%, which can probably be ignored.
Measurements of the capture SACS are usually subject to large corrections for room-return and multiple scattering, which might be the cause of some discrepancy in the U238g reaction. Poorer (but acceptable) agreement is also noted for $^{63}$Cu(n,2n), $^{238}$U(n,2n) and $^{204}$Pb(n,n') reactions. New extensive measurements of SACS in the reference $^{235}$U(n$_{th}$,f) neutron field are on-going. % at \v{R}e\v{z} near Prague, and will be published elsewhere.

\begin{table*}[htbp]
\vspace{-2mm}
\caption{Measured and calculated SACS for the IRDFF-II cross sections in the $^{235}$U(n$_{th}$,f) neutron benchmark field.}
\label{TableLongSACSu235}
\begin{tabular}{l r l r c l c c c c}
\hline\hline
Reaction & E$_{50\%}$ & \multicolumn{3}{p{1.9in}}{Measured SACS}       & \multicolumn{4}{p{1.8in}}{Calculated SACS} & C/E \\
                        \cline{3-5}                                      \cline{6-9}
Notation & MeV   & Value, mb & Unc.,\% & Reference                     & Value, mb & XS Unc.,\% & Sp.Unc.,\% & Tot.Unc.,\% &     \\
\hline
\T
Au197g  & 0.705 & 7.400E+01 & 4.05 & \cite{Fab68}                     & 7.8011E+01 &  0.51 &  0.80 &  0.95 & 1.054 $\pm$ 4.16 \% \cellcolor{green} \\
% Mn55g   & 0.726 & 3.820E+00 & 10.00& \cite{Hug53}                     & 2.9111E+00 & 27.08 &  0.96 & 27.09 & 0.762 $\pm$ 28.88\% \cellcolor{red}   \\
% Co59g   & 0.878 & 1.100E+01 & 10.00& \cite{Hug53}                     & 5.0259E+00 &  4.00 &  0.76 &  4.07 & 0.457 $\pm$ 10.80\% \cellcolor{red}   \\
U238g   & 0.907 & 8.500E+01 & 9.41 & \cite{Fab68}                     & 7.0197E+01 &  1.97 &  0.69 &  2.09 & 0.826 $\pm$ 9.64 \% \cellcolor{yellow}\\
Cu63g   & 0.932 & 1.080E+01 & 23.15& \cite{Fab68}                     & 1.0693E+01 &  8.57 &  0.64 &  8.60 & 0.990 $\pm$ 24.70\% \cellcolor{green} \\
% Na23g   & 0.944 & 2.600E-01 & 10.00& \cite{Hug53}                     & 2.8090E-01 &  3.88 &  0.82 &  3.96 & 1.080 $\pm$ 10.76\% \cellcolor{green} \\
% La139g  & 1.261 & 5.300E+00 & 10.00& \cite{Zol02}                     & 6.7979E+00 &  5.08 &  0.52 &  5.11 & 1.283 $\pm$ 11.23\% \cellcolor{red}   \\
In115gm & 1.095 & 1.245E+02 & 4.25 & \cite{Zol13}                     & 1.2805E+02 &  2.61 &  0.58 &  2.67 & 1.029 $\pm$ 5.02 \% \cellcolor{green} \\
B10He4  & 1.329 & 5.410E+02 & 4.44 & \cite{Oli82}                     & 5.5077E+02 &  8.35 &  0.54 &  8.37 & 1.018 $\pm$ 9.47 \% \cellcolor{green} \\
Li6He4  & 1.472 & 4.560E+02 & 4.39 & \cite{Oli82}                     & 4.7641E+02 &  6.03 &  0.60 &  6.06 & 1.045 $\pm$ 7.48 \% \cellcolor{green} \\
U235f   & 1.618 & 1.217E+03 & 1.12 & \cite{Man08}                     & 1.2253E+03 &  1.21 &  0.36 &  1.26 & 1.007 $\pm$ 1.69 \% \cellcolor{green} \\
Pu239f  & 1.691 & 1.831E+03 & 1.65 & \cite{Man08}                     & 1.7933E+03 &  1.25 &  0.37 &  1.30 & 0.979 $\pm$ 2.10 \% \cellcolor{green} \\
Np237f  & 1.965 & 1.350E+03 & 1.78 & \cite{Man08}                     & 1.3413E+03 &  1.58 &  0.47 &  1.75 & 0.994 $\pm$ 2.43 \% \cellcolor{green} \\
Rh103nm & 2.267 & 7.103E+02 & 3.38 & \cite{Kob92,Hor89,Gri83}         & 7.0643E+02 &  3.95 &  0.53 &  3.99 & 0.995 $\pm$ 5.23 \% \cellcolor{green} \\
Nb93n   & 2.589 & 1.476E+02 & 4.74 & \cite{Zol99}                     & 1.4109E+02 &  2.63 &  0.66 &  2.71 & 0.956 $\pm$ 5.46 \% \cellcolor{green} \\
In115nm & 2.589 & 1.878E+02 & 1.23 & \cite{Man08}                     & 1.8383E+02 &  1.68 &  0.69 &  1.82 & 0.979 $\pm$ 2.20 \% \cellcolor{green} \\
In113nm & 2.651 & 1.668E+02 & 5.13 & \cite{Zol13}                     & 1.5217E+02 &  1.19 &  0.72 &  1.39 & 0.912 $\pm$ 5.31 \% \cellcolor{green} \\
U238f   & 2.659 & 3.094E+02 & 1.13 & \cite{Man08}                     & 3.0618E+02 &  1.22 &  0.75 &  1.43 & 0.990 $\pm$ 1.82 \% \cellcolor{green} \\
Th232f  & 2.867 & 8.186E+01 & 3.10 & \cite{Gil84}                     & 7.8267E+01 &  5.79 &  0.78 &  5.84 & 0.956 $\pm$ 6.61 \% \cellcolor{green} \\
Hg199nm & 2.972 & 2.787E+02 & 5.30 & \cite{Zol08}                     & 2.8024E+02 &  3.68 &  0.77 &  3.76 & 1.006 $\pm$ 6.50 \% \cellcolor{green} \\
B10H3   & 3.533 & 3.023E+01 & 6.70 & \cite{Wym58}                     & 4.7915E+01 & 12.37 &  0.85 & 12.40 & 1.585 $\pm$ 14.09\% \cellcolor{lred}   \\
P31p    & 3.633 & 3.586E+01 & 2.68 & \cite{Zol14}                     & 3.4842E+01 &  3.47 &  1.06 &  3.63 & 0.972 $\pm$ 4.51 \% \cellcolor{green} \\
Ti47p   & 3.647 & 1.784E+01 & 1.99 & \cite{Man08}                     & 1.7765E+01 &  2.79 &  0.99 &  2.96 & 0.996 $\pm$ 3.57 \% \cellcolor{green} \\
S32p    & 3.967 & 6.908E+01 & 1.97 & \cite{Man08}                     & 6.6846E+01 &  2.53 &  1.14 &  2.78 & 0.968 $\pm$ 3.41 \% \cellcolor{green} \\
Zn64p   & 4.036 & 3.889E+01 & 7.25 & \cite{Zol08}                     & 3.8133E+01 &  1.74 &  1.20 &  2.12 & 0.981 $\pm$ 7.55 \% \cellcolor{green} \\
% Zn64p   & 4.036 & 3.539E+01 & 3.02 & \cite{Man02}                     & 3.8133E+01 &  1.74 &  1.20 &  2.12 & 1.078 $\pm$ 3.69 \% \cellcolor{green} \\
Ni58p   & 4.051 & 1.082E+02 & 1.30 & \cite{Man08}                     & 1.0519E+02 &  1.75 &  1.14 &  2.09 & 0.972 $\pm$ 2.46 \% \cellcolor{green} \\
Fe54p   & 4.294 & 7.809E+01 & 1.50 & \cite{Zol13}                     & 7.6659E+01 &  3.16 &  1.24 &  3.39 & 0.982 $\pm$ 3.71 \% \cellcolor{green} \\
Zn67p   & 4.421 & 9.660E-01 & 5.59 & \cite{Zol13}                     & 9.5344E-01 &  5.52 &  1.19 &  5.64 & 0.987 $\pm$ 7.94 \% \cellcolor{green} \\
Pb204nm & 4.839 & 1.946E+01 & 6.27 & \cite{Kim71,Bro77}               & 1.7364E+01 &  4.66 &  1.42 &  4.87 & 0.892 $\pm$ 7.94 \% \cellcolor{yellow}\\
Mo92pm  & 5.190 & 6.687E+00 & 2.53 & \cite{Zol13}                     & 6.5723E+00 &  3.83 &  1.49 &  4.11 & 0.983 $\pm$ 4.83 \% \cellcolor{green} \\
Al27p   & 5.700 & 3.902E+00 & 1.77 & \cite{Man08}                     & 3.8527E+00 &  3.49 & 12.23 &  2.66 & 0.987 $\pm$ 3.20 \% \cellcolor{green} \\
Co59p   & 5.703 & 1.396E+00 & 2.36 & \cite{Man08}                     & 1.3812E+00 &  3.59 &  1.68 &  3.97 & 0.989 $\pm$ 4.62 \% \cellcolor{green} \\
Ti46p   & 5.862 & 1.151E+01 & 1.99 & \cite{Man08}                     & 1.1118E+01 &  3.19 &  1.74 &  3.64 & 0.966 $\pm$ 4.15 \% \cellcolor{green} \\
Ni60p   & 6.811 & 2.180E+00 & 4.77 & \cite{Man11}                     & 2.1062E+00 &  1.98 &  2.10 &  2.88 & 0.966 $\pm$ 5.57 \% \cellcolor{green} \\
Cu63a   & 7.019 & 5.173E-01 & 2.74 & \cite{Man08,Kos19_1}             & 5.1459E-01 &  3.09 &  2.16 &  3.77 & 0.995 $\pm$ 4.66 \% \cellcolor{green} \\
Si28p   & 7.079 & 5.470E+00 & 3.07 & \cite{Zol14}                     & 5.2831E+00 &  2.53 &  2.28 &  3.41 & 0.966 $\pm$ 4.59 \% \cellcolor{green} \\
Fe54a   & 7.205 & 8.650E-01 & 4.02 & \cite{Sua97,Hor89,Bus87}         & 8.1560E-01 &  4.02 &  2.25 &  4.59 & 0.943 $\pm$ 6.10 \% \cellcolor{green} \\
Fe56p   & 7.362 & 1.079E+00 & 1.54 & \cite{Man08}                     & 1.0583E+00 &  2.68 &  2.38 &  3.59 & 0.981 $\pm$ 3.91 \% \cellcolor{green} \\
U2382   & 8.087 & 1.600E+01 & 4.34 & \cite{Kob76,Has78,Sha83}         & 1.4604E+01 &  5.10 &  2.92 &  5.87 & 0.913 $\pm$ 7.30 \% \cellcolor{green}\\
Ti48p   & 8.103 & 3.014E-01 & 1.46 & \cite{Man08,Kos19_1}             & 2.9347E-01 &  5.57 &  2.84 &  6.25 & 0.974 $\pm$ 6.42 \% \cellcolor{green} \\
Mg24p   & 8.125 & 1.449E+00 & 1.43 & \cite{Man08,Kos19_1}             & 1.4512E+00 &  0.83 &  2.85 &  2.97 & 1.002 $\pm$ 3.30 \% \cellcolor{green} \\
Co59a   & 8.127 & 1.563E-01 & 2.25 & \cite{Man08}                     & 1.5260E-01 &  3.99 &  2.81 &  4.88 & 0.976 $\pm$ 5.37 \% \cellcolor{green} \\
Al27a   & 8.471 & 7.005E-01 & 1.19 & \cite{Man08,Kos19_2}             & 6.8720E-01 &  0.74 &  3.16 &  3.24 & 0.981 $\pm$ 3.45 \% \cellcolor{green} \\
V51a    & 9.737 & 2.429E-02 & 2.29 & \cite{Man08}                     & 2.3918E-02 &  3.17 &  4.46 &  5.48 & 0.985 $\pm$ 5.94 \% \cellcolor{green} \\
Tm1692  &10.265 & 3.735E+00 & 4.23 & \cite{Zol10}                     & 3.7685E+00 &  3.34 &  6.01 &  6.88 & 1.009 $\pm$ 8.08 \% \cellcolor{green} \\
Au1972  &10.414 & 3.387E+00 & 1.67 & \cite{Man08,Kos19_1}             & 3.2969E+00 &  1.93 &  6.11 &  6.41 & 0.973 $\pm$ 6.62 \% \cellcolor{green} \\
Nb932   &11.210 & 4.348E-01 & 2.32 & \cite{Zol13}                     & 4.5189E-01 &  0.86 &  6.65 &  6.71 & 1.039 $\pm$ 7.10 \% \cellcolor{green} \\
I1272   &11.459 & 1.199E+00 & 2.79 & \cite{Zol08,Bur19}               & 1.1860E+00 &  3.16 &  6.60 &  7.32 & 0.989 $\pm$ 7.83 \% \cellcolor{green} \\
Mn552   &12.796 & 2.356E-01 & 2.81 & \cite{Man08,Bur19}               & 2.4852E-01 &  2.48 &  6.84 &  7.28 & 1.055 $\pm$ 7.80 \% \cellcolor{green} \\
As752   &12.797 & 3.207E-01 & 3.47 & \cite{Zol99,Kos18,Kos19_1}       & 3.2561E-01 &  6.07 &  6.87 &  9.16 & 1.015 $\pm$ 9.80 \% \cellcolor{green} \\
Co592   &12.974 & 2.028E-01 & 2.51 & \cite{Man08}                     & 2.1220E-01 &  1.65 &  6.98 &  7.17 & 1.046 $\pm$ 7.60 \% \cellcolor{green} \\
Cu632   &13.721 & 1.160E-01 & 5.88 & \cite{Gru68}                     & 9.9640E-02 &  1.44 &  7.69 &  7.82 & 0.859 $\pm$ 9.80 \% \cellcolor{yellow}\\
Y892    &13.797 & 1.697E-01 & 2.34 & \cite{Rau68,Ste70,Boj92,Kos18,Kos19_1}& 1.7274E-01 &  1.30 &  7.87 &  7.97 & 1.015 $\pm$ 8.37 \% \cellcolor{green}\\
F192    &13.911 & 8.056E-03 & 3.26 & \cite{Man08,Kos16,Kos19_2}       & 8.1151E-03 &  2.99 &  7.86 &  8.41 & 1.007 $\pm$ 9.02 \% \cellcolor{green} \\
Zr902   &14.320 & 1.041E-01 & 2.02 & \cite{Man08,Kos18,Kos19_2}       & 1.0607E-01 &  0.92 &  8.84 &  8.89 & 1.019 $\pm$ 9.12 \% \cellcolor{green} \\
Ni582   &14.879 & 4.076E-03 & 2.98 & \cite{Kob80a,Hor91,Sek81,Man85_2}& 4.1047E-03 &  1.29 & 10.23 & 10.31 & 1.007 $\pm$ 10.73\% \cellcolor{green} \\
Na232   &15.483 & 3.860E-03 & 3.39 & \cite{Kos16,Kos18_2}             & 3.9599E-03 &  1.27 & 11.65 & 11.72 & 1.026 $\pm$ 12.20\% \cellcolor{green} \\
Al272   &16.023 & 3.760E-03 & 13.30& \cite{Zol17}                     & 3.4823E-03 &  2.06 &  1.69 & 12.72 & 0.926 $\pm$ 18.40\% \cellcolor{green} \\
\hline\hline
\end{tabular}
\vspace{-2mm}
\end{table*}

\subsection{Ratio of $^{252}$Cf SFNS to $^{235}$U(n$_{th}$,f) PFNS Spectrum-averaged Cross Sections} \label{Sec_VII_C}

\begin{figure*}[htbp]
\vspace{-3mm}
   \includegraphics[width=0.99\textwidth]{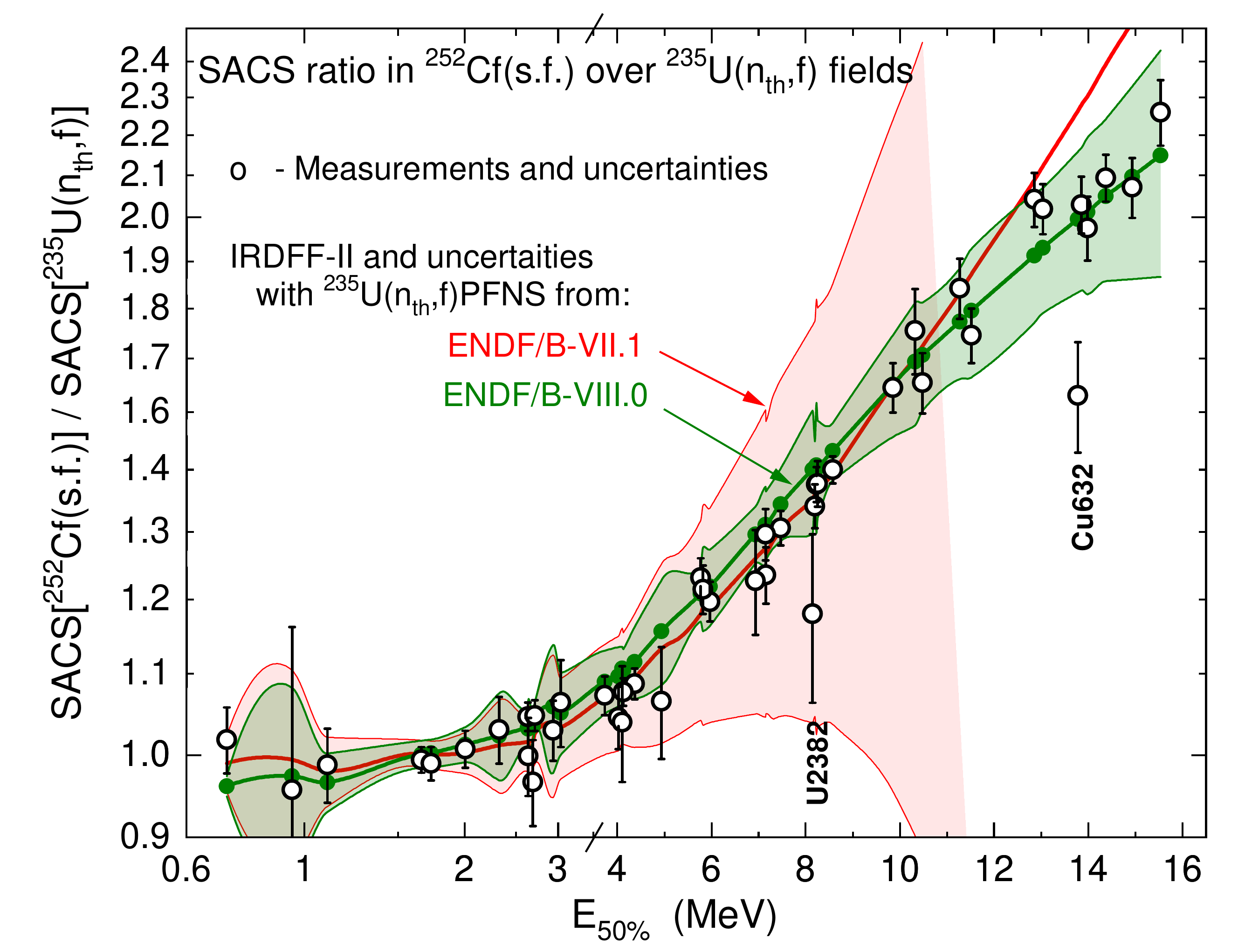}
\vspace{-2mm}
   \caption{(Color online) Ratio of SACS for the $^{252}$Cf SFNS and $^{235}$U(n$_{th}$,f) PFNS fields as a function of median energy $E_{50\%}$ : circles - measurements, crosses - computed with the \mbox{IRDFF-II} cross sections. Plotted values are calculated as ratios of SACS from Tables~\ref{tab:TableLongSACScf252} and \ref{tab:TableLongSACScf252_Rez} to SACS listed in Table~\ref{TableLongSACSu235}.}
   \label{fig:SACS_Cf252_U235}
\vspace{-2mm}
\end{figure*}

The activation cross sections in the $^{252}$Cf SFNS and $^{235}$U(n$_{th}$,f) PFNS benchmark neutron fields discussed in Secs.~\ref{Sec_VII_A} and~\ref{Sec_VII_B} form the foundation for the \mbox{IRDFF-II} validation. In addition to these two separate sets of validation data for a fission neutron spectrum, the smooth behaviour, \ie, the systematics, of the ratios of these spectrum-averaged cross sections, as a function of the $E_{50}$ response energy for a given reaction, provides another good indicator of the fidelity of the cross sections. This ratio is independent of the gamma emission probability, which is a significant source of measured SACS uncertainties. The application of this metric and a systematic of this ratio is addressed in Ref.~\cite{Cap18}. The systematic can be used also to assess the quality of experimental data in one of the reference fields~\cite{Cap18}. The actual status of C/E ratios computed with \mbox{IRDFF-II} is depicted in Fig.~\ref{fig:SACS_Cf252_U235}. The ratio for the reaction $^{63}$Cu(n,2n) is an outlier indicating that one or both of the SACS measurements used to calculate the ratio are likely wrong. Indeed, it can be seen that the corresponding C/E of this particular reaction in the $^{235}$U(n$_{th}$,f) PFNS neutron field is 0.859(9.8\%) pointing out to a possible experimental issue. A similar situation is found for the $^{238}$U(n,2n) reaction, where the C/E in the $^{235}$U(n$_{th}$,f) PFNS neutron field is 0.913(7.3\%), which also points to a potential experimental problem. Corresponding C/E in the $^{252}$Cf SFNS field show good agreement for these two reactions.

\subsection{Measurements in the SPR-III Fast Burst Reactor Central Cavity} \label{Sec_VII_D}

The baseline measured quantities in this reference benchmark field are end-of-irradiation (EOI) spectrum-averaged cross sections measured in operation \#13127, which was a 16-minute steady-state operation that took place on 1/27/2006 and represented a total facility-reported reactor integrated power of 125~MJ.

Although a baseline set of activation foils were used in this reactor operation \#13127, it was desirable to add more reactions to the set of measured data, so additional reactor exposures of activation foils were conducted. In order to eliminate the effect of variations in the reactor power during the other exposures, a $^{58}$Ni(n,p)$^{58}$Co monitor activity was used to facilitate a renormalization of data gathered in the other operations to the baseline exposure. When required, corrections were made in the measured activities of other foils to convert them to represent the end-of-irradiation from the baseline 16-minute uniform power operation. The other reactor operations are detailed in Table~\ref{tab:X_D_4}.

\begin{table}[htbp]
\vspace{-4mm}
\caption{Reactor operations used to support the SPR-III fast burst reactor central cavity spectral characterization.}
\label{tab:X_D_4}
%\begin{tabular}{ l c  p{0.6in}  p{1.0in}  } \hline\hline
 \begin{tabular}{ l c  l         p{3cm}  } \hline\hline
 Date        & Oper.   & Reactor            & Comment                \\
  {[d/m/y]}  & No.     & Power              &                        \\ \hline
 19/01/2006  & 13090   & 10~kW, 250~s       & 3x3 grid Foils Set \#1 \\
 19/01/2006  & 13091   & 10~kW,             & $^{235}$U foil         \\
 19/01/2006  & 13092   & 10~kW,             & $^{239}$Pu foil        \\
 26/01/2006  & 13117   & 5~kW, 275~s        & $^{237}$Np foil        \\
 26/01/2006  & 13118   & ---                & $^{238}$U foil         \\
 27/01/2006  & 13127   & 125~MJ,    16 min, & Foil Set \#2           \\
 10/02/2006  & 13156   & 27.5~MJ,   40 min, & ---                    \\
 14/04/2006  & 13295   & 27.5~MJ,  138 min, & Sc in boron ball       \\
 14/04/2006  & 13296   & 127.5~MJ, 217 min, & Mn  in boron ball      \\
 19/05/2006  & 13368   & 25~MJ,     32 min, & Fe in boron ball       \\
 19/05/2006  & 13369   & ---                & Ag in boron ball       \\
 30/06/2006  & 13452   & 25 MJ, 190 min,    & 3x3 grid Foils Set \#1 \\ \hline\hline
\end{tabular}
\vspace{-4mm}
\end{table}

In some of the cases covers were used to shield the activation foil and shift the sensitivity of the measured response to higher energies. These measurements used the following cover geometries:

\begin{enumerate}
 \item  ``bare'' indicating a bare foil with a thickness between 5 and 10 mils;
 \item  ``Cd'' indicating a cylindrical cadmium cover with thickness of 2.587x10$^{-3}$ atoms/barn;
 \item  ``Cdtk'' indicating a cylindrical cadmium cover with thickness of 4.705x10$^{-3}$ atoms/barn;
 \item  ``Cdtk/B4C'' indicates an inner thick cadmium cover with an areal thickness of 4.705x10$^{-3}$ atoms/barn and an outer spherical boron ball.  The boron ball is a spherical shell with an outer diameter of 4.76 cm and thickness of 1.03 cm. The B$_{4}$C has a density of 2.5 g/cm$^{2}$ and used isotopically enriched Boron (91.67~\% $^{10}$B and 8.33~\% $^{11}$B).
\end{enumerate}

A detailed description of the covers and of how self-shielding effects are rigorously addressed in their use in the activation measurements can be found in Reference \cite{Gri90}. Data were gathered for a total of 40 combinations of reactions and covers in this reference field.

The fission foils were in the form of a sintered oxide (uranium oxide-to-metal ratio of 1.995; plutonium ratio of 1.886) measuring approximately 0.5 inch by 0.03-inch thick, and having an approximate mass of 1 gram. The isotopic mixtures of the various fission foils are given in Table~\ref{tab:X_D_5}.

\begin{table}[htbp]
\vspace{-4mm}
\caption{Isotopic composition of fission foils}
\label{tab:X_D_5}
\begin{tabular}{ p{1.0in} p{0.4in} l } \hline\hline
Foil               & Isotope    & Atom Fraction$^*$ \\ \hline
\T
Enriched $^{235}$U & $^{235}$U  & 0.9300        \\
                   & $^{234}$U  & 0.00981       \\
                   & $^{236}$U  & 0.00359       \\
                   & $^{238}$U  & 0.0566        \\ \hline
\T
Depleted $^{238}$U & $^{238}$U  & 0.9979        \\
                   & $^{234}$U  & 0.00001       \\
                   & $^{235}$U  & 0.00205       \\
                   & $^{236}$U  & 0.00004       \\  \hline
\T
$^{239}$Pu         & $^{239}$Pu & 0.869965      \\
                   & $^{238}$Pu & 0.0006798     \\ \
                   & $^{240}$Pu & 0.115968688   \\
                   & $^{241}$Pu & 0.010797      \\
                   & $^{242}$Pu & 0.00235936    \\
                   & $^{235}$U  & 0.000199946   \\
                   & $^{237}$Np & 0.00002999    \\ \hline\hline
\addlinespace[1mm]
\multicolumn{3}{ p{3.0in} }{$^*$~Atom fraction sum of the number of atoms of the prime sensor element is unity. Due to contaminants, the total sum may be greater than 1.0.} \\ % \hline\hline
\end{tabular}
\vspace{-4mm}
\end{table}

The baseline set of end-of-irradiation activity measurements, normalized using the $^{58}$Ni(n,p)$^{58}$Co monitor reaction used in the same reactor operation, are shown in Table~\ref{Table_Long_XXII}. We cannot compare these measured activities to the corresponding calculated dosimetry metrics, spectrum-averaged cross sections, in the absence of a known neutron fluence level for the reference operation. Since a spectral index (SI) is defined to be the ratio of the spectrum-averaged cross sections for two specified reactions -- where one is generally considered to be a reference monitor reaction -- this quantity is independent of the neutron fluence. Thus, we have adopted the C/E for spectral indices, using the $^{58}$Ni(n,p)$^{58}$Co reaction as the reference reaction, as the relevant comparison metric. The C/E spectral index metric, for the reaction ``$\alpha$'' and the monitor ``Ni58p'', is notated as the CoE$^{SI}$$_{\alpha}$$_{/Ni58p}$.

Table~\ref{Table_Long_XXII} documents the data in this field for the 40 measured reactions. The entries in the table start with the $^{58}$Ni(n,p) normalizing reaction but are then ordered with respect to the increasing $E_{50}$ mean response energy for the reactions. The table includes:
\begin{enumerate}
 \item  Reaction identifiers;
 \item  Median response energy in this neutron field for the indicated dosimeter/cover combination;
 \item  measured end-of-irradiation (EOI) activities with measurement uncertainty;
 \item  uncertainty contribution in the calculated spectrum-averaged cross sections due to the \textit{a priori} spectrum;
 \item  calculated spectral index and associated uncertainty, including the contributions from the cross section and spectrum;
 \item  calculated-over-experimental (C/E) ratio of the spectral index (SI) and associated uncertainty.
\end{enumerate}

\begin{table*}[hbtp]
\vspace{-3mm}
\caption{Measured EOI activities and calculated quantities supporting the SI C/E for SPR-III central cavity neutron benchmark field.}
\label{Table_Long_XXII}
\begin{tabular}{l c r r r r r r r}
\\ \hline \hline
             &             & \multicolumn{2}{c}{Measured Act.} & Calc.     & \multicolumn{2}{c}{Spectral}  & \multicolumn{2}{c}{SI C/E}  \\
Entry Label  & $E_{50\%}$  & Bq/atom or & Unc.                 & SACS Unc. & Index (SI)& Unc.      &  Value & Unc. \\
             & [MeV]       & Fiss./atom & [\%]                 & [\%]      &           & [\%]      &        & [\%]  \\
\hline
\T
 Sc45g-Cd       &  0.2060 & 1.3720E-20 & ~3.30 & 23.14 & 2.0093E-01 & ~35.84 & \cellcolor{yellow} 1.0832 & \cellcolor{yellow} 36.12 \\
 Mn55g-Cd       &  0.2348 & 6.6180E-18 & ~2.60 & 10.61 & 1.0265E-01 & ~35.45 & \cellcolor{yellow} 0.8926 & \cellcolor{yellow} 35.68 \\
 Au197g-bare    &  0.2385 & 7.4140E-18 & ~4.50 &  9.87 & 2.8463E+00 & ~36.52 & \cellcolor{yellow} 0.8827 & \cellcolor{yellow} 36.92 \\
 Au197g-Cd      &  0.2505 & 6.5740E-18 & ~4.50 & 10.11 & 2.7742E+00 & ~36.93 &  \cellcolor{green} 0.9706 &  \cellcolor{green} 37.33 \\
 Sc45g-Cdtk/B4C &  0.2515 & 1.1920E-20 & ~3.50 & 11.26 & 1.8599E-01 & ~39.01 & \cellcolor{yellow} 1.1540 & \cellcolor{yellow} 39.29 \\
 Na23g-bare     &  0.2628 & 7.0870E-20 & ~2.10 &  4.21 & 5.8091E-03 & ~37.54 & \cellcolor{yellow} 0.8146 & \cellcolor{yellow} 37.73 \\
 Fe58g-Cd       &  0.2883 & 7.9740E-21 & ~2.80 &  9.63 & 6.8808E-02 & ~35.13 & \cellcolor{yellow} 1.2018 & \cellcolor{yellow} 35.38 \\
 Na23g-Cdna     &  0.2887 & 7.1700E-20 & ~2.10 &  3.41 & 5.4893E-03 & ~32.59 &    \cellcolor{lred} 0.7609 &    \cellcolor{lred} 32.81 \\
 Mn55g-Cdtk/B4C &  0.2958 & 4.5570E-18 & ~2.70 &  9.18 & 9.0805E-02 & ~38.60 & \cellcolor{yellow} 1.1468 & \cellcolor{yellow} 38.82 \\
 W186g-bare     &  0.3215 & 7.4543E-18 & ~9.02 &  8.56 & 1.2806E+00 & ~34.08 & \cellcolor{yellow} 1.0769 & \cellcolor{yellow} 35.39 \\
 Cu63g-Cd       &  0.3395 & 3.3020E-18 & ~2.20 &  7.62 & 3.1820E-01 & ~35.77 & \cellcolor{yellow} 1.1281 & \cellcolor{yellow} 35.97 \\
 Na23g-Cdtk/B4C &  0.3589 & 5.9630E-20 & ~2.10 &  5.28 & 6.2782E-03 & ~33.33 & \cellcolor{yellow} 1.0464 & \cellcolor{yellow} 33.54 \\
 U235f-Cdtk     &  0.6877 & 1.7550E-11 & ~3.00 &  1.27 & 2.4819E+01 & ~29.38 & \cellcolor{yellow} 1.0925 & \cellcolor{yellow} 29.69 \\
 U235f-Cdtk/B4C &  0.7568 & 1.5000E-11 & ~3.00 &  0.37 & 2.2920E+01 & ~29.39 & \cellcolor{yellow} 1.1804 & \cellcolor{yellow} 29.71 \\
 Pu239f-Cdtk    &  0.8244 & 2.2330E-11 & ~2.70 &  1.44 & 3.3176E+01 & ~27.33 & \cellcolor{yellow} 1.1477 & \cellcolor{yellow} 27.64 \\
 Pu239f-Cdtk/B4C&  0.8812 & 1.9120E-11 & ~2.70 &  2.33 & 3.0745E+01 & ~27.56 & \cellcolor{yellow} 1.2422 & \cellcolor{yellow} 27.86 \\
 Np237f-Cdtk    &  1.3904 & 1.2340E-11 & ~2.80 & 11.85 & 1.7581E+01 & ~22.87 & \cellcolor{yellow} 1.1006 & \cellcolor{yellow} 23.25 \\
 Np237f-Cdtk/B4C&  1.3952 & 1.1820E-11 & ~2.80 & 12.22 & 1.6182E+01 & ~21.34 & \cellcolor{yellow} 1.0576 & \cellcolor{yellow} 21.74 \\
 In115nm-bare   &  2.1498 & 6.2040E-17 & ~4.70 & 18.12 & 2.0190E+00 & ~15.57 & \cellcolor{yellow} 1.0779 & \cellcolor{yellow} 16.55 \\
 U238f-Cd       &  2.2862 & 2.3170E-12 & ~3.20 & 19.29 & 3.2721E+00 & ~12.77 &  \cellcolor{green} 1.0909 &  \cellcolor{green} 13.53 \\
 U238f-Cdtk/B4C &  2.2909 & 2.2230E-12 & ~3.20 & 19.53 & 2.9664E+00 & ~11.44 &  \cellcolor{green} 1.0308 &  \cellcolor{green} 12.28 \\
 Ti47p-Cd       &  3.1728 & 2.8820E-19 & ~2.70 & 21.61 & 1.7116E-01 & ~~2.97 &  \cellcolor{green} 1.0991 &  \cellcolor{green} ~5.07 \\
 S32p-bare      &  3.5235 & 2.5080E-19 & ~3.00 & 23.78 & 5.9089E-01 & ~~1.36 &  \cellcolor{green} 1.0221 &  \cellcolor{green} ~4.52 \\
 Ni58p-Cd       &  3.5954 & 8.7520E-20 & ~3.10 & 23.51 & 1.0000E+00 & ~~0.00 &  \cellcolor{green} 1.0000 &  \cellcolor{green} ~4.38 \\
 Zn64p-Cd       &  3.6230 & 3.8820E-18 & ~2.20 & 23.89 & 3.5650E-01 & ~~1.36 &  \cellcolor{green} 1.0749 &  \cellcolor{green} ~4.04 \\
 Fe54p-Cd       &  3.6969 & 1.4000E-20 & ~3.20 & 24.14 & 7.3304E-01 & ~~1.27 &  \cellcolor{green} 1.0396 &  \cellcolor{green} ~4.63 \\
 Al27p-Cd       &  5.0886 & 2.8720E-17 & ~3.00 & 27.07 & 3.3676E-02 & ~11.64 &  \cellcolor{green} 1.0675 &  \cellcolor{green} 12.41 \\
 Co59p-bare     &  5.1337 & 1.8835E-21 & ~5.18 & 26.70 & 1.3281E-02 & ~11.12 &  \cellcolor{green} 0.9818 &  \cellcolor{green} 12.65 \\
 Ti46p-Cd       &  5.2634 & 7.3250E-21 & ~3.40 & 27.59 & 1.0288E-01 & ~12.29 &  \cellcolor{green} 1.0389 &  \cellcolor{green} 13.13 \\
 Cu63a-bare     &  6.3286 & 1.6723E-23 & ~5.14 & 28.41 & 5.1287E-03 & ~16.82 & \cellcolor{yellow} 0.9873 & \cellcolor{yellow} 17.86 \\
 Fe56p-Cd       &  6.6224 & 5.4080E-19 & ~2.40 & 29.70 & 1.0197E-02 & ~~0.01 & \cellcolor{yellow} 1.0851 & \cellcolor{yellow} ~3.92 \\
 Ti48p-Cd       &  7.3414 & 8.6840E-21 & ~1.20 & 30.16 & 2.9767E-03 & ~20.68 & \cellcolor{yellow} 1.1673 & \cellcolor{yellow} 20.94 \\
 Mg24p-Cd       &  7.4113 & 1.2420E-19 & ~3.00 & 30.77 & 1.4247E-02 & ~21.75 & \cellcolor{yellow} 1.1400 & \cellcolor{yellow} 22.18 \\
 Al27a-Cd       &  7.6839 & 6.1350E-20 & ~1.90 & 30.83 & 7.2208E-03 & ~22.52 & \cellcolor{yellow} 1.1696 & \cellcolor{yellow} 22.81 \\
 Nb932-bare     & 10.2890 & 2.4821E-21 & ~4.72 & 32.61 & 4.5229E-03 & ~30.32 & \cellcolor{yellow} 1.1127 & \cellcolor{yellow} 30.84 \\
 Co592-bare     & 12.1390 & 1.6686E-22 & ~7.10 & 33.11 & 1.9395E-03 & ~33.58 & \cellcolor{yellow} 1.0172 & \cellcolor{yellow} 34.46 \\
 Mn552-bare     & 12.1420 & 4.7562E-23 & ~8.13 & 32.36 & 2.0609E-03 & ~32.34 & \cellcolor{yellow} 0.8601 & \cellcolor{yellow} 33.49 \\
 Mn552-Cdtk/B4C & 12.1450 & 4.0170E-23 & ~5.01 & 32.40 & 1.9708E-03 & ~32.53 & \cellcolor{yellow} 0.9740 & \cellcolor{yellow} 33.06 \\
 Zr902-Cd       & 13.5890 & 1.6160E-21 & ~3.40 & 33.12 & 9.9648E-04 & ~34.08 & \cellcolor{yellow} 1.1699 & \cellcolor{yellow} 34.39 \\
 Ni582-bare     & 14.0830 & 1.5485E-22 & ~5.87 & 33.06 & 4.1111E-05 & ~36.92 & \cellcolor{yellow} 1.0997 & \cellcolor{yellow} 37.51 \\
\hline \hline
\end{tabular}
\vspace{-3mm}
\end{table*}

\begin{figure}[htbp]
\vspace{-5mm}
\includegraphics[width=0.99\columnwidth]{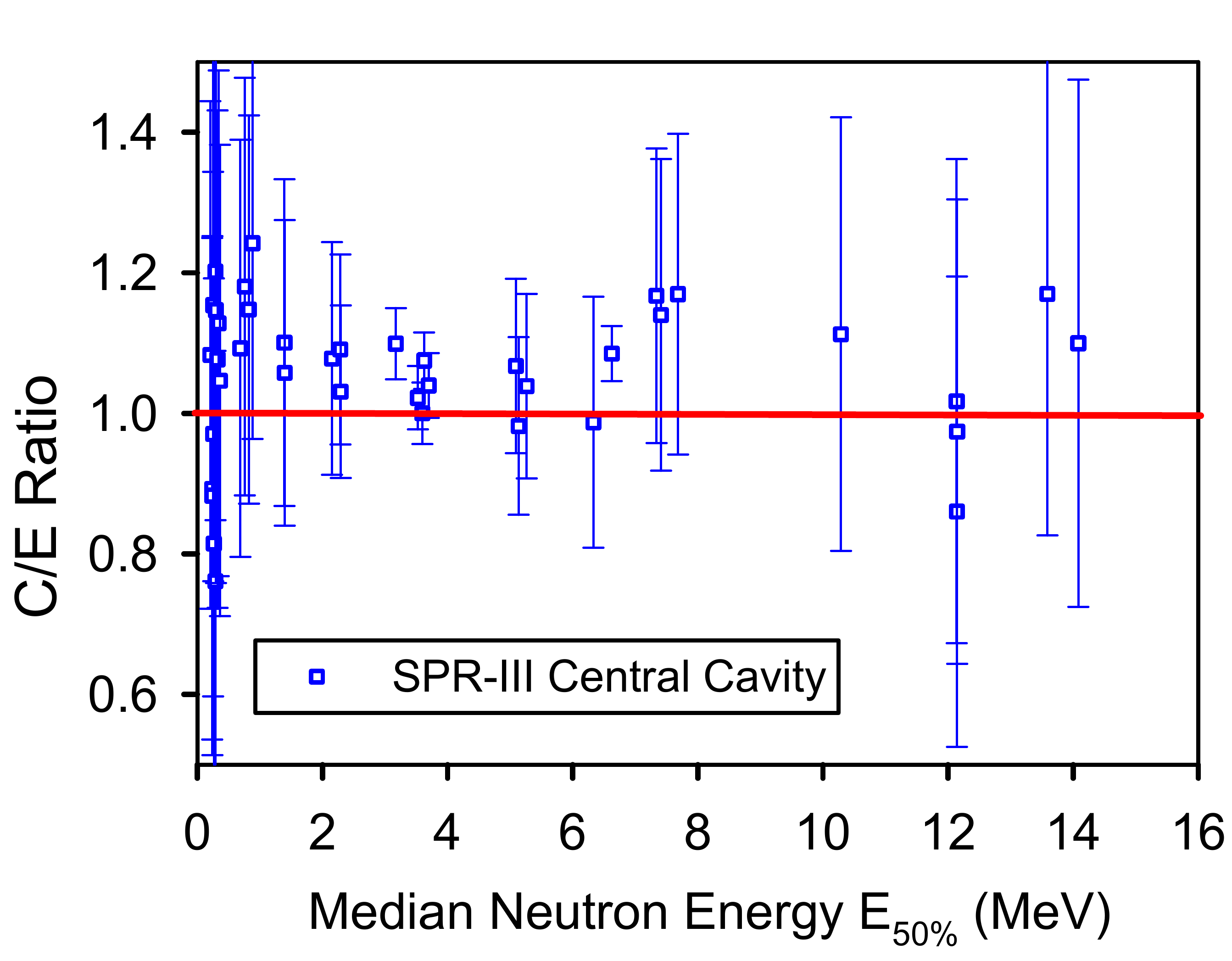}
\vspace{-4mm}
\caption{(Color online) CoE$^{SI}$$_{\alpha }$$_{/Ni58p}$ Ratios for the calculated spectrum in the SPR-III fast burst reactor central cavity reference benchmark field.}
\label{fig:image37}
\vspace{-2mm}
\end{figure}

Column 5 of Table~\ref{Table_Long_XXII} shows the contribution from the \textit{a priori} spectrum to the uncertainty in the spectrum-averaged cross sections. The \textit{a priori} spectrum is based solely on a calculation and, while the statistical uncertainty in this Monte Carlo calculation was typically less than 1~\% in a given energy bin, the uncertainty in the modeling and in the transport cross sections result is a very large overall spectrum uncertainty - as was seen in Fig.~\ref{fig:image37}. Since the uncertainty in the calculated spectral index has the spectrum appearing in both the numerator and the denominator of the metric, a Monte Carlo approach was used to propagate the spectrum uncertainty through this ratio of integrals. This methodology is described in reference~\cite{Gri18}. The use of spectral indices to make the C/E comparison means that the spectrum contribution to the SI uncertainty can be much smaller for reactions that have the dominant response in the same spectral region as the reference $^{58}$Ni(n,p)$^{58}$Co reaction. This is clearly seen in the column 7 spectrum contribution to the SI uncertainty. Here the spectrum-contribution to the SI uncertainty becomes zero for the reference reaction, and is much reduced for reactions where the dominant sensor response is similar to that from the reference reaction. For regions where the dosimeter dominant energy-dependent response differs from that of the reference reaction, the spectrum contribution to the SI uncertainty can be more than twice as large as that seen in the calculated SACS.

Columns 8 and 9 show the overall calculated-to-experimental (C/E) ratios, CoE$^{SI}$$\alpha$$_{/Ni58p}$, and associated uncertainties. The CoE$^{SI}$$_{\alpha}$$_{/Ni58p}$ value is the primary metric reported in this benchmark field. The uncertainty shown for this C/E metric is taken to be the root-mean-square of the measurement uncertainty and the spectrum and cross section contributions to the calculated spectrum-averaged cross section uncertainty. The color-coding in the CoE$^{SI}$$_{\alpha}$$_{/Ni58p}$ columns corresponds to the criteria appearing in Table~\ref{tab:X_A_3}. Good or acceptable validation evidence was seen for 39 of the 40 measured activities in this reference field. The only discrepant reaction among this set of data were for the $^{23}$Na(n,$\gamma$)$^{24}$Na reaction and, even in this case, the CoE$^{SI}$$_{\alpha}$$_{/Ni58p}$ was within one standard deviation of unity -- but the ratio fell outside the criteria adopted in Table~\ref{tab:X_A_3} and there was a large spectrum contribution to the CoE$^{SI}$$_{\alpha}$$_{/Ni58p}$ uncertainty.

Fig.~\ref{fig:image37} shows the resulting CoE$^{SI}$$_{\alpha}$$_{/Ni58p}$ ratios for this benchmark neutron field. The x-axis represents the $E_{50}$ mean response energy for the foil/cover combination in that calculated reference field spectrum. The large uncertainties, due to the \textit{a priori} spectrum uncertainty, are clearly seen here. A small positive bias in the C/E is also seen. The usefulness of this CoE$^{SI}$$_{\alpha}$$_{/Ni58p}$ metric in the validation process is limited by the large uncertainty due to the spectrum in this neutron field. Sec.~\ref{Sec_VIII} will address another metric that can more rigorously address the uncertainty in the spectrum and thus provide more powerful validation evidence.

\subsection{ACRR Pool-type Reactor Central Cavity} \label{Sec_VII_E}

The baseline measured quantities in this reference benchmark field are end-of-irradiation (EOI) spectrum-averaged cross sections measured in operation \#10639, which was a 153.2~MJ pulsed operation that took place on 9/19/2013. In order to eliminate the effect of variations in the reactor power during the other exposures, a $^{58}$Ni(n,p)$^{58}$Co monitor activity was used to facilitate a renormalization of data gathered in the other operations to the baseline exposure. When required, corrections were made in the measured activities of other foils to convert them to represent the end-of-irradiation from the baseline 16-minute uniform power operation. The other reactor operations are detailed in Table~\ref{Table_X_E_6}. As discussed in Sec.~\ref{Sec_VII_D}, in some of the cases covers were used to shield the activation foil and shift the sensitivity of the measured response to higher energies.

\begin{table}[htbp]
\caption{Reactor operations used to support the ACRR central cavity spectral characterization.}
\label{Table_X_E_6}
 \begin{tabular}{ l c  p{1.5cm}  p{4cm}  } \hline\hline
 Date       & Oper.   & Operation              & Comment                \\
 {[d/m/y]}  & No.     & Type                   &                        \\ \hline
\T
18/09/2013  & 10636   & Steady 172.4~MJ        & Activation foil set 2: Co, Fe, Sc, Ni, and Ti \\
19/09/2013  & 10638   & Pulse  157.7~MJ        & Activation foil set 3: Zr, Mo, Ag, Au, Au, Au-dilute, and W \\
02/10/2013  & 10649   & Pulse  148.0~MJ        & Activation foil sets 4 and 5: Na, Mg, Mn, Fe, Cu, Zn, In, Nb, and Al \\
19/09/2013  & 10637   & Pulse  151.5~MJ        & Cd covered activation foil sets 6 and 7: Co, Fe, Sc, Ni, Mo, Ag, Au, and Au-dilute \\
03/10/2013  & 10650   & Pulse  152.8~MJ        & Cd covered activation foil sets 8 and 9: Na, Mg, Mn, Fe, In, Cu \\
08/04/2014  & 10870   & Steady $\approx$150~MJ & Depleted U activation foil in Cd cover and B$_{4}$C ball \\
08/04/2014  & 10871   & Steady $\approx$150~MJ & Enriched U activation foil in Cd cover and B$_{4}$C ball \\
08/04/2014  & 10872   & Steady $\approx$150~MJ & Np activation foil in Cd cover and B$_{4}$C ball \\
08/04/2014  & 10873   & Steady $\approx$150~MJ & Pu activation foil in Cd cover and B$_{4}$C ball \\ \hline\hline
\end{tabular}
\end{table}
Data were gathered for a total of 33 combinations of reactions and covers in this reference field. The baseline set of end-of-irradiation activity measurements, normalized using the $^{58}$Ni(n,p)$^{58}$Co monitor reaction used in the same reactor operation, are shown in Table~\ref{Table_Long_XXIII}. As in Sec.~\ref{Sec_VII_D}, a spectral index is defined to be the ratio of the spectrum-averaged cross sections for two specified reactions, where one is generally considered to be a reference monitor reaction. This quantity is independent of the neutron fluence. Table~\ref{Table_Long_XXIII} provides the calculated spectrum-averaged cross sections for the measured reactions, with covers, and reports both the spectrum and cross section uncertainty contributions to the calculated spectrum-averaged cross sections. The entries in the table are ordered with respect to the increasing $E_{50}$ mean response energy for the reactions. The overall calculated-to-experimental (C/E) ratios, CoE$^{SI}$$_{\alpha}$$_{/Ni58p}$, are also shown in the table. The uncertainty shown for this C/E metric is taken to be the root-mean-square of the measurement uncertainty and the spectrum and cross section contributions to the calculated spectrum-averaged cross section uncertainty. Several of the table entries have such a large uncertainty that they do not qualify as acceptable validation cases, but they have a good C/E value, so they are also not designated as discrepant. These table entries correspond to measurements that are not of validation quality, typically due to the spectrum uncertainty, and are left uncoloured.

\begin{table*}[hbtp]
\caption{Measured EOI activities and calculated quantities supporting the SI C/E for ACRR central cavity neutron benchmark field.}
\label{Table_Long_XXIII}
\begin{tabular}{l c r r r r r r r}
\\ \hline \hline
             &             & \multicolumn{2}{c}{Measured Act.} & Calc.     & \multicolumn{2}{c}{Spectral}  & \multicolumn{2}{c}{SI C/E}  \\
Entry Label  & $E_{50\%}$  & Bq/atom or & Unc.                 & SACS      & Index (SI)& Unc.      &  Value & Unc. \\
             & [MeV]       & Fiss./atom & [\%]                 & Unc.[\%]  &           & [\%]      &        & [\%]  \\
\hline
\T
 Sc45g-bare      & 4.7701E-08  & 7.3161E-16 & 2.97  & 43.79 & 8.9065E+01 & 50.05 &  \cellcolor{white} 1.0063 &  \cellcolor{white} 50.26 \\
 Na23g-bare      & 6.0353E-08  & 1.8587E-15 & 2.97  & 38.98 & 1.7629E+00 & 45.73 &  \cellcolor{white} 1.0532 &  \cellcolor{white} 45.96 \\
 Mn55g-bare      & 6.8033E-08  & 3.4622E-13 & 2.76  & 36.16 & 5.1304E+01 & 42.67 &  \cellcolor{white} 0.9531 &  \cellcolor{white} 42.91 \\
 Fe58g-bare      & 7.2933E-08  & 8.3270E-17 & 2.69  & 33.19 & 5.4350E+00 & 39.79 &  \cellcolor{white} 1.0160 &  \cellcolor{white} 40.03 \\
 Co59g-bare      & 1.2520E-07  & 5.6707E-17 & 4.83  & 27.78 & 1.7960E+02 & 34.44 & \cellcolor{yellow} 1.1392 & \cellcolor{yellow} 25.58 \\
 Sc45g-Cd        & 1.7900E-06  & 1.1446E-16 & 2.97  & 28.59 & 1.3490E+01 & 34.29 & \cellcolor{yellow} 0.9746 & \cellcolor{yellow} 34.59 \\
 Au197g-bare     & 3.0900E-06  & 5.6538E-13 & 3.77  & 25.63 & 2.4586E+03 & 32.11 & \cellcolor{yellow} 1.1177 & \cellcolor{yellow} 32.52 \\
 Au197g-Cd       & 3.3100E-06  & 4.7342E-13 & 3.77  & 30.74 & 2.1328E+03 & 36.15 & \cellcolor{yellow} 1.1580 & \cellcolor{yellow} 36.51 \\
 Ag109g-bare     & 3.5260E-06  & 8.7524E-17 & 4.30  & 20.00 & 2.9347E+01 & 28.40 & \cellcolor{yellow} 0.9297 & \cellcolor{yellow} 28.94 \\
 Ag109g-Cd       & 4.8240E-06  & 5.7839E-17 & 4.29  & 25.61 & 2.3331E+01 & 31.29 & \cellcolor{yellow} 1.1184 & \cellcolor{yellow} 31.77 \\
 Na23g-Cd        & 6.4680E-06  & 4.2940E-16 & 2.97  & 18.52 & 3.6902E-01 & 24.23 & \cellcolor{yellow} 0.9544 & \cellcolor{yellow} 24.67 \\
 W186g-bare      & 1.1850E-05  & 2.2104E-13 & 2.50  & 18.33 & 3.2344E+02 & 26.52 & \cellcolor{yellow} 1.0251 & \cellcolor{yellow} 26.87 \\
 Co59g-Cd        & 9.5720E-05  & 1.8963E-17 & 4.83  & 13.66 & 6.9245E+01 & 16.41 &    \cellcolor{lred} 1.3135 &    \cellcolor{lred} 17.46 \\
 Mn55g-Cd        & 1.3490E-04  & 9.1556E-14 & 2.76  & 13.37 & 1.2863E+01 & 19.04 &  \cellcolor{green} 0.9036 &  \cellcolor{green} 19.56 \\
 Fe58g-Cd        & 1.7710E-04  & 2.5208E-17 & 2.69  & 11.82 & 1.6630E+00 & 16.38 &  \cellcolor{green} 1.0270 &  \cellcolor{green} 16.96 \\
 rmleu-Cdtk/B4C  & 3.3292E-01  & 3.1399E-09 & 4.03  & ~8.11 & 3.3913E+01 & ~4.79 &  \cellcolor{green} 0.9324 &  \cellcolor{green} ~7.18 \\
 rmlpu-Cdtk/B4C  & 6.9648E-01  & 3.1557E-09 & 4.03  & ~7.94 & 3.8531E+01 & ~5.16 &  \cellcolor{green} 1.0540 &  \cellcolor{green} ~7.43 \\
 Np237f-Cdtk/B4C & 1.4277E+00  & 1.5815E-09 & 4.03  & ~8.89 & 1.8067E+01 & ~6.30 &  \cellcolor{green} 0.9865 &  \cellcolor{green} ~8.26 \\
 rmldu-Cdtk/B4C  & 2.1768E+00  & 2.8531E-10 & 4.03  & ~9.03 & 3.4254E+00 & ~3.78 &  \cellcolor{green} 1.0365 &  \cellcolor{green} ~6.55 \\
 Ti47p-bare      & 3.0036E+00  & 3.8055E-17 & 3.69  & ~9.07 & 1.7495E-01 & ~1.13 &  \cellcolor{green} 0.9505 &  \cellcolor{green} ~5.22 \\
 S32p-bare       & 3.4419E+00  & 2.7894E-17 & 4.12  & ~9.29 & 5.8302E-01 & ~0.61 &  \cellcolor{green} 1.0133 &  \cellcolor{green} ~5.45 \\
 Ni58p-bare      & 3.4929E+00  & 9.7808E-18 & 3.52  & ~9.14 & 1.0000E+00 & ~0.00 &  \cellcolor{green} 1.0000 &  \cellcolor{green} ~4.98 \\
 Zn64p-bare      & 3.5683E+00  & 4.7963E-16 & 3.69  & ~9.35 & 3.5292E-01 & ~0.59 &  \cellcolor{green} 0.9626 &  \cellcolor{green} ~5.13 \\
 Fe54p-bare      & 3.6245E+00  & 1.6954E-18 & 2.97  & ~9.18 & 7.2790E-01 & ~0.47 &  \cellcolor{green} 0.9526 &  \cellcolor{green} ~4.63 \\
 Co59p-Cd        & 5.0698E+00  & 1.7965E-19 & 6.13  & ~9.49 & 1.2094E-02 & ~2.81 &  \cellcolor{green} 1.0478 &  \cellcolor{green} ~7.61 \\
 Ti46p-bare      & 5.1988E+00  & 7.8515E-19 & 3.05  & ~9.21 & 9.6822E-02 & ~3.25 &  \cellcolor{green} 1.0193 &  \cellcolor{green} ~5.68 \\
 Ni60p-bare      & 6.0778E+00  & 6.3203E-21 & 5.66  & ~9.36 & 1.8502E-02 & ~4.11 &  \cellcolor{green} 1.0529 &  \cellcolor{green} ~7.83 \\
 Fe56p-bare      & 6.5505E+00  & 5.8136E-17 & 2.69  & ~9.23 & 9.1059E-03 & ~4.65 &  \cellcolor{green} 1.0074 &  \cellcolor{green} ~6.42 \\
 Ti48p-bare      & 7.2201E+00  & 8.8424E-19 & 2.44  & ~9.38 & 2.5831E-03 & ~5.16 & \cellcolor{yellow} 1.1116 & \cellcolor{yellow} ~6.71 \\
 Mg24p-bare      & 7.3367E+00  & 1.2890E-17 & 3.05  & ~9.38 & 1.2267E-02 & ~5.49 &  \cellcolor{green} 1.0569 &  \cellcolor{green} ~7.20 \\
 Al27a-bare      & 7.6106E+00  & 6.1585E-18 & 3.05  & ~9.57 & 6.1899E-03 & ~5.62 & \cellcolor{yellow} 1.1163 & \cellcolor{yellow} ~7.30 \\
 Nb932-bare      & 1.0266E+01  & 2.5526E-19 & 2.83  & ~9.94 & 3.5065E-03 & ~7.08 &  \cellcolor{green} 0.9372 &  \cellcolor{green} ~8.40 \\
 Zr902-bare      & 1.3553E+01  & 1.7102E-19 & 5.48  & 13.77 & 8.2414E-04 & 12.09 &  \cellcolor{green} 1.0217 &  \cellcolor{green} 13.73 \\
\hline \hline
\end{tabular}
\end{table*}

\begin{figure}[htbp]
\includegraphics[width=\columnwidth]{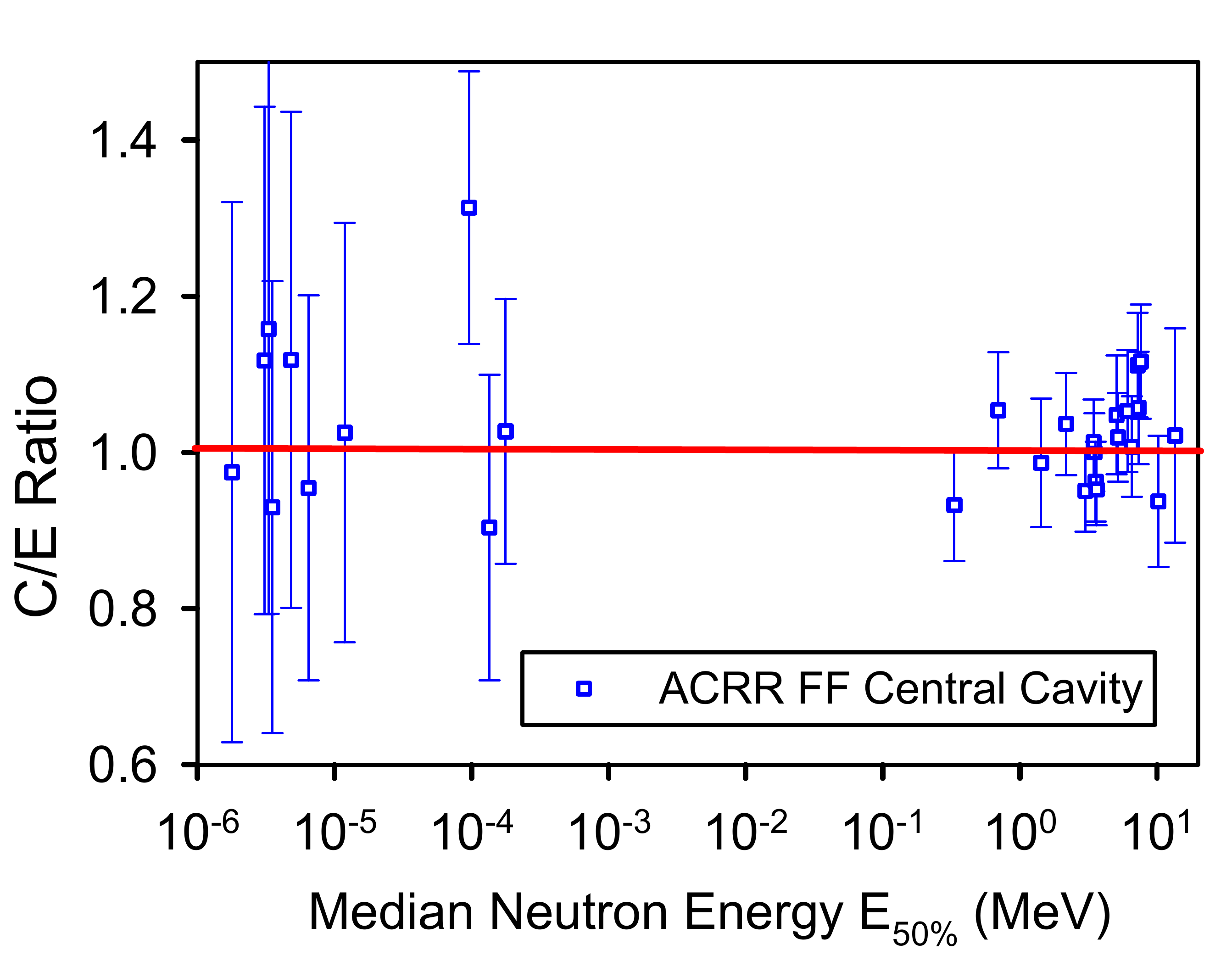}
\caption{(Color online) CoE$^{SI}$$_{\alpha}$$_{/Ni58p}$ Ratios for the Calculated Spectrum in the ACRR Central Cavity Reference Benchmark Field.}
\label{fig:image38}
\end{figure}

Fig.~\ref{fig:image38} shows the resulting C/E ratios for this benchmark neutron field. The reactions with a dominant high energy response, similar to that of the $^{58}$Ni(n,p)$^{58}$Co reference reaction, show smaller uncertainty bounds and general good agreement. The low energy dominant response reactions show much larger uncertainty bounds and considerable scatter in their central values.

\subsection{Other ``Bucket''-Modifying Environments in the ACRR Reference Reactor Fields} \label{Sec_VII_F}

C/E ratios for the spectral indices in several other reference reactor fields have been considered. These neutron fields and activity measurements include:
\begin{enumerate}
 \item  ACRR LB44; 35 measured activities
 \item  ACRR PLG; 35 measured activities
 \item  ACRR Cd-poly; 31 measured activities
 \item  FREC-II; 33 measured activities
%\item  FBR 6-in leakage; 31 measured activities
\end{enumerate}

The utility of these data is best addressed in the context of the consistency of the measured activities in Sec.~\ref{Sec_VIII_F} rather than as isolated spectral indices.

\subsection{Mol-BR1 Mark-III Irradiation Facility} \label{Sec_VII_G}
\begin{figure}[htbp]
  \includegraphics[width=\columnwidth]{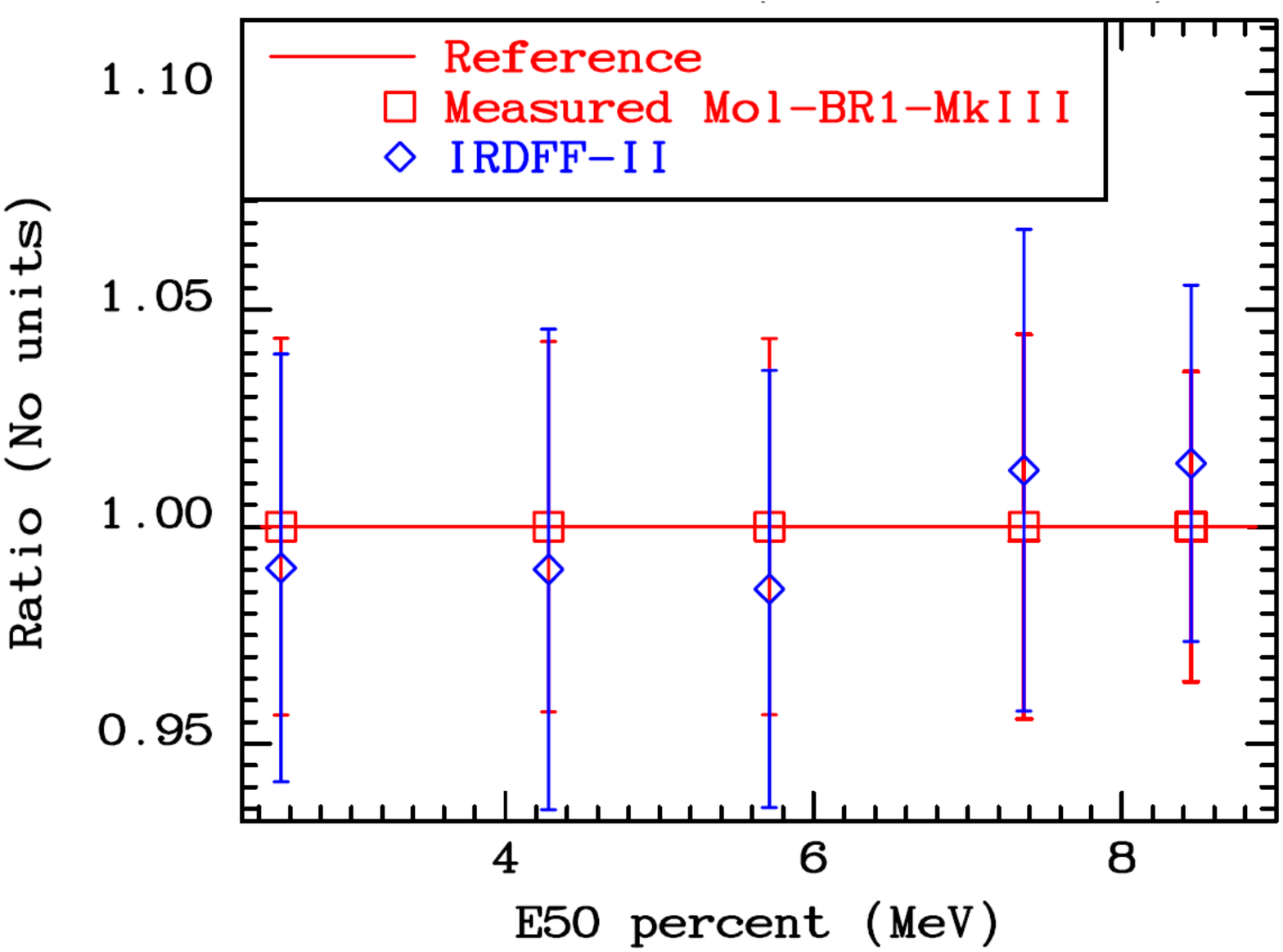}
  \caption{(Color online) C/E Spectral Index Ratios Relative to $^{58}$Ni(n,p) for the Calculated Spectrum in the Mol-BR1-Mark-III Reference Benchmark Field.}
  \label{Figure_Mol-BR1-MkIII}
\end{figure}

New measurements from the BR1 reactor at the Belgian Nuclear Research Centre SCK$\bullet$CEN in Mol were published recently~\cite{Wag16}.
The aim of the work was to make an update of the Mark-III spectrum and flux, using state of the art computing tools and certified neutron activation measurements. The \mbox{IRDFF-v1.02} library, which was available at the time, was used. The measured flux calibration factors were scaled by the corresponding calculated cross sections listed in the same publication to recover the measured reaction rates, which in turn were normalized by the $^{58}$Ni(n,p) reaction rate to remove the absolute flux dependence. The ratios were compared to the calculated values using the \mbox{IRDFF-II} cross sections, as shown in Table~\ref{Table_Mol-BR1-MkIII} and Fig.~\ref{Figure_Mol-BR1-MkIII}. The figure undoubtedly displays excellent agreement between measurements and calculations.

\begin{table}[htbp]
\vspace{-2mm}
\caption{Measured and calculated SI in the Mark-III facility of the BR1 reactor in Mol.}
\label{Table_Mol-BR1-MkIII}
\small
\begin{tabular}{ l c rr rr rr } \hline\hline
 Entry Label   &$E_{50\%}$& \multicolumn{2}{c}{Measured SI } & \multicolumn{2}{c}{Calculated SI} & \multicolumn{2}{c}{SI C/E} \&   \\
 and monitor   & [MeV]    & \multicolumn{2}{c}{\& Unc.~[\%]} & \multicolumn{2}{c}{\& Unc.~[\%] } & \multicolumn{2}{c}{Unc.~[\%]  } \\ \hline
\T
 In115nm/Ni58p & 2.54    & 1.816   & 4.34 & 1.798  & 2.43  & \cellcolor{green} 0.990 &\cellcolor{green} 5.0 \\
 Fe54p/Ni58p   & 4.28    & 0.733   & 4.27 & 0.726  & 3.61  & \cellcolor{green} 0.990 &\cellcolor{green} 5.6 \\
 Al27p/Ni58p   & 5.71    & 0.037   & 4.34 & 0.036  & 2.70  & \cellcolor{green} 0.986 &\cellcolor{green} 5.1 \\
 Fe56p/Ni58p   & 7.36    & 9.93E-3 & 4.43 & 0.010  & 3.22  & \cellcolor{green} 1.013 &\cellcolor{green} 5.5 \\
 Al27a/Ni58p   & 8.45    & 6.44E-3 & 3.57 &6.53E-3 & 1.90  & \cellcolor{green} 1.015 &\cellcolor{green} 4.0 \\
 \hline \hline
\end{tabular}
\vspace{-2mm}
\end{table}

\subsection{LR-0 Reactor at \v{R}e\v{z}} \label{Sec_VII_H}
The measured quantities in the experiments are reaction rate ratios and neutron spectra. Irradiations were carried out at an atmospheric pressure and room temperature, using fairly large samples due to the low neutron flux. It was shown that the neutron energy spectrum above 6~MeV is indistinguishable from the $^{235}$U~PFNS~\cite{Kos18_5}. Therefore, when corrected, the cross sections of reactions with the threshold over 6~MeV averaged in the LR-0 spectrum correspond to spectrum-averaged cross sections in $^{235}$U~PFNS~\cite{Kos18_3}. Corrections are based on the ratio between the share of neutrons over 6~MeV in LR-0 spectrum and in the $^{235}$U~PFNS. Another correction, taking into account the difference between LR-0 and PFNS, which is defined as a ratio between cross section averaged in reactor field and the $^{235}$U~PFNS field, is applied as well~\cite{Kos17_2}.

The neutron transport in large samples for correct evaluation of significant flux loss effect, as well as correction to flux level, are determined using well validated mathematical model of the irradiation assembly.

The efficiencies of the semiconductor detector were determined using a precise mathematical model of the used HPGe~\cite{Kos17_2}, which has been compiled using experimentally determined HPGe crystal parameters and its dead layer~\cite{Dry06}. The same arrangement of the gamma spectrum measurement was used in case of $^{252}$Cf irradiation experiments as well (see Table~\ref{tab:TableLongSACScf252_Rez}, \cite{Sch18_1}). Thanks to the usage of the same arrangement in case of $^{235}$U as well as $^{252}$Cf experiments, reaction rates ratio obtained in the $^{235}$U and $^{252}$Cf irradiations can be obtained with low uncertainties. Results are presented in the section on the $^{235}$U PFNS and $^{252}$Cf SFNS spectrum-averaged cross sections.

\subsection{TRIGA-JSI Pneumatic Tube Irradiation Channel} \label{Sec_VII_I}
The measured quantities in the two measurement campaigns were the induced activities after irradiation of bare samples and samples with covers of cadmium, boron nitride, boron carbide and enriched boron carbide. The reported unfiltered relative to filtered reaction rate ratios were derived from gamma spectrometry measurements of samples with the same geometry and material composition, performed with the same detector at the same geometry, and of the same gamma line. Therefore, the uncertainties in the derived unfiltered relative to filtered reaction rate ratios are unaffected by uncertainties in the sample material composition, detection efficiency, coincidence correction factors, or gamma emission probabilities. The uncertainties in the unfiltered relative to filtered reaction rate ratios reflect the uncertainties in the measured peak areas, sample mass and the irradiation, cooling and measurement timing. The uncertainties in the ratios of one reaction rate relative to another, on the other hand, are affected by all the above mentioned sources of uncertainty and indeed are reflected in the magnitude of the uncertainties.

For the JSI TRIGA case, a neutron spectrum adjustment procedure was carried out using the GRUPINT code, developed at the JSI~\cite{ISRD16}. The code enables the parametrization of the spectrum with a 19-parameter analytical function, and subsequent fitting of selected function parameters to measured reaction-rate ratios. Fine structure present in the spectra due to neutron resonances is taken into account through a form function, which is defined as the ratio of the calculated and the fitted analytical function. Measured reaction rate ratios can be labelled as active in the spectrum parameter fitting, or excluded. The active reaction rate ratios in the JSI TRIGA case were cadmium ratios $R_{Cd}$ (defined as the ratios of the bare relative to cadmium filtered reaction rates), for the $^{197}$Au$(n,\gamma)$ and $^{238}$U$(n,\gamma)$ reactions, and the ratios of the $^{27}$Al$(n,p)$ and $^{27}$Al$(n,\alpha)$ reaction rates relative to the $^{197}$Au$(n,\gamma)$ reaction rate. Neutron spectrum adjustment is required in order to improve the thermal to fission peak strength in the spectra resulting from the Monte Carlo calculations. In the present case the adjustment procedure resulted in an increase of the thermal spectrum by about 10~\%.

In addition to cadmium filters, filters made from boron nitride (BN), boron carbide (B$_4$C) and enriched boron carbide ($^{10}$B$_4$C) were employed as well. The filter effect on the neutron spectrum, \ie, the transmission function $t(E)$ is modelled in the GRUPINT code as a exponential attenuation function governed by two parameters:

\begin{equation}
    t(E) = e^{-\sigma_{tr}(E) n d_{eff}},
    \label{Transmission_function}
\end{equation}

\noindent where $n$ is the atom density in the filter material and $d_{eff}$ is the effective filter thickness, the first parameter, and $\sigma_{tr}(E)$ is the filter transport cross-section, defined as:

\begin{equation}
    \sigma_{tr}(E) = \sigma_{a}(E) + \xi \sigma_{s}(E),
\end{equation}

\noindent where $\sigma_{a}(E)$ is the filter absorption cross-section, $\sigma_{s}(E)$ is the filter scattering cross-section and $\xi$ is the effective scattering fraction, the second parameter. Filtered reaction rates are calculated as:

\begin{equation}
    R = \int_0^\infty \sigma(E) t(E)\varphi(E)dE,
\end{equation}

\noindent where $\sigma(E)$ is the reaction cross-section, $t(E)$ is the filter transmission function and $\varphi(E)$ is the neutron spectrum, in the present case the adjusted neutron spectrum, resulting from the adjustment procedure described above.

\begin{table*}[thbp]
%\vspace{-4mm}
  \caption{Comparison between experimental and calculated BN, B$_4$C and $^{10}$B$_4$C reaction rate ratios. The experimental results for the $^{197}$Au(n,$\gamma$) and $^{238}$U(n,$\gamma$) reactions have been used for adjustment of the BN, B$_4$C and $^{10}$B$_4$C transmission function parameters.}
  \footnotesize
    \begin{tabular}{|c|c|c|c|c|c|c|c|c|c|}
    \hline
\multirow{2}{*}{React.} & $R_{BN}$, & $R_{BN}$, & \textbf{Diff.} & $R_{B_4C}$, & $R_{B_4C}$, & \textbf{Diff.}  & $R_{^{10}B_4C}$,  & $R_{^{10}B_4C}$, & \textbf{Diff.} \\
 & \textbf{exp.} & \textbf{calc.} & [\%] & \textbf{exp.} & \textbf{calc.} & [\%] & \textbf{exp.} & \textbf{calc.} & [\%] \\
 \hline
\T
 $^{197}$Au & 8.17$\pm$1.2\%&8.28$\pm$8.52\% &                   1.3 &  27.81$\pm$1.2\% &  28.76$\pm$13.8\% &                    3.4 & 256.04$\pm$1.1\% & 258.65$\pm$4.52\% &  1.0  \\
 $^{238}$U  & 2.60$\pm$1.1\%&2.53$\pm$7.01\% &                  -2.7 &   5.44$\pm$1.2\% &   5.18$\pm$6.00\% &                   -4.7 &  51.39$\pm$1.0\% &  48.28$\pm$4.77\% & -6.0  \\ \hline
\T
 $^{238}$U  & 2.63$\pm$1.4\%&2.53$\pm$7.01\% &\cellcolor{green} -3.7 &   5.32$\pm$1.3\% &   5.18$\pm$6.00\% &\cellcolor{green}  -2.5 &  45.47$\pm$1.3\% &  48.28$\pm$4.77\%&\cellcolor{green} 6.2\\
 $^{238}$U  & 2.46$\pm$1.1\%&2.53$\pm$7.01\% &\cellcolor{green}  3.1 &   5.33$\pm$1.1\% &   5.18$\pm$6.00\% &\cellcolor{green}  -2.8 &  49.66$\pm$1.0\% &  48.28$\pm$4.77\%&\cellcolor{green}-2.8\\
 $^{55}$Mn  &39.13$\pm$2.5\%&35.83$\pm$5.17\% &\cellcolor{green} -8.4 &  48.26$\pm$2.6\% & 47.04$\pm$5.54\% &\cellcolor{green}  -2.5 & 183.11$\pm$3.0\% & 147.43$\pm$6.64\%&\cellcolor{lred} -19.5\\
 $^{232}$Th & 4.07$\pm$2.2\%& 4.19$\pm$18.9\% &\cellcolor{green}  2.7 &   6.50$\pm$2.2\% &  6.42$\pm$18.2\% &\cellcolor{green}  -1.2 &  31.13$\pm$2.1\% &  30.39$\pm$11.3\%&\cellcolor{green}-2.4\\
 $^{23}$Na  &92.77$\pm$2.3\%&87.38$\pm$3.99\% &\cellcolor{green} -5.8 & 136.03$\pm$2.2\% &122.64$\pm$4.07\% &\cellcolor{yellow} -9.8 & 283.94$\pm$2.2\% & 229.53$\pm$4.46\%&\cellcolor{lred} -19.2\\
\hline\hline
  \end{tabular}
  \label{Table_JSI_epithermal}
%\vspace{-3mm}
\end{table*}

\begin{figure*}[htbp]
\vspace{-3mm}
\centering
    \includegraphics[width=\textwidth]{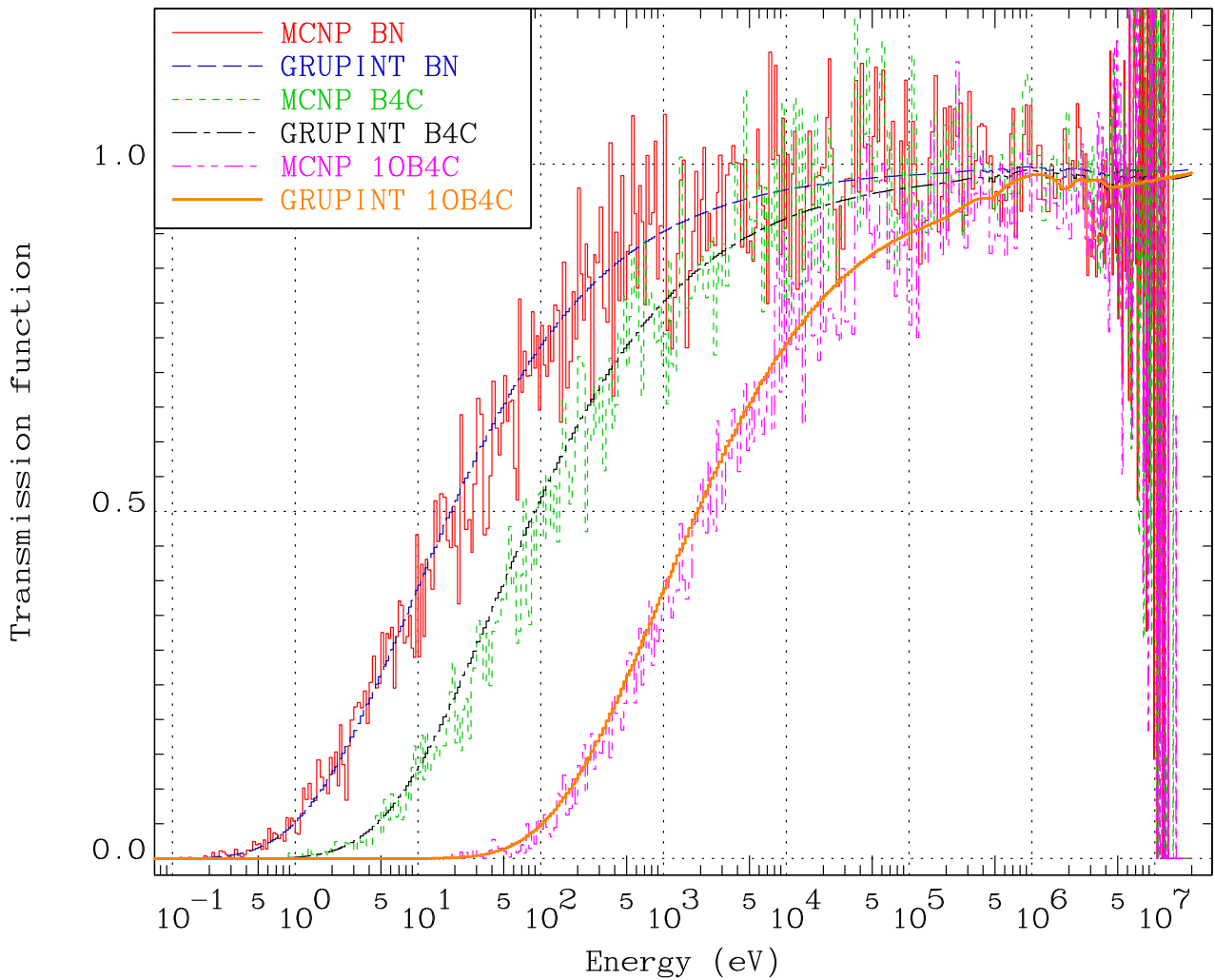}
\vspace{-2mm}
    \caption{(Color online) Comparison between the filter transmission functions $t(E)$ for the BN, B$_4$C and $^{10}$B$_4$C filters calculated with MCNP and as modelled in the GRUPINT code, with the function parameters adjusted to measured $^{197}$Au$(n,\gamma)$ and $^{238}$U$(n,\gamma)$ reaction rate ratios.}
    \label{Filter_t(E)}
\vspace{-3mm}
\end{figure*}

The BN, B$_4$C and $^{10}$B$_4$C filter cross sections were generated on the basis of \mbox{ENDF/B-VIII.0} nuclear data, the filter chemical compositions and the measured filter densities, using the \mbox{PREPRO-2018} nuclear data processing code package. The effective filter thicknesses $d_{eff}$ were first adjusted so that the transmission function shapes approximately matched the shapes calculated using MCNP, and second, finely adjusted on the basis of the experimental BN, B$_4$C and $^{10}$B$_4$C ratios for the $^{197}$Au$(n,\gamma)$ and $^{238}$U$(n,\gamma)$ reactions. The effective scattering fractions $\xi$, were adjusted on the basis of the filter transmission functions calculated with MCNP; however, the fractions $\xi$ were seen to be very close to zero, implying that neutron scattering has a low impact on the overall filter transmission function. Fig.~\ref{Filter_t(E)} displays the filter transmission functions calculated using MCNP and adjusted on the basis of experimental data.

It is very important to note, that since in the TRIGA-JSI case a subset of the experimental data is used for neutron spectrum adjustment or the adjustment of the parameters of the filter transmission functions, these data cannot be part of the TRIGA-JSI data validation case. The relevance of the TRIGA-JSI case is in the comparison of calculated and measured reaction rate ratios for other measured nuclear reactions in the experimental campaigns, which are included in the \mbox{IRDFF-II} library.
%Additionally, we report a comparison of experimental and calculated BN, B$_4$C and $^{10}$B$_4$C ratios for a specific selection of $(n,\gamma)$ reactions not included in the \mbox{IRDFF-II} library, but of potential interest for epithermal neutron dosimetry~\cite{Radulovic_NENE2018}.
Additional measurements of a specific selection of $(n,\gamma)$ reactions of potential interest for epithermal neutron dosimetry were made~\cite{Radulovic_NENE2018}, but they are outside the scope of the present analysis.
Table~\ref{Table_JSI_epithermal} reports the experimental and calculated BN, B$_4$C and $^{10}$B$_4$C reaction rate ratios.

\subsection{Uranium Fueled ICSBEP Assemblies from the ICSBEP Handbook} \label{Sec_VII_J}
In this section the C/E results for the spectral indices (SI) for the uranium-fueled ICSBEP assemblies (HMF001, HMF028 and IMF007) are summarized.  As indicated earlier the simulation models are derived from the specifications provided in the ICSBEP Handbook.  However the Handbook does not provide a summary of experimental reaction rate data.  The reaction rate data cited here are derived from a variety of sources (\eg, References~\cite{CIELO:2014,ENDF8,You07,Fra99,ENDF-202}).

\begin{figure}[htbp]
\vspace{-2mm}
\includegraphics[width=\columnwidth]{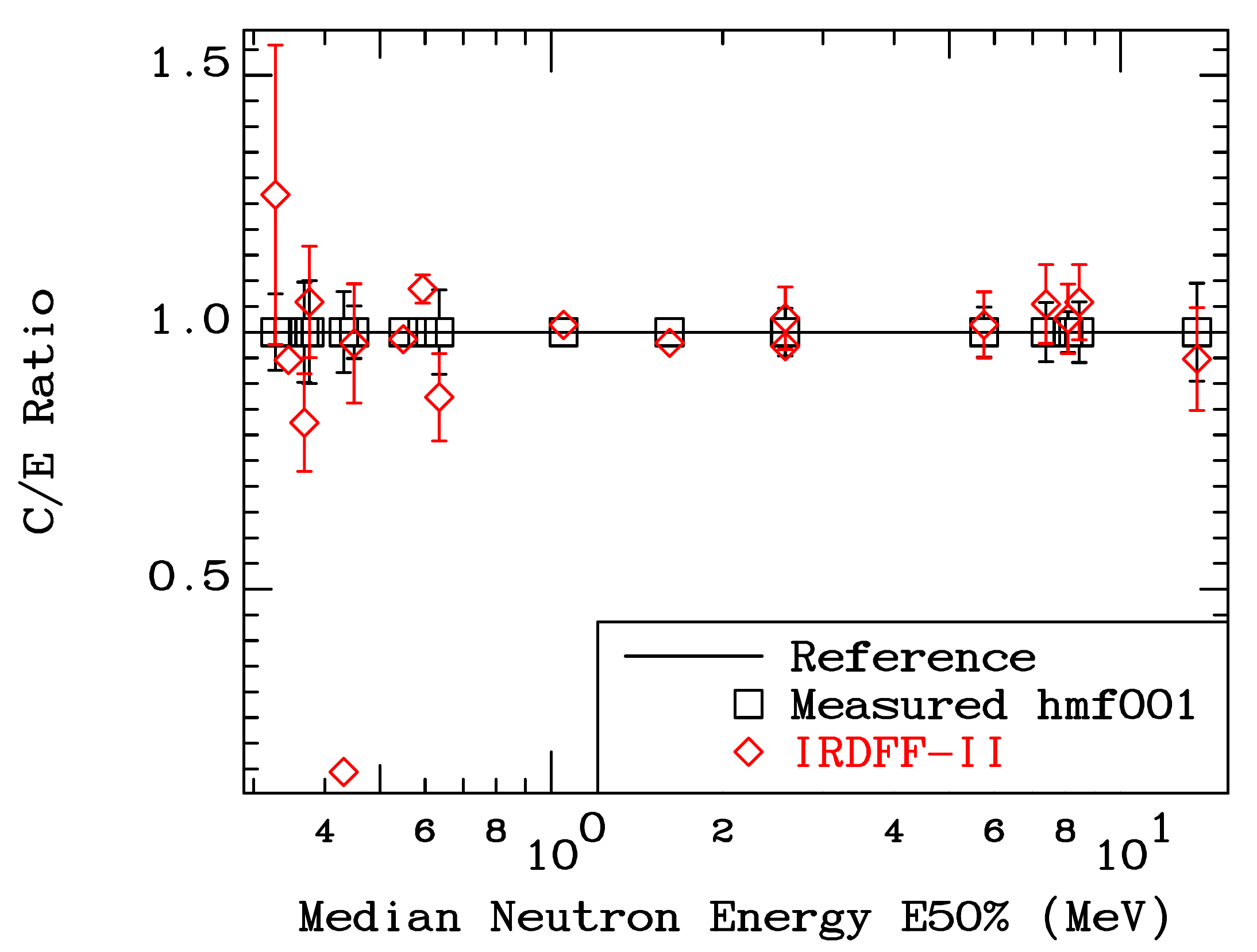}
\caption{(Color online) Ratio of Calculated and Measured SI in the central region of HMF001 (Godiva).}
\label{FigureICSBEP_AAA}
%\vspace{-2mm}
\end{figure}

Table~\ref{TableICSBEP_AAA} provides results for HMF001, with the corresponding SI C/E values shown in Fig.~\ref{FigureICSBEP_AAA}.  Aside from some scatter in the low energy activation foil capture reactions, where C/E values oscillate between 0.80 and 1.30, the calculated SI C/E are generally in good agreement with experiment, with all values within 10~\% of unity for $E_{50\%}$ values up to 8~MeV.  Even SI C/E for the high energy $^{63}$Cu(n,2n) reaction is in good agreement with experiment, given the large measurement and stochastic uncertainties for this reaction.

\begin{table*}[htbp]
\vspace{-4mm}
\caption{Measured and calculated SI in the central region of HMF001 (Godiva).}
\label{TableICSBEP_AAA}
\begin{tabular}{ l c rl rl rl r} \hline \hline
 Entry Label   &          & \multicolumn{2}{c}{Measured   } & \multicolumn{2}{c}{Calculated } & \multicolumn{2}{c}{SI C/E}         & Diff \\
 and monitor   &$E_{50\%}$& \multicolumn{2}{c}{SI and Unc.} & \multicolumn{2}{c}{SI and Unc.} & \multicolumn{2}{c}{Value and Unc.} & [\%] \\
               & MeV      &           & [\%]  &           & [\%]  &                            & [\%]                    &                         \\ \hline
\T
 Mn55g/U5f     & 0.328    & ~ 2.70E-3 & ~7.41 & ~ 3.42E-3 & 21.79 &\cellcolor{yellow} ~ 1.2673 &\cellcolor{yellow} 23.02 &\cellcolor{yellow} 26.73 \\
 Au197g/U5f    & 0.345    & ~ ~0.1000 & ~2.00 & ~ ~0.0946 & ~1.29 &\cellcolor{yellow} ~ 0.9458 &\cellcolor{yellow} ~2.38 &\cellcolor{yellow} -5.42 \\
 Ta181g/U5f    & 0.368    & ~ ~0.1230 & ~9.76 & ~ ~0.1013 & ~6.21 &\cellcolor{green}  ~ 0.8238 &\cellcolor{green}  11.56 &\cellcolor{green} -17.62 \\
 Nb93g/U5f     & 0.376    & ~ ~0.0300 & 10.00 & ~ ~0.0318 & ~2.11 &\cellcolor{green}  ~ 1.0588 &\cellcolor{green}  10.22 &\cellcolor{green}   5.88 \\
 Co59g/U5f     & 0.432    & ~ ~0.0380 & ~7.89 & ~ 5.48E-3 & ~2.76 &\cellcolor{lred}    ~ 0.1442 &\cellcolor{lred}    ~8.36 &\cellcolor{lred}   -85.58 \\
 Cu63g/U5f     & 0.450    & ~ ~0.0117 & ~5.13 & ~ ~0.0114 & 10.70 &\cellcolor{green}  ~ 0.9780 &\cellcolor{green}  11.86 &\cellcolor{green}  -2.20 \\
 U8g/U8f       & 0.549    & ~ ~0.4760 &       & ~ ~0.4695 & ~2.20 &\cellcolor{green}  ~ 0.9863 &\cellcolor{green}  ~2.20 &\cellcolor{green}  -1.37 \\
 In115gm/U5f   & 0.594    & ~ ~0.1168 & ~0.68 & ~ ~0.1266 & ~2.48 &\cellcolor{lred}    ~ 1.0841 &\cellcolor{lred}    ~2.57 &\cellcolor{lred}     8.41 \\
 La139g/U5f    & 0.635    & ~ 7.30E-3 & ~8.22 & ~ 6.38E-3 & ~5.20 &\cellcolor{green}  ~ 0.8736 &\cellcolor{green}  ~9.73 &\cellcolor{green} -12.64 \\
 Pu9f/U5f      & 1.051    & ~ ~1.3650 &       & ~ ~1.3841 & ~1.73 &\cellcolor{green}  ~ 1.0140 &\cellcolor{green}  ~1.73 &\cellcolor{green}   1.40 \\
 Np7f/U5f      & 1.614    & ~ ~0.8516 & ~1.41 & ~ ~0.8338 & ~2.12 &\cellcolor{green}  ~ 0.9791 &\cellcolor{green}  ~2.53 &\cellcolor{green}  -2.09 \\
 U8f/U5f       & 2.576    & ~ ~0.1629 &       & ~ ~0.1584 & ~1.71 &\cellcolor{green}  ~ 0.9726 &\cellcolor{green}  ~1.71 &\cellcolor{green}  -2.74 \\
 U8f/P31p      & 2.576    & ~ ~8.9460 & ~4.61 & ~ ~9.1893 & ~3.67 &\cellcolor{green}  ~ 1.0272 &\cellcolor{green}  ~5.89 &\cellcolor{green}   2.72 \\
 Al27p/P31p    & 5.757    & ~ ~0.1126 & ~4.88 & ~ ~0.1142 & ~4.03 &\cellcolor{green}  ~ 1.0143 &\cellcolor{green}  ~6.33 &\cellcolor{green}   1.43 \\
 Fe56p/P31p    & 7.397    & ~ ~0.0310 & ~5.80 & ~ ~0.0327 & ~4.39 &\cellcolor{green}  ~ 1.0545 &\cellcolor{green}  ~7.27 &\cellcolor{green}   5.45 \\
 U82/U5f       & 8.082    & ~ 7.73E-3 & ~4.00 & ~ 7.93E-3 & ~5.25 &\cellcolor{green}  ~ 1.0260 &\cellcolor{green}  ~6.60 &\cellcolor{green}   2.60 \\
 Al27a/P31p    & 8.461    & ~ ~0.0203 & ~5.91 & ~ ~0.0215 & ~3.55 &\cellcolor{green}  ~ 1.0584 &\cellcolor{green}  ~6.89 &\cellcolor{green}   5.84 \\
 Cu632/P31p    &13.619    & ~ 2.94E-3 & ~9.52 & ~ 2.79E-3 & ~4.49 &\cellcolor{green}  ~ 0.9477 &\cellcolor{green}  10.53 &\cellcolor{green}  -5.23 \\ \hline \hline
\end{tabular}
\vspace{-3mm}
\end{table*}

\begin{figure}[!htbp]
\vspace{-2mm}
\includegraphics[width=\columnwidth]{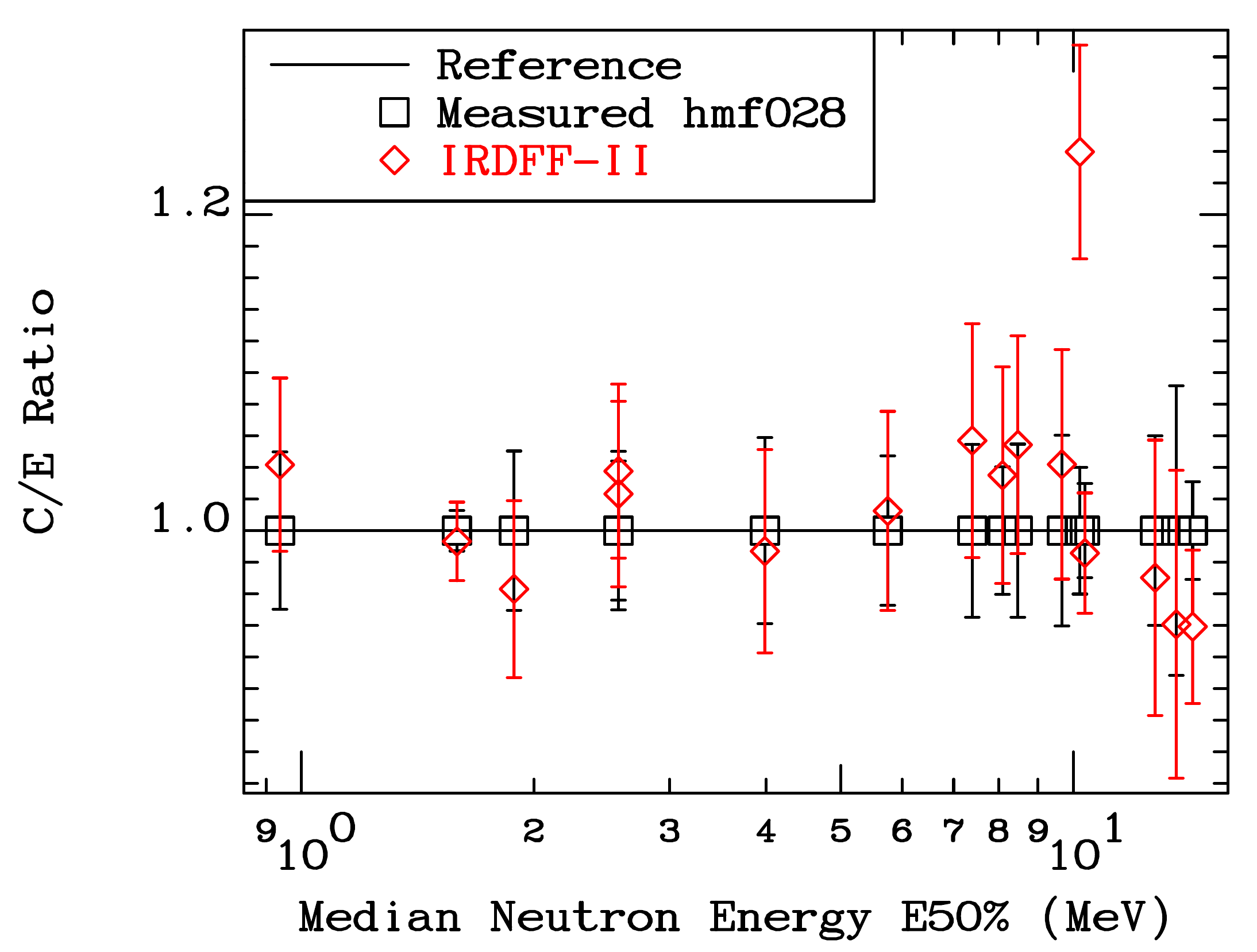}
\caption{(Color online) Ratio of Calculated and Measured SI in the central region of HMF028 (Flattop-25).}
\label{FigureICSBEP_BBB}
%\vspace{-2mm}
\end{figure}

Table~\ref{TableICSBEP_BBB} provides results for HMF028, with the corresponding SI C/E values shown in Fig.~\ref{FigureICSBEP_BBB}.  Aside from the $^{169}$Tm(n,2n) result (at E50\%$\approx 10$~MeV), which seems anomalous given the excellent $^{197}$Au(n,2n) result at nearly the same $E_{50\%}$ energy, these C/E results are excellent.

\begin{table*}[htbp]
\vspace{-4mm}
\caption{Measured and calculated SI in the central region of HMF028 (Flattop-25).}
\label{TableICSBEP_BBB}
\begin{tabular}{ l c rl rl rl r} \hline \hline
 Entry Label   &          & \multicolumn{2}{c}{Measured   } & \multicolumn{2}{c}{Calculated } & \multicolumn{2}{c}{SI C/E}         & Diff \\
 and monitor   &$E_{50\%}$& \multicolumn{2}{c}{SI and Unc.} & \multicolumn{2}{c}{SI and Unc.} & \multicolumn{2}{c}{Value and Unc.} & [\%] \\
               & MeV      &           & [\%]  &           & [\%]  &                            & [\%]                    &                         \\ \hline
\T
 Pu9f/U5f      & 0.938    & ~  1.3070 & 4.97  & ~  1.3615 & 1.73  &\cellcolor{green} ~ 1.0417 &\cellcolor{green}  5.27 &\cellcolor{green}   4.17  \\
 Np7f/U5f      & 1.590    & ~  0.7804 & 1.28  & ~  0.7751 & 2.12  &\cellcolor{green} ~ 0.9932 &\cellcolor{green}  2.48 &\cellcolor{green}  -0.68  \\
 Am1f/U5f      & 1.886    & ~  0.7540 & 5.04  & ~  0.7261 & 2.92  &\cellcolor{green} ~ 0.9631 &\cellcolor{green}  5.82 &\cellcolor{green}  -3.69  \\
 U8f/U5f       & 2.575    & ~  0.1397 & 5.01  & ~  0.1450 & 1.72  &\cellcolor{green} ~ 1.0377 &\cellcolor{green}  5.30 &\cellcolor{green}   3.77  \\
 U8f/P31p      & 2.575    & ~  8.9810 & 4.41  & ~  9.1903 & 3.67  &\cellcolor{green} ~ 1.0233 &\cellcolor{green}  5.74 &\cellcolor{green}   2.33  \\
 S32p/U5f      & 3.985    & ~  0.0306 & 5.88  & ~  0.0302 & 2.79  &\cellcolor{green} ~ 0.9870 &\cellcolor{green}  6.51 &\cellcolor{green}  -1.30  \\
 Al27p/P31p    & 5.749    & ~  0.1128 & 4.74  & ~  0.1135 & 4.03  &\cellcolor{green} ~ 1.0125 &\cellcolor{green}  6.22 &\cellcolor{green}   1.25  \\
 Fe56p/P31p    & 7.398    & ~  0.0306 & 5.47  & ~  0.0324 & 4.39  &\cellcolor{green} ~ 1.0570 &\cellcolor{green}  7.02 &\cellcolor{green}   5.70  \\
 U82/U5f       & 8.097    & ~ 6.94E-3 & 4.03  & ~ 7.18E-3 & 5.25  &\cellcolor{green} ~ 1.0351 &\cellcolor{green}  6.62 &\cellcolor{green}   3.51  \\
 Al27a/P31p    & 8.476    & ~  0.0202 & 5.48  & ~  0.0213 & 3.56  &\cellcolor{green} ~ 1.0534 &\cellcolor{green}  6.54 &\cellcolor{green}   5.43  \\
 V51a/U5f      & 9.665    & ~ 1.11E-5 & 6.04  & ~ 1.16E-5 & 3.50  &\cellcolor{green} ~ 1.0420 &\cellcolor{green}  6.98 &\cellcolor{green}   4.20  \\
 Tm1692/U5f    &10.200    & ~ 1.47E-3 & 4.01  & ~ 1.82E-3 & 3.71  &\cellcolor{lred}   ~ 1.2398 &\cellcolor{lred}    5.47 &\cellcolor{lred}    23.98  \\
 Au1972/U5f    &10.351    & ~ 1.61E-3 & 2.98  & ~ 1.59E-3 & 2.46  &\cellcolor{green} ~ 0.9858 &\cellcolor{green}  3.87 &\cellcolor{green}  -1.42  \\
 As752/U5f     &12.767    & ~ 1.50E-4 & 6.00  & ~ 1.46E-4 & 6.69  &\cellcolor{green} ~ 0.9702 &\cellcolor{green}  8.99 &\cellcolor{green}  -2.98  \\
 Cu632/P31p    &13.597    & ~ 2.94E-3 & 9.17  & ~ 2.76E-3 & 4.82  &\cellcolor{green} ~ 0.9407 &\cellcolor{green} 10.36 &\cellcolor{green}  -5.93  \\
 Zr902/U5f     &14.279    & ~ 4.85E-5 & 3.09  & ~ 4.56E-5 & 4.14  &\cellcolor{green} ~ 0.9392 &\cellcolor{green}  5.17 &\cellcolor{green}  -6.08  \\ \hline \hline
\end{tabular}
\vspace{-3mm}
\end{table*}

\begin{figure}[!htbp]
\vspace{-2mm}
\includegraphics[width=\columnwidth]{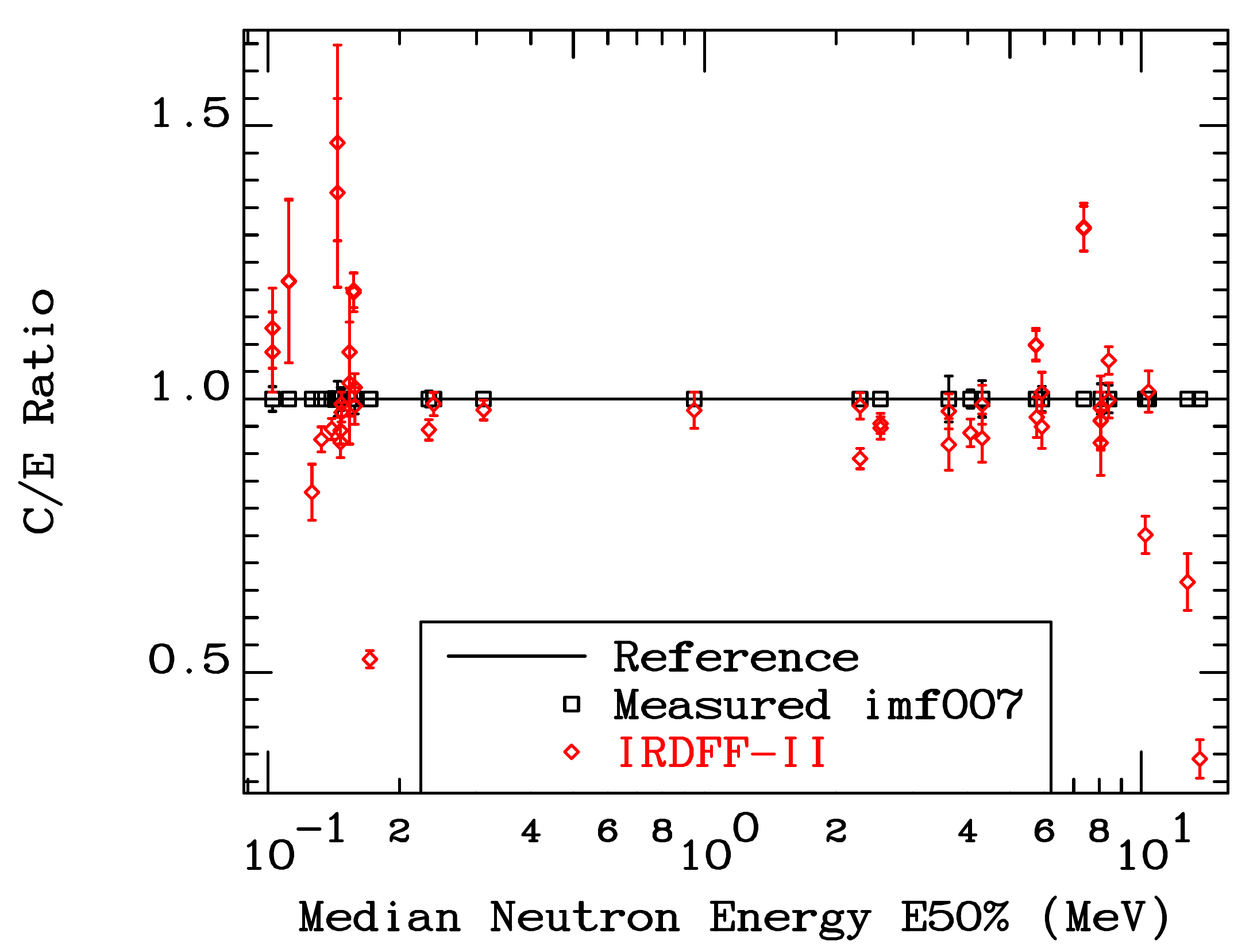}
\caption{(Color online) Ratio of Calculated and Measured SI in the central region of HMF028 (Big-Ten).}
\label{FigureICSBEP_CCC}
\vspace{-2mm}
\end{figure}

%\newpage
Table~\ref{TableICSBEP_CCC} provides results for IMF007, with the corresponding SI C/E values shown in Fig.~\ref{FigureICSBEP_CCC}. As with HMF001 there is a large scatter in the low energy activation foil capture reaction C/E values.  This large variation, both below and above unity, suggests a deficiency in the underlying measurements rather than in the source spectrum.  Unfortunately many of these measurements were made decades ago making it difficult to know whether details such as gamma transition intensity, decay half-life, possible capture branch to isomers, \textit{etc} were properly accounted for.  Once again the $^{169}$Tm(n,2n) SI C/E, near 10.3~MeV, stands out for its marked deviation from unity.% although in contrast to HMF028, where the result was too high, it is now too low.
Also once again the $^{197}$Au(n,2n) result is near unity suggesting the issue is with the measured $^{169}$Tm SACS (or a decay data deficiency) rather than problems in the IMF007 spectrum.  In general we observe good C/E results with most values being consistent with unity to within their respective uncertainties.

\begin{table*}[htbp]
\vspace{-2mm}
\caption{Measured and calculated SI in the central region of IMF007 (Big-Ten).}
\label{TableICSBEP_CCC}
\begin{tabular}{ l c rl rl rl r} \hline\hline
 Entry Label   &          & \multicolumn{2}{c}{Measured   } & \multicolumn{2}{c}{Calculated } & \multicolumn{2}{c}{SI C/E}         & Diff \\
 and monitor   &$E_{50\%}$& \multicolumn{2}{c}{SI and Unc.} & \multicolumn{2}{c}{SI and Unc.} & \multicolumn{2}{c}{Value and Unc.} & [\%] \\
               & MeV      &           & [\%]  &           & [\%]  &                            & [\%]                    &                         \\ \hline
\T
 Sc45g/U5f     & 0.102    & ~ 0.0132 & ~2.27& ~  0.0143 & ~6.36 &\cellcolor{green}  ~ 1.0857 &\cellcolor{green}  ~6.75 &\cellcolor{green}    8.57 \\
 Sc45g/Ni58p   & 0.102    & ~ 1.1000 &      & ~  1.2424 & ~6.49 &\cellcolor{green}  ~ 1.1295 &\cellcolor{green}  ~6.49 &\cellcolor{green}   12.95 \\
 Mn55g/U5f     & 0.112    & ~5.37E-3 &      & ~ 6.52E-3 & 12.25 &\cellcolor{green}  ~ 1.2147 &\cellcolor{green}  12.25 &\cellcolor{green}   21.47 \\
 Mn55g/Ni58p   & 0.112    & ~ 0.4650 &      & ~  0.5655 & 12.32 &\cellcolor{green}  ~ 1.2161 &\cellcolor{green}  12.32 &\cellcolor{green}   21.61 \\
 Ta181g/U5f    & 0.126    & ~ 0.2160 &      & ~  0.1791 & ~6.18 &\cellcolor{yellow} ~ 0.8293 &\cellcolor{yellow} ~6.18 &\cellcolor{yellow} -17.07 \\
 W186g/U5f     & 0.133    & ~ .05688 &      & ~  0.0527 & ~2.48 &\cellcolor{green}  ~ 0.9259 &\cellcolor{green}  ~2.48 &\cellcolor{green}   -7.41 \\
 B10a/U5f      & 0.136    & ~ 1.0110 & ~1.38& ~  0.9557 & ~1.50 &\cellcolor{green}  ~ 0.9453 &\cellcolor{green}  ~2.04 &\cellcolor{green}   -5.47 \\
 Fe58g/U5f     & 0.144    & ~3.10E-3 & ~3.23& ~ 4.27E-3 & 12.13 &\cellcolor{lred}    ~ 1.3773 &\cellcolor{lred}    12.55 &\cellcolor{lred}     37.73 \\
 Fe58g/Ni58p   & 0.144    & ~ 0.2520 &      & ~  0.3701 & 12.20 &\cellcolor{lred}    ~ 1.4688 &\cellcolor{lred}    12.20 &\cellcolor{lred}     46.88 \\
 Co59g/U5f     & 0.147    & ~9.50E-3 & ~2.11& ~ 8.75E-3 & ~2.27 &\cellcolor{yellow} ~ 0.9209 &\cellcolor{yellow} ~3.10 &\cellcolor{yellow}  -7.91 \\
 Co59g/Ni58p   & 0.147    & ~ 0.8050 &      & ~  0.7585 & ~2.61 &\cellcolor{yellow} ~ 0.9422 &\cellcolor{yellow} ~2.61 &\cellcolor{yellow}  -5.78 \\
 Au197g/U5f    & 0.147    & ~ 0.1670 & ~1.80& ~  0.1655 & ~1.29 &\cellcolor{green}  ~ 0.9907 &\cellcolor{green}  ~2.21 &\cellcolor{green}   -0.93 \\
 Au197g/Ni58p  & 0.147    & ~ 14.700 &      & ~ 14.3431 & ~1.83 &\cellcolor{green}  ~ 0.9757 &\cellcolor{green}  ~1.83 &\cellcolor{green}   -2.43 \\
 Cu63g/U5f     & 0.154    & ~ 0.0164 & ~0.61& ~  0.0178 & 10.75 &\cellcolor{green}  ~ 1.0858 &\cellcolor{green}  10.77 &\cellcolor{green}    8.58 \\
 Cu63g/Ni58p   & 0.154    & ~ 1.5000 &      & ~  1.5437 & 10.83 &\cellcolor{green}  ~ 1.0291 &\cellcolor{green}  10.83 &\cellcolor{green}    2.91 \\
 In115gm/U5f   & 0.157    & ~ 0.1460 &      & ~  0.1750 & ~2.64 &\cellcolor{lred}    ~ 1.1990 &\cellcolor{lred}    ~2.64 &\cellcolor{lred}     19.90 \\
 In115gm/Ni58p & 0.157    & ~ 12.700 &      & ~ 15.1752 & ~2.94 &\cellcolor{lred}    ~ 1.1949 &\cellcolor{lred}    ~2.94 &\cellcolor{lred}     19.49 \\
 U8g/U5f       & 0.158    & ~ 0.1100 & ~2.73& ~  0.1086 & ~2.09 &\cellcolor{green}  ~ 0.9873 &\cellcolor{green}  ~3.43 &\cellcolor{green}   -1.27 \\
 U8g/Ni58p     & 0.158    & ~ 9.2200 &      & ~  9.4151 & ~2.45 &\cellcolor{green}  ~ 1.0212 &\cellcolor{green}  ~2.45 &\cellcolor{green}    2.12 \\
 In113gm/U5f   & 0.171    & ~ 0.4220 &      & ~  0.2210 & ~2.95 &\cellcolor{lred}    ~ 0.5238 &\cellcolor{lred}    ~2.95 &\cellcolor{lred}    -47.62 \\
 Li6a/U5f      & 0.231    & ~ 0.7100 & ~1.41& ~  0.6544 & ~1.35 &\cellcolor{yellow} ~ 0.9217 &\cellcolor{yellow} ~1.95 &\cellcolor{yellow}  -5.64 \\
 U5f/Ni58p     & 0.240    & ~87.5000 &      & ~ 86.6913 & ~2.14 &\cellcolor{green}  ~ 0.9908 &\cellcolor{green}  ~2.14 &\cellcolor{green}   -0.92 \\
 Pu9f/U5f      & 0.312    & ~ 1.1936 & ~0.70& ~  1.1697 & ~1.74 &\cellcolor{green}  ~ 0.9799 &\cellcolor{green}  ~1.84 &\cellcolor{green}   -2.01 \\
 Np7f/U5f      & 0.946    & ~ 0.3223 & ~1.20& ~  0.3156 & ~3.14 &\cellcolor{green}  ~ 0.9794 &\cellcolor{green}  ~4.87 &\cellcolor{green}   -2.06 \\
 In115nm/U5f   & 2.269    & ~ 0.0271 &      & ~  0.0241 & ~2.09 &\cellcolor{lred}    ~ 0.8909 &\cellcolor{lred}    ~2.09 &\cellcolor{lred}    -10.91 \\
 In115nm/Ni58p & 2.269    & ~ 2.1200 &      & ~  2.0931 & ~2.46 &\cellcolor{green}  ~ 0.9873 &\cellcolor{green}  ~2.46 &\cellcolor{green}   -1.27 \\
 U8f/U5f       & 2.528    & ~ 0.0374 & ~0.90& ~  0.0357 & ~1.73 &\cellcolor{green}  ~ 0.9551 &\cellcolor{green}  24.13 &\cellcolor{green}   -4.49 \\
 U8f/Ni58p     & 2.528    & ~ 3.2700 &      & ~  3.0957 & ~2.16 &\cellcolor{yellow} ~ 0.9467 &\cellcolor{yellow} ~2.16 &\cellcolor{yellow}  -5.33 \\
 Ti47p/U5f     & 3.622    & ~2.15E-3 & ~4.19& ~ 1.97E-3 & ~3.01 &\cellcolor{green}  ~ 0.9166 &\cellcolor{green}  ~5.16 &\cellcolor{green}   -8.34 \\
 Ti47p/Ni58p   & 3.622    & ~ 0.1748 &      & ~  0.1708 & ~3.28 &\cellcolor{green}  ~ 0.9773 &\cellcolor{green}  ~3.28 &\cellcolor{green}   -2.27 \\
 Ni58p/U5f     & 4.061    & ~ 0.0123 & ~1.63& ~  0.0115 & ~2.14 &\cellcolor{green}  ~ 0.9378 &\cellcolor{green}  ~2.69 &\cellcolor{green}   -6.22 \\
 Fe54p/U5f     & 4.318    & ~9.00E-3 & ~3.33& ~ 8.35E-3 & ~3.38 &\cellcolor{green}  ~ 0.9282 &\cellcolor{green}  ~4.75 &\cellcolor{green}   -7.18 \\
 Fe54p/Ni58p   & 4.318    & ~ 0.7317 &      & ~  0.7242 & ~3.62 &\cellcolor{green}  ~ 0.9920 &\cellcolor{green}  ~3.62 &\cellcolor{green}   -1.03 \\
 Al27p/U5f     & 5.732    & ~3.88E-4 &      & ~ 4.26E-4 & ~2.45 &\cellcolor{lred}    ~ 1.0981 &\cellcolor{lred}    ~2.45 &\cellcolor{lred}      9.81 \\
 Al27p/Ni58p   & 5.732    & ~ 0.0336 &      & ~  0.0369 & ~2.77 &\cellcolor{lred}    ~ 1.0993 &\cellcolor{lred}    ~2.77 &\cellcolor{lred}      9.93 \\
 Co59p/U5f     & 5.757    & ~1.58E-4 &      & ~ 1.53E-4 & ~3.81 &\cellcolor{green}  ~ 0.9667 &\cellcolor{green}  ~3.81 &\cellcolor{green}   -3.33 \\
 Ti46p/U5f     & 5.916    & ~1.30E-3 & ~2.31& ~ 1.23E-3 & ~3.43 &\cellcolor{green}  ~ 0.9491 &\cellcolor{green}  ~4.13 &\cellcolor{green}   -5.09 \\
 Ti46p/Ni58p   & 5.916    & ~ 0.1057 &      & ~  0.1070 & ~3.67 &\cellcolor{green}  ~ 0.9813 &\cellcolor{green}  ~3.67 &\cellcolor{green}    1.20 \\
 Fe56p/U5f     & 7.378    & ~9.12E-5 &      & ~ 1.20E-4 & ~3.11 &\cellcolor{lred}    ~ 1.3117 &\cellcolor{lred}    ~3.11 &\cellcolor{lred}     31.17 \\
 Fe56p/Ni58p   & 7.378    & ~7.89E-3 &      & ~  0.0104 & ~3.36 &\cellcolor{lred}    ~ 1.3144 &\cellcolor{lred}    ~3.36 &\cellcolor{lred}     31.44 \\
 U82/U5f       & 8.067    & ~1.74E-3 &      & ~ 1.67E-3 & ~5.39 &\cellcolor{green}  ~ 0.9607 &\cellcolor{green}  ~5.39 &\cellcolor{green}   -3.93 \\
 U82/Ni58p     & 8.067    & ~ 0.1510 &      & ~  0.1449 & ~5.54 &\cellcolor{green}  ~ 0.9597 &\cellcolor{green}  ~5.54 &\cellcolor{green}   -4.03 \\
 Ti48p/U5f     & 8.068    & ~3.60E-5 & ~2.78& ~ 3.31E-5 & ~5.87 &\cellcolor{green}  ~ 0.9198 &\cellcolor{green}  ~6.49 &\cellcolor{green}   -8.02 \\
 Ti48p/Ni58p   & 8.068    & ~2.92E-3 &      & ~ 2.87E-3 & ~6.01 &\cellcolor{green}  ~ 0.9831 &\cellcolor{green}  ~6.01 &\cellcolor{green}   -1.69 \\
 Al27a/U5f     & 8.417    & ~7.80E-5 & ~2.56& ~ 7.78E-5 & ~1.97 &\cellcolor{green}  ~ 0.9974 &\cellcolor{green}  ~3.24 &\cellcolor{green}   -0.26 \\
 Al27a/Ni58p   & 8.417    & ~6.30E-3 &      & ~ 6.74E-3 & ~2.36 &\cellcolor{green}  ~ 1.0705 &\cellcolor{green}  ~2.36 &\cellcolor{green}    7.05 \\
 Tm1692/U5f    &10.244    & ~5.45E-4 &      & ~ 4.10E-4 & ~4.58 &\cellcolor{lred}    ~ 0.7515 &\cellcolor{lred}    ~4.58 &\cellcolor{lred}    -24.85 \\
 Au1972/U5f    &10.380    & ~3.52E-4 &      & ~ 3.57E-4 & ~3.74 &\cellcolor{green}  ~ 1.0135 &\cellcolor{green}  ~3.74 &\cellcolor{green}    1.35 \\
 Co592/U5f     &12.753    & ~3.14E-5 &      & ~ 2.09E-5 & ~7.77 &\cellcolor{lred}    ~ 0.6650 &\cellcolor{lred}    ~7.77 &\cellcolor{lred}    -33.50 \\
 Y892/U5f      &13.602    & ~4.67E-5 &      & ~ 1.59E-5 & 10.33 &\cellcolor{lred}    ~ 0.3414 &\cellcolor{lred}    10.33 &\cellcolor{lred}    -65.86 \\ \hline \hline
\end{tabular}
\vspace{-2mm}
\end{table*}

%\clearpage
\subsection{Plutonium Fueled ICSBEP Assemblies from the ICSBEP Handbook} \label{Sec_VII_K}
In contrast to the good SI C/E results for the uranium fueled ICSBEP assemblies that were noted in Sec.~\ref{Sec_VII_J} the results for plutonium fueled ICSBEP assemblies are poor.  Measured reaction rates, our calculated values and the corresponding SI C/E results for the PMF001 (Jezebel), PMF006 (Flattop-Pu) and PMF008 (Thor) critical assemblies are given in Table~\ref{TableICSBEP_DDD}.

\begin{figure}[!htbp]
\vspace{-2mm}
\includegraphics[width=\columnwidth]{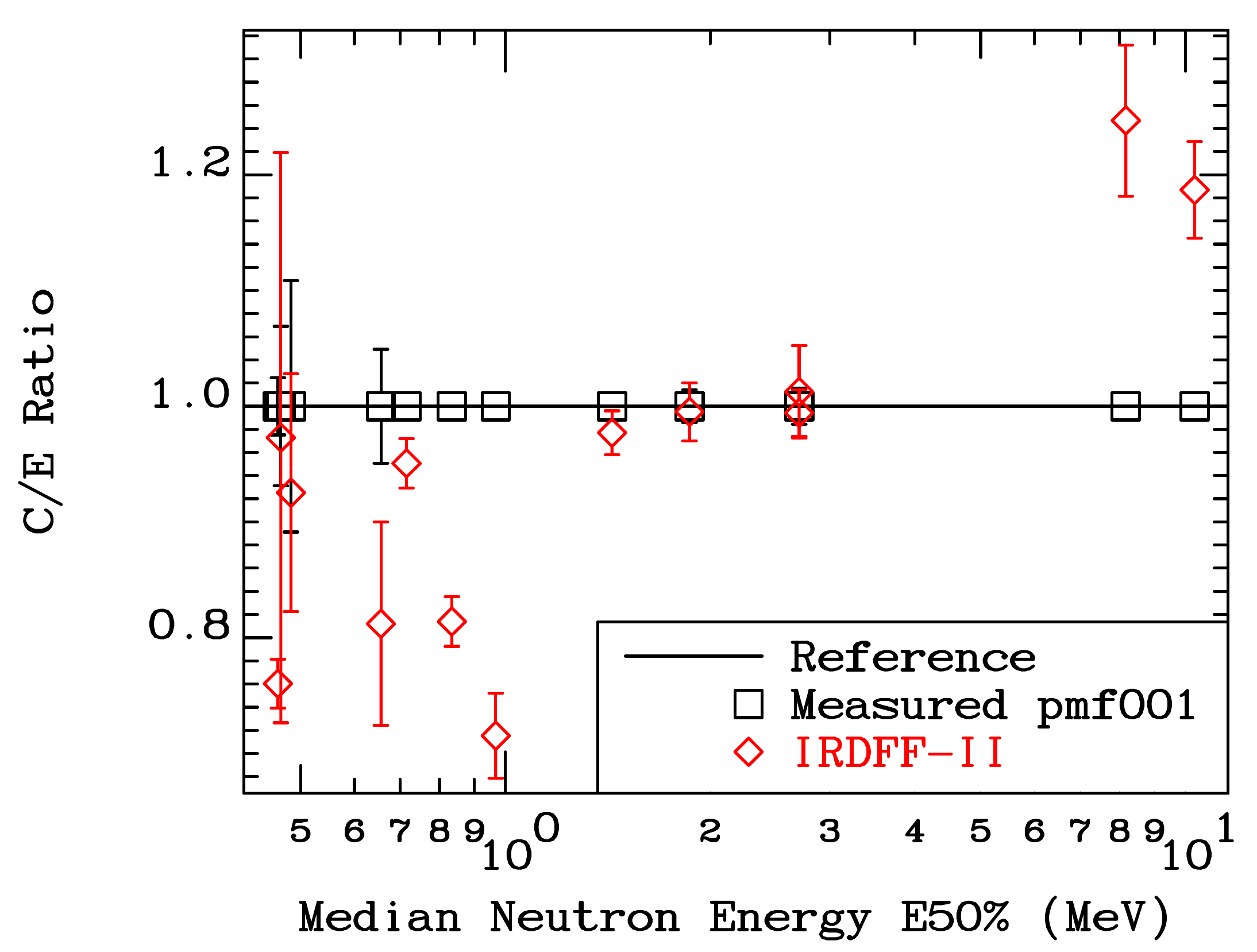}
\caption{(Color online) Ratio of Calculated and Measured SI in the central region of PMF001 (Jezebel) assembly.}
\label{Fig.ICSBEP_DDD}
\vspace{-2mm}
\end{figure}

\begin{figure}[!htbp]
%\vspace{-2mm}
\includegraphics[width=\columnwidth]{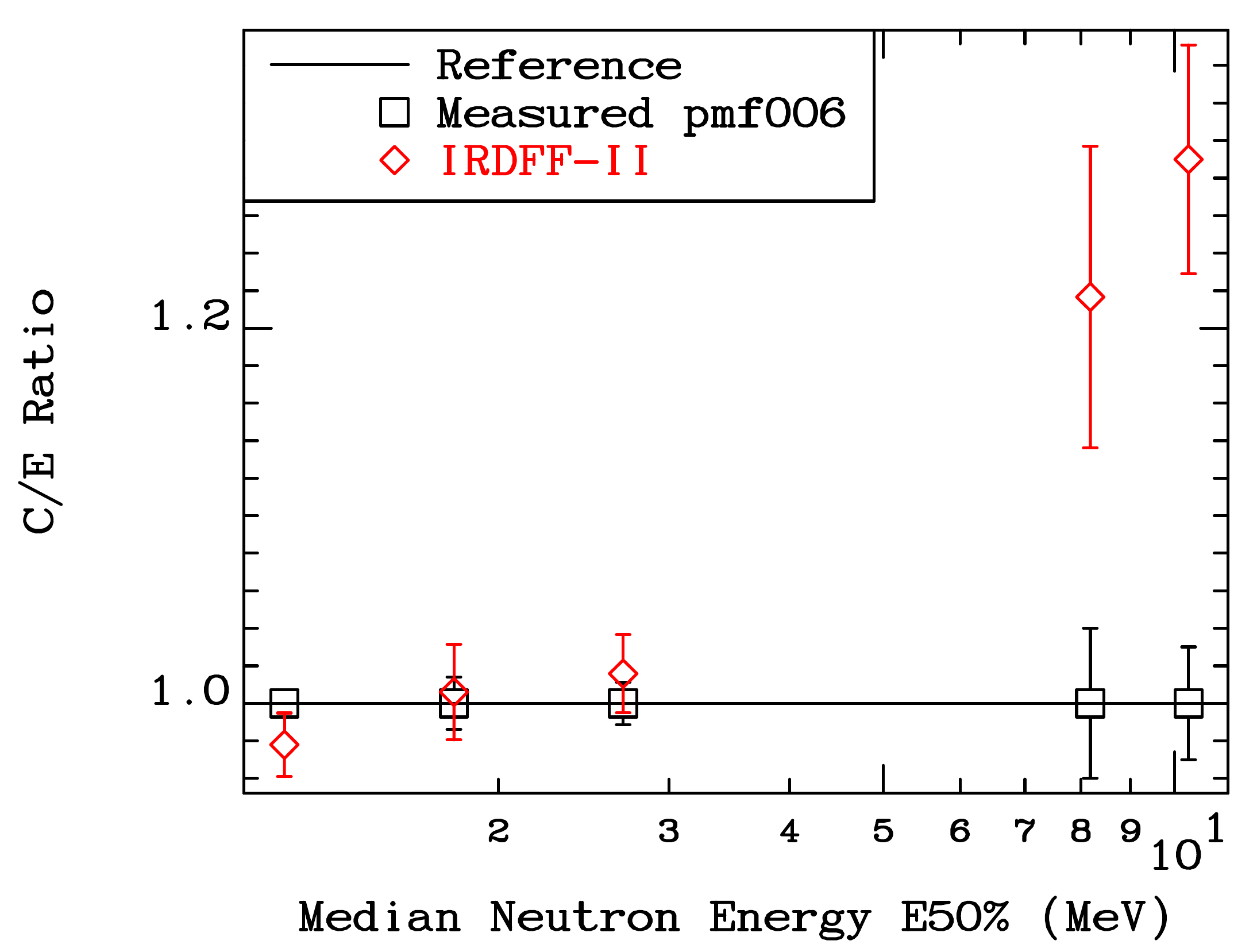}
\caption{(Color online) Ratio of Calculated and Measured SI in the central region of PMF006 (Flattop-Pu) assembly.}
\label{Fig.ICSBEP_EEE}
\vspace{-2mm}
\end{figure}

\begin{figure}[!htbp]
%\vspace{-2mm}
\includegraphics[width=\columnwidth]{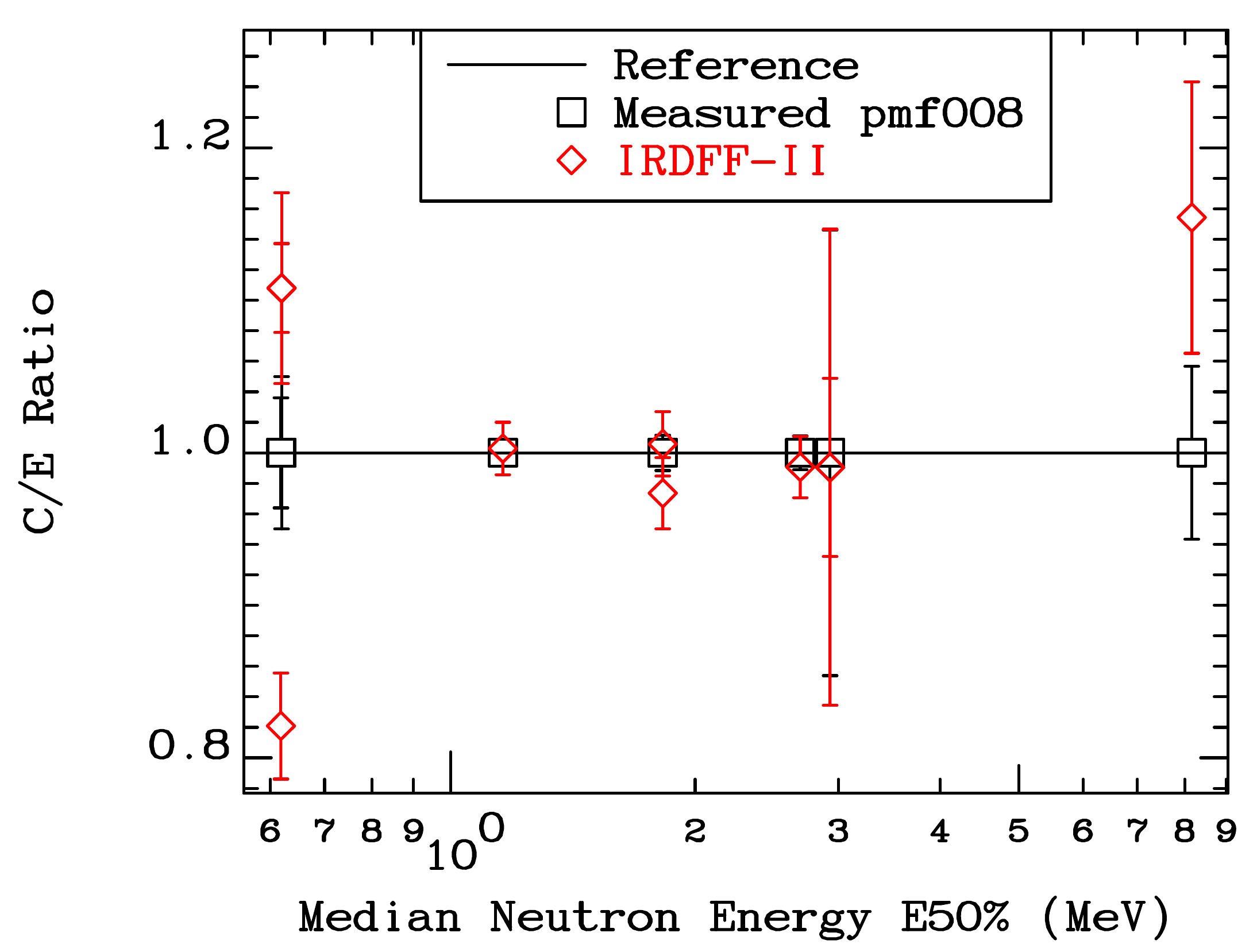}
\caption{(Color online) Ratio of Calculated and Measured SI in the central region of PMF008 (Thor) assembly.}
\label{Fig.ICSBEP_FFF}
\vspace{-2mm}
\end{figure}

\begin{figure}[htbp]
\vspace{-2mm}
\includegraphics[width=\columnwidth]{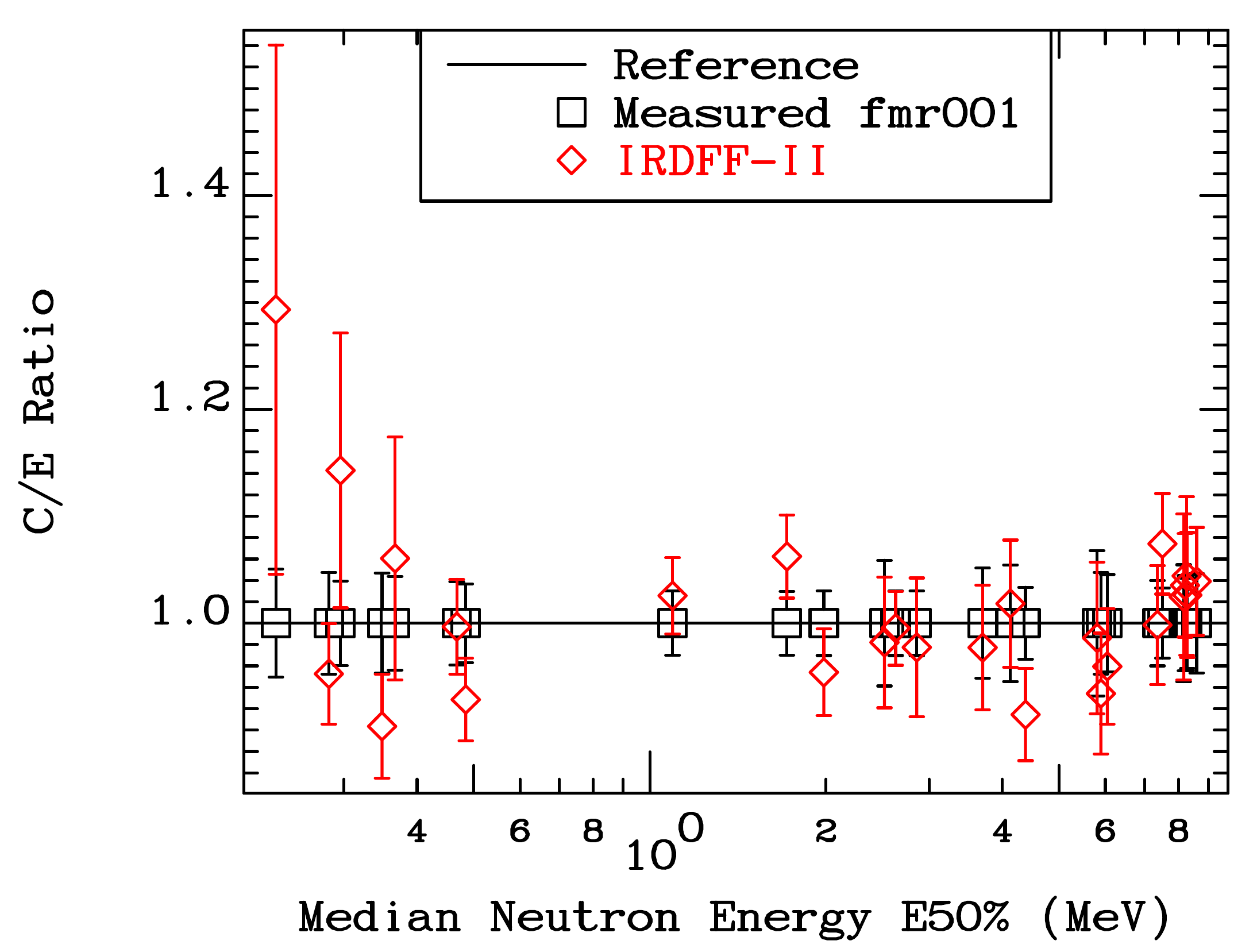}
\caption{(Color online) Ratio of Calculated and Measured SI in the central region of FMR001 assembly (formally designated FUND-IPPE-FR-MULT-RRR-001 in ICSBEP).}
\label{Fig.ICSBEP_GGG}
\vspace{-2mm}
\end{figure}

The SI C/E values, plotted against $E_{50\%}$ as has been done with all other results, are shown in Figs.~\ref{Fig.ICSBEP_DDD}, \ref{Fig.ICSBEP_EEE} and \ref{Fig.ICSBEP_FFF}.  These figures display a clear trend of increasing C/E with increasing energy.  With a similar trend seen for multiple assemblies it seems likely that the source for this trend is in the high energy tail of the $^{239}$Pu prompt fission spectrum.  However, we have also calculated the ICSBEP FMR001 assembly (more formally designated FUND-IPPE-FR-MULT-RRR-001 in the ICSBEP Handbook), a Pu metal rodded critical assembly with copper and uranium reflectors.  The SI measured data, our calculations and the corresponding C/E values are given in Table~\ref{TableICSBEP_GGG} and the C/E values are shown in Fig.~\ref{Fig.ICSBEP_GGG}.  In contrast to the other Pu fueled assemblies these results are quite good, with minimal variation in C/E out to the highest available energy.  The large C/E variation seen at low energies is for a variety of capture reactions and mimics the pattern seen in the results for uranium fueled assemblies.

\begin{table*}[htbp]
\vspace{-4mm}
\caption{Measured and calculated SI in the central region of PMF001 (Jezebel), PMF006 (Flattop-Pu) and PMF008 (Thor) assemblies.}
\label{TableICSBEP_DDD}
\begin{tabular}{ l c rl rl rl r} \hline \hline
 Entry Label   &          & \multicolumn{2}{c}{Measured   } & \multicolumn{2}{c}{Calculated } & \multicolumn{2}{c}{SI C/E}         & Diff \\
 and monitor   &$E_{50\%}$& \multicolumn{2}{c}{SI and Unc.} & \multicolumn{2}{c}{SI and Unc.} & \multicolumn{2}{c}{Value and Unc.} & [\%] \\
               & MeV      &           & [\%]  &           & [\%]  &                            & [\%]                    &                         \\ \hline
               & \multicolumn{8}{l}{Jezebel}                                         \\ \hline
\T
 Au197g/U5f    & 0.463    & ~  0.1012 & ~2.47 & ~  0.0769 & ~1.29 &\cellcolor{lred}    ~ 0.7603 &\cellcolor{lred}    ~2.79 &\cellcolor{lred}    -23.97 \\
 Mn55g/U5f     & 0.467    & ~ 2.90E-3 & ~6.90 & ~ 2.82E-3 & 24.36 &\cellcolor{green}  ~ 0.9728 &\cellcolor{green}  25.32 &\cellcolor{green}   -2.72 \\
 Nb93g/U5f     & 0.484    & ~  0.0276 & 10.87 & ~  0.0255 & ~2.26 &\cellcolor{green}  ~ 0.9252 &\cellcolor{green}  11.10 &\cellcolor{green}   -7.48 \\
 Cu63g/U5f     & 0.656    & ~  0.0122 & ~4.92 & ~ 9.91E-3 & ~9.64 &\cellcolor{yellow} ~ 0.8120 &\cellcolor{yellow} 10.82 &\cellcolor{yellow} -18.80 \\
 U8g/U5f       & 0.715    & ~  0.0677 &       & ~  0.0644 & ~2.24 &\cellcolor{yellow} ~ 0.9506 &\cellcolor{yellow} ~2.24 &\cellcolor{yellow}  -4.94 \\
 In115gm/U5f   & 0.834    & ~  0.1390 &       & ~  0.1131 & ~2.63 &\cellcolor{lred}    ~ 0.8140 &\cellcolor{lred}    ~2.63 &\cellcolor{lred}    -18.60 \\
 La139g/U5f    & 0.968    & ~ 8.20E-3 &       & ~ 5.87E-3 & ~5.13 &\cellcolor{lred}    ~ 0.7153 &\cellcolor{lred}    ~5.13 &\cellcolor{lred}    -28.47 \\
 Pu9f/U5f      & 1.435    & ~  1.4609 & ~0.89 & ~  1.4274 & ~1.73 &\cellcolor{green}  ~ 0.9771 &\cellcolor{green}  ~1.95 &\cellcolor{green}   -2.29 \\
 Np7f/U5f      & 1.865    & ~  0.9835 & ~1.42 & ~  0.9787 & ~2.06 &\cellcolor{green}  ~ 0.9951 &\cellcolor{green}  ~2.50 &\cellcolor{green}   -0.49 \\
 U8f/U5f       & 2.704    & ~  0.2133 & ~1.08 & ~  0.2121 & ~1.71 &\cellcolor{green}  ~ 0.9942 &\cellcolor{green}  ~2.02 &\cellcolor{green}   -0.58 \\
 U82/U5f       & 8.172    & ~  0.0106 &       & ~  0.0132 & ~5.24 &\cellcolor{lred}    ~ 1.2469 &\cellcolor{lred}    ~5.24 &\cellcolor{lred}     24.69 \\
 Tm1692/U5f    &10.317    & ~ 3.13E-3 &       & ~ 3.71E-3 & ~3.50 &\cellcolor{lred}    ~ 1.1868 &\cellcolor{lred}    ~3.50 &\cellcolor{lred}     18.68 \\ \hline
               & \multicolumn{8}{l}{Flattop-Pu}                               \\ \hline
 \T
 Pu9f/U5f      & 1.202    & ~  1.4150 &       & ~  1.3837 & ~1.73 &\cellcolor{green} ~ 0.9779 &\cellcolor{green} ~1.73 &\cellcolor{green}   -2.21 \\
 Np7f/U5f      & 1.799    & ~  0.8561 & ~1.40 & ~  0.8612 & ~2.08 &\cellcolor{green} ~ 1.0060 &\cellcolor{green} ~2.51 &\cellcolor{green}    0.60 \\
 U8f/U5f       & 2.691    & ~  0.1772 & ~1.13 & ~  0.1800 & ~1.72 &\cellcolor{green} ~ 1.0158 &\cellcolor{green} ~2.06 &\cellcolor{green}    1.58 \\
 U82/U5f       & 8.178    & ~ 9.02E-3 & ~3.99 & ~  0.0110 & ~5.27 &\cellcolor{lred}   ~ 1.2166 &\cellcolor{lred}   ~6.61 &\cellcolor{lred}     21.66 \\
 Tm1692/U5f    &10.333    & ~ 2.43E-3 & ~3.00 & ~ 3.13E-3 & ~3.65 &\cellcolor{lred}   ~ 1.2900 &\cellcolor{lred}   ~4.73 &\cellcolor{lred}     29.00 \\ \hline
               & \multicolumn{8}{l}{Thor}                                     \\ \hline
 \T
 U8g/U5f       & 0.619    & ~  0.0830 & ~3.61 & ~  0.0681 & ~2.21 &\cellcolor{lred}    ~ 0.8209 &\cellcolor{lred}    ~4.24 &\cellcolor{lred}    -17.91 \\
%Th2g/U8g\_w   & 0.651    & ~  1.2000 & ~5.00 & ~  1.3586 & ~3.81 &\cellcolor{green}  ~ 1.1322 &\cellcolor{green}  ~6.28 &\cellcolor{green}   10.80 \\
 Th2g/U8g      & 0.651    & ~  1.2000 &       & ~  1.3586 & ~3.81 &\cellcolor{yellow} ~ 1.1322 &\cellcolor{yellow} ~3.81 &\cellcolor{yellow}  10.80 \\
 U5f/U8f      & 1.160    & ~  5.1300 &       & ~  5.1445 & ~1.71 &\cellcolor{green}  ~ 1.0028 &\cellcolor{green}  ~1.71 &\cellcolor{green}    0.28 \\
 Np7f/U5f      & 1.824    & ~  0.9419 & ~1.17 & ~  0.9169 & ~2.07 &\cellcolor{green}  ~ 0.9735 &\cellcolor{green}  ~2.38 &\cellcolor{green}   -2.65 \\
 Np7f/U8f      & 1.824    & ~  4.6900 &       & ~  4.7172 & ~2.08 &\cellcolor{green}  ~ 1.0058 &\cellcolor{green}  ~2.08 &\cellcolor{green}    0.58 \\
 U8f/U5f       & 2.693    & ~  0.1962 & ~1.10 & ~  0.1944 & ~1.71 &\cellcolor{green}  ~ 0.9907 &\cellcolor{green}  ~2.03 &\cellcolor{green}   -0.93 \\
%Th2f/U8f\_w   & 2.931    & ~  0.2600 & 14.62 & ~  0.2575 & ~5.91 &\cellcolor{green}  ~ 0.9904 &\cellcolor{green}  15.76 &\cellcolor{green}   -0.96 \\
 Th2f/U8f      & 2.931    & ~  0.2600 &       & ~  0.2575 & ~5.91 &\cellcolor{green}  ~ 0.9904 &\cellcolor{green}  ~5.91 &\cellcolor{green}   -0.96 \\
 U82/U8f       & 8.165    & ~  0.0530 & ~5.66 & ~  0.0612 & ~5.24 &\cellcolor{green}  ~ 1.1543 &\cellcolor{green}  ~7.72 &\cellcolor{green}   15.43 \\ \hline \hline
\end{tabular}
\vspace{-4mm}
\end{table*}

\begin{table*}[htbp]
\vspace{-4mm}
\caption{Measured and calculated SI in the central region of FMR001 assembly (formally designated FUND-IPPE-FR-MULT-RRR-001 in ICSBEP).}
\label{TableICSBEP_GGG}
\begin{tabular}{ l c rl rl rl r} \hline \hline
 Entry Label   &          & \multicolumn{2}{c}{Measured   } & \multicolumn{2}{c}{Calculated } & \multicolumn{2}{c}{SI C/E}         & Diff \\
 and monitor   &$E_{50\%}$& \multicolumn{2}{c}{SI and Unc.} & \multicolumn{2}{c}{SI and Unc.} & \multicolumn{2}{c}{Value and Unc.} & [\%] \\
               & MeV      &            & [\%]  &           & [\%]  &                            & [\%]                    &                         \\ \hline
 \T
 Mn55g/U5f     & 0.230    & ~ 2.97E-3  & ~5.05 & ~ 3.84E-3 & 18.46 &\cellcolor{green} 1.2933 &\cellcolor{green} 19.14 &\cellcolor{green} 29.33 \\
 Au197g/U5f    & 0.283    & ~  0.1050  & ~4.76 & ~  0.1000 &  1.29 &\cellcolor{green} 0.9526 &\cellcolor{green}  4.93 &\cellcolor{green} -4.74 \\
 Fe58g/U5f     & 0.296    & ~ 2.28E-3  & ~3.95 & ~ 2.61E-3 & 10.53 &\cellcolor{green} 1.1428 &\cellcolor{green} 11.24 &\cellcolor{green} 14.28 \\
 Co59g/U5f     & 0.349    & ~ 6.40E-3  & ~4.69 & ~ 5.78E-3 &  2.63 &\cellcolor{green} 0.9036 &\cellcolor{green}  5.37 &\cellcolor{green} -9.64 \\
 Cu63g/U5f     & 0.367    & ~  0.0114  & ~4.39 & ~  0.0121 &  9.76 &\cellcolor{green} 1.0606 &\cellcolor{green} 10.70 &\cellcolor{green}  6.06 \\
 U8g/U5f       & 0.468    & ~  0.0770  & ~3.90 & ~  0.0767 &  2.14 &\cellcolor{green} 0.9967 &\cellcolor{green}  4.44 &\cellcolor{green} -0.33 \\
 Th2g/U5f      & 0.518    & ~  0.1090  & ~3.67 & ~  0.1012 &  1.94 &\cellcolor{green} 0.9287 &\cellcolor{green}  4.15 &\cellcolor{green} -7.13 \\
 Pu9f/U5f      & 1.093    & ~  1.3300  & ~3.01 & ~  1.3642 &  1.73 &\cellcolor{green} 1.0257 &\cellcolor{green}  3.47 &\cellcolor{green}  2.57 \\
 Np7f/U5f      & 1.714    & ~  0.7710  & ~2.98 & ~  0.8191 &  2.11 &\cellcolor{green} 1.0624 &\cellcolor{green}  3.66 &\cellcolor{green}  6.24 \\
 Am1f/U5f      & 1.982    & ~  0.8250  & ~3.03 & ~  0.7871 &  2.96 &\cellcolor{green} 0.9541 &\cellcolor{green}  4.24 &\cellcolor{green} -4.59 \\
 In115nm/U5f   & 2.516    & ~  0.1020  & ~5.88 & ~  0.1002 &  2.06 &\cellcolor{green} 0.9821 &\cellcolor{green}  6.23 &\cellcolor{green} -1.79 \\
 U8f/U5f       & 2.629    & ~  0.1650  & ~3.03 & ~  0.1642 &  1.71 &\cellcolor{green} 0.9953 &\cellcolor{green}  3.48 &\cellcolor{green} -0.47 \\
 Th2f/U5f      & 2.857    & ~  0.0430  & ~3.02 & ~  0.0420 &  5.91 &\cellcolor{green} 0.9774 &\cellcolor{green}  6.64 &\cellcolor{green} -2.26 \\
 Ti47p/U5f     & 3.702    & ~ 9.70E-3  & ~5.15 & ~ 9.48E-3 &  2.99 &\cellcolor{green} 0.9772 &\cellcolor{green}  5.96 &\cellcolor{green} -2.28 \\
 Ni58p/U5f     & 4.129    & ~  0.0550  & ~5.45 & ~  0.0560 &  2.11 &\cellcolor{green} 1.0183 &\cellcolor{green}  5.85 &\cellcolor{green}  1.83 \\
 Fe54p/U5f     & 4.383    & ~  0.0447  & ~3.36 & ~  0.0409 &  3.31 &\cellcolor{green} 0.9145 &\cellcolor{green}  4.72 &\cellcolor{green} -8.55 \\
 Al27p/U5f     & 5.806    & ~ 2.21E-3  & ~6.79 & ~ 2.18E-3 &  2.38 &\cellcolor{green} 0.9862 &\cellcolor{green}  7.19 &\cellcolor{green} -1.38 \\
 Co59p/U5f     & 5.895    & ~ 8.40E-4  & ~4.76 & ~ 7.85E-4 &  3.70 &\cellcolor{green} 0.9343 &\cellcolor{green}  6.03 &\cellcolor{green} -6.57 \\
 Ti46p/U5f     & 6.044    & ~ 6.60E-3  & ~4.55 & ~ 6.33E-3 &  3.31 &\cellcolor{green} 0.9596 &\cellcolor{green}  5.63 &\cellcolor{green} -4.04 \\
 Fe54a/U5f     & 7.364    & ~ 5.00E-4  & ~4.00 & ~ 4.99E-4 &  3.87 &\cellcolor{green} 0.9984 &\cellcolor{green}  5.57 &\cellcolor{green} -0.16 \\
 Fe56p/U5f     & 7.509    & ~ 6.10E-4  & ~3.28 & ~ 6.55E-4 &  2.90 &\cellcolor{green} 1.0743 &\cellcolor{green}  4.38 &\cellcolor{green}  7.43 \\
 U82/U5f       & 8.161    & ~ 9.16E-3  & ~5.46 & ~ 9.39E-3 &  5.24 &\cellcolor{green} 1.0246 &\cellcolor{green}  7.57 &\cellcolor{green}  2.46 \\
 Mg24p/U5f     & 8.207    & ~ 9.00E-4  & ~4.44 & ~ 9.32E-4 &  1.45 &\cellcolor{green} 1.0354 &\cellcolor{green}  4.68 &\cellcolor{green}  3.54 \\
 Ti48p/U5f     & 8.258    & ~ 1.80E-4  & ~4.44 & ~ 1.88E-4 &  5.53 &\cellcolor{green} 1.0442 &\cellcolor{green}  7.10 &\cellcolor{green}  4.42 \\
 Co59a/U5f     & 8.281    & ~ 9.50E-5  & ~4.21 & ~ 9.75E-5 &  3.83 &\cellcolor{green} 1.0263 &\cellcolor{green}  5.69 &\cellcolor{green}  2.63 \\
 Al27a/U5f     & 8.591    & ~ 4.30E-4  & ~4.65 & ~ 4.47E-4 &  1.40 &\cellcolor{green} 1.0391 &\cellcolor{green}  4.86 &\cellcolor{green}  3.91 \\ \hline \hline
\end{tabular}
\vspace{-4mm}
\end{table*}

In view of these generally poorer, and certainly conflicting, Pu fueled system results the reader is cautioned that SI results for plutonium fueled systems obtained with current evaluated nuclear data files is subject to larger than expected biases.

\subsection{Legacy Reference Reactor Fields} \label{Sec_VII_L}

We have revisited the older reference SACS data obtained at the following reactor facilities: coupled thermal/fast uranium and boron carbide spherical assembly (Sigma-Sigma), Coupled Fast Reactivity Measurement Facility (CFRMF), Intermediate-energy Standard Neutron Fields (ISNF) and Glory hole of the Tokyo University research reactor (YAYOI) \cite{Gru86}. Their neutron spectra are available in the IRDF-2002 database \cite{Spectra02} which in turn were adopted from the IRDF-90 library~\cite{IRDF90}; the latter was assembled in 1993. It is important to notice that the measurements in these neutron fields were not previously incorporated into the evidence package supporting validation of the IRDFF cross section library.

For the present analysis, we selected the facilities and spectra for ISNF and Sigma-Sigma after their spectra were re-calculated based on the transport cross sections from the \mbox{ENDF/B-VIII.0} library and represented in 725 groups up to 60~MeV, as described in Secs.~\ref{sssect:ISNF} and~\ref{sssect:Sigma}. The CFRMF spectrum in 460 energy groups was adopted directly from \mbox{IRDF-2002} as we were not able to reproduce it,  only the material number was changed from MAT=5 into MAT=9006.

We have examined the use of the YAYOI spectrum from IRDF-2002 (MAT= 9, in 100 energy groups) with the SACS measured by K. Kobayashi \etals\cite{Kob76, Kob80} for validation with this neutron environment. However practically all C/E ratios turned out to be below unity by 20-50~\%. The difference is too large to be attributed to the rather coarse 100-groups representation of the energy spectrum. Most likely there are some unidentified differences in the interpretation of the published measured spectrum-averaged cross section values and the neutron field characterization data. Attempts to use an alternative YAYOI spectrum received through private communications did not improve the agreement, so work in this benchmark field remains for a future study.

The experimental spectrum-averaged reaction cross sections were taken from EXFOR and were also checked against the original publications. Some experimental data are in the form of spectral indices, \ie, the reaction rate ratios measured relative to $^{235}$U(n,f), $^{238}$U(n,f) or $^{197}$Au(n,$\gamma$) -- and these data were used in this form. The calculated energy-averaged cross sections were obtained from the \mbox{IRDFF-II} cross sections and facility spectra with the help of the RR\_UNC code.

\begin{figure}[!htbp]
\vspace{-2mm}
  \includegraphics[width=\columnwidth]{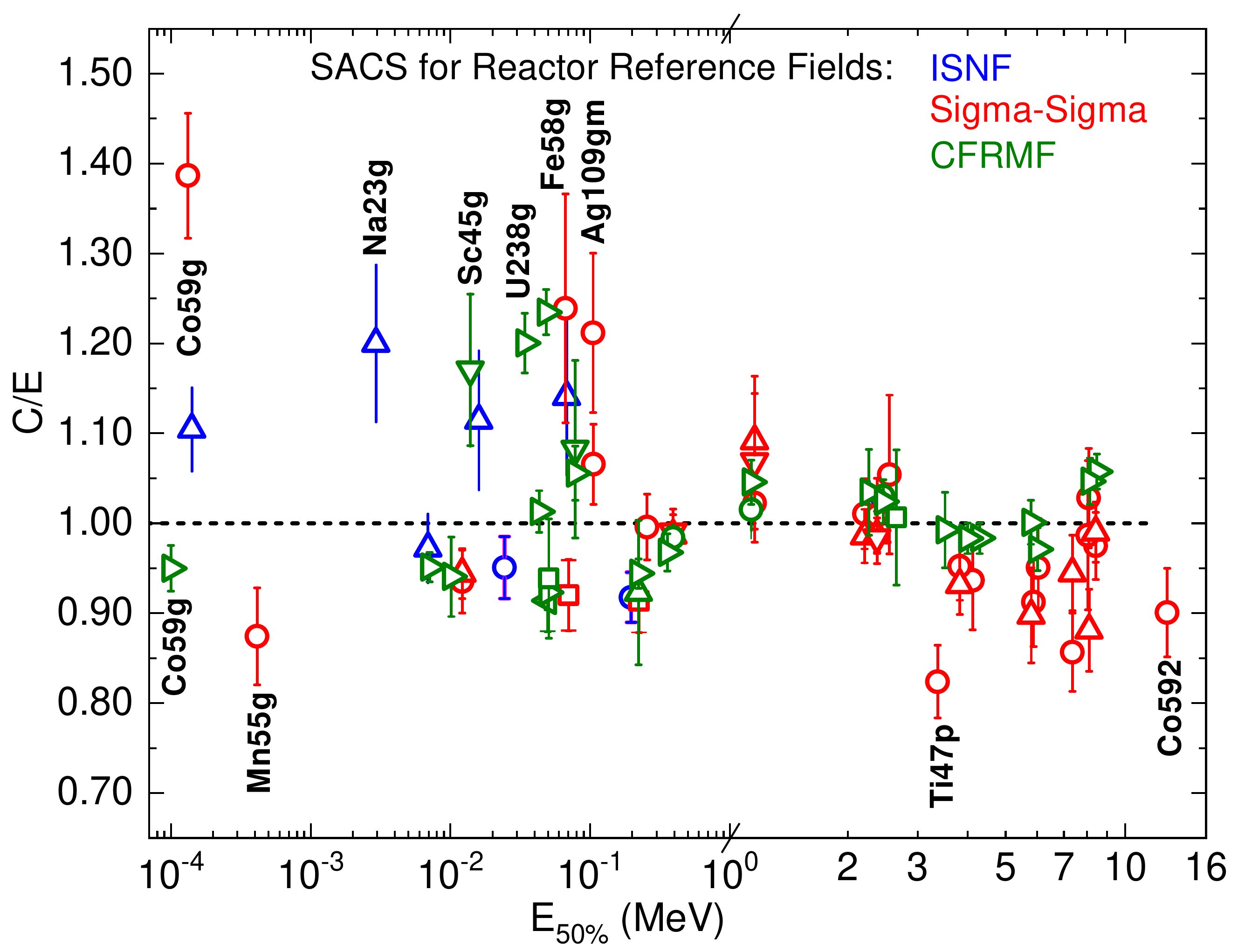}
  \caption{(Color online) C/E ratios for SACS and spectral indices (SI) measured in the reference reactor fields: ISNF (blue symbols), CFRMF (green), Sigma-Sigma (red). Plotted C/E values are listed in Table~\ref{TableLongSACSreactors}.}
  \label{fig:SACS_Reactors}
\vspace{-2mm}
\end{figure}

The C/E ratios for SACS or spectral indices are presented in Table~\ref{TableLongSACSreactors} and Fig.~\ref{fig:SACS_Reactors} as a function of the mean response energy $E_{50\%}$. The error bars displayed in Fig.~\ref{fig:SACS_Reactors} correspond to the sum of the experimental, \mbox{IRDFF-II} cross sections and the spectra uncertainties when the latter were available. It is seen that deviation of C/E from unity lies within two to three-sigma for most reactions and facilities, however with several exceptions which are explicitly labelled in the figure.
The agreement seems to be worse for E$_{50\%}<0.1$~MeV, \ie, for evaluated IRDFF-II cross sections in the epithermal region.

It is worth noting that the research reactor benchmarks have notably softer spectra than the fission sources $^{252}$Cf SFNS or $^{235}$U(n$_{th}$,f) PFNS. Because of this, they are capable of delivering validation data for reactions that augments those obtained in the fission source neutron fields. This is particularly true for the (n,$\gamma$) reaction on targets, such as $^{45}$Sc, $^{55}$Mn, $^{58}$Fe, and $^{109}$Ag, where the observed C/E agreement is seen to be within 2--3 standard deviations of unity.

In the case of reactions sensitive to neutron energies above 10~MeV, we see agreement between the C/E values in the Sigma-Sigma and $^{235}$U(n$_{th}$,f) PFNS benchmark neutron fields for the $^{59}$Co(n,2n) reaction (E$_{50\%}$ = 12.7~MeV): C/E = 0.900$\pm$0.049 and C/E = 0.995$\pm$0.044, respectively.

\LTcapwidth=\textwidth
\clearpage
\begin{longtable*}[!tbp]{l l l l l l l l c}
\caption{Measured and calculated SACS or SI ratio in the reference reactor neutron benchmark fields ISNF, Sigma-Sigma, and CFRMF.
For the calculated SACS uncertainties the contributions from the dosimetry cross sections (Sigma Unc.) and neutron spectra (Spect. Unc.) are given separately.
The C/E uncertainty is the sum of experimental and calculated ones.}
\label{TableLongSACSreactors} \\
\hline\hline
              &           & \multicolumn{3}{l}{Experimental SACS or Spectral Index (SI)} & \multicolumn{3}{l}{Calculated SACS or SI} &  \\
                            \cline{3-5}                                      \cline{6-8}
              &           &             & ~Exp.~  & Ref. &            & Sigma  & Spect. & C/E \\
 Reaction     & E$_{50\%}$& SACS [mb]   & ~Unc.~  &      & SACS [mb]  & ~Unc.  & ~Unc.   &     \\
 or SI        &  [MeV]    & or SI ratio & ~~[\%]~ &      & or SI ratio& ~~[\%]  & ~~[\%]   &     \\
\hline
\endfirsthead
\caption{continued} \\
\hline\hline
              &           & \multicolumn{3}{l}{Experimental SACS or Spectral Index (SI)} & \multicolumn{3}{l}{Calculated SACS or SI} &  \\
                            \cline{3-5}                                      \cline{6-8}
              &           &             & ~Exp.~  & Ref. &            & Sigma  & Spect. & C/E \\
 Reaction     & E$_{50\%}$& SACS [mb]   & ~Unc.~  &      & SACS [mb]  & ~Unc.  & ~Unc.   &     \\
 or SI        &  [MeV]    & or SI ratio & ~~[\%]~ &      & or SI ratio& ~~[\%]  & ~~[\%]   &     \\
\hline\hline
\endhead
%\endfoot
%\endlastfoot
\multicolumn{9}{c}{Intermediate Energy Standard Neutron Field (ISNF)}                                                             \\ \hline
 Co59g          & 1.416E-04 & 3.630E+01 & 4.13 & \cite{Lam88}  & 4.009E+00 & 0.85 & 1.13 & 1.104 $\pm$  4.22\% \cellcolor{yellow}\\
 Na23g          & 2.944E-03 & 1.570E+00 & 6.40 & \cite{Lam88}  & 1.884E+00 & 9.31 & 1.01 & 1.200 $\pm$  11.2\% \cellcolor{yellow}\\
 Au197g         & 6.946E-03 & 4.110E+02 & 2.68 & \cite{Lam88}  & 3.995E+02 & 2.86 & 0.26 & 0.972 $\pm$  4.91\% \cellcolor{green} \\
 Sc45g          & 1.598E-02 & 2.440E+01 & 3.30 & \cite{Lam88}  & 2.719E+01 & 6.13 & 0.26 & 1.115 $\pm$  6.95\% \cellcolor{yellow}\\
 B10He4         & 2.451E-02 & 1.831E+03 & 3.28 & \cite{Oli84}  & 1.741E+03 & 1.60 & 0.16 & 0.951 $\pm$  3.65\% \cellcolor{green} \\
 Ag109gm        & 6.809E-02 & 2.380E+01 & 3.78 & \cite{Lam88}  & 2.715E+01 & 6.67 & 0.12 & 1.141 $\pm$  7.67\% \cellcolor{green} \\
 Li6He4         & 1.963E-01 & 8.310E+02 & 3.01 & \cite{Oli84}  & 7.625E+02 & 0.40 & 0.10 & 0.918 $\pm$  3.03\% \cellcolor{yellow}\\
 In115nm        & 2.262E+00 & 9.700E+01 & 2.58 & \cite{Lam88}  & 9.846E+01 & 1.74 & 0.07 & 1.015 $\pm$  3.11\% \cellcolor{green} \\ \hline
\multicolumn{9}{c}{Secondary Intermediate-Energy Standard Neutron Field (Sigma-Sigma)}                                           \\ \hline
 Co59g          & 1.330E-04 & 3.640E+01 & 4.95 & \cite{Gar81}  & 5.047E+01 & 0.78 & 0.10 & 1.387 $\pm$  5.01\% \cellcolor{lred}   \\
 Mn55g          & 4.153E-04 & 3.550E+01 & 3.38 & \cite{Gar81}  & 3.103E+01 & 5.18 & 0.10 & 0.874 $\pm$  6.19\% \cellcolor{yellow}\\
 Au197g         & 1.219E-02 & 4.103E+02 & 3.41 & \cite{Gar78}  & 3.837E+02 & 1.56 & 0.10 & 0.935 $\pm$  3.75\% \cellcolor{yellow}\\
 Au197g/U5f     & 1.325E-02 & 2.699E-01 & 2.50 & \cite{Han76}  & 2.548E-01 & 1.98 & 0.10 & 0.944 $\pm$  3.15\% \cellcolor{green} \\
 Fe58g          & 6.688E-02 & 5.890E+00 & 4.92 & \cite{Gar81}  & 7.298E+00 & 9.03 & 0.10 & 1.239 $\pm$  10.2\% \cellcolor{yellow}\\
 B10He4         & 7.038E-02 & 1.695E+03 & 4.19 & \cite{Oli82}  & 1.560E+03 & 1.06 & 0.20 & 0.920 $\pm$  4.32\% \cellcolor{yellow}\\
 Ag109gm        & 1.052E-01 & 2.430E+01 & 3.70 & \cite{Gar81}  & 2.944E+01 & 6.31 & 0.10 & 1.212 $\pm$  7.32\% \cellcolor{yellow}\\
 In115gm        & 1.061E-01 & 2.602E+02 & 3.46 & \cite{Gar78}  & 2.773E+02 & 2.35 & 0.10 & 1.066 $\pm$  4.18\% \cellcolor{green} \\
 Li6He4         & 2.236E-01 & 9.310E+02 & 3.76 & \cite{Oli82}  & 8.503E+02 & 0.49 & 0.12 & 0.913 $\pm$  3.79\% \cellcolor{yellow}\\
 U235f          & 2.558E-01 & 1.512E+03 & 3.51 & \cite{Gar78}  & 1.506E+03 & 1.11 & 0.10 & 0.996 $\pm$  3.68\% \cellcolor{green} \\
 Pu9f/U5f       & 3.930E-01 & 1.179E+00 & 2.00 & \cite{Han76}  & 1.164E+00 & 0.43 & 0.10 & 0.987 $\pm$  2.57\% \cellcolor{green} \\
 Pu9f/U5f       & 3.933E-01 & 1.175E+00 & 2.30 & \cite{Fab78}  & 1.164E+00 & 0.43 & 0.10 & 0.993 $\pm$  2.58\% \cellcolor{green} \\
 Np237f         & 1.163E+00 & 6.030E+02 & 3.65 & \cite{Gar78}  & 6.164E+02 & 0.00 & 0.10 & 1.022 $\pm$  3.65\% \cellcolor{green} \\
 Np7f/U5f       & 1.163E+00 & 3.830E-01 & 7.06 & \cite{Fab78}  & 4.094E-01 & 1.67 & 0.10 & 1.071 $\pm$  7.06\% \cellcolor{green} \\
 Np7f/U5f       & 1.165E+00 & 3.751E-01 & 6.29 & \cite{Han76}  & 4.094E-01 & 1.67 & 0.10 & 1.091 $\pm$  6.38\% \cellcolor{green} \\
 In115m         & 2.201E+00 & 5.480E+01 & 3.47 & \cite{Gar78}  & 5.537E+01 & 1.72 & 0.10 & 1.010 $\pm$  3.87\% \cellcolor{green} \\
 In115nm/U5f    & 2.202E+00 & 3.731E-02 & 2.50 & \cite{Han76}  & 3.677E-02 & 1.72 & 0.10 & 0.986 $\pm$  3.23\% \cellcolor{green} \\
 U238f          & 2.368E+00 & 8.346E+01 & 3.95 & \cite{Gar78}  & 8.417E+01 & 1.22 & 0.10 & 1.009 $\pm$  4.14\% \cellcolor{green} \\
 U8f/U5f        & 2.368E+00 & 5.680E-02 & 2.70 & \cite{Fab78}  & 5.590E-02 & 0.53 & 0.10 & 0.974 $\pm$  2.96\% \cellcolor{green} \\
 U8f/U5f        & 2.383E+00 & 5.664E-02 & 1.50 & \cite{Han76}  & 5.590E-02 & 0.53 & 0.10 & 0.987 $\pm$  2.23\% \cellcolor{green} \\
 Th232f         & 2.544E+00 & 1.980E+01 & 6.06 & \cite{Gar78}  & 2.087E+01 & 5.80 & 0.10 & 1.054 $\pm$  8.39\% \cellcolor{green} \\
 Ti47p          & 3.363E+00 & 5.200E+00 & 4.04 & \cite{Gar81}  & 4.284E+00 & 2.76 & 0.10 & 0.824 $\pm$  4.89\% \cellcolor{lred}   \\
 Ni58p          & 3.828E+00 & 2.532E+01 & 3.55 & \cite{Gar78}  & 2.410E+01 & 1.72 & 0.10 & 0.952 $\pm$  3.95\% \cellcolor{green} \\
 Ni58p/U5f      & 3.828E+00 & 1.718E-02 & 3.10 & \cite{Han76}  & 1.601E-02 & 1.72 & 0.10 & 0.932 $\pm$  3.71\% \cellcolor{green} \\
 Fe54p          & 4.123E+00 & 1.820E+01 & 4.95 & \cite{Gar81}  & 1.705E+01 & 3.19 & 0.10 & 0.937 $\pm$  5.88\% \cellcolor{green} \\
 Al27p/U5f      & 5.793E+00 & 6.148E-04 & 5.50 & \cite{Han76}  & 5.518E-04 & 2.07 & 0.10 & 0.898 $\pm$  5.98\% \cellcolor{yellow}\\
 Co59p          & 5.870E+00 & 3.250E-01 & 4.00 & \cite{Gar81}  & 2.965E-01 & 3.64 & 0.10 & 0.912 $\pm$  5.41\% \cellcolor{green} \\
 Ti46p          & 6.040E+00 & 2.510E+00 & 4.78 & \cite{Gar81}  & 2.387E+00 & 3.21 & 0.10 & 0.951 $\pm$  5.76\% \cellcolor{green} \\
 Fe56p          & 7.361E+00 & 2.800E-01 & 4.29 & \cite{Gar81}  & 2.398E-01 & 2.72 & 0.10 & 0.856 $\pm$  5.08\% \cellcolor{lred}   \\
 Fe56p/U5f      & 7.361E+00 & 1.686E-04 & 3.50 & \cite{Han76}  & 1.593E-04 & 2.72 & 0.10 & 0.945 $\pm$  4.57\% \cellcolor{green} \\
 Ti48p          & 8.051E+00 & 6.700E-02 & 6.27 & \cite{Gar81}  & 6.610E-02 & 5.59 & 0.10 & 0.987 $\pm$  8.40\% \cellcolor{green} \\
 Co59a          & 8.090E+00 & 3.340E-02 & 3.89 & \cite{Gar81}  & 3.433E-02 & 3.71 & 0.10 & 1.028 $\pm$  5.38\% \cellcolor{green} \\
 Mg24p/U5f      & 8.112E+00 & 2.472E-04 & 5.10 & \cite{Han76}  & 2.178E-04 & 0.84 & 0.10 & 0.881 $\pm$  5.28\% \cellcolor{yellow}\\
 Al27a          & 8.433E+00 & 1.600E-01 & 3.75 & \cite{Gar81}  & 1.552E-01 & 0.74 & 0.10 & 0.975 $\pm$  3.82\% \cellcolor{green} \\
 Al27a/U5f      & 8.433E+00 & 1.041E-04 & 3.30 & \cite{Han76}  & 1.031E-04 & 0.74 & 0.10 & 0.990 $\pm$  3.56\% \cellcolor{green} \\
 Co592          & 1.275E+01 & 4.330E-02 & 5.08 & \cite{Gar81}  & 3.899E-02 & 2.01 & 0.10 & 0.900 $\pm$  5.46\% \cellcolor{yellow}\\ \hline
\multicolumn{9}{c}{Coupled Fast Reactivity Measurement Facility (CFRMF)}                                                         \\ \hline
 Co59g/U5f      & 1.000E-04 & 5.890E-02 & 2.30 & \cite{Rog78}  & 5.594E-02 & 0.75 & 0.10 & 0.950 $\pm$  2.80\% \cellcolor{green} \\
 Au197g/U5f     & 7.100E-03 & 2.730E-01 & 1.00 & \cite{Rog78}  & 2.596E-01 & 2.52 & 0.10 & 0.951 $\pm$  3.05\% \cellcolor{green} \\
 Cu63g/U5f      & 1.020E-02 & 2.920E-02 & 4.50 & \cite{Rog78}  & 2.747E-02 & 6.70 & 0.10 & 0.941 $\pm$  8.19\% \cellcolor{green} \\
 La139g/Au197ng & 1.391E-02 & 4.600E-02 & 0.00 & \cite{Har72}  & 5.384E-02 & 7.25 & 0.10 & 1.170 $\pm$  7.20\% \cellcolor{yellow}\\
 U8g/U5f        & 3.400E-02 & 1.190E-01 & 2.40 & \cite{Rog78}  & 1.429E-01 & 1.13 & 0.10 & 1.201 $\pm$  3.00\% \cellcolor{lred}   \\
 Sc45g/U5f      & 4.320E-02 & 1.510E-02 & 1.80 & \cite{Rog78}  & 1.530E-02 & 5.79 & 0.10 & 1.013 $\pm$  6.22\% \cellcolor{green} \\
 Fe58g/U5f      & 4.850E-02 & 3.940E-03 & 1.50 & \cite{Rog78}  & 4.865E-03 & 8.71 & 0.10 & 1.235 $\pm$  8.95\% \cellcolor{yellow}\\
 B10He4         & 4.963E-02 & 1.900E+03 & 3.68 & \cite{Far75}  & 1.737E+03 & 0.91 & 0.10 & 0.914 $\pm$  3.79\% \cellcolor{yellow}\\
 B10He4a/U5f    & 4.963E-02 & 1.102E+00 & 1.60 & \cite{Rog78}  & 1.102E+00 & 0.91 & 0.10 & 0.923 $\pm$  2.31\% \cellcolor{lred}   \\
 Th232g         & 5.080E-02 & 2.910E+02 & 3.10 & \cite{And79}  & 2.731E+02 & 6.37 & 0.10 & 0.938 $\pm$  7.08\% \cellcolor{green} \\
 In115gm/U5f    & 7.810E-02 & 1.810E-01 & 2.50 & \cite{Rog78}  & 1.911E-01 & 2.36 & 0.10 & 1.056 $\pm$  3.71\% \cellcolor{green} \\
 In115gm/Au197g & 7.810E-02 & 6.800E-01 & 8.82 & \cite{Har72}  & 7.360E-01 & 2.36 & 0.10 & 1.082 $\pm$  9.12\% \cellcolor{green} \\
 Li6He4         & 2.007E-01 & 1.004E+03 & 8.67 & \cite{Lip73}  & 9.268E+02 & 0.84 & 0.10 & 0.923 $\pm$  8.71\% \cellcolor{green} \\
 Li6He4/U5f     & 2.222E-01 & 5.881E-01 & 1.00 & \cite{Rog78}  & 5.679E-01 & 0.84 & 0.10 & 0.944 $\pm$  1.91\% \cellcolor{lred}   \\
 Pu9f/U5f       & 3.563E-01 & 1.165E+00 & 1.60 & \cite{Rog78}  & 1.127E+00 & 1.14 & 0.10 & 0.968 $\pm$  2.41\% \cellcolor{green} \\
 Pu9f/U5f       & 3.930E-01 & 1.145E+00 & 2.30 & \cite{Gru75}  & 1.127E+00 & 0.42 & 0.10 & 0.984 $\pm$  1.56\% \cellcolor{green} \\
 Np7f/U5f       & 1.139E+00 & 3.560E-01 & 1.90 & \cite{Rog78}  & 3.724E-01 & 2.23 & 0.10 & 1.046 $\pm$  2.36\% \cellcolor{green} \\
 Np7f/U5f       & 1.139E+00 & 7.340E+00 & 7.06 & \cite{Gru75}  & 7.451E+00 & 2.23 & 0.10 & 1.015 $\pm$  3.20\% \cellcolor{green} \\
 In115nm/U5f    & 2.258E+00 & 3.190E-02 & 4.40 & \cite{Rog78}  & 3.300E-02 & 1.70 & 0.10 & 1.034 $\pm$  4.92\% \cellcolor{green} \\
 U8f/U5f        & 2.451E+00 & 4.880E-02 & 1.90 & \cite{Rog78}  & 4.997E-02 & 1.22 & 0.10 & 1.024 $\pm$  2.66\% \cellcolor{green} \\
 U8f/U5f        & 2.451E+00 & 4.850E-02 & 2.70 & \cite{Gru75}  & 4.997E-02 & 1.22 & 0.10 & 1.030 $\pm$  1.86\% \cellcolor{green} \\
 Th232f         & 2.651E+00 & 1.960E+01 & 4.70 & \cite{And79}  & 1.973E+01 & 5.79 & 0.10 & 1.006 $\pm$  7.46\% \cellcolor{green} \\
 Ti47p/U5f      & 3.512E+00 & 2.690E-03 & 4.00 & \cite{Rog78}  & 2.670E-03 & 2.74 & 0.10 & 0.992 $\pm$  5.05\% \cellcolor{green} \\
 B10H3          & 3.641E+00 & 7.780E+00 & 8.61 & \cite{Lip73}  & 1.707E+01 & 16.5 & 0.10 & 2.194 $\pm$  18.6\% \cellcolor{lred}   \\
 Ni58p/U5f      & 4.006E+00 & 1.560E-02 & 1.00 & \cite{Rog78}  & 1.534E-02 & 1.73 & 0.10 & 0.983 $\pm$  2.44\% \cellcolor{green} \\
 Fe54p/U5f      & 4.297E+00 & 1.120E-02 & 1.00 & \cite{Rog78}  & 1.101E-02 & 3.13 & 0.10 & 0.983 $\pm$  3.57\% \cellcolor{green} \\
 Al27p/U5f      & 5.782E+00 & 5.630E-04 & 2.00 & \cite{Rog78}  & 5.635E-04 & 2.07 & 0.10 & 1.001 $\pm$  3.20\% \cellcolor{green} \\
 Ti46p/U5f      & 6.009E+00 & 1.680E-03 & 2.00 & \cite{Rog78}  & 1.631E-03 & 3.13 & 0.10 & 0.971 $\pm$  3.97\% \cellcolor{green} \\
 Ti48p/U5f      & 8.153E+00 & 4.430E-05 & 2.00 & \cite{Rog78}  & 4.638E-05 & 5.54 & 0.10 & 1.047 $\pm$  6.05\% \cellcolor{green} \\
 Al27a/U5f      & 8.488E+00 & 1.040E-04 & 1.20 & \cite{Rog78}  & 1.100E-04 & 0.73 & 0.10 & 1.057 $\pm$  1.98\% \cellcolor{yellow}\\
\hline\hline
% \end{tabular}
% \end{table*}
\end{longtable*}

%The gamma-emission probabilities of the capture product of $^186}$W are claimed with relatively small uncertainties, but the change from an older evaluation was about 20~\%, far in excess of the experimental uncertainties. We believe the assigned uncertainties are too small, considering that $^{187}$W has many small gamma transitions, which are very difficult to observe experimentally.
\begin{table}[!htbp]
\vspace{-2mm}
\caption{Comparison of the thermal cross sections by Mughabghab~\cite{Mug16} and \mbox{IRDFF-II} to the values derived from the Kayzero library~\cite{Kayzero}.
Relative uncertainties and $\sigma_0$ differences are given in \%.}
\label{tab:Kayzero_sig0}
\begin{tabular}{ c c | c c | c c r | c c r } \hline\hline
             &             & \multicolumn{2}{c}{Kayzero}   & \multicolumn{3}{c}{Mughabghab} & \multicolumn{3}{c}{IRDFF-II} \\
 Target      & Prod.       & $\sigma_0$ & Unc.  & $\sigma_0$ & Unc. & Diff. & $\sigma_0$ & Unc. & Diff. \\
             &             & [b] & [\%] & [b]   & [\%]& [\%]& [b] & [\%]  & [\%] \\ \hline
 \T
 ~$^{23}$Na  & ~~$^{24}$Na &0.513 & 0.6 & 0.525 & 1.0 &{\cellcolor{yellow} 2.3} & 0.536 & 1.6 &{\cellcolor{lred}    4.4} \\
 ~$^{45}$Sc  & ~~$^{46}$Sc & 26.3 & 0.4 &  27.2 & 0.7 &{\cellcolor{lred}    3.5} &  27.2 & 6.9 &{\cellcolor{green}  3.6} \\
 ~$^{50}$Cr  & ~~$^{51}$Cr &15.18 & 0.6 &  14.7 & 2.7 &{\cellcolor{green} -3.2} &  14.4 & 2.0 &{\cellcolor{lred}   -5.0} \\
 ~$^{55}$Mn  & ~~$^{56}$Mn &13.21 & 0.5 & 13.36 & 0.4 &{\cellcolor{yellow} 1.2} & 13.28 & 1.1 &{\cellcolor{green}  0.5} \\
 ~$^{58}$Fe  & ~~$^{59}$Fe & 1.29 & 1.5 &  1.32 & 2.3 &{\cellcolor{green}  1.9} &  1.31 & 5.1 &{\cellcolor{green}  1.5} \\
 ~$^{59}$Co  & ~~$^{60}$Co &37.92 & 0.3 & 37.18 & 0.2 &{\cellcolor{lred}   -2.0} & 37.18 & 0.7 &{\cellcolor{lred}   -2.0} \\
 ~$^{63}$Cu  & ~~$^{64}$Cu &  4.6 & 0.8 &   4.5 & 0.4 &{\cellcolor{yellow}-2.6} &  4.47 & 4.1 &{\cellcolor{green} -3.3} \\
 ~$^{93}$Nb  & ~~$^{94}$Nb &1.150 & 22. &  1.15 & 4.3 &{\cellcolor{green}  0.0} & 1.149 & 4.0 &{\cellcolor{green}  1.1} \\
 ~$^{93}$Nb  & ~$^{94m}$Nb &0.864 & 12. & 0.853 & 0.9 &{\cellcolor{green}  0.0} & 0.862$   $    &{\cellcolor{green}  0.9} \\
 $^{109}$Ag  & $^{110m}$Ag &3.909 & 1.3 &  3.95 & 1.3 &{\cellcolor{green}  1.1} & 4.211 & 5.1 &{\cellcolor{green}  7.7} \\
 $^{113}$In  & $^{114m}$In &8.733 & 1.7 &   8.1 & 3.7 &{\cellcolor{green} -7.2} & 8.134 & 4.6 &{\cellcolor{green} -6.9} \\
 $^{115}$In  & $^{116m}$In &157.5 & 0.7 & 162.3 & 0.4 &{\cellcolor{lred}    3.1} & 159.8 & 1.0 &{\cellcolor{green}  1.5} \\
 $^{139}$La  & ~$^{140}$La & 9.29 & 0.6 &  9.21 & 0.4 &{\cellcolor{green} -0.8} &  9.04 & 3.9 &{\cellcolor{green} -2.7} \\
 $^{181}$Ta  & ~$^{182}$Ta & 20.3 & 4.7 &  20.4 & 1.5 &{\cellcolor{green}  0.3} &  20.7 & 3.0 &{\cellcolor{green}  1.6} \\
 $^{186}$W   & ~$^{187}$ W & 34.7 & 3.0 &  38.1 & 1.3 &{\cellcolor{lred}    9.9} &  38.1 & 5.0 &{\cellcolor{green}  9.9} \\
 $^{197}$Au  & ~$^{198}$Au &98.65 & 0.1 & 98.65 & 0.1 &{\cellcolor{green}  0.0} & 98.70 & 0.5 &{\cellcolor{green}  0.0} \\
 $^{232}$Th  & ~$^{233}$Th & 7.34 & 0.5 &  7.35 & 0.4 &{\cellcolor{green}  0.1} &  7.34 & 5.4 &{\cellcolor{green}  0.0} \\
 $^{238}$U   & ~$^{239}$ U & 2.67 & 0.6 & 2.682 & 0.7 &{\cellcolor{green}  0.5} &  2.68 & 1.0 &{\cellcolor{green}  0.5} \\ \hline\hline
\end{tabular}
\vspace{-2mm}
\end{table}

\begin{table*}[!htbp]
\vspace{-4mm}
\caption{Comparison of the resonance integrals $I_{0}$ from \mbox{IRDFF-II} to the values derived from the Kayzero library~\cite{Kayzero} and the values by Mughabghab~\cite{Mug16}.
$I_{0}$ is calculated by \eqref{eq:I0} using $E1=0.55$~eV and $E2=2$~MeV for the Kayzero comparison and $E1=0.5$~eV and $E2=20$~MeV for the Mughabghab comparison.
Relative uncertainties and $I_{0}$ differences are given in \%.}
\label{tab:Kayzero_RI}
\begin{tabular}{ c c | c c c c r | c c c c r } \hline\hline
             &             & \multicolumn{2}{c}{Kayzero} & \multicolumn{3}{c}{IRDFF-II}& \multicolumn{2}{c}{Mughabghab-2016} & \multicolumn{3}{c}{IRDFF-II} \\
 Target      & Prod.       & $I_0$  & Unc.  & $I_0$  & Unc.   & Diff.& $I_0$  & Unc.    & $I_0$  & Unc.   & Diff  \\
             &             &  [b]   & [\%]  & [b]    & [\%]   & [\%] & [b]    & [\%]    & [b]    & [\%]   & [\%]  \\ \hline
\T
 ~$^{23}$Na  & ~~$^{24}$Na &0.3029 & 4.7  & 0.3041 & 4.5  &{\cellcolor{green}  0.4} &0.3120 & 3.2  & 0.3158 & 4.5  &{\cellcolor{green}  1.2}  \\
 ~$^{45}$Sc  & ~~$^{46}$Sc &  11.3 & 0.4  &  11.32 & 7.6  &{\cellcolor{green}  0.2} &  12.1 & 4.1  &  11.89 & 7.6  &{\cellcolor{green} -1.8}  \\
 ~$^{50}$Cr  & ~~$^{51}$Cr & 8.047 & 2.5  &  6.528 & 3.9  &{\cellcolor{lred}   -19.}&   7.2 & 2.8  &  6.834 & 3.9  &{\cellcolor{green} -5.1}  \\
 ~$^{55}$Mn  & ~~$^{56}$Mn & 13.91 & 3.0  &  13.25 & 3.8  &{\cellcolor{green} -4.7} &  13.3 & 3.8  &  13.53 & 3.8  &{\cellcolor{green}  1.7}  \\
 ~$^{58}$Fe  & ~~$^{59}$Fe &  1.26 & 1.8  &  1.246 & 4.9  &{\cellcolor{green} -1.3} &  1.34 & 12.  &  1.275 & 4.9  &{\cellcolor{green} -4.8}  \\
 ~$^{59}$Co  & ~~$^{60}$Co & 75.58 & 3.0  &  75.01 & 0.8  &{\cellcolor{green} -0.8} &  75.8 & 2.6  &  75.80 & 0.8  &{\cellcolor{green}  0.0}  \\
 ~$^{63}$Cu  & ~~$^{64}$Cu &   5.3 & 0.8  &  4.885 & 4.2  &{\cellcolor{green} -7.3} &  4.97 & 1.6  &  4.988 & 4.2  &{\cellcolor{green}  0.4}  \\
 ~$^{93}$Nb  & ~~$^{94}$Nb &  8.45 & 12.  &  8.725 & 3.4  &{\cellcolor{green}  3.2} &  8.37 & 4.4  &  8.753 & 3.4  &{\cellcolor{green}  4.6}  \\
 ~$^{93}$Nb  & ~~$^{94m}$Nb&  6.35 & 12.  &  6.546 & 3.4  &{\cellcolor{green}  3.1} &  6.21 & 6.4  &  6.567 & 3.4  &{\cellcolor{green}  5.8}  \\
 $^{109}$Ag  & $^{110m}$Ag & 65.27 & 4.4  &  68.39 & 7.0  &{\cellcolor{green}  4.8} &  65.1 & 4.5  &  68.50 & 7.0  &{\cellcolor{green}  5.3}  \\
 $^{113}$In  & $^{114m}$In &211.33 & 2.4  &  224.7 & 8.7  &{\cellcolor{green}  6.3} & 182.9 & 5.5  &  225.1 & 8.7  &{\cellcolor{yellow}23.0}  \\
 $^{115}$In  & $^{116m}$In &2645.4 & 2.0  &  2539. & 3.1  &{\cellcolor{green} -4.0} &  2650 & 3.8  &  2647. & 3.1  &{\cellcolor{green} -0.1}  \\
 $^{139}$La  & ~$^{140}$La & 11.52 & 0.6  &  11.92 & 5.5  &{\cellcolor{green}  3.5} &  12.1 & 5.0  &  12.11 & 5.5  &{\cellcolor{green}  0.1}  \\
 $^{181}$Ta  & ~$^{182}$Ta & 677.6 & 4.7  &  659.5 & 3.8  &{\cellcolor{green} -2.7} &   655 & 3.1  &  660.0 & 3.8  &{\cellcolor{green}  0.8}  \\
 $^{186}$W   & ~$^{187}$ W & 474.9 & 3.5  &  483.4 & 4.5  &{\cellcolor{green}  1.8} &   480 & 3.1  &  484.2 & 4.5  &{\cellcolor{green}  0.9}  \\
 $^{197}$Au  & ~$^{198}$Au & 1550. & 1.8  &  1542. & 2.1  &{\cellcolor{green} -0.5} &  1550 & 1.8  &  1545. & 2.1  &{\cellcolor{green} -0.3}  \\
 $^{232}$Th  & ~$^{233}$Th & 84.42 & 3.6  &  84.14 & 19.  &{\cellcolor{green} -0.3} &  83.3 & 1.8  &  84.29 & 19.  &{\cellcolor{green}  1.2}  \\
 $^{238}$U   & ~$^{239}$ U & 276.0 & 1.4  &  274.9 & 1.1  &{\cellcolor{green} -0.4} &   277 & 1.1  &  275.0 & 1.1  &{\cellcolor{green} -0.7}  \\ \hline\hline
\end{tabular}
\vspace{-2mm}
\end{table*}

\subsection{Thermal Reactor Neutron Fields - Thermal Cross Sections and Resonance Integrals} \label{Sec_VII_M}
There is a direct relation between average cross sections with 1/$v$ shape and the 2200~m/s value in a Maxwellian thermal reactor spectrum. The resonance integral also represents an idealized epithermal reactor spectrum. These quantities can be compared to compilations such as the Atlas of Resonance Parameters~\cite{Mug16}. The values of the thermal capture cross sections reported in the Atlas are based on decades of experience with the interpretation of different kinds of experimental data, resonance theory and systematics. There could also be an element of subjective judgement involved, in the case of inconsistent or lacking data. Since the values in the Atlas were often used in the evaluation process, such comparison should be considered as data verification, rather than validation.

The $k_{0}$ and $Q_{0}$ constants for neutron activation analysis by the k$_{0}$ standardization method in the Kayzero library were measured independently over the years. The thermal cross sections and the resonance integrals derived from them are representative of an integral measurement in an idealized thermal reactor system. They can be used for data validation purposes.

The k$_{0}$ factors can be measured quite accurately in irradiation facilities with strongly enhanced thermal spectra. They provide the ratio to the gold standard of the product of the cross section and the gamma-emission probability. In some cases the uncertainty in the gamma-emission probabilities introduces a dominant contribution to the uncertainty in the derived cross section. Other sources of uncertainty originate from detector calibration, measured masses and other details related to the analysis of the measured gamma spectra. The methodology was intended mainly for use with 1/\textit{v} absorbers. The Westcott \textit{g}-factors and the corrections for the non-ideal cadmium filter were seldom used in the interpretation of the measured data, but for the nuclides relevant for dosimetry these corrections are generally smaller than the overall experimental uncertainties.

In Table~\ref{tab:Kayzero_sig0} the thermal cross sections from the compilation by Mughabghab and in the \mbox{IRDFF-II} library are compared to the reference values derived from the Kayzero library. Generally, the values from \mbox{IRDFF-II} lie close to the reference values, except for the $^{186}$W capture cross section, which differs by nearly 10~\%. However, according to the consistency criteria defined in Table~\ref{tab:X_A_3}, the thermal capture cross sections of $^{23}$Na and $^{59}$Co also show poor agreement, mainly because the uncertainties assigned to the cross sections are small. Comparison of the cross sections in the Atlas by Mughabghab shows poor agreement for $^{45}$Sc, $^{59}$Co and $^{115}$In and marginally acceptable agreement for $^{23}$Na and $^{63}$Cu for the same reason, even though the values are comparably close to the reference. It must be mentioned that the main contribution to the uncertainties of the thermal cross sections of $^{93}$Nb come from the gamma-emission probabilities.

The case of $^{186}$W is an interesting one, where we believe the gamma-emission probabilities are mainly responsible for the difference. In the recent re-evaluation of the gamma-emission probabilities the values changed by about 20~\%, but the resulting derived thermal capture cross section could not be fitted in a rigorous resonance analysis performed at JRC in Geel~\cite{Sch14}. The case of the thermal capture cross section of $^{186}$W remains an open issue.

Resonance integrals can be derived from the Q$_{0}$ factors, which are usually measured by the cadmium ratio method. The method is convenient because it practically eliminates the dependence on the detector efficiency and other features related to gamma-spectrum analysis. Its drawback is that it requires two sample irradiations, bare and under Cd cover, where the presence of Cd may distort the local spectrum in the irradiation facility close to the Cd-covered sample. The other source of ambiguity is the definition of the resonance integral. The common definition:

\begin{equation}
\label{eq:I0}
 I_{0} = \int_{E1}^{E2} \sigma(E) \frac{dE}{E},
\end{equation}
where $E1 = 0.5$~eV and $E2$ goes to infinity is not what is measured. Mughaghab used $E1=0.5$~eV and $E2=20$~MeV. The effective Cd cutoff energy for a 1 mm thick Cd cover is around 0.55~eV and it is not sharp. Secondly, measurements are made in thermal-reactor spectra, which fall off rapidly above 2~MeV. A better approximation to compare with measured data is to use $E1$ = 0.55~eV and $E2$ = 2~MeV. An additional refinement is to define the Cd transmission correction factor $F_{cd}$ to correct for the difference between the idealized sharp cutoff at 0.55~eV and the actual cutoff by the rising Cd cross section with decreasing neutron energy. The Cd factor is defined as

\begin{equation}
F_{cd}=\frac{\int _{0}^{\infty } t(E)\sigma (E)\phi(E)dE }{\int _{Ecd}^{E3}\sigma(E)\phi (E)dE}
\end{equation}

From the definition it follows that measured values by the Cd ratio method must be divided by $F_{cd}$ for comparison with the idealized values. A compilation of cadmium factors is available in the IAEA document \href{https://www-nds.iaea.org/publications/indc/indc-nds-0693/}{INDC(NDS)-0693} for all nuclides of interest for NAA.

%In Table~\ref{tab:Kayzero_sig0} the thermal cross sections in the the compilation by Mughabghab and in the \mbox{IRDFF-II} library are compared to the reference values derived from the Kayzero library. Differences from the reference values are also shown.

The $F_{cd}$ factors were seldom used in the determination of the $Q_{0}$ factors for the Kayzero library. In the current Kayzero library an attempt was made to measure the $F_{cd}$ factor of  $^{186}$W by making measurements in three irradiation channels with different spectral characteristics, but no uncertainty was specified. We believe that the value 0.908 adopted by DeCorte~\cite{Dec87} is unrealistic and that good agreement of the resonance integral with Atlas and with \mbox{IRDFF-II} in Table~\ref{tab:Kayzero_RI} is a coincidence, stemming from error compensation between $F_{cd}$ and the gamma-emission probabilities in deriving the thermal capture cross section.

\begin{table*}[t]
\vspace{-4mm}
\caption{Measured and calculated SACS for the \mbox{IRDFF-II} cross sections in High Temperature Maxwellian Neutron Fields. Data from KADoNiS 1.0 \cite{KADoNiS10} are the stellar Maxwellian averaged cross sections in the centre-of-mass system.}
\label{tab:SACSmaxwel}
% \begin{tabular}{|p{0.3in}|p{0.6in}|p{0.6in}|p{0.6in}|p{0.7in}|p{0.7in}|p{0.5in}|p{0.5in}|p{0.7in}|}
\begin{tabular}{ l r l l r c l c c c}
\hline\hline
&  &  & \multicolumn{3}{p{2.0in}}{Measured SACS} & \multicolumn{3}{p{1.8in}}{Calculated SACS} \\
\cline{4-6} \cline{7-9}
Reaction& \textit{kT$_{lab}$} & $E_{50\%}$ & Value & Unc. & Reference & Value & Unc. & Spec. Unc. & C/E \\
Notation& [keV] & [keV] & [mb] & [\%] & & [mb] & [\%] & [\%] & \\
\hline
\T
Mn55g   & 30.55 & 25.78 & 3.315E+01 & 9.6  & \cite{KADoNiS10} & 3.283E+01 & 6.05 & 0.00 & 0.990 $\pm$ 11.3\% \cellcolor{green}  \\
Cu63g   & 30.48 & 25.84 & 6.093E+01 & 10.4 & \cite{KADoNiS10} & 7.144E+01 & 9.70 & 0.00 & 1.173 $\pm$ 14.2\% \cellcolor{yellow} \\
Au197g  & 25.00 & 26.64 & 6.450E+02 & 3.9  & \cite{Jim14}     & 6.899E+02 & 0.69 & 0.00 & 1.070 $\pm$ 4.1\%  \cellcolor{green}  \\
Co59g   & 30.51 & 27.05 & 4.379E+01 & 4.6  & \cite{KADoNiS10} & 3.444E+01 & 2.57 & 0.00 & 0.787 $\pm$ 5.3\%  \cellcolor{lred}    \\
U238g   & 25.30 & 29.00 & 3.910E+02 & 4.35 & \cite{Wal14}     & 3.899E+02 & 1.44 & 0.00 & 0.997 $\pm$ 4.6\%  \cellcolor{green}  \\
Nb93g   & 30.32 & 29.46 & 2.807E+02 & 2.4  & \cite{KADoNiS10} & 2.638E+02 & 0.78 & 0.00 & 0.940 $\pm$ 2.5\%  \cellcolor{green}  \\
La139g  & 30.22 & 29.52 & 3.284E+01 & 9.7  & \cite{KADoNiS10} & 4.159E+01 & 10.8 & 0.00 & 1.266 $\pm$ 14.5\% \cellcolor{yellow} \\
Au197g  & 28.00 & 30.25 & 6.160E+02 & 2.8  & \cite{Fei12}     & 6.451E+02 & 0.64 & 0.00 & 1.047 $\pm$ 2.8\%  \cellcolor{green}  \\
Au197g  & 28.00 & 30.25 & 6.200E+02 & 2.6  & \cite{Fei12}     & 6.451E+02 & 0.64 & 0.00 & 1.040 $\pm$ 2.7\%  \cellcolor{green}  \\
Ta181g  & 30.17 & 31.71 & 7.765E+02 & 2.5  & \cite{KADoNiS10} & 7.506E+02 & 3.87 & 0.00 & 0.967 $\pm$ 4.6\%  \cellcolor{green}  \\
Sc45g   & 30.67 & 32.58 & 6.234E+01 & 3.9  & \cite{KADoNiS10} & 6.797E+01 & 6.99 & 0.00 & 1.090 $\pm$ 8.0\%  \cellcolor{green}  \\
Au197g  & 30.15 & 32.65 & 6.200E+02 & 1.8  & \cite{KADoNiS10} & 6.181E+02 & 0.61 & 0.00 & 0.997 $\pm$ 1.9\%  \cellcolor{green}  \\
In113gm & 30.26 & 35.03 & 1.561E+02 & 29.9 & \cite{KADoNiS10} & 1.616E+02 & 4.05 & 0.00 & 1.035 $\pm$ 30.2\% \cellcolor{green}  \\
In115gm & 30.26 & 35.04 & 6.305E+02 & 30.1 & \cite{KADoNiS10} & 6.080E+02 & 4.05 & 0.00 & 0.964 $\pm$ 30.4\% \cellcolor{green}  \\
In113gg & 30.27 & 35.04 & 7.867E+02 & 10.7 & \cite{KADoNiS10} & 7.696E+02 & 0.00 & 0.00 & 0.978 $\pm$ 10.7\% \cellcolor{green}  \\
Na23g   & 31.30 & 35.27 & 1.825E+00 & 3.1  & \cite{KADoNiS10} & 2.023E+00 & 8.49 & 0.00 & 1.109 $\pm$ 9.1\%  \cellcolor{yellow} \\
In113g  & 30.27 & 35.91 & 8.201E+02 & 14.0 & \cite{KADoNiS10} & 9.539E+02 & 0.00 & 0.00 & 1.163 $\pm$ 14.0\% \cellcolor{yellow} \\
Fe58g   & 30.52 & 37.38 & 1.450E+01 & 6.5  & \cite{KADoNiS10} & 1.506E+01 & 16.3 & 0.00 & 1.039 $\pm$ 17.5\% \cellcolor{green}  \\
W186g   & 30.16 & 37.46 & 2.291E+02 & 5.5  & \cite{KADoNiS10} & 1.920E+02 & 1.93 & 0.00 & 0.838 $\pm$ 5.8\%  \cellcolor{lred}    \\
Ag109g  & 30.28 & 42.23 & 4.075E+01 & 3.8  & \cite{KADoNiS10} & 5.785E+01 & 6.76 & 0.00 & 1.420 $\pm$ 7.7\%  \cellcolor{lred}    \\
U238g   & 426.0 & 632.2 & 1.080E+02 & 3.70 & \cite{Wal14}     & 3.899E+02 & 1.82 & 0.00 & 1.011 $\pm$ 4.1\%  \cellcolor{green}  \\
\hline\hline
\end{tabular}
\vspace{-4mm}
\end{table*}

Another possible source of discrepancy in the measured resonance integrals stems from the fast-fission spectrum contribution. This depends on the irradiation facility. It may become significant in the Cd-ratio measurements, since these have to be made in conditions, where the epithermal contribution to the reaction rate is not so small. By a minor extension to the governing equations the fast fission effect can be treated explicitly \cite{Trkov:NAA}. The effect is usually small, but it has been demonstrated that in some cases it can amount to a few percent in the measured $Q_{0}$ \cite{Trkov:Al}. However, this methodology is not generally used by the NAA community.

The Atlas value of the resonance integrals of $^{45}$Sc and $^{55}$Mn are within two-sigma uncertainty compared to the values derived from the $Q_{0}$ factors in the Kayzero library, but the resonance integrals of $^{50}$Cr, $^{63}$Cu and $^{113}$In disagree with the values derived from the Kayzero library, as seen from Table~\ref{tab:Kayzero_RI}. The value for $^{113}$In was derived from the total capture resonance integral, assuming the same branching ratio as for the thermal cross section, which might be inappropriate. The values calculated from the \mbox{IRDFF-II} library generally agree to within two-sigma with Kayzero, except $^{50}$Cr. The neutron-induced resonance data of these nuclides and particularly for n+$^{50}$Cr remain an open issue that requires further investigation.

%\newpage
\subsection{High Temperature Maxwellian ($kT\approx$~25--430~keV)}\label{Sec_VII_N}

Most of the experimental Maxwellian spectrum-averaged cross sections, MACS, were taken from the latest version 1.0 of KADoNiS~\cite{KADoNiS10} based on the gold standard $^{197}$Au(n,$\gamma$) = 611.6$~\pm~$6.0~mb. This database presents the so-called stellar SACS which were derived from the experimental by applying the factor 2/$\sqrt{\pi}$ and converting them in the centre-of-mass system to the temperature \textit{kT$_{cm}$} = 30~keV \cite{Bao00}. For comparison with these data we have computed the \mbox{IRDFF-II} SACS for the corresponding \textit{kT$_{lab}$ = kT$_{cm}$~*~(1 + A)/A}, where \textit{A} is the ratio of the mass of the target nucleus to that of a neutron.
Additionally, all KADoNiS cross sections and their uncertainties were renormalized to the newly introduced SACS gold standard $^{197}$Au(n,$\gamma$) = 620$~\pm~$11~mb \cite{Car18} which adds a +1.4\% correction.

\begin{figure*}[!thbp]
\vspace{-2mm}
\includegraphics[width=\textwidth]{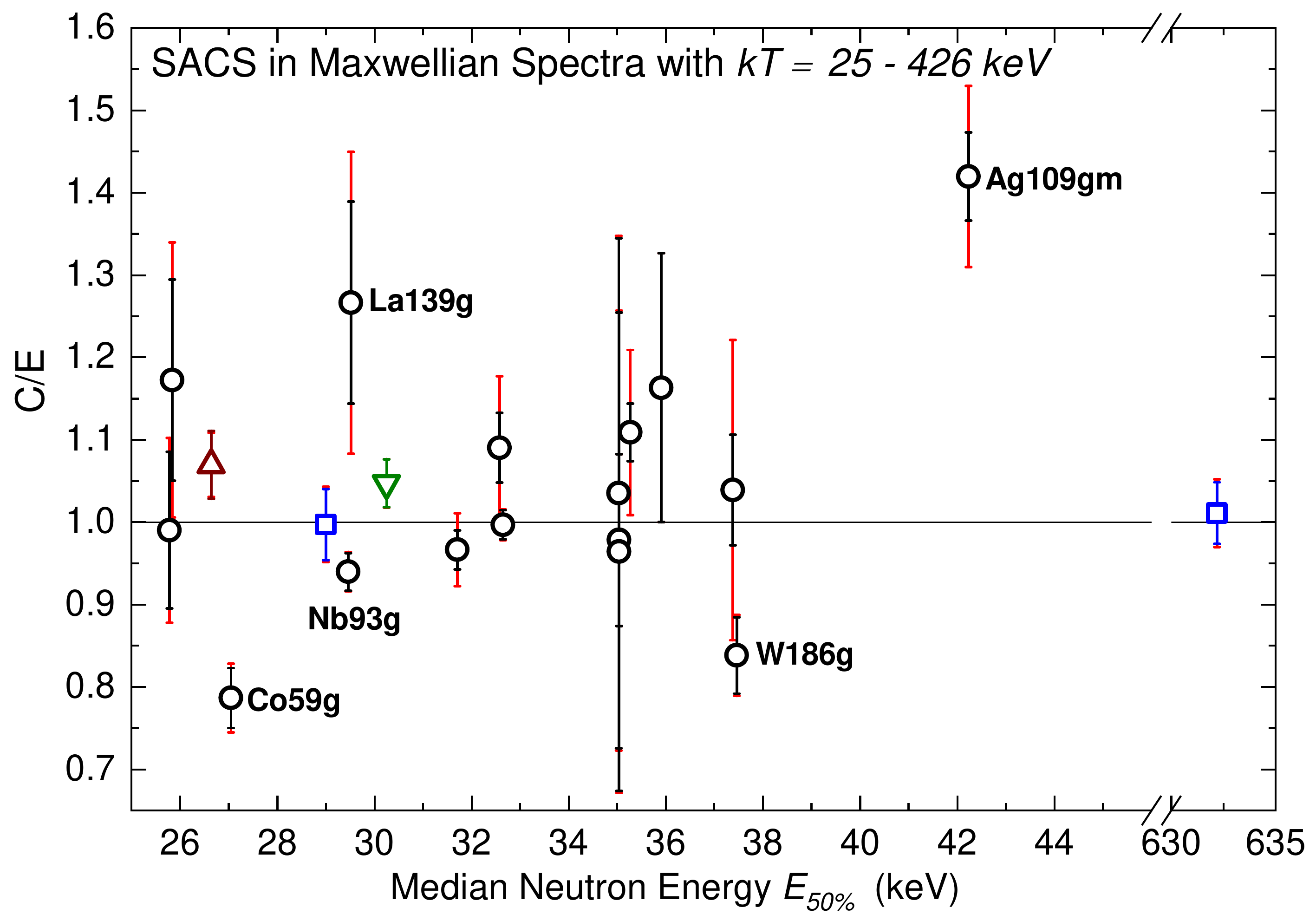}
\caption{(Color online) C/E ratios for SACS measured and calculated in the (25--426)~keV Maxwellian spectra as a function of the median energy. Experimental data: $\circ$ - KADoNiS-1.0~\cite{KADoNiS10}, other symbols -- G. Feinberg \etal\cite{Fei12}, A. Wallner \etal\cite{Wal14}, P. Jimenez-Bonilla \etal\cite{Jim14}. Uncertainties resulting from measurements and \mbox{IRDFF-II} cross sections are shown by black and red bars. Plotted values are listed in Table~\ref{tab:SACSmaxwel}.}
\label{fig:SACS_Maxw}
\vspace{-2mm}
\end{figure*}

Several other measurements \cite{Fei12,Jim14,Wal14} were carried out, but were not included in KADoNiS yet. In particular, A.~Wallner and co-workers~\cite{Wal14} have measured SACS, not by the traditional activation technique, but by the highly-accurate Accelerator Mass Spectroscopy technique (AMS). In addition to the measurement at typical neutron spectrum average energy of 25~keV they have measured SACS at higher $kT = 426$~keV. These experiments delivered their results in the laboratory system corresponding to different $kT$ values, so the \mbox{IRDFF-II} SACS were calculated at those temperatures.

The resulting C/E ratios with \mbox{IRDFF-II} for the 17 dosimetry reactions addressed within the KADoNiS database and several new experiments, along with the associated experimental and \mbox{IRDFF-II} cross section uncertainties, are listed in Table~\ref{tab:SACSmaxwel}. The trends in the C/E ratios are displayed in Fig.~\ref{fig:SACS_Maxw}. There the uncertainties of measurements and of the \mbox{IRDFF-II} cross sections are displayed separately allowing to observe their contributions to the total C/E uncertainty.

As one can see, most of the \mbox{IRDFF-II} Maxwellian averaged cross sections agree with measurements to within 1--2 sigma. However we do not observe a good agreement for (n,$\gamma$) reactions on  $^{59}$Co, $^{186}$W and $^{109}$Ag targets. Whether this is a problem in IRDFF-II evaluations or it is an experimental problem remains an open question. The reaction $^{232}$Th(n,$\gamma$) is a rare case among the \mbox{IRDFF-II} (n,$\gamma$) reactions that is not validated in the Maxwellian-30~keV benchmark field yet.

%\clearpage
\section{CONSISTENCY OF CROSS SECTIONS IN STANDARD AND REFERENCE FIELDS}  \label{Sec_VIII}

Because of the uncertainty in the neutron fluence for a given reactor operation, a better metric than a spectral index for a fission reactor field is the consistency of the set of measured activities in a given reactor field. The consistency is best captured by the overall chi-squared per degree of freedom ($\chi^2/dof$) metric as obtained from a least-squares-based spectrum adjustment in neutron benchmark fields. The relevant reaction-specific metric is then the resulting least-squares-based activity adjustment and the C/E for the adjusted activity.

The following subsections address this least-squares metric for specific neutron benchmark fields.

\subsection{$^{252}$Cf(s.f.)} \label{Sec_VIII_A}

Sec.~\ref{Sec_VII_A} reports on the validation evidence using individual measured values for the spectrum-averaged cross section in this standard neutron benchmark field. A more exacting validation metric, based on the consistency of the set of validation data gathered within this neutron field, is depicted in Table~\ref{tab:XI_A_9}.

The LSL-M2 code was used to perform the least-squares analysis. The inputs included the measurements, with uncertainties, the \mbox{IRDFF-II} cross sections, with covariance matrices, and the \textit{a priori }spectrum, in this case as provided by time-of-flight measurements, along with the corresponding covariance matrix. This approach permitted all of the input data to be adjusted, consistent with the input uncertainties, but the major result was to adjust the spectrum and to provide a high-fidelity test of the consistency of the data for the neutron field. A chi-squared per degree of freedom, $\chi^2/dof$, is reported and provides a metric on the consistency of the data.

\begin{table}[!htp]
 \caption{Least-squares-based consistency of data for the $^{252}$Cf(s.f.) neutron benchmark field}
 \label{tab:XI_A_9}
 \begin{tabular}{ l c c c c } \hline\hline
%\begin{tabular}{ p{0.6in}   p{0.6in}    p{0.7in}   c } \hline\hline
%\begin{tabular}{|p{0.4in} | p{0.4in} |  p{0.5in} |p{0.7in}|} \hline
  Reaction  & Act. Adj. & Spectrum Unc. &     \\
  Notation  & [\%]      &  [\%]         & C/E \\ \hline
\T\T
Au197g  & -0.33\%  & 0.42\% &\cellcolor{green}  0.991 $\mathrm{\pm}$~1.79\% \\
Ta181g  & -0.35\%  & 0.75\% &\cellcolor{green}  0.935 $\mathrm{\pm}$~5.65\% \\
Th232g  & ~0.43\%  & 0.43\% &\cellcolor{green}  1.032 $\mathrm{\pm}$~5.53\% \\
Cu63g   & ~0.20\%  & 0.61\% &\cellcolor{green}  1.010 $\mathrm{\pm}$~8.64\% \\
In115g  & -1.20\%  & 0.37\% &\cellcolor{green}  0.979 $\mathrm{\pm}$~2.45\% \\
U235f   & ~1.94\%  & 0.07\% &\cellcolor{green}  1.013 $\mathrm{\pm}$~1.21\% \\
Pu239f  & -0.12\%  & 0.07\% &\cellcolor{green}  0.992 $\mathrm{\pm}$~1.38\% \\
Np237f  & ~0.66\%  & 0.23\% &\cellcolor{green}  1.000 $\mathrm{\pm}$~1.69\% \\
In115n  & -2.28\%  & 0.38\% &\cellcolor{yellow} 0.964 $\mathrm{\pm}$~1.57\% \\
Nb93n   & ~1.47\%  & 0.36\% &\cellcolor{green}  1.010 $\mathrm{\pm}$~3.73\% \\
In113n  & -1.39\%  & 0.40\% &\cellcolor{green}  0.978 $\mathrm{\pm}$~2.15\% \\
U238f   & -1.45\%  & 0.40\% &\cellcolor{green}  0.978 $\mathrm{\pm}$~1.70\% \\
Th232f  & -5.66\%  & 0.43\% &\cellcolor{green}  0.935 $\mathrm{\pm}$~5.08\% \\
Hg199n  & -0.54\%  & 0.44\% &\cellcolor{green}  0.985 $\mathrm{\pm}$~2.22\% \\
BH3     & -1.95\%  & 0.51\% &\cellcolor{green}  0.931 $\mathrm{\pm}$11.58\% \\
Ti47p   & ~1.31\%  & 0.62\% &\cellcolor{green}  1.015 $\mathrm{\pm}$~2.16\% \\
S32p    & ~2.68\%  & 0.73\% &\cellcolor{green}  1.023 $\mathrm{\pm}$~3.65\% \\
Zn64p   & ~1.27\%  & 0.78\% &\cellcolor{green}  1.009 $\mathrm{\pm}$~3.31\% \\
Ni58p   & ~0.35\%  & 0.75\% &\cellcolor{green}  1.000 $\mathrm{\pm}$~1.66\% \\
Fe54p   & ~1.36\%  & 0.78\% &\cellcolor{green}  1.015 $\mathrm{\pm}$~1.79\% \\
Pb204n  & -1.58\%  & 0.97\% &\cellcolor{green}  0.971 $\mathrm{\pm}$~2.74\% \\
Al27p   & -4.51\%  & 1.14\% &\cellcolor{green}  0.950 $\mathrm{\pm}$~2.51\% \\
Co59p   & ~1.12\%  & 1.15\% &\cellcolor{green}  1.015 $\mathrm{\pm}$~3.16\% \\
Ti46p   & -1.44\%  & 1.18\% &\cellcolor{green}  0.974 $\mathrm{\pm}$~2.63\% \\
Si28p   & ~2.18\%  & 1.41\% &\cellcolor{green}  1.031 $\mathrm{\pm}$~2.67\% \\
Cu63a   & ~0.22\%  & 1.37\% &\cellcolor{green}  1.006 $\mathrm{\pm}$~2.93\% \\
Fe56p   & -0.23\%  & 1.45\% &\cellcolor{green}  0.998 $\mathrm{\pm}$~2.91\% \\
Mg24p   & ~4.59\%  & 1.56\% &\cellcolor{green}  1.050 $\mathrm{\pm}$~2.92\% \\
U2382   & ~3.92\%  & 1.56\% &\cellcolor{green}  1.045 $\mathrm{\pm}$11.23\% \\
Ti48p   & ~0.07\%  & 1.55\% &\cellcolor{green}  1.005 $\mathrm{\pm}$~3.53\% \\
Co59a   & -0.14\%  & 1.54\% &\cellcolor{green}  1.000 $\mathrm{\pm}$~3.39\% \\
Al27a   & -0.11\%  & 1.59\% &\cellcolor{green}  1.003 $\mathrm{\pm}$~2.07\% \\
U51a    & -0.98\%  & 1.85\% &\cellcolor{green}  0.989 $\mathrm{\pm}$~3.23\% \\
Au1972  & ~0.04\%  & 1.95\% &\cellcolor{green}  1.005 $\mathrm{\pm}$~2.85\% \\
Nb932   & -0.44\%  & 2.20\% &\cellcolor{green}  1.000 $\mathrm{\pm}$~4.95\% \\
I1272   & ~1.10\%  & 2.30\% &\cellcolor{green}  1.018 $\mathrm{\pm}$~3.87\% \\
Cu652   & -0.84\%  & 2.90\% &\cellcolor{green}  0.993 $\mathrm{\pm}$~3.70\% \\
Co592   & ~0.36\%  & 3.24\% &\cellcolor{green}  1.006 $\mathrm{\pm}$~4.27\% \\
Cu632   & ~3.32\%  & 4.10\% &\cellcolor{green}  1.032 $\mathrm{\pm}$~5.74\% \\
F192    & ~1.09\%  & 4.39\% &\cellcolor{green}  1.007 $\mathrm{\pm}$~5.66\% \\
Zr902   & -1.39\%  & 4.99\% &\cellcolor{green}  0.981 $\mathrm{\pm}$~5.77\% \\
Ni582   & ~0.87\%  & 5.90\% &\cellcolor{green}  0.997 $\mathrm{\pm}$~7.07\% \\
Ni60p   & ~2.41\%  & 1.35\% &\cellcolor{green}  1.027 $\mathrm{\pm}$~5.03\% \\
Tm1692  & -1.75\%  & 1.91\% &\cellcolor{green}  0.986 $\mathrm{\pm}$~4.74\% \\
Mn552   & -1.49\%  & 3.04\% &\cellcolor{green}  0.958 $\mathrm{\pm}$15.63\% \\
Y892    & ~1.21\%  & 4.19\% &\cellcolor{green}  0.984 $\mathrm{\pm}$~5.64\% \\
Na232   & -4.46\%  & 6.79\% &\cellcolor{green}  0.990 $\mathrm{\pm}$~7.63\% \\
Tm1693  & -0.33\%  &14.75\% &\cellcolor{lred}    0.795 $\mathrm{\pm}$16.24\% \\ \hline\hline
 \end{tabular}
\end{table}

The least-squares calculation in this standard neutron benchmark field used 48 measured reactions. As expected from the consideration of the separate spectrum-averaged cross sections addressed in Sect.~\ref{Sec_VI_A}, the measured data were very consistent with calculations and the overall least-squares spectrum adjustment had a very small $\chi^2/dof$ of 0.7042. A least-squares process using data for which there was an accurate identification of the \textit{a priori} uncertainties should have a $\chi^2/dof$ metric of 1.00. This $\chi^2/dof$ less than unity suggests that some of the input data (spectrum, cross sections, or measured activities) actually may have had smaller uncertainties than were assigned to the input quantities.

The resulting adjustment in the spectrum-averaged activity, the reduced spectrum uncertainty contributions to the C/E, and the C/E values are tabulated in Table~\ref{tab:XI_A_9}. The C/E column is color-coded in a manner consistent with the discussion in Sec.~\ref{Sec_VI_A}. The discrepant reactions are shown in ``red'', the ``acceptable'' reactions in ``yellow'', and the ``good'' reactions in \ql green\qrs.  While a least-squares analysis can adjust all of the input data that have identified associated uncertainties, changing their contribution to the uncertainty in the resulting C/E value, this table does not  provide the details of the resulting cross section and measurement uncertainty contributions since these uncertainties are not changed significantly from those presented in the spectrum-averaged cross section comparison in Table~\ref{tab:XI_A_9}. While the spectrum contribution to the overall C/E uncertainty is reduced here relative to that seen in the spectrum-averaged cross section comparison, the difference is not large because of the very-well-characterized \textit{a priori} $^{252}$Cf(s.f.) spectrum, which was measured using time-of-flight techniques, used in the analysis. In less-well-characterized neutron fields, the least-squares analysis will significantly reduce the spectrum contribution to the resulting uncertainty in the C/E value.

\begin{figure}[!htbp]
\vspace{-3mm}
\includegraphics[width=\columnwidth]{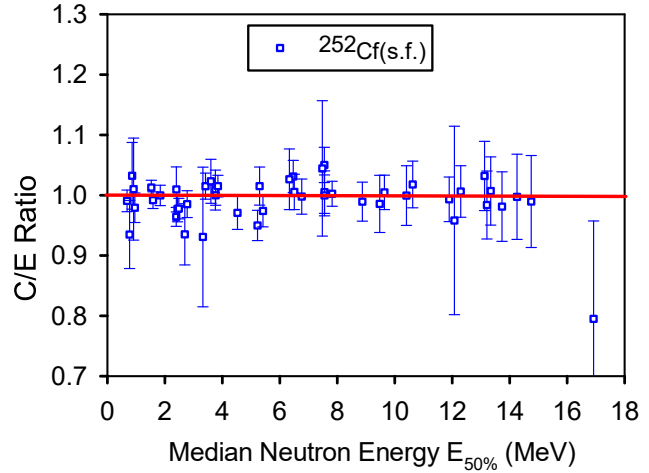}
\vspace{-5mm}
\caption{(Color online) Least-squares-based C/E as a function of median energy E$_{50\%}$ for the activity measurements in the $^{252}$Cf(s.f.) standard neutron benchmark Field. Uncertainties bars reflect the combined rms uncertainty contributions from the measurement, cross sections, and spectrum. Plotted values are listed in Table~\ref{tab:XI_A_9}.}
\label{fig:XI_A_40}
%\vspace{-2mm}
\end{figure}

Fig.~\ref{fig:XI_A_40} shows the C/E values in this $^{252}$Cf(s.f.) neutron benchmark field plotted versus the median response energy for the 48 reactions used in the least-squares adjustment. There is no significant trend seen in the C/E ratios versus the median response energy. The C/E uncertainties are clearly much larger in the low and high energy portion of the spectrum. The adjustment produced a 1.0\% adjusted scale factor. The scale factor adjustment is taken out of the C/E values in this figure.

The largest contributor to the overall $\chi^2/dof$ came from the Al27p reaction, which had an individual $\chi^2$ contribution of 4.02, but, based on the acceptance criteria presented in  Table~\ref{tab:X_A_3},  the Al27p reaction, with a C/E of 0.950 $\pm$ 2.51\%, still showed a \ql good\qr validation status in this neutron field.  The next largest contributor to the overall $\chi^2/dof$ came from the In115n reaction, which had an individual $\chi^2$ contribution of 3.23, with a C/E of 0.964 $\pm$ 1.57\%. The color coding in Table~\ref{tab:XI_A_9} shows that this reaction, because of its small overall C/E uncertainty, was just outside of the required two standard deviations and has a status of \ql acceptable\qr rather than \ql good\qr as validation evidence.  The only \ql poor\qr validation status seen in the data available in this benchmark field came from the Tm1693 reaction. This reaction, which had an individual $\chi^2$ contribution of 2.37, had a C/E of 0.795 $\pm$ 16.24\%, which, even though its large uncertainty gave it a C/E that was well within even the two standard deviations of unity required by Table~\ref{tab:X_A_3} criterion \#1 for a \ql good\qr status, it failed to meet the Table~\ref{tab:X_A_3} C/E interval criterion \#2 for even an \ql acceptable\qr status.

Note that the uncertainty of the TOF measured \textit{a priori} spectrum is larger than 15\% at the reaction threshold of ~16~MeV and quickly rises up to 75\% at 20~MeV, therefore the least-squares fit is not well constrained in the energy region where this particular dosimeter is sensitive. The least-squares analysis showed a very strong upward adjustment (+19.3\%) in the high energy ($>17$~MeV) portion of the spectrum and a downward shift in the reaction rate (-4.5\%), that is, the least-squares process made adjustments in both the reaction probability and the spectrum in order to minimize the resulting disagreement in this high energy region.

\subsection{$^{235}$U(n$_{th}$,f)} \label{Sec_VIII_B}

Sect.~\ref{Sec_VII_B} reports on the validation evidence using individual measured values for the spectrum-averaged cross sections in this reference neutron benchmark field. The more exacting validation metric, based on the consistency of the set of validation data gathered within this neutron field as assessed by the LSL-M2 least square code, is depicted in Table~\ref{tab:XI_B_10}. The least square inputs included the measurements, with uncertainties, the \mbox{IRDFF-II} cross sections, with covariance matrices, and the \textit{a priori }spectrum, in this case as provided from the GMA analysis and incorporated into the ENDF/B-VIII.0 evaluation as addressed in Sec.~\ref{Sec_VII_B}, along with the corresponding covariance matrix. The chi-squared per degree of freedom, $\chi^2/dof$, is reported from the analysis and provides a metric on the consistency of the input data.

\begin{table}[!htp]
 \caption{Least-squares-based consistency of data for the $^{235}$U(n$_{th}$,f) neutron reference benchmark field.}
 \label{tab:XI_B_10}
 \begin{tabular}{ l c  c  c }  \hline\hline
  Reaction  & Act. Adj. & Spectrum Unc. &     \\
  Notation  & [\%]      &  [\%]         & C/E \\ \hline
\T
Li6He4   & ~~4.69\%  & 0.65\% &\cellcolor{green}  1.074 ${\pm}$~6.02\% \\
Au197g   & -11.27\%  & 0.67\% &\cellcolor{yellow} 0.889 ${\pm}$~4.58\% \\
B10a     & ~-3.82\%  & 0.64\% &\cellcolor{yellow} 0.848 ${\pm}$~9.72\% \\
U238g    & -17.62\%  & 0.64\% &\cellcolor{yellow} 0.825 ${\pm}$~9.66\% \\
Cu63g    & ~~1.46\%  & 0.61\% &\cellcolor{green}  1.019 ${\pm}$24.61\% \\
In115gm  & ~~4.78\%  & 0.60\% &\cellcolor{green}  1.052 ${\pm}$~4.83\% \\
U235f    & ~~3.11\%  & 0.52\% &\cellcolor{green}  1.034 ${\pm}$~2.35\% \\
Pu239f   & ~~~0.4\%  & 0.52\% &\cellcolor{green}  1.007 ${\pm}$~2.65\% \\
Np237f   & ~~1.82\%  & 0.56\% &\cellcolor{green}  1.022 ${\pm}$~2.79\% \\
Rh103n   & ~~1.24\%  & 0.57\% &\cellcolor{green}  1.017 ${\pm}$~4.22\% \\
Nb93nm   & ~-1.81\%  & 0.62\% &\cellcolor{green}  0.983 ${\pm}$~5.38\% \\
In115nm  & ~~0.28\%  & 0.63\% &\cellcolor{green}  1.006 ${\pm}$~2.52\% \\
In113nm  & ~-6.94\%  & 0.65\% &\cellcolor{green}  0.933 ${\pm}$~5.57\% \\
U238f    & ~~~0.4\%  & 0.65\% &\cellcolor{green}  1.007 ${\pm}$~2.40\% \\
Th232f   & ~-8.05\%  & 0.65\% &\cellcolor{yellow} 0.918 ${\pm}$~3.83\% \\
Hg199n   & ~~2.34\%  & 0.65\% &\cellcolor{green}  1.027 ${\pm}$~5.84\% \\
P31p     & ~-0.49\%  & 0.78\% &\cellcolor{green}  0.997 ${\pm}$~3.62\% \\
Ti47p    & ~~1.78\%  & 0.73\% &\cellcolor{green}  1.024 ${\pm}$~3.19\% \\
S32p     & ~-0.67\%  & 0.82\% &\cellcolor{green}  0.995 ${\pm}$~3.04\% \\
Ni58p    & ~-0.19\%  & 0.80\% &\cellcolor{green}  1.000 ${\pm}$~2.62\% \\
Zn64p    & ~~~0.6\%  & 0.85\% &\cellcolor{green}  1.008 ${\pm}$~7.59\% \\
Fe54p    & ~~2.84\%  & 0.84\% &\cellcolor{green}  1.034 ${\pm}$~2.78\% \\
Zn67p    & ~~1.01\%  & 0.80\% &\cellcolor{green}  1.014 ${\pm}$~6.64\% \\
Pb204n   & ~-8.62\%  & 0.93\% &\cellcolor{green}  0.909 ${\pm}$~6.87\% \\
Mo92p    & ~~0.67\%  & 0.97\% &\cellcolor{green}  1.011 ${\pm}$~4.07\% \\
Al27p    & ~-3.01\%  & 1.08\% &\cellcolor{green}  0.968 ${\pm}$~2.96\% \\
Co59p    & ~~1.25\%  & 1.07\% &\cellcolor{green}  1.017 ${\pm}$~3.72\% \\
Ti46p    & ~~-0.5\%  & 1.13\% &\cellcolor{green}  0.993 ${\pm}$~3.45\% \\
Ni60p    & ~~-0.6\%  & 1.34\% &\cellcolor{green}  0.994 ${\pm}$~5.43\% \\
Cu63a    & ~~1.72\%  & 1.36\% &\cellcolor{green}  1.022 ${\pm}$~4.11\% \\
Si28p    & ~-0.45\%  & 1.56\% &\cellcolor{green}  0.993 ${\pm}$~4.16\% \\
Fe54a    & ~~-2.1\%  & 1.44\% &\cellcolor{green}  0.969 ${\pm}$~5.60\% \\
Fe56p    & ~~0.21\%  & 1.54\% &\cellcolor{green}  1.003 ${\pm}$~3.51\% \\
Ti48p    & ~-0.04\%  & 1.70\% &\cellcolor{green}  1.001 ${\pm}$~4.15\% \\
Co59a    & ~-0.03\%  & 1.68\% &\cellcolor{green}  1.001 ${\pm}$~4.50\% \\
U2382    & ~-5.03\%  & 1.87\% &\cellcolor{green}  0.936 ${\pm}$~5.63\% \\
Mg24p    & ~-0.08\%  & 1.83\% &\cellcolor{green}  0.999 ${\pm}$~3.08\% \\
Al27a    & ~~0.65\%  & 1.86\% &\cellcolor{green}  1.009 ${\pm}$~3.00\% \\
V51a     & ~~1.79\%  & 2.28\% &\cellcolor{green}  1.008 ${\pm}$~4.12\% \\
Tm1692   & ~~-1.5\%  & 2.74\% &\cellcolor{green}  1.036 ${\pm}$~5.55\% \\
Au1972   & ~~3.55\%  & 2.79\% &\cellcolor{green}  1.001 ${\pm}$~3.96\% \\
Nb932m   & ~-1.25\%  & 3.61\% &\cellcolor{green}  1.065 ${\pm}$~4.75\% \\
I1272    & ~~0.12\%  & 3.79\% &\cellcolor{green}  1.016 ${\pm}$~5.36\% \\
Mn552    & ~~0.55\%  & 4.72\% &\cellcolor{green}  1.057 ${\pm}$16.61\% \\
As752    & ~~3.25\%  & 4.81\% &\cellcolor{green}  1.040 ${\pm}$~6.88\% \\
Co592    & -19.81\%  & 4.95\% &\cellcolor{green}  1.069 ${\pm}$~6.05\% \\
Cu632    & ~~1.11\%  & 4.94\% &\cellcolor{green}  0.828 ${\pm}$~7.95\% \\
Y892     & ~-0.29\%  & 5.06\% &\cellcolor{green}  1.039 ${\pm}$~5.98\% \\
F192     & ~-0.31\%  & 4.60\% &\cellcolor{green}  1.023 ${\pm}$~6.12\% \\
Ni582    & ~~3.92\%  & 3.42\% &\cellcolor{green}  1.015 ${\pm}$~5.00\% \\
Na232    & ~-10.7\%  & 2.86\% &\cellcolor{green}  1.054 ${\pm}$~4.89\% \\
Al272g   & ~~4.69\%  & 2.30\% &\cellcolor{green}  0.902 ${\pm}$13.72\% \\ \hline \hline
 \end{tabular}
\end{table}

\begin{figure}[!htbp]
\vspace{-2mm}
\includegraphics[width=\columnwidth]{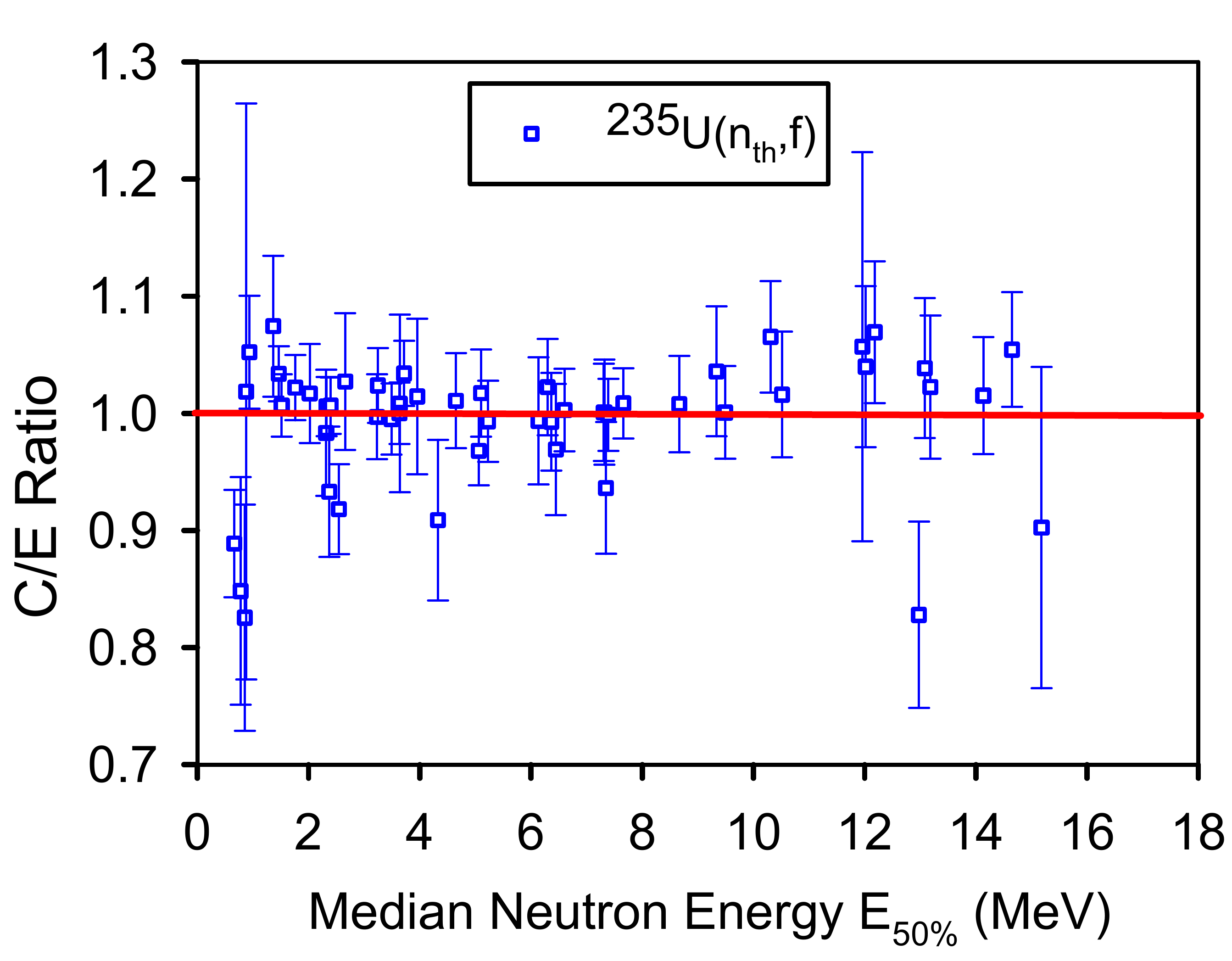}
\vspace{-5mm}
\caption{(Color online) Least-squares-based C/E as a function of median energy E$_{50\%}$ for the activity measurements in the $^{235}$U(n$_{th}$,f) reference neutron benchmark Field. Uncertainties bars reflect the combined rms uncertainty contributions from the measurement, cross sections, and spectrum.  Plotted values are listed in Table~\ref{tab:XI_B_10}.}
\label{fig:XI_B_41}
%\vspace{-2mm}
\end{figure}

The least-squares calculation used 52 measured reactions. Two reactions that were included in the SACS analysis were not included in this least-squares analysis: the B10t reaction was not addressed in the analysis because it was correlated with the B10a reaction that was used and the LSL-M2 code implementation used here did not handle strongly correlated cross sections; the Zr902 reaction was not addressed in the analysis because it was used in support of the determination of the \textit{a priori} $^{235}$U(n$_{th}$,f) spectrum and this would have introduced a strong correlation that the LSL-M2 code could not address. As expected from the consideration of the separate spectrum-averaged cross sections addressed in Sect.~\ref{Sec_VI_B1}, most of the reactions included in the least-squares adjustment proved to provide good or acceptable validation evidence, and the least-squares analysis yielded a very good $\chi^2/dof$ of 0.943.

There were no \ql poor\qr or discrepant reactions. Four reactions (Au197g with an individual $\chi^{2}$ contribution of 6.14 and a C/E of 0.889 $\pm$ 4.58\%; Th232f with a $\chi^2$ contribution of 4.84 and a C/E of 0.918 $\pm$ 3.83\%; U238g with a $\chi^2$ contribution of 3.33 and a C/E of 0.825 $\pm$ 9.66\%; and B10a with an $\chi^2$ contribution of 2.48 and a C/E of 0.848 $\pm$ 9.72\%) were labeled as \ql acceptable\qr validation evidence, and the remaining reactions met the criteria for \ql good\qr validation evidence. The U238g reaction was seen to be labeled as \ql acceptable\qr in both the SACS analysis shown in Table~\ref{TableLongSACSu235} and in this least-squares analysis. The Cu632 and Pb204n reactions were only labeled as \ql acceptable\qr in the SACS analysis but were both \ql good\qr in this least square’s analysis. The B10a, Au197g, and Th232f reactions are only labeled as \ql acceptable\qr in this least square analysis but met the criteria for \ql good\qr validation evidence in the SACS analysis. The fact that these three reactions also corresponded to the reactions with the lowest median response energy might indicate that it is the low energy portion of the \textit{a priori} $^{235}$U(n$_{th}$,f) spectrum, and not the cross sections or measurements, that has an issue.

The resulting adjustment in the spectrum-averaged activities, the reduced spectrum uncertainty contributions to the C/E, and C/E values are tabulated in Table~\ref{tab:XI_B_10}. The C/E column is again color-coded in a manner consistent with the discussion in Sec.~\ref{Sec_VI_B1}. Overall good or acceptable validation evidence, using these least-squares validation criteria, is found for all 52 \mbox{IRDFF-II} reactions in the $^{235}$U(n$_{th}$,f) reference neutron benchmark field.  It should be noted that, while the \textit{a priori} {$^{252}$Cf(s.f.)}  neutron spectrum showed a high uncertainty and a large least-squares adjustment in the high energy (15--20 MeV) region, neither this large uncertainty nor the large least-squares adjustment were seen for the \textit{a priori} $^{235}$U(n$_{th}$,f) spectrum. This observation is related to the role that the Zr902 reaction played in the determination of the \textit{a priori} $^{235}$U(n$_{th}$,f) spectrum by significantly reducing the uncertainty of the evaluated spectra due to the low uncertainty of the employed Zr902 SACS (~7\%) . Al272 was the reaction in this $^{235}$U(n$_{th}$,f) field that had the highest median response energy. While this reaction, with a median response energy of 15.18 MeV, had a significant measurement uncertainty (13.3~\%), a very small individual $\chi^2$ component, 0.58, and a good C/E of 0.902 $\pm$  13.72~\%, the least-squares analysis resulted in only a 0.3~\% adjustment in the neutron spectrum above 15~MeV.  This least-squares analysis is a very strong indication of the fidelity of the $>10$ MeV portion of the \textit{a priori} $^{235}$U(n$_{th}$,f) spectrum as defined in original PFNS evaluations \cite{PFNS,PFNS1,PFNS2} adopted for the ENDF/B-VIII.0 library \cite{ENDF8,CIELO:2018,Capote:2018}.

Fig.~\ref{fig:XI_B_41} shows the C/E values in this $^{235}$U(n$_{th}$,f) neutron benchmark field plotted versus the median response energy for the 52 reactions used in the final least-squares adjustment. The adjustment produced a 0.3~\% adjusted scale factor. The scale factor adjustment is taken out of the C/E values in this figure. There is no significant trend seen in the C/E ratios versus the median response energy. The C/E uncertainties are larger for several of the reactions in the low and high energy portion of the spectrum that reflects uncertainties of the \textit{a priori} spectrum in those energy regions.

\begin{table}[!htbp]
 \caption{Least-squares-based consistency of data for the SPR-III central cavity neutron reference benchmark field.}
 \label{tab:XI_C_11}
 \begin{tabularx} {\linewidth}{ X | cX | cX | cX } \hline\hline
  Reaction  & Act. Adj. & Spectrum Unc. &     \\
  Notation  & [\%]      &  [\%]         & C/E \\ \hline
\T
Ni58p-Cd     & -4.89 & 10.48 &\cellcolor{green}  1.000 $\mathrm{\pm}$ 10.96\% \\
Sc45g-Cd     &  0.2  &  5.93 &\cellcolor{green}  0.994 $\mathrm{\pm}$ ~8.90\% \\
Mn55g-Cd     & -3.43 &  5.64 &\cellcolor{yellow} 0.802 $\mathrm{\pm}$ ~8.31\% \\
Au197g-bare  &-15.58 &  5.58 &\cellcolor{yellow} 0.839 $\mathrm{\pm}$ ~7.17\% \\
Au197g-Cd    & -6.78 &  5.72 &\cellcolor{green}  0.929 $\mathrm{\pm}$ ~7.28\% \\
Sc45g-fiss   &  1.71 &  6.05 &\cellcolor{green}  1.051 $\mathrm{\pm}$ ~8.57\% \\
Na23g-bare   &  0.53 &  6.37 &\cellcolor{yellow} 1.129 $\mathrm{\pm}$ 11.26\% \\
Fe58g-Cd     &  0.66 &  5.86 &\cellcolor{yellow} 1.105 $\mathrm{\pm}$ 12.04\% \\
Na23g-Cdna   &  0.21 &  6.62 &\cellcolor{green}  1.075 $\mathrm{\pm}$ 11.78\% \\
Mn55g-fiss   &  1.14 &  5.69 &\cellcolor{green}  1.082 $\mathrm{\pm}$ ~8.69\% \\
W186g-bare   &-20.28 &  5.11 &\cellcolor{yellow} 0.818 $\mathrm{\pm}$ 10.41\% \\
Cu63g-Cd     &  0.04 &  5.52 &\cellcolor{green}  1.041 $\mathrm{\pm}$ 11.99\% \\
Na23g-fiss   &  0.59 &  6.24 &\cellcolor{yellow} 1.178 $\mathrm{\pm}$ 11.94\% \\
U235f-Cdtk   & -1.63 &  5.43 &\cellcolor{green}  1.045 $\mathrm{\pm}$ ~6.21\% \\
U235f-fiss   &  2.07 &  5.51 &\cellcolor{green}  1.089 $\mathrm{\pm}$ ~6.27\% \\
Pu239f-Cdtk  &  3.36 &  5.6  &\cellcolor{yellow} 1.109 $\mathrm{\pm}$ ~6.21\% \\
Pu239f-fiss  &  8.33 &  5.68 &\cellcolor{yellow} 1.165 $\mathrm{\pm}$ ~6.29\% \\
Np237f-Cdtk  & -0.01 &  6.99 &\cellcolor{yellow} 1.118 $\mathrm{\pm}$ ~7.55\% \\
Np237f-fiss  & -1.46 &  7.01 &\cellcolor{yellow} 1.103 $\mathrm{\pm}$ ~7.57\% \\
In115nm-bare & -0.84 &  9.03 &\cellcolor{green}  1.068 $\mathrm{\pm}$ 10.20\% \\
U238f-Cd     & -1.54 &  9.87 &\cellcolor{green}  1.046 $\mathrm{\pm}$ 10.37\% \\
U238f-fiss   & -2.6  &  9.85 &\cellcolor{green}  1.036 $\mathrm{\pm}$ 10.36\% \\
Ti47p-Cd     &  4.05 &  9.77 &\cellcolor{yellow} 1.102 $\mathrm{\pm}$ 10.22\% \\
S32p-bare    & -5.13 & 11.07 &\cellcolor{green}  0.990 $\mathrm{\pm}$ 11.50\% \\
Zn64p-Cd     &  2.35 & 11.19 &\cellcolor{green}  1.077 $\mathrm{\pm}$ 11.42\% \\
Fe54p-Cd     & -0.22 & 10.91 &\cellcolor{green}  1.049 $\mathrm{\pm}$ 11.40\% \\
Al27p-Cd     &  6.4  & 12.06 &\cellcolor{yellow} 1.138 $\mathrm{\pm}$ 12.45\% \\
Co59p-bare   & -7.44 & 11.83 &\cellcolor{green}  0.983 $\mathrm{\pm}$ 13.04\% \\
Ti46p-Cd     & -0.9  & 12.4  &\cellcolor{green}  1.054 $\mathrm{\pm}$ 12.96\% \\
Cu63a-bare   &-10.01 & 12.86 &\cellcolor{green}  0.966 $\mathrm{\pm}$ 13.97\% \\
Fe56p-Cd     & -0.35 & 14.11 &\cellcolor{green}  1.076 $\mathrm{\pm}$ 14.43\% \\
Ti48p-Cd     &  0.31 & 13.88 &\cellcolor{yellow} 1.105 $\mathrm{\pm}$ 14.22\% \\
Mg24p-Cd     &  3.48 & 15.41 &\cellcolor{yellow} 1.122 $\mathrm{\pm}$ 15.71\% \\
Al27a-Cd     & -1.46 & 14.78 &\cellcolor{green}  1.075 $\mathrm{\pm}$ 14.90\% \\
Nb932-bare   &  3.25 & 16.47 &\cellcolor{yellow} 1.113 $\mathrm{\pm}$ 17.14\% \\
Co592-bare   &  4.16 & 14.94 &\cellcolor{green}  1.087 $\mathrm{\pm}$ 16.60\% \\
Mn552-bare   &-16.09 & 15.06 &\cellcolor{green}  0.878 $\mathrm{\pm}$ 17.15\% \\
Mn552-fiss   & -3.18 & 15.05 &\cellcolor{green}  1.009 $\mathrm{\pm}$ 15.90\% \\
Zr902-Cd     &  5.45 & 15.17 &\cellcolor{yellow} 1.142 $\mathrm{\pm}$ 15.55\% \\
Ni582-bare   & -8.87 & 15.62 &\cellcolor{green}  1.014 $\mathrm{\pm}$ 16.69\% \\ \hline\hline
 \end{tabularx}
\end{table}

\subsection{SPR-III Fast Burst Reactor Central Cavity}\label{Sec_VIII_C}

\begin{figure}[!htbp]
\vspace{-2mm}
\includegraphics[width=\columnwidth]{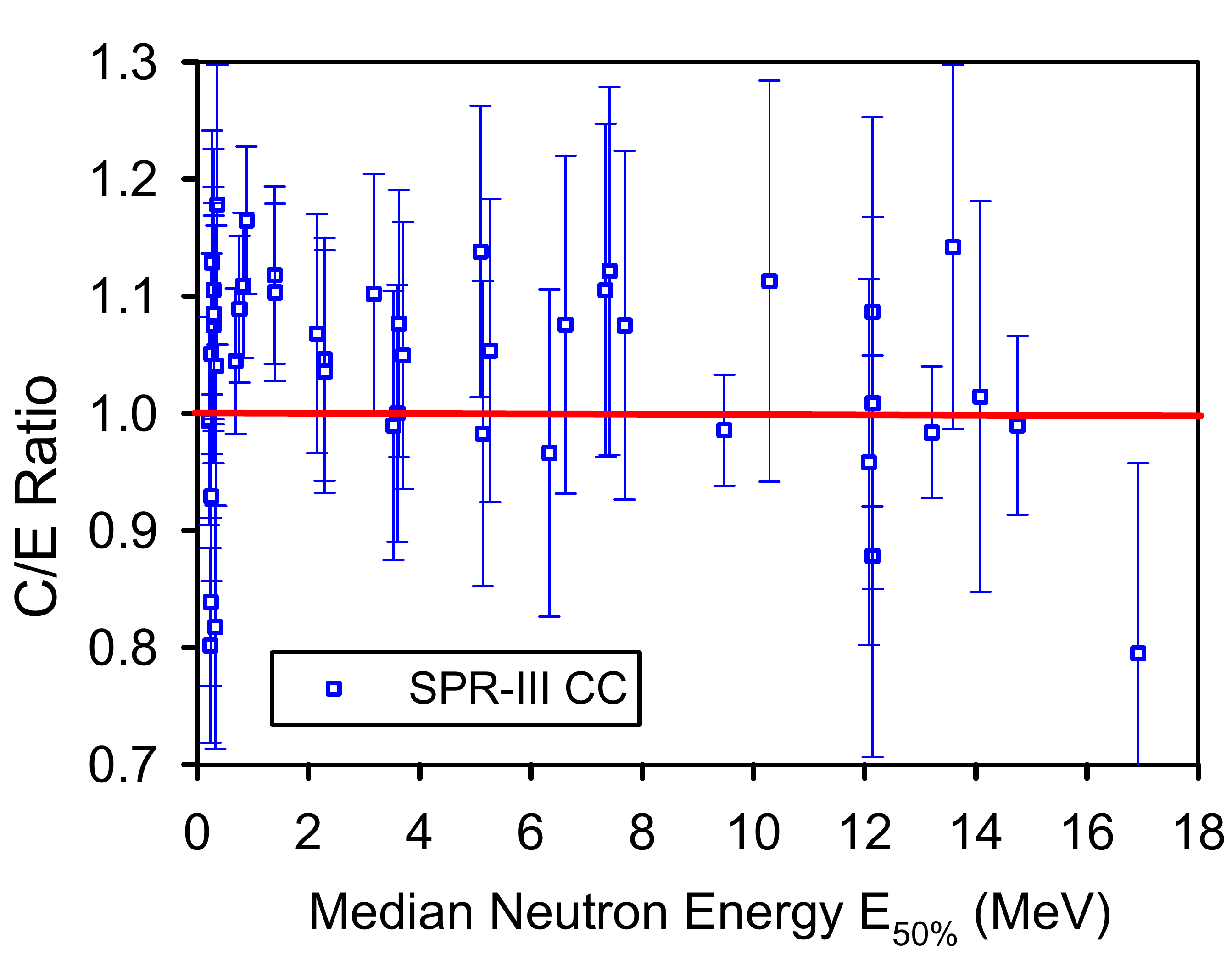}
\vspace{-5mm}
\caption{(Color online) Least-squares-based C/E as a function of median energy E$_{50\%}$ for the activity measurements in the SPR-III fast-burst reactor central cavity reference neutron benchmark field. Uncertainties bars reflect the combined rms  uncertainty contributions from the measurement, cross sections, and spectrum (remove bias). Plotted values are listed in Table~\ref{tab:XI_C_11}.}
\label{fig:XI_C_49}
%\vspace{-2mm}
\end{figure}

The least-squares analysis used 40 reaction/cover combinations and had a $\chi$$^{2}$/\textit{dof} of 2.409. Although this chi-squares is high, none of the resulting dosimeters are found to be discrepant when we use the acceptance criteria adopted in Table~\ref{tab:X_A_3}. Table~\ref{tab:XI_C_11} shows the resulting C/E along with the consistent least-squares spectrum uncertainty and adjustment in the activity measurement.
Fig.~\ref{fig:XI_C_49} shows the C/E values in the SPR-III reference neutron benchmark field plotted versus the median response energy for the 40 reactions used in the final least-squares
adjustment.

\subsection{ACRR Pool-type Reactor Central Cavity}\label{Sec_VIII_D}

The least-squares analysis for this neutron field used 33 reaction/cover combinations and had a $\chi$$^{2}$/\textit{dof} of 0.8984. This reflects an excellent consistency of the data (and associated input uncertainties). Table~\ref{tab:XI_D_12} shows the resulting C/E along with the consistent least-squares spectrum uncertainty and adjustment in the activity measurement. Two of the 33 reactions, Ag109g-bare and Co59g-Cd, were labeled as discrepant based on the criteria in Table~\ref{tab:X_A_3} -- due to the large C/E value even if this value was within two standard deviations of unity.
\begin{figure}[!htbp]
\vspace{-2mm}
\includegraphics[width=\columnwidth]{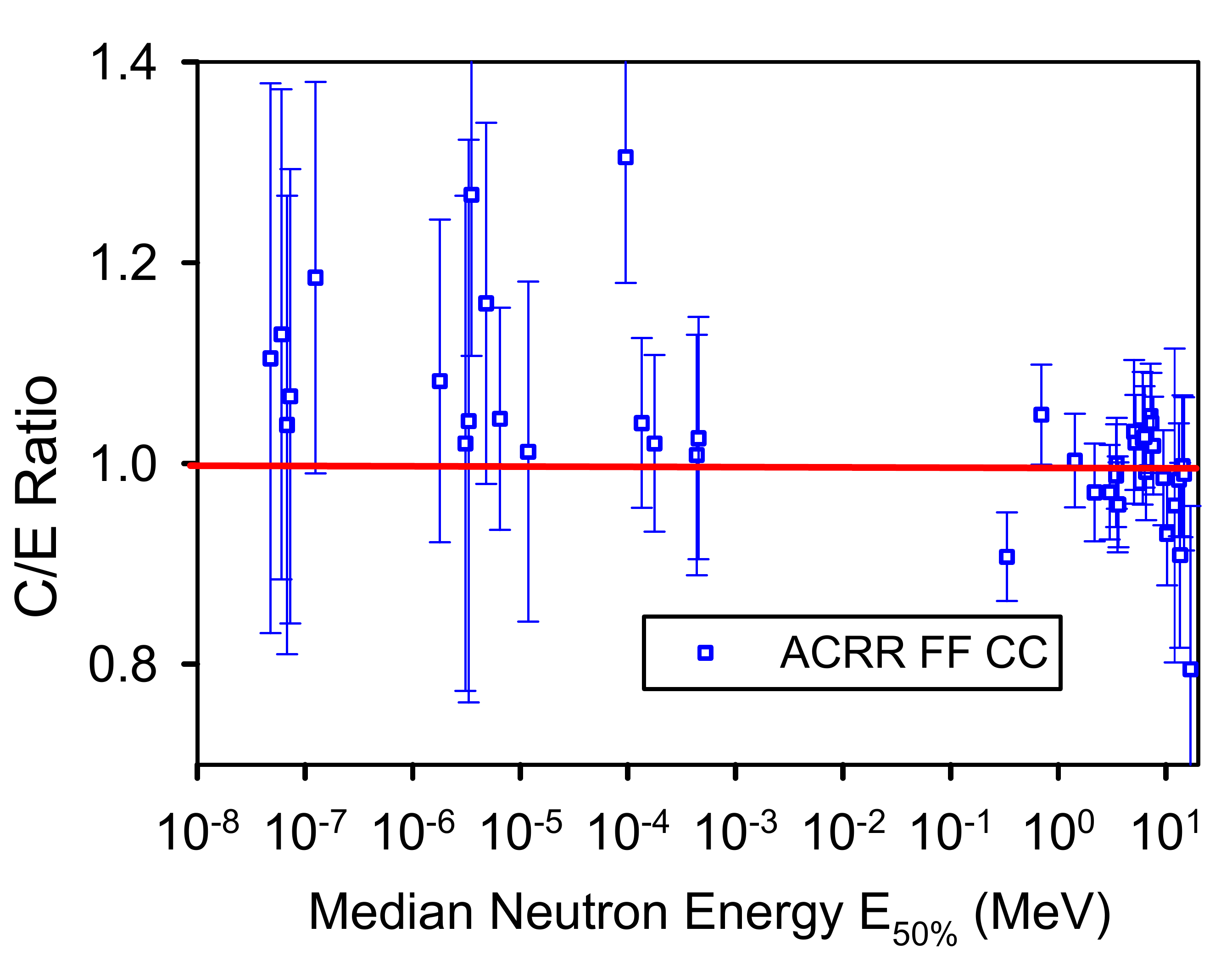}
\vspace{-5mm}
\caption{(Color online) Least-squares-based C/E as a function of median energy E$_{50\%}$ for the activity measurements in the ACRR pool-type reactor central cavity reference neutron benchmark field. Uncertainties bars reflect the combined rms uncertainty contributions from the measurement, cross sections, and spectrum. Plotted values are listed in Table~\ref{tab:XI_D_12}.}
\label{fig:XI_D_43}
%\vspace{-2mm}
\end{figure}
\begin{table}[hbtp]
\vspace{-3mm}
 \caption{Least-squares-based consistency of data for the ACRR central cavity neutron reference benchmark field.}
 \label{tab:XI_D_12}
 \begin{tabular}{p{0.9in}|p{0.6in}|p{0.5in}|p{0.9in}} \hline\hline
  Reaction  & Act. Adj. & Spectrum Unc. &     \\
  Notation  & [\%]      &  [\%]         & C/E \\ \hline
\T
Ni58p-bare  & 1.67  & 2.75  &\cellcolor{green}  1.000 $\pm$ ~4.53\% \\
Sc45g-bare  & 0.16  & 26.17 &\cellcolor{yellow} 1.105 $\pm$ 27.39\% \\
Na23g-bare  & 3.27  & 24.15 &\cellcolor{yellow} 1.129 $\pm$ 24.40\% \\
Mn55g-bare  & -3.83 & 22.67 &\cellcolor{green}  1.038 $\pm$ 22.85\% \\
Fe58g-bare  & -0.17 & 22.09 &\cellcolor{green}  1.067 $\pm$ 22.64\% \\
Co59g-bare  & 1.87  & 18.88 &\cellcolor{yellow} 1.185 $\pm$ 19.50\% \\
Sc45g-Cd    & 0.07  & 14.32 &\cellcolor{green}  1.082 $\pm$ 16.09\% \\
Au197g-bare & -1.92 & 24.32 &\cellcolor{green}  1.020 $\pm$ 24.66\% \\
Au197g-Cd   & 0.94  & 27.71 &\cellcolor{yellow} 1.042 $\pm$ 28.04\% \\
Ag109g-bare & 4.66  & 14.74 &\cellcolor{lred}    1.268 $\pm$ 16.04\% \\
Ag109g-Cd   & -0.8  & 16.34 &\cellcolor{yellow} 1.160 $\pm$ 17.98\% \\
Na23g-Cd    & -1.05 & 10.17 &\cellcolor{green}  1.045 $\pm$ 11.08\% \\
W186g-bare  & -0.12 & 16.5  &\cellcolor{green}  1.012 $\pm$ 16.93\% \\
Co59g-Cd    & 4.26  & 11.53 &\cellcolor{lred}    1.305 $\pm$ 12.53\% \\
Mn55g-Cd    & 0.66  & 7.73  &\cellcolor{green}  1.040 $\pm$ ~8.45\% \\
Fe58g-Cd    & -0.18 & 7.31  &\cellcolor{green}  1.020 $\pm$ ~8.80\% \\
rmleu-fiss  & -7.44 & 1.8   &\cellcolor{yellow} 0.907 $\pm$ ~4.42\% \\
rmlpu-fiss  & 5.04  & 1.87  &\cellcolor{green}  1.049 $\pm$ ~4.99\% \\
Np237f-fiss & 1.86  & 2.29  &\cellcolor{green}  1.003 $\pm$ ~4.67\% \\
rmldu-fiss  & -0.89 & 2.71  &\cellcolor{green}  0.971 $\pm$ ~4.89\% \\
Ti47p-bare  & -0.87 & 2.59  &\cellcolor{green}  0.971 $\pm$ ~4.71\% \\
S32p-bare   & 0.73  & 2.93  &\cellcolor{green}  0.988 $\pm$ ~5.12\% \\
Zn64p-bare  & -2.15 & 2.95  &\cellcolor{green}  0.959 $\pm$ ~4.77\% \\
Fe54p-bare  & -2.18 & 2.87  &\cellcolor{green}  0.959 $\pm$ ~4.23\% \\
Co59p-Cd &  2.95    & 3.12  &\cellcolor{green}  1.032 $\pm$ ~7.16\% \\
Ti46p-bare  & 1.43  & 3.23  &\cellcolor{green}  1.021 $\pm$ ~4.74\% \\
Ni60p-bare  & 1.39  & 3.28  &\cellcolor{green}  1.025 $\pm$ ~6.62\% \\
Fe56p-bare  & -1.65 & 3.49  &\cellcolor{green}  0.992 $\pm$ ~4.81\% \\
Ti48p-bare  & 1.35  & 3.5   &\cellcolor{green}  1.048 $\pm$ ~5.20\% \\
Nb932-bare  & -2.17 & 4.23  &\cellcolor{green}  0.930 $\pm$ ~5.10\% \\
Mg24p-bare  & 1.93  & 3.91  &\cellcolor{green}  1.041 $\pm$ ~4.98\% \\
Al27a-bare  & 0.19  & 3.8   &\cellcolor{green}  1.018 $\pm$ ~4.89\% \\
Zr902-bare  & -1.89 & 7.43  &\cellcolor{green}  0.909 $\pm$ ~9.23\% \\ \hline\hline
\end{tabular}
\vspace{-2mm}
\end{table}

The table reflects a large spectrum contribution to the uncertainty, and there is also seen to be a correlation in the C/E between the two configurations, bare and cover, -- both configurations showing a positive C/E value. It is very possible that the discrepancy reflects an issue with the \textit{a priori} neutron spectrum, or the associated uncertainty, rather than with the dosimetry cross section. It should also be noted that both of these basic reactions appear in the set of sensors in both, a bare configuration, and with a cadmium cover; only one of the sensor configurations was labelled as discrepant. It is important that we also note a limitation in our current least-squares treatment. While a least-squares analysis should capture any correlation between input quantities, the LSL code used here did not include any correlation between the dosimetry responses. While this is not an issue for most dosimetry reactions, it could be an issue when for the response functions that address two different configurations for the same basic reaction. This is an area for later tool development.
Fig.~\ref{fig:XI_D_43} shows the C/E values in the ACRR central cavity reference neutron benchmark field plotted versus the median response energy for the 33 reactions used in the final least-squares adjustment.

\begin{table*}[hbtp]
\vspace{-3mm}
%\tiny
\caption{Least-squares-based consistency of data for the ACRR LB44 neutron reference benchmark field.}
\label{ACRR_LB44_LSQ}
%\begin{tabular}{p{0.9in}|p{0.9in}|p{0.6in}|p{0.5in}|p{0.9in}} \hline\hline
 \begin{tabular}{l r c c c c} \hline\hline
  Reaction       &\multicolumn{2}{c}{Measured Activity}  & Act. Adj. & Spectrum  &  \\
  Notation       &[Bq/at or & Unc.                       & ~~Unc.    & ~~Unc.    &   \\
                 &~Fiss/at.] & ~~[\%]                    & ~~[\%]    & ~~[\%]    & C/E $\pm$ Unc. [\%]\\ \hline
\T
 Co59g-bare      &6.2470E-18 & ~0.8                      &  0.84     & 14.15     &\cellcolor{lred}   ~ 1.485$\pm$14.33 \\[1.4mm]
 Au197g-bare     &2.4293E-14 & ~2.8                      &  -0.87    & 9.67      &\cellcolor{green} ~ 1.039$\pm$10.59 \\[1.4mm]
 Mn55g-bare      &4.0558E-14 & ~2.0                      &  2.08     & 4.81      &\cellcolor{yellow}~ 1.195$\pm$~6.37 \\[1.4mm]
 Fe58g-bare      &1.2442E-17 & ~2.3                      &  -0.81    & 4.54      &\cellcolor{green} ~ 1.095$\pm$~8.08 \\[1.4mm]
 Mo98g-bare      &1.6478E-15 & ~1.8                      &  0.18     & 4.02      &\cellcolor{yellow}~ 1.110$\pm$~9.93 \\[1.4mm]
 Cu63g-bare      &6.6548E-15 & ~5.7                      &  -11.28   & 4.34      &\cellcolor{yellow}~ 0.859$\pm$~8.91 \\[1.4mm]
 Sc45g-bare      &1.6744E-17 & ~0.7                      &  0.11     & 2.8       &\cellcolor{yellow}~ 1.171$\pm$15.18 \\[1.4mm]
 Co59g-Cdtk/B4C  &2.0990E-19 & ~0.8                      &  0.27     & 3.65      &\cellcolor{yellow}~ 1.167$\pm$~4.38 \\[1.4mm]
 Au197g- Cdtk/B4C&2.7153E-15 & ~4.2                      &  -1.53    & 2.15      &\cellcolor{green} ~ 1.056$\pm$~6.37 \\[1.4mm]
 Fe58g-Cdtk/B4C  &2.7876E-18 & ~7.2                      &  3.06     & 2.9       &\cellcolor{yellow}~ 1.180$\pm$13.44 \\[1.4mm]
 Mo98g-Cdtk/B4C  &3.9782E-16 & ~2.2                      &  3.27     & 2.82      &\cellcolor{lred}   ~ 1.286$\pm$~8.07 \\[1.4mm]
 rmleu-Cdtk/B4C  &3.8036E-09 & ~2.1                      &  -1.51    & 4.39      &\cellcolor{green} ~ 1.046$\pm$~5.27 \\[1.4mm]
 rmlpu-Cdtk/B4C  &4.5468E-09 & ~2.0                      &  -2.73    & 5.69      &\cellcolor{green} ~ 1.007$\pm$~6.76 \\[1.4mm]
 Np237f-Cdtk/B4C &1.9753E-09 & ~3.0                      &  0.7      & 10.2      &\cellcolor{green} ~ 1.032$\pm$10.83 \\[1.4mm]
 In115n-bare     &9.8199E-15 & ~5.6                      &  -7.47    & 13.09     &\cellcolor{green} ~ 0.956$\pm$14.39 \\[1.4mm]
 rmldu-Cdtk/B4C  &2.8865E-10 & ~2.0                      &  2.44     & 14.44     &\cellcolor{green} ~ 1.072$\pm$14.72 \\[1.4mm]
 Ti47p-bare      &3.6124E-17 & ~2.6                      &  0.99     & 13.07     &\cellcolor{green} ~ 1.046$\pm$13.55 \\[1.4mm]
 S32p-bare       &2.6391E-17 & ~2.9                      &  -0.76    & 14.54     &\cellcolor{green} ~ 1.016$\pm$14.98 \\[1.4mm]
 Ni58p-bare      &9.5960E-18 & ~1.6                      &  -2.78    & 13.63     &\cellcolor{green} ~ 1.000$\pm$13.89 \\[1.4mm]
 Fe54p-bare      &1.4959E-18 & ~1.1                      &  1.64     & 14.18     &\cellcolor{green} ~ 1.048$\pm$14.39 \\[1.4mm]
 Co59p-bare      &1.6482E-19 & ~2.4                      &  -1.49    & 16.66     &\cellcolor{green} ~ 1.019$\pm$17.08 \\[1.4mm]
 Ti46p-bare      &6.7205E-19 & ~0.8                      &  0.93     & 17.24     &\cellcolor{green} ~ 1.057$\pm$17.45 \\[1.4mm]
 Ni60p-bare      &5.4018E-21 & ~6.0                      &  -0.2     & 17.11     &\cellcolor{green} ~ 1.053$\pm$18.27 \\[1.4mm]
 Fe56p-bare      &4.7933E-17 & ~0.9                      &  -0.22    & 17.89     &\cellcolor{green} ~ 1.056$\pm$18.13 \\[1.4mm]
 Ti48p-bare      &7.7602E-19 & ~0.6                      &  0.01     & 18.16     &\cellcolor{green} ~ 1.061$\pm$18.51 \\[1.4mm]
 Al27a-bare      &5.2690E-18 & ~3.7                      &  0.14     & 20.11     &\cellcolor{green} ~ 1.059$\pm$20.54 \\[1.4mm]
 Nb932-bare      &2.0258E-19 & ~1.0                      &  0        & 16.1      &\cellcolor{green} ~ 1.097$\pm$16.26 \\[1.4mm]
 Co592-bare      &1.2528E-20 & 12.0                      & 8.36      & 20.27     &\cellcolor{yellow}~ 1.206$\pm$23.68 \\[1.4mm]
 Mn552-bare      &3.5010E-21 & 16.0                      & -12.24    & 20.46     &\cellcolor{yellow}~ 0.999$\pm$26.07 \\[1.4mm]
 \hline\hline
 \end{tabular}
\vspace{-4mm}
\end{table*}

\begin{table*}[hbtp]
\vspace{-4mm}
%\tiny
 \caption{Least-squares-based consistency of data for the ACRR PLG neutron reference benchmark field.}
 \label{ACRR_PLG_LSQ}
 \begin{tabular}{l r c c c c} \hline\hline
  Reaction       &\multicolumn{2}{c}{Measured Activity}  & Act. Adj. & Spectrum  &     \\
  Notation       &[Bq/at. or & Unc.                      & ~~Unc.    & ~~Unc.    &     \\
                 &~Fiss/at.] & ~~[\%]                    & ~~[\%]    & ~~[\%]    & C/E $\pm$ Unc. [\%]\\ \hline
\T
 Sc45g-bare      &1.6579E-15 & 2.38                      &  0.06     & 30.63     &\cellcolor{yellow}~ 1.216$\pm$31.62 \\[1.3mm]
 Na23g-bare      &3.9324E-15 & 2.38                      &  1.7      & 29.74     &\cellcolor{yellow}~ 1.245$\pm$29.90 \\[1.3mm]
 Fe58g-bare      &1.6831E-16 & 2.90                      &   -0.1    & 28.23     &\cellcolor{yellow}~ 1.205$\pm$28.68 \\[1.3mm]
 Cu63g-bare      &5.0078E-14 & 3.69                      &  -2.61    & 28.17     &\cellcolor{yellow}~ 1.164$\pm$28.52 \\[1.3mm]
 Co59g-bare      &1.1350E-16 & 2.50                      &   -0.56   & 26.6      &\cellcolor{yellow}~ 1.205$\pm$26.72 \\[1.3mm]
 Sc45g-Cd        &1.2948E-16 & 2.38                      &  -0.65    & 14.74     &\cellcolor{green} ~ 1.054$\pm$16.43 \\[1.3mm]
 Au197g-bare     &7.1560E-13 & 2.69                      &  -0.35    & 22.13     &\cellcolor{green} ~ 1.078$\pm$22.33 \\[1.3mm]
 Au197g-Cd       &5.3238E-13 & 2.69                      &  0.21     & 28.07     &\cellcolor{green} ~ 1.026$\pm$28.26 \\[1.3mm]
 Na23g-Cd        &4.1746E-16 & 2.38                      &  0.71     & 10.97     &\cellcolor{yellow}~ 1.114$\pm$11.60 \\[1.3mm]
 Co59g-Cd        &2.2206E-17 & 2.50                      &   0.86    & 11.24     &\cellcolor{yellow}~ 1.169$\pm$11.54 \\[1.3mm]
 Fe58g-Cd        &2.6925E-17 & 2.90                      &   0.89    & 7.34      &\cellcolor{green} ~ 1.072$\pm$~8.90 \\[1.3mm]
 Mn55g-Cd        &1.1075E-13 & 2.62                      &  -1.39    & 7.57      &\cellcolor{green} ~ 1.022$\pm$~8.32 \\[1.3mm]
 Cu63g-Cd        &7.5389E-15 & 3.69                      &  2.37     & 7.1       &\cellcolor{green} ~ 1.095$\pm$~8.48 \\[1.3mm]
 rmleu-Cdtk/B4C  &2.5730E-09 & 2.83                      &  -6.39    & 1.77      &\cellcolor{green} ~ 0.927$\pm$~3.35 \\[1.3mm]
 rmlpu-Cdtk/B4C  &2.5696E-09 & 2.83                      &  4.06     & 1.84      &\cellcolor{green} ~ 1.059$\pm$~4.08 \\[1.3mm]
 In115n-bare     &7.0790E-15 & 5.95                      &  -7.32    & 2.57      &\cellcolor{green} ~ 0.921$\pm$~6.51 \\[1.3mm]
 rmldu-Cdtk/B4C  &2.1936E-10 & 2.90                      &   1.78    & 2.72      &\cellcolor{green} ~ 1.014$\pm$~4.01 \\[1.3mm]
 Ti47p-bare      &2.1537E-17 & 4.12                      &  -2.23    & 2.58      &\cellcolor{green} ~ 0.974$\pm$~4.71 \\[1.3mm]
 S32p-bare       &7.7936E-18 & 2.62                      &  1.49     & 2.92      &\cellcolor{green} ~ 1.017$\pm$~5.11 \\[1.3mm]
 Ni58p-bare      &3.7146E-16 & 3.61                      &  -0.1     & 2.74      &\cellcolor{green} ~ 1.000$\pm$~3.87 \\[1.3mm]
 Zn64p-bare      &1.2772E-18 & 3.36                      &  -1.68    & 2.94      &\cellcolor{green} ~ 0.985$\pm$~4.70 \\[1.3mm]
 Fe54p-bare      &1.4707E-19 & 3.36                      &  1.13     & 2.87      &\cellcolor{green} ~ 1.014$\pm$~4.51 \\[1.3mm]
 Co59p-Cd        &6.1277E-19 & 2.56                      &  -0.04    & 3.12      &\cellcolor{green} ~ 1.013$\pm$~4.98 \\[1.3mm]
 Ti46p-bare      &5.2239E-21 & 3.69                      &  2.57     & 3.23      &\cellcolor{green} ~ 1.051$\pm$~4.44 \\[1.3mm]
 Ni60p-Cd        &1.4983E-21 & 8.15                      &  -3.24    & 3.28      &\cellcolor{green} ~ 0.989$\pm$~5.04 \\[1.3mm]
 Cu63a-bare      &4.6190E-17 & 2.69                      &  -17.86   & 3.22      &\cellcolor{yellow}~ 0.839$\pm$~8.96 \\[1.3mm]
 Fe56p-bare      &7.1657E-19 & 2.24                      &  -1.64    & 3.49      &\cellcolor{green} ~ 1.009$\pm$~4.81 \\[1.3mm]
 Ti48p-bare      &9.8561E-18 & 2.69                      &  0.23     & 3.51      &\cellcolor{green} ~ 1.046$\pm$~5.11 \\[1.3mm]
 Mg24p-bare      &4.8547E-18 & 2.97                      &  4.87     & 3.92      &\cellcolor{green} ~ 1.098$\pm$~4.77 \\[1.3mm]
 Al27a-bare      &1.9054E-19 & 2.33                      &  -0.28    & 3.8       &\cellcolor{green} ~ 1.044$\pm$~4.84 \\[1.3mm]
 Nb932-bare      &3.5325E-21 & 20.3                      &  0.26     & 4.23      &\cellcolor{green} ~ 1.005$\pm$~4.85 \\[1.3mm]
 Mn552-bare      &1.2657E-20 & 18.2                      & -18.09    & 5.4       &\cellcolor{lred}   ~ 0.764$\pm$21.04 \\[1.3mm]
 Co592-Cd        &1.4115E-19 & 3.61                      &  -0.84    & 5.38      &\cellcolor{green} ~ 0.937$\pm$19.04 \\[1.3mm]
 Zr902-bare      &1.1537E-20 & 6.51                      &  -1.98    & 7.49      &\cellcolor{yellow}~ 0.882$\pm$ 8.32 \\[1.3mm]
 Ni582-Cd        &1.6579E-15 & 2.38                      &  -0.36    & 8.73      &\cellcolor{yellow}~ 0.885$\pm$10.91 \\[1.3mm]
\hline \hline
\end{tabular}
\vspace{-4mm}
\end{table*}

\begin{table*}[hbtp]
\vspace{-3mm}
%\tiny
\caption{Least-squares-based consistency of data for the ACRR CdPoly neutron reference benchmark field.}
\label{ACRR_CdPoly_LSQ}
 \begin{tabular}{l r c c c c} \hline\hline
  Reaction       &\multicolumn{2}{c}{Measured Activity}  & Act. Adj. & Spectrum  &     \\
  Notation       &[Bq/at. or & Unc.                      & ~~Unc.    & ~~Unc.    &     \\
                 &~Fiss/at.] & ~~[\%]                    & ~~[\%]    & ~~[\%]    & C/E $\pm$ Unc. [\%]\\ \hline
\T
 Sc45g-bare      &9.8273E-17 &~1.30                      &  -0.63    & 14.76     &\cellcolor{green} 0.940$\pm$16.44 \\[1.4mm]
 Au197g-bare     &3.7488E-13 &~1.70                      &  -0.05    & 28.43     &\cellcolor{green} 1.060$\pm$28.62 \\[1.4mm]
 Na23g-bare      &3.0951E-16 &~1.80                      &  1.31     & 11.12     &\cellcolor{green} 1.039$\pm$11.78 \\[1.4mm]
 Ag109g-bare     &4.8737E-17 &~0.60                      &  0.13     & 17.07     &\cellcolor{green} 1.088$\pm$18.33 \\[1.4mm]
 W186g-bare      &1.0211E-13 &~2.10                      &  -0.02    & 19.55     &\cellcolor{green} 1.014$\pm$20.09 \\[1.4mm]
 Co59g-bare      &1.3482E-17 &~1.40                      &  1.4      & 11.22     &\cellcolor{green} 1.274$\pm$11.51 \\[1.4mm]
 Fe58g-bare      &2.0046E-17 &~1.90                      &  -0.91    & 7.39      &\cellcolor{green} 0.965$\pm$~8.88 \\[1.4mm]
 Mn55g-bare      &7.9671E-14 &~1.60                      &  -1.38    & 7.58      &\cellcolor{green} 0.968$\pm$~8.32 \\[1.4mm]
 Mo98g-bare      &1.4967E-15 &~1.00                      &  -0.01    & 7.94      &\cellcolor{green} 1.017$\pm$12.50 \\[1.4mm]
 Cu63g-bare      &5.3525E-15 &~2.80                      &  0.56     & 7.22      &\cellcolor{green} 1.033$\pm$~8.45 \\[1.4mm]
 Nb93g-Cd        &6.0332E-22 &17.50                      & 11.49     & 3.17      &\cellcolor{yellow}1.169$\pm$17.96 \\[1.4mm]
 rmleu-Cdtk/B4C  &1.5452E-09 &~2.20                      &  0.8      & 1.74      &\cellcolor{green} 1.047$\pm$~3.46 \\[1.4mm]
 Np237f-Cdtk/B4C &7.9845E-10 &~3.60                      &  10.64    & 2.28      &\cellcolor{lred}  1.148$\pm$~4.74 \\[1.4mm]
 In115n-bare     &5.1109E-15 &~5.60                      &  -7.17    & 2.56      &\cellcolor{green} 0.954$\pm$~6.51 \\[1.4mm]
 Ti47p-bare      &2.3439E-17 &~3.10                      &  -2.8     & 2.58      &\cellcolor{green} 0.976$\pm$~4.70 \\[1.4mm]
 S32p-bare       &1.7162E-17 &~3.60                      &  0.59     & 2.93      &\cellcolor{green} 1.004$\pm$~5.12 \\[1.4mm]
 Ni58p-bare      &6.1122E-18 &~3.00                      &  0.05     & 2.76      &\cellcolor{green} 1.000$\pm$~4.60 \\[1.4mm]
 Zn64p-bare      &2.9304E-16 &~2.80                      &  -1.02    & 2.95      &\cellcolor{green} 0.986$\pm$~4.58 \\[1.4mm]
 Fe54p-bare      &1.0366E-18 &~1.70                      &  -1.01    & 2.88      &\cellcolor{green} 0.985$\pm$~4.00 \\[1.4mm]
 Al27p-bare      &2.3141E-15 &~2.20                      &  -4.11    & 3.17      &\cellcolor{green} 0.951$\pm$~4.40 \\[1.4mm]
 Co59p-bare      &1.2296E-19 &~3.10                      &  -0.19    & 3.1       &\cellcolor{green} 0.993$\pm$~5.17 \\[1.4mm]
 Ti46p-bare      &4.9967E-19 &~1.50                      &  3.28     & 3.21      &\cellcolor{green} 1.043$\pm$~4.38 \\[1.4mm]
 Cu63a-Cd        &1.1497E-21 &~4.40                      &  -10.22   & 3.19      &\cellcolor{green} 0.894$\pm$~6.08 \\[1.4mm]
 Fe56p-bare      &3.8669E-17 &~1.60                      &  -1.57    & 3.46      &\cellcolor{green} 0.995$\pm$~4.71 \\[1.4mm]
 Ti48p-bare      &6.0515E-19 &~1.00                      &  0.45     & 3.47      &\cellcolor{green} 1.040$\pm$~5.02 \\[1.4mm]
 Mg24p-bare      &8.6405E-18 &~1.90                      &  1.55     & 3.88      &\cellcolor{green} 1.053$\pm$~4.78 \\[1.4mm]
 Al27a-bare      &4.0228E-18 &~1.80                      &  2.65     & 3.77      &\cellcolor{green} 1.062$\pm$~4.65 \\[1.4mm]
 Nb932-bare      &1.7793E-19 &~1.90                      &  -1.88    & 4.19      &\cellcolor{green} 0.986$\pm$~5.04 \\[1.4mm]
 Mn552-bare      &2.8166E-21 &~4.50                      &  -2.34    & 5.46      &\cellcolor{green} 0.990$\pm$~7.43 \\[1.4mm]
 Co592-bare      &1.1389E-20 &18.30                      &  3.94     & 5.44      &\cellcolor{green} 1.054$\pm$19.24 \\[1.4mm]
 Zr902-bare      &1.2630E-19 &~2.20                      &  0.97     & 7.59      &\cellcolor{green} 1.050$\pm$~8.15 \\[1.4mm]
\hline\hline
\end{tabular}
\vspace{-3mm}
\end{table*}

\begin{table*}[hbtp]
\vspace{-4mm}
%\tiny
\caption{Least-squares-based consistency of data for the ACRR FREC-II neutron reference benchmark field.}
\label{ACRR_FREC_LSQ}
\begin{tabular}{l r c c c c} \hline\hline
  Reaction       &\multicolumn{2}{c}{Measured Activity}  & Act. Adj. & Spectrum  &                  \\
  Notation       &[Bq/at. or & ~~Unc.                    & ~~Unc.    & ~~Unc.    &                  \\
                 &~Fiss/at.] & ~~[\%]                    & ~~[\%]    & ~~[\%]    & C/E $\pm$ Unc. [\%]\\ \hline
\T
Sc45g-bare       &4.6441E-16 & 2.20                      &  2.67     & 3.79      &\cellcolor{yellow}~ 1.148$\pm$~4.66 \\[1.4mm]
Mn55g-bare       &1.9858E-13 & 1.70                      &  0.98     & 28.06     &\cellcolor{yellow}~ 0.894$\pm$28.37 \\[1.4mm]
Fe58g-bare       &4.7655E-17 & 2.00                      &  -1.83    & 21.94     &\cellcolor{yellow}~ 0.854$\pm$22.29 \\[1.4mm]
Cu63g-bare       &1.3579E-14 & 2.90                      &  -0.64    & 28.15     &\cellcolor{yellow}~ 0.886$\pm$28.26 \\[1.4mm]
Co59g-bare       &3.1311E-17 & 2.20                      &  0.72     & 11.26     &\cellcolor{yellow}~ 1.119$\pm$11.54 \\[1.4mm]
Sc45g-Cd         &2.4468E-17 & 2.20                      &  -4.65    & 3.07      &\cellcolor{green} ~ 0.975$\pm$~5.32 \\[1.4mm]
Au197g-bare      &1.6471E-13 & 3.10                      &  0.29     & 29.56     &\cellcolor{yellow}~ 0.872$\pm$29.87 \\[1.4mm]
Au197g-Cd        &1.0643E-13 & 1.80                      &  0.18     & 6.93      &\cellcolor{green} ~ 1.002$\pm$~8.25 \\[1.4mm]
Na23g-Cd         &7.8216E-17 & 1.80                      &  -1.2     & 2.88      &\cellcolor{green} ~ 0.980$\pm$~4.13 \\[1.4mm]
Co59g-Cd         &4.1486E-18 & 1.30                      &  -3.48    & 3.46      &\cellcolor{green} ~ 1.041$\pm$~5.08 \\[1.4mm]
Fe58g-Cd         &5.4352E-18 & 1.90                      &  0.6      & 29.64     &\cellcolor{yellow}~ 0.869$\pm$30.07 \\[1.4mm]
Mo98g-bare       &5.1417E-16 & 1.00                      &  -1.22    & 7.22      &\cellcolor{green} ~ 0.955$\pm$~8.76 \\[1.4mm]
Mo98g-Cd         &4.6865E-16 & 1.00                      &  -0.62    & 2.52      &\cellcolor{green} ~ 0.960$\pm$~6.49 \\[1.4mm]
Cu63g-Cd         &1.4968E-15 & 2.80                      &  7.43     & 3.9       &\cellcolor{yellow}~ 1.196$\pm$~5.01 \\[1.4mm]
rmle-Cdtk/B4C    &5.7944E-10 & 2.10                      &  -0.64    & 29.81     &\cellcolor{yellow}~ 0.858$\pm$29.93 \\[1.4mm]
rmlpu-Cdtk/B4C   &5.8279E-10 & 2.10                      &  -4.68    & 5.4       &\cellcolor{green} ~ 0.948$\pm$~9.83 \\[1.4mm]
Np237f- Cdtk/B4C &3.1382E-10 & 2.10                      &  2.3      & 7.66      &\cellcolor{green} ~ 0.920$\pm$12.31 \\[1.4mm]
In115n-bare      &1.6796E-15 & 5.60                      &  -2.46    & 7.31      &\cellcolor{yellow}~ 0.883$\pm$12.02 \\[1.4mm]
rmldu-Cdtk/B4C   &5.8963E-11 & 2.20                      &  1.62     & 10.81     &\cellcolor{green} ~ 1.063$\pm$11.54 \\[1.4mm]
Ti47p-bare       &8.0144E-18 & 3.20                      &  2.4      & 4.22      &\cellcolor{green} ~ 1.098$\pm$~5.02 \\[1.4mm]
S32p-bare        &6.3767E-18 & 3.60                      &  0.87     & 2.76      &\cellcolor{green} ~ 1.000$\pm$~4.47 \\[1.4mm]
Ni58p-bare       &2.0926E-18 & 2.80                      &  0.67     & 2.25      &\cellcolor{green} ~ 0.974$\pm$~3.71 \\[1.4mm]
Zn64p-bare       &9.8155E-17 & 4.80                      &  -3.16    & 2.65      &\cellcolor{green} ~ 0.936$\pm$~4.02 \\[1.4mm]
Fe54p-bare       &3.5797E-19 & 2.00                      &  -9.91    & 1.74      &\cellcolor{yellow}~ 0.864$\pm$~3.39 \\[1.4mm]
Co59p-bare       &4.3896E-20 & 3.40                      &  3.95     & 1.83      &\cellcolor{green} ~ 1.026$\pm$~4.09 \\[1.4mm]
Ti46p-bare       &1.7181E-19 & 1.70                      &  5.27     & 2.92      &\cellcolor{green} ~ 1.044$\pm$~5.11 \\[1.4mm]
Fe56p-bare       &1.3315E-17 & 2.50                      &  0.56     & 31.89     &\cellcolor{yellow}~ 0.899$\pm$32.97 \\[1.4mm]
Ti48p-bare       &2.0869E-19 & 1.00                      &  -0.89    & 14.65     &\cellcolor{green} ~ 0.985$\pm$16.44 \\[1.4mm]
Mg24p-bare       &2.7379E-18 & 2.40                      &  2.02     & 3.19      &\cellcolor{green} ~ 1.063$\pm$~4.43 \\[1.4mm]
Al27a-bare       &1.3531E-18 & 1.80                      &  -1.74    & 2.58      &\cellcolor{green} ~ 0.966$\pm$~4.76 \\[1.4mm]
Nb932-bare       &5.8125E-20 & 1.80                      &  -0.44    & 3.49      &\cellcolor{green} ~ 1.094$\pm$~5.00 \\[1.4mm]
Mn552-bare       &1.0289E-21 & 7.90                      &  2.14     & 2.95      &\cellcolor{green} ~ 1.015$\pm$~6.01 \\[1.4mm]
Zr902-bare       &4.9566E-20 & 3.00                      &  -1.54    & 7.62      &\cellcolor{green} ~ 0.929$\pm$~8.44 \\[1.4mm]
\hline\hline
\end{tabular}
\vspace{-3mm}
\end{table*}

\subsection{Other ``Bucket''-Modifying Environments in Reference Reactor Fields} \label{Sec_VIII_E}

The least-squares analysis for the SPR-III fast burst reactor central cavity and the pool-type ACRR central cavity reactor were good validation cases for the range of activities that are easily measured in research reactors, as shown in the previous section. The consistency of activity measurements for other reactor spectra produced using spectrum modifying buckets has also been considered in our validation. Tables~\ref{ACRR_LB44_LSQ}, \ref{ACRR_PLG_LSQ}, \ref{ACRR_CdPoly_LSQ} and~\ref{ACRR_FREC_LSQ} present the measured activities and the C/E ratios obtained from the least-squares adjustment. The C/E ratios are color-coded to indicate the status of the consistency.

Inspection of these tables shows that, while providing a larger quantity of validation evidence, the consideration of these other Bucket-Modifying Environments did not significantly increase the scope of reactions for which we have validation evidence over what is seen in the SPR-III Central Cavity and ACRR Central Cavity least-squares analyses. Consideration of the consistency of dosimeters in the least-squares analysis for these other research reactor configurations, addressing the following neutron benchmark fields, produced the following results:
\begin{enumerate}
  \item  ACRR LB44 with 29 sensors,    a $\chi^{2}$/\textit{dof }= 1.279, and 2 discrepant dosimeter/cover configurations (Co59g-bare, Mo98g-Cdtk-B4C);
  \item  ACRR PLG with 35 sensors,     a $\chi^{2}$/\textit{dof }= 1.025, and 1 discrepant dosimeter/cover configurations (Mn552-bare);
  \item  ACRR Cd-poly with 31 sensors, a $\chi^{2}$/\textit{dof }= 1.183, and 1 discrepant dosimeter/cover configuration (Np237f-Cdtkj-B4C);
  \item  FREC-II with 33 sensors,      a $\chi^{2}$/\textit{dof }= 1.711, and 0 discrepant dosimeter/cover configurations.
\end{enumerate}

There are also several reaction/cover combinations that are newly addressed in these neutron fields, such as:
\begin{enumerate}
  \item  Fe58g-Cdtk-B4C (configuration not used in ACRR-FF or SPR-III)
  \item  Co59g-Cdtk-B4C (configuration not used in ACRR-FF or SPR-III)
  \item  Co59g-Cd (which was discrepant in ACRR-FF)
  \item  Cu63g-bare (Cd cover was good in SPR-III)
  \item  Ag109g-bare (which was discrepant in ACRR-FF)
  \item  W186g-Cd
\end{enumerate}

\subsection{$^{9}$Be(d,n) Reaction Neutron Fields} \label{Sec_VIII_F}

\begin{table}[htb]
\vspace{-2mm}
%\tiny
 \caption{Be(d,n) Integral Testing of Natural Element Cross Sections in \mbox{IRDFF-II} (725 group calculations), $E_{d}$ = 16~MeV}
 \label{tab:XII_E_1}
%\begin{tabular}{ C{0.3in} R{0.3in} R{0.6in} R{0.2in} R{0.6in} R{0.3in} R{0.3in} R{0.3in} } \hline\hline
\begin{tabular}{ c r r r r r r r} \hline\hline
Elem. & Prod.      & \multicolumn{2}{c}{Measured}& \multicolumn{2}{c}{Calculated} & \multicolumn{2}{c}{$E$-Range} \\
      &            & at/at-s  & $\pm$ & at/at-s  & Diff.                  & \multicolumn{2}{c}{MeV} \\
      &            &          & [\%]  &          & [\%]                   &  5\% & 95\%  \\ \hline
\T
Ti    & $^{46}$Sc  & 9.74E-18 &  3    & 1.04E-17 &\cellcolor{yellow} 6.8  &  5.0 & 13.8  \\
      & $^{47}$Sc  & 5.77E-18 & 10    & 5.86E-18 &\cellcolor{green}  1.6  &  3.1 & 17.7  \\
      & $^{48}$Sc  & 8.33E-18 & 10    & 7.90E-18 &\cellcolor{green} -5.2  &  7.1 & 15.2  \\ \hline
\T
Fe    & $^{54}$Mn  & 1.92E-17 &  3    & 1.94E-17 &\cellcolor{green}  0.8  &  3.5 & 12.4  \\
      & $^{56}$Mn  & 2.51E-17 &  3    & 2.43E-17 &\cellcolor{green} -3.2  &  6.4 & 14.4  \\
      & $^{51}$Cr  & 1.17E-18 & 10    & 1.12E-18 &\cellcolor{green} -3.9  &  6.3 & 14.9  \\ \hline
\T
Ni    & $^{58}$Co  & 2.77E-16 &  3    & 2.81E-16 &\cellcolor{green}  1.5  &  3.3 & 12.2  \\
      & $^{57}$Ni  & 7.90E-19 &  3    & 8.28E-19 &\cellcolor{green}  4.8  & 13.2 & 19.6  \\
      & $^{60}$Co  & 1.40E-17 &  7    & 1.08E-17 &\cellcolor{lred}   -23.  &  6.0 & 14.0  \\ \hline
\T
Zr    & $^{89}$Zr  & 1.36E-17 &  3    & 1.41E-17 &\cellcolor{green}  4.1  & 12.8 & 19.3  \\ \hline \hline
\end{tabular}
\vspace{-3mm}
\end{table}

The \mbox{IRDFF-II} library contains cross sections up to 60~MeV for neutron dosimetry at accelerator-based neutron sources which require measurements for validation in high-energy neutron fields. Above 20 MeV there are many additional reaction channels that must be included to calculate the production of an activation product from elements that contain multiple isotopes.  For example, the production of $^{54}$Mn from Fe must include $^{56}$Fe(n,X)$^{54}$Mn that contains reactions from other Fe isotopes that create more $^{54}$Mn than the traditional $^{54}$Fe(n,p)$^{54}$Mn reaction cross section used below 20~MeV.  The impact of this effect on high energy neutron dosimetry was discussed in Ref.~\cite{INDC(NDS)-0731}. The IAEA Nuclear Data Section addressed this issue in the newly released \mbox{IRDFF-II} library.

\begin{table}[htb]
%\tiny
\vspace{-2mm}
 \caption{Be(d,n) Integral Testing of Natural Element Cross Sections in \mbox{IRDFF-II} (725 group calculations), $E_{d}$ = 40~MeV}
 \label{tab:XII_E_2}
% \begin{tabular}{ C{0.3in} R{0.3in} R{0.6in} R{0.2in} R{0.6in} R{0.3in} R{0.3in} R{0.3in} } \hline\hline
\begin{tabular}{ c r r r r r r r} \hline\hline
Elem. & Prod.      & \multicolumn{2}{c}{Measured} & \multicolumn{2}{c}{Calculated} & \multicolumn{2}{c}{$E$-Range} \\
      &            & at/at-s  & $\pm$ & at/at-s  & Diff.                  & \multicolumn{2}{c}{MeV} \\
      &            &          & [\%]  &          & [\%]                   &  5\% & 95\%  \\ \hline
\T
Al    & $^{24}$Na  & 3.14e-15 &  3    & 3.06E-15 &\cellcolor{green} -2.5  &  8.9 & 22.0  \\ \hline
\T
Ti    & $^{46}$Sc  & 1.42E-15 &  3    & 1.56E-15 &\cellcolor{lred}    9.7  &  8.3 & 37.0  \\
      & $^{47}$Sc  & 3.02E-15 &  3    & 2.94E-15 &\cellcolor{green} -2.8  & 12.9 & 33.0  \\
      & $^{48}$Sc  & 1.35E-15 &  3    & 1.49E-15 &\cellcolor{lred}   10.7  &  9.7 & 28.5  \\ \hline
\T
Fe    & $^{54}$Mn  & 1.68E-15 &  3    & 1.38E-15 &\cellcolor{green}-17.9  &  5.3 & 38.0  \\
      & $^{56}$Mn  & 2.80E-15 &  3    & 2.98E-15 &\cellcolor{green}  6.4  &  8.5 & 26.0  \\
      & $^{51}$Cr  & 2.30E-16 &  3    & 1.83E-16 &\cellcolor{green}-20.4  &  9.1 & 39.0  \\ \hline
\T
Ni    & $^{58}$Co  & 1.12E-14 &  3    & 1.09E-14 &\cellcolor{green} -2.2  &  3.8 & 23.0  \\
      & $^{57}$Ni  & 1.34E-15 & 10    & 1.46E-15 &\cellcolor{green}  8.7  & 14.8 & 29.5  \\
      & $^{60}$Co  & 1.04E-15 & 10    & 1.07E-15 &\cellcolor{green}  2.5  &  8.1 & 26.0  \\ \hline
\T
Zr    & $^{89}$Zr  & 1.70E-14 &  3    & 1.78E-14 &\cellcolor{green}  4.7  & 14.2 & 28.5  \\ \hline\hline
\end{tabular}
\vspace{-3mm}
\end{table}
New evaluated cross section files for natural Al, Ti, Fe, Ni,and Zr activation to various activation products were especially developed for \mbox{IRDFF-II}. While this file could be processed with the PREPRO codes, the format of the new file was not compatible with processing by the current version of NJOY. While interim steps have been taken to support processing this file with the NJOY code, we expect that NJOY compatibility with a legitimate \mbox{ENDF-6} format in \mbox{IRDFF-II} will soon be addressed.

The new natural element cross sections and covariances were processed into multi-group cross section libraries used by the STAYSL-PNNL computer code \cite{Gre18}.  Integral testing was then performed using activation data measured in several $^{9}$Be(d,n) neutron fields that have been characterized by neutron time-of-flight (TOF) measurements at deuteron energies of 16 and 40~MeV~\cite{Gre77, Gre78, Gre79}.  Measurements at $E_d=30$~MeV and $E_d=40$~MeV were presented in INDC(NDS)-0731~\cite{CRP17}, prior to the creation of the new Fe activation file. The plot of the spectra is shown in Fig.~\ref{Fig:XI_E}.

Eleven of the new natural element activation cross sections were tested in this report using previously published TOF neutron spectra to calculate the activation rates for direct comparison with the measured activation rates.  These integral tests are independent of prior work since the new cross sections and covariances were not used in any spectral adjustments of the TOF neutron spectra using STAYSL-PNNL.  The calculations were performed using 725-group cross section and covariance libraries.  The measured activation rates were updated using the recommended gamma branching ratios listed in Table~\ref{Table_Long_XIV}. Integral testing results are presented in Table~\ref{tab:XII_E_1} - Table~\ref{tab:XII_E_2} below. The neutron spectrum for $^{9}$Be(d,n) at $E_{d} = 16$~MeV does not exceed 20~MeV so it is a good check on the lower energy range. The spectra at $E_{d} = 40$~MeV provides a good test of the cross sections at progressively higher energies up to about 50~MeV. Tables~\ref{tab:XII_E_1} and~\ref{tab:XII_E_2} show generally good agreement between the measured activation rates and the calculated activation rates using the \mbox{IRDFF-II} natural element cross sections within the uncertainties of the neutron time-of-flight spectra as well as the activation rate measurements.

The new cross section and cross section libraries were also tested by performing neutron spectral adjustments with the \mbox{STAYSL-PNNL} computer code using the 725-group cross section and covariance libraries.  At 16~MeV, we used 24 reactions for the small spectral adjustment and 25 reactions at 40~MeV. Both the TOF and adjusted spectral activation rate calculations are shown in Fig.~\ref{Fig:XI_E}. In both cases we achieved acceptable $\chi^{2}$ values and overall standard deviations less than 10~\% attesting to the high quality of the new \mbox{IRDFF-II} file for neutron spectral adjustment up to 50~MeV using the new natural element cross sections.
% ------ Chapter XI: End --------

\section{OTHER VALIDATION EVIDENCE} \label{Sec_IX}

A large volume of additional data exists that supports the validation of the \mbox{IRDFF-II} library, as has been reported at other venues~\cite{CRP17}. This additional data includes measurements such as:
\begin{itemize}
%\item JAEA/FNS benchmark testing with graphite assembly and a DT source;
 \item JAEA/FNS benchmark testing with Li2O assembly and a DT source;
 \item ENEA Frascati copper block assembly with 14-MeV neutrons;
 \item CEA reactor data measured at MINERVE, EOLE, MASURCA, and CALIBAN reactor facilities;
 \item ASPIS-IRON88 benchmark described in the SINBAD package NEA-1517/75;
 \item activation experiments at iThemba Laboratory for Accelerator Based Sciences using 90 and 140 MeV quasi-monoenergetic neutron beams;
 \item Kyoto University Research Reactor 80 and 140 MeV p-Li quasi-monoenergetic neutron beams;
%\item Activation experiments at the NPI 20-35 MeV p+7Li quasi-monoenergetic beams.
\end{itemize}

This evidence is not explicitly captured in this paper because the results, typically, exhibited a strong influence from other uncertainty contributors or they did not significantly expand the strength of the validation evidence for the \mbox{IRDFF-II} library.

%\newpage
\section{VALIDATION STATUS}  \label{Sec_X}

Secs.~\ref{Sec_VII}, \ref{Sec_VIII}, and~\ref{Sec_IX} detail the validation evidence that has been assembled to support the \mbox{IRDFF-II} recommended cross sections in integral neutron fields. It is not sufficient for users of the library to just ask if a reaction cross section has been validated, users must also consider if the reaction cross section has been validated within the relevant energy response region addressed by their application. Table~\ref{Table_Long_XXIV} summarizes the validation data in a format where users can more readily assess the validation status for a given reaction for a given class of neutron field. Results are presented for selected integral benchmarks that cover the complete range of average spectrum energies. For a more detailed information the reader is referred to Secs.~\ref{Sec_VII}, \ref{Sec_VIII}, and~\ref{Sec_IX}.

Whereas all of the \mbox{IRDFF-II} metrology reactions have nuclear data evaluations that are based on some experimental data and trends informed by nuclear model calculations, some of the metrology reactions within the library are seen in Table~\ref{Table_Long_XXIV} to lack any “additional” validation evidence in integral benchmark neutron fields. Many of the validation deficiencies are for high threshold energy reactions where measurements are not easily acquired in fission neutron fields.
\newpage
\LTcapwidth=\textwidth
%

\section{LIBRARY DISTRIBUTIONS}  \label{Sec_XI}

The development of the \mbox{IRDFF-II} library was the result of an international collaboration, organized through a series of Research Projects at the International Atomic Energy Agency Nuclear Data Section (NDS). The library can be obtained from \href{https://www-nds.iaea.org/IRDFF/}{www-nds.iaea.org/IRDFF/}. It has been incorporated into the IAEA nuclear data viewers, \href{https://www-nds.iaea.org/exfor/endf.htm}{www-nds.iaea.org/exfor/endf.htm}. Using that online system, comparisons can be readily made between different nuclear data libraries, measured data available in EXFOR~\cite{EXF08}, %\footnote{And renormalized to consider the latest reference cross sections using the online EXFOR correction system.\newline},
and the \mbox{IRDFF-II} recommended cross section with uncertainties.

\section{CONCLUSIONS} \label{Sec_XII}

This paper reports on the assembly of a new neutron metrology library - the \mbox{IRDFF-II} - that represents the recommendations of an international collaboration and supports neutron metrology application in fission and fusion neutron fields. The elements of the \mbox{IRDFF-II} library are described and the basis for the recommended nuclear data values is provided. The \mbox{IRDFF-II} library, which now extends up to incident neutron energies of 60~MeV, contains a comprehensive, complete and consistent set of nuclear data that can be used for the characterization of neutron radiation fields and to support the neutron metrology used with experiments conducted in these neutron fields. In addition to providing the recommended nuclear data, this document summarizes the set of validation evidence gathered from integral neutron benchmark fields that confirms the consistency of the recommended nuclear data.

\section*{Acknowledgements}
Our sincere thanks to all colleagues who have contributed to and worked on this project during the last thirteen years since the release of the IRDF-2002 library. The preparation of this paper would not have been possible without the support, hard work and endless efforts of a large number of individuals and institutions.

The authors would also like to acknowledge the significant contributions by V. Chechev, V. G. Khlopin Radium Institute, St. Petersburg, Russia, in the area of the nuclear decay data that have been incorporated into the \mbox{IRDFF-II} library. V. Chechev, who collaborated closely with the IAEA Nuclear Data Section and the Decay Data Evaluation Project (DDEP), passed away on March 5, 2018.

%MS institutes
The IAEA is grateful to all participating laboratories for their assistance in the work and for support of the research meetings and activities.
The work described in this paper would not have been possible without IAEA Member State contributions.
%Work at ANL was supported by the U.S. Department of Energy, Office of Science, Office of Nuclear Physics, under contract no. DE-AC-06CH11357.
%Sandia National Laboratories is a multi-mission laboratory managed and operated by National Technology and Engineering Solutions of Sandia, LLC, a wholly owned subsidiary of Honeywell International, %Inc., for the U.S. Department of Energy's National Nuclear Security Administration under contract DE-NA0003525.
Work at SNL was supported by the U.S. Department of Energy's National Nuclear Security Administration under contract DE-NA0003525.

% other contributors
We acknowledge many fruitful exchanges of information with our colleagues A.J.~Koning, D. Rochman and J.-C.~Sublet on details of TENDL libraries. Without their efforts in producing TENDL we could not achieve a successful extension of the IRDFF up to 60 MeV.

The IAEA staff acknowledges the very valuable contribution to the preparation of this paper made by our colleague V. Zerkin by customizing and improving his online plotting package, and uploading the preliminary $\beta$-version of the \mbox{IRDFF-II} to the servers allowing for a comprehensive visual checking of the evaluated cross sections.

The IAEA appreciate the valuable contributions made by A.J.~Plompen, P.F.~Mastinu, I.~Kodeli, R. Nchodu, M. Stefanik, F. Wissmann, P. Maleka, N.B. Ndlovu and N. Kornilov during the various project meetings. %% Special Ack ...

Last but not least we are deeply indebted to W. Mannhart, who passed away on November 11, 2018. Mannhart contributions to many aspects of the field of neutron metrology over a period of several decades were very significant and are acknowledged as a source of inspiration.

%We would like to thank close collaborators ... who contributed to the overall success of the project.

% The bibliography is sorted in a separate file
%\input{NDS-IRDFF-II_Bib.tex}
% reinserted

\end{document}